\documentclass[]{book}
\usepackage{lmodern}
\usepackage{amssymb,amsmath}
\usepackage{ifxetex,ifluatex}
\usepackage{fixltx2e} 
\ifnum 0\ifxetex 1\fi\ifluatex 1\fi=0 
  \usepackage[T1]{fontenc}
  \usepackage[utf8]{inputenc}
\else 
  \ifxetex
    \usepackage{mathspec}
  \else
    \usepackage{fontspec}
  \fi
  \defaultfontfeatures{Ligatures=TeX,Scale=MatchLowercase}
\fi
\IfFileExists{upquote.sty}{\usepackage{upquote}}{}
\IfFileExists{microtype.sty}{%
\usepackage{microtype}
\UseMicrotypeSet[protrusion]{basicmath} 
}{}
\usepackage[margin=1in]{geometry}
\usepackage{hyperref}
\hypersetup{unicode=true,
            pdftitle={Loss Data Analytics},
            pdfauthor={An open text authored by the Actuarial Community},
            pdfborder={0 0 0},
            breaklinks=true}
\usepackage{natbib}
\usepackage{color}
\usepackage{fancyvrb}

\DefineVerbatimEnvironment{Highlighting}{Verbatim}{commandchars=\\\{\}}
\usepackage{framed}
\definecolor{shadecolor}{RGB}{248,248,248}
\newenvironment{Shaded}{\begin{snugshade}}{\end{snugshade}}
\newcommand{\KeywordTok}[1]{\textcolor[rgb]{0.13,0.29,0.53}{\textbf{#1}}}
\newcommand{\DataTypeTok}[1]{\textcolor[rgb]{0.13,0.29,0.53}{#1}}
\newcommand{\DecValTok}[1]{\textcolor[rgb]{0.00,0.00,0.81}{#1}}

\newcommand{\FloatTok}[1]{\textcolor[rgb]{0.00,0.00,0.81}{#1}}

\newcommand{\CharTok}[1]{\textcolor[rgb]{0.31,0.60,0.02}{#1}}

\newcommand{\StringTok}[1]{\textcolor[rgb]{0.31,0.60,0.02}{#1}}

\newcommand{\SpecialStringTok}[1]{\textcolor[rgb]{0.31,0.60,0.02}{#1}}

\newcommand{\CommentTok}[1]{\textcolor[rgb]{0.56,0.35,0.01}{\textit{#1}}}

\newcommand{\OtherTok}[1]{\textcolor[rgb]{0.56,0.35,0.01}{#1}}

\newcommand{\ControlFlowTok}[1]{\textcolor[rgb]{0.13,0.29,0.53}{\textbf{#1}}}
\newcommand{\OperatorTok}[1]{\textcolor[rgb]{0.81,0.36,0.00}{\textbf{#1}}}

\newcommand{\NormalTok}[1]{#1}
\usepackage{longtable,booktabs}
\usepackage{graphicx,grffile}
\makeatletter
\def\maxwidth{\ifdim\Gin@nat@width>\linewidth\linewidth\else\Gin@nat@width\fi}
\def\maxheight{\ifdim\Gin@nat@height>\textheight\textheight\else\Gin@nat@height\fi}
\makeatother
\setkeys{Gin}{width=\maxwidth,height=\maxheight,keepaspectratio}
\IfFileExists{parskip.sty}{%
\usepackage{parskip}
}{
\setlength{\parindent}{0pt}
\setlength{\parskip}{6pt plus 2pt minus 1pt}
}
\providecommand{\tightlist}{%
  \setlength{\itemsep}{0pt}\setlength{\parskip}{0pt}}
\ifx\paragraph\undefined\else
\let\oldparagraph\paragraph
\renewcommand{\paragraph}[1]{\oldparagraph{#1}\mbox{}}
\fi
\ifx\subparagraph\undefined\else
\let\oldsubparagraph\subparagraph
\renewcommand{\subparagraph}[1]{\oldsubparagraph{#1}\mbox{}}
\fi

\let\rmarkdownfootnote\footnote%
\def\footnote{\protect\rmarkdownfootnote}

\usepackage{titling}

  \title{Loss Data Analytics}
  \pretitle{\vspace{\droptitle}\centering\huge}
  \posttitle{\par}
  \author{An open text authored by the Actuarial Community}
  \preauthor{\centering\large\emph}
  \postauthor{\par}
  \predate{\centering\large\emph}
  \postdate{\par}
  \date{2018-07-28}

\usepackage{booktabs}
\usepackage{amsthm}

\theoremstyle{definition}

\theoremstyle{definition}

\theoremstyle{definition}

\theoremstyle{remark}

\let\BeginKnitrBlock\begin \let\EndKnitrBlock\end
\begin{document}
\maketitle

{
\setcounter{tocdepth}{2}
\tableofcontents
}
\chapter*{Preface}\label{preface}
\addcontentsline{toc}{chapter}{Preface}

\subsubsection*{Book Description}\label{book-description}
\addcontentsline{toc}{subsubsection}{Book Description}

\textbf{Loss Data Analytics} is an interactive, online, freely available
text.

\begin{itemize}
\item
  The online version contains many interactive objects (quizzes,
  computer demonstrations, interactive graphs, video, and the like) to
  promote \emph{deeper learning}.
\item
  A subset of the book is available for \emph{offline reading} in pdf
  and EPUB formats.
\item
  The online text will be available in multiple languages to promote
  access to a \emph{worldwide audience}.
\end{itemize}

\subsubsection*{What will success look
like?}\label{what-will-success-look-like}
\addcontentsline{toc}{subsubsection}{What will success look like?}

The online text will be freely available to a worldwide audience. The
online version will contain many interactive objects (quizzes, computer
demonstrations, interactive graphs, video, and the like) to promote
deeper learning. Moreover, a subset of the book will be available in pdf
format for low-cost printing. The online text will be available in
multiple languages to promote access to a worldwide audience.

\subsubsection*{How will the text be
used?}\label{how-will-the-text-be-used}
\addcontentsline{toc}{subsubsection}{How will the text be used?}

This book will be useful in actuarial curricula worldwide. It will cover
the loss data learning objectives of the major actuarial organizations.
Thus, it will be suitable for classroom use at universities as well as
for use by independent learners seeking to pass professional actuarial
examinations. Moreover, the text will also be useful for the continuing
professional development of actuaries and other professionals in
insurance and related financial risk management industries.

\subsubsection*{Why is this good for the
profession?}\label{why-is-this-good-for-the-profession}
\addcontentsline{toc}{subsubsection}{Why is this good for the
profession?}

An online text is a type of open educational resource (OER). One
important benefit of an OER is that it equalizes access to knowledge,
thus permitting a broader community to learn about the actuarial
profession. Moreover, it has the capacity to engage viewers through
active learning that deepens the learning process, producing analysts
more capable of solid actuarial work. Why is this good for students and
teachers and others involved in the learning process?

Cost is often cited as an important factor for students and teachers in
textbook selection (see a recent post on the
\href{https://www.aei.org/publication/the-new-era-of-the-400-college-textbook-which-is-part-of-the-unsustainable-higher-education-bubble/}{\$400
textbook}). Students will also appreciate the ability to ``carry the
book around'' on their mobile devices.

\subsubsection*{Why loss data analytics?}\label{why-loss-data-analytics}
\addcontentsline{toc}{subsubsection}{Why loss data analytics?}

Although the intent is that this type of resource will eventually
permeate throughout the actuarial curriculum, one has to start
somewhere. Given the dramatic changes in the way that actuaries treat
data, loss data seems like a natural place to start. The idea behind the
name \emph{loss data analytics} is to integrate classical loss data
models from applied probability with modern analytic tools. In
particular, we seek to recognize that big data (including social media
and usage based insurance) are here and high speed computation s readily
available.

\subsubsection*{Project Goal}\label{project-goal}
\addcontentsline{toc}{subsubsection}{Project Goal}

The project goal is to have the actuarial community author our textbooks
in a collaborative fashion.

To get involved, please visit our
\href{https://sites.google.com/a/wisc.edu/loss-data-analytics/}{Loss
Data Analytics Project Site}.

\chapter*{Contributor List}\label{contributor-list}
\addcontentsline{toc}{chapter}{Contributor List}

\begin{itemize}
\item
  \textbf{Zeinab Amin} American University in Cairo
\item
  \textbf{Katrien Antonio}, KU Leuven
\item
  \textbf{Jan Beirlant}, KU Leuven
\item
  \textbf{Carolina Castro} - University of Buenos Aires
\item
  \textbf{Gary Dean}, Ball State University
\item
  \textbf{Edward W. (Jed) Frees} is an emeritus professor, formerly the
  Hickman-Larson Chair of Actuarial Science at the University of
  Wisconsin-Madison. He is a Fellow of both the Society of Actuaries and
  the American Statistical Association. He has published extensively (a
  four-time winner of the Halmstad and Prize for best paper published in
  the actuarial literature) and has written three books. He also is a
  co-editor of the two-volume series \emph{Predictive Modeling
  Applications in Actuarial Science} published by Cambridge University
  Press.
\item
  \textbf{Guojun Gan} - University of Connecticut
\item
  \textbf{Lisa Gao} is a doctoral student at the University of
  Wisconsin-Madison.
\item
  \textbf{José Garrido}, Concordia University
\item
  \textbf{Noriszura Ismail}, University Kebangsaan Malaysia
\item
  \textbf{Joseph Kim}, Yonsei University
\item
  \textbf{Shyamalkumar Nariankadu} - University of Iowa
\item
  \textbf{Nii-Armah Okine} is a doctoral student at the University of
  Wisconsin-Madison.
\item
  \textbf{Margie Rosenberg} - University of Wisconsin
\item
  \textbf{Emine Selin Sarıdaş}, Mimar Sinan University
\item
  \textbf{Peng Shi} - University of Wisconsin
\item
  \textbf{Jianxi Su}, Purdue University
\item
  \textbf{Tim Verdonck}, KU Leuven
\item
  \textbf{Krupa Viswanathan} - Temple University
\end{itemize}

\chapter*{Acknowledgements}\label{acknowledgements}
\addcontentsline{toc}{chapter}{Acknowledgements}

Edward Frees acknowledges the John and Anne Oros Distinguished Chair for
Inspired Learning in Business which provided seed money to support the
project. Frees and his Wisconsin colleagues also acknowledge a Society
of Actuaries Center of Excellence Grant that provided funding to support
work in dependence modeling and health initiatives.

We acknowledge the Society of Actuaries for permission to use problems
from their examinations.

We also wish to acknowledge the support and sponsorship of the
\href{http://www.blackactuaries.org/}{International Association of Black
Actuaries} in our joint efforts to provide actuarial educational content
to all.

\begin{figure}
\centering
\includegraphics[width=0.25000\textwidth]{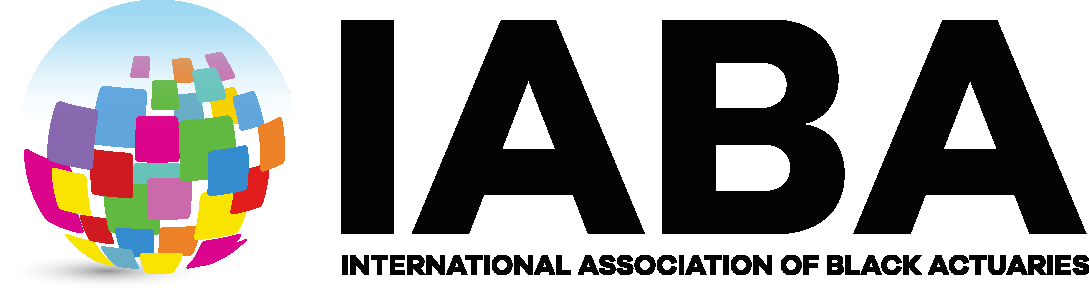}
\caption{}
\end{figure}

\section*{Reviewer Acknowledgment}\label{reviewer-acknowledgment}
\addcontentsline{toc}{section}{Reviewer Acknowledgment}

\begin{itemize}
\tightlist
\item
  Hirokazu (Iwahiro) Iwasawa
\end{itemize}

\chapter{Introduction to Loss Data Analytics}\label{C:Intro}

\emph{Chapter Preview}. This book introduces readers to methods of
analyzing insurance data. Section \ref{S:Intro} begins with a discussion
of why the use of data is important in the insurance industry. Although
obvious, the importance of data is critical - it is the whole premise of
the book. Next, Section \ref{S:PredModApps} gives a general overview of
the purposes of analyzing insurance data which is reinforced in the
Section \ref{S:LGPIF} case study. Naturally, there is a huge gap between
these broads goals and a case study application; this gap is covered
through the methods and techniques of data analysis covered in the rest
of the text.

\section{Relevance of Analytics}\label{S:Intro}

In this section, you learn how to:

\begin{itemize}
\tightlist
\item
  Motivate the relevance of insurance
\item
  Describe analytics
\item
  Describe data generating events associated with the timeline of a
  typical insurance contract
\end{itemize}

This book introduces the process of using data to make decisions in an
insurance context. It does not assume that readers are familiar with
insurance but introduces insurance concepts as needed. Insurance may not
be as entertaining as the sports industry nor as widely familiar as the
agricultural industry but it does affect the financial livelihoods of
many. By almost any measure, insurance is a major economy activity. On a
global level, insurance premiums comprised about 6.3\% of the world
gross domestic product (GDP) in 2013, \citep{III2015}. To illustrate,
premiums accounted for 17.6\% of GDP in Taiwan (the highest in the
study) and represented 7.5\% of GDP in the United States. On a personal
level, almost everyone owning a home has insurance to protect themselves
in the event of a fire, hailstorm, or some other calamitous event.
Almost every country requires insurance for those driving a car. So,
although not particulary entertaining nor widely familiar, insurance is
an important piece of the economy and relevant to individual
livelihoods.

Insurance is a data-driven industry. Like other major corporations,
insurers use data when trying to decide how much to pay employees, how
many employees to retain, how to market their services, how to forecast
financial trends, and so on. Although each industry retains its own
nuances, these represent general areas of activities that are not
specific to the insurance industry. You will find that the data methods
and tools introduced in this text relevant for these general areas.

Moreover, when introducing data methods, we will focus on losses that
potentially arise from obligations in insurance contracts. This could be
the amount of damage to one's apartment under a renter's insurance
agreement, the amount needed to compensate someone that you hurt in a
driving accident, and the like. We will call these \emph{insurance
claims} or \emph{loss amounts}. With this focus, we will be able to
introduce generally applicable statistical tools in techniques in
real-life situations where the tools can be used directly.

\subsection{What is Analytics?}\label{what-is-analytics}

Insurance is a data-driven industry and analytics is a key to deriving
information from data. But what is analytics? Making data-driven
business decisions has been described as business analytics, business
intelligence, and data science. These terms, among others, are sometimes
used interchangeably and sometimes used separately, referring to
distinct domains of applications. As an example of such distinctions,
\emph{business intelligence} may focus on processes of collecting data,
often through databases and data warehouses, whereas \emph{business
analytics} utilizes tools and methods for statistical analyses of data.
In contrast to these two terms that emphasize business applications, the
term \emph{data science} can encompass broader applications in many
scientific domains. For our purposes, we use the term \emph{analytics}
to refer to the process of using data to make decisions. This process
involves gathering data, understanding models of uncertainty, making
general inferences, and communicating results.

\subsection{Short-term Insurance}\label{short-term-insurance}

This text will focus on short-term insurance contracts. By short-term,
we mean contracts where the insurance coverage is typically provided for
six months or a year. If you are new to insurance, then it is probably
easiest to think about an insurance policy that covers the contents of
an apartment or house that you are renting (known as renters insurance)
or the contents and property of a building that is owned by you or a
friend (known as homeowners insurance). Another easy example is
automobile insurance. In the event of an accident, this policy may cover
damage to your vehicle, damage to other vehicles in the accident, as
well as medical expenses of those injured in the accident.

In the US, policies such as renters and homeowners are known as property
insurance whereas a policy such as auto that covers medical damages to
people is known as casualty insurance. In the rest of the world, these
are both known as nonlife or general insurance, to distinguish them from
life insurance.

Both life and nonlife insurances are important. To illustrate,
\citep{III2015} estimates that direct insurance premiums in the world
for 2013 was 2,608,091 for life and 2,032,850 for nonlife; these figures
are in millions of US dollars. As noted earlier, the total represents
6.3\% of the world GDP. Put another way, life accounts for 56.2\% of
insurance premiums and 3.5\% of world GDP, nonlife accounts for 43.8\%
of insurance premiums and 2.7\% of world GDP. Both life and nonlife
represent important economic activities and are worthy of study in their
own right.

Yet, life insurance considerations differ from nonlife. In life
insurance, the default is to have a multi-year contract. For example, if
a person 25 years old purchases a whole life policy that pays upon death
of the insured and that person does not die until age 100, then the
contract is in force for 75 years. We think of this as a long-term
contract.

Further, in life insurance, the benefit amount is often stipulated in
the contract provisions. In contrast, most short-term contracts provide
for reimbursement of insured losses which are unknown before the
accident. (Of course, there are usually limits placed on the
reimbursement amounts.) In a multi-year life insurance contract, the
time value of money plays a prominent role. In contrast, in a short-term
nonlife contract, the random amount of reimbursement takes priority.

In both life and nonlife insurances, the frequency of claims is very
important. For many life insurance contracts, the insured event (such as
death) happens only once. In contrast, for nonlife insurances such as
automobile, it is common for individuals (especially young male drivers)
to get into more than one accident during a year. So, our models need to
reflect this observation; we will introduce different frequency models
than you may have seen when studying life insurance.

For short-term insurance, the framework of the probabilistic model is
straightforward. We think of a one-period model (the period length,
e.g., six months, will be specified in the situation).

\begin{itemize}
\item
  At the beginning of the period, the insured pays the insurer a known
  premium that is agreed upon by both parties to the contract.
\item
  At the end of the period, the insurer reimburses the insured for a
  (possibly multivariate) random loss that we will denote as \(y\).
\end{itemize}

This framework will be developed as we proceed but we first focus on
integrating this framework with concerns about how the data may arise
and what we can accomplish with this framework.

\subsection{Insurance Processes}\label{S:InsProcesses}

\begin{figure}

{\centering \includegraphics[width=1\linewidth]{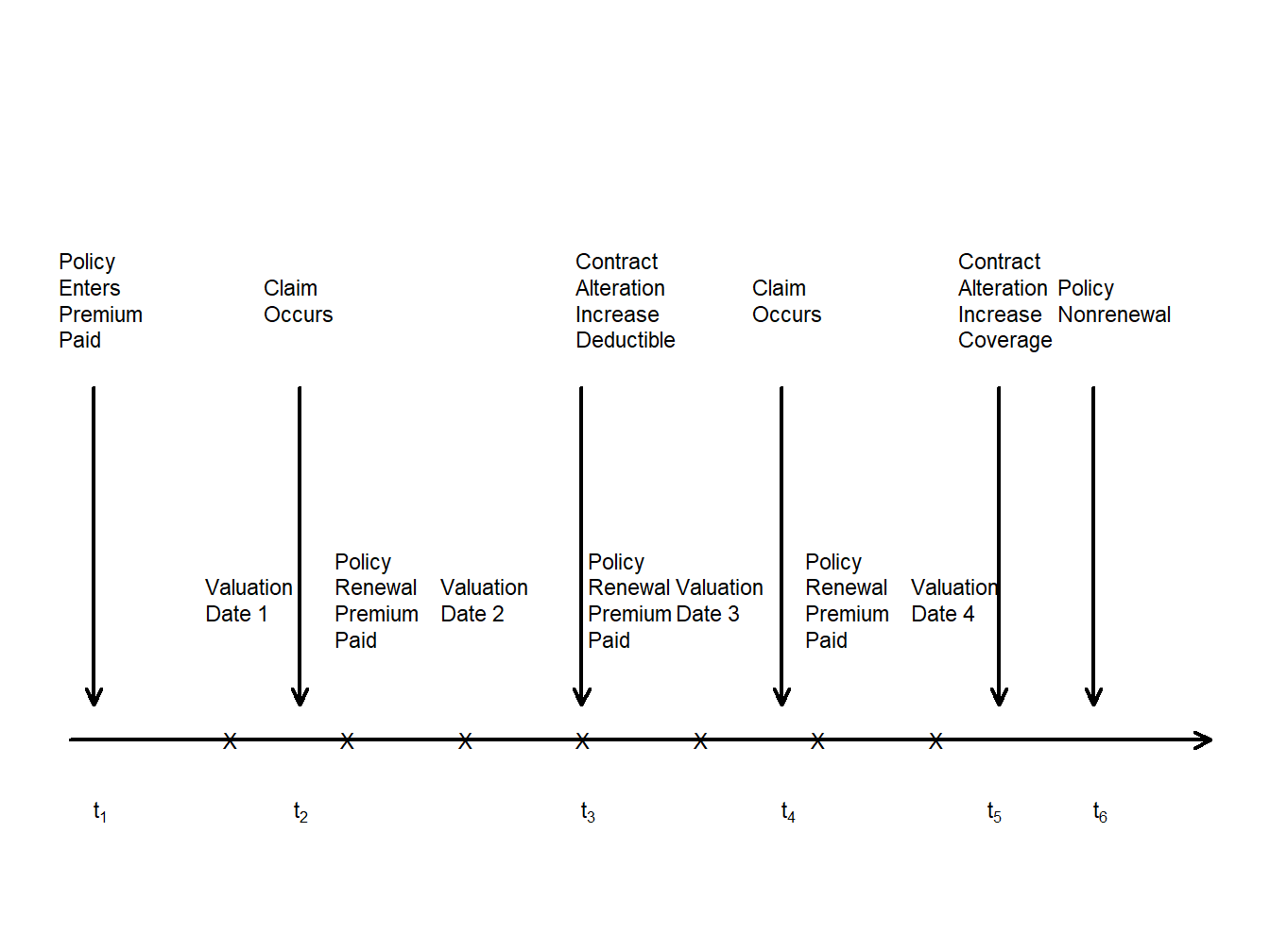}

}

\caption{Timeline of a Typical Insurance Policy. Arrows
mark the occurrences of random events. Each x marks the time of
scheduled events that are typically non-random.}\label{fig:StochOperations}
\end{figure}

One way to describe the data arising from operations of a company that
sells insurance products is to adopt a granular approach. In this micro
oriented view, we can think specifically about what happens to a
contract at various stages of its existence. Consider Figure
\ref{fig:StochOperations} that traces a timeline of a typical insurance
contract. Throughout the existence of the contract, the company
regularly processes events such as premium collection and valuation,
described in Section \ref{S:PredModApps}; these are marked with an
\textbf{x} on the timeline. Further, non-regular and unanticipated
events also occur. To illustrate, times \(\mathrm{t}_2\) and
\(\mathrm{t}_4\) mark the event of an insurance claim (some contracts,
such as life insurance, can have only a single claim). Times
\(\mathrm{t}_3\) and \(\mathrm{t}_5\) mark the events when a
policyholder wishes to alter certain contract features, such as the
choice of a deductible or the amount of coverage. Moreover, from a
company perspective, one can even think about the contract initiation
(arrival, time \(\mathrm{t}_1\)) and contract termination (departure,
time \(\mathrm{t}_6\)) as uncertain events.

\section{Insurance Company Operations}\label{S:PredModApps}

In this section, you learn how to:

\begin{itemize}
\tightlist
\item
  Describe five major operational areas of insurance companies.
\item
  Identify the role of data and analytics opportunities within each
  operational area.
\end{itemize}

Armed with insurance data and a method of organizing the data into
variable types, the end goal is to use data to make decisions. Of
course, we will need to learn more about methods of analyzing and
extrapolating data but that is the purpose of the remaining chapters in
the text. To begin, let us think about why we wish to do the analysis.
To provide motivation, we take the insurer's viewpoint (not a person)
and introduce ways of bringing money in, paying it out, managing costs,
and making sure that we have enough money to meet obligations.

Specifically, in many insurance companies, it is customary to aggregate
detailed insurance processes into larger operational units; many
companies use these functional areas to segregate employee activities
and areas of responsibilities. Actuaries and other financial analysts
work within these units and use data for the following activities:

\begin{enumerate}
\def\labelenumi{\arabic{enumi}.}
\item
  \textbf{Initiating Insurance}. At this stage, the company makes a
  decision as to whether or not to take on a risk (the
  \emph{underwriting} stage) and assign an appropriate premium (or
  rate). Insurance analytics has its actuarial roots in
  \emph{ratemaking}, where analysts seek to determine the right price
  for the right risk.
\item
  \textbf{Renewing Insurance}. Many contracts, particularly in general
  insurance, have relatively short durations such as 6 months or a year.
  Although there is an implicit expectation that such contracts will be
  renewed, the insurer has the opportunity to decline coverage and to
  adjust the premium. Analytics is also used at this policy renewal
  stage where the goal is to retain profitable customers.
\item
  \textbf{Claims Management}. Analytics has long been used in (1)
  detecting and preventing claims fraud, (2) managing claim costs,
  including identifying the appropriate support for claims handling
  expenses, as well as (3) understanding excess layers for reinsurance
  and retention.
\item
  \textbf{Loss Reserving}. Analytic tools are used to provide management
  with an appropriate estimate of future obligations and to quantify the
  uncertainty of the estimates.
\item
  \textbf{Solvency and Capital Allocation}. Deciding on the requisite
  amount of capital and ways of allocating capital to alternative
  investment activities represent other important analytics activities.
  Companies must understand how much capital is needed so that they will
  have sufficient flow of cash available to meet their obligations. This
  is an important question that concerns not only company managers but
  also customers, company shareholders, regulatory authorities, as well
  as the public at large. Related to issues of how much capital is the
  question of how to allocate capital to differing financial projects,
  typically to maximize an investor's return. Although this question can
  arise at several levels, insurance companies are typically concerned
  with how to allocate capital to different lines of business within a
  firm and to different subsidiaries of a parent firm.
\end{enumerate}

Although data is a critical component of solvency and capital
allocation, other components including an economic framework and
financial investments environment are also important. Because of the
background needed to address these components, we will not address
solvency and capital allocation issues further in this text.

Nonetheless, for all operating functions, we emphasize that analytics in
the insurance industry is not an exercise that a small group of analysts
can do by themselves. It requires an insurer to make significant
investments in their information technology, marketing, underwriting,
and actuarial functions. As these areas represent the primary end goals
of the analysis of data, additional background on each operational unit
is provided in the following subsections.

\subsection{Initiating Insurance}\label{initiating-insurance}

Setting the price of an insurance good can be a perplexing problem. In
manufacturing, the cost of a good is (relatively) known and provides a
benchmark for assessing a market demand price. In other areas of
financial services, market prices are available and provide the basis
for a market-consistent pricing structure of products. In contrast, for
many lines of insurance, the cost of a good is uncertain and market
prices are unavailable. Expectations of the random cost is a reasonable
place to start for a price, as this is the optimal price for a
risk-neutral insurer. Thus, it has been traditional in insurance pricing
to begin with the expected cost and to add to this so-called margins to
account for the product's riskiness, expenses incurred in servicing the
product, and a profit/surplus allowance for the insurance company.

For some lines of business, especially automobile and homeowners
insurance, analytics has served to sharpen the market by making the
calculation of the good's expectation more precise. The increasing
availability of the internet among consumers has promoted transparency
in pricing. Insurers seek to increase their market share by refining
their risk classification systems and employing skimming the cream
underwriting strategies. Recent surveys (e.g., \citep{survey2013})
indicate that pricing is the most common use of analytics among
insurers.

\emph{Underwriting}, the process of classifying risks into homogenous
categories and assigning policyholders to these categories, lies at the
core of ratemaking. Policyholders within a class have similar risk
profiles and so are charged the same insurance price. This is the
concept of an actuarially fair premium; it is fair to charge different
rates to policyholders only if they can be separated by identifiable
risk factors. To illustrate, an early contribution, Two Studies in
Automobile Insurance Ratemaking, by \citep{bailey1960} provided a
catalyst to the acceptance of analytic methods in the insurance
industry. This paper addresses the problem of classification ratemaking.
It describes an example of automobile insurance that has five use
classes cross-classified with four merit rating classes. At that time,
the contribution to premiums for use and merit rating classes were
determined independently of each other. Thinking about the interacting
effects of different classification variables is a more difficult
problem.

\subsection{Renewing Insurance}\label{renewing-insurance}

Insurance is a type of financial service and, like many service
contracts, insurance coverage is often agreed upon for a limited time
period, such as six months or a year, at which time commitments are
complete. Particularly for general insurance, the need for coverage
continues and so efforts are made to issue a new contract providing
similar coverage. Renewal issues can also arise in life insurance, e.g.,
term (temporary) life insurance, although other contracts, such as life
annuities, terminate upon the insured's death and so issues of
renewability are irrelevant.

In absence of legal restrictions, at renewal the insurer has the
opportunity to:

\begin{itemize}
\item
  accept or decline to underwrite the risk and
\item
  determine a new premium, possibly in conjunction with a new
  classification of the risk.
\end{itemize}

Risk classification and rating at renewal is based on two types of
information. First, as at the initial stage, the insurer has available
many rating variables upon which decisions can be made. Many variables
will not change, e.g., sex, whereas others are likely to have changed,
e.g., age, and still others may or may not change, e.g., credit score.
Second, unlike the initial stage, at renewal the insurer has available a
history of policyholder's loss experience, and this history can provide
insights into the policyholder that are not available from rating
variables. Modifying premiums with claims history is known as
\emph{experience rating}, also sometimes referred to as \emph{merit
rating}.

Experience rating methods are either applied retrospectively or
prospectively. With retrospective methods, a refund of a portion of the
premium is provided to the policyholder in the event of favorable (to
the insurer) experience. Retrospective premiums are common in life
insurance arrangements (where policyholders earned dividends in the U.S.
and bonuses in the U.K.). In general insurance, prospective methods are
more common, where favorable insured experience is rewarded through a
lower renewal premium.

Claims history can provide information about a policyholder's risk
appetite. For example, in personal lines it is common to use a variable
to indicate whether or not a claim has occurred in the last three years.
As another example, in a commercial line such as worker's compensation,
one may look to a policyholder's average claim over the last three
years. Claims history can reveal information that is hidden (to the
insurer) about the policyholder.

\subsection{Claims and Product
Management}\label{claims-and-product-management}

In some of areas of insurance, the process of paying claims for insured
events is relatively straightforward. For example, in life insurance, a
simple death certificate is all that is needed as the benefit amount is
provided in the contract terms. However, in non-life areas such as
property and casualty insurance, the process is much more complex. Think
about even a relatively simple insured event such as automobile
accident. Here, it is often helpful to determine which party is at
fault, one needs to assess damage to all of the vehicles and people
involved in the incident, both insured and non-insured, the expenses
incurred in assessing the damages, and so forth. The process of
determining coverage, legal liability, and settling claims is known as
\emph{claims adjustment}.

Insurance managers sometimes use the phrase \emph{claims leakage} to
mean dollars lost through claims management inefficiencies. There are
many ways in which analytics can help manage the claims process,
\citep{SASsurvey}. Historically, the most important has been fraud
detection. The claim adjusting process involves reducing information
asymmetry (the claimant knows exactly what happened; the company knows
some of what happened). Mitigating fraud is an important part of claims
management process.

One can think about the management of claims severity as consisting of
the following components:

\begin{itemize}
\item
  \textbf{Claims triaging}. Just as in the medical world, early
  identification and appropriate handling of high cost claims (patients,
  in the medical world), can lead to dramatic company savings. For
  example, in workers compensation, insurers look to achieve early
  identification of those claims that run the risk of high medical costs
  and a long payout period. Early intervention into those cases could
  give insurers more control over the handling of the claim, the medical
  treatment, and the overall costs with an earlier return-to-work.
\item
  \textbf{Claims processing}. The goal is to use analytics to identify
  situations suitable for small claims handling processes and those for
  adjuster assignment to complex claims.
\item
  \textbf{Adjustment decisions}. Once a complex claim has been
  identified and assigned to an adjuster, analytic driven routines can
  be established to aid subsequent decision-making processes. Such
  processes can also be helpful for adjusters in developing case
  reserves, an important input to the insurer's loss reserves, Section
  \ref{S:Reserving}.
\end{itemize}

In addition to the insured's reimbursement for insured losses, the
insurer also needs to be concerned with another source of revenue
outflow, expenses. Loss adjustment expenses are part of an insurer's
cost of managing claims. Analytics can be used to reduce expenses
directly related to claims handling (allocated) as well as general staff
time for overseeing the claims processes (unallocated). The insurance
industry has high operating costs relative to other portions of the
financial services sectors.

In addition to claims payments, there are many other ways in which
insurers use to data to manage their products. We have already discussed
the need for analytics in underwriting, that is, risk classification at
the initial acquisition stage. Insurers are also interested in which
policyholders elect to renew their contract and, as with other products,
monitor customer loyalty.

Analytics can also be used to manage the portfolio, or collection, of
risks that an insurer has acquired. When the risk is initially obtained,
the insurer's risk can be managed by imposing contract parameters that
modify contract payouts. In Chapter xx introduces common modifications
including coinsurance, deductibles, and policy upper limits.

After the contract has been agreed upon with an insured, the insurer may
still modify its net obligation by entering into a reinsurance
agreement. This type of agreement is with a reinsurer, an insurer of an
insurer. It is common for insurance companies to purchase insurance on
its portfolio of risks to gain protection from unusual events, just as
people and other companies do.

\subsection{Loss Reserving}\label{S:Reserving}

An important feature that distinguishes insurance from other sectors of
the economy is the timing of the exchange of considerations. In
manufacturing, payments for goods are typically made at the time of a
transaction. In contrast, for insurance, money received from a customer
occurs in advance of benefits or services; these are rendered at a later
date. This leads to the need to hold a reservoir of wealth to meet
future obligations in respect to obligations made. The size of this
reservoir of wealth, and the importance of ensuring its adequacy in
regard to liabilities already assumed, is a major concern for the
insurance industry.

Setting aside money for unpaid claims is known as \emph{loss reserving};
in some jurisdictions, reserves are also known as \emph{technical
provisions}. We saw in Figure \ref{fig:StochOperations} how future
obligations arise naturally at a specific (valuation) date; a company
must estimate these outstanding liabilities when determining its
financial strength. Accurately determining loss reserves is important to
insurers for many reasons.

\begin{enumerate}
\def\labelenumi{\arabic{enumi}.}
\item
  Loss reserves represent a loan that the insurer owes its customers.
  Under-reserving may result in a failure to meet claim liabilities.
  Conversely, an insurer with excessive reserves may present a weaker
  financial position than it truly has and lose market share.
\item
  Reserves provide an estimate for the unpaid cost of insurance that can
  be used for pricing contracts.
\item
  Loss reserving is required by laws and regulations. The public has a
  strong interest in the financial strength of insurers.
\item
  In addition to the insurance company management and regulators, other
  stakeholders such as investors and customers make decisions that
  depend on company loss reserves.
\end{enumerate}

Loss reserving is a topic where there are substantive differences
between life and general (also known as property and casualty, or
non-life), insurance. In life insurance, the severity (amount of loss)
is often not a source of concern as payouts are specified in the
contract. The frequency, driven by mortality of the insured, is a
concern. However, because of the length of time for settlement of life
insurance contracts, the time value of money uncertainty as measured
from issue to date of death can dominate frequency concerns. For
example, for an insured who purchases a life contract at age 20, it
would not be unusual for the contract to still be open in 60 years time.
See, for example, \citep{bowers1986actuarial} or
\citep{dickson2013actuarial} for introductions to reserving for life
insurance.

\section{Case Study: Wisconsin Property Fund}\label{S:LGPIF}

In this section, for a real case study such as the Wisconsin Property
Fund, you learn how to:

\begin{itemize}
\tightlist
\item
  Describe how data generating events can produce data of interest to
  insurance analysts.
\item
  Identify the type of each variable.
\item
  Produce relevant summary statistics for each variable.
\item
  Describe how these summary statistcs can be used in each of the major
  operational areas of an insurance company.
\end{itemize}

Let us illustrate the kind of data under consideration and the goals
that we wish to achieve by examining the Local Government Property
Insurance Fund (LGPIF), an insurance pool administered by the Wisconsin
Office of the Insurance Commissioner. The LGPIF was established to
provide property insurance for local government entities that include
counties, cities, towns, villages, school districts, and library boards.
The fund insures local government property such as government buildings,
schools, libraries, and motor vehicles. The fund covers all property
losses except those resulting from flood, earthquake, wear and tear,
extremes in temperature, mold, war, nuclear reactions, and embezzlement
or theft by an employee.

The property fund covers over a thousand local government entities who
pay approximately \$25 million in premiums each year and receive
insurance coverage of about \$75 billion. State government buildings are
not covered; the LGPIF is for local government entities that have
separate budgetary responsibilities and who need insurance to moderate
the budget effects of uncertain insurable events. Coverage for local
government property has been made available by the State of Wisconsin
since 1911.

\subsection{Fund Claims Variables}\label{S:OutComes}

At a fundamental level, insurance companies accept premiums in exchange
for promises to indemnify a policyholder upon the uncertain occurrence
of an insured event. This indemnification is known as a \emph{claim}. A
positive amount, also known as the \emph{severity} of the claim, is a
key financial expenditure for an insurer. So, knowing only the claim
amount summarizes the reimbursement to the policyholder.

Ignoring expenses, an insurer that examines only amounts paid would be
indifferent to two claims of 100 when compared to one claim of 200, even
though the number of claims differ. Nonetheless, it is common for
insurers to study how often claims arise, known as the \emph{frequency}
of claims. The frequency is important for expenses, but it also
influences contractual parameters (such as deductibles and policy
limits) that are written on a per occurrence basis, is routinely
monitored by insurance regulators, and is often a key driven in the
overall indemnification obligation of the insurer. We shall consider the
two claims variables, the severity and frequency, as the two main
outcome variables that we wish to understand, model, and manage.

To illustrate, in 2010 there were 1,110 policyholders in the property
fund. Table \ref{tab:Frequency2010} shows the distribution of the 1,377
claims. Almost two-thirds (0.637) of the policyholders did not have any
claims and an additional 18.8\% only had one claim. The remaining 17.5\%
(=1 - 0.637 - 0.188) had more than one claim; the policyholder with the
highest number recorded 239 claims. The average number of claims for
this sample was 1.24 (=1377/1110).

\begin{longtable}[]{@{}llllllllllll@{}}
\caption{\label{tab:Frequency2010} 2010 Claims Frequency
Distribution}\tabularnewline
\toprule
Type & & & & & & & & & & &\tabularnewline
\midrule
\endfirsthead
\toprule
Type & & & & & & & & & & &\tabularnewline
\midrule
\endhead
Number & 0 & 1 & 2 & 3 & 4 & 5 & 6 & 7 & 8 & 9 or more &
Sum\tabularnewline
Count & 707 & 209 & 86 & 40 & 18 & 12 & 9 & 4 & 6 & 19 &
1,110\tabularnewline
Proportion & 0.637 & 0.188 & 0.077 & 0.036 & 0.016 & 0.011 & 0.008 &
0.004 & 0.005 & 0.017 & 1.000\tabularnewline
\bottomrule
\end{longtable}

R Code for Frequency Table

\hypertarget{display.T:Frequency.2}{}
\begin{verbatim}
Insample <- read.csv("Insample.csv", header=T,  na.strings=c("."), stringsAsFactors=FALSE)
Insample2010 <- subset(Insample, Year==2010)
table(Insample2010$Freq)
\end{verbatim}

For the severity distribution, one common approach is to examine the
distribution of the sample of 1,377 claims. However, another common
approach is to examine the distribution of the average claims of those
policyholders with claims. In our 2010 sample, there were 403
(=1110-707) such policyholders. For 209 of these policyholders with one
claim, the average claim equals the only claim they experienced. For the
policyholder with highest frequency, the average claim is an average
over 239 separately reported claim events. The total severity divided by
the number of claims is also known as the \emph{pure premium} or
\emph{loss cost}.

\begin{figure}

{\centering \includegraphics[width=0.8\linewidth]{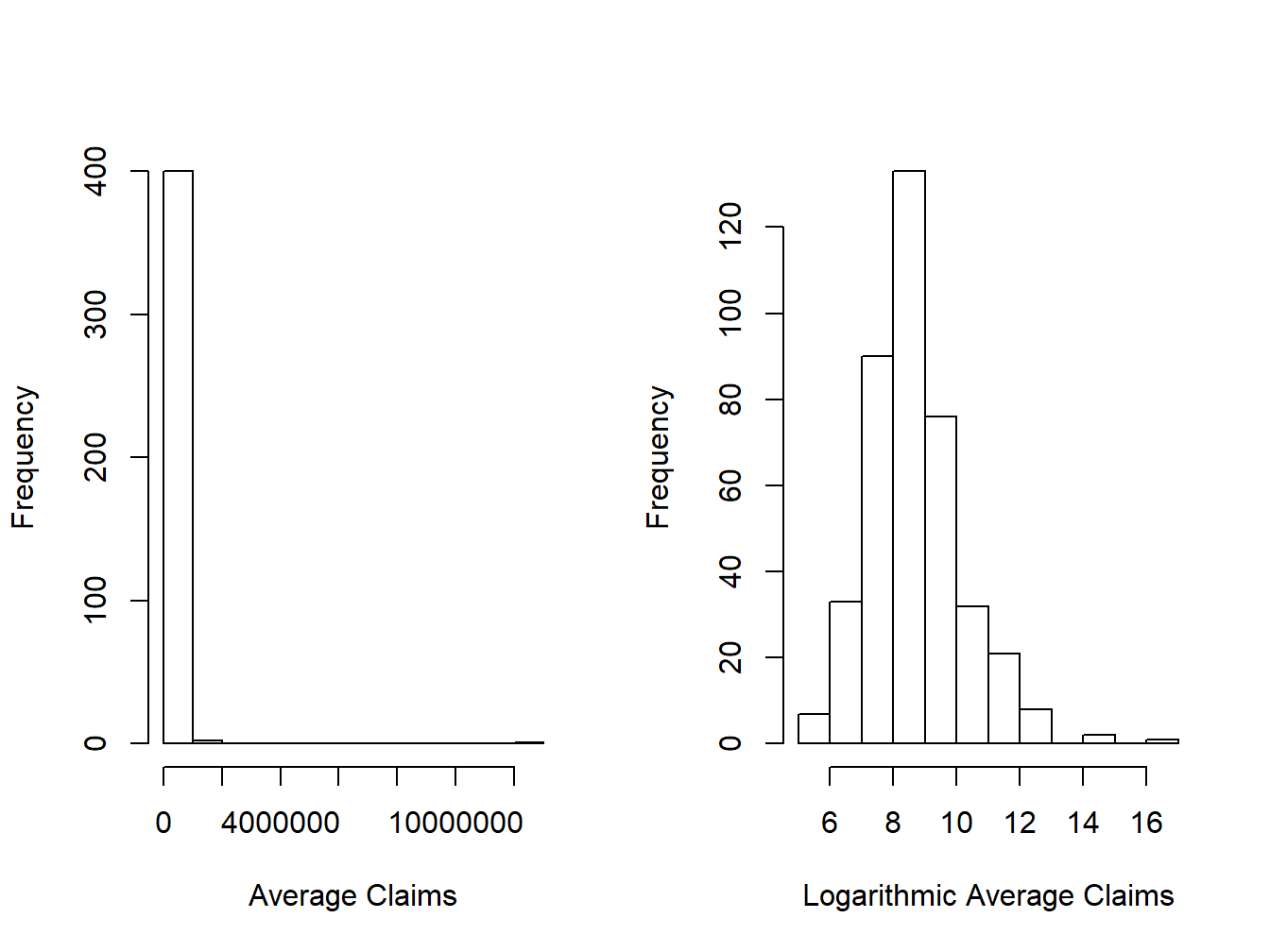}

}

\caption{Distribution of Positive Average Severities}\label{fig:SeverityFig}
\end{figure}

Table \ref{tab:Severity2010} summarizes the sample distribution of
average severities from the 403 policyholders; it shows that the average
claim amount was 56,330 (all amounts are in US Dollars). However, the
average gives only a limited look at the distribution. More information
can be gleaned from the summary statistics which show a very large claim
in the amount of 12,920,000. Figure \ref{fig:SeverityFig} provides
further information about the distribution of sample claims, showing a
distribution that is dominated by this single large claim so that the
histogram is not very helpful. Even when removing the large claim, you
will find a distribution that is skewed to the right. A generally
accepted technique is to work with claims in logarithmic units
especially for graphical purposes; the corresponding figure in the
right-hand panel is much easier to interpret.

\begin{longtable}[]{@{}rrrrrr@{}}
\caption{\label{tab:Severity2010} 2010 Average Severity
Distribution}\tabularnewline
\toprule
\begin{minipage}[b]{0.12\columnwidth}\raggedleft\strut
Minimum\strut
\end{minipage} & \begin{minipage}[b]{0.13\columnwidth}\raggedleft\strut
First Quartile\strut
\end{minipage} & \begin{minipage}[b]{0.10\columnwidth}\raggedleft\strut
Median\strut
\end{minipage} & \begin{minipage}[b]{0.10\columnwidth}\raggedleft\strut
Mean\strut
\end{minipage} & \begin{minipage}[b]{0.13\columnwidth}\raggedleft\strut
Third Quartile\strut
\end{minipage} & \begin{minipage}[b]{0.14\columnwidth}\raggedleft\strut
Maximum\strut
\end{minipage}\tabularnewline
\midrule
\endfirsthead
\toprule
\begin{minipage}[b]{0.12\columnwidth}\raggedleft\strut
Minimum\strut
\end{minipage} & \begin{minipage}[b]{0.13\columnwidth}\raggedleft\strut
First Quartile\strut
\end{minipage} & \begin{minipage}[b]{0.10\columnwidth}\raggedleft\strut
Median\strut
\end{minipage} & \begin{minipage}[b]{0.10\columnwidth}\raggedleft\strut
Mean\strut
\end{minipage} & \begin{minipage}[b]{0.13\columnwidth}\raggedleft\strut
Third Quartile\strut
\end{minipage} & \begin{minipage}[b]{0.14\columnwidth}\raggedleft\strut
Maximum\strut
\end{minipage}\tabularnewline
\midrule
\endhead
\begin{minipage}[t]{0.12\columnwidth}\raggedleft\strut
167\strut
\end{minipage} & \begin{minipage}[t]{0.13\columnwidth}\raggedleft\strut
2,226\strut
\end{minipage} & \begin{minipage}[t]{0.10\columnwidth}\raggedleft\strut
4,951\strut
\end{minipage} & \begin{minipage}[t]{0.10\columnwidth}\raggedleft\strut
56,330\strut
\end{minipage} & \begin{minipage}[t]{0.13\columnwidth}\raggedleft\strut
11,900\strut
\end{minipage} & \begin{minipage}[t]{0.14\columnwidth}\raggedleft\strut
12,920,000\strut
\end{minipage}\tabularnewline
\bottomrule
\end{longtable}

R Code for Severity Distribution Table and Figures

\hypertarget{display.SeverityFig.2}{}
\begin{verbatim}
Insample <- read.csv("Data/PropertyFundInsample.csv", header=T, na.strings=c("."), stringsAsFactors=FALSE)
Insample2010 <- subset(Insample, Year==2010)
InsamplePos2010 <- subset(Insample2010, yAvg>0)
# Table
summary(InsamplePos2010$yAvg)
length(InsamplePos2010$yAvg)
# Figures
par(mfrow=c(1, 2))
hist(InsamplePos2010$yAvg, main="", xlab="Average Claims")
hist(log(InsamplePos2010$yAvg), main="", xlab="Logarithmic Average Claims")
\end{verbatim}

\subsection{Fund Rating Variables}\label{S:FundVariables}

Developing models to represent and manage the two outcome variables,
frequency and severity, is the focus of the early chapters of this text.
However, when actuaries and other financial analysts use those models,
they do so in the context of externally available variables. In general
statistical terminology, one might call these explanatory or predictor
variables; there are many other names in statistics, economics,
psychology, and other disciplines. Because of our insurance focus, we
call them \emph{rating variables} as they will be useful in setting
insurance rates and premiums.

We earlier considered a sample of 1,110 observations which may seem like
a lot. However, as we will seen in our forthcoming applications, because
of the preponderance of zeros and the skewed nature of claims, actuaries
typically yearn for more data. One common approach that we adopt here is
to examine outcomes from multiple years, thus increasing the sample
size. We will discuss the strengths and limitations of this strategy
later but, at this juncture, just want to show the reader how it works.

Specifically, Table \ref{tab:CoverageBCIM} shows that we now consider
policies over five years of data, years 2006, \ldots{}, 2010, inclusive.
The data begins in 2006 because there was a shift in claim coding in
2005 so that comparisons with earlier years are not helpful. To mitigate
the effect of open claims, we consider policy years prior to 2011. An
open claim means that all of the obligations are not known at the time
of the analysis; for some claims, such an injury to a person in an auto
accident or in the workplace, it can take years before costs are fully
known.

Table \ref{tab:CoverageBCIM} shows that the average claim varies over
time, especially with the high 2010 value due to a single large claim.
The total number of policyholders is steadily declining and, conversely,
the coverage is steadily increasing. The coverage variable is the amount
of coverage of the property and contents. Roughly, you can think of it
as the maximum possible payout of the insurer. For our immediate
purposes, it is our first rating variable. Other things being equal, we
would expect that policyholders with larger coverage will have larger
claims. We will make this vague idea much more precise as we proceed.

\begin{longtable}[]{@{}lrrrr@{}}
\caption{\label{tab:CoverageBCIM} Building and Contents Claims
Summary}\tabularnewline
\toprule
\begin{minipage}[b]{0.25\columnwidth}\raggedright\strut
Year\strut
\end{minipage} & \begin{minipage}[b]{0.14\columnwidth}\raggedleft\strut
Average Frequency\strut
\end{minipage} & \begin{minipage}[b]{0.13\columnwidth}\raggedleft\strut
Average Severity\strut
\end{minipage} & \begin{minipage}[b]{0.15\columnwidth}\raggedleft\strut
Average Coverage\strut
\end{minipage} & \begin{minipage}[b]{0.17\columnwidth}\raggedleft\strut
Number of Policyholders\strut
\end{minipage}\tabularnewline
\midrule
\endfirsthead
\toprule
\begin{minipage}[b]{0.25\columnwidth}\raggedright\strut
Year\strut
\end{minipage} & \begin{minipage}[b]{0.14\columnwidth}\raggedleft\strut
Average Frequency\strut
\end{minipage} & \begin{minipage}[b]{0.13\columnwidth}\raggedleft\strut
Average Severity\strut
\end{minipage} & \begin{minipage}[b]{0.15\columnwidth}\raggedleft\strut
Average Coverage\strut
\end{minipage} & \begin{minipage}[b]{0.17\columnwidth}\raggedleft\strut
Number of Policyholders\strut
\end{minipage}\tabularnewline
\midrule
\endhead
\begin{minipage}[t]{0.25\columnwidth}\raggedright\strut
2006\strut
\end{minipage} & \begin{minipage}[t]{0.14\columnwidth}\raggedleft\strut
0.951\strut
\end{minipage} & \begin{minipage}[t]{0.13\columnwidth}\raggedleft\strut
9,695\strut
\end{minipage} & \begin{minipage}[t]{0.15\columnwidth}\raggedleft\strut
32,498,186\strut
\end{minipage} & \begin{minipage}[t]{0.17\columnwidth}\raggedleft\strut
1,154\strut
\end{minipage}\tabularnewline
\begin{minipage}[t]{0.25\columnwidth}\raggedright\strut
2007\strut
\end{minipage} & \begin{minipage}[t]{0.14\columnwidth}\raggedleft\strut
1.167\strut
\end{minipage} & \begin{minipage}[t]{0.13\columnwidth}\raggedleft\strut
6,544\strut
\end{minipage} & \begin{minipage}[t]{0.15\columnwidth}\raggedleft\strut
35,275,949\strut
\end{minipage} & \begin{minipage}[t]{0.17\columnwidth}\raggedleft\strut
1,138\strut
\end{minipage}\tabularnewline
\begin{minipage}[t]{0.25\columnwidth}\raggedright\strut
2008\strut
\end{minipage} & \begin{minipage}[t]{0.14\columnwidth}\raggedleft\strut
0.974\strut
\end{minipage} & \begin{minipage}[t]{0.13\columnwidth}\raggedleft\strut
5,311\strut
\end{minipage} & \begin{minipage}[t]{0.15\columnwidth}\raggedleft\strut
37,267,485\strut
\end{minipage} & \begin{minipage}[t]{0.17\columnwidth}\raggedleft\strut
1,125\strut
\end{minipage}\tabularnewline
\begin{minipage}[t]{0.25\columnwidth}\raggedright\strut
2009\strut
\end{minipage} & \begin{minipage}[t]{0.14\columnwidth}\raggedleft\strut
1.219\strut
\end{minipage} & \begin{minipage}[t]{0.13\columnwidth}\raggedleft\strut
4,572\strut
\end{minipage} & \begin{minipage}[t]{0.15\columnwidth}\raggedleft\strut
40,355,382\strut
\end{minipage} & \begin{minipage}[t]{0.17\columnwidth}\raggedleft\strut
1,112\strut
\end{minipage}\tabularnewline
\begin{minipage}[t]{0.25\columnwidth}\raggedright\strut
2010\strut
\end{minipage} & \begin{minipage}[t]{0.14\columnwidth}\raggedleft\strut
1.241\strut
\end{minipage} & \begin{minipage}[t]{0.13\columnwidth}\raggedleft\strut
20,452\strut
\end{minipage} & \begin{minipage}[t]{0.15\columnwidth}\raggedleft\strut
41,242,070\strut
\end{minipage} & \begin{minipage}[t]{0.17\columnwidth}\raggedleft\strut
1,110\strut
\end{minipage}\tabularnewline
\bottomrule
\end{longtable}

R Code for Building and Contents Claims Summary

\hypertarget{display.CoverageBC.2}{}
\begin{verbatim}
Insample <- read.csv("Data/PropertyFundInsample.csv", header=T, na.strings=c("."), stringsAsFactors=FALSE)
library(doBy)
T1A <- summaryBy(Freq ~ Year, data = Insample,
   FUN = function(x) { c(m = mean(x), num=length(x)) } )
T1B <- summaryBy(yAvg    ~ Year, data = Insample,
   FUN = function(x) { c(m = mean(x), num=length(x)) } )
T1C <- summaryBy(BCcov    ~ Year, data = Insample,
   FUN = function(x) { c(m = mean(x), num=length(x)) } )
Table1In <- cbind(T1A[1],T1A[2],T1B[2],T1C[2],T1A[3])
names(Table1In) <- c("Year", "Average Frequency","Average Severity", "Average","Number of Policyholders")
Table1In
\end{verbatim}

For a different look at this five-year sample, Table \ref{tab:DeductCov}
summarizes the distribution of our two outcomes, frequency and claims
amount. In each case, the average exceeds the median, suggesting that
the two distributions are right-skewed. In addition, the table
summarizes our continuous rating variables, coverage and deductible
amount. The table also suggests that these variables also have
right-skewed distributions.

\begin{longtable}[]{@{}lrrrr@{}}
\caption{\label{tab:DeductCov} Summary of Claim Frequency and Severity,
Deductibles, and Coverages}\tabularnewline
\toprule
\begin{minipage}[b]{0.23\columnwidth}\raggedright\strut
\strut
\end{minipage} & \begin{minipage}[b]{0.12\columnwidth}\raggedleft\strut
Minimum\strut
\end{minipage} & \begin{minipage}[b]{0.11\columnwidth}\raggedleft\strut
Median\strut
\end{minipage} & \begin{minipage}[b]{0.12\columnwidth}\raggedleft\strut
Average\strut
\end{minipage} & \begin{minipage}[b]{0.14\columnwidth}\raggedleft\strut
Maximum\strut
\end{minipage}\tabularnewline
\midrule
\endfirsthead
\toprule
\begin{minipage}[b]{0.23\columnwidth}\raggedright\strut
\strut
\end{minipage} & \begin{minipage}[b]{0.12\columnwidth}\raggedleft\strut
Minimum\strut
\end{minipage} & \begin{minipage}[b]{0.11\columnwidth}\raggedleft\strut
Median\strut
\end{minipage} & \begin{minipage}[b]{0.12\columnwidth}\raggedleft\strut
Average\strut
\end{minipage} & \begin{minipage}[b]{0.14\columnwidth}\raggedleft\strut
Maximum\strut
\end{minipage}\tabularnewline
\midrule
\endhead
\begin{minipage}[t]{0.23\columnwidth}\raggedright\strut
Claim Frequency\strut
\end{minipage} & \begin{minipage}[t]{0.12\columnwidth}\raggedleft\strut
0\strut
\end{minipage} & \begin{minipage}[t]{0.11\columnwidth}\raggedleft\strut
0\strut
\end{minipage} & \begin{minipage}[t]{0.12\columnwidth}\raggedleft\strut
1.109\strut
\end{minipage} & \begin{minipage}[t]{0.14\columnwidth}\raggedleft\strut
263\strut
\end{minipage}\tabularnewline
\begin{minipage}[t]{0.23\columnwidth}\raggedright\strut
Claim Severity\strut
\end{minipage} & \begin{minipage}[t]{0.12\columnwidth}\raggedleft\strut
0\strut
\end{minipage} & \begin{minipage}[t]{0.11\columnwidth}\raggedleft\strut
0\strut
\end{minipage} & \begin{minipage}[t]{0.12\columnwidth}\raggedleft\strut
9,292\strut
\end{minipage} & \begin{minipage}[t]{0.14\columnwidth}\raggedleft\strut
12,922,218\strut
\end{minipage}\tabularnewline
\begin{minipage}[t]{0.23\columnwidth}\raggedright\strut
Deductible\strut
\end{minipage} & \begin{minipage}[t]{0.12\columnwidth}\raggedleft\strut
500\strut
\end{minipage} & \begin{minipage}[t]{0.11\columnwidth}\raggedleft\strut
1,000\strut
\end{minipage} & \begin{minipage}[t]{0.12\columnwidth}\raggedleft\strut
3,365\strut
\end{minipage} & \begin{minipage}[t]{0.14\columnwidth}\raggedleft\strut
100,000\strut
\end{minipage}\tabularnewline
\begin{minipage}[t]{0.23\columnwidth}\raggedright\strut
Coverage (000's)\strut
\end{minipage} & \begin{minipage}[t]{0.12\columnwidth}\raggedleft\strut
8.937\strut
\end{minipage} & \begin{minipage}[t]{0.11\columnwidth}\raggedleft\strut
11,354\strut
\end{minipage} & \begin{minipage}[t]{0.12\columnwidth}\raggedleft\strut
37,281\strut
\end{minipage} & \begin{minipage}[t]{0.14\columnwidth}\raggedleft\strut
2,444,797\strut
\end{minipage}\tabularnewline
\bottomrule
\end{longtable}

R Code for Summary of Claim Frequency and Severity, Deductibles, and
Coverages

\hypertarget{display.DeductCov.2}{}
\begin{verbatim}
Insample <- read.csv("Data/PropertyFundInsample.csv", header=T, na.strings=c("."), stringsAsFactors=FALSE)
t1<- summaryBy(Insample$Freq ~ 1, data = Insample,
   FUN = function(x) { c(ma=min(x), m1=median(x),m=mean(x),mb=max(x)) } )
names(t1) <- c("Minimum", "Median","Average", "Maximum")
t2 <- summaryBy(Insample$yAvg ~ 1, data = Insample,
   FUN = function(x) { c(ma=min(x), m1=median(x), m=mean(x),mb=max(x)) } )
names(t2) <- c("Minimum", "Median","Average", "Maximum")
t3 <- summaryBy(Deduct ~ 1, data = Insample,
   FUN = function(x) { c(ma=min(x), m1=median(x), m=mean(x),mb=max(x)) } )
names(t3) <- c("Minimum", "Median","Average", "Maximum")
t4 <- summaryBy(BCcov/1000 ~ 1, data = Insample,
   FUN = function(x) { c(ma=min(x), m1=median(x), m=mean(x),mb=max(x)) } )
names(t4) <- c("Minimum", "Median","Average", "Maximum")
Table2 <- rbind(t1,t2,t3,t4)
Table2a <- round(Table2,3)
Rowlable <- rbind("Claim Frequency","Claim Severity","Deductible","Coverage (000's)")
Table2aa <- cbind(Rowlable,as.matrix(Table2a))
Table2aa
\end{verbatim}

The following display describes the rating variables considered in this
chapter. To handle the skewness, we henceforth focus on logarithmic
transformations of coverage and deductibles. To get a sense of the
relationship between the non-continuous rating variables and claims,
Table \ref{tab:ClaimRateVar} relates the claims outcomes to these
categorical variables. Table \ref{tab:ClaimRateVar} suggests substantial
variation in the claim frequency and average severity of the claims by
entity type. It also demonstrates higher frequency and severity for the
\({\tt Fire5}\) variable and the reverse for the \({\tt NoClaimCredit}\)
variable. The relationship for the \({\tt Fire5}\) variable is
counter-intuitive in that one would expect lower claim amounts for those
policyholders in areas with better public protection (when the
protection code is five or less). Naturally, there are other variables
that influence this relationship. We will see that these background
variables are accounted for in the subsequent multivariate regression
analysis, which yields an intuitive, appealing (negative) sign for the
\({\tt Fire5}\) variable.

Description of Rating Variables \[{\small \begin{matrix}
\begin{array}{ l | l}
\hline
Variable    & Description \\
\hline
\text{EntityType}   & \text{Categorical variable that is one of six types:  (Village, City,} \\
& ~~~~ \text{County, Misc, School, or Town)} \\
\text{LnCoverage}   & \text{Total building and content coverage, in logarithmic millions of dollars}\\
\text{LnDeduct}     & \text{Deductible, in logarithmic dollars} \\
\text{AlarmCredit}  & \text{Categorical variable that is one of four types:  (0, 5, 10, or 15)} \\
 &  ~~~~   \text{for automatic smoke alarms in main rooms} \\
\text{NoClaimCredit}    & \text{Binary variable to indicate no claims in the past two years} \\
\text{Fire5 }           & \text{Binary variable to indicate the fire class is below 5} \\
& ~~~~ \text{(The range of fire class is 0 to 10} \\
\hline
\end{array}
\end{matrix}}\]

\begin{longtable}[]{@{}lrrr@{}}
\caption{\label{tab:ClaimRateVar} Claims Summary by Entity Type, Fire Class,
and No Claim Credit}\tabularnewline
\toprule
\begin{minipage}[b]{0.27\columnwidth}\raggedright\strut
Variable\strut
\end{minipage} & \begin{minipage}[b]{0.15\columnwidth}\raggedleft\strut
Number of Policies\strut
\end{minipage} & \begin{minipage}[b]{0.15\columnwidth}\raggedleft\strut
Claim Frequency\strut
\end{minipage} & \begin{minipage}[b]{0.15\columnwidth}\raggedleft\strut
Average Severity\strut
\end{minipage}\tabularnewline
\midrule
\endfirsthead
\toprule
\begin{minipage}[b]{0.27\columnwidth}\raggedright\strut
Variable\strut
\end{minipage} & \begin{minipage}[b]{0.15\columnwidth}\raggedleft\strut
Number of Policies\strut
\end{minipage} & \begin{minipage}[b]{0.15\columnwidth}\raggedleft\strut
Claim Frequency\strut
\end{minipage} & \begin{minipage}[b]{0.15\columnwidth}\raggedleft\strut
Average Severity\strut
\end{minipage}\tabularnewline
\midrule
\endhead
\begin{minipage}[t]{0.27\columnwidth}\raggedright\strut
\emph{EntityType}\strut
\end{minipage} & \begin{minipage}[t]{0.15\columnwidth}\raggedleft\strut
\strut
\end{minipage} & \begin{minipage}[t]{0.15\columnwidth}\raggedleft\strut
\strut
\end{minipage} & \begin{minipage}[t]{0.15\columnwidth}\raggedleft\strut
\strut
\end{minipage}\tabularnewline
\begin{minipage}[t]{0.27\columnwidth}\raggedright\strut
Village\strut
\end{minipage} & \begin{minipage}[t]{0.15\columnwidth}\raggedleft\strut
1,341\strut
\end{minipage} & \begin{minipage}[t]{0.15\columnwidth}\raggedleft\strut
0.452\strut
\end{minipage} & \begin{minipage}[t]{0.15\columnwidth}\raggedleft\strut
10,645\strut
\end{minipage}\tabularnewline
\begin{minipage}[t]{0.27\columnwidth}\raggedright\strut
City\strut
\end{minipage} & \begin{minipage}[t]{0.15\columnwidth}\raggedleft\strut
793\strut
\end{minipage} & \begin{minipage}[t]{0.15\columnwidth}\raggedleft\strut
1.941\strut
\end{minipage} & \begin{minipage}[t]{0.15\columnwidth}\raggedleft\strut
16,924\strut
\end{minipage}\tabularnewline
\begin{minipage}[t]{0.27\columnwidth}\raggedright\strut
County\strut
\end{minipage} & \begin{minipage}[t]{0.15\columnwidth}\raggedleft\strut
328\strut
\end{minipage} & \begin{minipage}[t]{0.15\columnwidth}\raggedleft\strut
4.899\strut
\end{minipage} & \begin{minipage}[t]{0.15\columnwidth}\raggedleft\strut
15,453\strut
\end{minipage}\tabularnewline
\begin{minipage}[t]{0.27\columnwidth}\raggedright\strut
Misc\strut
\end{minipage} & \begin{minipage}[t]{0.15\columnwidth}\raggedleft\strut
609\strut
\end{minipage} & \begin{minipage}[t]{0.15\columnwidth}\raggedleft\strut
0.186\strut
\end{minipage} & \begin{minipage}[t]{0.15\columnwidth}\raggedleft\strut
43,036\strut
\end{minipage}\tabularnewline
\begin{minipage}[t]{0.27\columnwidth}\raggedright\strut
School\strut
\end{minipage} & \begin{minipage}[t]{0.15\columnwidth}\raggedleft\strut
1,597\strut
\end{minipage} & \begin{minipage}[t]{0.15\columnwidth}\raggedleft\strut
1.434\strut
\end{minipage} & \begin{minipage}[t]{0.15\columnwidth}\raggedleft\strut
64,346\strut
\end{minipage}\tabularnewline
\begin{minipage}[t]{0.27\columnwidth}\raggedright\strut
Town\strut
\end{minipage} & \begin{minipage}[t]{0.15\columnwidth}\raggedleft\strut
971\strut
\end{minipage} & \begin{minipage}[t]{0.15\columnwidth}\raggedleft\strut
0.103\strut
\end{minipage} & \begin{minipage}[t]{0.15\columnwidth}\raggedleft\strut
19,831\strut
\end{minipage}\tabularnewline
\begin{minipage}[t]{0.27\columnwidth}\raggedright\strut
Fire5=0\strut
\end{minipage} & \begin{minipage}[t]{0.15\columnwidth}\raggedleft\strut
2,508\strut
\end{minipage} & \begin{minipage}[t]{0.15\columnwidth}\raggedleft\strut
0.502\strut
\end{minipage} & \begin{minipage}[t]{0.15\columnwidth}\raggedleft\strut
13,935\strut
\end{minipage}\tabularnewline
\begin{minipage}[t]{0.27\columnwidth}\raggedright\strut
Fire5=1\strut
\end{minipage} & \begin{minipage}[t]{0.15\columnwidth}\raggedleft\strut
3,131\strut
\end{minipage} & \begin{minipage}[t]{0.15\columnwidth}\raggedleft\strut
1.596\strut
\end{minipage} & \begin{minipage}[t]{0.15\columnwidth}\raggedleft\strut
41,421\strut
\end{minipage}\tabularnewline
\begin{minipage}[t]{0.27\columnwidth}\raggedright\strut
NoClaimCredit=0\strut
\end{minipage} & \begin{minipage}[t]{0.15\columnwidth}\raggedleft\strut
3,786\strut
\end{minipage} & \begin{minipage}[t]{0.15\columnwidth}\raggedleft\strut
1.501\strut
\end{minipage} & \begin{minipage}[t]{0.15\columnwidth}\raggedleft\strut
31,365\strut
\end{minipage}\tabularnewline
\begin{minipage}[t]{0.27\columnwidth}\raggedright\strut
NoClaimCredit=1\strut
\end{minipage} & \begin{minipage}[t]{0.15\columnwidth}\raggedleft\strut
1,853\strut
\end{minipage} & \begin{minipage}[t]{0.15\columnwidth}\raggedleft\strut
0.310\strut
\end{minipage} & \begin{minipage}[t]{0.15\columnwidth}\raggedleft\strut
30,499\strut
\end{minipage}\tabularnewline
\begin{minipage}[t]{0.27\columnwidth}\raggedright\strut
Total\strut
\end{minipage} & \begin{minipage}[t]{0.15\columnwidth}\raggedleft\strut
5,639\strut
\end{minipage} & \begin{minipage}[t]{0.15\columnwidth}\raggedleft\strut
1.109\strut
\end{minipage} & \begin{minipage}[t]{0.15\columnwidth}\raggedleft\strut
31,206\strut
\end{minipage}\tabularnewline
\bottomrule
\end{longtable}

R Code for Claims Summary by Entity Type, Fire Class, and No Claim
Credit

\hypertarget{display.ClaimRateVar.2}{}
\begin{verbatim}
ByVarSumm<-function(datasub){
  tempA <- summaryBy(Freq    ~ 1 , data = datasub,
     FUN = function(x) { c(m = mean(x), num=length(x)) } )
  datasub1 <-  subset(datasub, yAvg>0)
  tempB <- summaryBy(yAvg   ~ 1, data = datasub1,FUN = function(x) { c(m = mean(x)) } )
  tempC <- merge(tempA,tempB,all.x=T)[c(2,1,3)]
  tempC1 <- as.matrix(tempC)
  return(tempC1)
  }
datasub <-  subset(Insample, TypeVillage == 1);
t1 <- ByVarSumm(datasub)
datasub <-  subset(Insample, TypeCity == 1);
t2 <- ByVarSumm(datasub)
datasub <-  subset(Insample, TypeCounty == 1);
t3 <- ByVarSumm(datasub)
datasub <-  subset(Insample, TypeMisc == 1);
t4 <- ByVarSumm(datasub)
datasub <-  subset(Insample, TypeSchool == 1);
t5 <- ByVarSumm(datasub)
datasub <-  subset(Insample, TypeTown == 1);
t6 <- ByVarSumm(datasub)
datasub <-  subset(Insample, Fire5 == 0);
t7 <- ByVarSumm(datasub)
datasub <-  subset(Insample, Fire5 == 1);
t8 <- ByVarSumm(datasub)
datasub <-  subset(Insample, Insample$NoClaimCredit == 0);
t9 <- ByVarSumm(datasub)
datasub <-  subset(Insample, Insample$NoClaimCredit == 1);
t10 <- ByVarSumm(datasub)
t11 <- ByVarSumm(Insample)

Tablea <- rbind(t1,t2,t3,t4,t5,t6,t7,t8,t9,t10,t11)
Tableaa <- round(Tablea,3)
Rowlable <- rbind("Village","City","County","Misc","School",
          "Town","Fire5--No","Fire5--Yes","NoClaimCredit--No",
        "NoClaimCredit--Yes","Total")
Table4 <- cbind(Rowlable,as.matrix(Tableaa))
Table4
\end{verbatim}

Table \ref{tab:RateAlarmCredit} shows the claims experience by alarm
credit. It underscores the difficulty of examining variables
individually. For example, when looking at the experience for all
entities, we see that policyholders with no alarm credit have on average
lower frequency and severity than policyholders with the highest (15\%,
with 24/7 monitoring by a fire station or security company) alarm
credit. In particular, when we look at the entity type School, the
frequency is 0.422 and the severity 25,257 for no alarm credit, whereas
for the highest alarm level it is 2.008 and 85,140. This may simply
imply that entities with more claims are the ones that are likely to
have an alarm system. Summary tables do not examine multivariate
effects; for example, Table \ref{tab:ClaimRateVar} ignores the effect of
size (as we measure through coverage amounts) that affect claims.

\begin{longtable}[]{@{}lrrrrrr@{}}
\caption{\label{tab:RateAlarmCredit} Claims Summary by Entity Type and Alarm
Credit Category}\tabularnewline
\toprule
\begin{minipage}[b]{0.10\columnwidth}\raggedright\strut
Entity Type\strut
\end{minipage} & \begin{minipage}[b]{0.12\columnwidth}\raggedleft\strut
Claim Frequency\strut
\end{minipage} & \begin{minipage}[b]{0.11\columnwidth}\raggedleft\strut
Avg. Severity\strut
\end{minipage} & \begin{minipage}[b]{0.11\columnwidth}\raggedleft\strut
Num. Policies\strut
\end{minipage} & \begin{minipage}[b]{0.12\columnwidth}\raggedleft\strut
Claim Frequency\strut
\end{minipage} & \begin{minipage}[b]{0.11\columnwidth}\raggedleft\strut
Avg. Severity\strut
\end{minipage} & \begin{minipage}[b]{0.11\columnwidth}\raggedleft\strut
Num. Policies\strut
\end{minipage}\tabularnewline
\midrule
\endfirsthead
\toprule
\begin{minipage}[b]{0.10\columnwidth}\raggedright\strut
Entity Type\strut
\end{minipage} & \begin{minipage}[b]{0.12\columnwidth}\raggedleft\strut
Claim Frequency\strut
\end{minipage} & \begin{minipage}[b]{0.11\columnwidth}\raggedleft\strut
Avg. Severity\strut
\end{minipage} & \begin{minipage}[b]{0.11\columnwidth}\raggedleft\strut
Num. Policies\strut
\end{minipage} & \begin{minipage}[b]{0.12\columnwidth}\raggedleft\strut
Claim Frequency\strut
\end{minipage} & \begin{minipage}[b]{0.11\columnwidth}\raggedleft\strut
Avg. Severity\strut
\end{minipage} & \begin{minipage}[b]{0.11\columnwidth}\raggedleft\strut
Num. Policies\strut
\end{minipage}\tabularnewline
\midrule
\endhead
\begin{minipage}[t]{0.10\columnwidth}\raggedright\strut
Village\strut
\end{minipage} & \begin{minipage}[t]{0.12\columnwidth}\raggedleft\strut
0.326\strut
\end{minipage} & \begin{minipage}[t]{0.11\columnwidth}\raggedleft\strut
11,078\strut
\end{minipage} & \begin{minipage}[t]{0.11\columnwidth}\raggedleft\strut
829\strut
\end{minipage} & \begin{minipage}[t]{0.12\columnwidth}\raggedleft\strut
0.278\strut
\end{minipage} & \begin{minipage}[t]{0.11\columnwidth}\raggedleft\strut
8,086\strut
\end{minipage} & \begin{minipage}[t]{0.11\columnwidth}\raggedleft\strut
54\strut
\end{minipage}\tabularnewline
\begin{minipage}[t]{0.10\columnwidth}\raggedright\strut
City\strut
\end{minipage} & \begin{minipage}[t]{0.12\columnwidth}\raggedleft\strut
0.893\strut
\end{minipage} & \begin{minipage}[t]{0.11\columnwidth}\raggedleft\strut
7,576\strut
\end{minipage} & \begin{minipage}[t]{0.11\columnwidth}\raggedleft\strut
244\strut
\end{minipage} & \begin{minipage}[t]{0.12\columnwidth}\raggedleft\strut
2.077\strut
\end{minipage} & \begin{minipage}[t]{0.11\columnwidth}\raggedleft\strut
4,150\strut
\end{minipage} & \begin{minipage}[t]{0.11\columnwidth}\raggedleft\strut
13\strut
\end{minipage}\tabularnewline
\begin{minipage}[t]{0.10\columnwidth}\raggedright\strut
County\strut
\end{minipage} & \begin{minipage}[t]{0.12\columnwidth}\raggedleft\strut
2.140\strut
\end{minipage} & \begin{minipage}[t]{0.11\columnwidth}\raggedleft\strut
16,013\strut
\end{minipage} & \begin{minipage}[t]{0.11\columnwidth}\raggedleft\strut
50\strut
\end{minipage} & \begin{minipage}[t]{0.12\columnwidth}\raggedleft\strut
-\strut
\end{minipage} & \begin{minipage}[t]{0.11\columnwidth}\raggedleft\strut
-\strut
\end{minipage} & \begin{minipage}[t]{0.11\columnwidth}\raggedleft\strut
1\strut
\end{minipage}\tabularnewline
\begin{minipage}[t]{0.10\columnwidth}\raggedright\strut
Misc\strut
\end{minipage} & \begin{minipage}[t]{0.12\columnwidth}\raggedleft\strut
0.117\strut
\end{minipage} & \begin{minipage}[t]{0.11\columnwidth}\raggedleft\strut
15,122\strut
\end{minipage} & \begin{minipage}[t]{0.11\columnwidth}\raggedleft\strut
386\strut
\end{minipage} & \begin{minipage}[t]{0.12\columnwidth}\raggedleft\strut
0.278\strut
\end{minipage} & \begin{minipage}[t]{0.11\columnwidth}\raggedleft\strut
13,064\strut
\end{minipage} & \begin{minipage}[t]{0.11\columnwidth}\raggedleft\strut
18\strut
\end{minipage}\tabularnewline
\begin{minipage}[t]{0.10\columnwidth}\raggedright\strut
School\strut
\end{minipage} & \begin{minipage}[t]{0.12\columnwidth}\raggedleft\strut
0.422\strut
\end{minipage} & \begin{minipage}[t]{0.11\columnwidth}\raggedleft\strut
25,523\strut
\end{minipage} & \begin{minipage}[t]{0.11\columnwidth}\raggedleft\strut
294\strut
\end{minipage} & \begin{minipage}[t]{0.12\columnwidth}\raggedleft\strut
0.410\strut
\end{minipage} & \begin{minipage}[t]{0.11\columnwidth}\raggedleft\strut
14,575\strut
\end{minipage} & \begin{minipage}[t]{0.11\columnwidth}\raggedleft\strut
122\strut
\end{minipage}\tabularnewline
\begin{minipage}[t]{0.10\columnwidth}\raggedright\strut
Town\strut
\end{minipage} & \begin{minipage}[t]{0.12\columnwidth}\raggedleft\strut
0.083\strut
\end{minipage} & \begin{minipage}[t]{0.11\columnwidth}\raggedleft\strut
25,257\strut
\end{minipage} & \begin{minipage}[t]{0.11\columnwidth}\raggedleft\strut
808\strut
\end{minipage} & \begin{minipage}[t]{0.12\columnwidth}\raggedleft\strut
0.194\strut
\end{minipage} & \begin{minipage}[t]{0.11\columnwidth}\raggedleft\strut
3,937\strut
\end{minipage} & \begin{minipage}[t]{0.11\columnwidth}\raggedleft\strut
31\strut
\end{minipage}\tabularnewline
\begin{minipage}[t]{0.10\columnwidth}\raggedright\strut
Total\strut
\end{minipage} & \begin{minipage}[t]{0.12\columnwidth}\raggedleft\strut
0.318\strut
\end{minipage} & \begin{minipage}[t]{0.11\columnwidth}\raggedleft\strut
15,118\strut
\end{minipage} & \begin{minipage}[t]{0.11\columnwidth}\raggedleft\strut
2,611\strut
\end{minipage} & \begin{minipage}[t]{0.12\columnwidth}\raggedleft\strut
0.431\strut
\end{minipage} & \begin{minipage}[t]{0.11\columnwidth}\raggedleft\strut
10,762\strut
\end{minipage} & \begin{minipage}[t]{0.11\columnwidth}\raggedleft\strut
239\strut
\end{minipage}\tabularnewline
\bottomrule
\end{longtable}

\begin{longtable}[]{@{}lrrrrrr@{}}
\caption{Claims Summary by Entity Type and Alarm Credit
Category}\tabularnewline
\toprule
\begin{minipage}[b]{0.10\columnwidth}\raggedright\strut
Entity Type\strut
\end{minipage} & \begin{minipage}[b]{0.12\columnwidth}\raggedleft\strut
Claim Frequency\strut
\end{minipage} & \begin{minipage}[b]{0.11\columnwidth}\raggedleft\strut
Avg. Severity\strut
\end{minipage} & \begin{minipage}[b]{0.11\columnwidth}\raggedleft\strut
Num. Policies\strut
\end{minipage} & \begin{minipage}[b]{0.12\columnwidth}\raggedleft\strut
Claim Frequency\strut
\end{minipage} & \begin{minipage}[b]{0.11\columnwidth}\raggedleft\strut
Avg. Severity\strut
\end{minipage} & \begin{minipage}[b]{0.11\columnwidth}\raggedleft\strut
Num. Policies\strut
\end{minipage}\tabularnewline
\midrule
\endfirsthead
\toprule
\begin{minipage}[b]{0.10\columnwidth}\raggedright\strut
Entity Type\strut
\end{minipage} & \begin{minipage}[b]{0.12\columnwidth}\raggedleft\strut
Claim Frequency\strut
\end{minipage} & \begin{minipage}[b]{0.11\columnwidth}\raggedleft\strut
Avg. Severity\strut
\end{minipage} & \begin{minipage}[b]{0.11\columnwidth}\raggedleft\strut
Num. Policies\strut
\end{minipage} & \begin{minipage}[b]{0.12\columnwidth}\raggedleft\strut
Claim Frequency\strut
\end{minipage} & \begin{minipage}[b]{0.11\columnwidth}\raggedleft\strut
Avg. Severity\strut
\end{minipage} & \begin{minipage}[b]{0.11\columnwidth}\raggedleft\strut
Num. Policies\strut
\end{minipage}\tabularnewline
\midrule
\endhead
\begin{minipage}[t]{0.10\columnwidth}\raggedright\strut
Village\strut
\end{minipage} & \begin{minipage}[t]{0.12\columnwidth}\raggedleft\strut
0.500\strut
\end{minipage} & \begin{minipage}[t]{0.11\columnwidth}\raggedleft\strut
8,792\strut
\end{minipage} & \begin{minipage}[t]{0.11\columnwidth}\raggedleft\strut
50\strut
\end{minipage} & \begin{minipage}[t]{0.12\columnwidth}\raggedleft\strut
0.725\strut
\end{minipage} & \begin{minipage}[t]{0.11\columnwidth}\raggedleft\strut
10,544\strut
\end{minipage} & \begin{minipage}[t]{0.11\columnwidth}\raggedleft\strut
408\strut
\end{minipage}\tabularnewline
\begin{minipage}[t]{0.10\columnwidth}\raggedright\strut
City\strut
\end{minipage} & \begin{minipage}[t]{0.12\columnwidth}\raggedleft\strut
1.258\strut
\end{minipage} & \begin{minipage}[t]{0.11\columnwidth}\raggedleft\strut
8,625\strut
\end{minipage} & \begin{minipage}[t]{0.11\columnwidth}\raggedleft\strut
31\strut
\end{minipage} & \begin{minipage}[t]{0.12\columnwidth}\raggedleft\strut
2.485\strut
\end{minipage} & \begin{minipage}[t]{0.11\columnwidth}\raggedleft\strut
20,470\strut
\end{minipage} & \begin{minipage}[t]{0.11\columnwidth}\raggedleft\strut
505\strut
\end{minipage}\tabularnewline
\begin{minipage}[t]{0.10\columnwidth}\raggedright\strut
County\strut
\end{minipage} & \begin{minipage}[t]{0.12\columnwidth}\raggedleft\strut
2.125\strut
\end{minipage} & \begin{minipage}[t]{0.11\columnwidth}\raggedleft\strut
11,688\strut
\end{minipage} & \begin{minipage}[t]{0.11\columnwidth}\raggedleft\strut
8\strut
\end{minipage} & \begin{minipage}[t]{0.12\columnwidth}\raggedleft\strut
5.513\strut
\end{minipage} & \begin{minipage}[t]{0.11\columnwidth}\raggedleft\strut
15,476\strut
\end{minipage} & \begin{minipage}[t]{0.11\columnwidth}\raggedleft\strut
269\strut
\end{minipage}\tabularnewline
\begin{minipage}[t]{0.10\columnwidth}\raggedright\strut
Misc\strut
\end{minipage} & \begin{minipage}[t]{0.12\columnwidth}\raggedleft\strut
0.077\strut
\end{minipage} & \begin{minipage}[t]{0.11\columnwidth}\raggedleft\strut
3,923\strut
\end{minipage} & \begin{minipage}[t]{0.11\columnwidth}\raggedleft\strut
26\strut
\end{minipage} & \begin{minipage}[t]{0.12\columnwidth}\raggedleft\strut
0.341\strut
\end{minipage} & \begin{minipage}[t]{0.11\columnwidth}\raggedleft\strut
87,021\strut
\end{minipage} & \begin{minipage}[t]{0.11\columnwidth}\raggedleft\strut
179\strut
\end{minipage}\tabularnewline
\begin{minipage}[t]{0.10\columnwidth}\raggedright\strut
School\strut
\end{minipage} & \begin{minipage}[t]{0.12\columnwidth}\raggedleft\strut
0.488\strut
\end{minipage} & \begin{minipage}[t]{0.11\columnwidth}\raggedleft\strut
11,597\strut
\end{minipage} & \begin{minipage}[t]{0.11\columnwidth}\raggedleft\strut
168\strut
\end{minipage} & \begin{minipage}[t]{0.12\columnwidth}\raggedleft\strut
2.008\strut
\end{minipage} & \begin{minipage}[t]{0.11\columnwidth}\raggedleft\strut
85,140\strut
\end{minipage} & \begin{minipage}[t]{0.11\columnwidth}\raggedleft\strut
1,013\strut
\end{minipage}\tabularnewline
\begin{minipage}[t]{0.10\columnwidth}\raggedright\strut
Town\strut
\end{minipage} & \begin{minipage}[t]{0.12\columnwidth}\raggedleft\strut
0.091\strut
\end{minipage} & \begin{minipage}[t]{0.11\columnwidth}\raggedleft\strut
2,338\strut
\end{minipage} & \begin{minipage}[t]{0.11\columnwidth}\raggedleft\strut
44\strut
\end{minipage} & \begin{minipage}[t]{0.12\columnwidth}\raggedleft\strut
0.261\strut
\end{minipage} & \begin{minipage}[t]{0.11\columnwidth}\raggedleft\strut
9,490\strut
\end{minipage} & \begin{minipage}[t]{0.11\columnwidth}\raggedleft\strut
88\strut
\end{minipage}\tabularnewline
\begin{minipage}[t]{0.10\columnwidth}\raggedright\strut
Total\strut
\end{minipage} & \begin{minipage}[t]{0.12\columnwidth}\raggedleft\strut
0.517\strut
\end{minipage} & \begin{minipage}[t]{0.11\columnwidth}\raggedleft\strut
10,194\strut
\end{minipage} & \begin{minipage}[t]{0.11\columnwidth}\raggedleft\strut
327\strut
\end{minipage} & \begin{minipage}[t]{0.12\columnwidth}\raggedleft\strut
2.093\strut
\end{minipage} & \begin{minipage}[t]{0.11\columnwidth}\raggedleft\strut
41,458\strut
\end{minipage} & \begin{minipage}[t]{0.11\columnwidth}\raggedleft\strut
2,462\strut
\end{minipage}\tabularnewline
\bottomrule
\end{longtable}

R Code for Claims Summary by Entity Type and Alarm Credit Category

\hypertarget{display.RateAlarmCredit.2}{}
\begin{verbatim}
#Claims Summary by Entity Type and Alarm Credit
ByVarSumm<-function(datasub){
  tempA <- summaryBy(Freq    ~ AC00 , data = datasub,
                     FUN = function(x) { c(m = mean(x), num=length(x)) } )
  datasub1 <-  subset(datasub, yAvg>0)
  if(nrow(datasub1)==0) { n<-nrow(datasub)
    return(c(0,0,n))
  } else
  {
    tempB <- summaryBy(yAvg   ~ AC00, data = datasub1,
                       FUN = function(x) { c(m = mean(x)) } )
    tempC <- merge(tempA,tempB,all.x=T)[c(2,4,3)]
    tempC1 <- as.matrix(tempC)
    return(tempC1)
  }
}
AlarmC <- 1*(Insample$AC00==1) + 2*(Insample$AC05==1)+ 3*(Insample$AC10==1)+ 4*(Insample$AC15==1)
ByVarCredit<-function(ACnum){
datasub <-  subset(Insample, TypeVillage == 1 & AlarmC == ACnum);
  t1 <- ByVarSumm(datasub)
datasub <-  subset(Insample, TypeCity == 1 & AlarmC == ACnum);
  t2 <- ByVarSumm(datasub)
datasub <-  subset(Insample, TypeCounty == 1 & AlarmC == ACnum);
  t3 <- ByVarSumm(datasub)
datasub <-  subset(Insample, TypeMisc == 1 & AlarmC == ACnum);
  t4 <- ByVarSumm(datasub)
datasub <-  subset(Insample, TypeSchool == 1 & AlarmC == ACnum);
  t5 <- ByVarSumm(datasub)
datasub <-  subset(Insample, TypeTown == 1 & AlarmC ==ACnum);
  t6 <- ByVarSumm(datasub)
datasub <-  subset(Insample, AlarmC == ACnum);
  t7 <- ByVarSumm(datasub)
Tablea <- rbind(t1,t2,t3,t4,t5,t6,t7)
Tableaa <- round(Tablea,3)
Rowlable <- rbind("Village","City","County","Misc","School",
                  "Town","Total")
Table4 <- cbind(Rowlable,as.matrix(Tableaa))
}
Table4a <- ByVarCredit(1)    #Claims Summary by Entity Type and Alarm Credit==00
Table4b <- ByVarCredit(2)    #Claims Summary by Entity Type and Alarm Credit==05
Table4c <- ByVarCredit(3)    #Claims Summary by Entity Type and Alarm Credit==10
Table4d <- ByVarCredit(4)    #Claims Summary by Entity Type and Alarm Credit==15
\end{verbatim}

\subsection{Fund Operations}\label{fund-operations}

We have now seen the Fund's two outcome variables, a count variable for
the number of claims and a continuous variable for the claims amount. We
have also introduced a continuous rating variable, coverage, discrete
quantitative variable, (logarithmic) deductibles, two binary rating
variable, no claims credit and fire class, as well as two categorical
rating variables, entity type and alarm credit. Subsequent chapters will
explain how to analyze and model the distribution of these variables and
their relationships. Before getting into these technical details, let us
first think about where we want to go. General insurance company
functional areas are described in Section \ref{S:PredModApps}; let us
now think about how these areas might apply in the context of the
property fund.

\subsubsection*{Initiating Insurance}\label{initiating-insurance-1}
\addcontentsline{toc}{subsubsection}{Initiating Insurance}

Because this is a government sponsored fund, we do not have to worry
about selecting good or avoiding poor risks; the fund is not allowed to
deny a coverage application from a qualified local government entity. If
we do not have to underwrite, what about how much to charge?

We might look at the most recent experience in 2010, where the total
fund claims were approximately 28.16 million USD
(\(=1377 \text{ claims} \times 20452 \text{ average severity}\)).
Dividing that among 1,110 policyholders, that suggests a rate of 24,370
( \(\approx\) 28,160,000/1110). However, 2010 was a bad year; using the
same method, our premium would be much lower based on 2009 data. This
swing in premiums would defeat the primary purpose of the fund, to allow
for a steady charge that local property managers could utilize in their
budgets.

Having a single price for all policyholders is nice but hardly seems
fair. For example, Table \ref{tab:ClaimRateVar} suggests that Schools
have much higher claims than other entities and so should pay more.
However, simply doing the calculation on an entity by entity basis is
not right either. For example, we saw in Table \ref{tab:RateAlarmCredit}
that had we used this strategy, entities with a 15\% alarm credit (for
good behavior, having top alarm systems) would actually wind up paying
more.

So, we have the data for thinking about the appropriate rates to charge
but will need to dig deeper into the analysis. We will explore this
topic further in Chapter 6 on \emph{premium calculation fundamentals}.
Selecting appropriate risks is introduced in Chapter 7 on \emph{risk
classification}.

\subsubsection*{Renewing Insurance}\label{renewing-insurance-1}
\addcontentsline{toc}{subsubsection}{Renewing Insurance}

Although property insurance is typically a one-year contract, Table
\ref{tab:CoverageBCIM} suggests that policyholders tend to renew; this
is typical of general insurance. For renewing policyholders, in addition
to their rating variables we have their claims history and this claims
history can be a good predictor of future claims. For example, Table
\ref{tab:CoverageBCIM} shows that policyholders without a claim in the
last two years had much lower claim frequencies than those with at least
one accident (0.310 compared to 1.501); a lower predicted frequency
typically results in a lower premium. This is why it is common for
insurers to use variables such as \({\tt NoClaimCredit}\) in their
rating. We will explore this topic further in Chapter 8 on
\emph{experience rating}.

\subsubsection*{Claims Management}\label{claims-management}
\addcontentsline{toc}{subsubsection}{Claims Management}

Of course, the main story line of 2010 experience was the large claim of
over 12 million USD, nearly half the claims for that year. Are there
ways that this could have been prevented or mitigated? Are their ways
for the fund to purchase protection against such large unusual events?
Another unusual feature of the 2010 experience noted earlier was the
very large frequency of claims (239) for one policyholder. Given that
there were only 1,377 claims that year, this means that a single
policyholder had 17.4 \% of the claims. This also suggestions
opportunities for managing claims, the subject of Chapter 9.

\subsubsection*{Loss Reserving}\label{loss-reserving}
\addcontentsline{toc}{subsubsection}{Loss Reserving}

In our case study, we look only at the one year outcomes of closed
claims (the opposite of open). However, like many lines of insurance,
obligations from insured events to buildings such as fire, hail, and the
like, are not known immediately and may develop over time. Other lines
of business, including those were there are injuries to people, take
much longer to develop. Chapter 10 introduces this concern and
\emph{loss reserving}, the discipline of determining how much the
insurance company should retain to meet its obligations.

\section{Further Resources and
Contributors}\label{Intro-further-reading-and-resources}

\subsubsection*{Contributor}\label{contributor}
\addcontentsline{toc}{subsubsection}{Contributor}

\begin{itemize}
\tightlist
\item
  \textbf{Edward W. (Jed) Frees}, University of Wisconsin-Madison, is
  the principal author of the initital version of this chapter. Email:
  \href{mailto:jfrees@bus.wisc.edu}{\nolinkurl{jfrees@bus.wisc.edu}} for
  chapter comments and suggested improvements.
\end{itemize}

This book introduces loss data analytic tools that are most relevant to
actuaries and other financial risk analysts. Here are a few reference
cited in the chapter.

\chapter{Frequency Modeling}\label{C:Frequency-Modeling}

\emph{Chapter Preview.} A primary focus for insurers is estimating the
magnitude of aggregate claims it must bear under its insurance
contracts. Aggregate claims are affected by both the frequency of
insured events and the severity of the insured event. Decomposing
aggregate claims into these two components, which each warrant
significant attention, is essential for analysis and pricing. This
chapter discusses frequency distributions, measures, and parameter
estimation techniques.

\section{Frequency Distributions}\label{S:frequency-distributions}

\subsection{How Frequency Augments Severity
Information}\label{S:how-frequency-augments-severity-information}

\subsubsection{Basic Terminology}\label{S:basic-terminology}

We use \textbf{claim} to denote the indemnification upon the occurrence
of an insured event. While some authors use claim and loss
interchangeably, others think of loss as the amount suffered by the
insured whereas claim is the amount paid by the insurer.
\textbf{Frequency} represents how often an insured event occurs,
typically within a policy contract. Here, we focus on count random
variables that represent the number of claims, that is, how frequently
an event occurs. \textbf{Severity} denotes the amount, or size, of each
payment for an insured event. In future chapters, the aggregate model,
which combines frequency models with severity models, is examined.

\subsubsection{The Importance of
Frequency}\label{S:the-importance-of-frequency}

Recall from Chapter \ref{C:Intro} that setting the price of an insurance
good can be a complex problem. In manufacturing, the cost of a good is
(relatively) known. In other financial service areas, market prices are
available. In insurance, we can generalize the price setting as follows:
start with an expected cost. Add ``margins'' to account for the
product's riskiness, expenses incurred in servicing the product, and a
profit/surplus allowance for the insurer.

That expected cost for insurance can be defined as the expected number
of claims times the expected amount per claim, that is, expected
\emph{frequency times severity}. The focus on claim count allows the
insurer to consider those factors which directly affect the occurrence
of a loss, thereby potentially generating a claim. The frequency process
can then be modeled.

\subsubsection{Why Examine Frequency
Information}\label{S:why-examine-frequency-information}

Insurers and other stakeholders, including governmental organizations,
have various motivations for gathering and maintaining frequency
datasets.

\begin{itemize}
\tightlist
\item
  \textbf{Contractual} - In insurance contracts, it is common for
  particular deductibles and policy limits to be listed and invoked for
  each occurrence of an insured event. Correspondingly, the claim count
  data generated would indicate the number of claims which meet these
  criteria, offering a unique claim frequency measure. Extending this,
  models of total insured losses would need to account for deductibles
  and policy limits for each insured event.
\item
  \textbf{Behaviorial} - In considering factors that influence loss
  frequency, the risk-taking and risk-reducing behavior of individuals
  and companies should be considered. Explanatory (rating) variables can
  have different effects on models of how often an event occurs in
  contrast to the size of the event.

  \begin{itemize}
  \tightlist
  \item
    In healthcare, the decision to utilize healthcare by individuals,
    and minimize such healthcare utilization through preventive care and
    wellness measures, is related primarily to his or her personal
    characteristics. The cost per user is determined by those personal,
    the medical condition, potential treatment measures, and decisions
    made by the healthcare provider (such as the physician) and the
    patient. While there is overlap in those factors and how they affect
    total healthcare costs, attention can be focused on those separate
    drivers of healthcare visit frequency and healthcare cost severity.
  \item
    In personal lines, prior claims history is an important underwriting
    factor. It is common to use an indicator of whether or not the
    insured had a claim within a certain time period prior to the
    contract.
  \item
    In homeowners insurance, in modeling potential loss frequency, the
    insurer could consider loss prevention measures that the homeowner
    has adopted, such as visible security systems. Separately, when
    modeling loss severity, the insurer would examine those factors that
    affect repair and replacement costs.
  \end{itemize}
\item
  \textbf{Databases}. Many insurers keep separate data files that
  suggest developing separate frequency and severity models. For
  example, a policyholder file is established when a policy is written.
  This file records much underwriting information about the insured(s),
  such as age, gender, and prior claims experience, policy information
  such as coverage, deductibles and limitations, as well as the
  insurance claims event. A separate file, known as the ``claims'' file,
  records details of the claim against the insurer, including the
  amount. (There may also be a ``payments'' file that records the timing
  of the payments although we shall not deal with that here.) This
  recording process makes it natural for insurers to model the frequency
  and severity as separate processes.
\item
  \textbf{Regulatory and Administrative} Insurance is a highly regulated
  and monitored industry, given its importance in providing financial
  security to individuals and companies facing risk. As part of its
  duties, regulators routinely require the reporting of claims numbers
  as well as amounts. This may be due to the fact that there can be
  alternative definitions of an ``amount,'' e.g., paid versus incurred,
  and there is less potential error when reporting claim numbers. This
  continual monitoring helps ensure financial stability of these
  insurance companies.
\end{itemize}

\section{Basic Frequency
Distributions}\label{S:basic-frequency-distributions}

In this section, we will introduce the distributions that are commonly
used in actuarial practice to model count data. The claim count random
variable is denoted by \(N\); by its very nature it assumes only
non-negative integral values. Hence the distributions below are all
discrete distributions supported on the set of non-negative integers
(\(\mathbb{Z}^+\)).

\subsection{Foundations}\label{S:foundations}

Since \(N\) is a discrete random variable taking values in
\(\mathbb{Z}^+\), the most natural full description of its distribution
is through the specification of the probabilities with which it assumes
each of the non-negative integral values. This leads us to the concept
of the \textbf{probability mass function} (\emph{pmf}) of \(N\), denoted
as \(p_N(\cdot)\) and defined as follows:

\begin{equation}
p_N(k)=\Pr(N=k), \quad \hbox{for } k=0,1,\ldots
\end{equation}

We note that there are alternate complete descriptions, or
characterizations, of the distribution of \(N\); for example, the
\textbf{distribution function} of \(N\) denoted by \(F_N(\cdot)\) and
defined below is another such:

\begin{equation}
F_N(x):=\begin{cases}
\sum\limits_{k=0}^{\lfloor x \rfloor } \Pr(N=k), &x\geq 0;\\
0, & \hbox{otherwise}.
\end{cases}
\end{equation}

In the above, \(\lfloor \cdot \rfloor\) denotes the floor function;
\(\lfloor x \rfloor\) denotes the greatest integer less than or equal to
\(x\). We note that the \textbf{survival function} of \(N\), denoted by
\(S_N(\cdot)\), is defined as the ones'-complement of \(F_N(\cdot)\),
\emph{i.e.} \(S_N(\cdot):=1-F_N(\cdot)\). Clearly, the latter is another
characterization of the distribution of \(N\).

Often one is interested in quantifying a certain aspect of the
distribution and not in its complete description. This is particularly
useful when comparing distributions. A \emph{center of location} of the
distribution is one such aspect, and there are many different measures
that are commonly used to quantify it. Of these, the \textbf{mean} is
the most popular; the mean of \(N\), denoted by \(\mu_N\)\footnote{For
  example, if there are 3 risk factors each of which the number of
  levels are 2, 3 and 4, respectively, we have
  \(k=(2-1)\times(3-1)\times (4-1)=6\).}, is defined as

\begin{equation}
\mu_N=\sum_{k=0}^\infty kp_N(k).
\end{equation}

We note that \(\mu_N\) is the expected value of the random variable
\(N\), \emph{i.e.} \(\mu_N=\mathrm{E}~N\). This leads to a general class
of measures, the **moments*8 of the distribution; the \(r\)-th moment of
\(N\), for \(r> 0\), is defined as \(\mathrm{E}{N^r}\) and denoted by
\(\mu_N'(r)\). Hence, for \(r>0\), we have\\

\begin{equation}
\mu_N(r)= \mathrm{E}{N^r}= \sum_{k=0}^\infty k^r p_N(k).
\end{equation}

We note that \(\mu_N'(\cdot)\) is a well-defined non-decreasing function
taking values in \([0,\infty)\), as \(\Pr(N\in\mathbb{Z}^+)=1\); also,
note that \(\mu_N=\mu_N'(1)\).

Another basic aspect of a distribution is its dispersion, and of the
various measures of dispersion studied in the literature, the
\textbf{standard deviation} is the most popular. Towards defining it, we
first define the \textbf{variance} of \(N\), denoted by
\(\mathrm{Var}~N\), as \(\mathrm{Var}~N:=\mathrm{E}{(N-\mu_N)^2}\) when
\(\mu_N\) is finite. By basic properties of the expected value of a
random variable, we see that
\(\mathrm{Var}~N:=\mathrm{E}~{N^2}-(\mathrm{E}~N)^2\). The standard
deviation of \(N\), denoted by \(\sigma_N\), is defined as the
square-root of \(\mathrm{Var}~N\). Note that the latter is well-defined
as \(\mathrm{Var}~N\), by its definition as the average squared
deviation from the mean, is non-negative; \(\mathrm{Var}~N\) is denoted
by \(\sigma_N^2\). Note that these two measures take values in
\([0,\infty)\).

\subsection{Moment and Probability Generating
Functions}\label{S:generating-functions}

Now we will introduce two generating functions that are found to be
useful when working with count variables. Recall that the \textbf{moment
generating function} (mgf) of \(N\), denoted as \(M_N(\cdot)\), is
defined as \[
M_N(t) = \mathrm{E}~{e^{tN}} = \sum^{\infty}_{k=0} ~e^{tk}~ p_N(k), \quad t\in \mathbb{R}.
\] We note that while \(M_N(\cdot)\) is well defined as it is the
expectation of a non-negative random variable (\(e^{tN}\)), though it
can assume the value \(\infty\). Note that for a count random variable,
\(M_N(\cdot)\) is finite valued on \((-\infty,0]\) with \(M_N(0)=1\).
The following theorem, whose proof can be found in \citep{billingsley}
(pages 285-6), encapsulates the reason for its name.

\BeginKnitrBlock{theorem}
\protect\hypertarget{thm:freq-thm1}{}{\label{thm:freq-thm1} }Let \(N\) be a
count random variable such that \(\mathrm{E}~{e^{t^\ast N}}\) is finite
for some \(t^\ast>0\). We have the following:

All moment of \(N\) are finite, \emph{i.e.} \[
\mathrm{E}{N^r}<\infty, \quad r\geq 0.
\] The \emph{mgf} can be used to \emph{generate} its moments as follows:
\[
\left.\frac{{\rm d}^m}{{\rm d}t^m} M_N(t)\right\vert_{t=0}=\mathrm{E}{N^m}, \quad m\geq 1.
\] The \emph{mgf} \(M_N(\cdot)\) characterizes the distribution; in
other words it uniquely specifies the distribution.
\EndKnitrBlock{theorem}

Another reason that the \emph{mgf} is very useful as a tool is that for
two independent random variables \(X\) and \(Y\), with their mgfs
existing in a neighborhood of \(0\), the \emph{mgf} of \(X+Y\) is the
product of their respective mgfs.

A related generating function to the \emph{mgf} is called the
\textbf{probability generating function} (\emph{pgf}), and is a useful
tool for random variables taking values in \(\mathbb{Z}^+\). For a
random variable \(N\), by \(P_N(\cdot)\) we denote its \emph{pgf} and we
define it as follows:

\begin{equation}
P_N(s):=\mathrm{E}~{s^N}, \quad s\geq 0.
\end{equation}

It is straightforward to see that if the \emph{mgf} \(M_N(\cdot)\)
exists on \((-\infty,t^\ast)\) then \[
P_N(s)=M_N(\log(s)), \quad s<e^{t^\ast}.
\] Moreover, if the \emph{pgf} exists on an interval \([0,s^\ast)\) with
\(s^\ast>1\), then the \emph{mgf} \(M_N(\cdot)\) exists on
\((-\infty,\log(s^\ast))\), and hence uniquely specifies the
distribution of \(N\) by Theorem \ref{thm:freq-thm1}. The following
result for \emph{pgf} is an analog of Theorem \ref{thm:freq-thm1}, and
in particular justifies its name.

\BeginKnitrBlock{theorem}
\protect\hypertarget{thm:pgfthm}{}{\label{thm:pgfthm} }Let \(N\) be a count
random variable such that \(\mathrm{E}~{(s^{\ast})^N}\) is finite for
some \(s^\ast>1\). We have the following:

All moment of \(N\) are finite, \emph{i.e.} \[
\mathrm{E}~{N^r}<\infty, \quad r\geq 0.
\] The \emph{pmf} of \(N\) can be derived from the \emph{pgf} as
follows: \[
p_N(m)=\begin{cases}
P_N(0), &m=0;\cr
&\cr
\left(\frac{1}{m!}\right) \left.\frac{{\rm d}^m}{{\rm d}s^m} P_N(s)\right\vert_{s=0}\;, &m\geq 1.\cr
\end{cases}
\] The factorial moments of \(N\) can be derived as follows: \[
\left.\frac{{\rm d}^m}{{\rm d}s^m} P_N(s)\right\vert_{s=1}=\mathrm{E}~{\prod\limits_{i=0}^{m-1} (N-i)}, \quad m\geq 1.
\] The \emph{pgf} \(P_N(\cdot)\) characterizes the distribution; in
other words it uniquely specifies the distribution.
\EndKnitrBlock{theorem}

\subsection{Important Frequency
Distributions}\label{S:important-frequency-distributions}

In this sub-section we will study three important frequency
distributions used in Statistics, namely the Binomial, the Negative
Binomial and the Poisson distributions. In the following, a risk denotes
a unit covered by insurance. A risk could be an individual, a building,
a company, or some other identifier for which insurance coverage is
provided. For context, imagine an insurance data set containing the
number of claims by risk or stratified in some other manner. The above
mentioned distributions also happen to be the most commonly used in
insurance practice for reasons, some of which we mention below.

\begin{itemize}
\tightlist
\item
  These distributions can be motivated by natural random experiments
  which are good approximations to real life processes from which many
  insurance data arise. Hence, not surprisingly, they together offer a
  reasonable fit to many insurance data sets of interest. The
  appropriateness of a particular distribution for the set of data can
  be determined using standard statistical methodologies, as we discuss
  later in this chapter.
\item
  They provide a rich enough basis for generating other distributions
  that even better approximate or well cater to more real situations of
  interest to us.

  \begin{itemize}
  \tightlist
  \item
    The three distributions are either one-parameter or two-parameter
    distributions. In fitting to data, a parameter is assigned a
    particular value. The set of these distributions can be enlarged to
    their convex hulls by treating the parameter(s) as a random variable
    (or vector) with its own probability distribution, with this larger
    set of distributions offering greater flexibility. A simple example
    that is better addressed by such an enlargement is a portfolio of
    claims generated by insureds belonging to many different risk
    classes.
  \item
    In insurance data, we may observe either a marginal or inordinate
    number of zeros, \emph{i.e.} zero claims by risk. When fitting to
    the data, a frequency distribution in its standard specification
    often fails to reasonably account for this occurrence. The natural
    modification of the above three distributions, however, accommodate
    this phenomenon well towards offering a better fit.
  \item
    In insurance we are interested in total claims paid, whose
    distribution results from compounding the fitted frequency
    distribution with a severity distribution. These three distributions
    have properties that make it easy to work with the resulting
    aggregate severity distribution.
  \end{itemize}
\end{itemize}

\subsubsection{Binomial Distribution}\label{S:binomial-distribution}

We begin with the binomial distribution which arises from any finite
sequence of identical and independent experiments with binary outcomes.
The most canonical of such experiments is the (biased or unbiased) coin
tossing experiment with the outcome being heads or tails. So if \(N\)
denotes the number of heads in a sequence of \(m\) independent coin
tossing experiments with an identical coin which turns heads up with
probability \(q\), then the distribution of \(N\) is called the binomial
distribution with parameters \((m,q)\), with \(m\) a positive integer
and \(q\in[0,1]\). Note that when \(q=0\) (resp., \(q=1\)) then the
distribution is degenerate with \(N=0\) (resp., \(N=m\)) with
probability \(1\). Clearly, its support when \(q\in(0,1)\) equals
\(\{0,1,\ldots,m\}\) with \emph{pmf} given by \footnote{Preferring the
  multiplicative form to others (e.g., additive one) was already hinted
  in \eqref{eq:log-lin-mu}.}

\begin{equation*}
p_k:= \binom{m}{k} q^k (1-q)^{m-k}, \quad k=0,\ldots,m.
\end{equation*}

The reason for its name is that the \emph{pmf} takes values among the
terms that arise from the binomial expansion of \((q +(1-q))^m\). This
realization then leads to the the following expression for the
\emph{pgf} of the binomial distribution: \[
P(z):= \sum_{k=0}^m z^k \binom{m}{k} q^k (1-q)^{m-k} = \sum_{k=0}^m  \binom{m}{k} (zq)^k (1-q)^{m-k} = (qz+(1-q))^m = (1+q(z-1))^m.
\] Note that the above expression for the \emph{pgf} confirms the fact
that the binomial distribution is the \(m\)-convolution of the Bernoulli
distribution, which is the binomial distribution with \(m=1\) and
\emph{pgf} \((1+q(z-1))\). Also, note that the \emph{mgf} of the
binomial distribution is given by \((1+q(e^t-1))^m\).

The central moments of the binomial distribution can be found in a few
different ways. To emphasize the key property that it is a
\(m\)-convolution of the Bernoulli distribution, we derive below the
moments using this property. We begin by observing that the Bernoulli
distribution with parameter \(q\) assigns probability of \(q\) and
\(1-q\) to \(1\) and \(0\), respectively. So its mean equals \(q\)
(\(=0\times (1-q) + 1\times q\)); note that its raw second moment equals
its mean as \(N^2=N\) with probability \(1\). Using these two facts we
see that the variance equals \(q(1-q)\). Moving on to the Binomial
distribution with parameters \(m\) and \(q\), using the fact that it is
the \(m\)-convolution of the Bernoulli distribution, we write \(N\) as
the sum of \(N_1,\ldots,N_m\), where \(N_i\) are \emph{iid} Bernoulli
variates. Now using the moments of Bernoulli and linearity of the
expectation, we see that \[
\mathrm{E}~{N}=\mathrm{E}~{\sum_{i=1}^m N_i} = \sum_{i=1}^m ~\mathrm{E}~{N_i} = mq.
\] Also, using the fact that the variance of the sum of independent
random variables is the sum of their variances, we see that\\
\[
\mathrm{Var}~{N}=\mathrm{Var}~\left({\sum_{i=1}^m N_i}\right)=\sum_{i=1}^m \mathrm{Var}~{N_i} = mq(1-q).
\] Alternate derivations of the above moments are suggested in the
exercises. One important observation, especially from the point of view
of applications, is that the mean is greater than the variance unless
\(q=0\).

\subsubsection{Poisson Distribution}\label{S:poisson-distribution}

After the Binomial distribution, the Poisson distribution (named after
the French polymath Sim'eon Denis Poisson) is probably the most well
known of discrete distributions. This is partly due to the fact that it
arises naturally as the distribution of the count of the random
occurrences of a type of event in a certain time period, if the rate of
occurrences of such events is a constant. Relatedly, it also arises as
the asymptotic limit of the Binomial distribution with
\(m\rightarrow \infty\) and \(mq\rightarrow \lambda\).

The Poisson distribution is parametrized by a single parameter usually
denoted by \(\lambda\) which takes values in \((0,\infty)\). Its
\emph{pmf} is given by \[
p_k = \frac{e^{-\lambda}\lambda^k}{k!}, k=0,1,\ldots
\] It is easy to check that the above specifies a \emph{pmf} as the
terms are clearly non-negative, and that they sum to one follows from
the infinite Taylor series expansion of \(e^\lambda\). More generally,
we can derive its \emph{pgf}, \(P(\cdot)\), as follows: \[
P(z):= \sum_{k=0}^\infty p_k z^k = \sum_{k=0}^\infty  \frac{e^{-\lambda}\lambda^kz^k}{k!} = e^{-\lambda} e^{\lambda z}
= e^{\lambda(z-1)}, \forall z\in\mathbb{R}.
\] From the above, we derive its \emph{mgf} as follows: \[
M(t)=P(e^t)=e^{\lambda(e^t-1)}, t\in \mathbb{R}.
\] Towards deriving its mean, we note that for the Poisson distribution
\[
kp_k=\begin{cases}
0,  &k=0;\cr
\lambda p_{k-1}, &k\geq1;
\end{cases}
\] this can be checked easily. In particular, this implies that \[
\mathrm{E}~{N}=\sum_{k\geq 0} k~p_k =\lambda\sum_{k\geq 1} p_{k-1} = \lambda\sum_{j\geq 0} p_{j} =\lambda.
\] In fact, more generally, using either a generalization of the above
or using Theorem \ref{thm:pgfthm}, we see that \[
\mathrm{E}{\prod\limits_{i=0}^{m-1} (N-i)}=\left.\frac{{\rm d}^m}{{\rm d}s^m} P_N(s)\right\vert_{s=1}=\lambda^m, \quad m\geq 1.
\] This, in particular, implies that \[
\mathrm{Var}~{N}=\mathrm{E}~{N^2}-(\mathrm{E}~{N})^2 = \mathrm{E}~{N(N-1)}+\mathrm{E}~{N}-(\mathrm{E}~{N})^2=\lambda^2+\lambda-\lambda^2=\lambda.
\] Note that interestingly for the Poisson distribution
\(\mathrm{Var}~{N}=\mathrm{E}~{N}\).

\subsubsection{Negative Binomial
Distribution}\label{S:negative-binomial-distribution}

The third important count distribution is the Negative Binomial
distribution. Recall that the Binomial distribution arose as the
distribution of the number of \emph{successes} in \(m\) independent
repetition of an experiment with binary outcomes. If we instead consider
the number of \emph{successes} until we observe the \(r\)-th
\emph{failure} in independent repetitions of an experiment with binary
outcomes, then its distribution is a Negative Binomial distribution. A
particular case, when \(r=1\), is the geometric distribution. In the
following we will allow the parameter \(r\) to be any positive real, and
unfortunately when \(r\) in not an integer the above random experiment
would not be applicable. To then motivate the distribution more
generally, and in the process explain its name, we recall the binomial
series, \emph{i.e.} \[
(1+x)^s= 1 + s x + \frac{s(s-1)}{2!}x^2 + \ldots..., \quad s\in\mathbb{R}; \vert x \vert<1.
\] If we define \(\binom{s}{k}\), the generalized binomial coefficient,
by \[
\binom{s}{k}=\frac{s(s-1)\cdots(s-k+1)}{k!},
\] then we have \[
(1+x)^s= \sum_{k=0}^{\infty} \binom{s}{k} x^k, \quad s\in\mathbb{R}; \vert x \vert<1.
\] If we let \(s=-r\), then we see that the above yields \[
(1-x)^{-r}= 1 + r x + \frac{(r+1)r}{2!}x^2 + \ldots...= \sum_{k=0}^\infty \binom{r+k-1}{k} x^k, \quad r\in\mathbb{R}; \vert x \vert<1.
\] This implies that if we define \(p_k\) as \[
p_k = \binom{k+r-1}{k} \left(\frac{1}{1+\beta}\right)^r \left(\frac{\beta}{1+\beta}\right)^k, \quad k=0,1,\ldots
\] for \(r>0\) and \(\beta>=0\), then it defines a valid \emph{pmf}.
Such defined distribution is called the negative binomial distribution
with parameters \((r,\beta)\) with \(r>0\) and \(\beta\geq 0\).
Moreover, the binomial series also implies that the \emph{pgf} of this
distribution is given by \[
\begin{aligned}
  P(z) &= (1-\beta(z-1))^{-r}, \quad \vert z \vert \leq 1+\frac{1}{\beta}, \beta\geq0.
\end{aligned}
\] The above implies that the \emph{mgf} is given by \[
\begin{aligned}
  M(t) &= (1-\beta(e^t-1))^{-r}, \quad t \leq \log\left(1+\frac{1}{\beta}\right), \beta\geq0.
\end{aligned}
\] We derive its moments using Theorem \ref{thm:freq-thm1} as follows:

\begin{eqnarray*}
\mathrm{E}~{N}&=&M'(0)= \left. r\beta e^t (1-\beta(e^t-1))^{-r-1}\right\vert_{t=0}=r\beta;\\
\mathrm{E}~{N^2}&=&M''(0)= \left.\left[ r\beta e^t (1-\beta(e^t-1))^{-r-1} + r(r+1)\beta^2 e^{2t} (1-\beta(e^t-1))^{-r-2}\right]\right\vert_{t=0}\\
&=&r\beta(1+\beta)+r^2\beta^2;\\
\hbox{and }\mathrm{E}{N}&=&\mathrm{E}{N^2}-(\mathrm{E}{N})^2=r\beta(1+\beta)+r^2\beta^2-r^2\beta^2=r\beta(1+\beta)
\end{eqnarray*}

We note that when \(\beta>0\), we have
\(\mathrm{Var}~{N} >\mathrm{E}~{N}\). In other words, this distribution
is \textbf{overdispersed} (relative to the Poisson); similarly, when
\(q>0\) the binomial distribution is said to be \textbf{underdispersed}
(relative to the Poisson).

Finally, we observe that the Poisson distribution also emerges as a
limit of negative binomial distributions. Towards establishing this, let
\(\beta_r\) be such that as \(r\) approaches infinity \(r\beta_r\)
approaches \(\lambda>0\). Then we see that the mgfs of negative binomial
distributions with parameters \((r,\beta_r)\) satisfies \[
\lim_{r\rightarrow 0} (1-\beta_r(e^t-1))^{-r}=\exp\{\lambda(e^t-1)\},
\] with the right hand side of the above equation being the \emph{mgf}
of the Poisson distribution with parameter \(\lambda\)\footnote{corresponding
  to \(\texttt{VAgecat1}\)}

\section{The (a, b, 0) Class}\label{S:the-a-b-0-class}

In the previous section we studied three distributions, namely the
binomial, the Poisson and the negative binomial distributions. In the
case of the Poisson, to derive its mean we used the the fact that \[
kp_k=\lambda p_{k-1}, \quad k\geq 1,
\] which can be expressed equivalently as \[
\frac{p_k}{p_{k-1}}=\frac{\lambda}{k}, \quad k\geq 1.
\] Interestingly, we can similarly show that for the binomial
distribution \[
\frac{p_k}{p_{k-1}}=\frac{-q}{1-q}+\left(\frac{(m+1)q}{1-q}\right)\frac{1}{k}, \quad k=1,\ldots,m,
\] and that for the negative binomial distribution \[
\frac{p_k}{p_{k-1}}=\frac{\beta}{1+\beta}+\left(\frac{(r-1)\beta}{1+\beta}\right)\frac{1}{k}, \quad k\geq 1.
\] The above relationships are all of the form

\begin{equation}
\frac{p_k}{p_{k-1}}=a+\frac{b}{k}, \quad k\geq 1;
\label{eq:ab0}
\end{equation}

this raises the question if there are any other distributions which
satisfy this seemingly general recurrence relation.

To begin with let \(a<0\). In this case as \((a+b/k)\rightarrow a<0\) as
\(k\rightarrow \infty\), and the ratio on the left is non-negative, it
follows that if \(a<0\) then \(b\) should satisfy \(b=-ka\), for some
\(k\geq 1\). Any such pair \((a,b)\) can be written as \[
\left(\frac{-q}{1-q},\frac{(m+1)q}{1-q}\right), \quad q\in(0,1), m\geq 1;
\] note that the case \(a<0\) with \(a+b=0\) yields the degenerate at
\(0\) distribution which is the binomial distribution with \(q=0\) and
arbitrary \(m\geq 1\).

In the case of \(a=0\), again by non-negativity of the ratio
\(p_k/p_{k-1}\), we have \(b\geq 0\). If \(b=0\) the distribution is
degenerate at \(0\), which is a binomial with \(q=0\) or a Poisson
distribution with \(\lambda=0\) or a negative binomial distribution with
\(\beta=0\). If \(b>0\), then clearly such a distribution is a Poisson
distribution with mean (\emph{i.e.} \(\lambda\)) equal to \(b\).

In the case of \(a>0\), again by non-negativity of the ratio
\(p_k/p_{k-1}\), we have \(a+b/k\geq 0\) for all \(k\geq 1\). The most
stringent of these is the inequality \(a+b\geq 0\). Note that \(a+b=0\)
again results in degeneracy at \(0\); excluding this case we have
\(a+b>0\) or equivalently \(b=(r-1)a\) with \(r>0\). Some algebra easily
yields the following expression for \(p_k\): \[
p_k = \binom{k+r-1}{k} p_0 a^k, \quad k=1,2,\ldots.
\] The above series converges for \(a<1\) when \(r>0\), with the sum
given by \(p_0*((1-a)^{(-r)}-1)\). Hence, equating the latter to
\(1-p_0\) we get \(p_0=(1-a)^{(-r)}\). So in this case the pair
\((a,b)\) is of the form \((a,(r-1)a)\), for \(r>0\) and \(0<a<1\);
since a equivalent parametrization is
\((\beta/(1+\beta),(r-1)\beta/(1+\beta))\), for \(r>0\) and \(0<\beta\),
we see from above that such distributions are negative binomial
distributions.

From the above development we see that not only does the recurrence
\eqref{eq:ab0} tie these three distributions together, but also it
characterizes them. For this reason these three distributions are
collectively referred to in the actuarial literature as \((a,b,0)\)
class of distributions, with \(0\) referring to the starting point of
the recurrence. Note that the value of \(p_0\) is implied by \((a,b)\)
since the probabilities have to sum to one. Of course, \eqref{eq:ab0} as a
recurrence relation for \(p_k\) makes the computation of the \emph{pmf}
efficient by removing redundancies. Later, we will see that it does so
even in the case of compound distributions with the frequency
distribution belonging to the \((a,b,0)\) class - this fact is the more
important motivating reason to study these three distribution from this
viewpoint.

\textbf{Example 2.3.1.} A discrete probability distribution has the
following properties \[
\begin{aligned}
p_k&=c\left( 1+\frac{2}{k}\right) p_{k-1} \:\:\: k=1,2,3,\ldots\\
p_1&= \frac{9}{256}
\end{aligned}
\] Determine the expected value of this discrete random variable.

Show Example Solution

\hypertarget{toggleExampleFreq.3.1}{}
\textbf{Solution:} Since the \emph{pmf} satisfies the \((a,b,0)\)
recurrence relation we know that the underlying distribution is one
among the binomial, Poisson and negative binomial distributions. Since
the ratio of the parameters (\emph{i.e.} \(b/a\)) equals \(2\), we know
that it is negative binomial and that \(r=3\). Moreover, since for a
negative binomial \(p_1=r(1+\beta)^{-(r+1)}\beta\), we have \(\beta=3\).
Finally, since the mean of a negative binomial is \(r\beta\) we have the
mean of the given distribution equals \(9\).

\begin{center}\rule{0.5\linewidth}{\linethickness}\end{center}

\section{Estimating Frequency
Distributions}\label{S:estimating-frequency-distributions}

\subsection{Parameter estimation}\label{S:parameter-estimation}

In Section \ref{S:basic-frequency-distributions} we introduced three
distributions of importance in modeling various types of count data
arising from insurance. Let us now suppose that we have a set of count
data to which we wish to fit a distribution, and that we have determined
that one of these \((a,b,0)\) distributions is more appropriate than the
others. Since each one of these forms a class of distributions if we
allow its parameter(s) to take any permissible value, there remains the
task of determining the \textbf{best} value of the parameter(s) for the
data at hand. This is a statistical point estimation problem, and in
parametric inference problems the statistical inference paradigm of
\emph{maximum likelihood} usually yields efficient estimators. In this
section we will describe this paradigm and derive the maximum likelihood
estimators (\emph{mle}s).

\begin{figure}

{\centering \includegraphics[width=0.45\linewidth]{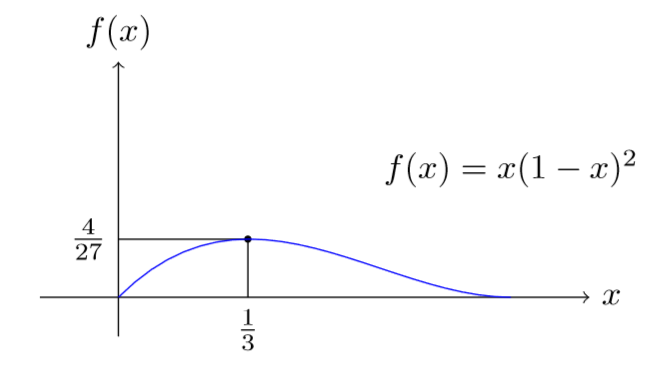}

}

\caption{Likelihood of a $(0,1,0)$ $3$-sample from Bernoulli}\label{fig:berlik}
\end{figure}

Let us suppose that we observe the \emph{iid} random variables
\(X_1,X_2,\ldots,X_n\) from a distribution with \emph{pmf} \(p_\theta\),
where \(\theta\) is an unknown value in
\(\Theta\subseteq \mathbb{R}^d\). For example, in the case of the
Poisson distribution \[
p_\theta(x)=e^{-\theta}\frac{\theta^x}{x!}, \quad x=0,1,\ldots,
\] with \(\Theta=(0,\infty)\). In the case of the binomial distribution
we have \[
p_\theta(x)=\binom{m}{x} q^x(1-q)^{m-x}, \quad x=0,1,\ldots,m,
\] with \(\theta:=(m,q)\in \{0,1,2,\ldots\}\times(0,1]\). Let us suppose
that the observations are \(x_1,\ldots,x_n\); in this case the
probability of observing this sample from \(p_\theta\) equals \[
\prod_{i=1}^n p_\theta(x_i).
\] The above, denoted by \(L(\theta)\), viewed as a function of
\(\theta\) is called the \emph{likelihood}. Note that we suppressed its
dependence on the data, to emphasize that we are viewing it as a
function of the parameter. For example, in the case of the Poisson
distribution we have \[
L(\lambda)=e^{-n\lambda} \lambda^{\sum_{i=1}^n x_i} \left(\prod_{i=1}^n x_i!\right)^{-1};
\] in the case of the binomial distribution we have \[
L(m,q)=\left(\prod_{i=1}^n \binom{m}{x_i}\right) q^{\sum_{i=1}^n x_i} (1-q)^{nm-\sum_{i=1}^n x_i} .
\] The \textbf{maximum likelihood estimator} (\emph{mle}) for \(\theta\)
is any maximizer of the likelihood; in a sense the \emph{mle} chooses
the parameter value that best explains the observed observations.
Consider a sample of size \(3\) from a Bernoulli distribution (binomial
with \(m=1\)) with values \(0,1,0\). The likelihood in this case is
easily checked to equal \[
L(q)=q(1-q)^2,
\] and the plot of the likelihood is given in Figure \ref{fig:berlik}.
As shown in the plot, the maximum value of the likelihood equals
\(4/27\) and is attained at \(q=1/3\), and hence the \emph{mle} for
\(q\) is \(1/3\) for the given sample. In this case one can resort to
algebra to show that \[
q(1-q)^2=\left(q-\frac{1}{3}\right)^2\left(q-\frac{4}{3}\right)+\frac{4}{27},
\] and conclude that the maximum equals \(4/27\), and is attained at
\(q=1/3\) (using the fact that the first term is non-positive in the
interval \([0,1]\)). But as is apparent, this way of deriving the
\emph{mle} using algebra does not generalize. In general, one resorts to
calculus to derive the \emph{mle} - note that for some likelihoods one
may have to resort to other optimization methods, especially when the
likelihood has many local extrema. It is customary to equivalently
maximize the logarithm of the likelihood\footnote{We use matrix
  derivative here.} \(L(\cdot)\), denoted by \(l(\cdot)\), and look at
the set of zeros of its first derivative\footnote{A slight benefit of
  working with \(l(\cdot)\) is that constant terms in \(L(\cdot)\) do
  not appear in \(l'(\cdot)\) whereas they do in \(L'(\cdot)\).}
\(l'(\cdot)\). In the case of the above likelihood,
\(l(q)=\log(q)+2\log(1-q)\), and \[
l'(q):=\frac{\rm d}{{\rm d}q}l(q)=\frac{1}{q}-\frac{2}{1-q}.
\] The unique zero of \(l'(\cdot)\) equals \(1/3\), and since
\(l''(\cdot)\) is negative, we have \(1/3\) is the unique maximizer of
the likelihood and hence its \emph{mle}.

\subsection{Frequency Distributions
MLE}\label{S:frequency-distributions-mle}

In the following, we derive the \emph{mle} for the three members of the
\((a,b,0)\) class. We begin by summarizing the discussion above. In the
setting of observing \emph{iid} random variables \(X_1,X_2,\ldots,X_n\)
from a distribution with \emph{pmf} \(p_\theta\), where \(\theta\) is an
unknown value in \(\Theta\subseteq \mathbb{R}^d\), the likelihood
\(L(\cdot)\), a function on \(\Theta\) is defined as \[
L(\theta):=\prod_{i=1}^n p_\theta(x_i),
\] where \(x_1,\ldots,x_n\) are the observed values. The maximum
likelihood estimator (\emph{mle}) of \(\theta\), denoted as
\(\hat{\theta}_{\rm MLE}\) is a function which maps the observations to
an element of the set of maximizers of \(L(\cdot)\), namely \[
\{\theta \vert L(\theta)=\max_{\eta\in\Theta}L(\eta)\}.
\] Note the above set is a function of the observations, even though
this dependence is not made explicit. In the case of the three
distributions that we will study, and quite generally, the above set is
a singleton with probability tending to one (with increasing sample
size). In other words, for many commonly used distributions and when the
sample size is large, the \emph{mle} is uniquely defined with high
probability.

In the following, we will assume that we have observed \(n\) \emph{iid}
random variables \(X_1,X_2,\ldots,X_n\) from the distribution under
consideration, even though the parametric value is unknown. Also,
\(x_1,x_2,\ldots,x_n\) will denote the observed values. We note that in
the case of count data, and data from discrete distributions in general,
the likelihood can alternately be represented as \[
L(\theta):=\prod_{k\geq 0} \left(p_\theta(k)\right)^{m_k},
\] where \[
m_k:= \left\vert \{i\vert x_i=k, 1\leq i \leq n\} \right\vert=\sum_{i= 1}^n I(x_i=k), \quad k\geq 0.
\] Note that this is an information loss-less transformation of the
data. For large \(n\) it leads to compression of the data in the sense
of \emph{sufficiency}. Below, we present expressions for the \emph{mle}
in terms of \(\{m_k\}_{k\geq 1}\) as well.

\textbf{MLE - Poisson Distribution:} In this case, as noted above, the
likelihood is given by \[
L(\lambda)=\left(\prod_{i=1}^n x_i!\right)^{-1}e^{-n\lambda}\lambda^{\sum_{i=1}^n x_i},
\] which implies that \[
l(\lambda)= -\sum_{i=1}^n \log(x_i!) -n\lambda +\log(\lambda) \cdot \sum_{i=1}^n x_i,
\] and \[
l'(\lambda)= -n +\frac{1}{\lambda}\sum_{i=1}^n x_i.
\] Since \(l''< 0\) if \(\sum_{i=1}^n x_i>0\), the maximum is attained
at the sample mean. In the contrary, the maximum is attained at the
least possible parameter value, that is the \emph{mle} equals zero.
Hence, we have\\
\[
\hat{\lambda}_{\rm MLE} = \frac{1}{n}\sum_{i=1}^n X_i.
\] Note that the sample mean can be computed also as \[
\frac{1}{n} \sum_{k\geq 1} km_k.
\] It is noteworthy that in the case of the Poisson, the exact
distribution of \(\hat{\lambda}_{\rm MLE}\) is available in closed form
- it is a scaled Poisson - when the underlying distribution is a
Poisson. This is so as the sum of independent Poisson random variables
is a Poisson as well. Of course, for large sample size one can use the
ordinary Central Limit Theorem (CLT) to derive a normal approximation.
Note that the latter approximation holds even if the underlying
distribution is any distribution with a finite second moment.

\textbf{MLE - Binomial Distribution:} Unlike the case of the Poisson
distribution, the parameter space in the case of the binomial is
\(2\)-dimensional. Hence the optimization problem is a bit more
challenging. We begin by observing that the likelihood is given by \[
L(m,q)= \left(\prod_{i=1}^n \binom{m}{x_i}\right) q^{\sum_{i=1}^n x_i} (1-q)^{nm-\sum_{i=1}^n x_i},
\] and the log-likelihood by \[
l(m,q)= \sum_{i=1}^n \log\left(\binom{m}{x_i}\right) + \left({\sum_{i=1}^n x_i}\right)\log(q)+ \left({nm-\sum_{i=1}^n x_i}\right)\log(1-q).
\] Note that since \(m\) takes only non-negative integral values, we
cannot use multivariate calculus to find the optimal values.
Nevertheless, we can use single variable calculus to show that

\begin{equation}
\hat{q}_{\rm MLE}\times \hat{m}_{\rm MLE}= \frac{1}{n}\sum_{i=1}^n X_i.
\label{eq:binmle}
\end{equation}

Towards this we note that for a fixed value of \(m\), \[
\frac{\delta}{\delta q} l(m,q) = \left({\sum_{i=1}^n x_i}\right)\frac{1}{q}- \left({nm-\sum_{i=1}^n x_i}\right)\frac{1}{1-q},
\] and that \[
\frac{\delta^2}{\delta q^2} l(m,q) = -\left[\left({\sum_{i=1}^n x_i}\right)\frac{1}{q^2} + \left({nm-\sum_{i=1}^n x_i}\right)\frac{1}{(1-q)^2}\right]\leq 0.
\] The above implies that for any fixed value of \(m\), the maximizing
value of \(q\) satisfies \[
mq=\frac{1}{n}\sum_{i=1}^n X_i,
\] and hence we establish equation \eqref{eq:binmle}. The above reduces
the task to the search for \(\hat{m}_{\rm MLE}\), which is member of the
set of the maximizers of

\begin{equation}
L\left(m,\frac{1}{nm}\sum_{i=1}^n x_i\right).
\label{eq:binlikm}
\end{equation}

Note the likelihood would be zero for values of \(m\) smaller than
\(\max\limits_{1\leq i \leq n}x_i\), and hence \[
\hat{m}_{\rm MLE}\geq \max_{1\leq i \leq n}x_i.
\] Towards specifying an algorithm to compute \(\hat{m}_{\rm MLE}\), we
first point out that for some data sets \(\hat{m}_{\rm MLE}\) could
equal \(\infty\), indicating that a Poisson distribution would render a
better fit than any binomial distribution. This is so as the binomial
distribution with parameters \((m,\overline{x}/m)\) approaches the
Poisson distribution with parameter \(\overline{x}\) with \(m\)
approaching infinity. The fact that some data sets will \textbf{prefer}
a Poisson distribution should not be surprising since in the above sense
the set of Poisson distribution is on the boundary of the set of
binomial distributions.

\begin{figure}

{\centering \includegraphics[width=0.8\linewidth]{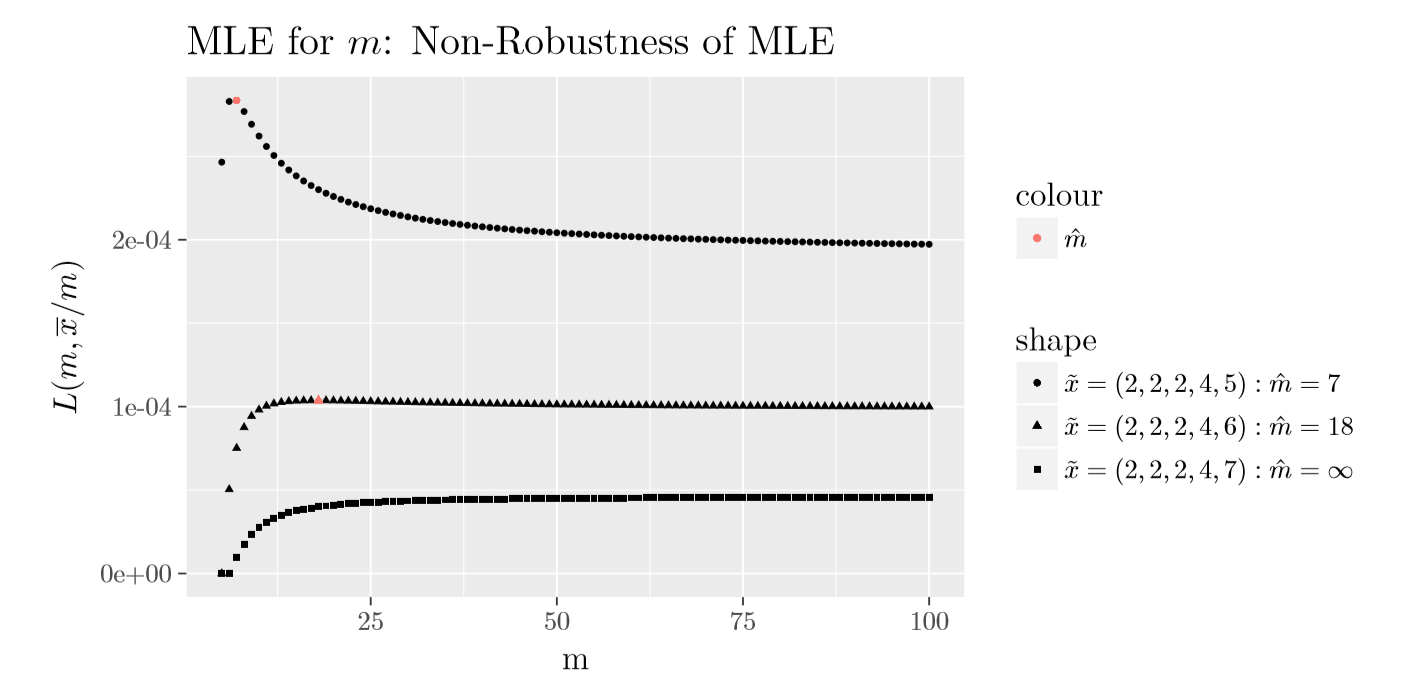}

}

\caption{Plot of $L(m,\overline{x}/m)$ for binomial distribution}\label{fig:MLEm}
\end{figure}

Interestingly, in \citep{olkin1981} they show that if the sample mean is
less than or equal to the sample variance then
\(\hat{m}_{\rm MLE}=\infty\); otherwise, there exists a finite \(m\)
that maximizes equation \eqref{eq:binlikm}. In Figure \ref{fig:MLEm} below
we display the plot of \(L\left(m,\frac{1}{nm}\sum_{i=1}^n x_i\right)\)
for three different samples of size \(5\); they differ only in the value
of the sample maximum. The first sample of \((2,2,2,4,5)\) has the ratio
of sample mean to sample variance greater than \(1\) (\(1.875\)), the
second sample of \((2,2,2,4,6)\) has the ratio equal to \(1.25\) which
is closer to \(1\), and the third sample of \((2,2,2,4,7)\) has the
ratio less than \(1\) (\(0.885\)). For these three samples, as shown in
Figure \ref{fig:MLEm}, \(\hat{m}_{\rm MLE}\) equals \(7\), \(18\) and
\(\infty\), respectively. Note that the limiting value of
\(L\left(m,\frac{1}{nm}\sum_{i=1}^n x_i\right)\) as \(m\) approaches
infinity equals

\begin{equation}
\left(\prod_{i=1}^n x_i! \right)^{-1} \exp\left\{-\sum_{i=1}^n x_i\right\} \overline{x}^{n\overline{x}}.
\label{eq:Poilik}
\end{equation}

Also, note that Figure \ref{fig:MLEm} shows that the \emph{mle} of \(m\)
is non-robust, \emph{i.e.} changes in a small proportion of the data set
can cause large changes in the estimator.

The above discussion suggests the following simple algorithm:

\begin{itemize}
\tightlist
\item
  \emph{Step 1}. If the sample mean is less than or equal to the sample
  variance, \(\hat{m}_{MLE}=\infty\). The \emph{mle} suggested
  distribution is a Poisson distribution with
  \(\hat{\lambda}=\overline{x}\).
\item
  \emph{Step 2}. If the sample mean is greater than the sample variance,
  then compute \(L(m,\overline{x}/m)\) for \(m\) values greater than or
  equal to the sample maximum until \(L(m,\overline{x}/m)\) is close to
  the value of the Poisson likelihood given in \eqref{Poilik}. The value
  of \(m\) that corresponds to the maximum value of
  \(L(m,\overline{x}/m)\) among those computed equals \(\hat{m}_{MLE}\).
\end{itemize}

We note that if the underlying distribution is the binomial distribution
with parameters \((m,q)\) (with \(q>0\)) then \(\hat{m}_{MLE}\) will
equal \(m\) for large sample sizes. Also, \(\hat{q}_{MLE}\) will have an
asymptotically normal distribution and converge with probability one to
\(q\).

\textbf{MLE - Negative Binomial Distribution:} The case of the negative
binomial distribution is similar to that of the binomial distribution in
the sense that we have two parameters and the MLEs are not be available
in closed form. A difference between them is that unlike the binomial
parameter \(m\) which takes positive integral values, the parameter
\(r\) of the negative binomial can assume any positive real value. This
makes the optimization problem a tad more complex. We begin by observing
that the likelihood can be expressed in the following form: \[
L(r,\beta)=\left(\prod_{i=1}^n \binom{r+x_i-1}{x_i}\right) (1+\beta)^{-n(r+\overline{x})} \beta^{n\overline{x}}.
\] The above implies that log-likelihood is given by \[
l(r,\beta)=\sum_{i=1}^n \log\binom{r+x_i-1}{x_i} -n(r+\overline{x}) \log(1+\beta) +n\overline{x}\log\beta,
\] and hence \[
\frac{\delta}{\delta\beta} l(r,\beta) = -\frac{n(r+\overline{x})}{1+\beta} + \frac{n\overline{x}}{\beta}.
\] Equating the above to zero, we get \[
\hat{r}_{MLE}\times \hat{\beta}_{MLE} = \overline{x}.
\] The above reduces the two dimensional optimization problem to a
one-dimensional problem - we need to maximize \[
l(r,\overline{x}/r)=\sum_{i=1}^n \log\binom{r+x_i-1}{x_i} -n(r+\overline{x}) \log(1+\overline{x}/r) +n\overline{x}\log(\overline{x}/r),
\] with respect to \(r\), with the maximizing \(r\) being its \emph{mle}
and \(\hat{\beta}_{MLE}=\overline{x}/\hat{r}_{MLE}\). In
\citep{levin1977} it is show that if the sample variance is greater than
the sample mean then there exists a unique \(r>0\) that maximizes
\(l(r,\overline{x}/r)\) and hence a unique MLE for \(r\) and \(\beta\).
Also, they show that if \(\hat{\sigma}^2\leq \overline{x}\), then the
negative binomial likelihood will be dominated by the Poisson likelihood
with \(\hat{\lambda}=\overline{x}\) - in other words, a Poisson
distribution offers a better fit to the data. The guarantee in the case
of \(\hat{\sigma}^2>\hat{\mu}\) permits us to use any algorithm to
maximize \(l(r,\overline{x}/r)\). Towards an alternate method of
computing the likelihood, we note that \[
l(r,\overline{x}/r)=\sum_{i=1}^n \sum_{j=1}^{x_i}\log(r-1+j) - \sum_{i=1}^n\log(x_i!) - n(r+\overline{x}) \log(r+\overline{x}) + nr\log(r) + n\overline{x}\log(\overline{x}),
\] which yields \[
\left(\frac{1}{n}\right)\frac{\delta}{\delta r}l(r,\overline{x}/r)=\left(\frac{1}{n}\right)\sum_{i=1}^n \sum_{j=1}^{x_i}\frac{1}{r-1+j} - \log(r+\overline{x}) + \log(r).
\] We note that, in the above expressions, the inner sum equals zero if
\(x_i=0\). The \emph{mle} for \(r\) is a zero of the last expression,
and hence a root finding algorithm can be used to compute it. Also, we
have \[
\left(\frac{1}{n}\right)\frac{\delta^2}{\delta r^2}l(r,\overline{x}/r)=\frac{\overline{x}}{r(r+\overline{x})}-\left(\frac{1}{n}\right)\sum_{i=1}^n \sum_{j=1}^{x_i}\frac{1}{(r-1+j)^2}.
\] A simple but quickly converging iterative root finding algorithm is
the Newton's method, which incidentally the Babylonians are believed to
have used for computing square roots. Applying the Newton's method to
our problem results in the following algorithm:\\
 \emph{Step i}. Choose an approximate solution, say \(r_0\). Set \(k\)
to \(0\).\\
\emph{Step ii}. Define \(r_{k+1}\) as \[
r_{k+1}:= r_k - \frac{\left(\frac{1}{n}\right)\sum_{i=1}^n \sum_{j=1}^{x_i}\frac{1}{r_k-1+j} - \log(r_k+\overline{x}) + \log(r_k)}{\frac{\overline{x}}{r_k(r_k+\overline{x})}-\left(\frac{1}{n}\right)\sum_{i=1}^n \sum_{j=1}^{x_i}\frac{1}{(r_k-1+j)^2}}
\]\\
\emph{Step iii}. If \(r_{k+1}\sim r_k\), then report \(r_{k+1}\) as MLE;
else increment \(k\) by \(1\) and repeat \emph{Step ii}.

For example, we simulated a \(5\) sample of \(41, 49, 40, 27, 23\) from
the negative binomial with parameters \(r=10\) and \(\beta=5\). Choosing
the starting value of \(r\) such that \[
r\beta=\hat{\mu} \quad \hbox{and} \quad r\beta(1+\beta)=\hat{\sigma}^2
\] leads to the starting value of \(23.14286\). The iterates of \(r\)
from the Newton's method are \[
21.39627, 21.60287, 21.60647, 21.60647;
\] the rapid convergence seen above is typical of the Newton's method.
Hence in this example, \(\hat{r}_{MLE}\sim21.60647\) and
\(\hat{\beta}_{MLE}=8.3308\)

\emph{R Implementation of Newton's Method - Negative Binomial MLE for
\(r\)}

Show R Code

\hypertarget{toggleCodeFreq.1}{}
\begin{verbatim}
Newton<-function(x,abserr){
mu<-mean(x);
sigma2<-mean(x^2)-mu^2;
r<-mu^2/(sigma2-mu);
b<-TRUE;
iter<-0;
while (b) {
tr<-r;
m1<-mean(c(x[x==0],sapply(x[x>0],function(z){sum(1/(tr:(tr-1+z)))})));
m2<-mean(c(x[x==0],sapply(x[x>0],function(z){sum(1/(tr:(tr-1+z))^2)})));
r<-tr-(m1-log(1+mu/tr))/(mu/(tr*(tr+mu))-m2);
b<-!(abs(tr-r)<abserr);
iter<-iter+1;
}
c(r,iter)
}
\end{verbatim}

\begin{center}\rule{0.5\linewidth}{\linethickness}\end{center}

\begin{figure}

{\centering \includegraphics[width=0.8\linewidth]{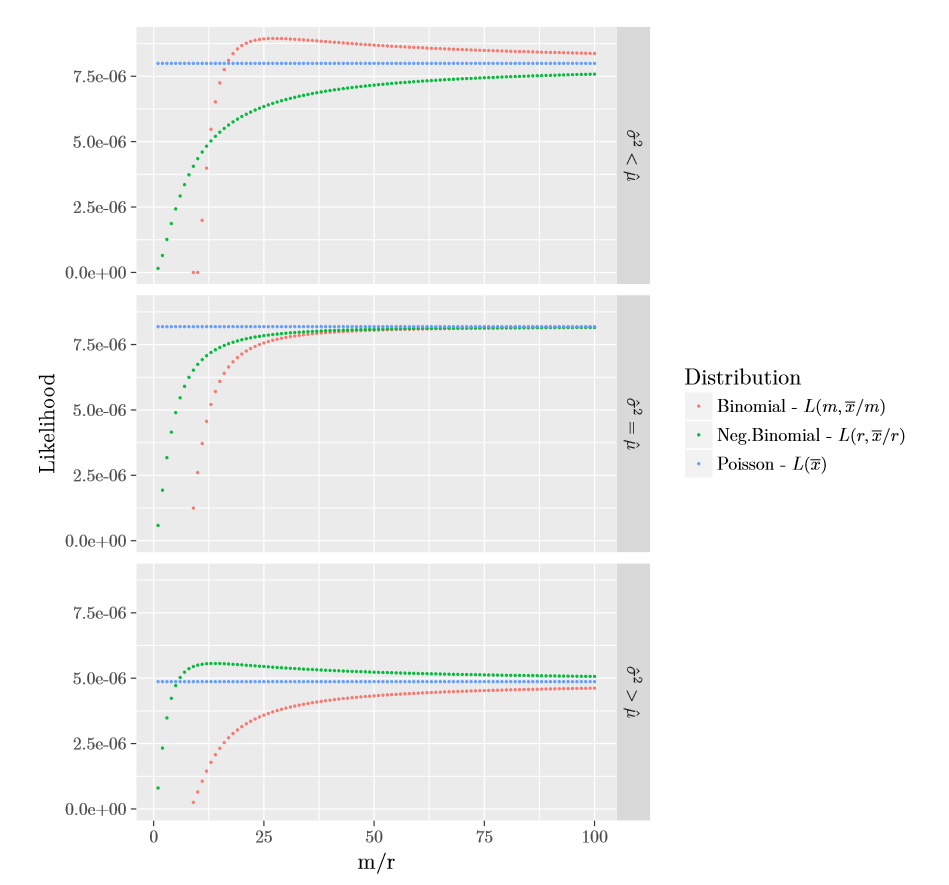}

}

\caption{Plot of $(a,b,0)$ Partially Maximized Likelihoods}\label{fig:MLEab0}
\end{figure}

To summarize our discussion of \emph{mle} for the \((a,b,0)\) class of
distributions, in Figure \ref{fig:MLEab0} below we plot the maximum
value of the Poisson likelihood, \(L(m,\overline{x}/m)\) for the
binomial, and \(L(r,\overline{x}/r)\) for the negative binomial, for the
three samples of size \(5\) given in \protect\hyperlink{tab:2.1}{Table
2.1}. The data was constructed to cover the three orderings of the
sample mean and variance. As show in the Figure \ref{fig:MLEab0}, and
supported by theory, if \(\hat{\mu}<\hat{\sigma}^2\) then the negative
binomial will result in a higher maximum likelihood value; if
\(\hat{\mu}=\hat{\sigma}^2\) the Poisson will have the highest
likelihood value; and finally in the case that
\(\hat{\mu}>\hat{\sigma}^2\) the binomial will give a better fit than
the others. So before fitting a frequency data with an \((a,b,0,)\)
distribution, it is best to start with examining the ordering of
\(\hat{\mu}\) and \(\hat{\sigma}^2\). We again emphasize that the
Poisson is on the \textbf{boundary} of the negative binomial and
binomial distributions. So in the case that
\(\hat{\mu}\geq\hat{\sigma}^2\) (\(\hat{\mu}\leq\hat{\sigma}^2\), resp.)
the Poisson will yield a better fit than the negative binomial
(binomial, resp.), which will also be indicated by \(\hat{r}=\infty\)
(\(\hat{m}=\infty\), resp.).

\[\begin{matrix}
\begin{array}{c|c|c}
\hline
\text{Data} & \text{Mean }(\hat{\mu}) & \text{Variance }(\hat{\sigma}^2) \\
\hline
(2,3,6,8,9) & 5.60 & 7.44 \\
(2,5,6,8,9) & 6 & 6\\
(4,7,8,10,11) & 8 & 6\\\hline
\end{array}
\end{matrix}\]

\protect\hyperlink{tab:2.1}{Table 2.1} : Three Samples of Size \(5\)

\section{Other Frequency
Distributions}\label{S:other-frequency-distributions}

In the above we discussed three distributions with supports contained in
the set of non-negative integers, which well cater to many insurance
applications. Moreover, typically by allowing the parameters to be a
function of known (to the insurer) explanatory variables such as age,
sex, geographic location (territory), and so forth, these distributions
allow us to explain claim probabilities in terms of these variables. The
field of statistical study that studies such models is known as
\textbf{regression analysis} - it is an important topic of actuarial
interest that we will not pursue in this book; see
\citep{freesregression}.

There are clearly infinitely many other count distributions, and more
importantly the above distributions by themselves do not cater to all
practical needs. In particular, one feature of some insurance data is
that the proportion of zero counts can be out of place with the
proportion of other counts to be explainable by the above distributions.
In the following we modify the above distributions to allow for
arbitrary probability for zero count irrespective of the assignment of
relative probabilities for the other counts. Another feature of a data
set which is naturally comprised of homogeneous subsets is that while
the above distributions may provide good fits to each subset, they may
fail to do so to the whole data set. Later we naturally extend the
\((a,b,0)\) distributions to be able to cater to, in particular, such
data sets.

\subsection{Zero Truncation or
Modification}\label{S:zero-truncation-or-modification}

Let us suppose that we are looking at auto insurance policies which
appear in a database of auto claims made in a certain period. If one is
to study the number of claims that these policies have made during this
period, then clearly the distribution has to assign a probability of
zero to the count variable assuming the value zero. In other words, by
restricting attention to count data from policies in the database of
claims, we have in a sense zero-truncated the count data of all
policies. In personal lines (like auto), policyholders may not want to
report that first claim because of fear that it may increase future
insurance rates - this behavior will inflate the proportion of zero
counts. Examples such as the latter modify the proportion of zero
counts. Interestingly, natural modifications of the three distributions
considered above are able to provide good fits to
zero-modified/truncated data sets arising in insurance.

In the below we modify the probability assigned to zero count by the
\((a,b,0)\) class while maintaining the relative probabilities assigned
to non-zero counts - zero modification. Note that since the \((a,b,0)\)
class of distribution satisfies the recurrence \eqref{eq:ab0}, maintaining
relative probabilities of non-zero counts implies that recurrence
\eqref{eq:ab0} is satisfied for \(k\geq 2\). This leads to the definition
of the following class of distributions.

\textbf{Definition}. A count distribution is a member of the
\((a, b, 1)\) class if for some constants \(a\) and \(b\) the
probabilities \(p_k\) satisfy

\begin{equation}
\frac{p_k}{p_{k-1}}=a+\frac{b}{k},\quad k\geq 2.
\label{eq:ab1}
\end{equation}

Note that since the recursion starts with \(p_1\), and not \(p_0\), we
refer to this super-class of \((a,b,0)\) distributions by \((a,b,1)\).
To understand this class, recall that each valid pair of values for
\(a\) and \(b\) of the \((a,b,0)\) class corresponds to a unique vector
of probabilities \(\{p_k\}_{k\geq 0}\). If we now look at the
probability vector \(\{\tilde{p}_k\}_{k\geq 0}\) given by \[
\tilde{p}_k= \frac{1-\tilde{p}_0}{1-p_0}\cdot p_k, \quad k\geq 1,
\] where \(\tilde{p}_0\in[0,1)\) is arbitrarily chosen, then since the
relative probabilities for positive values according to
\(\{p_k\}_{k\geq 0}\) and \(\{\tilde{p}_k\}_{k\geq 0}\) are the same, we
have \(\{\tilde{p}_k\}_{k\geq 0}\) satisfies recurrence \eqref{eq:ab1}.
This, in particular, shows that the class of \((a,b,1)\) distributions
is strictly wider than that of \((a,b,0)\).

In the above, we started with a pair of values for \(a\) and \(b\) that
led to a valid \((a,b,0)\) distribution, and then looked at the
\((a,b,1)\) distributions that corresponded to this \((a,b,0)\)
distribution. We will now argue that the \((a,b,1)\) class allows for a
larger set of permissible values for \(a\) and \(b\) than the
\((a,b,0)\) class. Recall from Section \ref{S:the-a-b-0-class} that in
the case of \(a<0\) we did not use the fact that the recurrence
\eqref{eq:ab0} started at \(k=1\), and hence the set of pairs \((a,b)\)
with \(a<0\) that are permissible for the \((a,b,0)\) class is identical
to those that are permissible for the \((a,b,1)\) class. The same
conclusion is easily drawn for pairs with \(a=0\). In the case that
\(a>0\), instead of the constraint \(a+b>0\) for the \((a,b,0)\) class
we now have the weaker constraint of \(a+b/2>0\) for the \((a,b,1)\)
class. With the parametrization \(b=(r-1)a\) as used in Section
\ref{S:the-a-b-0-class}, instead of \(r>0\) we now have the weaker
constraint of \(r>-1\). In particular, we see that while zero modifying
a \((a,b,0)\) distribution leads to a distribution in the \((a,b,1)\)
class, the conclusion does not hold in the other direction.

Zero modification of a count distribution \(F\) such that it assigns
zero probability to zero count is called a zero truncation of \(F\).
Hence, the zero truncated version of probabilities \(\{p_k\}_{k\geq 0}\)
is given by \[
\tilde{p}_k=\begin{cases}
0, & k=0;\\
\frac{p_k}{1-p_0}, & k\geq 1.
\end{cases}
\]

In particular, we have that a zero modification of a count distribution
\(\{p_k\}_{k\geq 0}\), denoted by \(\{p^M_k\}_{k\geq 0}\), can be
written as a convex combination of the degenerate distribution at \(0\)
and the zero truncation of \(\{p_k\}_{k\geq 0}\), denoted by
\(\{p^T_k\}_{k\geq 0}\). That is we have \[
p^M_k= p^M_0 \cdot \delta_{0}(k) + (1-p^M_0) \cdot p^T_k, \quad k\geq 0.
\]

\textbf{Example 2.5.1. Zero Truncated/Modified Poisson}. Consider a
Poisson distribution with parameter \(\lambda=2\). Calculate
\(p_k, k=0,1,2,3\), for the usual (unmodified), truncated and a modified
version with \((p_0^M=0.6)\).

Show Example Solution

\hypertarget{toggleExampleFreq.5.1}{}
\textbf{Solution.} For the Poisson distribution as a member of the
(\(a,b\),0) class, we have \(a=0\) and \(b=\lambda=2\). Thus, we may use
the recursion \(p_k = \lambda p_{k-1}/k= 2 p_{k-1}/k\) for each type,
after determining starting probabilities. The calculation of
probabilities for \(k\leq 3\) is shown in
\protect\hyperlink{tab:2.2}{Table 2.2}.

\[\begin{matrix}
\begin{array}{c|c|c|c}
\hline
k & p_k & p_k^T & p_k^M\\\hline
0 & p_0=e^{-\lambda}=0.135335 & 0 & 0.6\\\hline
1 & p_1=p_0(0+\frac{\lambda}{1})=0.27067 &
\frac{p_1}{1-p_0}=0.313035 &
\frac{1-p_0^M}{1-p_0}~p_1=0.125214\\\hline
2 & p_2=p_1\left( \frac{\lambda}{2}\right)=0.27067 &
p_2^T=p_1^T\left(\frac{\lambda}{2}\right)=0.313035 &
p_2^M=0.125214\\\hline
3 & p_3=p_2\left(\frac{\lambda}{3}\right)=0.180447 &
p_3^T=p_2^T\left(\frac{\lambda}{3}\right)=0.208690 &
p_3^M=p_2^M\left(\frac{\lambda}{3}\right)=0.083476\\\hline
\end{array}
\end{matrix}\]

\protect\hyperlink{tab:2.2}{Table 2.2} : Calculation of probabilities
for \(k\leq 3\)

\begin{center}\rule{0.5\linewidth}{\linethickness}\end{center}

\section{Mixture Distributions}\label{S:mixture-distributions}

In many applications, the underlying population consists of naturally
defined sub-groups with some homogeneity within each sub-group. In such
cases it is convenient to model the individual sub-groups, and in a
ground-up manner model the whole population. As we shall see below,
beyond the aesthetic appeal of the approach, it also extends the range
of applications that can be catered to by standard parametric
distributions.

Let \(k\) denote the number of defined sub-groups in a population, and
let \(F_i\) denote the distribution of an observation drawn from the
\(i\)-th subgroup. If we let \(\alpha_i\) denote the proportion of the
population in the \(i\)-th subgroup, then the distribution of a randomly
chosen observation from the population, denoted by \(F\), is given by

\begin{equation}
F(x)=\sum_{i=1}^n \alpha_i \cdot F_i(x).
\label{eq:mixdefn}
\end{equation}

The above expression can be seen as a direct application of Bayes
theorem. As an example, consider a population of drivers split broadly
into two sub-groups, those with less than \(5\)-years of driving
experience and those with more than \(5\)-years experience. Let
\(\alpha\) denote the proportion of drivers with less than \(5\) years
experience, and \(F_{\leq 5}\) and \(F_{> 5}\) denote the distribution
of the count of claims in a year for a driver in each group,
respectively. Then the distribution of claim count of a randomly
selected driver is given by \[
\alpha\cdot F_{\leq 5} + (1-\alpha)F_{> 5}.
\]

An alternate definition of a mixture distribution is as follows. Let
\(N_i\) be a random variable with distribution \(F_i\),
\(i=1,\ldots, k\). Let \(I\) be a random variable taking values
\(1,2,\ldots,k\) with probabilities \(\alpha_1,\ldots,\alpha_k\),
respectively. Then the random variable \(N_I\) has a distribution given
by equation \eqref{eq:mixdefn}\footnote{This in particular lays out a way
  to simulate from a mixture distribution that makes use of efficient
  simulation schemes that may exist for the component distributions.}.

In \eqref{eq:mixdefn} we see that the distribution function is a convex
combination of the component distribution functions. This result easily
extends to the density function, the survival function, the raw moments,
and the expectation as these are all linear functionals of the
distribution function. We note that this is not true for central moments
like the variance, and conditional measures like the hazard rate
function. In the case of variance it is easily seen as \[
\mathrm{E}{N_I}=\mathrm{E}{\mathrm{E}{N_I\vert I}} + \mathrm{E}{\mathrm{E}{N_I|I}}=\sum_{i=1}^k \alpha_i \mathrm{E}{N_i} + \mathrm{E}{\mathrm{E}{N_I|I}},
\] and hence is not a convex function of the variances unless the group
means are all equal.

\textbf{Example 2.6.1. SOA Exam Question}. In a certain town the number
of common colds an individual will get in a year follows a Poisson
distribution that depends on the individual's age and smoking status.
The distribution of the population and the mean number of colds are as
follows:

\[\begin{matrix}
\begin{array}{l|c|c}
\hline
 & \text{Proportion of population} &
\text{Mean number of colds}\\\hline
\text{Children} & 0.3 & 3\\
\text{Adult Non-Smokers} & 0.6 & 1\\
\text{Adult Smokers} & 0.1 & 4\\\hline
\end{array}
\end{matrix}\]

\protect\hyperlink{tab:2.3}{Table 2.3} : The distribution of the
population and the mean number of colds

\begin{enumerate}
\def\labelenumi{\arabic{enumi}.}
\tightlist
\item
  Calculate the probability that a randomly drawn person has 3 common
  colds in a year.
\item
  Calculate the conditional probability that a person with exactly 3
  common colds in a year is an adult smoker.
\end{enumerate}

Show Example Solution

\hypertarget{toggleExampleFreq.6.1}{}
\textbf{Solution.}

\begin{enumerate}
\def\labelenumi{\arabic{enumi}.}
\tightlist
\item
  Using development above, we can write the required probability as
  \(\Pr(N_I=3)\), with \(I\) denoting the group of the randomly selected
  individual with \(1,2\) and \(3\) signifying the groups
  \emph{Children}, \emph{Adult Non-Smoker}, and \emph{Adult Smoker},
  respectively. Now by conditioning we get \[
  \Pr(N_I=3)=0.3\cdot\Pr(N_1=3)+0.6\cdot\Pr(N_2=3)+0.1\cdot\Pr(N_3=3),
  \] with \(N_1,N_2\) and \(N_3\) following Poisson distributions with
  means \(3,1\) and \(4\), respectively. Using the above, we get
  \(\Pr(N_I=3)\sim0.1235\)
\item
  The required probability can be written as \(\Pr(I=3\vert N_I=3)\),
  which equals \[
  \Pr(I=3\vert N_I=3)=\frac{\Pr(I=3;N_3=3)}{\Pr(N_I=3)}\sim\frac{0.1 \times 0.1954}{0.1235}\sim 0.1581.
  \]
\end{enumerate}

\begin{center}\rule{0.5\linewidth}{\linethickness}\end{center}

In the above example, the number of subgroups \(k\) was equal to three.
In general, \(k\) can be any natural number, but when \(k\) is large it
is parsimonious from a modeling point of view to take the following
\emph{infinitely many subgroup} approach. To motivate this approach, let
the \(i\)-th subgroup be such that its component distribution \(F_i\) is
given by \(G_{\tilde{\theta_i}}\), where \(G_\cdot\) is a parametric
family of distributions with parameter space
\(\Theta\subseteq \mathbb{R}^d\). With this assumption, the distribution
function \(F\) of a randomly drawn observation from the population is
given by \[
F(x)=\sum_{i=1}^k \alpha_i G_{\tilde{\theta_i}}(x),\quad \forall x\in\mathbb{R}.
\] which can be alternately written as\\
\[
F(x)=\mathbf{R}{G_{\tilde{\vartheta}}(x)},\quad \forall x\in\mathbb{R},
\] where \(\tilde{\vartheta}\) takes values \(\tilde{\theta_i}\) with
probability \(\alpha_i\), for \(i=1,\ldots,k\). The above makes it clear
that when \(k\) is large, one could model the above by treating
\(\tilde{\vartheta}\) as continuous random variable.

To illustrate this approach, suppose we have a population of drivers
with the distribution of claims for an individual driver being
distributed as a Poisson. Each person has their own (personal) expected
number of claims \(\lambda\) - smaller values for good drivers, and
larger values for others. There is a distribution of \(\lambda\) in the
population; a popular and convenient choice for modeling this
distribution is a gamma distribution with parameters
\((\alpha, \theta)\). With these specifications it turns out that the
resulting distribution of \(N\), the claims of a randomly chosen driver,
is a negative binomial with parameters \((r=\alpha,\beta=\theta)\). This
can be shown in many ways, but a straightforward argument is as follows:

\begin{align*}
\Pr(N=k)&= \int_0^\infty \frac{e^{-\lambda}\lambda^k}{k!} \frac{\lambda^{\alpha-1}e^{-\lambda/\theta}}{\Gamma{(\alpha)}\theta^{\alpha}} {\rm d}\lambda =
\frac{1}{k!\Gamma(\alpha)\theta^\alpha}\int_0^\infty \lambda^{\alpha+k-1}e^{-\lambda(1+1/\theta)}{\rm d}\lambda=\frac{\Gamma{(\alpha+k)}}{k!\Gamma(\alpha)\theta^\alpha(1+1/\theta)^{\alpha+k}} \\
&=\binom{\alpha+k-1}{k}\left(\frac{1}{1+\theta}\right)^\alpha\left(\frac{\theta}{1+\theta}\right)^k, \quad k=0,1,\ldots
\end{align*}

It is worth mention that by considering mixtures of a parametric class
of distributions we increase the richness of the class, resulting in the
mixture class being able to cater well to more applications that the
parametric class we started with. In the above case, this is seen as we
have observed earlier that in a sense the Poisson distributions are on
the boundary of negative binomial distributions and by mixing Poisson we
get the interior distributions as well. Mixture modeling is a very
important modeling technique in insurance applications, and later
chapters will cover more aspects of this modeling technique.

\textbf{Example 2.6.2.} Suppose that \(N|\Lambda \sim\)
Poisson\((\Lambda)\) and that \(\Lambda \sim\) gamma with mean of 1 and
variance of 2. Determine the probability that \(N=1\).

Show Example Solution

\hypertarget{toggleExampleFreq.6.2}{}
\textbf{Solution.} For a gamma distribution with parameters
\((\alpha, \theta)\), we have that the mean is \(\alpha \theta\) and the
variance is \(\alpha \theta^2\). Using these expressions we have \[
\begin{aligned}
\alpha &= \frac{1}{2} \text{   and   } \theta =2.
\end{aligned}
\] Now, one can directly use the above result to conclude that \(N\) is
distributed as a negative binomial with \(r = \alpha = \frac{1}{2}\) and
\(\beta= \theta =2\). Thus \[
\begin{aligned}
\Pr(N=1)  &= \binom{1+r-1}{1}(\frac{1}{(1+\beta)^r})\left(\frac{\beta}{1+\beta}\right)^1 \\
&=                 \binom{1+\frac{1}{2}-1}{1}{\frac{1}{(1+2)^{1/2}}}\left(\frac{2}{1+2}\right)^1\\
&=  \frac{1}{3^{3/2}} = 0.19245 .
\end{aligned}
\]

\begin{center}\rule{0.5\linewidth}{\linethickness}\end{center}

\section{Goodness of Fit}\label{S:goodness-of-fit}

In the above we have discussed three basic frequency distributions,
along with their ways to enhance the reach of these classes through zero
modification/truncation and by looking at mixtures of these
distributions. Nevertheless, these classes still remain parametric and
hence by their very nature a small subset of the class of all possible
frequency distributions (\emph{i.e.} the set of distributions on
non-negative integers.) Hence, even though we have talked about methods
for estimating the unknown parameters, the \emph{fitted} distribution
will not be a good representation of the underlying distribution if the
latter is \textbf{far} from the class of distribution used for modeling.
In fact, it can be shown that the \emph{mle} estimate will converge to a
value such that the fitted distribution will be a certain
\emph{projection} of the underlying distribution on the class of
distributions used for modeling. Below we present one testing method -
Pearson's chi-square statistic - to check for the \emph{goodness of fit}
of the fitted distribution.

In \(1993\), a portfolio of \(n=7,483\) automobile insurance policies
from a major Singaporean insurance company had the distribution of auto
accidents per policyholder as given in \protect\hyperlink{tab:2.4}{Table
2.4}.

\[\begin{matrix}
\begin{array}{c|c|c|c|c|c|c}
\hline
\text{Count }(k) & 0 & 1 & 2 & 3 & 4 & \text{Total}\\
\hline
\text{No. of Policies with }k\text{ accidents }(m_k) & 6,996 & 455 & 28 & 4 & 0 & 7483\\
\hline
\end{array}
\end{matrix}\]

\protect\hyperlink{tab:2.4}{Table 2.4} : Singaporean Automobile Accident
Data

If we a fit a Poisson distribution, then the \emph{mle} for \(\lambda\),
the Poisson mean, is the sample mean which is given by \[
\overline{N} = \frac{0\cdot 6996 + 1 \cdot 455 + 2 \cdot 28 + 3 \cdot 4 + 4 \cdot 0}{7483} = 0.06989.
\] Now if we use Poisson(\(\hat{\lambda}_{MLE}\)) as the fitted
distribution, then a tabular comparison of the fitted counts and
observed counts is given by \protect\hyperlink{tab:2.5}{Table 2.5}
below, where \(\hat{p}_k\) represents the estimated probabilities under
the fitted Poisson distribution.

\[\begin{matrix}
\begin{array}{c|c|c}
\hline
\text{Count}  & \text{Observed}  & \text{Fitted Counts}\\
(k) & (m_k) & \text{Using Poisson }(n\hat{p}_k)\\
\hline
0 & 6,996 & 6,977.86 \\
1 & 455 & 487.70 \\
2 & 28 & 17.04 \\
3 & 4 & 0.40 \\
\geq4 & 0 & 0.01\\
\hline
\text{Total} & 7,483 & 7,483.00\\
\hline
\end{array}
\end{matrix}\]

\protect\hyperlink{tab:2.5}{Table 2.5} : Comparison of Observed to
Fitted Counts: Singaporean Auto Data

While the fit seems \emph{reasonable}, a tabular comparison falls short
of a statistic test of the hypothesis that the underlying distribution
is indeed Poisson. The \emph{Pearson's chi-square statistic} is a
goodness of fit statistical test that can be used for this purpose. To
explain this statistic let use suppose that a dataset of size \(n\) is
grouped into \(k\) cells with \(m_k/n\) and \(\hat{p}_k\), for
\(k=1\ldots,K\) being the observed and estimated probabilities of an
observation belonging to the \(k\)-th cell, respectively. The Pearson's
chi-square test statistic is then given by \[
\sum_{k=1}^K\frac{\left( m_k-n\widehat{p}_k \right) ^{2}}{n\widehat{p}_k}.
\] The motivation for the above statistic derives from the fact that \[
\sum_{k=1}^K\frac{\left( m_k-n{p}_k \right) ^{2}}{n{p}_k}
\] has a limiting chi-square distribution with \(K-1\) degrees of
freedom if \(p_k\), \(k=1,\ldots,K\) are the true cell probabilities.
Now suppose that only the summarized data represented by \(m_k\),
\(k=1,\ldots,K\) is available. Further, if \(p_k\)'s are functions of
\(s\) parameters, replacing \(p_k\)'s by any \emph{efficiently}
estimated probabilities \(\widehat{p}_k\)'s results in the statistic
continuing to have a limiting chi-square distribution but with degrees
of freedom given by \(K-1-s\). Such efficient estimates can be derived
for example by using the \emph{mle} method (with a multinomial
likelihood) or by estimating the \(s\) parameters which minimizes the
Pearson's chi-square statistic above. For example, the \texttt{R} code
below does calculate an estimate for \(\lambda\) doing the latter and
results in the estimate \(0.06623153\), close but different from the
\emph{mle} of \(\lambda\) using the full data:

\begin{verbatim}
m<-c(6996,455,28,4,0);
op<-m/sum(m);
g<-function(lam){sum((op-c(dpois(0:3,lam),1-ppois(3,lam)))^2)};
optim(sum(op*(0:4)),g,method="Brent",lower=0,upper=10)$par
\end{verbatim}

When one uses the full-data to estimate the probabilities the asymptotic
distribution is \emph{in between} chi-square distributions with
parameters \(K-1\) and \(K-1-s\). In practice it is common to ignore
this subtlety and assume the limiting chi-square has \(K-1-s\) degrees
of freedom. Interestingly, this practical shortcut works quite well in
the case of the Poisson distribution.

For the Singaporean auto data the Pearson's chi-square statistic equals
\(41.98\) using the full data \emph{mle} for \({\lambda}\). Using the
limiting distribution of chi-square with \(5-1-1=3\) degrees of freedom,
we see that the value of \(41.98\) is way out in the tail (\(99\)-th
percentile is below \(12\)). Hence we can conclude that the Poisson
distribution provides an inadequate fit for the data.

In the above we started with the cells as given in the above tabular
summary. In practice, a relevant question is how to define the cells so
that the chi-square distribution is a good approximation to the finite
sample distribution of the statistic. A rule of thumb is to define the
cells in such a way to have at least \(80\%\) if not all of the cells
having expected counts greater than \(5\). Also, it is clear that a
larger number of cells results in a higher power of the test, and hence
a simple rule of thumb is to maximize the number of cells such that each
cell has at least 5 observations.

\section{Exercises}\label{S:exercises}

\textbf{Theoretical Exercises:}

\textbf{Exercise 2.1.} Derive an expression for \(p_N(\cdot)\) in terms
of \(F_N(\cdot)\) and \(S_N(\cdot)\).

\textbf{Exercise 2.2.} A measure of center of location must be
\textbf{equi-variant} with respect to shifts. In other words, if \(N_1\)
and \(N_2\) are two random variables such that \(N_1+c\) has the same
distribution as \(N_2\), for some constant \(c\), then the difference
between the measures of the center of location of \(N_2\) and \(N_1\)
must equal \(c\). Show that the mean satisfies this property.

\textbf{Exercise 2.3.} Measures of dispersion should be invariant w.r.t.
shifts and scale equi-variant. Show that standard deviation satisfies
these properties by doing the following:

\begin{itemize}
\tightlist
\item
  Show that for a random variable \(N\), its standard deviation equals
  that of \(N+c\), for any constant \(c\).
\item
  Show that for a random variable \(N\), its standard deviation equals
  \(1/c\) times that of \(cN\), for any positive constant \(c\).
\end{itemize}

\textbf{Exercise 2.4.} Let \(N\) be a random variable with probability
mass function given by \[
p_N(k):= \begin{cases}
\left(\frac{6}{\pi^2}\right)\left(\frac{1}{k^{2}}\right), & k\geq 1;\\
0, &\hbox{otherwise}.
\end{cases}
\] Show that the mean of \(N\) is \(\infty\).

\textbf{Exercise 2.5.} Let \(N\) be a random variable with a finite
second moment. Show that the function \(\psi(\cdot)\) defined by \[
\psi(x):=\mathrm{E}{(N-x)^2}. \quad x\in\mathbb{R}
\] is minimized at \(\mu_N\) without using calculus. Also, give a proof
of this fact which uses calculus. Conclude that the minimum value equals
the variance of \(N\).

\textbf{Exercise 2.6.} Derive the first two central moments of the
\((a,b,0)\) distributions using the methods mentioned below:

\begin{itemize}
\tightlist
\item
  For the Binomial distribution derive the moments using only its
  \emph{pmf}, its \emph{mgf} and its \emph{pgf}.
\item
  For the Poisson distribution derive the moments using only its mgf.
\item
  For the Negative-Binomial distribution derive the moments using only
  its \emph{pmf}, and its \emph{pgf}.
\end{itemize}

\textbf{Exercise 2.7.} Let \(N_1\) and \(N_2\) be two independent
Poisson random variables with means \(\lambda_1\) and \(\lambda_2\),
respectively. Identify the conditional distribution of \(N_1\) given
\(N_1+N_2\).

\textbf{Exercise 2.8.} (\textbf{Non-Uniqueness of the MLE}) Consider the
following parametric family of densities indexed by the parameter \(p\)
taking values in \([0,1]\): \[
f_p(x)=p\cdot\phi(x+2)+(1-p)\cdot\phi(x-2), \quad x\in\mathbb{R},
\] where \(\phi(\cdot)\) represents the standard normal density.

\begin{itemize}
\tightlist
\item
  Show that for all \(p\in[0,1]\), \(f_p(\cdot)\) above is a valid
  density function.
\item
  Find an expression in \(p\) for the mean and the variance of
  \(f_p(\cdot)\).
\item
  Let us consider a sample of size one consisting of \(x\). Show that
  when \(x\) equals \(0\), the set of MLEs for \(p\) equals \([0,1]\);
  also show that the \emph{mle} is unique otherwise.
\end{itemize}

\textbf{Exercise 2.9.} Graph the region of the plane corresponding to
values of \((a,b)\) that give rise to valid \((a,b,0)\) distributions.
Do the same for \((a,b,1)\) distributions.

\textbf{Exercise 2.10.} (\textbf{Computational Complexity}) For the
\((a,b,0)\) class of distributions, count the number of basic math
operations needed to compute the \(n\) probabilities
\(p_0\ldots p_{n-1}\) using the recurrence relationship. For the
negative binomial distribution with non-integral \(r\), count the number
of such operations using the brute force approach. What do you observe?

\textbf{Exercises with a Practical Focus:}

\textbf{Exercise 2.11. SOA Exam Question.} You are given:

\begin{enumerate}
\def\labelenumi{\arabic{enumi}.}
\tightlist
\item
  \(p_k\) denotes the probability that the number of claims equals \(k\)
  for \(k=0,1,2,\ldots\)
\item
  \(\frac{p_n}{p_m}=\frac{m!}{n!}, m\ge 0, n\ge 0\)
\end{enumerate}

Using the corresponding zero-modified claim count distribution with
\(p_0^M=0.1\), calculate \(p_1^M\).

\textbf{Exercise 2.12. SOA Exam Question.} During a one-year period, the
number of accidents per day was distributed as follows:

\[
\begin{matrix}
\begin{array}{c|c|c|c|c|c|c}
\hline
\text{No. of Accidents} & 0 & 1 & 2 & 3 & 4 & 5\\
\hline
\text{No. of Days} & 209 & 111 & 33 & 7 & 5 & 2\\
\hline
\end{array}
\end{matrix}
\]

You use a chi-square test to measure the fit of a Poisson distribution
with mean 0.60. The minimum expected number of observations in any group
should be 5. The maximum number of groups should be used. Determine the
value of the chi-square statistic.

A discrete probability distribution has the following properties \[
\begin{aligned}
\Pr(N=k) = \left( \frac{3k+9}{8k}\right) \Pr(N=k-1), \quad k=1,2,3,\ldots
\end{aligned}
\] Determine the value of \(\Pr(N=3)\). (Ans: 0.1609)

\subsubsection*{Exercises}\label{exercises}
\addcontentsline{toc}{subsubsection}{Exercises}

Here are a set of exercises that guide the viewer through some of the
theoretical foundations of \textbf{Loss Data Analytics}. Each tutorial
is based on one or more questions from the professional actuarial
examinations -- typically the Society of Actuaries Exam C.

\href{https://www.ssc.wisc.edu/~jfrees/loss-data-analytics/loss-data-analytics-problems/}{Frequency
Distribution Guided Tutorials}

\section{R Code for Plots in this
Chapter}\label{r-code-for-plots-in-this-chapter}

\textbf{Code for Figure \ref{fig:MLEab0}:}

Show R Code

\hypertarget{toggleCodeFreq.2}{}
\begin{verbatim}
likbinm<-function(m){
  prod((dbinom(x,m,mean(x)/m)))
}
liknbinm<-function(r){
  prod(dnbinom(x,r,1-mean(x)/(mean(x)+r)))
}
x<-c(2,5,6,8,9)+2;
n<-(9:100);
r<-(1:100);
ll<-unlist(lapply(n,likbinm));
n[ll==max(ll[!is.na(ll)])]
y<-cbind(n,ll);
z<-cbind(rep("$\\hat{\\sigma}^2<\\hat{\\mu}$",length(n)),rep("Binomial - $L(m,\\overline{x}/m)$",length(n)));
ll<-unlist(lapply(r,liknbinm));
ll[is.na(ll)]=0;
r[ll==max(ll[!is.na(ll)])];
y<-rbind(y,cbind(r,ll));
z<-rbind(z,cbind(rep("$\\hat{\\sigma}^2<\\hat{\\mu}$",length(r)),rep("Neg.Binomial - $L(r,\\overline{x}/r)$",length(r))));
y<-rbind(y,cbind(r,rep(prod(dpois(x,mean(x))),length(r))));
z<-rbind(z,cbind(rep("$\\hat{\\sigma}^2<\\hat{\\mu}$",length(r)),rep("Poisson - $L(\\overline{x})$",length(r))));
x<-c(2,5,6,8,9);
ll<-unlist(lapply(n,likbinm));
n[ll==max(ll[!is.na(ll)])]
y<-rbind(y,cbind(n,ll));
z<-rbind(z,cbind(rep("$\\hat{\\sigma}^2=\\hat{\\mu}$",length(n)),rep("Binomial - $L(m,\\overline{x}/m)$",length(n))));
ll<-unlist(lapply(r,liknbinm));
ll[is.na(ll)]=0;
r[ll==max(ll[!is.na(ll)])];
y<-rbind(y,cbind(r,ll));
z<-rbind(z,cbind(rep("$\\hat{\\sigma}^2=\\hat{\\mu}$",length(r)),rep("Neg.Binomial - $L(r,\\overline{x}/r)$",length(r))));
y<-rbind(y,cbind(r,rep(prod(dpois(x,mean(x))),length(r))));
z<-rbind(z,cbind(rep("$\\hat{\\sigma}^2=\\hat{\\mu}$",length(r)),rep("Poisson - $L(\\overline{x})$",length(r))));
x<-c(2,3,6,8,9);
ll<-unlist(lapply(n,likbinm));
n[ll==max(ll[!is.na(ll)])]
y<-rbind(y,cbind(n,ll));
z<-rbind(z,cbind(rep("$\\hat{\\sigma}^2>\\hat{\\mu}$",length(n)),rep("Binomial - $L(m,\\overline{x}/m)$",length(n))));
ll<-unlist(lapply(r,liknbinm));
ll[is.na(ll)]=0;
r[ll==max(ll[!is.na(ll)])];
y<-rbind(y,cbind(r,ll));
z<-rbind(z,cbind(rep("$\\hat{\\sigma}^2>\\hat{\\mu}$",length(r)),rep("Neg.Binomial - $L(r,\\overline{x}/r)$",length(r))));
y<-rbind(y,cbind(r,rep(prod(dpois(x,mean(x))),length(r))));
z<-rbind(z,cbind(rep("$\\hat{\\sigma}^2>\\hat{\\mu}$",length(r)),rep("Poisson - $L(\\overline{x})$",length(r))));
colnames(y)<-c("x","lik");
colnames(z)<-c("dataset","Distribution");
dy<-cbind(data.frame(y),data.frame(z));

library(tikzDevice);
library(ggplot2);
options(tikzMetricPackages = c("\\usepackage[utf8]{inputenc}","\\usepackage[T1]{fontenc}", "\\usetikzlibrary{calc}",
                               "\\usepackage{amssymb}","\\usepackage{amsmath}","\\usepackage[active]{preview}"))
tikz(file = "plot_test_2.tex", width = 6.25, height = 6.25);
ggplot(data=dy,aes(x=x,y=lik,col=Distribution)) + geom_point(size=0.25) + facet_grid(dataset~.)+
  labs(x="m/r",y="Likelihood",title="");
dev.off();
\end{verbatim}

\begin{center}\rule{0.5\linewidth}{\linethickness}\end{center}

\textbf{Code for Figure \ref{fig:MLEm}:}

Show R Code

\hypertarget{toggleCodeFreq.3}{}
\begin{verbatim}
likm<-function(m){
  prod((dbinom(x,m,mean(x)/m)))
}
x<-c(2,2,2,4,5);
n<-(5:100);
ll<-unlist(lapply(n,likm));
n[ll==max(ll)]
y<-cbind(n,ll);
x<-c(2,2,2,4,6);
ll<-unlist(lapply(n,likm));
n[ll==max(ll)]
y<-cbind(y,ll);
x<-c(2,2,2,4,7);
ll<-unlist(lapply(n,likm));
n[ll==max(ll)]
y<-cbind(y,ll);
colnames(y)<-c("m","$\\tilde{x}=(2,2,2,4,5)$","$\\tilde{x}=(2,2,2,4,6)$","$\\tilde{x}=(2,2,2,4,7)$");
dy<-data.frame(y);
library(tikzDevice);
library(ggplot2);
options(tikzMetricPackages = c("\\usepackage[utf8]{inputenc}","\\usepackage[T1]{fontenc}", "\\usetikzlibrary{calc}",
                               "\\usepackage{amssymb}","\\usepackage{amsmath}","\\usepackage[active]{preview}"))
tikz(file = "plot_test.tex", width = 6.25, height = 3.125);
ggplot(dy) +
  geom_point(aes(x=m, y=(X..tilde.x...2.2.2.4.5..),shape="$\\tilde{x}=(2,2,2,4,5):\\hat{m}=7$"),size=0.75) +
  geom_point(aes(x=m, y=(X..tilde.x...2.2.2.4.6..),shape="$\\tilde{x}=(2,2,2,4,6):\\hat{m}=18$"),size=0.75) +
  geom_point(aes(x=m, y=(X..tilde.x...2.2.2.4.7..),shape="$\\tilde{x}=(2,2,2,4,7):\\hat{m}=\\infty$"),size=0.75) +
  geom_point(aes(x=c(7),y=dy$X..tilde.x...2.2.2.4.5..[3],colour="$\\hat{m}$",shape="$\\tilde{x}=(2,2,2,4,5):\\hat{m}=7$"),size=0.75)+
  geom_point(aes(x=c(18),y=dy$X..tilde.x...2.2.2.4.6..[14],colour="$\\hat{m}$",shape="$\\tilde{x}=(2,2,2,4,6):\\hat{m}=18$"),size=0.75)+
  labs(x="m",y="$L(m,\\overline{x}/m)$",title="MLE for $m$: Non-Robustness of MLE ");
dev.off();
\end{verbatim}

\begin{center}\rule{0.5\linewidth}{\linethickness}\end{center}

\section{Further Resources and
Contributors}\label{Freq-further-reading-and-resources}

\subsubsection*{Contributors}\label{contributors}
\addcontentsline{toc}{subsubsection}{Contributors}

\begin{itemize}
\tightlist
\item
  \textbf{N.D. Shyamalkumar}, The University of Iowa, is the principal
  author of the initital version of this chapter. Email:
  \href{mailto:shyamal-kumar@uiowa.edu}{\nolinkurl{shyamal-kumar@uiowa.edu}}
  for chapter comments and suggested improvements.
\item
  \textbf{Krupa Viswanathan}, Temple University,
  \href{mailto:ksubrama@temple.edu}{\nolinkurl{ksubrama@temple.edu}},
  provided substantial improvements.
\end{itemize}

Here are a few reference cited in the chapter.

\chapter{Modeling Loss Severity}\label{C:Severity}

\emph{Chapter Preview.} The traditional loss distribution approach to
modeling aggregate losses starts by separately fitting a frequency
distribution to the number of losses and a severity distribution to the
size of losses. The estimated aggregate loss distribution combines the
loss frequency distribution and the loss severity distribution by
convolution. Discrete distributions often referred to as counting or
frequency distributions were used in Chapter \ref{C:Frequency-Modeling}
to describe the number of events such as number of accidents to the
driver or number of claims to the insurer. Lifetimes, asset values,
losses and claim sizes are usually modeled as continuous random
variables and as such are modeled using continuous distributions, often
referred to as loss or severity distributions. Mixture distributions are
used to model phenomenon investigated in a heterogeneous population,
such as modelling more than one type of claims in liability insurance
(small frequent claims and large relatively rare claims). In this
chapter we explore the use of continuous as well as mixture
distributions to model the random size of loss. We present key
attributes that characterize continuous models and means of creating new
distributions from existing ones. In this chapter we explore the effect
of coverage modifications, which change the conditions that trigger a
payment, such as applying deductibles, limits, or adjusting for
inflation, on the distribution of individual loss amounts.

\section{Basic Distributional Quantities}\label{S:BasicQuantities}

In this section we calculate the basic distributional quantities:
moments, percentiles and generating functions.

\subsection{Moments}\label{moments}

Let \(X\) be a continuous random variable with probability density
function \(f_{X}\left( x \right)\). The \emph{k}-th raw moment of \(X\),
denoted by \(\mu_{k}^{\prime}\), is the expected value of the
\emph{k}-th power of \(X\), provided it exists. The first raw moment
\(\mu_{1}^{\prime}\) is the mean of \(X\) usually denoted by \(\mu\).
The formula for \(\mu_{k}^{\prime}\) is given as
\[\mu_{k}^{\prime} = \mathrm{E}\left( X^{k} \right) = \int_{0}^{\infty}{x^{k}f_{X}\left( x \right)dx } .\]
The support of the random variable \(X\) is assumed to be nonnegative
since actuarial phenomena are rarely negative.

The \emph{k}-th central moment of \(X\), denoted by \(\mu_{k}\), is the
expected value of the \emph{k}-th power of the deviation of \(X\) from
its mean \(\mu\). The formula for \(\mu_{k}\) is given as
\[\mu_{k} = \mathrm{E}\left\lbrack {(X - \mu)}^{k} \right\rbrack = \int_{0}^{\infty}{\left( x - \mu \right)^{k}f_{X}\left( x \right) dx }.\]
The second central moment \(\mu_{2}\) defines the variance of \(X\),
denoted by \(\sigma^{2}\). The square root of the variance is the
standard deviation \(\sigma\). A further characterization of the shape
of the distribution includes its degree of symmetry as well as its
flatness compared to the normal distribution. The ratio of the third
central moment to the cube of the standard deviation
\(\left( \mu_{3} / \sigma^{3} \right)\) defines the coefficient of
skewness which is a measure of symmetry. A positive coefficient of
skewness indicates that the distribution is skewed to the right
(positively skewed). The ratio of the fourth central moment to the
fourth power of the standard deviation
\(\left(\mu_{4} / \sigma^{4} \right)\) defines the coefficient of
kurtosis. The normal distribution has a coefficient of kurtosis of 3.
Distributions with a coefficient of kurtosis greater than 3 have heavier
tails and higher peak than the normal, whereas distributions with a
coefficient of kurtosis less than 3 have lighter tails and are flatter.

\textbf{Example 3.1.1. SOA Exam Question.} Assume that the \emph{rv}
\(X\) has a gamma distribution with mean 8 and skewness 1. Find the
variance of \(X\).

Show Example Solution

\hypertarget{toggleExampleLoss.1.1}{}
\textbf{Solution.} The probability density function of \(X\) is given by
\[f_{X}\left( x \right) = \frac{\left( x / \theta \right)^{\alpha}}{x\Gamma\left( \alpha \right)} e^{- x / \theta} \]
for \(x > 0\). For \(\alpha>0\), the \emph{k}-th raw moment is
\[\mu_{k}^{\prime} = E\left( X^{k} \right) = \int_{0}^{\infty}{\frac{1}{\left( \alpha - 1 \right)!\theta^{\alpha}}x^{k + \alpha - 1}e^{- x / \theta} dx} = \frac{\Gamma\left( k + \alpha \right)}{\Gamma\left( \alpha \right)}\theta^{k}\]
Given \(\Gamma\left( r + 1 \right) = r\Gamma\left( r \right)\) and
\(\Gamma\left( 1 \right) = 1\), then
\(\mu_{1}^{\prime} = E\left( X \right) = \alpha\theta\),
\(\mu_{2}^{\prime} = E\left( X^{2} \right) = \left( \alpha + 1 \right)\alpha\theta^{2}\),
\(\mu_{3}^{\prime} = E\left( X^{3} \right) = \left( \alpha + 2 \right)\left( \alpha + 1 \right)\alpha\theta^{3}\),
and
\(\mathrm{Var}\left( X \right) = (\alpha + 1)\alpha\theta^2 - (\alpha\theta)^2 = \alpha\theta^{2}\).

\[\text{Skewness}  = \frac{E\left\lbrack {(X - \mu_{1}^{\prime})}^{3} \right\rbrack}{{\mathrm{Var}\left( X \right)}^{3/2}} = \frac{\mu_{3}^{\prime} - 3\mu_{2}^{\prime}\mu_{1}^{\prime} + 2{\mu_{1}^{\prime}}^{3}}{{\mathrm{Var}\left( X \right)}^{3/2}} \\
 = \frac{\left( \alpha + 2 \right)\left( \alpha + 1 \right)\alpha\theta^{3} - 3\left( \alpha + 1 \right)\alpha^{2}\theta^{3} + 2\alpha^{3}\theta^{3}}{\left( \alpha\theta^{2} \right)^{3/2}} = \frac{2}{\alpha^{1/2}} = 1\]

Hence, \(\alpha = 4\). Since, \(E\left( X \right) = \alpha\theta = 8\),
then \(\theta = 2\) and finally,
\(\mathrm{Var}\left( X \right) = \alpha\theta^{2} = 16\).

\begin{center}\rule{0.5\linewidth}{\linethickness}\end{center}

\subsection{Quantiles}\label{quantiles}

Percentiles can also be used to describe the characteristics of the
distribution of \(X\). The 100p\emph{th} percentile of the distribution
of \(X\), denoted by \(\pi_{p}\), is the value of \(X\) which satisfies
\[F_{X}\left( {\pi_{p}}^{-} \right) \leq p \leq F\left( \pi_{p} \right) ,\]
for \(0 \leq p \leq 1\).

The 50-th percentile or the middle point of the distribution,
\(\pi_{0.5}\), is the median. Unlike discrete random variables,
percentiles of continuous variables are distinct.

\textbf{Example 3.1.1. SOA Exam Question.} Let \(X\) be a continuous
random variable with density function
\(f_{X}\left( x \right) = \theta e^{- \theta x}\), for \(x > 0\) and 0
elsewhere. If the median of this distribution is \(\frac{1}{3}\), find
\(\theta\).

Show Example Solution

\hypertarget{toggleExampleLoss.1.2}{}
\textbf{Solution.}

The distribution function is
\(F_{X}\left( x \right) = 1 - e^{- \theta x}\). So,
\(F_{X}\left( \pi_{0.5} \right) = 1 - e^{- \theta\pi_{0.5}} = 0.5\). As,
\(\pi_{0.5} = \frac{1}{3}\), we have
\(F_X\left(\frac{1}{3}\right) = 1 - e^{-\theta / 3} = 0.5\) and
\(\theta = 3 \ln 2\).

\begin{center}\rule{0.5\linewidth}{\linethickness}\end{center}

\subsection{Moment Generating
Function}\label{moment-generating-function}

The moment generating function, denoted by \(M_{X}\left( t \right)\)
uniquely characterizes the distribution of \(X\). While it is possible
for two different distributions to have the same moments and yet still
differ, this is not the case with the moment generating function. That
is, if two random variables have the same moment generating function,
then they have the same distribution. The moment generating is a real
function whose \emph{k}-th derivative at zero is equal to the
\emph{k}-th raw moment of \(X\). The moment generating function is given
by
\[M_{X}\left( t \right) = \mathrm{E}\left( e^{\text{tX}} \right) = \int_{0}^{\infty}{e^{\text{tx}}f_{X}\left( x \right) dx }\]
for all \(t\) for which the expected value exists.

\textbf{Example 3.1.3. SOA Exam Question.} The random variable \(X\) has
an exponential distribution with mean \(\frac{1}{b}\). It is found that
\(M_{X}\left( - b^{2} \right) = 0.2\). Find \(b\).

Show Example Solution

\hypertarget{toggleExampleLoss.1.3}{}
\textbf{Solution.}

With \(X \sim Exp \left( \frac{1}{b}\right)\),
\[M_{X}\left( t \right) = E\left( e^{\text{tX}} \right) = \int_{0}^{\infty}{e^{\text{tx}}be^{- bx} dx} = \int_{0}^{\infty}{be^{- x\left( b - t \right)} dx} = \frac{b}{\left( b - t \right)}.\]

Then,
\[M_{X}\left( - b^{2} \right) = \frac{b}{\left( b + b^{2} \right)} = \frac{1}{\left( 1 + b \right)} = 0.2.\]
Thus, \(b = 4\).

\begin{center}\rule{0.5\linewidth}{\linethickness}\end{center}

\textbf{Example 3.1.4. SOA Exam Question.} Let \(X_{1}, \ldots, X_{n}\)
be independent \(\text{Ga}\left( \alpha_{i},\theta \right)\) random
variables. Find the distribution of \(S = \sum_{i = 1}^{n}X_{i}\).

Show Example Solution

\hypertarget{toggleExampleLoss.1.4}{}
\textbf{Solution.}

The moment generating function of \(S\) is
\[M_{S}\left( t \right) = \text{E}\left( e^{\text{tS}} \right) = E\left( e^{t\sum_{i = 1}^{n}X_{i}} \right) \\
= E\left( \prod_{i = 1}^{n}e^{tX_{i}} \right)\] using independence we
get\\
\[= \prod_{i = 1}^{n}{E\left( e^{tX_{i}} \right) = \prod_{i = 1}^{n}{M_{X_{i}}\left( t \right)}} .\]

The moment generating function of \(X_{i}\) is
\(M_{X_{i}}\left( t \right) = \left( 1 - \theta t \right)^{- \alpha_{i}}\).
Then,
\[M_{S}\left( t \right) = \prod_{i = 1}^{n}\left( 1 - \theta t \right)^{- \alpha_{i}} = \left( 1 - \theta t \right)^{- \sum_{i = 1}^{n}\alpha_{i}}, \]
indicating that
\(S\sim Ga\left( \sum_{i = 1}^{n}\alpha_{i},\theta \right)\).

By finding the first and second derivatives of \(M_{S}\left( t \right)\)
at zero, we can show that
\(E\left( S \right) = \left. \ \frac{\partial M_{S}\left( t \right)}{\partial t} \right|_{t = 0} = \alpha\theta\)
where \(\alpha = \sum_{i = 1}^{n}\alpha_{i}\), and
\[E\left( S^{2} \right) = \left. \ \frac{\partial^{2}M_{S}\left( t \right)}{\partial t^{2}} \right|_{t = 0} = \left( \alpha + 1 \right)\alpha\theta^{2}.\]
Hence, \(\mathrm{Var}\left( S \right) = \alpha\theta^{2}\).

\begin{center}\rule{0.5\linewidth}{\linethickness}\end{center}

\subsection{Probability Generating
Function}\label{probability-generating-function}

The probability generating function, denoted by
\(P_{X}\left( z \right)\), also uniquely characterizes the distribution
of \(X\). It is defined as
\[P_{X}\left( z \right) = \mathrm{E}\left( z^{X} \right) = \int_{0}^{\infty}{z^{x}f_{X}\left( x \right) dx}\]
for all \(z\) for which the expected value exists.

We can also use the probability generating function to generate moments
of \(X\). By taking the \emph{k}-th derivative of
\(P_{X}\left( z \right)\) with respect to \(z\) and evaluate it at
\(z\  = \ 1\), we get
\(\mathrm{E}\left\lbrack X\left( X - 1 \right)\ldots\left( X - k + 1 \right) \right\rbrack .\)

The probability generating function is more useful for discrete
\emph{rv}s and was introduced in Section \ref{S:generating-functions}.

\section{Continuous Distributions for Modeling Loss
Severity}\label{S:ContinuousDistn}

In this section we explain the characteristics of distributions suitable
for modeling severity of losses, including gamma, Pareto, Weibull and
generalized beta distribution of the second kind. Applications for which
each distribution may be used are identified.

\subsection{Gamma Distribution}\label{gamma-distribution}

The gamma distribution is commonly used in modeling claim severity. The
traditional approach in modelling losses is to fit separate models for
claim frequency and claim severity. When frequency and severity are
modeled separately it is common for actuaries to use the Poisson
distribution for claim count and the gamma distribution to model
severity. An alternative approach for modelling losses that has recently
gained popularity is to create a single model for pure premium (average
claim cost) that will be described in Chapter 4.

The continuous variable \(X\) is said to have the gamma distribution
with shape parameter \(\alpha\) and scale parameter \(\theta\) if its
probability density function is given by
\[f_{X}\left( x \right) = \frac{\left( x/ \theta  \right)^{\alpha}}{x\Gamma\left( \alpha \right)}\exp \left( -x/ \theta \right) \ \ \ \text{for } x > 0 .\]
Note that \(\ \alpha > 0,\ \theta > 0\).

\begin{figure}

{\centering \includegraphics[width=1.2\linewidth]{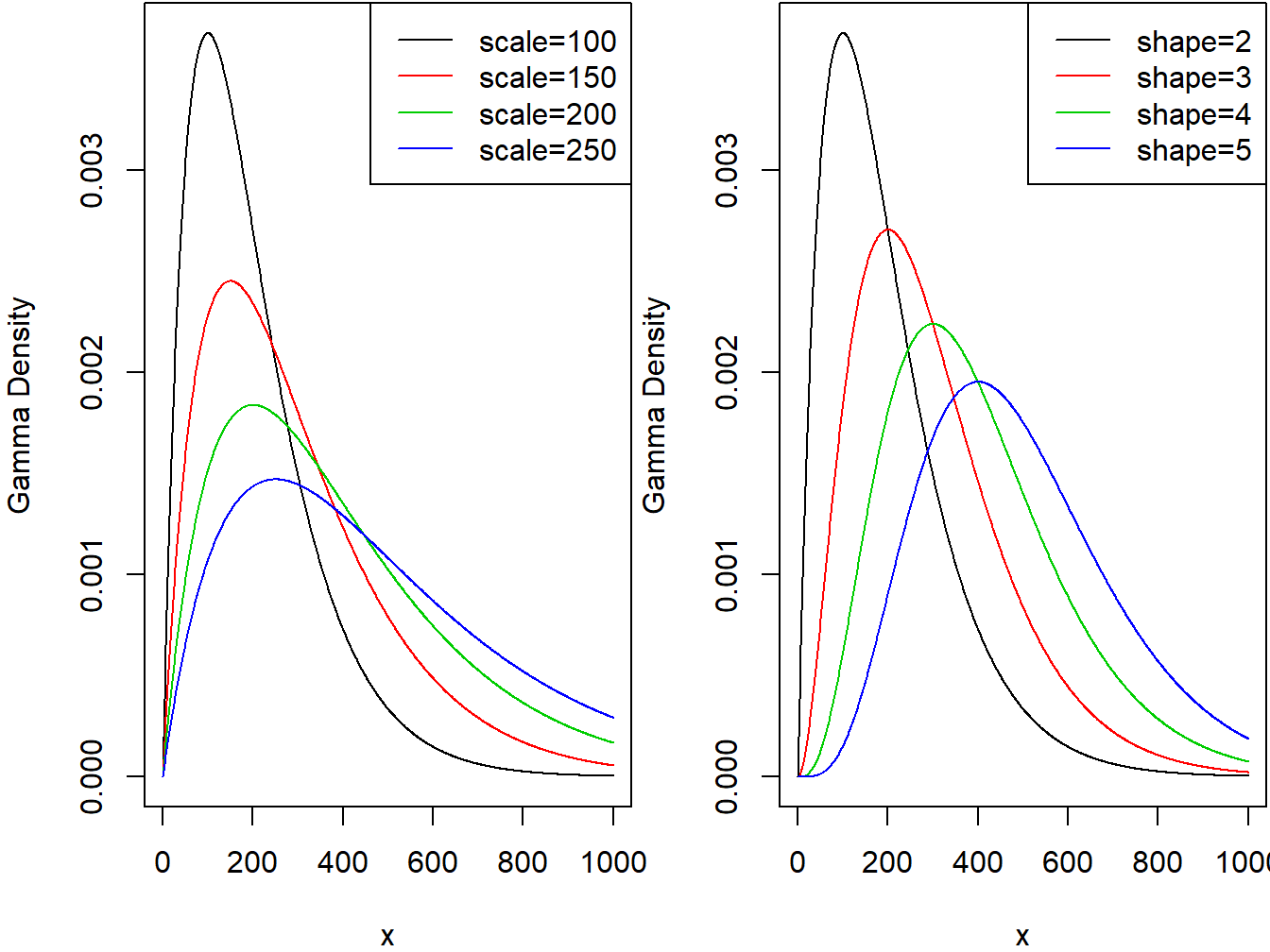}

}

\caption{Gamma Densities. The left-hand panel is with shape=2 and Varying Scale.
 The right-hand panel is with scale=100 and Varying Shape.}\label{fig:gammapdf}
\end{figure}

The two panels in Figure \ref{fig:gammapdf} demonstrate the effect of
the scale and shape parameters on the gamma density function.

R Code for Gamma Density Plots

\hypertarget{display.gammascale.2}{}
\begin{verbatim}
par(mfrow=c(1, 2), mar = c(4, 4, .1, .1))

# Varying Scale Gamma Densities
scaleparam <- seq(100, 250, by = 50)
shapeparam <- 2:5
x <- seq(0, 1000, by = 1)
fgamma <- dgamma(x, shape = 2, scale = scaleparam[1])
plot(x, fgamma, type = "l", ylab = "Gamma Density")
for(k in 2:length(scaleparam)){
  fgamma <- dgamma(x,shape = 2, scale = scaleparam[k])
  lines(x,fgamma, col = k)
}
legend("topright", c("scale=100", "scale=150", "scale=200", "scale=250"), lty=1, col = 1:4)

# Varying Shape Gamma Densities
fgamma <- dgamma(x, shape = shapeparam[1], scale = 100)
plot(x, fgamma, type = "l", ylab = "Gamma Density")
for(k in 2:length(shapeparam)){
  fgamma <- dgamma(x,shape = shapeparam[k], scale = 100)
  lines(x,fgamma, col = k)
}
legend("topright", c("shape=2", "shape=3", "shape=4", "shape=5"), lty=1, col = 1:4)
\end{verbatim}

When \(\alpha = 1\) the gamma reduces to an exponential distribution and
when \(\alpha = \frac{n}{2}\) and \(\theta = 2\) the gamma reduces to a
chi-square distribution with \(n\) degrees of freedom. As we will see in
Section \ref{MLEGrouped}, the chi-square distribution is used
extensively in statistical hypothesis testing.

The distribution function of the gamma model is the incomplete gamma
function, denoted by \(\Gamma\left( \frac{\alpha;x}{\theta} \right)\),
and defined as
\[F_{X}\left( x \right) = \Gamma\left( \alpha; \frac{x}{\theta} \right) = \frac{1}{\Gamma\left( \alpha \right)}\int_{0}^{x /\theta}t^{\alpha - 1}e^{- t}\text{dt}\]
\(\alpha > 0,\ \theta > 0\).

The \(k\)-th moment of the gamma distributed random variable for any
positive \(k\) is given by
\[\mathrm{E}\left( X^{k} \right) = \theta^{k} \frac{\Gamma\left( \alpha + k \right)}{\Gamma\left( \alpha \right)}  \ \ \ \text{for } k > 0 .\]
The mean and variance are given by
\(\mathrm{E}\left( X \right) = \alpha\theta\) and
\(\mathrm{Var}\left( X \right) = \alpha\theta^{2}\), respectively.

Since all moments exist for any positive \(k\), the gamma distribution
is considered a light tailed distribution, which may not be suitable for
modeling risky assets as it will not provide a realistic assessment of
the likelihood of severe losses.

\subsection{Pareto Distribution}\label{pareto-distribution}

The Pareto distribution, named after the Italian economist Vilfredo
Pareto (1843-1923), has many economic and financial applications. It is
a positively skewed and heavy-tailed distribution which makes it
suitable for modeling income, high-risk insurance claims and severity of
large casualty losses. The survival function of the Pareto distribution
which decays slowly to zero was first used to describe the distribution
of income where a small percentage of the population holds a large
proportion of the total wealth. For extreme insurance claims, the tail
of the severity distribution (losses in excess of a threshold) can be
modelled using a Pareto distribution.

\begin{figure}

{\centering \includegraphics[width=1.2\linewidth]{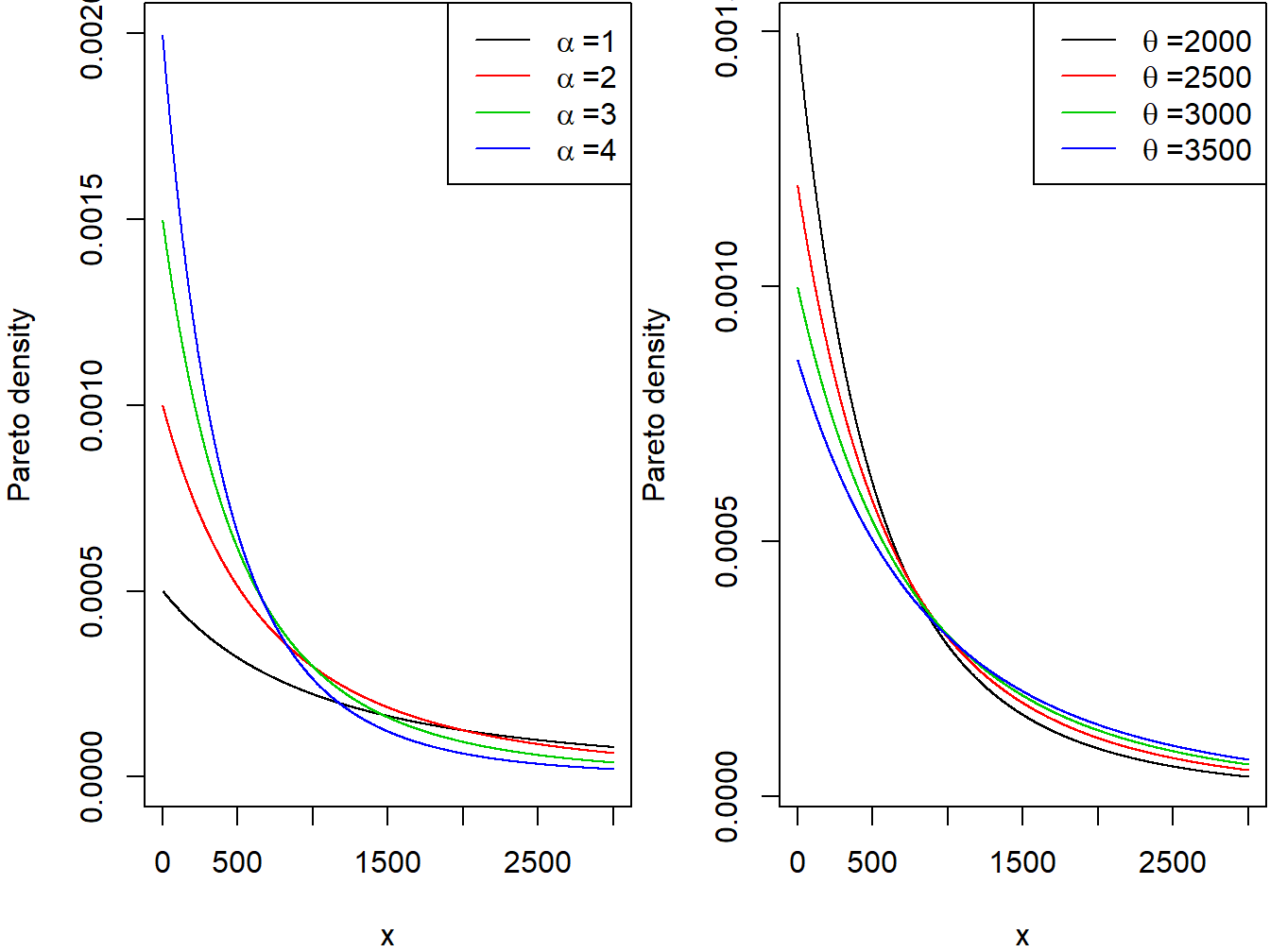}

}

\caption{Pareto Densities. The left-hand panel is with scale=2000 and Varying Shape.  The right-hand panel is with shape=3 and Varying Scale}\label{fig:Paretopdf}
\end{figure}

The continuous variable \(X\) is said to have the Pareto distribution
with shape parameter \(\alpha\) and scale parameter \(\theta\) if its
pdf is given by
\[f_{X}\left( x \right) = \frac{\alpha\theta^{\alpha}}{\left( x + \theta \right)^{\alpha + 1}} \ \ \  x  >  0,\ \alpha >  0,\ \theta > 0.\]
The two panels in Figure \ref{fig:Paretopdf} demonstrate the effect of
the scale and shape parameters on the Pareto density function.

\begin{verbatim}
## Loading required package: stats4
\end{verbatim}

\begin{verbatim}
## Loading required package: splines
\end{verbatim}

R Code for Pareto Density Plots

\hypertarget{display.Paretoscale.2}{}
\begin{verbatim}
par(mfrow=c(1, 2), mar = c(4, 4, .1, .1))

# Varying Shape Pareto Densities
x <- seq(1, 3000, by = 1)
scaleparam <- seq(2000, 3500, 500)
shapeparam <- 1:4

# varying the shape parameter
plot(x, dparetoII(x, loc=0, shape = shapeparam[1], scale = 2000), ylim=c(0,0.002),type = "l", ylab = "Pareto density")
for(k in 2:length(shapeparam)){
  lines(x, dparetoII(x, loc=0, shape = shapeparam[k], scale = 2000), col = k)
}
legend("topright", c(expression(alpha~'=1'), expression(alpha~'=2'), expression(alpha~'=3'), expression(alpha~'=4')), lty=1, col = 1:4)

# Varying Scale Pareto Densities
plot(x, dparetoII(x, loc=0, shape = 3, scale = scaleparam[1]), type = "l", ylab = "Pareto density")
for(k in 2:length(scaleparam)){
  lines(x, dparetoII(x, loc=0, shape = 3, scale = scaleparam[k]), col = k)
}
legend("topright", c(expression(theta~'=2000'), expression(theta~'=2500'), expression(theta~'=3000'), expression(theta~'=3500')), lty=1, col = 1:4)
\end{verbatim}

The distribution function of the Pareto distribution is given by
\[F_{X}\left( x \right) = 1 - \left( \frac{\theta}{x + \theta} \right)^{\alpha}  \ \ \ x > 0,\ \alpha > 0,\ \theta > 0.\]
It can be easily seen that the hazard function of the Pareto
distribution is a decreasing function in \(x\), another indication that
the distribution is heavy tailed.

The \(k\)-th moment of the Pareto distributed random variable exists, if
and only if, \(\alpha > k\). If \(k\) is a positive integer then
\[\mathrm{E}\left( X^{k} \right) = \frac{k!\theta^{k}}{\left( \alpha - 1 \right)\cdots\left( \alpha - k \right)} \ \ \ \alpha > k.\]
The mean and variance are given by
\[\mathrm{E}\left( X \right) = \frac{\theta}{\alpha - 1} \ \ \ \text{for } \alpha > 1\]
and
\[\mathrm{Var}\left( X \right) = \frac{\alpha\theta^{2}}{\left( \alpha - 1 \right)^{2}\left( \alpha - 2 \right)} \ \ \ \text{for } \alpha > 2,\]respectively.

\textbf{Example 3.2.1. } The claim size of an insurance portfolio
follows the Pareto distribution with mean and variance of 40 and 1800
respectively. Find

\begin{enumerate}
\def\labelenumi{\alph{enumi}.}
\tightlist
\item
  The shape and scale parameters.

  \begin{enumerate}
  \def\labelenumii{\alph{enumii}.}
  \setcounter{enumii}{1}
  \tightlist
  \item
    The 95-th percentile of this distribution.
  \end{enumerate}
\end{enumerate}

Show Example Solution

\hypertarget{toggleExampleLoss.2.1}{}
\textbf{Solution.}

\textbf{a.} As, \(X\sim Pa(\alpha,\theta)\), we have
\(\mathrm{E}\left( X \right) = \frac{\theta}{\alpha - 1} = 40\) and
\(\mathrm{Var}\left( X \right) = \frac{\alpha\theta^{2}}{\left( \alpha - 1 \right)^{2}\left( \alpha - 2 \right)} = 1800\).
By dividing the square of the first equation by the second we get
\(\frac{\alpha - 2}{\alpha} = \frac{40^{2}}{1800}\). Thus,
\(\alpha = 18.02\) and \(\theta = 680.72\).

\textbf{b.} The 95-th percentile, \(\pi_{0.95}\), satisfies the equation
\[F_{X}\left( \pi_{0.95} \right) = 1 - \left( \frac{680.72}{\pi_{0.95} + 680.72} \right)^{18.02} = 0.95.\]
Thus, \(\pi_{0.95} = 122.96\).

\begin{center}\rule{0.5\linewidth}{\linethickness}\end{center}

\subsection{Weibull Distribution}\label{weibull-distribution}

The Weibull distribution, named after the Swedish physicist Waloddi
Weibull (1887-1979) is widely used in reliability, life data analysis,
weather forecasts and general insurance claims. Truncated data arise
frequently in insurance studies. The Weibull distribution is
particularly useful in modeling left-truncated claim severity
distributions. Weibull was used to model excess of loss treaty over
automobile insurance as well as earthquake inter-arrival times.

\begin{figure}

{\centering \includegraphics[width=1.2\linewidth]{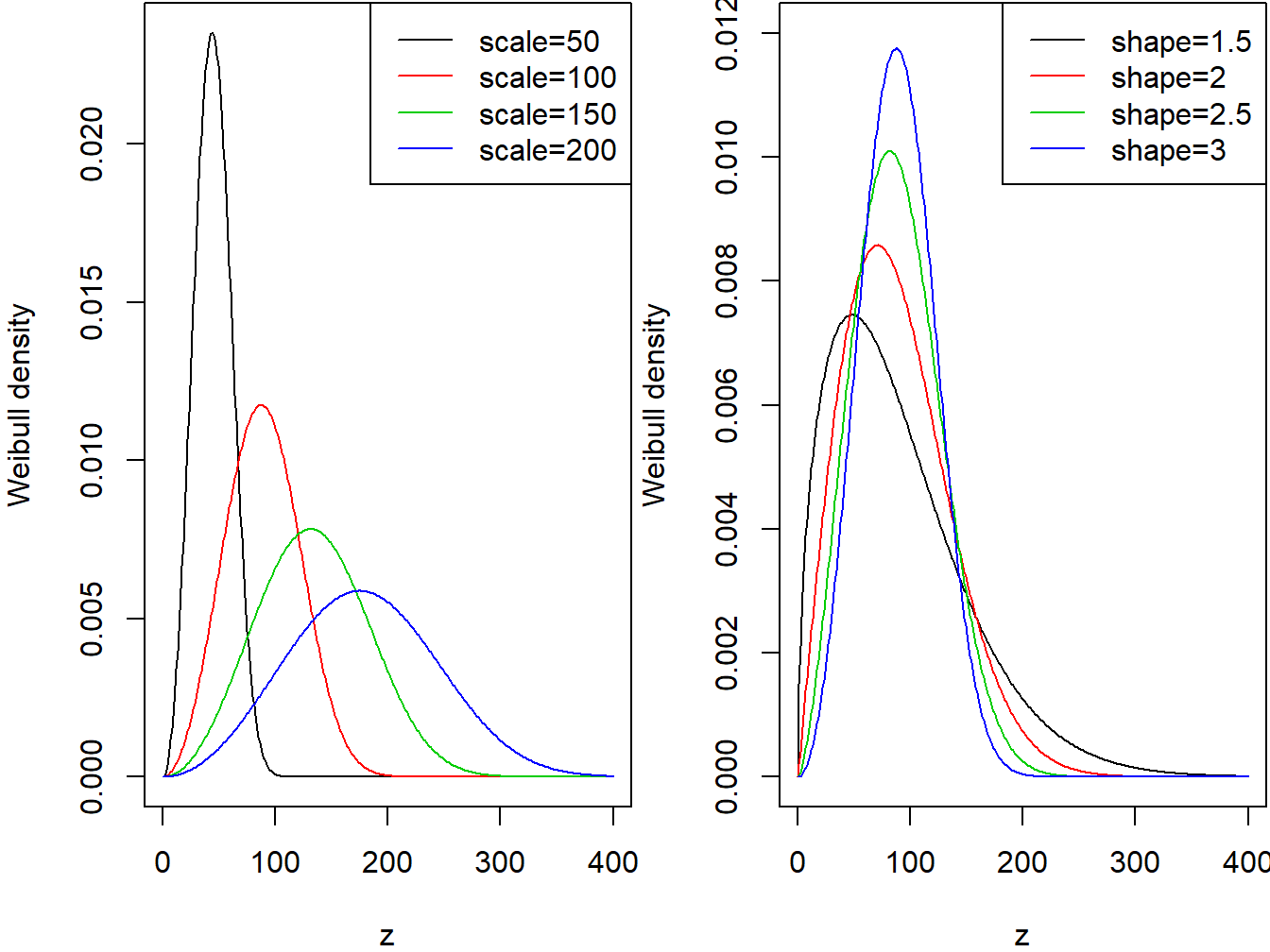}

}

\caption{Weibull Densities. The left-hand panel is with shape=3 and Varying Scale. The right-hand panel is with scale=100 and Varying Shape.}\label{fig:Weibullpdf}
\end{figure}

The continuous variable \(X\) is said to have the Weibull distribution
with shape parameter \(\alpha\) and scale parameter \(\theta\) if its
probability density function is given by
\[f_{X}\left( x \right) = \frac{\alpha}{\theta}\left( \frac{x}{\theta} \right)^{\alpha - 1} \exp \left(- \left( \frac{x}{\theta} \right)^{\alpha}\right) \ \ \ x > 0,\ \alpha > 0,\ \theta > 0.\]
The two panels Figure \ref{fig:Weibullpdf} demonstrate the effects of
the scale and shape parameters on the Weibull density function.

R Code for Weibull Density Plots

\hypertarget{display.ux20Weibullscale.2}{}
\begin{verbatim}
par(mfrow=c(1, 2), mar = c(4, 4, .1, .1))

# Varying Scale Weibull Densities
z<- seq(0,400,by=1)
scaleparam <- seq(50,200,50)
shapeparam <- seq(1.5,3,0.5)
plot(z, dweibull(z, shape = 3, scale = scaleparam[1]), type = "l", ylab = "Weibull density")
for(k in 2:length(scaleparam)){
  lines(z,dweibull(z,shape = 3, scale = scaleparam[k]), col = k)}
legend("topright", c("scale=50", "scale=100", "scale=150", "scale=200"), lty=1, col = 1:4)

# Varying Shape Weibull Densities
plot(z, dweibull(z, shape = shapeparam[1], scale = 100), ylim=c(0,0.012), type = "l", ylab = "Weibull density")
for(k in 2:length(shapeparam)){
  lines(z,dweibull(z,shape = shapeparam[k], scale = 100), col = k)}
legend("topright", c("shape=1.5", "shape=2", "shape=2.5", "shape=3"), lty=1, col = 1:4)
\end{verbatim}

The distribution function of the Weibull distribution is given by
\[F_{X}\left( x \right) = 1 - e^{- \left( x / \theta \right)^{\alpha}}  \ \ \ x >  0,\ \alpha >  0,\ \theta > 0.\]

It can be easily seen that the shape parameter \(\alpha\) describes the
shape of the hazard function of the Weibull distribution. The hazard
function is a decreasing function when \(\alpha < 1\), constant when
\(\alpha = 1\) and increasing when \(\alpha > 1\). This behavior of the
hazard function makes the Weibull distribution a suitable model for a
wide variety of phenomena such as weather forecasting, electrical and
industrial engineering, insurance modeling and financial risk analysis.

The \(k\)-th moment of the Weibull distributed random variable is given
by
\[\mathrm{E}\left( X^{k} \right) = \theta^{k}\Gamma\left( 1 + \frac{k}{\alpha} \right) .\]

The mean and variance are given by
\[\mathrm{E}\left( X \right) = \theta\Gamma\left( 1 + \frac{1}{\alpha} \right)\]
and
\[\mathrm{Var}(X)= \theta^{2}\left( \Gamma\left( 1 + \frac{2}{\alpha} \right)  - \left\lbrack \Gamma\left( 1 + \frac{1}{\alpha} \right) \right\rbrack  ^{2}\right),\]
respectively.

\textbf{Example 3.2.2.} Suppose that the probability distribution of the
lifetime of AIDS patients (in months) from the time of diagnosis is
described by the Weibull distribution with shape parameter 1.2 and scale
parameter 33.33.

\begin{enumerate}
\def\labelenumi{\alph{enumi}.}
\tightlist
\item
  Find the probability that a randomly selected person from this
  population survives at least 12 months,

  \begin{enumerate}
  \def\labelenumii{\alph{enumii}.}
  \setcounter{enumii}{1}
  \tightlist
  \item
    A random sample of 10 patients will be selected from this
    population. What is the probability that at most two will die within
    one year of diagnosis.

    \begin{enumerate}
    \def\labelenumiii{\alph{enumiii}.}
    \setcounter{enumiii}{2}
    \tightlist
    \item
      Find the 99-th percentile of this distribution.
    \end{enumerate}
  \end{enumerate}
\end{enumerate}

Show Example Solution

\hypertarget{toggleExampleLoss.2.2}{}
\textbf{Solution.}

\textbf{a.} \$Let \(X \sim Wei \left( 1.2,33.33 \right)\) be the
lifetime of AIDS patients (in months). We have,
\[{\Pr\left( X \geq 12 \right) = S}_{X}\left( 12 \right) = e^{- \left( \frac{12}{33.33} \right)^{1.2}} = 0.746.\]
\textbf{b.} \$Let \(Y\) be the number of patients who die within one
year of diagnosis. Then, \(Y\sim Bin\left( 10,\ 0.254 \right)\) and
\(\Pr\left( Y \leq 2 \right) = 0.514.\)

\textbf{c.} \$Let \(\pi_{0.99}\) denote the 99-th percentile of this
distribution. Then,
\[S_{X}\left( \pi_{0.99} \right) = \exp\left\{- \left( \frac{\pi_{0.99}}{33.33} \right)^{1.2}\right\} = 0.01.\]
Solving for \(\pi_{0.99}\), we get \(\pi_{0.99} = 118.99\).

\begin{center}\rule{0.5\linewidth}{\linethickness}\end{center}

\subsection{The Generalized Beta Distribution of the Second
Kind}\label{the-generalized-beta-distribution-of-the-second-kind}

The Generalized Beta Distribution of the Second Kind (GB2) was
introduced by \citet{venter1983transformed} in the context of insurance
loss modeling and by \citet{mcdonald1984some} as an income and wealth
distribution. It is a four-parameter very flexible distribution that can
model positively as well as negatively skewed distributions.

The continuous variable \(X\) is said to have the GB2 distribution with
parameters \(a\), \(b\), \(\alpha\) and \(\beta\) if its probability
density function is given by
\[f_{X}\left( x \right) = \frac{ax^{a \alpha - 1}}{b^{a \alpha}B\left( \alpha,\beta \right)\left\lbrack 1 + \left( x/b \right)^{a} \right\rbrack^{\alpha + \beta}} \ \ \ \text{for } x > 0,\]
\(a,b,\alpha,\beta > 0\), and where the beta function
\(B\left( \alpha,\beta \right)\) is defined as
\[B\left( \alpha,\beta \right) = \int_{0}^{1}{t^{\alpha - 1}\left( 1 - t \right)^{\beta - 1}}\text{dt}.\]

The GB2 provides a model for heavy as well as light tailed data. It
includes the exponential, gamma, Weibull, Burr, Lomax, F, chi-square,
Rayleigh, lognormal and log-logistic as special or limiting cases. For
example, by setting the parameters \(a = \alpha = \beta = 1\), then the
GB2 reduces to the log-logistic distribution. When \(a = 1\) and
\(\beta \rightarrow \infty\), it reduces to the gamma distribution and
when \(\alpha = 1\) and \(\beta \rightarrow \infty\), it reduces to the
Weibull distribution.

The \(k\)-th moment of the GB2 distributed random variable is given by
\[\mathrm{E}\left( X^{k} \right) = \frac{b^{k}\left( \alpha + \frac{k}{a},\beta - \frac{k}{a} \right)}{\left( \alpha,\beta \right)}, \ \ \ k > 0.\]
Earlier applications of the GB2 were on income data and more recently
have been used to model long-tailed claims data. GB2 was used to model
different types of automobile insurance claims, severity of fire losses
as well as medical insurance claim data.

\section{Methods of Creating New Distributions}\label{MethodsCreation}

In this section we

understand connections among the distributions;

give insights into when a distribution is preferred when compared to
alternatives;

provide foundations for creating new distributions.

\subsection{Functions of Random Variables and their
Distributions}\label{functions-of-random-variables-and-their-distributions}

In Section \ref{S:ContinuousDistn} we discussed some elementary known
distributions. In this section we discuss means of creating new
parametric probability distributions from existing ones. Let \(X\) be a
continuous random variable with a known probability density function
\(f_{X}(x)\) and distribution function \(F_{X}(x)\). Consider the
transformation \(Y = g\left( X \right)\), where \(g(X)\) is a one-to-one
transformation defining a new random variable \(Y\). We can use the
distribution function technique, the change-of-variable technique or the
moment-generating function technique to find the probability density
function of the variable of interest \(Y\). In this section we apply the
following techniques for creating new families of distributions: (a)
multiplication by a constant (b) raising to a power, (c) exponentiation
and (d) mixing.

\subsection{Multiplication by a
Constant}\label{multiplication-by-a-constant}

If claim data show change over time then such transformation can be
useful to adjust for inflation. If the level of inflation is positive
then claim costs are rising, and if it is negative then costs are
falling. To adjust for inflation we multiply the cost \(X\) by 1+
inflation rate (negative inflation is deflation). To account for
currency impact on claim costs we also use a transformation to apply
currency conversion from a base to a counter currency.

Consider the transformation \(Y = cX\), where \(c > 0\), then the
distribution function of \(Y\) is given by
\[F_{Y}\left( y \right) = \Pr\left( Y \leq y \right) = \Pr\left( cX \leq y \right) = \Pr\left( X \leq \frac{y}{c} \right) = F_{X}\left( \frac{y}{c} \right).\]
Hence, the probability density function of interest \(f_{Y}(y)\) can be
written as
\[f_{Y}\left( y \right) = \frac{1}{c}f_{X}\left( \frac{y}{c} \right).\]
Suppose that \(X\) belongs to a certain set of parametric distributions
and define a rescaled version \(Y\  = \ cX\), \(c\  > \ 0\). If \(Y\) is
in the same set of distributions then the distribution is said to be a
scale distribution. When a member of a scale distribution is multiplied
by a constant \(c\) (\(c > 0\)), the scale parameter for this scale
distribution meets two conditions:

The parameter is changed by multiplying by \(c\);

All other parameter remain unchanged.

\textbf{Example 3.3.1. SOA Exam Question.} The aggregate losses of
Eiffel Auto Insurance are denoted in Euro currency and follow a
Lognormal distribution with \(\mu = 8\) and \(\sigma = 2\). Given that 1
euro \(=\) 1.3 dollars, find the set of lognormal parameters, which
describe the distribution of Eiffel's losses in dollars?

Show Example Solution

\hypertarget{toggleExampleLoss.3.1}{}
\textbf{Solution.}

Let \(X\) and \(Y\) denote the aggregate losses of Eiffel Auto Insurance
in euro currency and dollars respectively. As \(Y = 1.3X\), we have,
\[F_{Y}\left( y \right) = \Pr\left( Y \leq y \right) = \Pr\left( 1.3X \leq y \right) = \Pr\left( X \leq \frac{y}{1.3} \right) = F_{X}\left( \frac{y}{1.3} \right).\]

\(X\) follows a lognormal distribution with parameters \(\mu = 8\) and
\(\sigma = 2\). The probability density function of \(X\) is given by
\[f_{X}\left( x \right) = \frac{1}{x \sigma \sqrt{2\pi}}\exp \left\{- \frac{1}{2}\left( \frac{\ln x - \mu}{\sigma} \right)^{2}\right\} \ \ \ \text{for } x > 0.\]
As \(\left| \frac{dx}{dy} \right| = \frac{1}{1.3}\), the probability
density function of interest \(f_{Y}(y)\) is
\[f_{Y}\left( y \right) = \frac{1}{1.3}f_{X}\left( \frac{y}{1.3} \right) \\
= \frac{1}{1.3}\frac{1.3}{y \sigma \sqrt{2\pi}}\exp \left\{- \frac{1}{2}\left( \frac{\ln\left( y/1.3 \right) - \mu}{\sigma} \right)^{2}\right\} \\
= \frac{1}{y \sigma\sqrt{2\pi}}\exp \left\{- \frac{1}{2}\left( \frac{\ln y - \left( \ln 1.3 + \mu \right)}{\sigma} \right)^{2}\right\}.\]
Then \(Y\) follows a lognormal distribution with parameters
\(\ln 1.3 + \mu = 8.26\) and \(\sigma = 2.00\). If we let
\(\mu = ln(m)\) then it can be easily seen that \(m\)=\(e^{\mu}\) is the
scale parameter which was multiplied by 1.3 while \(\sigma\) is the
shape parameter that remained unchanged.

\begin{center}\rule{0.5\linewidth}{\linethickness}\end{center}

\textbf{Example 3.3.2. SOA Exam Question.} Demonstrate that the gamma
distribution is a scale distribution.

Show Example Solution

\hypertarget{toggleExampleLoss.3.2}{}
\textbf{Solution.}

Let \(X\sim Ga(\alpha,\theta)\) and \(Y = cX\). As
\(\left| \frac{dx}{dy} \right| = \frac{1}{c}\), then
\[f_{Y}\left( y \right) = \frac{1}{c}f_{X}\left( \frac{y}{c} \right) = \frac{\left( \frac{y}{c\theta} \right)^{\alpha}}{y\Gamma\left( \alpha \right)}\exp \left( - \frac{y}{c\theta} \right)  .\]
We can see that \(Y\sim Ga(\alpha,c\theta)\) indicating that gamma is a
scale distribution and \(\theta\) is a scale parameter.

\begin{center}\rule{0.5\linewidth}{\linethickness}\end{center}

\subsection{Raising to a Power}\label{raising-to-a-power}

In the previous section we have talked about the flexibility of the
Weibull distribution in fitting reliability data. Looking to the origins
of the Weibull distribution, we recognize that the Weibull is a power
transformation of the exponential distribution. This is an application
of another type of transformation which involves raising the random
variable to a power.

Consider the transformation \(Y = X^{\tau}\), where \(\tau > 0\), then
the distribution function of \(Y\) is given by
\[F_{Y}\left( y \right) = \Pr\left( Y \leq y \right) = \Pr\left( X^{\tau} \leq y \right) = \Pr\left( X \leq y^{1/ \tau} \right) = F_{X}\left( y^{1/ \tau} \right).\]

Hence, the probability density function of interest \(f_{Y}(y)\) can be
written as
\[f_{Y}(y) = \frac{1}{\tau} y^{1/ \tau - 1} f_{X}\left( y^{1/ \tau} \right).\]
On the other hand, if \(\tau < 0\), then the distribution function of
\(Y\) is given by
\[F_{Y}\left( y \right) = \Pr\left( Y \leq y \right) = \Pr\left( X^{\tau} \leq y \right) = \Pr\left( X \geq y^{1/ \tau} \right) = 1 - F_{X}\left( y^{1/ \tau} \right), \]
and
\[f_{Y}(y) = \left| \frac{1}{\tau} \right|{y^{1/ \tau - 1}f}_{X}\left( y^{1/ \tau} \right).\]

\textbf{Example 3.3.3.} We assume that \(X\) follows the exponential
distribution with mean \(\theta\) and consider the transformed variable
\(Y = X^{\tau}\). Show that \(Y\) follows the Weibull distribution when
\(\tau\) is positive and determine the parameters of the Weibull
distribution.

Show Example Solution

\hypertarget{toggleExampleLoss.3.3}{}
\textbf{Solution.}

As \(\ X\sim Exp(\theta)\), we have
\[f_{X}(x) = \frac{1}{\theta}e^{- x/ \theta} \ \ \ \, x > 0.\] Solving
for \emph{x} yields \(x = y^{1/\tau}\). Taking the derivative, we have
\[\left| \frac{dx}{dy} \right| = \frac{1}{\tau}{y^{\frac{1}{\tau}} - 1}.\]
Thus,
\[f_{Y}\left( y \right) = \frac{1}{\tau}{y^{\frac{1}{\tau} - 1}f}_{X}\left( y^{\frac{1}{\tau}} \right) \\
= \frac{1}{\tau \theta }y^{\frac{1}{\tau} - 1}e^{- \frac{y^{\frac{1}{\tau}}}{\theta}} = \frac{\alpha}{\beta}\left( \frac{y}{\beta} \right)^{\alpha - 1}e^{- \left( y/ \beta \right)^{\alpha}}.\]
where \(\alpha = \frac{1}{\tau}\) and \(\beta = \theta^{\tau}\). Then,
\(Y\) follows the Weibull distribution with shape parameter \(\alpha\)
and scale parameter \(\beta\).

\begin{center}\rule{0.5\linewidth}{\linethickness}\end{center}

\subsection{Exponentiation}\label{exponentiation}

The normal distribution is a very popular model for a wide number of
applications and when the sample size is large, it can serve as an
approximate distribution for other models. If the random variable \(X\)
has a normal distribution with mean \(\mu\) and variance \(\sigma^{2}\),
then \(Y = e^{X}\) has lognormal distribution with parameters \(\mu\)
and \(\sigma^{2}\). The lognormal random variable has a lower bound of
zero, is positively skewed and has a long right tail. A lognormal
distribution is commonly used to describe distributions of financial
assets such as stock prices. It is also used in fitting claim amounts
for automobile as well as health insurance. This is an example of
another type of transformation which involves exponentiation.

Consider the transformation \(Y = e^{X}\), then the distribution
function of \(Y\) is given by
\[F_{Y}\left( y \right) = \Pr\left( Y \leq y \right) = \Pr\left( e^{X} \leq y \right) = \Pr\left( X \leq \ln y \right) = F_{X}\left( \ln y \right).\]
Hence, the probability density function of interest \(f_{Y}(y)\) can be
written as \[f_{Y}(y) = \frac{1}{y}f_{X}\left( \ln y \right).\]

\textbf{Example 3.3.4. SOA Exam Question.} \(X\) has a uniform
distribution on the interval \((0,\ c)\). \(Y = e^{X}\). Find the
distribution of \(Y\).

Show Example Solution

\hypertarget{toggleExampleLoss.3.4}{}
\textbf{Solution.}

We begin with the \emph{cdf} of \(Y\),
\[F_{Y}\left( y \right) = \Pr\left( Y \leq y \right) = \Pr\left( e^{X} \leq y \right) = \Pr\left( X \leq \ln y \right) = F_{X}\left( \ln y \right).\]
Taking the derivative, we have,
\[f_{Y}\left( y \right) = \frac{1}{y}f_{X}\left(\ln y \right) = \frac{1}{\text{cy}}. \]
Since \(0 < x < c\), then \(1 < y < e^{c}\).

\begin{center}\rule{0.5\linewidth}{\linethickness}\end{center}

\subsection{Finite Mixtures}\label{finite-mixtures}

Mixture distributions represent a useful way of modelling data that are
drawn from a heterogeneous population. This parent population can be
thought to be divided into multiple subpopulations with distinct
distributions.

\subsubsection{Two-point Mixture}\label{two-point-mixture}

If the underlying phenomenon is diverse and can actually be described as
two phenomena representing two subpopulations with different modes, we
can construct the two point mixture random variable \(X\). Given random
variables \(X_{1}\) and \(X_{2}\), with probability density functions
\(f_{X_{1}}\left( x \right)\) and \(f_{X_{2}}\left( x \right)\)
respectively, the probability density function of \(X\) is the weighted
average of the component probability density function
\(f_{X_{1}}\left( x \right)\) and \(f_{X_{2}}\left( x \right)\). The
probability density function and distribution function of \(X\) are
given by
\[f_{X}\left( x \right) = af_{X_{1}}\left( x \right) + \left( 1 - a \right)f_{X_{2}}\left( x \right),\]
and
\[F_{X}\left( x \right) = aF_{X_{1}}\left( x \right) + \left( 1 - a \right)F_{X_{2}}\left( x \right),\]

for \(0 < a <1\), where the mixing parameters \(a\) and \((1 - a)\)
represent the proportions of data points that fall under each of the two
subpopulations respectively. This weighted average can be applied to a
number of other distribution related quantities. The \emph{k}-th moment
and moment generating function of \(X\) are given by
\(\mathrm{E}\left( X^{k} \right) = a\mathrm{E}\left( X_{1}^{K} \right) + \left( 1 - a \right)\mathrm{E}\left( X_{2}^{k} \right)\),
and
\[M_{X}\left( t \right) = aM_{X_{1}}\left( t \right) + \left( 1 - a \right)M_{X_{2}}\left( t \right),\]
respectively.

\textbf{Example 3.3.5. SOA Exam Question.} The distribution of the
random variable \(X\) is an equally weighted mixture of two Poisson
distributions with parameters \(\lambda_{1}\) and \(\lambda_{2}\)
respectively. The mean and variance of \(X\) are 4 and 13, respectively.
Determine \(\Pr\left( X > 2 \right)\).

Show Example Solution

\hypertarget{toggleExampleLoss.3.5}{}
\textbf{Solution.}

\[\mathrm{E}\left( X \right) = 0.5\lambda_{1} + 0.5\lambda_{2} = 4\] For
a Poisson random variable \(X\) with parameter \(\lambda\), we have
\(\mathrm{E~}X^2 = \mathrm{Var}(X) + [\mathrm{E}(X)]^2 = \lambda + \lambda^2\).
Thus,

\[\mathrm{E}\left( X^{2} \right) = 0.5\left( \lambda_{1} + \lambda_{1}^{2} \right) + 0.5\left( \lambda_{2} + \lambda_{2}^{2} \right) = 13 + 16.\]

Substituting the first in the second equation, we get
\(\lambda_{1} + \lambda_{2} = 8\) and
\(\lambda_{1}^{2} + \lambda_{2}^{2} = 50\). After further substitution,
we get that the parameters of the two Poisson distributions are 1 and 7,
respectively. So,
\[\Pr\left( X > 2 \right) = 0.5\Pr\left( X_{1} > 2 \right) + 0.5\Pr\left( X_{2} > 2 \right) = 0.05 .\]

\begin{center}\rule{0.5\linewidth}{\linethickness}\end{center}

\subsubsection{\texorpdfstring{\emph{k}-point
Mixture}{k-point Mixture}}\label{k-point-mixture}

In case of finite mixture distributions, the random variable of interest
\(X\) has a probability \(p_{i}\) of being drawn from homogeneous
subpopulation \(i\), where \(i = 1,2,\ldots,k\) and \(k\) is the
initially specified number of subpopulations in our mixture. The mixing
parameter \(p_{i}\) represents the proportion of observations from
subpopulation \(i\). Consider the random variable \(X\) generated from
\(k\) distinct subpopulations, where subpopulation \(i\) is modeled by
the continuous distribution \(f_{X_{i}}\left( x \right)\). The
probability distribution of \(X\) is given by
\[f_{X}\left( x \right) = \sum_{i = 1}^{k}{p_{i}f_{X_{i}}\left( x \right)},\]
where \(0 < p_{i} < 1\) and \(\sum_{i = 1}^{k} p_{i} = 1\).

This model is often referred to as a \emph{finite mixture} or a \(k\)
point mixture. The distribution function, \(r\)-th moment and moment
generating functions of the \(k\)-th point mixture are given as

\[F_{X}\left( x \right) = \sum_{i = 1}^{k}{p_{i}F_{X_{i}}\left( x \right)},\]
\[\mathrm{E}\left( X^{r} \right) = \sum_{i = 1}^{k}{p_{i}\mathrm{E}\left( X_{i}^{r} \right)}, \text{and}\]
\[M_{X}\left( t \right) = \sum_{i = 1}^{k}{p_{i}M_{X_{i}}\left( t \right)},\]
respectively.

\textbf{Example 3.3.6. SOA Exam Question.} \(Y_{1}\) is a mixture of
\(X_{1}\) and \(X_{2}\) with mixing weights \(a\) and \((1 - a)\).
\(Y_{2}\) is a mixture of \(X_{3}\) and \(X_{4}\) with mixing weights
\(b\) and \((1 - b)\). \(Z\) is a mixture of \(Y_{1}\) and \(Y_{2}\)
with mixing weights \(c\) and \((1 - c)\).

Show that \(Z\) is a mixture of \(X_{1}\), \(X_{2}\), \(X_{3}\) and
\(X_{4}\), and find the mixing weights.

Show Example Solution

\hypertarget{toggleExampleLoss.3.6}{}
\textbf{Solution.} Applying the formula for a mixed distribution, we get
\[f_{Y_{1}}\left( x \right) = af_{X_{1}}\left( x \right) + \left( 1 - a \right)f_{X_{2}}\left( x \right)\]

\[f_{Y_{2}}\left( x \right) = bf_{X_{3}}\left( x \right) + \left( 1 - b \right)f_{X_{4}}\left( x \right)\]

\[f_{Z}\left( x \right) = cf_{Y_{1}}\left( x \right) + \left( 1 - c \right)f_{Y_{2}}\left( x \right)\]

Substituting the first two equations into the third, we get

\[f_{Z}\left( x \right) = c\left\lbrack af_{X_{1}}\left( x \right) + \left( 1 - a \right)f_{X_{2}}\left( x \right) \right\rbrack + \left( 1 - c \right)\left\lbrack bf_{X_{3}}\left( x \right) + \left( 1 - b \right)f_{X_{4}}\left( x \right) \right\rbrack\]

\[= caf_{X_{1}}\left( x \right) + c\left( 1 - a \right)f_{X_{2}}\left( x \right) + \left( 1 - c \right)bf_{X_{3}}\left( x \right) + (1 - c)\left( 1 - b \right)f_{X_{4}}\left( x \right)\].

Then, \(Z\) is a mixture of \(X_{1}\), \(X_{2}\), \(X_{3}\) and
\(X_{4}\), with mixing weights \(\text{ca}\), \(c\left( 1 - a \right)\),
\(\left( 1 - c \right)b\) and \((1 - c)\left( 1 - b \right)\),
respectively.

\begin{center}\rule{0.5\linewidth}{\linethickness}\end{center}

\subsection{Continuous Mixtures}\label{continuous-mixtures}

A mixture with a very large number of subpopulations (\(k\) goes to
infinity) is often referred to as a continuous mixture. In a continuous
mixture, subpopulations are not distinguished by a discrete mixing
parameter but by a continuous variable \(\theta\), where \(\theta\)
plays the role of \(p_{i}\) in the finite mixture. Consider the random
variable \(X\) with a distribution depending on a parameter \(\theta\),
where \(\theta\) itself is a continuous random variable. This
description yields the following model for \(X\)
\[f_{X}\left( x \right) = \int_{0}^{\infty}{f_{X}\left( x\left| \theta \right.\  \right)g\left( \theta \right)} d \theta ,\]
where \(f_{X}\left( x\left| \theta \right.\  \right)\) is the
conditional distribution of \(X\) at a particular value of \(\theta\)
and \(g\left( \theta \right)\) is the probability statement made about
the unknown parameter \(\theta\), known as the prior distribution of
\(\theta\) (the prior information or expert opinion to be used in the
analysis).

The distribution function, \(k\)-th moment and moment generating
functions of the continuous mixture are given as
\[F_{X}\left( x \right) = \int_{-\infty}^{\infty}{F_{X}\left( x\left| \theta \right.\  \right)g\left( \theta \right)} d \theta,\]
\[\mathrm{E}\left( X^{k} \right) = \int_{-\infty}^{\infty}{\mathrm{E}\left( X^{k}\left| \theta \right.\  \right)g\left( \theta \right)}d \theta,\]
\[M_{X}\left( t \right) = \mathrm{E}\left( e^{t X} \right) = \int_{-\infty}^{\infty}{\mathrm{E}\left( e^{ tx}\left| \theta \right.\  \right)g\left( \theta \right)}d \theta, \]
respectively.

The \(k\)-th moment of the mixture distribution can be rewritten as
\[\mathrm{E}\left( X^{k} \right) = \int_{-\infty}^{\infty}{\mathrm{E}\left( X^{k}\left| \theta \right.\  \right)g\left( \theta \right)}d\theta = \mathrm{E}\left\lbrack \mathrm{E}\left( X^{k}\left| \theta \right.\  \right) \right\rbrack .\]

In particular the mean and variance of \(X\) are given by
\[\mathrm{E}\left( X \right) = \mathrm{E}\left\lbrack \mathrm{E}\left( X\left| \theta \right.\  \right) \right\rbrack\]
and
\[\mathrm{Var}\left( X \right) = \mathrm{E}\left\lbrack \mathrm{Var}\left( X\left| \theta \right.\  \right) \right\rbrack + \mathrm{Var}\left\lbrack \mathrm{E}\left( X\left| \theta \right.\  \right) \right\rbrack .\]

\textbf{Example 3.3.7. SOA Exam Question.} \(X\) has a binomial
distribution with a mean of \(100q\) and a variance of
\(100q\left( 1 - q \right)\) and \(q\) has a beta distribution with
parameters \(a = 3\) and \(b = 2\). Find the unconditional mean and
variance of \(X\).

Show Example Solution

\hypertarget{toggleExampleLoss.3.7}{}
\textbf{Solution.}

As \(q\sim Beta(3,2)\), we have
\(\mathrm{E}\left( q \right) = \frac{a}{a + b} = \frac{3}{5}\) and
\(\mathrm{E}\left( q^{2} \right) = \frac{a\left( a + 1 \right)}{\left( a + b \right)\left( a + b + 1 \right)} = \frac{2}{5}\).

Now, using the formulas for the unconditional mean and variance, we have
\[\mathrm{E}\left( X \right) = \mathrm{E}\left\lbrack E\left( X\left| q \right.\  \right) \right\rbrack = \mathrm{E}\left( 100q \right) = 100\mathrm{E}\left( q \right) = 60\]
\[\mathrm{Var}\left( X \right) = \mathrm{E}\left\lbrack \mathrm{Var}\left( X\left| q \right.\  \right) \right\rbrack + \mathrm{Var}\left\lbrack E\left( X\left| q \right.\  \right) \right\rbrack = \mathrm{E}\left\lbrack 100q\left( 1 - q \right) \right\rbrack + \mathrm{Var}\left( 100q \right)\]

\[= 100\mathrm{E}\left( q \right) - 100\mathrm{E}\left( q^{2} \right) + 100^{2}\mathrm{V}\left( q \right) = 420\].

\begin{center}\rule{0.5\linewidth}{\linethickness}\end{center}

\textbf{Example 3.3.8. SOA Exam Question.} Claim sizes, \(X\), are
uniform on for each policyholder. varies by policyholder according to an
exponential distribution with mean 5. Find the unconditional
distribution, mean and variance of \(X\).

Show Example Solution

\hypertarget{toggleExampleLoss.3.8}{}
\textbf{Solution.}

The conditional distribution of \(X\) is
\(f_{X}\left( \left. \ x \right|\theta \right) = \frac{1}{10}\) for
\(\theta < x < \theta + 10\).

The prior distribution of \(\theta\) is
\(g\left( \theta \right) = \frac{1}{5}e^{- \frac{\theta}{5}}\) for
\(0 < \theta < \infty\).

The conditional mean and variance of \(X\) are given by
\[\mathrm{E}\left( \left. \ X \right|\theta \right) = \frac{\theta + \theta + 10}{2} = \theta + 5\]
and
\[\mathrm{Var}\left( \left. \ X \right|\theta \right) = \frac{\left\lbrack \left( \theta + 10 \right) - \theta \right\rbrack^{2}}{12} = \frac{100}{12}, \]
respectively.

Hence, the unconditional mean and variance of \(X\) are given by
\[\mathrm{E}\left( X \right) = \mathrm{E}\left\lbrack \mathrm{E}\left( X\left| \theta \right.\  \right) \right\rbrack = \mathrm{E}\left( \theta + 5 \right) = \mathrm{E}\left( \theta \right) + 5 = 5 + 5 = 10,\]
and
\[\mathrm{Var}\left( X \right) = \mathrm{E}\left\lbrack V\left( X\left| \theta \right.\  \right) \right\rbrack + \mathrm{Var}\left\lbrack \mathrm{E}\left( X\left| \theta \right.\  \right) \right\rbrack \\
= \mathrm{E}\left( \frac{100}{12} \right) + \mathrm{Var}\left( \theta + 5 \right) = 8.33 + \mathrm{Var}\left( \theta \right) = 33.33. \]
The unconditional distribution of \(X\) is
\[f_{X}\left( x \right) = \int_{}^{}{f_{X}\left( x |\theta \right) ~g\left( \theta \right) d \theta} .\]

\[f_{X}\left( x \right) = \left\{ \begin{matrix}
\int_{0}^{x}{\frac{1}{50}e^{- \frac{\theta}{5}}d\theta = \frac{1}{10}\left( 1 - e^{- \frac{x}{5}} \right)} & 0 \leq x \leq 10, \\
\int_{x - 10}^{x}{\frac{1}{50}e^{- \frac{\theta}{5}} d\theta} = \frac{1}{10}\left( e^{- \frac{\left( x - 10 \right)}{5}} - e^{- \frac{x}{5}} \right) & 10 < x < \infty. \\
\end{matrix} \right.\ \]

\begin{center}\rule{0.5\linewidth}{\linethickness}\end{center}

\section{Coverage Modifications}\label{S:CoverageModifications}

In this section we evaluate the impacts of coverage modifications: a)
deductibles, b) policy limit, c) coinsurance and inflation on insurer's
costs.

\subsection{Policy Deductibles}\label{S:PolicyDeduct}

Under an ordinary deductible policy, the insured (policyholder) agrees
to cover a fixed amount of an insurance claim before the insurer starts
to pay. This fixed expense paid out of pocket is called the deductible
and often denoted by \(d\). The insurer is responsible for covering the
loss \(X\) less the deductible \(d\). Depending on the agreement, the
deductible may apply to each covered loss or to a defined benefit period
(month, year, etc.)

Deductibles eliminate a large number of small claims, reduce costs of
handling and processing these claims, reduce premiums for the
policyholders and reduce moral hazard. Moral hazard occurs when the
insured takes more risks, increasing the chances of loss due to perils
insured against, knowing that the insurer will incur the cost (e.g.~a
policyholder with collision insurance may be encouraged to drive
recklessly). The larger the deductible, the less the insured pays in
premiums for an insurance policy.

Let \(X\) denote the loss incurred to the insured and \(Y\) denote the
amount of paid claim by the insurer. Speaking of the benefit paid to the
policyholder, we differentiate between two variables: The payment per
loss and the payment per payment. The payment per loss variable, denoted
by \(Y^{L}\), includes losses for which a payment is made as well as
losses less than the deductible and hence is defined as
\[Y^{L} = \left( X - d \right)_{+}
= \left\{ \begin{array}{cc}
0 & X < d, \\
X - d & X > d
\end{array} \right. .\] \(Y^{L}\) is often referred to as left censored
and shifted variable because the values below \(d\) are not ignored and
all losses are shifted by a value \(d\).

On the other hand, the payment per payment variable, denoted by
\(Y^{P}\), is not defined when there is no payment and only includes
losses for which a payment is made. The variable is defined as
\[Y^{P} = \left\{ \begin{matrix}
\text{Undefined} & X \le d \\
X - d & X > d
\end{matrix} \right. \] \(Y^{P}\) is often referred to as left truncated
and shifted variable or excess loss variable because the claims smaller
than \(d\) are not reported and values above \(d\) are shifted by \(d\).

Even when the distribution of \(X\) is continuous, the distribution of
\(Y^{L}\) is partly discrete and partly continuous. The discrete part of
the distribution is concentrated at \(Y = 0\) (when \(X \leq d\)) and
the continuous part is spread over the interval \(Y > 0\) (when
\(X > d\)). For the discrete part, the probability that no payment is
made is the probability that losses fall below the deductible; that is,
\[\Pr\left( Y^{L} = 0 \right) = \Pr\left( X \leq d \right) = F_{X}\left( d \right).\]
Using the transformation \(Y^{L} = X - d\) for the continuous part of
the distribution, we can find the probability density function of
\(Y^{L}\) given by \[f_{Y^{L}}\left( y \right) = \left\{ \begin{matrix}
F_{X}\left( d \right) & y = 0, \\
f_{X}\left( y + d \right) & y > 0
\end{matrix} \right. \]

We can see that the payment per payment variable is the payment per loss
variable conditioned on the loss exceeding the deductible; that is,
\(Y^{P} = \left. \ Y^{L} \right|X > d\). Hence, the probability density
function of \(Y^{P}\) is given by
\[f_{Y^{P}}\left( y \right) = \frac{f_{X}\left( y + d \right)}{1 - F_{X}\left( d \right)},\]
for \(y > 0\). Accordingly, the distribution functions of \(Y^{L}\)and
\(Y^{P}\) are given by
\[F_{Y^{L}}\left( y \right) = \left\{ \begin{matrix}
F_{X}\left( d \right) & y = 0, \\
F_{X}\left( y + d \right) & y > 0. \\
\end{matrix} \right.\ \] and
\[F_{Y^{P}}\left( y \right) = \frac{F_{X}\left( y + d \right) - F_{X}\left( d \right)}{1 - F_{X}\left( d \right)},\]
for \(y > 0\), respectively.

The raw moments of \(Y^{L}\) and \(Y^{P}\) can be found directly using
the probability density function of \(X\) as follows
\[\mathrm{E}\left\lbrack \left( Y^{L} \right)^{k} \right\rbrack = \int_{d}^{\infty}\left( x - d \right)^{k}f_{X}\left( x \right)dx ,\]
and
\[\mathrm{E}\left\lbrack \left( Y^{P} \right)^{k} \right\rbrack = \frac{\int_{d}^{\infty}\left( x - d \right)^{k}f_{X}\left( x \right) dx }{{1 - F}_{X}\left( d \right)} = \frac{\mathrm{E}\left\lbrack \left( Y^{L} \right)^{k} \right\rbrack}{{1 - F}_{X}\left( d \right)},\]
respectively.

We have seen that the deductible \(d\) imposed on an insurance policy is
the amount of loss that has to be paid out of pocket before the insurer
makes any payment. The deductible \(d\) imposed on an insurance policy
reduces the insurer's payment. The loss elimination ratio (\emph{LER})
is the percentage decrease in the expected payment of the insurer as a
result of imposing the deductible. \emph{LER} is defined as
\[LER = \frac{\mathrm{E}\left( X \right) - \mathrm{E}\left( Y^{L} \right)}{\mathrm{E}\left( X \right)}.\]

A little less common type of policy deductible is the franchise
deductible. The Franchise deductible will apply to the policy in the
same way as ordinary deductible except that when the loss exceeds the
deductible \(d\), the full loss is covered by the insurer. The payment
per loss and payment per payment variables are defined as
\[Y^{L} = \left\{ \begin{matrix}
0 & X \leq d, \\
X & X > d, \\
\end{matrix} \right.\ \] and \[Y^{P} = \left\{ \begin{matrix}
\text{Undefined} & X \leq d, \\
X & X > d, \\
\end{matrix} \right.\ \] respectively.

\textbf{Example 3.4.1. SOA Exam Question.} A claim severity distribution
is exponential with mean 1000. An insurance company will pay the amount
of each claim in excess of a deductible of 100. Calculate the variance
of the amount paid by the insurance company for one claim, including the
possibility that the amount paid is 0.

Show Example Solution

\hypertarget{toggleExampleLoss.4.1}{}
\textbf{Solution.}

Let \(Y^{L}\) denote the amount paid by the insurance company for one
claim. \[Y^{L} = \left( X - 100 \right)_{+} = \left\{ \begin{matrix}
0 & X \leq 100, \\
X - 100 & X > 100. \\
\end{matrix} \right.\ \] The first and second moments of \(Y^{L}\) are
\[E\left( Y^{L} \right) = \int_{100}^{\infty}\left( x - 100 \right)f_{X}\left( x \right)dx \\
= {\int_{100}^{\infty}{S_{X}\left( x \right)}dx = 1000e}^{- \frac{100}{1000}},\]
and
\[E\left\lbrack \left( Y^{L} \right)^{2} \right\rbrack = \int_{100}^{\infty}\left( x - 100 \right)^{2}f_{X}\left( x \right)dx \\\\
= 2 \times 1000^{2}e^{- \frac{100}{1000}}.\] So,
\[\mathrm{Var}\left( Y^{L} \right) = \left( 2 \times 1000^{2}e^{- \frac{100}{1000}} \right) - \left( {1000e}^{- \frac{100}{1000}} \right)^{2} = 990,944.\]

An arguably simpler path to the solution is to make use of the
relationship between \(X\) and \(Y^{P}\). If \(X\) is exponentially
distributed with mean 1000, then \(Y^{P}\) is also exponentially
distributed with the same mean, because of the memoryless property of
the exponential distribution. Hence, \(E\left( Y^{P} \right)\)=1000 and
\[E\left\lbrack \left( Y^{P} \right)^{2} \right\rbrack = 2 \times 1000^{2}.\]
Using the relationship between \(Y^{L}\) and \(Y^{P}\) we find
\[E\left( Y^{L} \right) = \ E\left( Y^{P} \right)S_{X}\left( 100 \right){= 1000e}^{- \frac{100}{1000}}\]

\[E\left\lbrack \left( Y^{L} \right)^{2} \right\rbrack = E\left\lbrack \left( Y^{P} \right)^{2} \right\rbrack S_{X}\left( 100 \right) = 2 \times 1000^{2}e^{- \frac{100}{1000}}.\]

\begin{center}\rule{0.5\linewidth}{\linethickness}\end{center}

\textbf{Example 3.4.2. SOA Exam Question.} For an insurance:

Losses have a density function
\[f_{X}\left( x \right) = \left\{ \begin{matrix}
    0.02x & 0 < x  < 10, \\
    0 & \text{elsewhere.} \\
    \end{matrix} \right. \]

The insurance has an ordinary deductible of 4 per loss.

\(Y^{P}\) is the claim payment per payment random variable.

Calculate \(\mathrm{E}\left( Y^{P} \right)\).

Show Example Solution

\hypertarget{toggleExampleLoss.4.2}{}
\textbf{Solution.}

We define \(Y^P\) as follows \[Y^{P} = \left\{ \begin{matrix}
\text{Undefined} & X \leq 4, \\
X - 4 & X > 4. \\
\end{matrix} \right.\ \] So,
\(E\left( Y^{P} \right) = \frac{\int_{4}^{10}\left( x - 4 \right)0.02xdx}{{1 - F}_{X}\left( 4 \right)} = \frac{2.88}{0.84} = 3.43\).

Note that we divide by \(S_X(4)=1-F_X(4)\), as this is the range where
the variable \(Y^P\) is defined.

\begin{center}\rule{0.5\linewidth}{\linethickness}\end{center}

\textbf{Example 3.4.3. SOA Exam Question.} You are given:

Losses follow an exponential distribution with the same mean in all
years.

The loss elimination ratio this year is 70\%.

The ordinary deductible for the coming year is 4/3 of the current
deductible.

Compute the loss elimination ratio for the coming year.

Show Example Solution

\hypertarget{toggleExampleLoss.4.3}{}
\textbf{Solution.}

Let the losses \(X\sim Exp(\theta)\) and the deductible for the coming
year \(d' = \frac{4}{3}d\), the deductible of the current year. The
\emph{LER} for the current year is
\[\frac{E\left( X \right) - E\left( Y^{L} \right)}{E\left( X \right)} = \frac{\theta - \theta e^{- d / \theta}}{\theta} = 1 - e^{- d / \theta} = 0.7.\]
Then, \(e^{- d / \theta} = 0.3\).

The \emph{LER} for the coming year is

\begin{align*}
&\frac{\theta - \theta \exp(- \frac{d'}{\theta})}{\theta}=\frac{\theta - \theta \exp(- \frac{\left( \frac{4}{3}d \right)}{\theta})}{\theta} \\
&= 1 - \exp\left(- \frac{ \frac{4}{3} d }{\theta}\right) = 1 - \left( e^{-d /\theta} \right)^{4/3} = 1 - {0.3}^{4/3} = 0.8 .
\end{align*}

\begin{center}\rule{0.5\linewidth}{\linethickness}\end{center}

\subsection{Policy Limits}\label{S:PolicyLimits}

Under a limited policy, the insurer is responsible for covering the
actual loss \(X\) up to the limit of its coverage. This fixed limit of
coverage is called the policy limit and often denoted by \(u\). If the
loss exceeds the policy limit, the difference \(X - u\) has to be paid
by the policyholder. While a higher policy limit means a higher payout
to the insured, it is associated with a higher premium.

Let \(X\) denote the loss incurred to the insured and \(Y\) denote the
amount of paid claim by the insurer. Then \(Y\) is defined as
\[Y = X \land u = \left\{ \begin{matrix}
X & X \leq u, \\
u & X > u. \\
\end{matrix} \right.\ \] It can be seen that the distinction between
\(Y^{L}\) and \(Y^{P}\) is not needed under limited policy as the
insurer will always make a payment.

Even when the distribution of \(X\) is continuous, the distribution of
\(Y\) is partly discrete and partly continuous. The discrete part of the
distribution is concentrated at \(Y = u\) (when \(X > u\)), while the
continuous part is spread over the interval \(Y < u\) (when
\(X \leq u\)). For the discrete part, the probability that the benefit
paid is \(u\), is the probability that the loss exceeds the policy limit
\(u\); that is,
\[\Pr \left( Y = u \right) = \Pr \left( X > u \right) = {1 - F}_{X}\left( u \right).\]
For the continuous part of the distribution \(Y = X\), hence the
probability density function of \(Y\) is given by
\[f_{Y}\left( y \right) = \left\{ \begin{matrix}
f_{X}\left( y \right) & 0 < y < u, \\
1 - F_{X}\left( u \right) & y = u. \\
\end{matrix} \right.\ \] Accordingly, the distribution function of \(Y\)
is given by \[F_{Y}\left( y \right) = \left\{ \begin{matrix}
F_{X}\left( x \right) & 0 < y < u, \\
1 & y \geq u. \\
\end{matrix} \right.\ \] The raw moments of \(Y\) can be found directly
using the probability density function of \(X\) as follows
\[\mathrm{E}\left( Y^{k} \right) = \mathrm{E}\left\lbrack \left( X \land u \right)^{k} \right\rbrack = \int_{0}^{u}x^{k}f_{X}\left( x \right)dx + \int_{u}^{\infty}{u^{k}f_{X}\left( x \right)} dx \\ \int_{0}^{u}x^{k}f_{X}\left( x \right)dx + u^{k}\left\lbrack {1 - F}_{X}\left( u \right) \right\rbrack dx.\]

\textbf{Example 3.4.4. SOA Exam Question.} Under a group insurance
policy, an insurer agrees to pay 100\% of the medical bills incurred
during the year by employees of a small company, up to a maximum total
of one million dollars. The total amount of bills incurred, \(X\), has
probability density function
\[f_{X}\left( x \right) = \left\{ \begin{matrix}
\frac{x\left( 4 - x \right)}{9} & 0 < x < 3, \\
0 & \text{elsewhere.} \\
\end{matrix} \right.\ \] where \(x\) is measured in millions. Calculate
the total amount, in millions of dollars, the insurer would expect to
pay under this policy.

Show Example Solution

\textbf{Solution.}

Define the total amount of bills payed by the insurer as
\[Y = X \land 1 = \left\{ \begin{matrix}
X & X \leq 1, \\
1 & X > 1. \\
\end{matrix} \right.\ \] So
\(\mathrm{E}\left( Y \right) = \mathrm{E}\left( X \land 1 \right) = \int_{0}^{1}\frac{x^{2}(4 - x)}{9}dx + 1 * \int_{1}^{3}\frac{x\left( 4 - x \right)}{9}dx = 0.935\).

\begin{center}\rule{0.5\linewidth}{\linethickness}\end{center}

\subsection{Coinsurance}\label{coinsurance}

As we have seen in Section \ref{S:PolicyDeduct}, the amount of loss
retained by the policyholder can be losses up to the deductible \(d\).
The retained loss can also be a percentage of the claim. The percentage
\(\alpha\), often referred to as the coinsurance factor, is the
percentage of claim the insurance company is required to cover. If the
policy is subject to an ordinary deductible and policy limit,
coinsurance refers to the percentage of claim the insurer is required to
cover, after imposing the ordinary deductible and policy limit. The
payment per loss variable, \(Y^{L}\), is defined as
\[Y^{L} = \left\{ \begin{matrix}
0 & X \leq d, \\
\alpha\left( X - d \right) & d <  X \leq u, \\
\alpha\left( u - d \right) & X > u. \\
\end{matrix} \right.\ \] The policy limit (the maximum amount paid by
the insurer) in this case is \(\alpha\left( u - d \right)\), while \(u\)
is the maximum covered loss.

The \(k\)-th moment of \(Y^{L}\) is given by
\[\mathrm{E}\left\lbrack \left( Y^{L} \right)^{k} \right\rbrack = \int_{d}^{u}\left\lbrack \alpha\left( x - d \right) \right\rbrack^{k}f_{X}\left( x \right)dx + \int_{u}^{\infty}\left\lbrack \alpha\left( u - d \right) \right\rbrack^{k}f_{X}\left( x \right) dx .\]

A growth factor \(\left( 1 + r \right)\) may be applied to \(X\)
resulting in an inflated loss random variable \(\left( 1 + r \right)X\)
(the prespecified \emph{d} and \emph{u} remain unchanged). The resulting
per loss variable can be written as \[Y^{L} = \left\{ \begin{matrix}
0 & X \leq \frac{d}{1 + r}, \\
\alpha\left\lbrack \left( 1 + r \right)X - d \right\rbrack & \frac{d}{1 + r} <  X \leq \frac{u}{1 + r}, \\
\alpha\left( u - d \right) & X > \frac{u}{1 + r}. \\
\end{matrix} \right.\ \] The first and second moments of \(Y^{L}\) can
be expressed as
\[\mathrm{E}\left( Y^{L} \right) = \alpha\left( 1 + r \right)\left\lbrack \mathrm{E}\left( X \land \frac{u}{1 + r} \right) - \mathrm{E}\left( X \land \frac{d}{1 + r} \right) \right\rbrack,\]
and \[\mathrm{E}\left\lbrack \left( Y^{L} \right)^{2}
\right\rbrack = \alpha^{2}\left( 1 + r \right)^{2}  \left\{ \mathrm{E}\left\lbrack \left( X \land \frac{u}{1 + r} \right)^{2} \right\rbrack - \mathrm{E}\left\lbrack \left( X \land \frac{d}{1 + r} \right)^{2} \right\rbrack  \right. \\
\left. \ \ \ \ \ - 2\left( \frac{d}{1 + r} \right)\left\lbrack \mathrm{E}\left( X \land \frac{u}{1 + r} \right) - \mathrm{E}\left( X \land \frac{d}{1 + r} \right) \right\rbrack \right\} ,\]
respectively.

The formulae given for the first and second moments of \(Y^{L}\) are
general. Under full coverage, \(\alpha = 1\), \(r = 0\), \(u = \infty\),
\(d = 0\) and \(\mathrm{E}\left( Y^{L} \right)\) reduces to
\(\mathrm{E}\left( X \right)\). If only an ordinary deductible is
imposed, \(\alpha = 1\), \(r = 0\), \(u = \infty\) and
\(\mathrm{E}\left( Y^{L} \right)\) reduces to
\(\mathrm{E}\left( X \right) - \mathrm{E}\left( X \land d \right)\). If
only a policy limit is imposed \(\alpha = 1\), \(r = 0\), \(d = 0\) and
\(\mathrm{E}\left( Y^{L} \right)\) reduces to
\(\mathrm{E}\left( X \land u \right)\).

\textbf{Example 3.4.5. SOA Exam Question.} The ground up loss random
variable for a health insurance policy in 2006 is modeled with \emph{X},
an exponential distribution with mean 1000. An insurance policy pays the
loss above an ordinary deductible of 100, with a maximum annual payment
of 500. The ground up loss random variable is expected to be 5\% larger
in 2007, but the insurance in 2007 has the same deductible and maximum
payment as in 2006. Find the percentage increase in the expected cost
per payment from 2006 to 2007.

Show Example Solution

\hypertarget{toggleExampleLoss.4.5}{}
\textbf{Solution.}

We define the amount per loss \(Y^L\) in both years as
\[Y_{2006}^{L} = \left\{ \begin{matrix}
0 & X \leq 100, \\
X - 100 & 100 <  X \leq 600, \\
500 & X > 600. \\
\end{matrix} \right.\ \]

\[Y_{2007}^{L} = \left\{ \begin{matrix}
0 & X \leq 95.24, \\
1.05X - 100 & 95.24 <  X \leq 571.43, \\
500 & X > 571.43. \\
\end{matrix} \right.\ \]

So,

\[E\left( Y_{2006}^{L} \right) = E\left( X \land 600 \right) - E\left( X \land 100 \right) = 1000\left( {1 - e}^{- \frac{600}{1000}} \right) - 1000\left( {1 - e}^{- \frac{100}{1000}} \right)\]

\[= 356.026\].

\[E\left( Y_{2007}^{L} \right) = 1.05\left\lbrack E\left( X \land 571.43 \right) - E\left( X \land 95.24 \right) \right\rbrack\]

\[= 1.05\left\lbrack 1000\left( {1 - e}^{- \frac{571.43}{1000}} \right) - 1000\left( {1 - e}^{- \frac{95.24}{1000}} \right) \right\rbrack\]

\[=361.659\].

\(E\left( Y_{2006}^{P} \right) = \frac{356.026}{e^{- \frac{100}{1000}}} = 393.469\).

\(E\left( Y_{2007}^{P} \right) = \frac{361.659}{e^{- \frac{95.24}{1000}}} = 397.797\).

Because
\(\frac{E\left( Y_{2007}^{P} \right)}{E\left( Y_{2006}^{P} \right)} -1 = 0.011,\)
there is an increase of 1.1\% from 2006 to 2007.

\begin{center}\rule{0.5\linewidth}{\linethickness}\end{center}

\subsection{Reinsurance}\label{reinsurance}

In Section \ref{S:PolicyDeduct} we introduced the policy deductible,
which is a contractual arrangement under which an insured transfers part
of the risk by securing coverage from an insurer in return for an
insurance premium. Under that policy, when the loss exceeds the
deductible, the insurer is not required to pay until the insured has
paid the fixed deductible. We now introduce reinsurance, a mechanism of
insurance for insurance companies. Reinsurance is a contractual
arrangement under which an insurer transfers part of the underlying
insured risk by securing coverage from another insurer (referred to as a
reinsurer) in return for a reinsurance premium. Although reinsurance
involves a relationship between three parties: the original insured, the
insurer (often referred to as cedent or cedant) and the reinsurer, the
parties of the reinsurance agreement are only the primary insurer and
the reinsurer. There is no contractual agreement between the original
insured and the reinsurer. The reinsurer is not required to pay under
the reinsurance contract until the insurer has paid a loss to its
original insured. The amount retained by the primary insurer in the
reinsurance agreement (the reinsurance deductible) is called retention.

Reinsurance arrangements allow insurers with limited financial resources
to increase the capacity to write insurance and meet client requests for
larger insurance coverage while reducing the impact of potential losses
and protecting the insurance company against catastrophic losses.
Reinsurance also allows the primary insurer to benefit from underwriting
skills, expertize and proficient complex claim file handling of the
larger reinsurance companies.

\textbf{Example 3.4.6. SOA Exam Question.} In 2005 a risk has a
two-parameter Pareto distribution with \(\alpha = 2\) and
\(\theta = 3000\). In 2006 losses inflate by 20\%. Insurance on the risk
has a deductible of 600 in each year. \(P_{i}\), the premium in year
\(i\), equals 1.2 times expected claims. The risk is reinsured with a
deductible that stays the same in each year. \(R_{i}\), the reinsurance
premium in year \(i\), equals 1.1 times the expected reinsured claims.
\(\frac{R_{2005}}{P_{2005} = 0.55}\). Calculate
\(\frac{R_{2006}}{P_{2006}}\).

Show Example Solution

\hypertarget{toggleExampleLoss.4.6}{}
\textbf{Solution.}

Let us use the following notation:

\(X_{i}:\) The risk in year \(i\)

\(Y_{i}:\) The insured claim in year \(i\)

\(P_{i}:\) The insurance premium in year \(i\)

\(Y_{i}^{R}:\) The reinsured claim in year \(i\)

\(R_{i}:\) The reinsurance premium in year \(i\)

\(d:\) The insurance deductible in year \(i\) (the insurance deductible
is fixed each year, equal to 600)

\(d^{R}:\) The reinsurance deductible or retention in year \(i\) (the
reinsurance deductible is fixed each year, but unknown) where
\(i = 2005,\ 2006\)

\[Y_{i} = \left\{ \begin{matrix}
0 & X_{i} \leq 600 \\
X_{i} - 600 & X_{i} > 600 \\
\end{matrix} \right.\ \] where \(i = 2005,\ 2006\)

\[X_{2005}\sim Pa\left( 2,3000 \right)\]

\[\mathrm{E}\left( Y_{2005} \right) = \mathrm{E}\left( X_{2005} - 600 \right)_{+} = \mathrm{E}\left( X_{2005} \right) - \mathrm{E}\left( X_{2005} \land 600 \right)\]

\[= 3000 - 3000\left( 1 - \frac{3000}{3600} \right) = 2500\]

\[P_{2005} = 1.2\mathrm{E}\left( Y_{2005} \right) = 3000\]

Since \(X_{2006} = 1.2X_{2005}\) and Pareto is a scale distribution with
scale parameter \(\theta\), then
\(X_{2006}\sim Pa\left( 2,3600 \right)\)

\[\mathrm{E}\left( Y_{2006} \right) = \mathrm{E}\left( X_{2006} - 600 \right)_{+} = \mathrm{E}\left( X_{2006} \right) - \mathrm{E}\left( X_{2006} \land 600 \right)\]

\[= 3600 - 3600\left( 1 - \frac{3600}{4200} \right) = 3085.714\]

\[P_{2006} = 1.2\mathrm{E}\left( Y_{2006} \right) = 3702.857\]

\[Y_{i}^{R} = \left\{ \begin{matrix}
0 & X_{i} - 600 \leq d^{R} \\
X_{i} - 600 - d^{R} & X_{i} - 600 > d^{R} \\
\end{matrix} \right.\ \]

Since \(\frac{R_{2005}}{P_{2005}} = 0.55\), then
\(R_{2005} = 3000 \times 0.55 = 1650\)

Since \(R_{2005} = 1.1\mathrm{E}\left( Y_{2005}^{R} \right)\), then
\(\mathrm{E}\left( Y_{2005}^{R} \right) = \frac{1650}{1.1} = 1500\)

\[\mathrm{E}\left( Y_{2005}^{R} \right) = \mathrm{E}\left( X_{2005} - 600 - d^{R} \right)_{+} = \mathrm{E}\left( X_{2005} \right) - \mathrm{E}\left( X_{2005} \land \left( 600 + d^{R} \right) \right)\]

\[= 3000 - 3000\left( 1 - \frac{3000}{3600 + d^{R}} \right) = 1500 \Rightarrow d^{R} = 2400\]

\[\mathrm{E}\left( Y_{2006}^{R} \right) = \mathrm{E}\left( X_{2006} - 600 - d^{R} \right)_{+} = \mathrm{E}\left( X_{2006} - 3000 \right)_{+} = \mathrm{E}\left( X_{2006} \right) - \mathrm{E}\left( X_{2006} \land 3000 \right)\]

\[= 3600 - 3600\left( 1 - \frac{3600}{6600} \right) = 1963.636\]

\[R_{2006} = 1.1\mathrm{E}\left( Y_{2006}^{R} \right) = 1.1 \times 1963.636 = 2160\]

Therefore \(\frac{R_{2006}}{P_{2006}} = \frac{2160}{3702.857} = 0.583\)

\begin{center}\rule{0.5\linewidth}{\linethickness}\end{center}

\section{Maximum Likelihood Estimation}\label{S:MaxLikeEstimation}

In this section we estimate statistical parameters using the method of
maximum likelihood. Maximum likelihood estimates in the presence of
grouping, truncation or censoring are calculated.

\subsection{Maximum Likelihood Estimators for Complete
Data}\label{maximum-likelihood-estimators-for-complete-data}

Pricing of insurance premiums and estimation of claim reserving are
among many actuarial problems that involve modeling the severity of loss
(claim size). The principles for using maximum likelihood to estimate
model parameters were introduced in Chapter \textbf{xxx}. In this
section, we present a few examples to illustrate how actuaries fit a
parametric distribution model to a set of claim data using maximum
likelihood. In these examples we derive the asymptotic variance of
maximum-likelihood estimators of the model parameters. We use the delta
method to derive the asymptotic variances of functions of these
parameters.

\textbf{Example 3.5.1. SOA Exam Question.} Consider a random sample of
claim amounts: 8,000 10,000 12,000 15,000. You assume that claim amounts
follow an inverse exponential distribution, with parameter \(\theta\).

\begin{enumerate}
\def\labelenumi{\alph{enumi}.}
\tightlist
\item
  Calculate the maximum likelihood estimator for \(\theta\).

  \begin{enumerate}
  \def\labelenumii{\alph{enumii}.}
  \setcounter{enumii}{1}
  \tightlist
  \item
    Approximate the variance of the maximum likelihood estimator.

    \begin{enumerate}
    \def\labelenumiii{\alph{enumiii}.}
    \setcounter{enumiii}{2}
    \tightlist
    \item
      Determine an approximate 95\% confidence interval for \(\theta\).

      \begin{enumerate}
      \def\labelenumiv{\alph{enumiv}.}
      \setcounter{enumiv}{3}
      \tightlist
      \item
        Determine an approximate 95\% confidence interval for
        \(\Pr \left( X \leq 9,000 \right).\)
      \end{enumerate}
    \end{enumerate}
  \end{enumerate}
\end{enumerate}

Show Example Solution

\hypertarget{toggleExampleLoss.5.1}{}
\textbf{Solution.}

The probability density function is
\[f_{X}\left( x \right) = \frac{\theta e^{- \frac{\theta}{x}}}{x^{2}}, \]
where \(x > 0\).

\textbf{a.} \$The likelihood function, \(L\left( \theta \right)\), can
be viewed as the probability of the observed data, written as a function
of the model's parameter \(\theta\)
\[L\left( \theta \right) = \prod_{i = 1}^{4}{f_{X_{i}}\left( x_{i} \right)} = \frac{\theta^{4}e^{- \theta\sum_{i = 1}^{4}\frac{1}{x_{i}}}}{\prod_{i = 1}^{4}x_{i}^{2}}.\]

The log-likelihood function, \(\ln L \left( \theta \right)\), is the sum
of the individual logarithms.
\[\ln L \left( \theta \right) = 4ln\theta - \theta\sum_{i = 1}^{4}\frac{1}{x_{i}} - 2\sum_{i = 1}^{4}\ln x_{i} .\]

\[\frac{d \ln L \left( \theta \right)}{d \theta} = \frac{4}{\theta} - \sum_{i = 1}^{4}\frac{1}{x_{i}}.\]
The maximum likelihood estimator of \(\theta\), denoted by
\(\hat{\theta}\), is the solution to the equation
\[\frac{4}{\hat{\theta}} - \sum_{i = 1}^{4}{\frac{1}{x_{i}} = 0}.\]
Thus,
\(\hat{\theta} = \frac{4}{\sum_{i = 1}^{4}\frac{1}{x_{i}}} = 10,667\)

The second derivative of \(\ln L \left( \theta \right)\) is given by
\[\frac{d^{2}\ln L\left( \theta \right)}{d\theta^{2}} = \frac{- 4}{\theta^{2}}.\]
Evaluating the second derivative of the loglikelihood function at
\(\hat{\theta} = 10,667\) gives a negative value, indicating
\(\hat{\theta}\) as the value that maximizes the loglikelihood function.

\textbf{b.} \$Taking reciprocal of negative expectation of the second
derivative of \(\ln L \left( \theta \right)\), we obtain an estimate of
the variance of \(\hat{\theta}\)
\(\widehat{Var}\left( \hat{\theta} \right) = \left. \ \left\lbrack E\left( \frac{d^{2}\ln L \left( \theta \right)}{d\theta^{2}} \right) \right\rbrack^{- 1} \right|_{\theta = \hat{\theta}} = \frac{{\hat{\theta}}^{2}}{4} = 28,446,222\).

It should be noted that as the sample size \(n \rightarrow \infty\), the
distribution of the maximum likelihood estimator \(\hat{\theta}\)
converges to a normal distribution with mean \(\theta\) and variance
\(\hat{V}\left( \hat{\theta} \right)\). The approximate confidence
interval in this example is based on the assumption of normality,
despite the small sample size, only for the purpose of illustration.

\textbf{c.} \$The 95\% confidence interval for \(\theta\) is given by
\[10,667 \pm 1.96\sqrt{28,446,222} = \left( 213.34,\ 21,120.66 \right).\]
\textbf{d.} \$The distribution function of \(X\) is
\(F\left( x \right) = 1 - e^{- \frac{x}{\theta}}\). Then, the maximum
likelihood estimate of
\(g\left( \theta \right) = F\left( 9,000 \right)\) is
\[g\left( \hat{\theta} \right) = 1 - e^{- \frac{9,000}{10,667}} = 0.57.\]
We use the delta method to approximate the variance of
\(g\left( \hat{\theta} \right)\).
\[\frac{\text{dg}\left( \theta \right)}{d \theta} = {- \frac{9,000}{\theta^{2}}e}^{- \frac{9,000}{\theta}}.\]

\(\widehat{Var}\left\lbrack g\left( \hat{\theta} \right) \right\rbrack = \left( - {\frac{9,000}{{\hat{\theta}}^{2}}e}^{- \frac{9,000}{\hat{\theta}}} \right)^{2}\hat{V}\left( \hat{\theta} \right) = 0.0329\).

The 95\% confidence interval for \(F\left( 9,000 \right)\) is given by
\[0.57 \pm 1.96\sqrt{0.0329} = \left( 0.214,\ 0.926 \right).\]

\begin{center}\rule{0.5\linewidth}{\linethickness}\end{center}

\textbf{Example 3.5.2. SOA Exam Question.} A random sample of size 6 is
from a lognormal distribution with parameters \(\mu\) and \(\sigma\).
The sample values are 200, 3,000, 8,000, 60,000, 60,000, 160,000.

\begin{enumerate}
\def\labelenumi{\alph{enumi}.}
\tightlist
\item
  Calculate the maximum likelihood estimator for \(\mu\) and \(\sigma\).

  \begin{enumerate}
  \def\labelenumii{\alph{enumii}.}
  \setcounter{enumii}{1}
  \tightlist
  \item
    Estimate the covariance matrix of the maximum likelihood estimator.

    \begin{enumerate}
    \def\labelenumiii{\alph{enumiii}.}
    \setcounter{enumiii}{2}
    \tightlist
    \item
      Determine approximate 95\% confidence intervals for \(\mu\) and
      \(\sigma\).

      \begin{enumerate}
      \def\labelenumiv{\alph{enumiv}.}
      \setcounter{enumiv}{3}
      \tightlist
      \item
        Determine an approximate 95\% confidence interval for the mean
        of the lognormal distribution.
      \end{enumerate}
    \end{enumerate}
  \end{enumerate}
\end{enumerate}

Show Example Solution

\hypertarget{toggleExampleLoss.5.2}{}
\textbf{Solution.}

The probability density function is
\[f_{X}\left( x \right) = \frac{1}{x \sigma \sqrt{2\pi}}\exp - \frac{1}{2}\left( \frac{\ln x - \mu}{\sigma} \right)^{2},\]
where \(x > 0\).

\textbf{a.} \$The likelihood function, \(L\left( \mu,\sigma \right)\),
is the product of the pdf for each data point.
\[L\left( \mu,\sigma \right) = \prod_{i = 1}^{6}{f_{X_{i}}\left( x_{i} \right)} = \frac{1}{\sigma^{6}\left( 2\pi \right)^{3}\prod_{i = 1}^{6}x_{i}}exp - \frac{1}{2}\sum_{i = 1}^{6}\left( \frac{\ln x_{i} - \mu}{\sigma} \right)^{2}.\]
The loglikelihood function, \(\ln L \left( \mu,\sigma \right)\), is the
sum of the individual logarithms.
\[\ln \left( \mu,\sigma \right) = - 6ln\sigma - 3ln\left( 2\pi \right) - \sum_{i = 1}^{6}\ln x_{i} - \frac{1}{2}\sum_{i = 1}^{6}\left( \frac{\ln x_{i} - \mu}{\sigma} \right)^{2}.\]
The first partial derivatives are
\[\frac{\partial lnL\left( \mu,\sigma \right)}{\partial\mu} = \frac{1}{\sigma^{2}}\sum_{i = 1}^{6}\left( \ln x_{i} - \mu \right).\]
\[\frac{\partial lnL\left( \mu,\sigma \right)}{\partial\sigma} = \frac{- 6}{\sigma} + \frac{1}{\sigma^{3}}\sum_{i = 1}^{6}\left( \ln x_{i} - \mu \right)^{2}.\]
The maximum likelihood estimators of \(\mu\) and \(\sigma\), denoted by
\(\hat{\mu}\) and \(\hat{\sigma}\), are the solutions to the equations
\[\frac{1}{{\hat{\sigma}}^{2}}\sum_{i = 1}^{6}\left( \ln x_{i} - \hat{\mu} \right) = 0.\]
\[\frac{- 6}{\hat{\sigma}} + \frac{1}{{\hat{\sigma}}^{3}}\sum_{i = 1}^{6}\left( \ln x_{i} - \hat{\mu} \right)^{2} = 0.\]
These yield the estimates

\(\hat{\mu} = \frac{\sum_{i = 1}^{6}{\ln x_{i}}}{6} = 9.38\) and
\({\hat{\sigma}}^{2} = \frac{\sum_{i = 1}^{6}\left( \ln x_{i} - \hat{\mu} \right)^{2}}{6} = 5.12\).

The second partial derivatives are

\(\frac{\partial^{2}\text{lnL}\left( \mu,\sigma \right)}{\partial\mu^{2}} = \frac{- 6}{\sigma^{2}}\),
\(\frac{\partial^{2}\text{lnL}\left( \mu,\sigma \right)}{\partial\mu\partial\sigma} = \frac{- 2}{\sigma^{3}}\sum_{i = 1}^{6}\left( \ln x_{i} - \mu \right)\)
and
\(\frac{\partial^{2}\text{lnL}\left( \mu,\sigma \right)}{\partial\sigma^{2}} = \frac{6}{\sigma^{2}} - \frac{3}{\sigma^{4}}\sum_{i = 1}^{6}\left( \ln x_{i} - \mu \right)^{2}\).

\textbf{b.} \$ To derive the covariance matrix of the MLE we need to
find the expectations of the second derivatives. Since the random
variable \(X\) is from a lognormal distribution with parameters \(\mu\)
and \(\sigma\), then \(\text{lnX}\) is normally distributed with mean
\(\mu\) and variance \(\sigma^{2}\).

\(\mathrm{E}\left( \frac{\partial^{2}\text{lnL}\left( \mu,\sigma \right)}{\partial\mu^{2}} \right) = \mathrm{E}\left( \frac{- 6}{\sigma^{2}} \right) = \frac{- 6}{\sigma^{2}}\),

\(\mathrm{E}\left( \frac{\partial^{2}\text{lnL}\left( \mu,\sigma \right)}{\partial\mu\partial\sigma} \right) = \frac{- 2}{\sigma^{3}}\sum_{i = 1}^{6}{\mathrm{E}\left( \ln x_{i} - \mu \right)} = \frac{- 2}{\sigma^{3}}\sum_{i = 1}^{6}\left\lbrack \mathrm{E}\left( \ln x_{i} \right) - \mu \right\rbrack\)=\(\frac{- 2}{\sigma^{3}}\sum_{i = 1}^{6}\left( \mu - \mu \right) = 0\),

and

\(\mathrm{E}\left( \frac{\partial^{2}\text{lnL}\left( \mu,\sigma \right)}{\partial\sigma^{2}} \right) = \frac{6}{\sigma^{2}} - \frac{3}{\sigma^{4}}\sum_{i = 1}^{6}{\mathrm{E}\left( \ln x_{i} - \mu \right)}^{2} = \frac{6}{\sigma^{2}} - \frac{3}{\sigma^{4}}\sum_{i = 1}^{6}{\mathrm{V}\left( \ln x_{i} \right) = \frac{6}{\sigma^{2}} - \frac{3}{\sigma^{4}}\sum_{i = 1}^{6}{\sigma^{2} = \frac{- 12}{\sigma^{2}}}}\).

Using the negatives of these expectations we obtain the Fisher
information matrix \[\begin{bmatrix}
\frac{6}{\sigma^{2}} & 0 \\
0 & \frac{12}{\sigma^{2}} \\
\end{bmatrix}\].

The covariance matrix, \(\Sigma\), is the inverse of the Fisher
information matrix \[\Sigma = \begin{bmatrix}
\frac{\sigma^{2}}{6} & 0 \\
0 & \frac{\sigma^{2}}{12} \\
\end{bmatrix}\].

The estimated matrix is given by \[\hat{\Sigma} = \begin{bmatrix}
0.8533 & 0 \\
0 & 0.4267 \\
\end{bmatrix}\].

\textbf{c.} \$ The 95\% confidence interval for \(\mu\) is given by
\(9.38 \pm 1.96\sqrt{0.8533} = \left( 7.57,\ 11.19 \right)\).

The 95\% confidence interval for \(\sigma^{2}\) is given by
\(5.12 \pm 1.96\sqrt{0.4267} = \left( 3.84,\ 6.40 \right)\).

\textbf{d.} \$ The mean of \emph{X} is
\(\exp\left( \mu + \frac{\sigma^{2}}{2} \right)\). Then, the maximum
likelihood estimate of
\[g\left( \mu,\sigma \right) = \exp\left( \mu + \frac{\sigma^{2}}{2} \right)\]
is
\[g\left( \hat{\mu},\hat{\sigma} \right) = \exp\left( \hat{\mu} + \frac{{\hat{\sigma}}^{2}}{2} \right) = 153,277.\]

We use the delta method to approximate the variance of the mle
\(g\left( \hat{\mu},\hat{\sigma} \right)\).

\(\frac{\partial g\left( \mu,\sigma \right)}{\partial\mu} = exp\left( \mu + \frac{\sigma^{2}}{2} \right)\)
and
\(\frac{\partial g\left( \mu,\sigma \right)}{\partial\sigma} = \sigma exp\left( \mu + \frac{\sigma^{2}}{2} \right)\).

Using the delta method, the approximate variance of
\(g\left( \hat{\mu},\hat{\sigma} \right)\) is given by

\[\left. \ \hat{V}\left( g\left( \hat{\mu},\hat{\sigma} \right) \right) = \begin{bmatrix}
\frac{\partial g\left( \mu,\sigma \right)}{\partial\mu} & \frac{\partial g\left( \mu,\sigma \right)}{\partial\sigma} \\
\end{bmatrix}\Sigma\begin{bmatrix}
\frac{\partial g\left( \mu,\sigma \right)}{\partial\mu} \\
\frac{\partial g\left( \mu,\sigma \right)}{\partial\sigma} \\
\end{bmatrix} \right|_{\mu = \hat{\mu},\sigma = \hat{\sigma}}\]

\[= \begin{bmatrix}
153,277 & 346,826 \\
\end{bmatrix}\begin{bmatrix}
0.8533 & 0 \\
0 & 0.4267 \\
\end{bmatrix}\begin{bmatrix}
153,277 \\
346,826 \\
\end{bmatrix} =\]71,374,380,000

The 95\% confidence interval for
\(\exp\left( \mu + \frac{\sigma^{2}}{2} \right)\) is given by

\(153,277 \pm 1.96\sqrt{71,374,380,000} = \left( - 370,356,\ 676,910 \right)\).

Since the mean of the lognormal distribution cannot be negative, we
should replace the negative lower limit in the previous interval by a
zero.

\begin{center}\rule{0.5\linewidth}{\linethickness}\end{center}

\subsection{Maximum Likelihood Estimators for Grouped
Data}\label{MLEGrouped}

In the previous section we considered the maximum likelihood estimation
of continuous models from complete (individual) data. Each individual
observation is recorded, and its contribution to the likelihood function
is the density at that value. In this section we consider the problem of
obtaining maximum likelihood estimates of parameters from grouped data.
The observations are only available in grouped form, and the
contribution of each observation to the likelihood function is the
probability of falling in a specific group (interval). Let \(n_{j}\)
represent the number of observations in the interval
\(\left( \left. \ c_{j - 1},c_{j} \right\rbrack \right.\ \) The grouped
data likelihood function is thus given by
\[L\left( \theta \right) = \prod_{j = 1}^{k}\left\lbrack F\left( \left. \ c_{j} \right|\theta \right) - F\left( \left. \ c_{j - 1} \right|\theta \right) \right\rbrack^{n_{j}},\]
where \(c_{0}\) is the smallest possible observation (often set to zero)
and \(c_{k}\) is the largest possible observation (often set to
infinity).

\textbf{Example 3.5.3. SOA Exam Question.} For a group of policies, you
are given that losses follow the distribution function
\(F\left( x \right) = 1 - \frac{\theta}{x}\), for
\(\theta < x < \infty.\) Further, a sample of 20 losses resulted in the
following:

\[
{\small
\begin{matrix}\hline
\text{Interval} & \text{Number of Losses}  \\ \hline
(\theta, 10] & 9 \\
(10, 25] & 6 \\
(25, \infty) & 5  \\ \hline
\end{matrix}
}
\]

Calculate the maximum likelihood estimate of \(\theta\).

Show Example Solution

\hypertarget{toggleExampleLoss.5.3}{}
\textbf{Solution.}

The contribution of each of the 9 observations in the first interval to
the likelihood function is the probability of \(X \leq 10\); that is,
\(\Pr\left( X \leq 10 \right) = F\left( 10 \right)\). Similarly, the
contributions of each of 6 and 5 observations in the second and third
intervals are
\(\Pr\left( 10 < X \leq 25 \right) = F\left( 25 \right) - F(10)\) and
\(P\left( X > 25 \right) = 1 - F(25)\), respectively. The likelihood
function is thus given by
\[L\left( \theta \right) = \left\lbrack F\left( 10 \right) \right\rbrack^{9}\left\lbrack F\left( 25 \right) - F(10) \right\rbrack^{6}\left\lbrack 1 - F(25) \right\rbrack^{5}\]
\[{= \left( 1 - \frac{\theta}{10} \right)}^{9}\left( \frac{\theta}{10} - \frac{\theta}{25} \right)^{6}\left( \frac{\theta}{25} \right)^{5}\]
\[{= \left( \frac{10 - \theta}{10} \right)}^{9}\left( \frac{15\theta}{250} \right)^{6}\left( \frac{\theta}{25} \right)^{5}.\]
Then,
\(\ln L \left( \theta \right) = 9ln\left( 10 - \theta \right) + 6ln\theta + 5ln\theta - 9ln10 + 6ln15 - 6ln250 - 5ln25\).
\[\frac{d \ln L \left( \theta \right)}{d \theta} = \frac{- 9}{\left( 10 - \theta \right)} + \frac{6}{\theta} + \frac{5}{\theta}.\]
The maximum likelihood estimator, \(\hat{\theta}\), is the solution to
the equation
\[\frac{- 9}{\left( 10 - \hat{\theta} \right)} + \frac{11}{\hat{\theta}} = 0\]
and \(\hat{\theta} = 5.5\).

\begin{center}\rule{0.5\linewidth}{\linethickness}\end{center}

\subsection{Maximum Likelihood Estimators for Censored
Data}\label{maximum-likelihood-estimators-for-censored-data}

Another distinguishing feature of data gathering mechanism is censoring.
While for some event of interest (losses, claims, lifetimes, etc.) the
complete data maybe available, for others only partial information is
available; information that the observation exceeds a specific value.
The limited policy introduced in Section \ref{S:PolicyLimits} is an
example of right censoring. Any loss greater than or equal to the policy
limit is recorded at the limit. The contribution of the censored
observation to the likelihood function is the probability of the random
variable exceeding this specific limit. Note that contributions of both
complete and censored data share the survivor function, for a complete
point this survivor function is multiplied by the hazard function, but
for a censored observation it is not.

\textbf{Example 3.5.4. SOA Exam Question.} The random variable has
survival function:
\[S_{X}\left( x \right) = \frac{\theta^{4}}{\left( \theta^{2} + x^{2} \right)^{2}}.\]
Two values of \(X\) are observed to be 2 and 4. One other value exceeds
4. Calculate the maximum likelihood estimate of \(\theta\).

Show Example Solution

\hypertarget{toggleExampleLoss.5.4}{}
\textbf{Solution.}

The contributions of the two observations 2 and 4 are
\(f_{X}\left( 2 \right)\) and \(f_{X}\left( 4 \right)\) respectively.
The contribution of the third observation, which is only known to exceed
4 is \(S_{X}\left( 4 \right)\). The likelihood function is thus given by
\[L\left( \theta \right) = f_{X}\left( 2 \right)f_{X}\left( 4 \right)S_{X}\left( 4 \right).\]
The probability density function of \(X\) is given by
\[f_{X}\left( x \right) = \frac{4x\theta^{4}}{\left( \theta^{2} + x^{2} \right)^{3}}.\]
Thus,
\[L\left( \theta \right) = \frac{8\theta^{4}}{\left( \theta^{2} + 4 \right)^{3}}\frac{16\theta^{4}}{\left( \theta^{2} + 16 \right)^{3}}\frac{\theta^{4}}{\left( \theta^{2} + 16 \right)^{2}} = \\
\frac{128\theta^{12}}{\left( \theta^{2} + 4 \right)^{3}\left( \theta^{2} + 16 \right)^{5}},\]

So,
\[\ln L\left( \theta \right) = ln128 + 12ln\theta - 3ln\left( \theta^{2} + 4 \right) - 5ln\left( \theta^{2} + 16 \right)\],

and

\(\frac{\text{dlnL}\left( \theta \right)}{d \theta} = \frac{12}{\theta} - \frac{6\theta}{\left( \theta^{2} + 4 \right)} - \frac{10\theta}{\left( \theta^{2} + 16 \right)}\).

The maximum likelihood estimator, \(\hat{\theta}\), is the solution to
the equation
\[\frac{12}{\hat{\theta}} - \frac{6\hat{\theta}}{\left( {\hat{\theta}}^{2} + 4 \right)} - \frac{10\hat{\theta}}{\left( {\hat{\theta}}^{2} + 16 \right)} = 0\]
or
\[12\left( {\hat{\theta}}^{2} + 4 \right)\left( {\hat{\theta}}^{2} + 16 \right) - 6{\hat{\theta}}^{2}\left( {\hat{\theta}}^{2} + 16 \right) - 10{\hat{\theta}}^{2}\left( {\hat{\theta}}^{2} + 4 \right) = \\
- 4{\hat{\theta}}^{4} + 104{\hat{\theta}}^{2} + 768 = 0,\] which yields
\({\hat{\theta}}^{2} = 32\) and \(\hat{\theta} = 5.7\).

\begin{center}\rule{0.5\linewidth}{\linethickness}\end{center}

\subsection{Maximum Likelihood Estimators for Truncated
Data}\label{maximum-likelihood-estimators-for-truncated-data}

This section is concerned with the maximum likelihood estimation of the
continuous distribution of the random variable \(X\) when the data is
incomplete due to truncation. If the values of \(X\) are truncated at
\(d\), then it should be noted that we would not have been aware of the
existence of these values had they not exceeded \(d\). The policy
deductible introduced in Section \ref{S:PolicyDeduct} is an example of
left truncation. Any loss less than or equal to the deductible is not
recorded. The contribution to the likelihood function of an observation
\(x\) truncated at \(d\) will be a conditional probability and the
\(f_{X}\left( x \right)\) will be replaced by
\(\frac{f_{X}\left( x \right)}{S_{X}\left( d \right)}\).

\textbf{Example 3.5.5. SOA Exam Question.} For the single parameter
Pareto distribution with \(\theta = 2\), maximum likelihood estimation
is applied to estimate the parameter \(\alpha\). Find the estimated mean
of the ground up loss distribution based on the maximum likelihood
estimate of \(\alpha\) for the following data set:

Ordinary policy deductible of 5, maximum covered loss of 25 (policy
limit 20)

8 insurance payment amounts: 2, 4, 5, 5, 8, 10, 12, 15

2 limit payments: 20, 20.

Show Example Solution

\hypertarget{toggleExampleLoss.5.5}{}
\textbf{Solution.}

The contributions of the different observations can be summarized as
follows:

For the exact loss: \(f_{X}\left( x \right)\)

For censored observations: \(S_{X}\left( 25 \right)\).

For truncated observations:
\(\frac{f_{X}\left( x \right)}{S_{X}\left( 5 \right)}\).

Given that ground up losses smaller than 5 are omitted from the data
set, the contribution of all observations should be conditional on
exceeding 5. The likelihood function becomes
\[L\left( \alpha \right) = \frac{\prod_{i = 1}^{8}{f_{X}\left( x_{i} \right)}}{\left\lbrack S_{X}\left( 5 \right) \right\rbrack^{8}}\left\lbrack \frac{S_{X}\left( 25 \right)}{S_{X}\left( 5 \right)} \right\rbrack^{2}.\]
For the single parameter Pareto the probability density and distribution
functions are given by

\[f_{X}\left( x \right) = \frac{\alpha\theta^{\alpha}}{x^{\alpha + 1}} \ \ \text{and} \ \ F_{X}\left( x \right) = 1 - \left( \frac{\theta}{x} \right)^{\alpha},\]
for \(x > \theta\), respectively. Then, the likelihood and loglikelihood
functions are given by
\[L\left( \alpha \right) = \frac{\alpha^{8}}{\prod_{i = 1}^{8}x_{i}^{\alpha + 1}}\frac{5^{10\alpha}}{25^{2\alpha}},\]
\[\ln L \left( \alpha \right) = 8ln\alpha - \left( \alpha + 1 \right)\sum_{i = 1}^{8}{\ln x_{i}} + 10\alpha ln5 - 2\alpha ln25.\]

\(\frac{\text{dlnL}\left( \alpha \right)}{d \theta} = \frac{8}{\alpha} - \sum_{i = 1}^{8}{\ln x_{i}} + 10ln5 - 2ln25\).

The maximum likelihood estimator, \(\hat{\alpha}\), is the solution to
the equation
\[\frac{8}{\hat{\alpha}} - \sum_{i = 1}^{8}{\ln x_{i}} + 10ln5 - 2ln25 = 0,\]which
yields
\[\hat{\alpha} = \frac{8}{\sum_{i = 1}^{8}{\ln x_{i}} - 10ln5 + 2ln25} = \frac{8}{(ln7 + ln9 + \ldots + ln20) - 10ln5 + 2ln25} = 0.785.\]
The mean of the Pareto only exists for \(\alpha > 1\). Since
\(\hat{\alpha} = 0.785 < 1\). Then, the mean does not exist.

\begin{center}\rule{0.5\linewidth}{\linethickness}\end{center}

\section{Further Resources and
Contributors}\label{LM-further-reading-and-resources}

\subsubsection*{Contributors}\label{contributors-1}
\addcontentsline{toc}{subsubsection}{Contributors}

\begin{itemize}
\tightlist
\item
  \textbf{Zeinab Amin}, The American University in Cairo, is the
  principal author of this chapter. Email:
  \href{mailto:zeinabha@aucegypt.edu}{\nolinkurl{zeinabha@aucegypt.edu}}
  for chapter comments and suggested improvements.
\item
  Many helpful comments have been provided by Hirokazu (Iwahiro)
  Iwasawa,
  \href{mailto:iwahiro@bb.mbn.or.jp}{\nolinkurl{iwahiro@bb.mbn.or.jp}} .
\end{itemize}

\subsubsection*{Exercises}\label{exercises-1}
\addcontentsline{toc}{subsubsection}{Exercises}

Here are a set of exercises that guide the viewer through some of the
theoretical foundations of \textbf{Loss Data Analytics}. Each tutorial
is based on one or more questions from the professional actuarial
examinations -- typically the Society of Actuaries Exam C.

\href{http://www.ssc.wisc.edu/~jfrees/loss-data-analytics/chapter-3-modeling-loss-severity/loss-data-analytics-severity-problems/}{Severity
Distribution Guided Tutorials}

\subsubsection*{Further Readings and
References}\label{further-readings-and-references}
\addcontentsline{toc}{subsubsection}{Further Readings and References}

Notable contributions include: \citet{cummins1991managing},
\citet{frees2008hierarchical}, \citet{klugman2012},
\citet{kreer2015goodness}, \citet{mcdonald1984some},
\citet{mcdonald1995generalization}, \citet{tevet2016applying}, and
\citet{venter1983transformed}.

\chapter{Model Selection and Estimation}\label{C:ModelSelection}

\emph{Chapter Preview}. Chapters \ref{C:Frequency-Modeling} and
\ref{C:Severity} have described how to fit parametric models to
frequency and severity data, respectively. This chapter describes
selection of models. To compare alternative parametric models, it is
helpful to introduce models that summarize data without reference to a
specific parametric distribution. Section \ref{S:MS:NonParInf} describes
nonparametric estimation, how we can use it for model comparisons and
how it can be used to provide starting values for parametric procedures.

The process of model selection is then summarized in Section
\ref{S:MS:ModelSelection}. Although our focus is on continuous data, the
same process can be used for discrete data or data that come from a
hybrid combination of discrete and continuous data. Further, Section
\ref{S:MS:ModifiedData} describes estimation for alternative sampling
schemes, included grouped, censored and truncated data, following the
introduction provided in Chapter \ref{C:Severity}. The chapter closes
with Section \ref{S:MS:BayesInference} on Bayesian inference, an
alternative procedure where the (typically unknown) parameters are
treated as random variables.

\section{Nonparametric Inference}\label{S:MS:NonParInf}

\begin{center}\rule{0.5\linewidth}{\linethickness}\end{center}

In this section, you learn how to:

\begin{itemize}
\tightlist
\item
  Estimate moments, quantiles, and distributions without reference to a
  parametric distribution
\item
  Summarize the data graphically without reference to a parametric
  distribution
\item
  Determine measures that summarize deviations of a parametric from a
  nonparametric fit
\item
  Use nonparametric estimators to approximate parameters that can be
  used to start a parametric estimation procedure
\end{itemize}

\begin{center}\rule{0.5\linewidth}{\linethickness}\end{center}

Consider \(X_1, \ldots, X_n\), a \textbf{random sample} (with
replacement) from an unknown underlying population distribution
\(F(\cdot)\). As independent draws from the same distribution, we say
that \(X_1, \ldots, X_n\) are \emph{independently and identically
distributed (iid)} random variables. Now say we have a data sample,
\(x_1, \dots, x_n\), which represents a realization of
\(X_1, \ldots, X_n\). Note that \(x_1, \ldots, x_n\) is non-random; it
is simply a particular set of data values, i.e.~an observation of the
random variables \(X_1, \ldots, X_n\). Using this sample, we will try to
estimate the population distribution function \(F(\cdot)\). We first
proceed with a \textbf{nonparametric} analysis, in which we do not
assume or rely on any explicit parametric distributional forms for
\(F(\cdot)\).

\subsection{Nonparametric Estimation}\label{nonparametric-estimation}

The population distribution \(F(\cdot)\) can be summarized in various
ways. These include moments, the distribution function \(F(\cdot)\)
itself, the quantiles or percentiles associated with the distribution,
and the corresponding mass or density function \(f(\cdot)\). Summary
statistics based on the sample, \(X_1, \ldots, X_n\), are known as
\textbf{nonparametric estimators} of the corresponding summary measures
of the distribution. We will examine moment estimators, distribution
function estimators, quantile estimators, and density estimators, as
well as their statistical properties such as expected value and
variance. Using our data observations \(x_1, \ldots, x_n\), we can put
numerical values to these estimators and compute \textbf{nonparametric
estimates}.

\subsubsection{Moment Estimators}\label{S:MS:MomentEstimator}

The \(k\)\textbf{-th moment}, \(\mathrm{E~}[X^k] = \mu^{\prime}_k\), is
our first example of a population summary measure. It is estimated with
the corresponding sample statistic \[\frac{1}{n} \sum_{i=1}^n X_i^k .\]
In typical applications, \(k\) is a positive integer, although it need
not be. For the first moment (\(k=1\)), the prime symbol (\(\prime\))
and the \(1\) subscript are usually dropped, using
\(\mu=\mu^{\prime}_1\) to denote the \textbf{mean}. The corresponding
sample estimator for \(\mu\) is called the \textbf{sample mean}, denoted
with a bar on top of the random variable:
\[\bar{X} =\frac{1}{n} \sum_{i=1}^n X_i .\]

Sometimes, \(\mu^{\prime}_k\) is called the \(k\)-th \emph{raw} moment
to distinguish it from the \(k\)\textbf{-th central moment},
\(\mathrm{E~} [(X-\mu)^k] = \mu_k\), which is estimated as
\[\frac{1}{n} \sum_{i=1}^n \left(X_i - \bar{X}\right)^k .\] The second
central moment (\(k=2\)) is an important case for which we typically
assign a new symbol, \(\sigma^2 = \mathrm{E~} [(X-\mu)^2]\), known as
the \textbf{variance}. The corresponding sample estimator for
\(\sigma^2\) is called the \textbf{sample variance}.

\subsubsection{Empirical Distribution
Function}\label{empirical-distribution-function}

To estimate the distribution function nonparametrically, we define the
\textbf{empirical distribution function} to be

\[\begin{aligned}
F_n(x) &=  \frac{\text{number of observations less than or equal to }x}{n} \\
&=  \frac{1}{n} \sum_{i=1}^n I\left(X_i \le x\right).
\end{aligned}\]

Here, the notation \(I(\cdot)\) is the indicator function; it returns 1
if the event \((\cdot)\) is true and 0 otherwise.

\textbf{Example 4.1.1. Toy Data Set}. To illustrate, consider a
fictitious, or ``toy,'' data set of \(n=10\) observations. Determine the
empirical distribution function.

\[\begin{array}{c|cccccccccc}
\hline
i &1&2&3&4&5&6&7&8&9&10 \\
X_i& 10 &15 &15 &15 &20 &23 &23 &23 &23 &30\\
\hline
\end{array}\]

Show Example Solution

\begin{figure}

{\centering \includegraphics[width=0.6\linewidth]{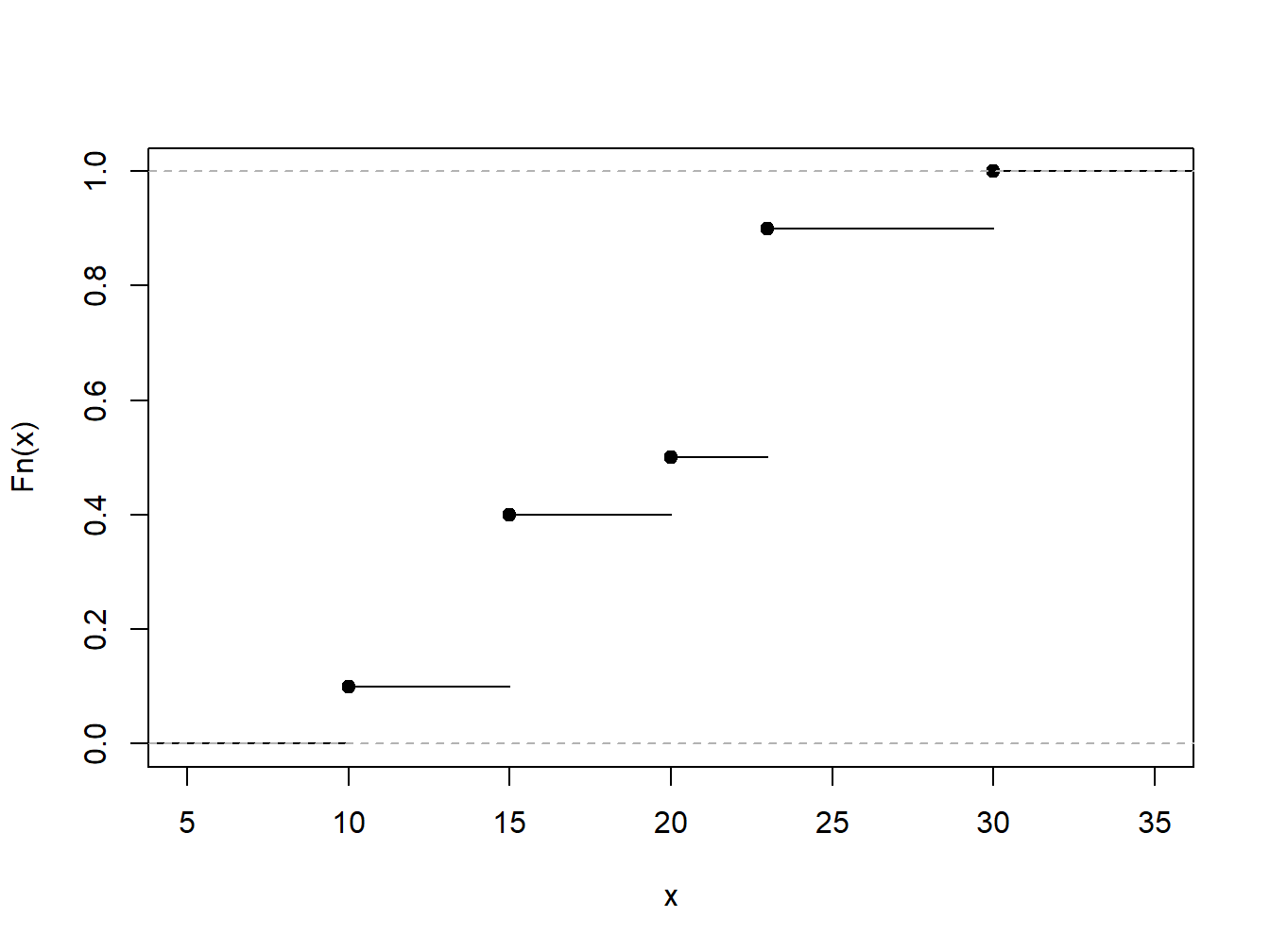}

}

\caption{Empirical Distribution Function of a Toy Example}\label{fig:EDFToy}
\end{figure}

\hypertarget{toggleExampleSelect.1.1}{}
You should check that the sample mean is \(\bar{x} = 19.7\) and that the
sample variance is \(34.45556\). The corresponding empirical
distribution function is

\[\begin{aligned}
F_n(x) &=
\left\{
\begin{array}{ll}
0 & \text{ for }\ x<10 \\
0.1 & \text{ for }\ 10 \leq x<15 \\
0.4 & \text{ for }\ 15 \leq x<20 \\
0.5 & \text{ for }\ 20 \leq x<23 \\
0.9 & \text{ for }\ 23 \leq x<30 \\
1 & \text{ for }\ x \geq 30,
\end{array}
\right.\end{aligned}\]

which is shown in the following graph in Figure \ref{fig:EDFToy}.

Show R Code

\hypertarget{toggleToy}{}
\begin{verbatim}
(xExample <- c(10,rep(15,3),20,rep(23,4),30))
PercentilesxExample <- ecdf(xExample)
plot(PercentilesxExample, main="",xlab="x")
\end{verbatim}

\begin{center}\rule{0.5\linewidth}{\linethickness}\end{center}

\subsubsection{Quantiles}\label{S:MS:QuantileEstimator}

We have already seen the \textbf{median}, which is the number such that
approximately half of a data set is below (or above) it. The
\textbf{first quartile} is the number such that approximately 25\% of
the data is below it and the \textbf{third quartile} is the number such
that approximately 75\% of the data is below it. A \(100p\)
\textbf{percentile} is the number such that \(100 \times p\) percent of
the data is below it.

To generalize this concept, consider a distribution function
\(F(\cdot)\), which may or may not be from a continuous variable, and
let \(q\) be a fraction so that \(0<q<1\). We want to define a quantile,
say \(q_F\), to be a number such that \(F(q_F) \approx q\). Notice that
when \(q = 0.5\), \(q_F\) is the median; when \(q = 0.25\), \(q_F\) is
the first quartile, and so on.

To be precise, for a given \(0<q<1\), define the \(q\)\textbf{th
quantile} \(q_F\) to be any number that satisfies

\begin{equation}
  F(q_F-) \le q \le F(q_F)
  \label{eq:Quantile}
\end{equation}

Here, the notation \(F(x-)\) means to evaluate the function \(F(\cdot)\)
as a left-hand limit.

\begin{figure}

{\centering \includegraphics[width=0.6\linewidth]{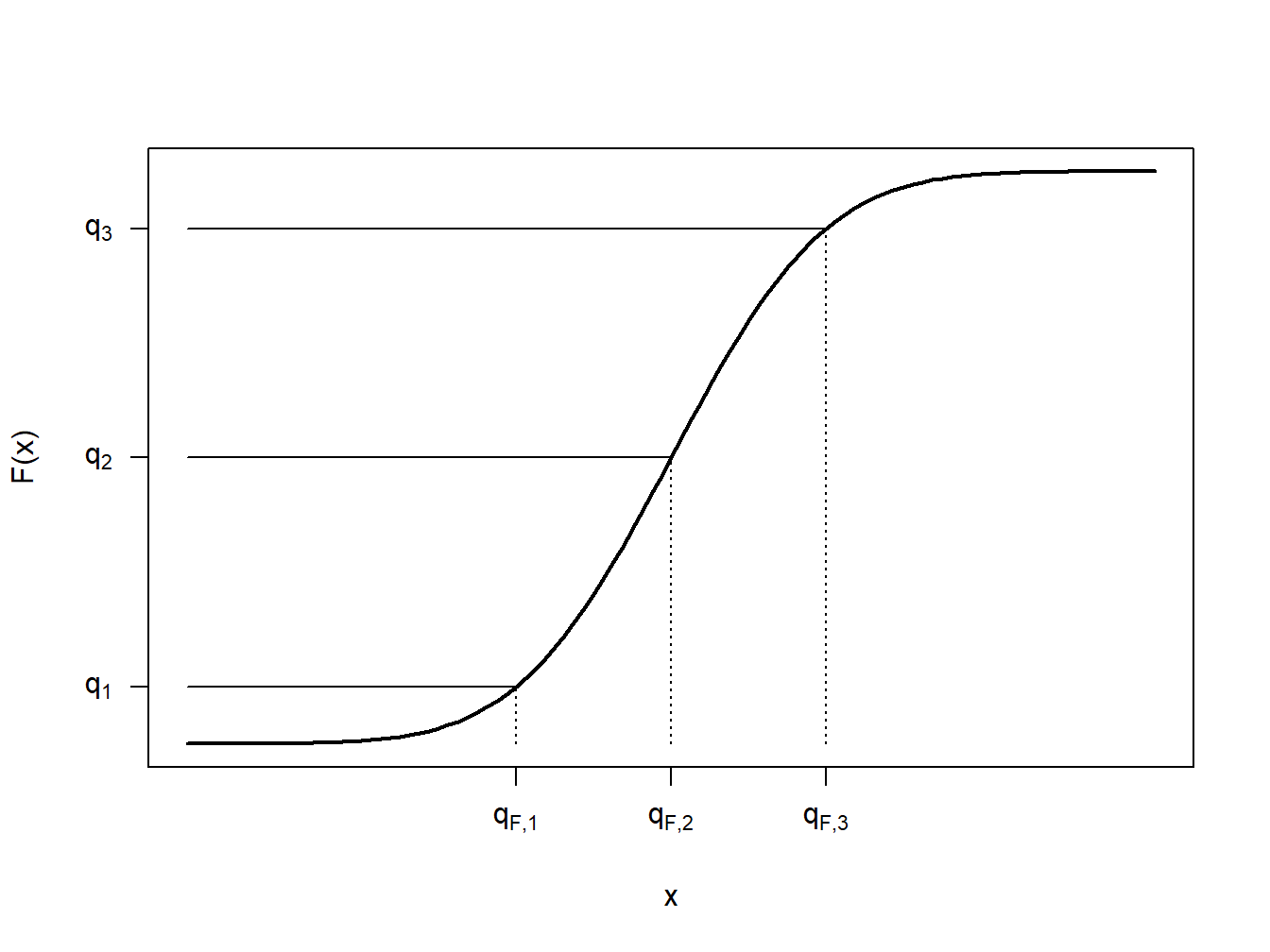}

}

\caption{Continuous Quantile Case}\label{fig:Quantile1}
\end{figure}

To get a better understanding of this definition, let us look at a few
special cases. First, consider the case where \(X\) is a continuous
random variable so that the distribution function \(F(\cdot)\) has no
jump points, as illustrated in Figure \ref{fig:Quantile1}. In this
figure, a few fractions, \(q_1\), \(q_2\), and \(q_3\) are shown with
their corresponding quantiles \(q_{F,1}\), \(q_{F,2}\), and \(q_{F,3}\).
In each case, it can be seen that \(F(q_F-)= F(q_F)\) so that there is a
unique quantile. Because we can find a unique inverse of the
distribution function at any \(0<q<1\), we can write \(q_F= F^{-1}(q)\).

\begin{figure}

{\centering \includegraphics[width=0.9\linewidth]{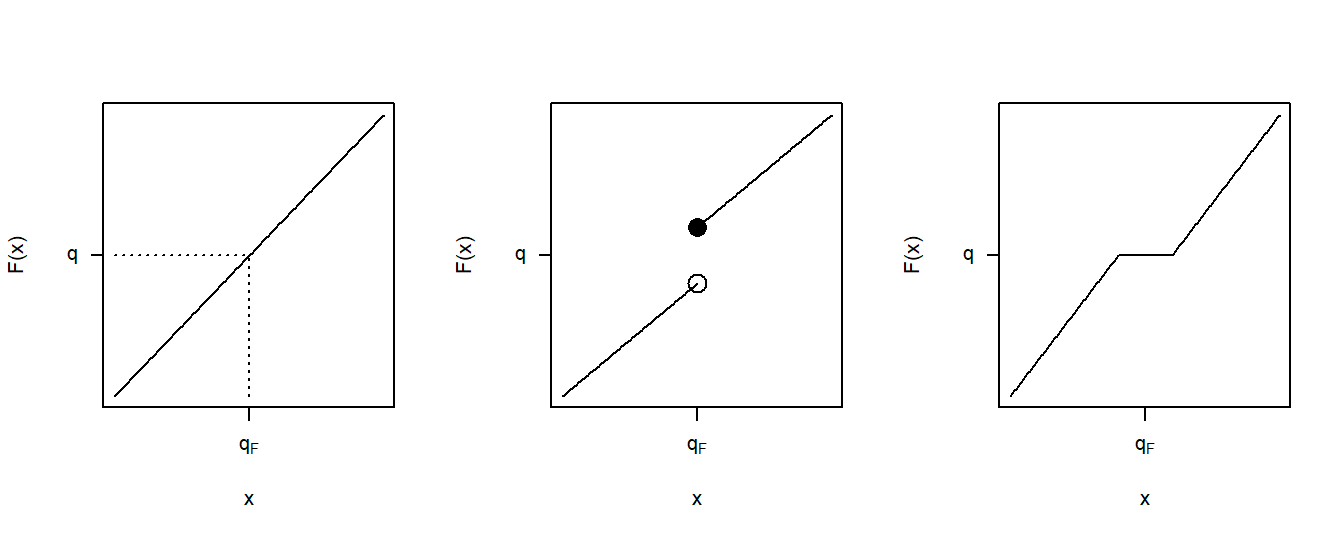}

}

\caption{Three Quantile Cases}\label{fig:Quantile2}
\end{figure}

Figure \ref{fig:Quantile2} shows three cases for distribution functions.
The left panel corresponds to the continuous case just discussed. The
middle panel displays a jump point similar to those we already saw in
the empirical distribution function of Figure \ref{fig:EDFToy}. For the
value of \(q\) shown in this panel, we still have a unique value of the
quantile \(q_F\). Even though there are many values of \(q\) such that
\(F(q_F-) \le q \le F(q_F)\), for a particular value of \(q\), there is
only one solution to equation \eqref{eq:Quantile}. The right panel depicts
a situation in which the quantile can not be uniquely determined for the
\(q\) shown as there is a range of \(q_F\)'s satisfying equation
\eqref{eq:Quantile}.

\begin{center}\rule{0.5\linewidth}{\linethickness}\end{center}

\textbf{Example 4.1.2. Toy Data Set: Continued.} Determine quantiles
corresponding to the 20th, 50th, and 95th percentiles.

Show Example Solution

\hypertarget{toggleExampleSelect.1.2}{}
\textbf{Solution}. Consider Figure \ref{fig:EDFToy}. The case of
\(q=0.20\) corresponds to the middle panel, so the 20th percentile is
15. The case of \(q=0.50\) corresponds to the right panel, so the median
is any number between 20 and 23 inclusive. Many software packages use
the average 21.5 (e.g. \texttt{R}, as seen below). For the 95th
percentile, the solution is 30. We can see from the graph that 30 also
corresponds to the 99th and the 99.99th percentiles.

\begin{Shaded}
\begin{Highlighting}[]
\KeywordTok{quantile}\NormalTok{(xExample, }\DataTypeTok{probs=}\KeywordTok{c}\NormalTok{(}\FloatTok{0.2}\NormalTok{, }\FloatTok{0.5}\NormalTok{, }\FloatTok{0.95}\NormalTok{), }\DataTypeTok{type=}\DecValTok{6}\NormalTok{)}
\end{Highlighting}
\end{Shaded}

\begin{verbatim}
##  20%  50%  95%
## 15.0 21.5 30.0
\end{verbatim}

\begin{center}\rule{0.5\linewidth}{\linethickness}\end{center}

By taking a weighted average between data observations, smoothed
empirical quantiles can handle cases such as the right panel in Figure
\ref{fig:Quantile2}. The \(q\)th \textbf{smoothed empirical quantile} is
defined as \[\hat{\pi}_q = (1-h) X_{(j)} + h X_{(j+1)}\] where
\(j=\lfloor(n+1)q\rfloor\), \(h=(n+1)q-j\), and
\(X_{(1)}, \ldots, X_{(n)}\) are the ordered values (the \textbf{order
statistics}) corresponding to \(X_1, \ldots, X_n\). Note that this is a
linear interpolation between \(X_{(j)}\) and \(X_{(j+1)}\).

\textbf{Example 4.1.3. Toy Data Set: Continued.} Determine the 50th and
20th smoothed percentiles.

Show Example Solution

\hypertarget{toggleExampleSelect.1.3}{}
\textbf{Solution} Take \(n=10\) and \(q=0.5\). Then,
\(j=\lfloor(11)0.5 \rfloor= \lfloor5.5 \rfloor=5\) and
\(h=(11)(0.5)-5=0.5\). Then the 0.5-th smoothed empirical quantile is
\[\hat{\pi}_{0.5} = (1-0.5) X_{(5)} + (0.5) X_{(6)} = 0.5 (20) + (0.5)(23) = 21.5.\]
Now take \(n=10\) and \(q=0.2\). In this case,
\(j=\lfloor(11)0.2\rfloor=\lfloor 2.2 \rfloor=2\) and
\(h=(11)(0.2)-2=0.2\). Then the 0.2-th smoothed empirical quantile is
\[\hat{\pi}_{0.2} = (1-0.2) X_{(2)} + (0.2) X_{(3)} = 0.2 (15) + (0.8)(15) = 15.\]

\begin{center}\rule{0.5\linewidth}{\linethickness}\end{center}

\subsubsection{Density Estimators}\label{density-estimators}

When the random variable is discrete, estimating the probability mass
function \(f(x) = \Pr(X=x)\) is straightforward. We simply use the
empirical average, defined to be
\[f_n(x) = \frac{1}{n} \sum_{i=1}^n I(X_i = x).\]

For a continuous random variable, consider a discretized formulation in
which the domain of \(F(\cdot)\) is partitioned by constants
\(\{c_0 < c_1 < \cdots < c_k\}\) into intervals of the form
\([c_{j-1}, c_j)\), for \(j=1, \ldots, k\). The data observations are
thus ``grouped'' by the intervals into which they fall. Then, we might
use the basic definition of the empirical mass function, or a variation
such as
\[f_n(x) = \frac{n_j}{n \times (c_j - c_{j-1})}  \ \ \ \ \ \ c_{j-1} \le x < c_j,\]
where \(n_j\) is the number of observations (\(X_i\)) that fall into the
interval \([c_{j-1}, c_j)\).

Extending this notion to instances where we observe individual data,
note that we can always create arbitrary groupings and use this formula.
More formally, let \(b>0\) be a small positive constant, known as a
\textbf{bandwidth}, and define a density estimator to be

\begin{equation}
  f_n(x) = \frac{1}{2nb} \sum_{i=1}^n I(x-b < X_i \le x + b)
  \label{eq:KDF}
\end{equation}

Show A Snippet of Theory

\hypertarget{Theorykerneldensity}{}
The idea is that the estimator \(f_n(x)\) in equation \eqref{eq:KDF} is
the average over \(n\) \emph{iid} realizations of a random variable with
mean

\[\begin{aligned}
\mathrm{E~ } \frac{1}{2b} I(x-b < X \le x + b) &=  \frac{1}{2b}\left(F(x+b)-F(x-b)\right) \\
&=  \frac{1}{2b} \left( \left\{ F(x) + b F^{\prime}(x) + b^2 C_1\right\}
\left\{ F(x) - b F^{\prime}(x) + b^2 C_2\right\} \right) \\
&=  F^{\prime}(x) + b \frac{C_1-C_2}{2} \rightarrow  F^{\prime}(x) = f(x),
\end{aligned}\]

as \(b\rightarrow 0\). That is, \(f_n(x)\) is an asymptotically unbiased
estimator of \(f(x)\) (its expectation approaches the true value as
sample size increases to infinity). This development assumes some
smoothness of \(F(\cdot)\), in particular, twice differentiability at
\(x\), but makes no assumptions on the form of the distribution function
\(F\). Because of this, the density estimator \(f_n\) is said to be
\emph{nonparametric}.

More generally, define the \textbf{kernel density estimator} as

\begin{equation}
  f_n(x) = \frac{1}{nb} \sum_{i=1}^n w\left(\frac{x-X_i}{b}\right)
  \label{eq:kernelDens}
\end{equation}

where \(w\) is a probability density function centered about 0. Note
that equation \eqref{eq:KDF} simply becomes the kernel density estimator
where \(w(x) = \frac{1}{2}I(-1 < x \le 1)\), also known as the
\textbf{uniform kernel}. Other popular choices are shown in
\protect\hyperlink{tab:41}{Table 4.1}.

\[\begin{matrix}
\text{Table 4.1: Popular Choices for the Kernel Density Estimator}\\
\begin{array}{l|cc}
\hline
\text{Kernel} &  w(x) \\
\hline
\text{Uniform } &  \frac{1}{2}I(-1 < x \le 1) \\
\text{Triangle} &  (1-|x|)\times I(|x| \le 1) \\
\text{Epanechnikov} & \frac{3}{4}(1-x^2) \times I(|x| \le 1) \\
\text{Gaussian} & \phi(x) \\
\hline
\end{array}\end{matrix}\]

Here, \(\phi(\cdot)\) is the standard normal density function. As we
will see in the following example, the choice of bandwidth \(b\) comes
with a \emph{bias-variance tradeoff} between matching local
distributional features and reducing the volatility.

\begin{center}\rule{0.5\linewidth}{\linethickness}\end{center}

\begin{figure}

{\centering \includegraphics[width=0.7\linewidth]{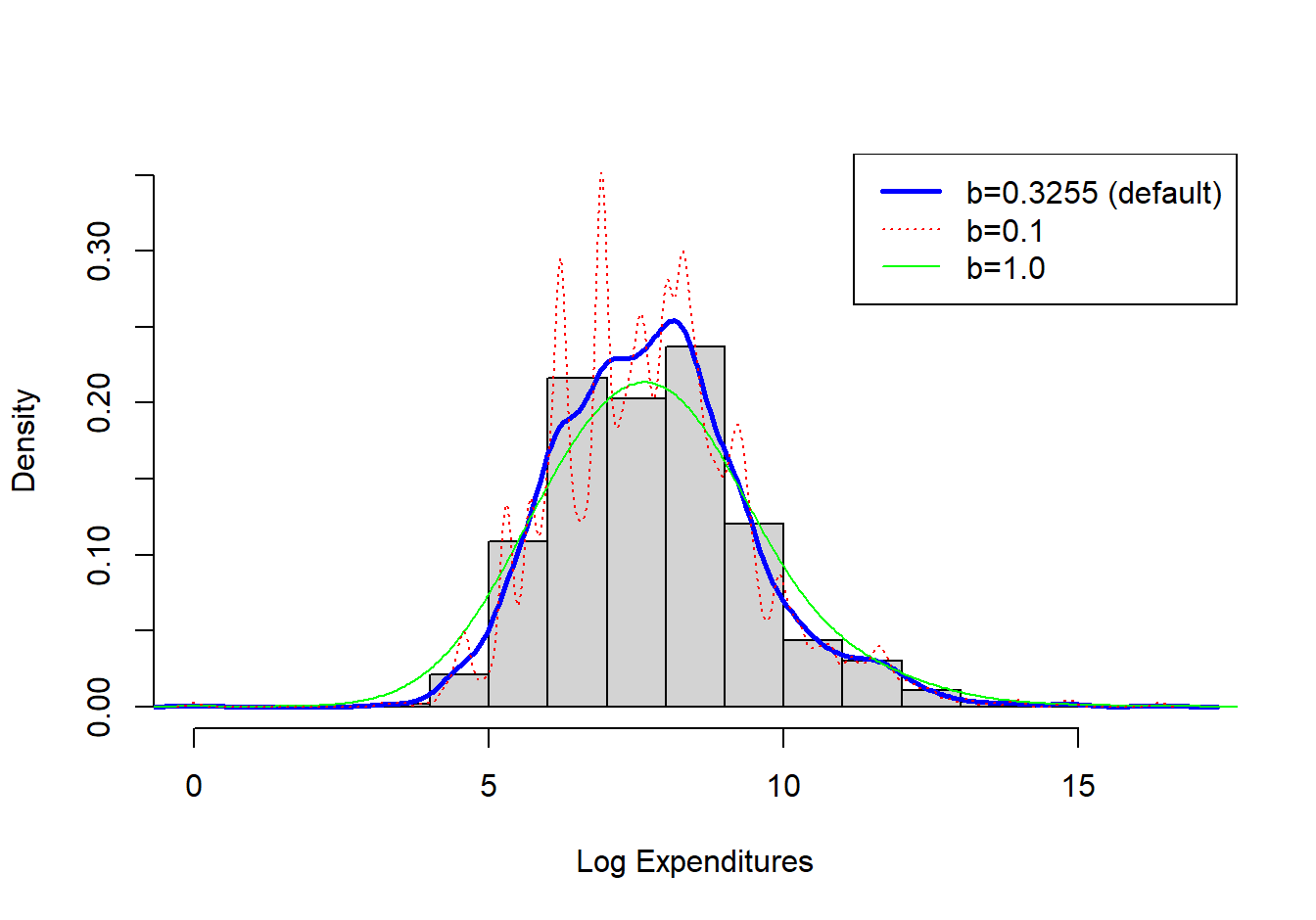}

}

\caption{Histogram of Logarithmic Property Claims with Superimposed Kernel Density Estimators}\label{fig:Density2}
\end{figure}

\textbf{Example 4.1.4. Property Fund.} Figure \ref{fig:Density2} shows a
histogram (with shaded gray rectangles) of logarithmic property claims
from 2010. The (blue) thick curve represents a Gaussian kernel density
where the bandwidth was selected automatically using an ad hoc rule
based on the sample size and volatility of the data. For this dataset,
the bandwidth turned out to be \(b=0.3255\). For comparison, the (red)
dashed curve represents the density estimator with a bandwidth equal to
0.1 and the green smooth curve uses a bandwidth of 1. As anticipated,
the smaller bandwidth (0.1) indicates taking local averages over less
data so that we get a better idea of the local average, but at the price
of higher volatility. In contrast, the larger bandwidth (1) smooths out
local fluctuations, yielding a smoother curve that may miss
perturbations in the local average. For actuarial applications, we
mainly use the kernel density estimator to get a quick visual impression
of the data. From this perspective, you can simply use the default ad
hoc rule for bandwidth selection, knowing that you have the ability to
change it depending on the situation at hand.

Show R Code

\hypertarget{togglekpdf}{}
\begin{verbatim}
#Density Comparison
hist(log(ClaimData$Claim), main="", ylim=c(0,.35),xlab="Log Expenditures", freq=FALSE, col="lightgray")
lines(density(log(ClaimData$Claim)), col="blue",lwd=2.5)
lines(density(log(ClaimData$Claim), bw=1), col="green")
lines(density(log(ClaimData$Claim), bw=.1), col="red", lty=3)
legend("topright", c("b=0.3255 (default)", "b=0.1", "b=1.0"), lty=c(1,3,1),
            lwd=c(2.5,1,1), col=c("blue", "red", "green"), cex=1)
\end{verbatim}

\begin{center}\rule{0.5\linewidth}{\linethickness}\end{center}

Nonparametric density estimators, such as the kernel estimator, are
regularly used in practice. The concept can also be extended to give
smooth versions of an empirical distribution function. Given the
definition of the kernel density estimator, the kernel estimator of the
distribution function can be found as \[\begin{aligned}
\hat{F}_n(x) = \frac{1}{n} \sum_{i=1}^n W\left(\frac{x-X_i}{b}\right).\end{aligned}\]

where \(W\) is the distribution function associated with the kernel
density \(w\). To illustrate, for the uniform kernel, we have
\(w(y) = \frac{1}{2}I(-1 < y \le 1)\), so \[\begin{aligned}
W(y) =
\begin{cases}
0 &            y<-1\\
\frac{y+1}{2}& -1 \le y < 1 \\
1 & y \ge 1 \\
\end{cases}\end{aligned}\]

\begin{center}\rule{0.5\linewidth}{\linethickness}\end{center}

\textbf{Example 4.1.5. SOA Exam Question.} You study five lives to
estimate the time from the onset of a disease to death. The times to
death are:

\[\begin{array}{ccccc}
2 & 3 & 3 & 3 & 7  \\
\end{array}\]

Using a triangular kernel with bandwith \(2\), calculate the density
function estimate at 2.5.

Show Example Solution

\hypertarget{toggleExampleSelect.1.5}{}
\textbf{Solution.} For the kernel density estimate, we have
\[f_n(x) = \frac{1}{nb} \sum_{i=1}^n w\left(\frac{x-X_i}{b}\right),\]
where \(n=5\), \(b=2\), and \(x=2.5\). For the triangular kernel,
\(w(x) = (1-|x|)\times I(|x| \le 1)\). Thus,

\[\begin{array}{c|c|c}
\hline
X_i & \frac{x-X_i}{b} & w\left(\frac{x-X_i}{b} \right) \\
\hline
2 & \frac{2.5-2}{2}=\frac{1}{4} &  (1-\frac{1}{4})(1) = \frac{3}{4} \\
\hline
3 & & \\
3 & \frac{2.5-3}{2}=\frac{-1}{4} & \left(1-\left| \frac{-1}{4} \right| \right)(1) = \frac{3}{4} \\
3 & & \\
\hline
7 & \frac{2.5-7}{2}=-2.25 & (1-|-2.25|)(0) = 0\\
\hline
\end{array}\]

Then the kernel density estimate is
\[f_n(x) = \frac{1}{5(2)}\left( \frac{3}{4} + (3) \frac{3}{4} + 0 \right) = \frac{3}{10}\]

\begin{center}\rule{0.5\linewidth}{\linethickness}\end{center}

\subsection{Tools for Model Selection}\label{S:MS:ToolsModelSelection}

The previous section introduced nonparametric estimators in which there
was no parametric form assumed about the underlying distributions.
However, in many actuarial applications, analysts seek to employ a
parametric fit of a distribution for ease of explanation and the ability
to readily extend it to more complex situations such as including
explanatory variables in a regression setting. When fitting a parametric
distribution, one analyst might try to use a gamma distribution to
represent a set of loss data. However, another analyst may prefer to use
a Pareto distribution. How does one know which model to select?

Nonparametric tools can be used to corroborate the selection of
parametric models. Essentially, the approach is to compute selected
summary measures under a fitted parametric model and to compare it to
the corresponding quantity under the nonparametric model. As the
nonparametric does not assume a specific distribution and is merely a
function of the data, it is used as a benchmark to assess how well the
parametric distribution/model represents the data. This comparison may
alert the analyst to deficiencies in the parametric model and sometimes
point ways to improving the parametric specification.

\subsubsection{Graphical Comparison of
Distributions}\label{graphical-comparison-of-distributions}

\begin{figure}

{\centering \includegraphics[width=0.8\linewidth]{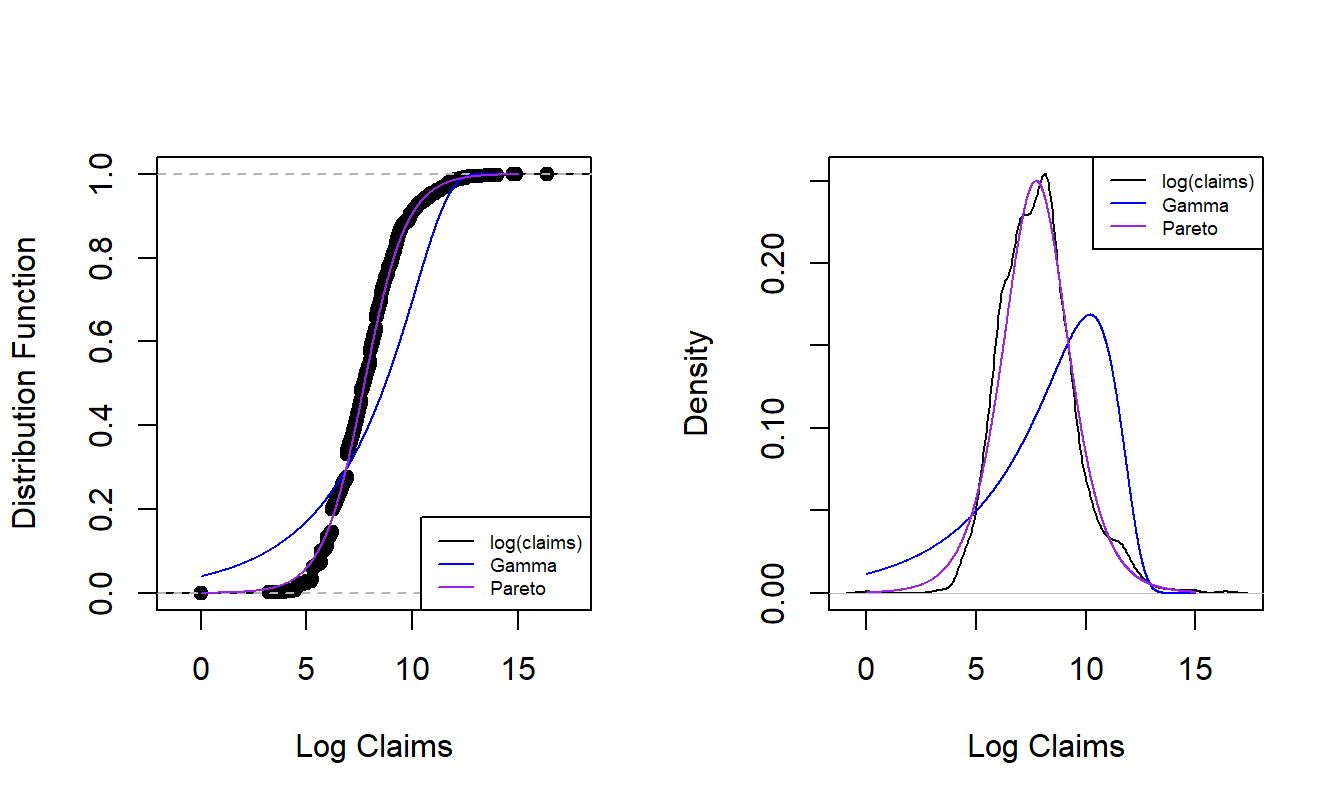}

}

\caption{Nonparametric Versus Fitted Parametric Distribution and Density Functions. The left-hand panel compares distribution functions, with the dots corresponding to the empirical distribution, the thick blue curve corresponding to the fitted gamma and the light purple curve corresponding to the fitted Pareto. The right hand panel compares these three distributions summarized using probability density functions.}\label{fig:ComparisonCDFPDF}
\end{figure}

We have already seen the technique of overlaying graphs for comparison
purposes. To reinforce the application of this technique, Figure
\ref{fig:ComparisonCDFPDF} compares the empirical distribution to two
parametric fitted distributions. The left panel shows the distribution
functions of claims distributions. The dots forming an ``S-shaped''
curve represent the empirical distribution function at each observation.
The thick blue curve gives corresponding values for the fitted gamma
distribution and the light purple is for the fitted Pareto distribution.
Because the Pareto is much closer to the empirical distribution function
than the gamma, this provides evidence that the Pareto is the better
model for this data set. The right panel gives similar information for
the density function and provides a consistent message. Based on these
figures, the Pareto distribution is the clear choice for the analyst.

\begin{figure}

{\centering \includegraphics[width=0.8\linewidth]{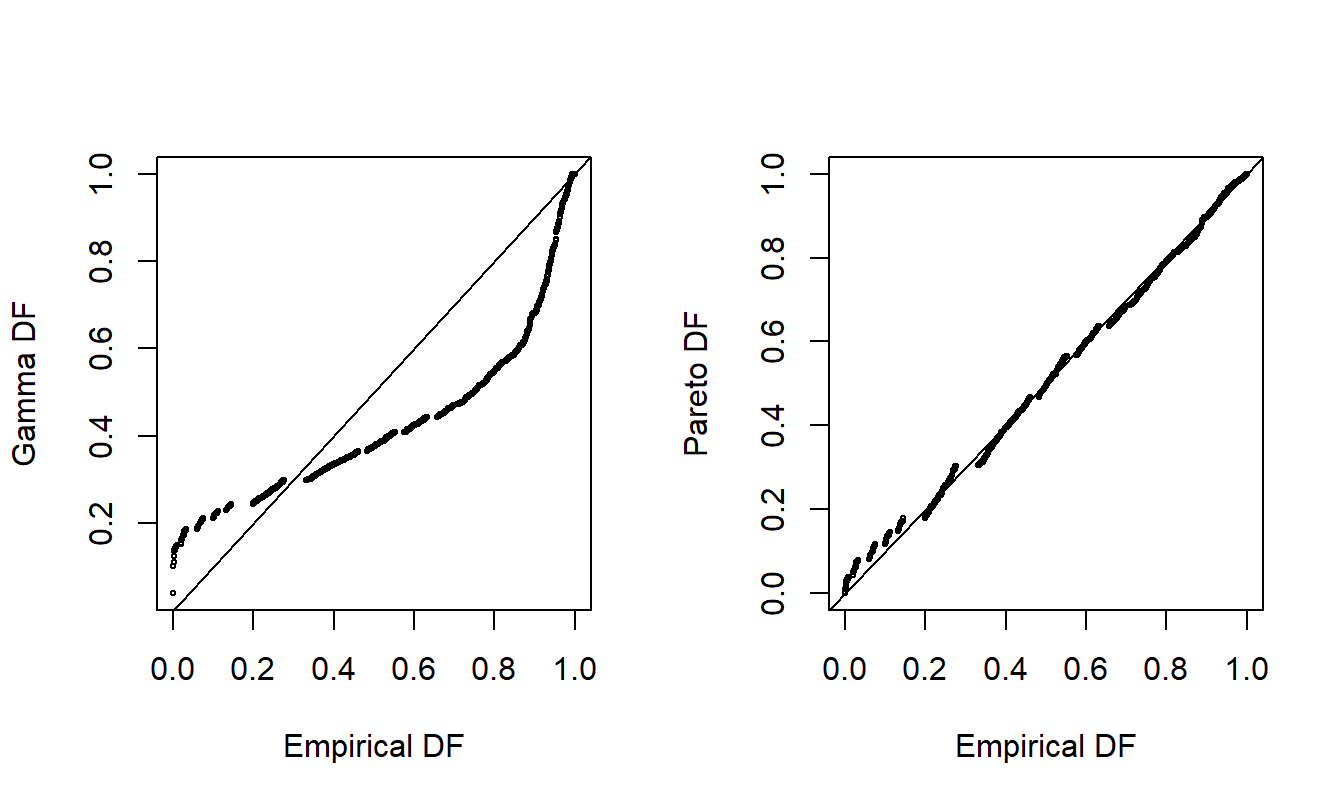}

}

\caption{Probability-Probability ($pp$) Plots. The horizontal axes gives the empirical distribution function at each observation. In the left-hand panel, the corresponding distribution function for the gamma is shown in the vertical axis. The right-hand panel shows the fitted Pareto distribution. Lines of $y=x$ are superimposed.}\label{fig:PPPlot}
\end{figure}

For another way to compare the appropriateness of two fitted models,
consider the \textbf{probability-probability (\(pp\)) plot}. A \(pp\)
plot compares cumulative probabilities under two models. For our
purposes, these two models are the nonparametric empirical distribution
function and the parametric fitted model. Figure \ref{fig:PPPlot} shows
\(pp\) plots for the Property Fund data. The fitted gamma is on the left
and the fitted Pareto is on the right, compared to the same empirical
distribution function of the data. The straight line represents equality
between the two distributions being compared, so points close to the
line are desirable. As seen in earlier demonstrations, the Pareto is
much closer to the empirical distribution than the gamma, providing
additional evidence that the Pareto is the better model.

\begin{figure}

{\centering \includegraphics[width=0.8\linewidth]{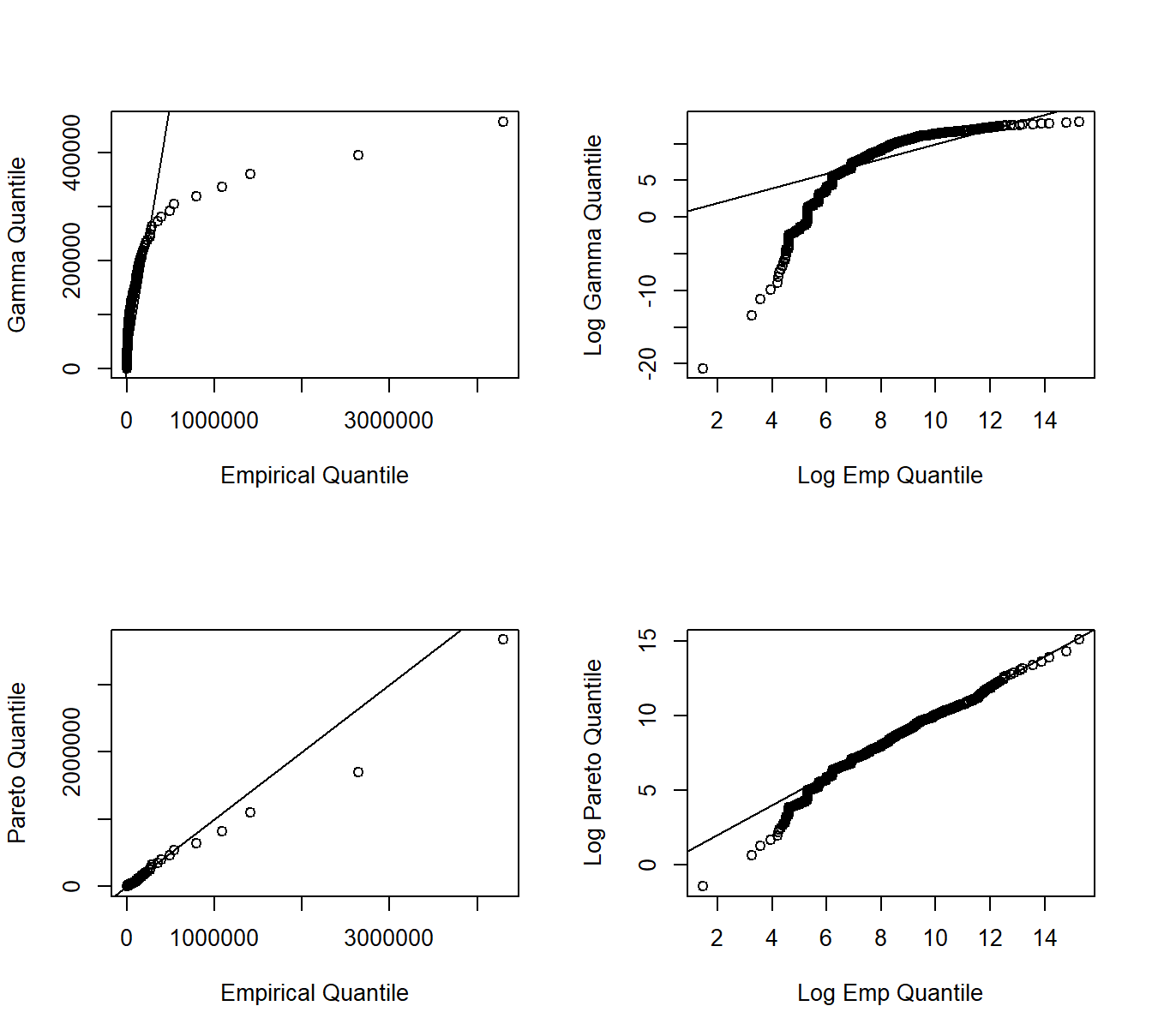}

}

\caption{Quantile-Quantile ($qq$) Plots. The horizontal axes gives the empirical quantiles at each observation. The right-hand panels they are graphed on a logarithmic basis. The vertical axis gives the quantiles from the fitted distributions; Gamma quantiles are in the upper panels, Pareto quantiles are in the lower panels.}\label{fig:QQPlot}
\end{figure}

A \(pp\) plot is useful in part because no artificial scaling is
required, such as with the overlaying of densities in Figure
\ref{fig:ComparisonCDFPDF}, in which we switched to the log scale to
better visualize the data. Furthermore, \(pp\) plots are available in
multivariate settings where more than one outcome variable is available.
However, a limitation of the \(pp\) plot is that, because they plot
\emph{cumulative} distribution functions, it can sometimes be difficult
to detect \emph{where} a fitted parametric distribution is deficient. As
an alternative, it is common to use a \textbf{quantile-quantile (\(qq\))
plot}, as demonstrated in Figure \ref{fig:QQPlot}.

The \(qq\) plot compares two fitted models through their quantiles. As
with \(pp\) plots, we compare the nonparametric to a parametric fitted
model. Quantiles may be evaluated at each point of the data set, or on a
grid (e.g., at \(0, 0.001, 0.002, \ldots, 0.999, 1.000\)), depending on
the application. In Figure \ref{fig:QQPlot}, for each point on the
aforementioned grid, the horizontal axis displays the empirical quantile
and the vertical axis displays the corresponding fitted parametric
quantile (gamma for the upper two panels, Pareto for the lower two).
Quantiles are plotted on the original scale in the left panels and on
the log scale in the right panels to allow us to see where a fitted
distribution is deficient. The straight line represents equality between
the empirical distribution and fitted distribution. From these plots, we
again see that the Pareto is an overall better fit than the gamma.
Furthermore, the lower-right panel suggests that the Pareto distribution
does a good job with large observations, but provides a poorer fit for
small observations.

\begin{center}\rule{0.5\linewidth}{\linethickness}\end{center}

\textbf{Example 4.1.6. SOA Exam Question.} The graph below shows a
\(pp\) plot of a fitted distribution compared to a sample.

\begin{figure}

{\centering \includegraphics[width=0.4\linewidth]{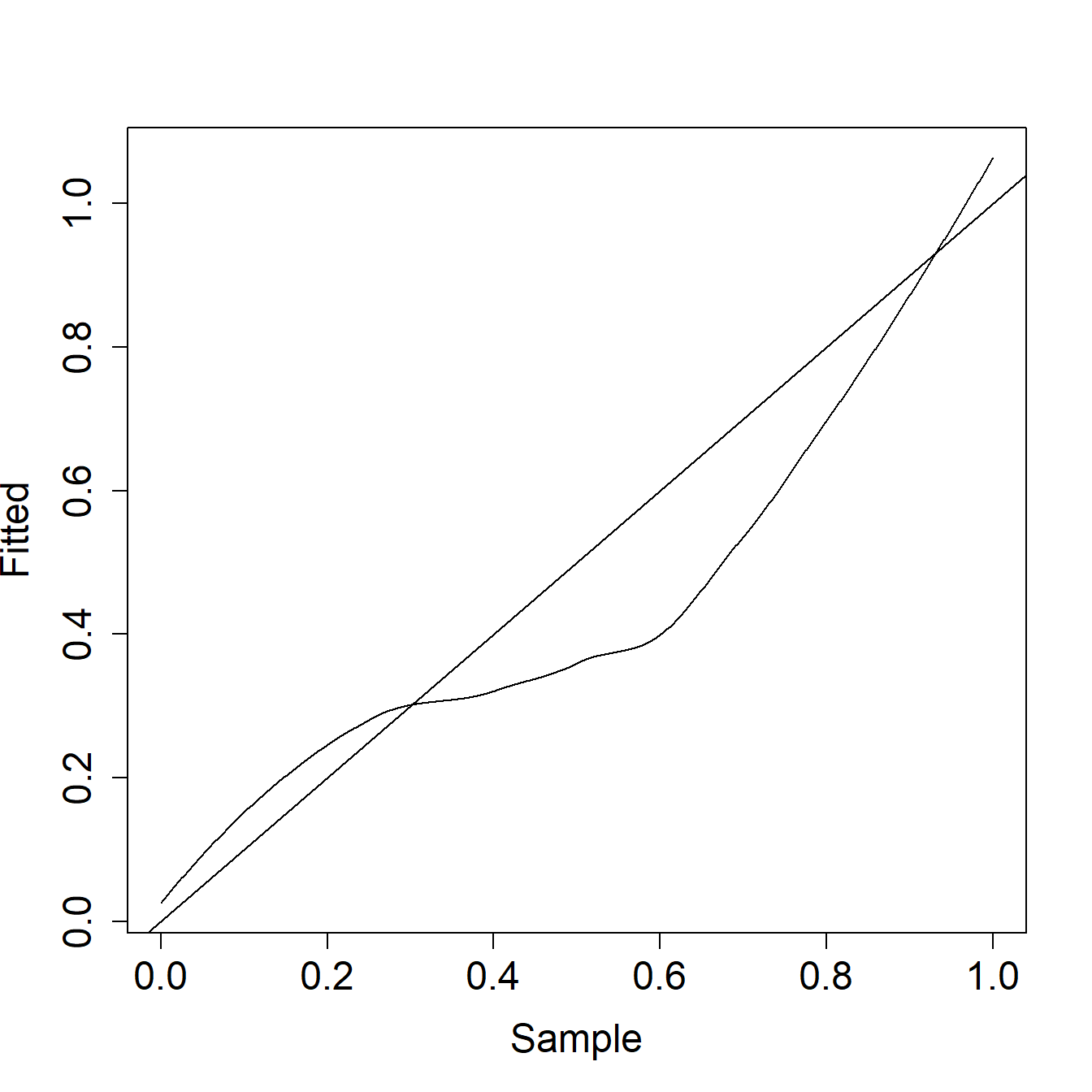}

}

\end{figure}

Comment on the two distributions with respect to left tail, right tail,
and median probabilities.

Show Example Solution

\hypertarget{toggleExampleSelect.1.6}{}
\textbf{Solution.} The tail of the fitted distribution is too thick on
the left, too thin on the right, and the fitted distribution has less
probability around the median than the sample. To see this, recall that
the \(pp\) plot graphs the cumulative distribution of two distributions
on its axes (empirical on the x-axis and fitted on the y-axis in this
case). For small values of \(x\), the fitted model assigns greater
probability to being below that value than occurred in the sample (i.e.
\(F(x) > F_n(x)\)). This indicates that the model has a heavier left
tail than the data. For large values of \(x\), the model again assigns
greater probability to being below that value and thus less probability
to being above that value (i.e. \(S(x) < S_n(x)\). This indicates that
the model has a lighter right tail than the data. In addition, as we go
from 0.4 to 0.6 on the horizontal axis (thus looking at the middle 20\%
of the data), the \(pp\) plot increases from about 0.3 to 0.4. This
indicates that the model puts only about 10\% of the probability in this
range.

\begin{center}\rule{0.5\linewidth}{\linethickness}\end{center}

\subsubsection{Statistical Comparison of
Distributions}\label{S:MS:Tools:Stats}

When selecting a model, it is helpful to make the graphical displays
presented. However, for reporting results, it can be effective to
supplement the graphical displays with selected statistics that
summarize model goodness of fit. \protect\hyperlink{tab:42}{Table 4.2}
provides three commonly used goodness of fit statistics. Here, \(F_n\)
is the empirical distribution and \(F\) is the fitted distribution.

\[\begin{matrix}
\text{Table 4.2: Three Goodness of Fit Statistics} \\
\begin{array}{l|cc}
\hline
\text{Statistic} & \text{Definition} & \text{Computational Expression} \\
\hline
\text{Kolmogorov-} & \max_x |F_n(x) - F(x)| & \max(D^+, D^-) \text{ where } \\
~~~\text{Smirnov} && D^+ = \max_{i=1, \ldots, n} \left|\frac{i}{n} - F_i\right| \\
&& D^- = \max_{i=1, \ldots, n} \left| F_i - \frac{i-1}{n} \right| \\
\text{Cramer-von Mises} & n \int (F_n(x) - F(x))^2 f(x) dx & \frac{1}{12n} + \sum_{i=1}^n \left(F_i - (2i-1)/n\right)^2 \\
\text{Anderson-Darling} & n \int \frac{(F_n(x) - F(x))^2}{F(x)(1-F(x))} f(x) dx & -n-\frac{1}{n} \sum_{i=1}^n (2i-1) \log\left(F_i(1-F_{n+1-i})\right)^2 \\
\hline
\end{array} \\
\text{where } F_i \text{ is defined to be } F(x_i).
\end{matrix}\]

The \textbf{Kolmogorov-Smirnov statistic} is the maximum absolute
difference between the fitted distribution function and the empirical
distribution function. Instead of comparing differences between single
points, the \textbf{Cramer-von Mises statistic} integrates the
difference between the empirical and fitted distribution functions over
the entire range of values. The \textbf{Anderson-Darling statistic} also
integrates this difference over the range of values, although weighted
by the inverse of the variance. It therefore places greater emphasis on
the tails of the distribution (i.e when \(F(x)\) or \(1-F(x)=S(x)\) is
small).

\begin{center}\rule{0.5\linewidth}{\linethickness}\end{center}

\textbf{Exaxmple 4.1.7. SOA Exam Question (modified).} A sample of claim
payments is:

\[\begin{array}{ccccc}
29 & 64 & 90 & 135 & 182  \\
\end{array}\]

Compare the empirical claims distribution to an exponential distribution
with mean \(100\) by calculating the value of the Kolmogorov-Smirnov
test statistic.

Show Example Solution

\hypertarget{toggleExampleSelect.1.7}{}
\textbf{Solution.} For an exponential distribution with mean \(100\),
the cumulative distribution function is \(F(x)=1-e^{-x/100}\). Thus,

\[\begin{array}{ccccc}
\hline
x & F(x) & F_n(x) & F_n(x-) & \max(|F(x)-F_n(x)|,|F(x)-F_n(x-)|) \\
\hline
29  & 0.2517 & 0.2 & 0   & \max(0.0517, 0.2517) = 0.2517 \\
64  & 0.4727 & 0.4 & 0.2 & \max(0.0727, 0.2727) = 0.2727 \\
90  & 0.5934 & 0.6 & 0.4 & \max(0.0066, 0.1934) = 0.1934 \\
135 & 0.7408 & 0.8 & 0.6 & \max(0.0592, 0.1408) = 0.1408 \\
182 & 0.8380 & 1   & 0.8 & \max(0.1620, 0.0380) = 0.1620 \\
\hline
\end{array}\]

The Kolmogorov-Smirnov test statistic is therefore
\(KS = \max(0.2517, 0.2727, 0.1934, 0.1408, 0.1620) = 0.2727\).

\begin{center}\rule{0.5\linewidth}{\linethickness}\end{center}

\subsection{Starting Values}\label{starting-values}

The method of moments and percentile matching are nonparametric
estimation methods that provide alternatives to maximum likelihood.
Generally, maximum likelihood is the preferred technique because it
employs data more efficiently. However, methods of moments and
percentile matching are useful because they are easier to interpret and
therefore allow the actuary or analyst to explain procedures to others.
Additionally, the numerical estimation procedure (e.g.~if performed in
\texttt{R}) for the maximum likelihood is iterative and requires
starting values to begin the recursive process. Although many problems
are robust to the choice of the starting values, for some complex
situations, it can be important to have a starting value that is close
to the (unknown) optimal value. Method of moments and percentile
matching are techniques that can produce desirable estimates without a
serious computational investment and can thus be used as a starting
value for computing maximum likelihood.

\subsubsection{Method of Moments}\label{method-of-moments}

Under the \textbf{method of moments}, we approximate the moments of the
parametric distribution using the empirical (nonparametric) moments
described in Section \ref{S:MS:MomentEstimator}. We can then
algebraically solve for the parameter estimates.

\begin{center}\rule{0.5\linewidth}{\linethickness}\end{center}

\textbf{Example 4.1.8. Property Fund.} For the 2010 property fund, there
are \(n=1,377\) individual claims (in thousands of dollars) with

\[m_1 = \frac{1}{n} \sum_{i=1}^n X_i = 26.62259 \ \ \ \
\text{and} \ \ \ \
 m_2 = \frac{1}{n} \sum_{i=1}^n X_i^2 = 136154.6 .\] Fit the parameters
of the gamma and Pareto distributions using the method of moments.

Show Example Solution

\hypertarget{toggleExampleSelect.1.8}{}
\textbf{Solution.} To fit a gamma distribution, we have
\(\mu_1 = \alpha \theta\) and
\(\mu_2^{\prime} = \alpha(\alpha+1) \theta^2\). Equating the two yields
the method of moments estimators, easy algebra shows that

\[\alpha = \frac{\mu_1^2}{\mu_2^{\prime}-\mu_1^2}  \ \ \ \text{and} \ \ \  \theta = \frac{\mu_2^{\prime}-\mu_1^2}{\mu_1}.\]

Thus, the method of moment estimators are

\[\begin{aligned}
\hat{\alpha} &=  \frac{26.62259^2}{136154.6-26.62259^2} = 0.005232809 \\
\hat{\theta} &=  \frac{136154.6-26.62259^2}{26.62259} = 5,087.629.
\end{aligned}\]

For comparison, the maximum likelihood values turn out to be
\(\hat{\alpha}_{MLE} = 0.2905959\) and
\(\hat{\theta}_{MLE} = 91.61378\), so there are big discrepancies
between the two estimation procedures. This is one indication, as we
have seen before, that the gamma model fits poorly.

In contrast, now assume a Pareto distribution so that
\(\mu_1 = \theta/(\alpha -1)\) and
\(\mu_2^{\prime} = 2\theta^2/((\alpha-1)(\alpha-2) )\). Easy algebra
shows

\[\alpha = 1+ \frac{\mu_2^{\prime}}{\mu_2^{\prime}-\mu_1^2} \ \ \ \
\text{and} \ \ \ \ \
 \theta = (\alpha-1)\mu_1.\]

Thus, the method of moment estimators are

\[\begin{aligned}
\hat{\alpha} &=  1+ \frac{136154.6}{136154.6-26,62259^2} = 2.005233 \\
\hat{\theta} &=  (2.005233-1) \cdot 26.62259 = 26.7619
\end{aligned}\]

The maximum likelihood values turn out to be
\(\hat{\alpha}_{MLE} = 0.9990936\) and
\(\hat{\theta}_{MLE} = 2.2821147\). It is interesting that
\(\hat{\alpha}_{MLE}<1\); for the Pareto distribution, recall that
\(\alpha <1\) means that the mean is infinite. This is another
indication that the property claims data set is a long tail
distribution.

\begin{center}\rule{0.5\linewidth}{\linethickness}\end{center}

As the above example suggests, there is flexibility with the method of
moments. For example, we could have matched the second and third moments
instead of the first and second, yielding different estimators.
Furthermore, there is no guarantee that a solution will exist for each
problem. You will also find that matching moments is possible for a few
problems where the data are censored or truncated, but in general, this
is a more difficult scenario. Finally, for distributions where the
moments do not exist or are infinite, method of moments is not
available. As an alternative for the infinite moment situation, one can
use the percentile matching technique.

\subsubsection{Percentile Matching}\label{percentile-matching}

Under percentile matching, we approximate the quantiles or percentiles
of the parametric distribution using the empirical (nonparametric)
quantiles or percentiles described in Section
\ref{S:MS:QuantileEstimator}.

\begin{center}\rule{0.5\linewidth}{\linethickness}\end{center}

\textbf{Example 4.1.9. Property Fund.} For the 2010 property fund, we
illustrate matching on quantiles. In particular, the Pareto distribution
is intuitively pleasing because of the closed-form solution for the
quantiles. Recall that the distribution function for the Pareto
distribution is
\[F(x) = 1 - \left(\frac{\theta}{x+\theta}\right)^{\alpha}.\] Easy
algebra shows that we can express the quantile as
\[F^{-1}(q) = \theta \left( (1-q)^{-1/\alpha} -1 \right).\] for a
fraction \(q\), \(0<q<1\).

Determine estimates of the Pareto distribution parameters using the 25th
and 95th empirical quantiles.

Show Example Solution

\hypertarget{toggleExampleSelect.1.9}{}
\textbf{Solution.}

The 25th percentile (the first quartile) turns out to be \(0.78853\) and
the 95th percentile is \(50.98293\) (both in thousands of dollars). With
two equations
\[0.78853 = \theta \left( 1- (1-.25)^{-1/\alpha} \right) \ \ \ \ \text{and} \ \ \ \ 50.98293 = \theta \left( 1- (1-.75)^{-1/\alpha} \right)\]
and two unknowns, the solution is
\[\hat{\alpha} = 0.9412076 \ \ \ \ \ \text{and} \ \ \ \
\hat{\theta} = 2.205617 .\] We remark here that a numerical routine is
required for these solutions as no analytic solution is available.
Furthermore, recall that the maximum likelihood estimates are
\(\hat{\alpha}_{MLE} = 0.9990936\) and
\(\hat{\theta}_{MLE} = 2.2821147\), so the percentile matching provides
a better approximation for the Pareto distribution than the method of
moments.

\begin{center}\rule{0.5\linewidth}{\linethickness}\end{center}

\textbf{Exercise 4.1.10. SOA Exam Question.} You are given:

\begin{enumerate}
\def\labelenumi{(\roman{enumi})}
\tightlist
\item
  Losses follow a loglogistic distribution with cumulative distribution
  function:
  \[F(x) = \frac{\left(x/\theta\right)^{\gamma}}{1+\left(x/\theta\right)^{\gamma}}\]
\item
  The sample of losses is:
\end{enumerate}

\[\begin{array}{ccccccccccc}
10 &35 &80 &86 &90 &120 &158 &180 &200 &210 &1500 \\
\end{array}\]

Calculate the estimate of \(\theta\) by percentile matching, using the
40th and 80th empirically smoothed percentile estimates.

Show Example Solution

\hypertarget{toggleExampleSelect.1.10}{}
\textbf{Solution.} With 11 observations, we have
\(j=\lfloor(n+1)q\rfloor = \lfloor 12(0.4) \rfloor = \lfloor 4.8\rfloor=4\)
and \(h=(n+1)q-j = 12(0.4)-4=0.8\). By interpolation, the 40th
empirically smoothed percentile estimate is
\(\hat{\pi}_{0.4} = (1-h) X_{(j)} + h X_{(j+1)} = 0.2(86)+0.8(90)=89.2\).

Similarly, for the 80th empirically smoothed percentile estimate, we
have \(12(0.8)=9.6\) so the estimate is
\(\hat{\pi}_{0.8} = 0.4(200)+0.6(210)=206\).

Using the loglogistic cumulative distribution, we need to solve the
following two equations for parameters \(\theta\) and \(gamma\):
\[0.4=\frac{(89.2/\theta)^\gamma}{1+(89.2/\theta)^\gamma} \ \ \ \text{and} \ \ \ \   0.8=\frac{(206/\theta)^\gamma}{1+(206+\theta)^\gamma}\]

Solving for each parenthetical expression gives
\(\frac{2}{3}=(89.2/\theta)^\gamma\) and \(4=(206/\theta)^\gamma\).
Taking the ratio of the second equation to the first gives
\(6=(206/89.2)^\gamma \Rightarrow \gamma=\frac{\ln(6)}{\ln(206/89.2)} = 2.1407\).
Then \(4^{1/2.1407}=206/\theta \Rightarrow \theta=107.8\)

\begin{center}\rule{0.5\linewidth}{\linethickness}\end{center}

\section{Model Selection}\label{S:MS:ModelSelection}

\begin{center}\rule{0.5\linewidth}{\linethickness}\end{center}

In this section, you learn how to:

\begin{itemize}
\tightlist
\item
  Describe the iterative model selection specification process
\item
  Outline steps needed to select a parametric model
\item
  Describe pitfalls of model selection based purely on insample data
  when compared to the advantages of out-of-sample model validation
\end{itemize}

\begin{center}\rule{0.5\linewidth}{\linethickness}\end{center}

This section underscores the idea that model selection is an iterative
process in which models are cyclically (re)formulated and tested for
appropriateness before using them for inference. After summarizing the
process of selecting a model based on the dataset at hand, we describe
model selection process based on:

\begin{itemize}
\item
  an in-sample or training dataset,
\item
  an out-of-sample or test dataset, and
\item
  a method that combines these approaches known as
  \textbf{cross-validation}.
\end{itemize}

\subsection{Iterative Model Selection}\label{iterative-model-selection}

\begin{figure}

{\centering \includegraphics[width=0.8\linewidth]{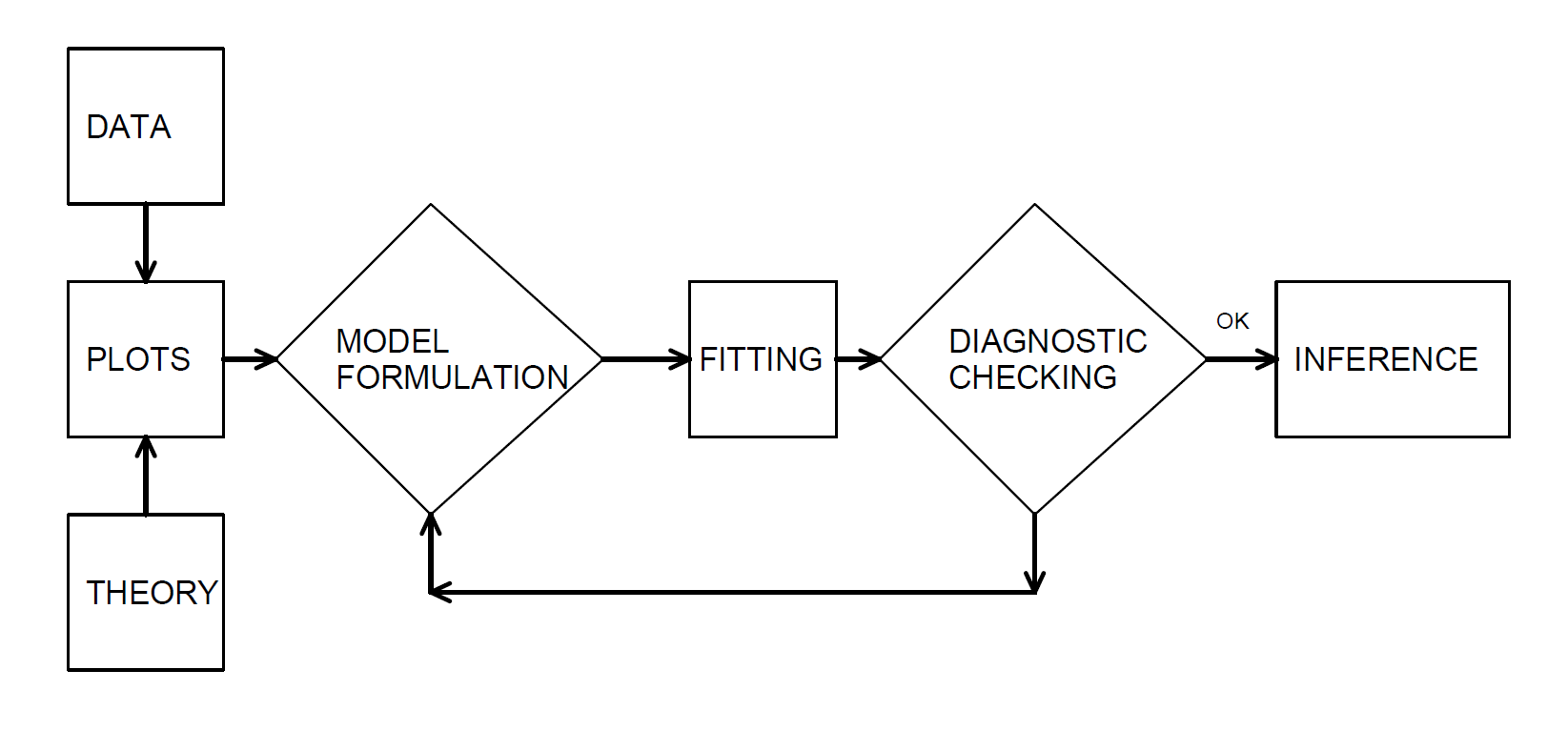}

}

\caption{The iterative model specification process.}\label{fig:Iterative}
\end{figure}

In our development, we examine the data graphically, hypothesize a model
structure, and compare the data to a candidate model in order to
formulate an improved model. \citet{box1980sampling} describes this as
an \emph{iterative process} which is shown in Figure
\ref{fig:Iterative}.

This iterative process provides a useful recipe for structuring the task
of specifying a model to represent a set of data. The first step, the
model formulation stage, is accomplished by examining the data
graphically and using prior knowledge of relationships, such as from
economic theory or industry practice. The second step in the iteration
is based on the assumptions of the specified model. These assumptions
must be consistent with the data to make valid use of the model. The
third step is diagnostic checking; the data and model must be consistent
with one another before additional inferences can be made. Diagnostic
checking is an important part of the model formulation; it can reveal
mistakes made in previous steps and provide ways to correct these
mistakes.

The iterative process also emphasizes the skills you need to make
analytics work. First, you need a willingness to summarize information
numerically and portray this information graphically. Second, it is
important to develop an understanding of model properties. You should
understand how a probabilistic model behaves in order to match a set of
data to it. Third, theoretical properties of the model are also
important for inferring general relationships based on the behavior of
the data.

\subsection{Model Selection Based on a Training
Dataset}\label{model-selection-based-on-a-training-dataset}

It is common to refer to a dataset used for analysis as an
\emph{in-sample} or \emph{training} dataset. Techniques available for
selecting a model depend upon whether the outcomes \(X\) are discrete,
continuous, or a hybrid of the two, although the principles are the
same.

\textbf{Graphical and other Basic Summary Measures.} Begin by
summarizing the data graphically and with statistics that do not rely on
a specific parametric form, as summarized in Section
\ref{S:MS:NonParInf}. Specifically, you will want to graph both the
empirical distribution and density functions. Particularly for loss data
that contain many zeros and that can be skewed, deciding on the
appropriate scale (e.g., logarithmic) may present some difficulties. For
discrete data, tables are often preferred. Determine sample moments,
such as the mean and variance, as well as selected quantiles, including
the minimum, maximum, and the median. For discrete data, the mode (or
most frequently occurring value) is usually helpful.

These summaries, as well as your familiarity of industry practice, will
suggest one or more candidate parametric models. Generally, start with
the simpler parametric models (for example, one parameter exponential
before a two parameter gamma), gradually introducing more complexity
into the modeling process.

Critique the candidate parametric model numerically and graphically. For
the graphs, utilize the tools introduced in Section
\ref{S:MS:ToolsModelSelection} such as \(pp\) and \(qq\) plots. For the
numerical assessments, examine the statistical significance of
parameters and try to eliminate parameters that do not provide
additional information.

\textbf{Likelihood Ratio Tests.} For comparing model fits, if one model
is a subset of another, then a likelihood ratio test may be employed;
see for example Sections \ref{S:AppA:HT:LRT} and
\ref{S:AppC:MLEModelVal}.

\textbf{Goodness of Fit Statistics.} Generally, models are not proper
subsets of one another so overall goodness of fit statistics are helpful
for comparing models. \emph{Information criteria} are one type of
goodness of statistic. The most widely used examples are Akaike's
Information Criterion (\emph{AIC}) and the Schwarz Bayesian Criterion
(\emph{BIC}); they are are widely cited because they can be readily
generalized to multivariate settings. Section \ref{S:AppA:HT:IC}
provides a summary of these statistics.

For selecting the appropriate distribution, statistics that compare a
parametric fit to a nonparametric alternative, summarized in Section
\ref{S:MS:Tools:Stats}, are useful for model comparison. For discrete
data, a \emph{chi-square goodness of fit statistic} (see Section 2.7) is
generally preferred as it is more intuitive and simpler to explain.

\subsection{Model Selection Based on a Test
Dataset}\label{model-selection-based-on-a-test-dataset}

Model validation is the process of confirming that the proposed model is
appropriate, especially in light of the purposes of the investigation.
An important criticism of the model selection process is that it can be
susceptible to \emph{data-snooping}, that is, fitting a great number of
models to a single set of data. By looking at a large number of models,
we may overfit the data and understate the natural variation in our
representation.

\textbf{Model Validation Process.} We can respond to this criticism by
using a technique sometimes known as \textbf{out-of-sample validation}.
The ideal situation is to have available two sets of data, one for
training, or model development, and one for testing, or model
validation. We initially develop one or several models on the first data
set that we call our \emph{candidate} models. Then, the relative
performance of the candidate models can be measured on the second set of
data. In this way, the data used to validate the model is unaffected by
the procedures used to formulate the model.

The model validation process not only addresses the problem of
overfitting the data but also supports the goal of \textbf{predictive
inference}. Particularly in actuarial applications, our goal is to make
statements about about new experience rather than a dataset at hand. For
example, we use claims experience from one year to develop a model that
can be used to price insurance contracts for the following year. As an
analogy, we can think about the training data set as experience from one
year that is used to predict the behavior of the next year's test data
set.

\begin{figure}

{\centering \includegraphics[width=0.6\linewidth]{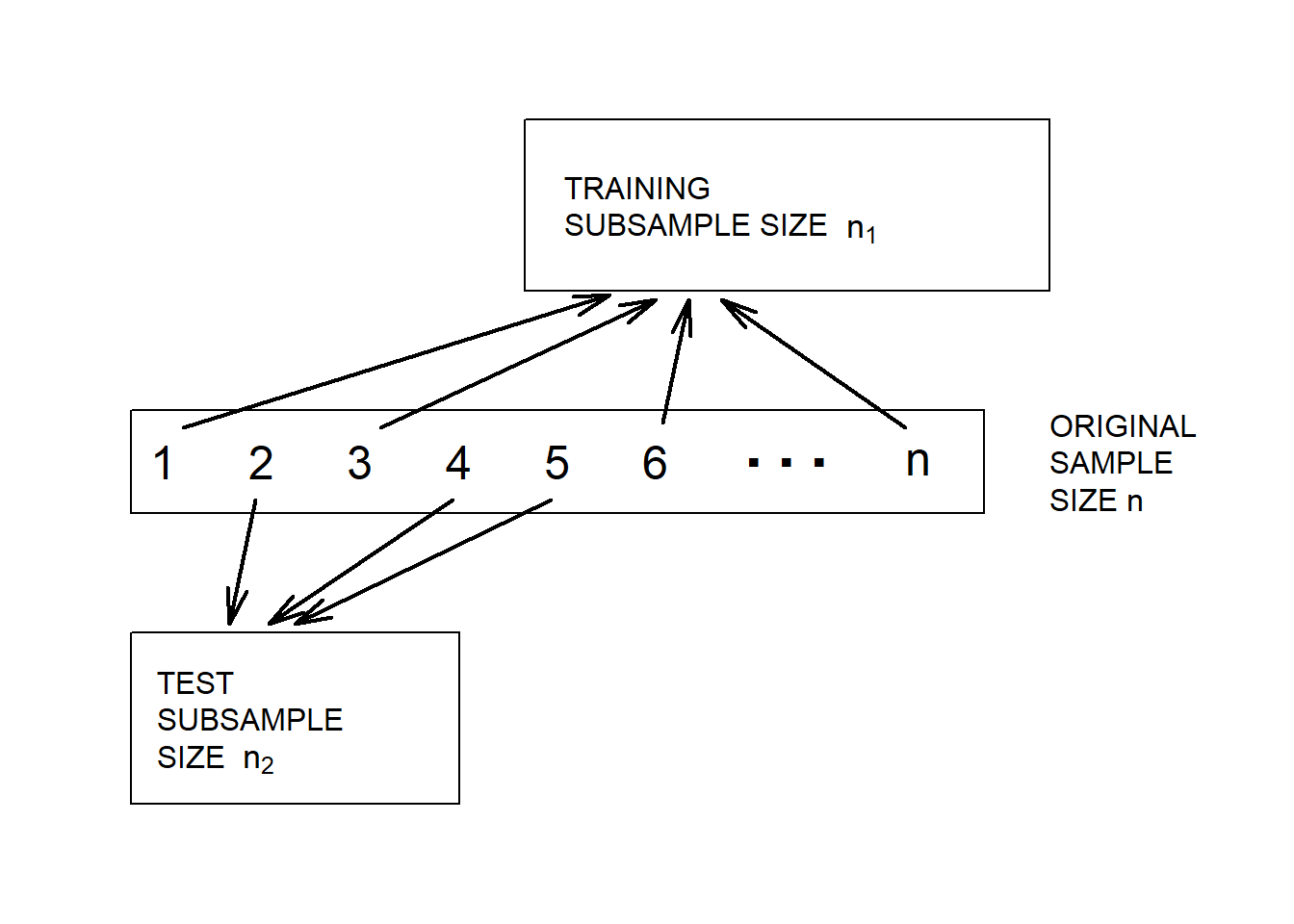}

}

\caption{Model Validation. A data set of size n is randomly split into two subsamples.}\label{fig:ModelValidation}
\end{figure}

\textbf{Random Split of the Data.} Unfortunately, rarely will two sets
of data be available to the investigator. However, we can implement the
validation process by splitting the data set into \textbf{training} and
\textbf{test} subsamples, respectively. Figure \ref{fig:ModelValidation}
illustrates this splitting of the data.

Various researchers recommend different proportions for the allocation.
\citet{snee1977validation} suggests that data-splitting not be done
unless the sample size is moderately large. The guidelines of
\citet{picard1990data} show that the greater the number of parameters to
be estimated, the greater the proportion of observations needed for the
model development subsample. As a rule of thumb, for data sets with 100
or fewer observations, use about 25-35\% of the sample for out-of-sample
validation. For data sets with 500 or more observations, use 50\% of the
sample for out-of-sample validation.

\textbf{Model Validation Statistics.} Much of the literature supporting
the establishment of a model validation process is based on regression
and classification models that you can think of as an
\emph{input-output} problem (\citet{james2013introduction}). That is, we
have several inputs \(x_1, \ldots, x_k\) that are related to an output
\(y\) through a function such as
\[y = \mathrm{g}\left(x_1, \ldots, x_k\right).\] One uses the training
sample to develop an estimate of \(\mathrm{g}\), say,
\(\hat{\mathrm{g}}\), and then calibrate the distance from the observed
outcomes to the predictions using a criterion of the form

\begin{equation}
\sum_i \mathrm{d}(y_i,\hat{\mathrm{g}}\left(x_{i1}, \ldots, x_{ik}\right) ) .
\label{eq:OutSampleCriter}
\end{equation}

Here, the sum \emph{i} is over the test data. In many regression
applications, it is common to use squared Euclidean distance of the form
\(\mathrm{d}(y_i,\mathrm{g}) = (y_i-\mathrm{g})^2\). In actuarial
applications, Euclidean distance
\(\mathrm{d}(y_i,\mathrm{g}) = |y_i-\mathrm{g}|\) is often preferred
because of the skewed nature of the data (large outlying values of \(y\)
can have a large effect on the measure). The Chapter 4 \emph{Technical
Supplement A} describes another measure, the \emph{Gini index} that is
useful in actuarial applications.

\textbf{Selecting a Distribution.} Still, our focus so far has been to
select a distribution for a data set that can be used for actuarial
modeling without additional inputs \(x_1, \ldots, x_k\). Even in this
more fundamental problem, the model validation approach is valuable. If
we base all inference on only in-sample data, then there is a tendency
to select more complicated models then needed. For example, we might
select a four parameter GB2, generalized beta of the second kind,
distribution when only a two parameter Pareto is needed. Information
criteria such as \emph{AIC} and \emph{BIC} included penalties for model
complexity and so provide some protection but using a test sample is the
best guarantee to achieve parsimonious models. From a quote often
attributed to Einstein, we want to ``use the simplest model as possible
but no simpler.''

\subsection{Model Selection Based on
Cross-Validation}\label{model-selection-based-on-cross-validation}

Although out-of-sample validation is the gold standard in predictive
modeling, it is not always practical to do so. The main reason is that
we have limited sample sizes and the out-of-sample model selection
criterion in equation \eqref{eq:OutSampleCriter} depends on a
\emph{random} split of the data. This means that different analysts,
even when working the same data set and same approach to modeling, may
select different models. This is likely in actuarial applications
because we work with skewed data sets where there is a large chance of
getting some very large outcomes and large outcomes may have a great
influence on the parameter estimates.

\textbf{Cross-Validation Procedure.} Alternatively, one may use
\textbf{cross-validation}, as follows.

\begin{itemize}
\item
  The procedure begins by using a random mechanism to split the data
  into \emph{K} subsets known as \emph{folds}, where analysts typcially
  use 5 to 10.
\item
  Next, one uses the first \emph{K}-1 subsamples to estimate model
  parameters. Then, ``predict'' the outcomes for the \emph{K}th
  subsample and use a measure such as in equation
  \eqref{eq:OutSampleCriter} to summarize the fit.
\item
  Now, repeat this by holding out each of the \emph{K} sub-samples,
  summarizing with a cumulative out-of-sample statistic.
\end{itemize}

Repeat these steps for several candidate models and choose the model
with the lowest cumulative out-of-sample statistic.

Cross-validation is widely used because it retains the predictive flavor
of the out-of-sample model validation process but, due to the re-use of
the data, is more stable over random samples.

\section{Estimation using Modified Data}\label{S:MS:ModifiedData}

\begin{center}\rule{0.5\linewidth}{\linethickness}\end{center}

In this section, you learn how to:

\begin{itemize}
\tightlist
\item
  Describe grouped, censored, and truncated data
\item
  Estimate parametric distributions based on grouped, censored, and
  truncated data
\item
  Estimate distributions nonparametrically based on grouped, censored,
  and truncated data
\end{itemize}

\begin{center}\rule{0.5\linewidth}{\linethickness}\end{center}

\subsection{Parametric Estimation using Modified
Data}\label{parametric-estimation-using-modified-data}

Basic theory and many applications are based on \emph{individual}
observations that are ``\emph{complete}'' and ``\emph{unmodified},'' as
we have seen in the previous section. Chapter 3 introduced the concept
of observations that are ``\emph{modified}'' due to two common types of
limitations: \textbf{censoring} and \textbf{truncation}. For example, it
is common to think about an insurance deductible as producing data that
are truncated (from the left) or policy limits as yielding data that are
censoreed (from the right). This viewpoint is from the primary insurer
(the seller of the insurance). However, as we will see in Chapter 10, a
reinsurer (an insurer of an insurance company) may not observe claims
smaller than an amount, only that a claim exists, an example of
censoring from the left. So, in this section, we cover the full gamut of
alternatives. Specifically, this section will address parametric
estmation methods for three alternatives to individual, complete, and
unmodified data: \textbf{interval-censored} data available only in
groups, data that are limited or \textbf{censored}, and data that may
not be observed due to \textbf{truncation}.

\subsubsection{Parametric Estimation using Grouped
Data}\label{S:MS:GroupedData}

Consider a sample of size \(n\) observed from the distribution
\(F(\cdot)\), but in groups so that we only know the group into which
each observation fell, not the exact value. This is referred to as
\textbf{grouped} or \textbf{interval-censored} data. For example, we may
be looking at two successive years of annual employee records. People
employed in the first year but not the second have left sometime during
the year. With an exact departure date (individual data), we could
compute the amount of time that they were with the firm. Without the
departure date (grouped data), we only know that they departed sometime
during a year-long interval.

Formalizing this idea, suppose there are \(k\) groups or intervals
delimited by boundaries \(c_0 < c_1< \cdots < c_k\). For each
observation, we only observe the interval into which it fell (e.g.
\((c_{j-1}, c_j)\)), not the exact value. Thus, we only know the number
of observations in each interval. The constants
\(\{c_0 < c_1 < \cdots < c_k\}\) form some partition of the domain of
\(F(\cdot)\). Then the probability of an observation \(X_i\) falling in
the \(j\)th interval is \[\Pr\left(X
_i \in (c_{j-1}, c_j]\right) = F(c_j) - F(c_{j-1}).\]

The corresponding probability mass function for an observation is
\[\begin{aligned}
f(x) &=
\begin{cases}
F(c_1) - F(c_{0}) &   \text{if }\ x \in (c_{0}, c_1]\\
\vdots & \vdots \\
F(c_k) - F(c_{k-1}) &   \text{if }\ x \in (c_{k-1}, c_k]\\
\end{cases} \\
&= \prod_{j=1}^k \left\{F(c_j) - F(c_{j-1})\right\}^{I(x \in (c_{j-1}, c_j])}
\end{aligned}\]

Now, define \(n_j\) to be the number of observations that fall in the
\(j\)th interval, \((c_{j-1}, c_j]\). Thus, the likelihood function
(with respect to the parameter(s) \(\theta\)) is \[\begin{aligned}
\mathcal{L}(\theta) = \prod_{j=1}^n f(x_i) = \prod_{j=1}^k \left\{F(c_j) - F(c_{j-1})\right\}^{n_j}
\end{aligned}\]

And the log-likelihood function is \[\begin{aligned}
L(\theta) = \ln \mathcal{L}(\theta) = \ln \prod_{j=1}^n f(x_i) = \sum_{j=1}^k n_j \ln \left\{F(c_j) - F(c_{j-1})\right\}
\end{aligned}\]

Maximizing the likelihood function (or equivalently, maximizing the
log-likelihood function) would then produce the maximum likelihood
estimates for grouped data.

\textbf{Example 4.3.1. SOA Exam Question.} You are given:

\begin{enumerate}
\def\labelenumi{(\roman{enumi})}
\tightlist
\item
  Losses follow an exponential distribution with mean \(\theta\).
\item
  A random sample of 20 losses is distributed as follows:
\end{enumerate}

\[\begin{array}{l|c}
\hline
\text{Loss Range} & \text{Frequency} \\
\hline
[0,1000] & 7 \\
(1000,2000] & 6 \\
(2000,\infty) & 7 \\
\hline
\end{array}\]

Calculate the maximum likelihood estimate of \(\theta\).

Show Example Solution

\hypertarget{toggleExampleSelect.3.1}{}
\textbf{Solution.} \[\begin{aligned}
\mathcal{L}(\theta) &= F(1000)^7[F(2000)-F(1000)]^6[1-F(2000)]^7 \\
&= (1-e^{-1000/\theta})^7(e^{-1000/\theta} - e^{-2000/\theta})^6(e^{-2000/\theta})^7 \\
&= (1-p)^7(p-p^2)^6(p^2)^7 \\
&= p^{20}(1-p)^{13}
\end{aligned}\]

where \(p=e^{-1000/\theta}\). Maximizing this expression with respect to
\(p\) is equivalent to maximizing the likelihood with respect to
\(\theta\). The maximum occurs at \(p=\frac{20}{33}\) and so
\(\hat{\theta}=\frac{-1000}{\ln(20/33)}= 1996.90\).

\begin{center}\rule{0.5\linewidth}{\linethickness}\end{center}

\subsubsection{Censored Data}\label{censored-data}

\textbf{Censoring} occurs when we observe only a limited value of an
observation. The most common form is \textbf{right-censoring}, in which
we record the smaller of the ``true'' dependent variable and a censoring
variable. Using notation, let \(X\) represent an outcome of interest,
such as the loss due to an insured event. Let \(C_U\) denote the
censoring time, such as \(C_U=5\). With right-censored observations, we
observe \(X\) if it is below censoring point \(C_U\); otherwise if \(X\)
is higher than the censoring point, we only observe the censored
\(C_U\). Therefore, we record \(X_U^{\ast}= \min(X, C_U)\). We also
observe whether or not censoring has occurred. Let
\(\delta_U= \mathrm{I}(X \geq C_U)\) be a binary variable that is 1 if
censoring occurs, \(y \geq C_U\), and 0 otherwise.

For example, \(C_U\) may represent the upper limit of coverage of an
insurance policy. The loss may exceed the amount \(C_U\), but the
insurer only has \(C_U\) in its records as the amount paid out and does
not have the amount of the actual loss \(X\) in its records.

Similarly, with \textbf{left-censoring}, we only observe \(X\) if \(X\)
is above censoring point (e.g.~time or loss amount) \(C_L\); otherwise
we observe \(C_L\). Thus, we record \(X_L^{\ast}= \max(X, C_L)\) along
with the censoring indicator \(\delta_L= \mathrm{I}(X \leq C_L)\).

For example, suppose a reinsurer will cover insurer losses greater than
\(C_L\). Let \(Y = X_L^{\ast} - C_L\) represent the amount that the
\emph{reinsurer} is responsible for. If the policyholder loss
\(X < C_L\), then the insurer will pay the entire claim and \(Y =0\), no
loss for the reinsurer. If the loss \(X \ge C_L\), then \(Y = X-C_L\)
represents the reinsurer's retained claims. If a loss occurs, the
reinsurer knows the actual amount if it exceeds the limit \(C_L\),
otherwise it only knows that it had a loss of \(0\).

As another example of a left-censored observation, suppose we are
conducting a study and interviewing a person about an event in the past.
The subject may recall that the event occurred before \(C_L\), but not
the exact date.

\subsubsection{Truncated Data}\label{truncated-data}

We just saw that censored observations are still available for study,
although in a limited form. In contrast, \textbf{truncated} outcomes are
a type of missing data. An outcome is potentially truncated when the
availability of an observation depends on the outcome.

In insurance, it is common for observations to be
\textbf{left-truncated} at \(C_L\) when tfhe amount is \[\begin{aligned}
Y &=
\left\{
\begin{array}{ll}
\text{we do not observe }X & X < C_L \\
X- C_L & X \geq C_L.
\end{array}
\right.\end{aligned}\]

In other words, if \(X\) is less than the threshold \(C_L\), then it is
not observed. FOr example, \(C_L\) may represent the deductible
associated with an insurance coverage. If the insured loss is less than
the deductible, then the insurer does not observe or record the loss at
all. If the loss exceeds the deductible, then the excess \(X-C_L\) is
the claim that the insurer covers.

Similarly for \textbf{right-truncated} data, if \(X\) exceeds a
threshold \(C_U\), then it is not observed. In this case, the amount is
\[\begin{aligned}
Y &=
\left\{
\begin{array}{ll}
X & X < C_U \\
\text{we do not observe }X & X \geq C_U.
\end{array}
\right.\end{aligned}\]

Classic examples of truncation from the right include \(X\) as a measure
of distance to a star. When the distance exceeds a certain level
\(C_U\), the star is no longer observable.

\begin{figure}

{\centering \includegraphics[width=0.6\linewidth]{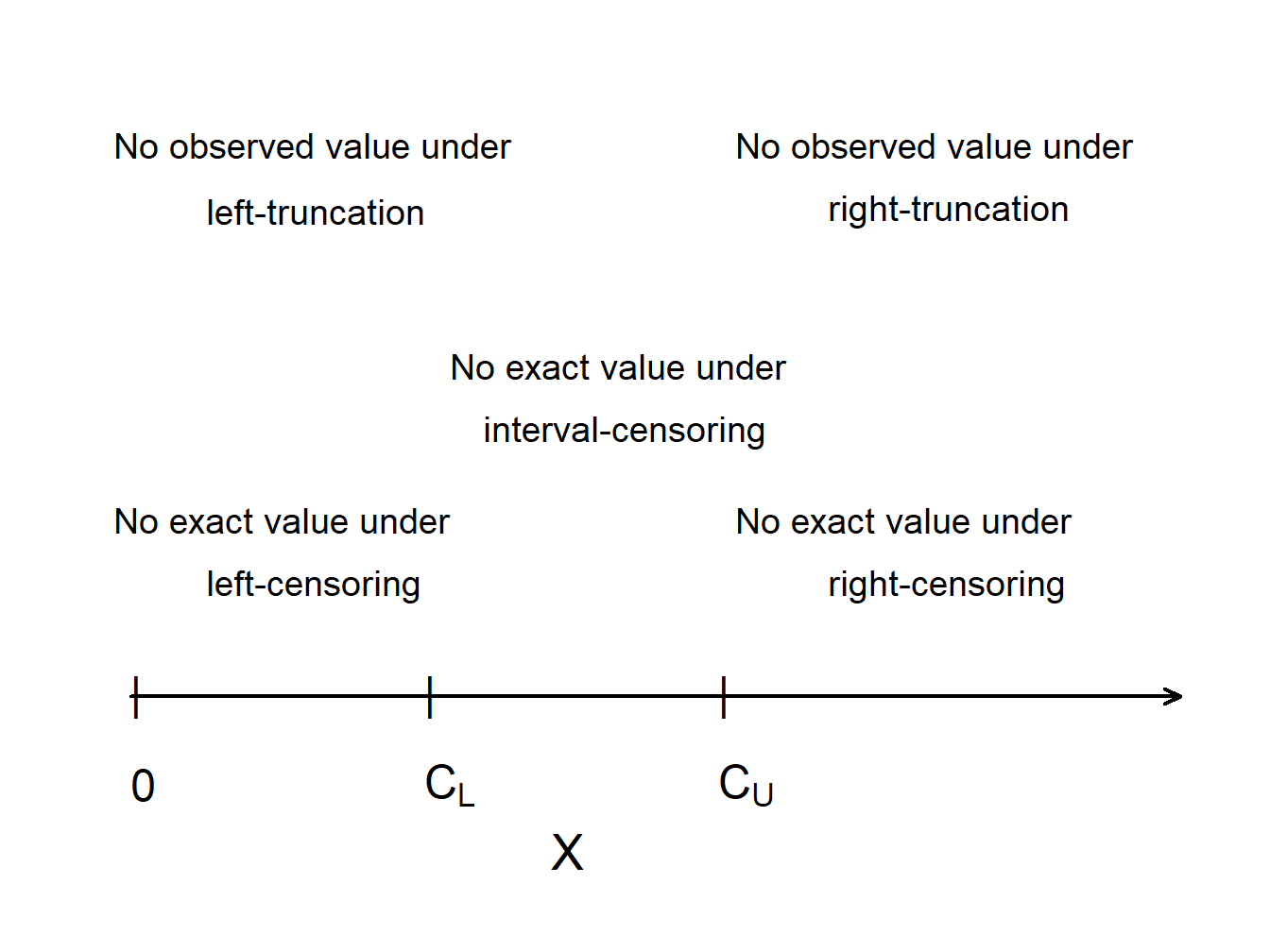}

}

\caption{Censoring and Truncation}\label{fig:CensorTrunc}
\end{figure}

Figure \ref{fig:CensorTrunc} compares truncated and censored
observations. Values of \(X\) that are greater than the ``upper''
censoring limit \(C_U\) are not observed at all (right-censored), while
values of \(X\) that are smaller than the ``lower'' truncation limit
\(C_L\) are observed, but observed as \(C_L\) rather than the actual
value of \(X\) (left-truncated).

\begin{center}\rule{0.5\linewidth}{\linethickness}\end{center}

Show Example

\begin{figure}

{\centering \includegraphics[width=0.6\linewidth]{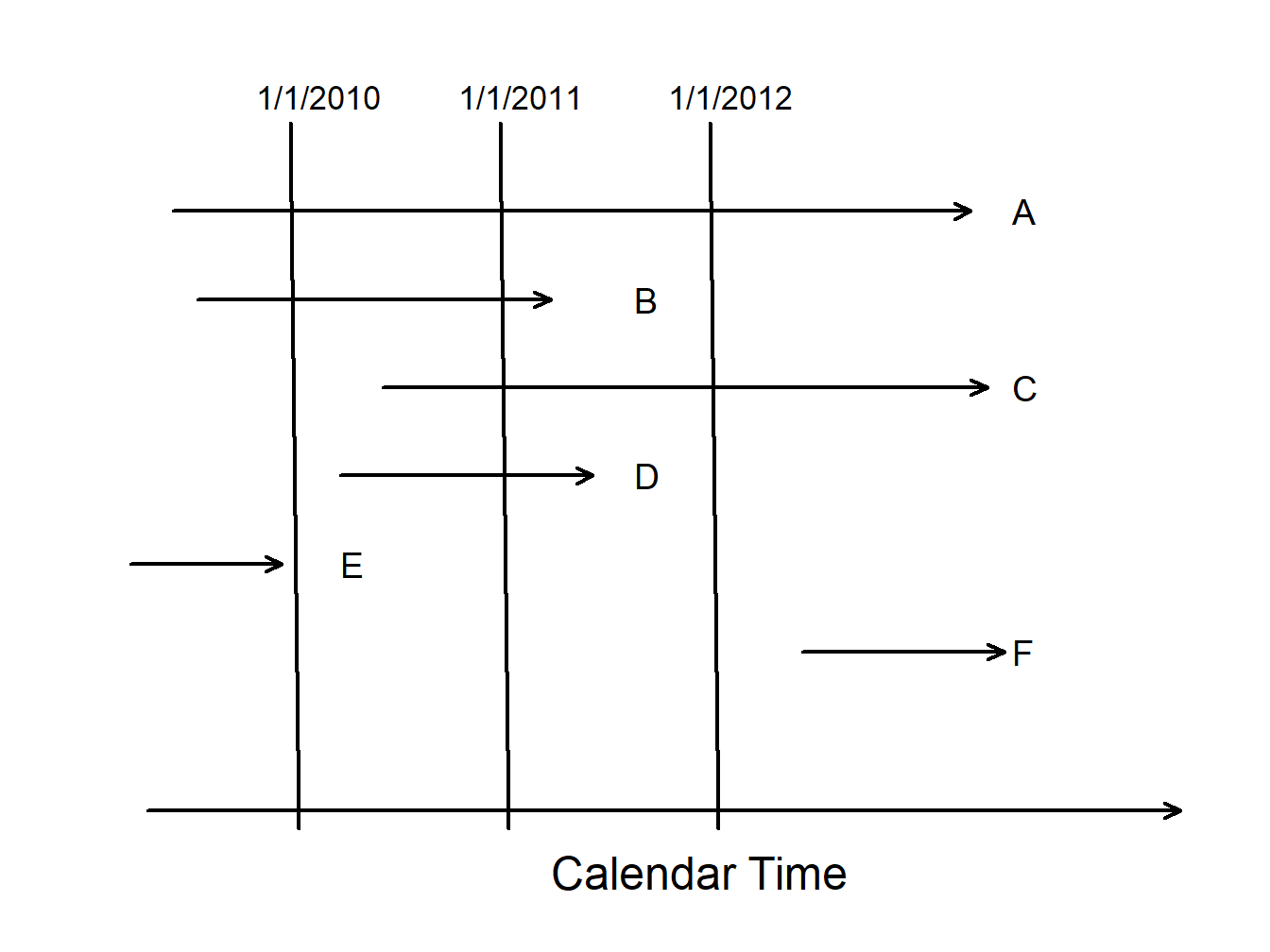}

}

\caption{Timeline for Several Subjects on Test in a Mortality Study}\label{fig:Mortality}
\end{figure}

\hypertarget{toggleExampleMort}{}
\textbf{Example -- Mortality Study.} Suppose that you are conducting a
two-year study of mortality of high-risk subjects, beginning January 1,
2010 and finishing January 1, 2012. Figure \ref{fig:Mortality}
graphically portrays the six types of subjects recruited. For each
subject, the beginning of the arrow represents that the the subject was
recruited and the arrow end represents the event time. Thus, the arrow
represents exposure time.

\begin{itemize}
\tightlist
\item
  \textbf{Type A - Right-censored.} This subject is alive at the
  beginning and the end of the study. Because the time of death is not
  known by the end of the study, it is right-censored. Most subjects are
  Type A.
\item
  \textbf{Type B - Complete} information is available for a type B
  subject. The subject is alive at the beginning of the study and the
  death occurs within the observation period.
\item
  \textbf{Type C - Right-censored and left-truncated.} A type C subject
  is right-censored, in that death occurs after the observation period.
  However, the subject entered after the start of the study and is said
  to have a \emph{delayed entry time}. Because the subject would not
  have been observed had death occurred before entry, it is
  left-truncated.
\item
  \textbf{Type D - Left-truncated.} A type D subject also has delayed
  entry. Because death occurs within the observation period, this
  subject is not right censored.
\item
  \textbf{Type E - Left-truncated.} A type E subject is not included in
  the study because death occurs prior to the observation period.
\item
  \textbf{Type F - Right-truncated.} Similarly, a type F subject is not
  included because the entry time occurs after the observation period.
\end{itemize}

\begin{center}\rule{0.5\linewidth}{\linethickness}\end{center}

To summarize, for outcome \(X\) and constants \(C_L\) and \(C_U\),

\begin{longtable}[]{@{}ccc@{}}
\toprule
Limitation Type & Limited Variable & Censoring
Information\tabularnewline
\midrule
\endhead
right censoring & \(X_U^{\ast}= \min(X, C_U)\) &
\(\delta_U= \mathrm{I}(X \geq C_U)\)\tabularnewline
left censoring & \(X_L^{\ast}= \max(y, C_L)\) &
\(\delta_L= \mathrm{I}(X \leq C_L)\)\tabularnewline
interval censoring & &\tabularnewline
right truncation & \(X\) & observe \(X\) if \(X < C_U\)\tabularnewline
left truncation & \(X\) & observe \(X\) if \(X < C_L\)\tabularnewline
\bottomrule
\end{longtable}

\subsubsection{Parametric Estimation using Censored and Truncated
Data}\label{parametric-estimation-using-censored-and-truncated-data}

For simplicity, we assume fixed censoring times and a continuous outcome
\(X\). To begin, consider the case of right-censored data where we
record \(X_U^{\ast}= \min(X, C_U)\) and censoring indicator
\(\delta_U= \mathrm{I}(X \geq C_U)\). If censoring occurs so that
\(\delta_U=1\), then \(X \geq C_U\) and the likelihood is
\(\Pr(X \geq C_U) = 1-F(C_U)\). If censoring does not occur so that
\(\delta_U=0\), then \(X < C_U\) and the likelihood is \(f(x)\).
Summarizing, we have the likelihood of a single observation as

\[\begin{aligned}
\left\{
\begin{array}{ll}
f(x) & \text{if } \delta = 0 \\
1-F(C_U) & \text{if }\delta=1
\end{array}
\right. = \left( f(x)\right)^{1-\delta} \left(1-F(C_U)\right)^{\delta} .
\end{aligned}\]

The right-hand expression allows us to present the likelihood more
compactly. Now, for an \emph{iid} sample of size \(n\),
\(\{ (x_{U1},\delta_1), \ldots,(x_{Un}, \delta_n) \}\), the likelihood
is

\[\mathcal{L}(\theta) = \prod_{i=1}^n \left( f(x_i)\right)^{1-\delta_i} \left(1-F(C_{Ui})\right)^{\delta_i} = \prod_{\delta_i=0} f(x_i) \prod_{\delta_i=1} \{1-F(C_{Ui})\},\]

with potential censoring times \(\{ C_{U1}, \ldots,C_{Un} \}\). Here,
the notation ``\(\prod_{\delta_i=0}\)'' means to take the product over
uncensored observations, and similarly for ``\(\prod_{\delta_i=1}\).''

On the other hand, truncated data are handled in likelihood inference
via conditional probabilities. Specifically, we adjust the likelihood
contribution by dividing by the probability that the variable was
observed. To summarize, we have the following contributions to the
likelihood function for six types of outcomes:

\[\begin{array}{lc}
\hline
\text{Outcome} & \text{Likelihood Contribution} \\
\hline
\text{exact value} & f(x) \\
\text{right-censoring} & 1-F(C_U) \\
\text{left-censoring} & F(C_L) \\
\text{right-truncation} & f(x)/F(C_U) \\
\text{left-truncation} & f(x)/(1-F(C_L)) \\
\text{interval-censoring} & F(C_U)-F(C_L) \\
\hline
\end{array}\]

For known outcomes and censored data, the likelihood is
\[\mathcal{L}(\theta) = \prod_{E} f(x_i) \prod_{R} \{1-F(C_{Ui})\} \prod_{L}
F(C_{Li}) \prod_{I} (F(C_{Ui})-F(C_{Li})),\] where ``\(\prod_{E}\)'' is
the product over observations with \emph{E}xact values, and similarly
for \emph{R}ight-, \emph{L}eft- and \emph{I}nterval-censoring.

For right-censored and left-truncated data, the likelihood is
\[\mathcal{L}(\theta) = \prod_{E} \frac{f(x_i)}{1-F(C_{Li})} \prod_{R} \frac{1-F(C_{Ui})}{1-F(C_{Li})},\]
and similarly for other combinations. To get further insights, consider
the following.

\begin{center}\rule{0.5\linewidth}{\linethickness}\end{center}

Show Example

\hypertarget{toggleExampleEXP}{}
\textbf{Special Case: Exponential Distribution.} Consider data that are
right-censored and left-truncated, with random variables \(X_i\) that
are exponentially distributed with mean \(\theta\). With these
specifications, recall that \(f(x) = \theta^{-1} \exp(-x/\theta)\) and
\(F(x) = 1-\exp(-x/\theta)\).

For this special case, the log-likelihood is

\[\begin{aligned}
L(\theta) &= \sum_{E} \left\{ \ln f(x_i) - \ln (1-F(C_{Li})) \right\} + \sum_{R}\left\{ \ln (1-F(C_{Ui}))- \ln (1-\mathrm{F}(C_{Li})) \right\}\\
&= \sum_{E} (-\ln \theta -(x_i-C_{Li})/\theta ) -\sum_{R} (C_{Ui}-C_{Li})/\theta .
\end{aligned}\]

To simplify the notation, define
\(\delta_i = \mathrm{I}(X_i \geq C_{Ui})\) to be a binary variable that
indicates right-censoring. Let
\(X_i^{\ast \ast} = \min(X_i, C_{Ui}) - C_{Li}\) be the amount that the
observed variable exceeds the lower truncation limit. With this, the
log-likelihood is

\begin{equation}
  L(\theta) =  - \sum_{i=1}^n ((1-\delta_i) \ln \theta + \frac{x_i^{\ast \ast}}{\theta})
  \label{eq:EXPloglik}
\end{equation}

Taking derivatives with respect to the parameter \(\theta\) and setting
it equal to zero yields the maximum likelihood estimator

\[\widehat{\theta}  = \frac{1}{n_u} \sum_{i=1}^n  x_i^{\ast \ast},\]

where \(n_u = \sum_i (1-\delta_i)\) is the number of uncensored
observations.

\begin{center}\rule{0.5\linewidth}{\linethickness}\end{center}

\textbf{Example 4.3.2. SOA Exam Question.} You are given:

\begin{enumerate}
\def\labelenumi{(\roman{enumi})}
\tightlist
\item
  A sample of losses is: 600 700 900
\item
  No information is available about losses of 500 or less.
\item
  Losses are assumed to follow an exponential distribution with mean
  \(\theta\).
\end{enumerate}

Calculate the maximum likelihood estimate of \(\theta\).

Show Example Solution

\hypertarget{toggleExampleSelect.3.2}{}
\textbf{Solution.} These observations are truncated at 500. The
contribution of each observation to the likelihood function is
\[\frac{f(x)}{1-F(500)} = \frac{\theta^{-1}e^{-x/\theta}}{e^{-500/\theta}}\]

Then the likelihood function is

\[\mathcal{L}(\theta)= \frac{\theta^{-1} e^{-600/\theta} \theta^{-1} e^{-700/\theta} \theta^{-1} e^{-900/\theta}}{(e^{-500/\theta})^3} = \theta^{-3}e^{-700/\theta}\]

The log-likelihood is

\[L(\theta) = \ln\mathcal{L}(\theta) = -3\ln \theta - 700\theta^{-1}\]

Maximizing this expression by setting the derivative with respect to
\(\theta\) equal to 0, we have

\[L'(\theta) = -3\theta^{-1} + 700\theta^{-2} = 0 \ \Rightarrow \ \hat{\theta} = \frac{700}{3} = 233.33\]

\begin{center}\rule{0.5\linewidth}{\linethickness}\end{center}

\textbf{Example 4.3.3. SOA Exam Question.} You are given the following
information about a random sample:

\begin{enumerate}
\def\labelenumi{(\roman{enumi})}
\tightlist
\item
  The sample size equals five.
\item
  The sample is from a Weibull distribution with \(\tau=2\).
\item
  Two of the sample observations are known to exceed 50, and the
  remaining three observations are 20, 30, and 45.
\end{enumerate}

Calculate the maximum likelihood estimate of \(\theta\).

Show Example Solution

\hypertarget{toggleExampleSelect.3.3}{}
\textbf{Solution.} The likelihood function is

\[\begin{aligned}
\mathcal{L}(\theta) &= f(20) f(30) f(45) [1-F(50)]^2 \\
&= \frac{2(20/\theta)^2 e^{-(20/\theta)^2}}{20} \frac{2(30/\theta)^2 e^{-(30/\theta)^2}}{30} \frac{2(45/\theta)^2 e^{-(45/\theta)^2}}{45}(e^{-(50/\theta)^2})^2 \\
&\propto \frac{1}{\theta^6} e^{-8325/\theta^2}
\end{aligned}\]

The natural logarithm of the above expression is
\(-6\ln\theta - \frac{8325}{\theta^2}\). Maximizing this expression by
setting its derivative to 0, we get

\[\frac{-6}{\theta} + \frac{16650}{\theta^3} = 0 \ \Rightarrow \ \hat{\theta} = \left(\frac{16650}{6}\right)^{\frac{1}{2}} = 52.6783\]

\begin{center}\rule{0.5\linewidth}{\linethickness}\end{center}

\subsection{Nonparametric Estimation using Modified
Data}\label{nonparametric-estimation-using-modified-data}

Nonparametric estimators provide useful benchmarks, so it is helpful to
understand the estimation procedures for grouped, censored, and
truncated data.

\subsubsection{Grouped Data}\label{grouped-data}

As we have seen in Section \ref{S:MS:GroupedData}, observations may be
grouped (also referred to as interval censored) in the sense that we
only observe them as belonging in one of \(k\) intervals of the form
\((c_{j-1}, c_j]\), for \(j=1, \ldots, k\). At the boundaries, the
empirical distribution function is defined in the usual way:
\[F_n(c_j) = \frac{\text{number of observations } \le c_j}{n}\]

For other values of \(x \in (c_{j-1}, c_j)\), we can estimate the
distribution function with the \textbf{ogive} estimator, which linearly
interpolates between \(F_n(c_{j-1})\) and \(F_n(c_j)\), i.e.~the values
of the boundaries \(F_n(c_{j-1}\) and \(F_n(c_j)\) are connected with a
straight line. This can formally be expressed as
\[F_n(x) = \frac{c_j-x}{c_j-c_{j-1}} F_n(c_{j-1}) + \frac{x-c_{j-1}}{c_j-c_{j-1}} F_n(c_j) \ \ \ \text{for } c_{j-1} \le x < c_j\]

The corresponding density is
\[f_n(x) = F^{\prime}_n(x) = \frac{F_n(c_j)-F_n(c_{j-1})}{c_j - c_{j-1}} \ \ \  \text{for } c_{j-1} \le x < c_j .\]

\begin{center}\rule{0.5\linewidth}{\linethickness}\end{center}

\textbf{Example 4.3.4. SOA Exam Question.} You are given the following
information regarding claim sizes for 100 claims:

\[
\begin{array}{r|c}
\hline
\text{Claim Size} &  \text{Number of Claims} \\
\hline
0 - 1,000 & 16 \\
1,000 - 3,000 & 22 \\
3,000 - 5,000 & 25 \\
5,000 - 10,000 & 18 \\
10,000 - 25,000 & 10 \\
25,000 - 50,000 & 5 \\
50,000 - 100,000 & 3 \\
\text{over  } 100,000 & 1 \\
\hline
\end{array}
\]

Using the ogive, calculate the estimate of the probability that a
randomly chosen claim is between 2000 and 6000.

Show Example Solution

\hypertarget{toggleExampleSelect.3.4}{}
\textbf{Solution.} At the boundaries, the empirical distribution
function is defined in the usual way, so we have
\[F_{100}(1000) = 0.16, \ F_{100}(3000)=0.38, \ F_{100}(5000)=0.63, \ F_{100}(10000)=0.81\]
For other claim sizes, the ogive estimator linearly interpolates between
these values:
\[F_{100}(2000) = 0.5F_{100}(1000) + 0.5F_{100}(3000) = 0.5(0.16)+0.5(0.38)=0.27\]
\[F_{100}(6000)=0.8F_{100}(5000)+0.2F_{100}(10000) = 0.8(0.63)+0.2(0.81)=0.666\]
Thus, the probability that a claim is between 2000 and 6000 is
\(F_{100}(6000) - F_{100}(2000) = 0.666-0.27 = 0.396\).

\begin{center}\rule{0.5\linewidth}{\linethickness}\end{center}

\subsubsection{Right-Censored Empirical Distribution
Function}\label{right-censored-empirical-distribution-function}

It can be useful to calibrate parametric likelihood methods with
nonparametric methods that do not rely on a parametric form of the
distribution. The product-limit estimator due to \citep{kaplan1958} is a
well-known estimator of the distribution in the presence of censoring.

To begin, first note that the empirical distribution function \(F_n(x)\)
is an \textbf{unbiased} estimator of the distribution function \(F(x)\)
(in the ``usual'' case in the absence of censoring). This is because
\(F_n(x)\) is the average of indicator variables that are also unbiased,
that is, \(\mathrm{E~} I(X \le x) = \Pr(X \le x) = F(x)\). Now suppose
the the random outcome is censored on the right by a limiting amount,
say, \(C_U\), so that we record the smaller of the two,
\(X^* = \min(X, C_U)\). For values of \(x\) that are smaller than
\(C_U\), the indicator variable still provides an unbiased estimator of
the distribution function before we reach the censoring limit. That is,
\(\mathrm{E~} I(X^* \le x) = F(x)\) because
\(I(X^* \le x) = I(X \le x)\) for \(x < C_U\). In the same way,
\(\mathrm{E~} I(X^* > x) = 1 -F(x) = S(x)\).

Now consider two random variables that have different censoring limits.
For illustration, suppose that we observe \(X_1^* = \min(X_1, 5)\) and
\(X_2^* = \min(X_2, 10)\) where \(X_1\) and \(X_2\) are independent
draws from the same distribution. For \(x \le 5\), the empirical
distribution function \(F_2(x)\) is an unbiased estimator of \(F(x)\).
However, for \(5 < x \le 10\), the first observation cannot be used for
the distribution function because of the censoring limitation. Instead,
the strategy developed by \citep{kaplan1958} is to use \(S_n(5)\) as an
estimator of \(S(5)\) and then to use the second observation to estimate
the conditional survivor function
\(\Pr(X > x | X >5) = \frac{S(x)}{S(5)}\). Specifically, for
\(5 < x \le 10\), the estimator of the survival function is \[
\hat{S}(x) = S_2(5) \times I(X_2^* > x ) .
\]

Extending this idea, for each observation \(i\), let \(u_i\) be the
upper censoring limit (\(=\infty\) if no censoring). Thus, the recorded
value is \(x_i\) in the case of no censoring and \(u_i\) if there is
censoring. Let \(t_{1} <\cdots< t_{k}\) be \(k\) distinct points at
which an uncensored loss occurs, and let \(s_j\) be the number of
uncensored losses \(x_i\)'s at \(t_{j}\). The corresponding \textbf{risk
set} is the number of observations that are active (not censored) at a
value \emph{less than} \(t_{j}\), denoted as
\(R_j = \sum_{i=1}^n I(x_i \geq t_{j}) + \sum_{i=1}^n I(u_i \geq t_{j})\).

\textbf{Kaplan-Meier Product Limit Estimator.} With this notation, the
\textbf{product-limit estimator} of the distribution function is

\begin{equation}
\hat{F}(x) =
\left\{
\begin{array}{ll}
0 & x<t_{1} \\
1-\prod_{j:t_{j} \leq x}\left( 1-\frac{s_j}{R_{j}}\right) & x \geq t_{1} .
\end{array}
\right. \label{eq:KaplanMeier}
\end{equation}

As usual, the corresponding estimate of the survival function is
\(\hat{S}(x) = 1 - \hat{F}(x)\).

\begin{center}\rule{0.5\linewidth}{\linethickness}\end{center}

\textbf{Example 4.3.5. SOA Exam Question.} The following is a sample of
10 payments:

\[\begin{array}{cccccccccc}
4 &4 &5+ &5+ &5+ &8 &10+ &10+ &12 &15 \\
\end{array}\]

where \(+\) indicates that a loss has exceeded the policy limit.

Using the Kaplan-Meier product-limit estimator, calculate the
probability that the loss on a policy exceeds 11, \(\hat{S}(11)\).

Show Example Solution

\hypertarget{toggleExampleSelect.3.5}{}
\textbf{Solution.} There are four event times (non-censored
observations). For each time \(t_j\), we can calcuate the number of
events \(s_j\) and the risk set \(R_j\) as the following:

\[\begin{array}{cccc}
\hline
j & t_j & s_j & R_j \\
\hline
1 & 4 & 2 & 10 \\
2 & 8 & 1 & 5 \\
3 & 12 & 1 & 2 \\
4 & 15 & 1 & 1 \\
\hline
\end{array}\]

Thus, the Kaplan-Meier estimate of \(S(11)\) is \[\begin{aligned}
\hat{S}(11) &= \prod_{j:t_j\leq 11} \left( 1- \frac{s_j}{R_j} \right) =  \prod_{j=1}^{2} \left( 1- \frac{s_j}{R_j} \right)\\
&= \left(1-\frac{2}{10} \right) \left(1-\frac{1}{5} \right) = (0.8)(0.8)= 0.64. \\
\end{aligned}\]

\begin{center}\rule{0.5\linewidth}{\linethickness}\end{center}

\subsubsection{Right-Censored, Left-Truncated Empirical Distribution
Function}\label{right-censored-left-truncated-empirical-distribution-function}

In addition to right-censoring, we now extend the framework to allow for
left-truncated data. As before, for each observation \(i\), let \(u_i\)
be the upper censoring limit (\(=\infty\) if no censoring). Further, let
\(d_i\) be the lower truncation limit (0 if no truncation). Thus, the
recorded value (if it is greater than \(d_i\)) is \(x_i\) in the case of
no censoring and \(u_i\) if there is censoring. Let
\(t_{1} <\cdots< t_{k}\) be \(k\) distinct points at which an event of
interest occurs, and let \(s_j\) be the number of recorded events
\(x_i\)'s at time point \(t_{j}\). The corresponding risk set is
\[R_j = \sum_{i=1}^n I(x_i \geq t_{j}) + \sum_{i=1}^n I(u_i \geq t_{j}) - \sum_{i=1}^n I(d_i \geq t_{j}).\]

With this new definition of the risk set, the product-limit estimator of
the distribution function is as in equation \eqref{eq:KaplanMeier}.

\textbf{Greenwood's Formula}. \citep{greenwood1926} derived the formula
for the estimated variance of the product-limit estimator to be

\[\widehat{Var}(\hat{F}(x)) = (1-\hat{F}(x))^{2} \sum _{j:t_{j} \leq x} \dfrac{s_j}{R_{j}(R_{j}-s_j)}.\]

\texttt{R}`s \texttt{survfit} method takes a survival data object and
creates a new object containing the Kaplan-Meier estimate of the
survival function along with confidence intervals. The Kaplan-Meier
method (\texttt{type=\textquotesingle{}kaplan-meier\textquotesingle{}})
is used by default to construct an estimate of the survival curve. The
resulting discrete survival function has point masses at the observed
event times (discharge dates) \(t_j\), where the probability of an event
given survival to that duration is estimated as the number of observed
events at the duration \(s_j\) divided by the number of subjects exposed
or 'at-risk' just prior to the event duration \(R_j\).

Two alternate types of estimation are also available for the
\texttt{survfit} method. The alternative
(\texttt{type=\textquotesingle{}fh2\textquotesingle{}}) handles ties, in
essence, by assuming that multiple events at the same duration occur in
some arbitrary order. Another alternative
(\texttt{type=\textquotesingle{}fleming-harrington\textquotesingle{}})
uses the Nelson-Äalen (see \citep{aalen1978}) estimate of the
\textbf{cumulative hazard function} to obtain an estimate of the
survival function. The estimated cumulative hazard \(\hat{H}(x)\) starts
at zero and is incremented at each observed event duration \(t_j\) by
the number of events \(s_j\) divided by the number at risk \(R_j\). With
the same notation as above, the \textbf{Nelson-Äalen} estimator of the
distribution function is

\[\begin{aligned}
\hat{F}_{NA}(x) &=
\left\{
\begin{array}{ll}
0 & x<t_{1} \\
1- \exp \left(-\sum_{j:t_{j} \leq x}\frac{s_j}{R_j} \right) & x \geq t_{1} .
\end{array}
\right.\end{aligned}\]

Note that the above expression is a result of the Nelson-Äalen estimator
of the cumulative hazard function
\[\hat{H}(x)=\sum_{j:t_j\leq x}  \frac{s_j}{R_j} \] and the relationship
between the survival function and cumulative hazard function,
\(\hat{S}_{NA}(x)=e^{-\hat{H}(x)}\).

\begin{center}\rule{0.5\linewidth}{\linethickness}\end{center}

\textbf{Example 4.3.6. SOA Exam Question.} For observation \(i\) of a
survival study:

\begin{itemize}
\tightlist
\item
  \(d_i\) is the left truncation point
\item
  \(x_i\) is the observed value if not right censored
\item
  \(u_i\) is the observed value if right censored
\end{itemize}

You are given:

\[\begin{array}{c|cccccccccc}
\hline
\text{Observation } (i) & 1 & 2 & 3 & 4 & 5 & 6 & 7 & 8 & 9 & 10\\ \hline
d_i & 0 & 0 & 0 & 0 & 0 & 0 & 0 & 1.3 & 1.5 & 1.6\\
x_i & 0.9 & - & 1.5 & - & - & 1.7 & - & 2.1 & 2.1 & - \\
u_i & - & 1.2 & - & 1.5 & 1.6 & - & 1.7 & - & - & 2.3 \\
\hline
\end{array}\]

Calculate the Kaplan-Meier product-limit estimate, \(\hat{S}(1.6)\)

Show Example Solution

\hypertarget{toggleExampleSelect.3.6}{}
\textbf{Solution.} Recall the risk set
\(R_j = \sum_{i=1}^n \left\{ I(x_i \geq t_{j}) + I(u_i \geq t_{j}) - I(d_i \geq t_{j}) \right\}\).
Then

\[\begin{array}{ccccc}
\hline
j & t_j & s_j & R_j & \hat{S}(t_j) \\
\hline
1  & 0.9   & 1   & 10-3 = 7 & 1-\frac{1}{7} = \frac{6}{7} \\
2  & 1.5   & 1   & 8-2 = 6  & \frac{6}{7}\left( 1 - \frac{1}{6} \right) = \frac{5}{7}\\
3  & 1.7   & 1   & 5-0 = 5  & \frac{5}{7}\left( 1 - \frac{1}{5} \right) = \frac{4}{7}\\
4  & 2.1   & 2   & 3        & \frac{4}{7}\left( 1 - \frac{2}{3}\right) = \frac{4}{21}\\
\hline
\end{array}\]

The Kaplan-Meier estimate is therefore \(\hat{S}(1.6) = \frac{5}{7}\).

\begin{center}\rule{0.5\linewidth}{\linethickness}\end{center}

\textbf{Exercise 4.3.7. SOA Exam Question. - Continued.}

\begin{enumerate}
\def\labelenumi{\alph{enumi})}
\tightlist
\item
  Using the Nelson-Äalen estimator, calculate the probability that the
  loss on a policy exceeds 11, \(\hat{S}_{NA}(11)\).
\item
  Calculate Greenwood's approximation to the variance of the
  product-limit estimate \(\hat{S}(11)\).
\end{enumerate}

Show Example Solution

\hypertarget{toggleExampleSelect.3.7}{}
\textbf{Solution.} As before, there are four event times (non-censored
observations). For each time \(t_j\), we can calcuate the number of
events \(s_j\) and the risk set \(R_j\) as the following:

\[\begin{array}{cccc}
\hline
j & t_j & s_j & R_j \\
\hline
1 & 4 & 2 & 10 \\
2 & 8 & 1 & 5 \\
3 & 12 & 1 & 2 \\
4 & 15 & 1 & 1 \\
\hline
\end{array}\]

The Nelson-Äalen estimate of \(S(11)\) is
\(\hat{S}_{NA}(11)=e^{-\hat{H}(11)} = e^{-0.4} = 0.67\), since
\[\begin{aligned}
\hat{H}(11) &= \sum_{j:t_j\leq 11} \frac{s_j}{R_j}  = \sum_{j=1}^{2} \frac{s_j}{R_j}  \\
&= \frac{2}{10} + \frac{1}{5}  = 0.2 + 0.2 = 0.4 .\\
\end{aligned}\]

From earlier work, the Kaplan-Meier estimate of \(S(11)\) is
\(\hat{S}(11) = 0.64\). Then Greenwood's estimate of the variance of the
product-limit estimate of \(S(11)\) is \[\begin{aligned}
\widehat{Var}(\hat{S}(11)) &= (\hat{S}(11))^2 \sum_{j:t_j\leq 11} \frac{s_j}{R_j(R_j-s_j)}
&= (0.64)^2 \left(\frac{2}{10(8)} + \frac{1}{5(4)} \right)  = 0.0307. \\
\end{aligned}\]

\begin{center}\rule{0.5\linewidth}{\linethickness}\end{center}

\section{Bayesian Inference}\label{S:MS:BayesInference}

\begin{center}\rule{0.5\linewidth}{\linethickness}\end{center}

In this section, you learn how to:

\begin{itemize}
\tightlist
\item
  Describe the Bayes model as an alternative to the frequentist approach
  and summarize the five components of this modeling approach.
\item
  Describe the Bayesian decision framework and its role in determining
  Bayesian predictions.
\item
  Determine posterior predictions.
\end{itemize}

\begin{center}\rule{0.5\linewidth}{\linethickness}\end{center}

Up to this point, our inferential methods have focused on the
\textbf{frequentist} setting, in which samples are repeatedly drawn from
a population. The vector of parameters \(\boldsymbol \theta\) is fixed
yet unknown, whereas the outcomes \(X\) are realizations of random
variables.

In contrast, under the \textbf{Bayesian} framework, we view both the
model parameters and the data as random variables. We are uncertain
about the parameters \(\boldsymbol \theta\) and use probability tools to
reflect this uncertainty.

There are several advantages of the Bayesian approach. First, we can
describe the entire distribution of parameters conditional on the data.
This allows us, for example, to provide probability statements regarding
the likelihood of parameters. Second, this approach allows analysts to
blend prior information known from other sources with the data in a
coherent manner. This topic is developed in detail in the credibility
chapter. Third, the Bayesian approach provides a unified approach for
estimating parameters. Some non-Bayesian methods, such as least squares,
require a separate approach to estimate variance components. In
contrast, in Bayesian methods, all parameters can be treated in a
similar fashion. This is convenient for explaining results to consumers
of the data analysis. Fourth, Bayesian analysis is particularly useful
for forecasting future responses.

\subsection{Bayesian Model}\label{bayesian-model}

As stated earlier, under the Bayesian perspective, the model parameters
and data are both viewed as random. Our uncertainty about the parameters
of the underlying data generating process is reflected in the use of
probability tools.

\textbf{Prior Distribution.} Specifically, think about
\(\boldsymbol \theta\) as a random vector and let
\(\pi(\boldsymbol \theta)\) denote the distribution of possible
outcomes. This is knowledge that we have before outcomes are observed
and is called the prior distribution. Typically, the prior distribution
is a regular distribution and so integrates or sums to one, depending on
whether \(\boldsymbol \theta\) is continuous or discrete. However, we
may be very uncertain (or have no clue) about the distribution of
\(\boldsymbol \theta\); the Bayesian machinery allows the following
situation \[\int \pi(\theta) d\theta = \infty,\] in which case
\(\pi(\cdot)\) is called an \textbf{improper prior}.

\textbf{Model Distribution.} The distribution of outcomes given an
assumed value of \(\boldsymbol \theta\) is known as the model
distribution and denoted as
\(f(x | \boldsymbol \theta) = f_{X|\boldsymbol \theta} (x|\boldsymbol \theta )\).
This is the usual frequentist mass or density function.

\textbf{Joint Distribution.} The distribution of outcomes and model
parameters is, unsurprisingly, known as the joint distribution and
denoted as
\(f(x , \boldsymbol \theta) = f(x|\boldsymbol \theta )\pi(\boldsymbol \theta)\).

\textbf{Marginal Outcome Distribution.} The distribution of outcomes can
be expressed as
\[f(x) = f(x | \boldsymbol \theta)\pi(\boldsymbol \theta) d\boldsymbol \theta.\]
This is analogous to a frequentist mixture distribution.

\textbf{Posterior Distribution of Parameters.} After outcomes have been
observed (hence the terminology ``posterior''), one can use Bayes
theorem to write the distribution as
\[\pi(\boldsymbol \theta | x) =\frac{f(x , \boldsymbol \theta)}{f(x)} =\frac{f(x|\boldsymbol \theta )\pi(\boldsymbol \theta)}{f(x)}\]
The idea is to update your knowledge of the distribution of
\(\boldsymbol \theta\) (\(\pi(\boldsymbol \theta)\)) with the data
\(x\).

We can summarize the distribution using a confidence interval type
statement.

\textbf{Definition.} \([a,b]\) is said to be a \(100(1-\alpha)\%\)
\textbf{credibility interval} for \(\boldsymbol \theta\) if
\[\Pr (a \le \theta \le b | \mathbf{x}) \ge 1- \alpha.\]

\begin{center}\rule{0.5\linewidth}{\linethickness}\end{center}

\textbf{Exercise 4.4.1. SOA Exam Question.} You are given:

\begin{enumerate}
\def\labelenumi{(\roman{enumi})}
\tightlist
\item
  In a portfolio of risks, each policyholder can have at most one claim
  per year.
\item
  The probability of a claim for a policyholder during a year is \(q\).
\item
  The prior density is \[\pi(q) = q^3/0.07, \ \ \ 0.6 < q < 0.8\]
\end{enumerate}

A randomly selected policyholder has one claim in Year 1 and zero claims
in Year 2. For this policyholder, calculate the posterior probability
that \(0.7 < q < 0.8\).

Show Example Solution

\hypertarget{toggleExampleSelect.4.1}{}
\textbf{Solution.} The posterior density is proportional to the product
of the likelihood function and prior density. Thus,
\[\pi(q|1,0) \propto f(1|q)\ f(0|q)\ \pi(q) \propto q(1-q)q^3 = q^4-q^5\]

To get the exact posterior density, we integrate the above function over
its range \((0.6, 0.8)\)

\[\int_{0.6}^{0.8} q^4-q^5 dq = \frac{q^5}{5} - \left. \frac{q^6}{6} \right|_{0.6}^{0.8} = 0.014069 \ \Rightarrow \ \pi(q|1,0)=\frac{q^4-q^5}{0.014069}\]

Then
\[P(0.7<q<0.8|1,0)= \int_{0.7}^{0.8} \frac{q^4-q^5}{0.014069}dq = 0.5572\]

\begin{center}\rule{0.5\linewidth}{\linethickness}\end{center}

\textbf{Example 4.4.2. SOA Exam Question.} You are given:

\begin{enumerate}
\def\labelenumi{(\roman{enumi})}
\tightlist
\item
  The prior distribution of the parameter \(\Theta\) has probability
  density function:
  \[\pi(\theta) = \frac{1}{\theta^2}, \ \ 1 < \theta < \infty\]
\item
  Given \(\Theta = \theta\), claim sizes follow a Pareto distribution
  with parameters \(\alpha=2\) and \(\theta\).
\end{enumerate}

A claim of 3 is observed. Calculate the posterior probability that
\(\Theta\) exceeds 2.

Show Example Solution

\hypertarget{toggleExampleSelect.4.2}{}
\emph{Solution:} The posterior density, given an observation of 3 is

\[\pi(\theta|3) =  \frac{f(3|\theta)\pi(\theta)}{\int_1^\infty f(3|\theta)\pi(\theta)d\theta} =
\frac{\frac{2\theta^2}{(3+\theta)^3}\frac{1}{\theta^2}}{\int_1^\infty 2(3+\theta)^{-3} d\theta} =
\frac{2(3+\theta)^{-3}}{\left. -(3+\theta)^{-2}\right|_1^\infty} = 32(3+\theta)^{-3}, \ \ \theta > 1\]

Then

\[P(\Theta>2|3) = \int_2^\infty 32(3+\theta)^{-3}d\theta = \left. -16(3+\theta)^{-2} \right|_2^\infty = \frac{16}{25} = 0.64\]

\begin{center}\rule{0.5\linewidth}{\linethickness}\end{center}

\subsection{Decision Analysis}\label{decision-analysis}

In classical decision analysis, the loss function
\(l(\hat{\theta}, \theta)\) determines the penalty paid for using the
estimate \(\hat{\theta}\) instead of the true \(\theta\).

The \textbf{Bayes estimate} is that value that minimizes the expected
loss \(\mathrm{E~}[ l(\hat{\theta}, \theta)]\).

Some important special cases include:

\[\begin{array}{cll}
\hline
\text{Loss function } l(\hat{\theta}, \theta) & \text{Descriptor} & \text{Bayes Estimate} \\
\hline
(\hat{\theta}- \theta)^2 & \text{squared error loss} & \mathrm{E}(\theta|X) \\
|\hat{\theta}- \theta| & \text{absolute deviation loss} & \text{median of } \pi(\theta|x) \\
I(\hat{\theta} =\theta) & \text{zero-one loss (for discrete probabilities)} & \text{mode of } \pi(\theta|x) \\
\hline
\end{array}\]

For new data \(y\), the predictive distribution is
\[f(y|x) = \int f(y|\theta) \pi(\theta|x) d\theta .\]

With this, the \textbf{Bayesian prediction} of \(y\) is

\[\begin{aligned}
\mathrm{E}(y|x) &=  \int y f(y|x) dy = \int y \left(\int f(y|\theta) \pi(\theta|x) d\theta \right) dy \\
&=  \int  \mathrm{E}(y|\theta) \pi(\theta|x) d\theta .
\end{aligned}\]

\begin{center}\rule{0.5\linewidth}{\linethickness}\end{center}

\textbf{Example 4.4.3. SOA Exam Question.} For a particular policy, the
conditional probability of the annual number of claims given
\(\Theta = \theta\), and the probability distribution of \(\Theta\) are
as follows:

\[\begin{array}{l|ccc}
\hline
\text{Number of Claims} & 0 & 1 & 2 \\
\text{Probability} & 2\theta & \theta & 1-3\theta \\
\hline
\end{array}\]

\[\begin{array}{l|cc}
\hline
\theta & 0.05 & 0.30 \\
\text{Probability} & 0.80 & 0.20 \\
\hline
\end{array}\]

Two claims are observed in Year 1. Calculate the Bayesian estimate
(Bühlmann credibility estimate) of the number of claims in Year 2.

Show Example Solution

\hypertarget{toggleExampleSelect.4.3}{}
\textbf{Solution.} Note that
\(\mathrm{E}(\theta) = 0.05(0.8) + 0.3(0.2) = 0.1\) and
\(\mathrm{E}(\theta^2) = 0.05^2(0.8) + 0.3^2(0.2)=0.02\)

We also have
\(\mu(\theta) = 0(2\theta) + 1(\theta) + 2(1-3\theta) = 2-5\theta\) and
\(v(\theta) = 0^2(2\theta) + 1^2(\theta) + 2^2(1-3\theta) - (2-5\theta)^2 = 9\theta-25\theta^2\).

Thus

\[\begin{aligned}
\mu &=  E(2-5\theta) = 2-5(0.1) = 1.5 \\
v   &=  EVPV = E(9\theta - 25\theta^2)=9(0.1)-25(0.02)=0.4 \\
a &= VHM = Var(2-5\theta) = 25Var(\theta) = 25(0.02-0.1^2) = 0.25 \\
\Rightarrow k &= \frac{v}{a} = \frac{0.4}{0.25} = 1.6 \\
\Rightarrow Z &= \frac{1}{1+1.6} = \frac{5}{13}
\end{aligned}\]

Therefore, \(P=\frac{5}{13}2 + \frac{8}{13}1.5 = 1.6923\).

\begin{center}\rule{0.5\linewidth}{\linethickness}\end{center}

\textbf{Example 4.4.4. SOA Exam Question.} You are given:

\begin{enumerate}
\def\labelenumi{(\roman{enumi})}
\tightlist
\item
  Losses on a company's insurance policies follow a Pareto distribution
  with probability density function:
  \[f(x|\theta) = \frac{\theta}{(x+\theta)^2}, \ \ 0 < x < \infty\]
\item
  For half of the company's policies \(\theta=1\) , while for the other
  half \(\theta=3\).
\end{enumerate}

For a randomly selected policy, losses in Year 1 were 5. Calculate the
posterior probability that losses for this policy in Year 2 will exceed
8.

Show Example Solution

\hypertarget{toggleExampleSelect.4.4}{}
\textbf{Solution.} We are given the prior distribution of \(\theta\) as
\(P(\theta=1)=P(\theta=3)=\frac{1}{2}\), the conditional distribution
\(f(x|\theta)\), and the fact that we observed \(X_1=5\). The goal is to
find the predictive probability \(P(X_2>8|X_1=5)\).

The posterior probabilities are

\[\begin{aligned}
P(\theta=1|X_1=5) &= \frac{f(5|\theta=1)P(\theta=1)}{f(5|\theta=1)P(\theta=1) + f(5|\theta=3)P(\theta=3)} \\
&= \frac{\frac{1}{36}(\frac{1}{2})}{\frac{1}{36}(\frac{1}{2})+\frac{3}{64}(\frac{1}{2})} = \frac{\frac{1}{72}}{\frac{1}{72} + \frac{3}{128}} = \frac{16}{43}
\end{aligned}\]

\[\begin{aligned}
P(\theta=3|X_1=5) &= \frac{f(5|\theta=3)P(\theta=3)}{f(5|\theta=1)P(\theta=1) + f(5|\theta=3)P(\theta=3)} \\
&= 1-P(\theta=1|X_1=5) = \frac{27}{43}
\end{aligned}\]

Note that the conditional probability that losses exceed 8 is

\[\begin{aligned}
P(X_2>8|\theta) &= \int_8^\infty f(x|\theta)dx \\
&= \int_8^\infty \frac{\theta}{(x+\theta)^2}dx = \left. -\frac{\theta}{x+\theta} \right|_8^\infty = \frac{\theta}{8 + \theta}
\end{aligned}\]

The predictive probability is therefore

\[\begin{aligned}
P(X_2>8|X_1=5) &= P(X_2>8|\theta=1) P(\theta=1|X_1=5) + P(X_2>8|\theta=3) P(\theta=3 | X_1=5) \\
&= \frac{1}{8+1}\left( \frac{16}{43}\right) + \frac{3}{8+3} \left( \frac{27}{43}\right) = 0.2126
\end{aligned}\]

\begin{center}\rule{0.5\linewidth}{\linethickness}\end{center}

\textbf{Exercise 4.4.5. SOA Exam Question.} You are given:

\begin{enumerate}
\def\labelenumi{(\roman{enumi})}
\tightlist
\item
  The probability that an insured will have at least one loss during any
  year is \(p\).
\item
  The prior distribution for \(p\) is uniform on \([0, 0.5]\).
\item
  An insured is observed for 8 years and has at least one loss every
  year.
\end{enumerate}

Calculate the posterior probability that the insured will have at least
one loss during Year 9.

Show Example Solution

\hypertarget{toggleExampleSelect.4.5}{}
\textbf{Solution.} The posterior probability density is
\[\begin{aligned}
\pi(p|1,1,1,1,1,1,1,1) &\propto Pr(1,1,1,1,1,1,1,1|p)\ \pi(p) = p^8(2) \propto p^8 \\
\Rightarrow \pi(p|1,1,1,1,1,1,1,1) &= \frac{p^8}{\int_0^5 p^8 dp} = \frac{p^8}{(0.5^9)/9} = 9(0.5^{-9})p^8
\end{aligned}\]

Thus, the posterior probability that the insured will have at least one
loss during Year 9 is

\[\begin{aligned}
P(X_9=1|1,1,1,1,1,1,1,1) &= \int_0^5 P(X_9=1|p) \pi(p|1,1,1,1,1,1,1,1) dp \\
&= \int_0^5 p(9)(0.5^{-9})p^8 dp = 9(0.5^{-9})(0.5^{10})/10 = 0.45
\end{aligned}\]

\begin{center}\rule{0.5\linewidth}{\linethickness}\end{center}

\textbf{Example 4.4.6. SOA Exam Question.} You are given:

\begin{enumerate}
\def\labelenumi{(\roman{enumi})}
\tightlist
\item
  Each risk has at most one claim each year. \[\begin{array}{ccc}
  \hline
  \text{Type of Risk} & \text{Prior Probability} & \text{Annual Claim Probability} \\
  \hline
  \text{I} & 0.7 & 0.1 \\
  \text{II} & 0.2 & 0.2 \\
  \text{III} & 0.1 & 0.4 \\
  \hline
  \end{array}\]
\end{enumerate}

One randomly chosen risk has three claims during Years 1-6. Calculate
the posterior probability of a claim for this risk in Year 7.

Show Example Solution

\hypertarget{toggleExampleSelect.4.6}{}
\textbf{Solution.} The probabilities are from a binomial distribution
with 6 trials in which 3 successes were observed.

\[\begin{aligned}
P(3|\text{I}) &= {6 \choose 3} (0.1^3)(0.9^3) = 0.01458 \\
P(3|\text{II}) &= {6 \choose 3} (0.2^3)(0.8^3) = 0.08192 \\
P(3|\text{III}) &= {6 \choose 3} (0.4^3)(0.6^3) = 0.27648
\end{aligned}\]

The probability of observing three successes is
\[\begin{aligned} P(3) &= P(3|\text{I})P(\text{I}) + P(3|\text{II})P(\text{II}) + P(3|\text{III})P(\text{III}) \\
&=  0.7(0.01458) + 0.2(0.08192) + 0.1(0.27648) = 0.054238
\end{aligned}\]

The three posterior probabilities are \[\begin{aligned}
P(\text{I}|3) &= \frac{P(3|\text{I})P(\text{I})}{P(3)} = \frac{0.7(0.01458)}{0.054238} = 0.18817 \\
P(\text{II}|3) &= \frac{P(3|\text{II})P(\text{II})}{P(3)} = \frac{0.2(0.08192)}{0.054238} = 0.30208 \\
P(\text{III}|3) &= \frac{P(3|\text{III})P(\text{III})}{P(3)} = \frac{0.1(0.27648)}{0.054238} = 0.50975
\end{aligned}\]

The posterior probability of a claim is then \[\begin{aligned}
P(\text{claim} | 3) &= P(\text{claim}|\text{I})P(\text{I} | 3) + P(\text{claim} | \text{II})P(\text{II} | 3) + P(\text{claim} | \text{III}) P(\text{III} | 3) \\
&= 0.1(0.18817) + 0.2(0.30208) + 0.4(0.50975) = 0.28313
\end{aligned}\]

\begin{center}\rule{0.5\linewidth}{\linethickness}\end{center}

\subsection{Posterior Distribution}\label{posterior-distribution}

How can we calculate the posterior distribution
\(\pi(\boldsymbol \theta | x) =\frac{f(x|\boldsymbol \theta )\pi(\boldsymbol \theta)}{f(x)}\)?

\begin{itemize}
\tightlist
\item
  \textbf{By hand:} we can do this in special cases
\item
  \textbf{Simulation:} use modern computational techniques such as
  Markov Chain Monte Carlo (MCMC) simulation
\item
  \textbf{Normal approximation:} !!! Theorem 12.39 of \textbf{KPW}
  provides a justification
\item
  \textbf{Conjugate distributions:} classical approach. Although this
  approach is available only for a limited number of distributions, it
  has the appeal that it provides closed-form expressions for the
  distributions, allowing for easy interpretations of results. We focus
  on this approach.
\end{itemize}

To relate the prior and posterior distributions of the parameters, we
have

\[\begin{array}{ccc}
\pi(\boldsymbol \theta | x) & = & \frac{f(x|\boldsymbol \theta )\pi(\boldsymbol \theta)}{f(x)}  \\
 & \propto  & f(x|\boldsymbol \theta ) \pi(\boldsymbol \theta) \\
\text{Posterior} & \text{is proportional to} & \text{likelihood} \times \text{prior} \\
\end{array}\]

For \textbf{conjugate distributions}, the posterior and the prior come
from the same family of distributions. The following illustration looks
at the Poisson-gamma special case, the most well-known in actuarial
applications.

\begin{center}\rule{0.5\linewidth}{\linethickness}\end{center}

Show Example

\hypertarget{toggleExampleConj}{}
\textbf{Special Case -- Poisson-Gamma} Assume a Poisson(\(\lambda\))
model distribution so that
\[f(\mathbf{x} | \lambda) = \prod_{i=1}^n \frac{\lambda^{x_i} e^{-\lambda}}{x_i!}\]
Assume \(\lambda\) follows a gamma(\(\alpha, \theta\)) prior
distribution so that
\[\pi(\lambda) = \frac{\left(\lambda/\theta\right)^{\alpha} \exp(-\lambda/\theta)}{\lambda \Gamma(\alpha)}.\]
The posterior distribution is proportional to \[\begin{aligned}
\pi(\lambda | \mathbf{x}) &\propto f(\mathbf{x}|\theta ) \pi(\lambda) \\
&= C \lambda^{\sum_i x_i + \alpha +1} \exp(-\lambda(n+1/\theta))
\end{aligned}\]

where \(C\) is a constant. We recognize this to be a gamma distribution
with new parameters \(\alpha_{new} = \sum_i x_i + \alpha\) and
\(\theta_{new} = 1/(n + 1/\theta)\). Thus, the gamma distribution is a
conjugate prior for the Poisson model distribution.

\begin{center}\rule{0.5\linewidth}{\linethickness}\end{center}

\textbf{Example 4.4.7. SOA Exam Question.} You are given:

\begin{enumerate}
\def\labelenumi{(\roman{enumi})}
\tightlist
\item
  The conditional distribution of the number of claims per policyholder
  is Poisson with mean \(\lambda\).
\item
  The variable \(\lambda\) has a gamma distribution with parameters
  \(\alpha\) and \(\theta\).
\item
  For policyholders with 1 claim in Year 1, the credibility estimate for
  the number of claims in Year 2 is 0.15.
\item
  For policyholders with an average of 2 claims per year in Year 1 and
  Year 2, the credibility estimate for the number of claims in Year 3 is
  0.20.
\end{enumerate}

Calculate \(\theta\).

Show Example Solution

\hypertarget{toggleExampleSelect.4.7}{}
\textbf{Solution.} Since the conditional distribution of the number of
claims per policyholder, \(E(X|\lambda)=Var(X|\lambda)=\lambda\)

Thus,

\[\begin{aligned}
\mu &= v = E(\lambda) = \alpha\theta \\
a &= Var(\lambda) = \alpha\theta^2 \\
k &= \frac{v}{a} = \frac{1}{\theta} \\
\Rightarrow Z &= \frac{n}{n+1/\theta} = \frac{n\theta}{n\theta+1}
\end{aligned}\]

Using the credibility estimates given,

\[\begin{aligned}
0.15 &= \frac{\theta}{\theta + 1}(1) + \frac{1}{\theta + 1}\mu = \frac{\theta + \mu}{\theta + 1} \\
0.20 &= \frac{2\theta}{2\theta+1}(2) + \frac{1}{2\theta+1}\mu = \frac{4\theta+\mu}{2\theta+1}
\end{aligned}\]

From the first equation,
\(0.15\theta + 0.15 = \theta + \mu \ \Rightarrow \ \mu = 0.15- 0.85\theta\).

Then the second equation becomes
\(0.4\theta + 0.2 = 4\theta + 0.15 - 0.85\theta \ \Rightarrow \ \theta=0.01818.\)

\begin{center}\rule{0.5\linewidth}{\linethickness}\end{center}

\section{Further Resources and
Contributors}\label{MS:further-reading-and-resources}

\subsubsection*{Exercises}\label{exercises-2}
\addcontentsline{toc}{subsubsection}{Exercises}

Here are a set of exercises that guide the viewer through some of the
theoretical foundations of \textbf{Loss Data Analytics}. Each tutorial
is based on one or more questions from the professional actuarial
examinations, typically the Society of Actuaries Exam C.

\href{http://www.ssc.wisc.edu/~jfrees/loss-data-analytics/loss-data-analytics-model-selection/}{Model
Selection Guided Tutorials}

\subsubsection*{Contributors}\label{contributors-2}
\addcontentsline{toc}{subsubsection}{Contributors}

\begin{itemize}
\tightlist
\item
  \textbf{Edward W. (Jed) Frees} and \textbf{Lisa Gao}, University of
  Wisconsin-Madison, are the principal authors of the initital version
  of this chapter. Email:
  \href{mailto:jfrees@bus.wisc.edu}{\nolinkurl{jfrees@bus.wisc.edu}} for
  chapter comments and suggested improvements.
\end{itemize}

\section*{Technical Supplement A. Gini
Statistic}\label{technical-supplement-a.-gini-statistic}
\addcontentsline{toc}{section}{Technical Supplement A. Gini Statistic}

\subsection*{TS A.1. The Classic Lorenz
Curve}\label{ts-a.1.-the-classic-lorenz-curve}
\addcontentsline{toc}{subsection}{TS A.1. The Classic Lorenz Curve}

In welfare economics, it is common to compare distributions via the
\textbf{Lorenz curve}, developed by Max Otto Lorenz
\citep{lorenz1905methods}. A Lorenz curve is a graph of the proportion
of a population on the horizontal axis and a distribution function of
interest on the vertical axis. It is typically used to represent income
distributions. When the income distribution is perfectly aligned with
the population distribution, the Lorenz curve results in a 45 degree
line that is known as the \textbf{line of equality}. The area between
the Lorenz curve and the line of equality is a measure of the
discrepancy between the income and population distributions. Two times
this area is known as the \textbf{Gini index}, introduced by Corrado
Gini in 1912.

\begin{figure}

{\centering \includegraphics[width=0.9\linewidth]{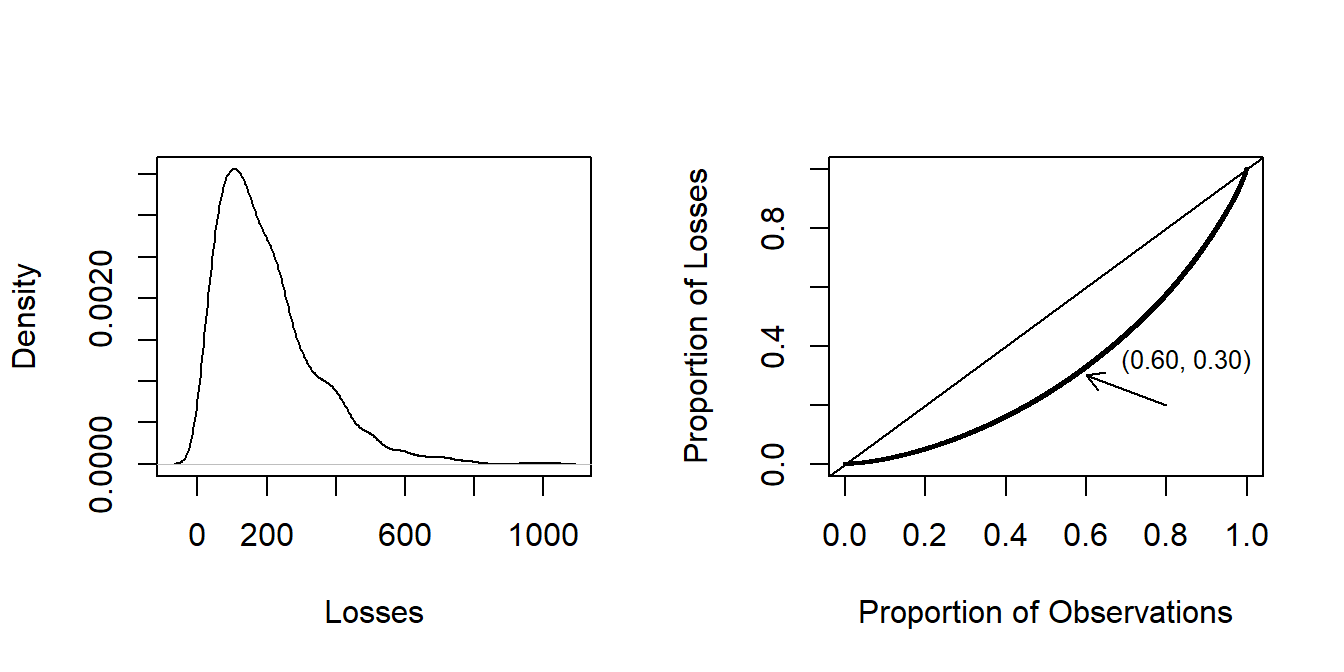}

}

\caption{Distribution of insurance losses.}\label{fig:ClassicLorenz}
\end{figure}

\textbf{Example -- Classic Lorenz Curve.} For an insurance example,
Figure \ref{fig:ClassicLorenz} shows a distribution of insurance losses.
This figure is based on a random sample of 2000 losses. The left-hand
panel shows a right-skewed histogram of losses. The right-hand panel
provides the corresponding Lorenz curve, showing again a skewed
distribution. For example, the arrow marks the point where 60 percent of
the policyholders have 30 percent of losses. The 45 degree line is the
line of equality; if each policyholder has the same loss, then the loss
distribution would be at this line. The Gini index, twice the area
between the Lorenz curve and the 45 degree line, is 37.6 percent for
this data set.

\subsection*{TS A.2. Ordered Lorenz Curve and the Gini
Index}\label{ts-a.2.-ordered-lorenz-curve-and-the-gini-index}
\addcontentsline{toc}{subsection}{TS A.2. Ordered Lorenz Curve and the
Gini Index}

We now introduce a modification of the classic Lorenz curve and Gini
statistic that is useful in insurance applications. Specifically, we
introduce an \emph{ordered} Lorenz curve which is a graph of the
distribution of losses versus premiums, where both losses and premiums
are ordered by relativities. Intuitively, the relativities point towards
aspects of the comparison where there is a mismatch between losses and
premiums. To make the ideas concrete, we first provide some notation. We
will consider \(i=1, \ldots, n\) policies. For the \(i\)th policy, let

\begin{itemize}
\tightlist
\item
  \(y_i\) denote the insurance loss,
\item
  \(\mathbf{x}_i\) be the set of policyholder characteristics known to
  the analyst,
\item
  \(P_i=P(\mathbf{x}_i)\) be the associated premium that is a function
  of \(\mathbf{x}_i\),
\item
  \(S_i = S(\mathbf{x}_i)\) be an insurance score under consideration
  for rate changes, and
\item
  \(R_i = R(\mathbf{x}_i)=S(\mathbf{x}_i)/P(\mathbf{x}_i)\) is the
  relativity, or relative premium.
\end{itemize}

Thus, the set of information used to calculate the ordered Lorenz curve
for the \(i\)th policy is \((y_i, P_i, S_i, R_i)\).

\subsubsection*{Ordered Lorenz Curve}\label{ordered-lorenz-curve}
\addcontentsline{toc}{subsubsection}{Ordered Lorenz Curve}

We now sort the set of policies based on relativities (from smallest to
largest) and compute the premium and loss distributions. Using notation,
the premium distribution is

\begin{equation}
\hat{F}_P(s) =  \frac{ \sum_{i=1}^n
P(\mathbf{x}_i) \mathrm{I}(R_i \leq s) }{\sum_{i=1}^n P(\mathbf{x}_i)} ,\label{eq:EmpPremDF}
\end{equation}

and the loss distribution is

\begin{equation}
\hat{F}_{L}(s) =  \frac{ \sum_{i=1}^n y_i \mathrm{I}(R_i
\leq s) }{\sum_{i=1}^n y_i} ,\label{eq:EmpLossDF}
\end{equation}

where \(\mathrm{I}(\cdot)\) is the indicator function, returning a 1 if
the event is true and zero otherwise. The graph
\(\left(\hat{F}_P(s),\hat{F}_{L}(s) \right)\) is an \textbf{ordered
Lorenz curve}.

The classic Lorenz curve shows the proportion of policyholders on the
horizontal axis and the loss distribution function on the vertical axis.
The ordered Lorenz curve extends the classical Lorenz curve in two ways,
(1) through the ordering of risks and prices by relativities and (2) by
allowing prices to vary by observation. We summarize the ordered Lorenz
curve in the same way as the classic Lorenz curve using a Gini index,
defined as twice the area between the curve and a 45 degree line. The
analyst seeks ordered Lorenz curves that approach passing through the
southeast corner (1,0); these have greater separation between the loss
and premium distributions and therefore larger Gini indices.

\textbf{Example -- Loss Distribution.}

Suppose we have \(n=5\) policyholders with experience as:

\begin{longtable}[]{@{}lcllllll@{}}
\toprule
Variable & \(i\) & 1 & 2 & 3 & 4 & 5 & Sum\tabularnewline
\midrule
\endhead
Loss & \(y_i\) & 5 & 5 & 5 & 4 & 6 & 25\tabularnewline
Premium & \(P(\mathbf{x}_i)\) & 4 & 2 & 6 & 5 & 8 & 25\tabularnewline
Relativity & \(R(\mathbf{x}_i)\) & 5 & 4 & 3 & 2 & 1 &\tabularnewline
\bottomrule
\end{longtable}

Determine the Lorenz curve and the ordered Lorenz curve.

Show Example Solution

\hypertarget{toggleLorenz}{}
\begin{figure}

{\centering \includegraphics[width=0.9\linewidth]{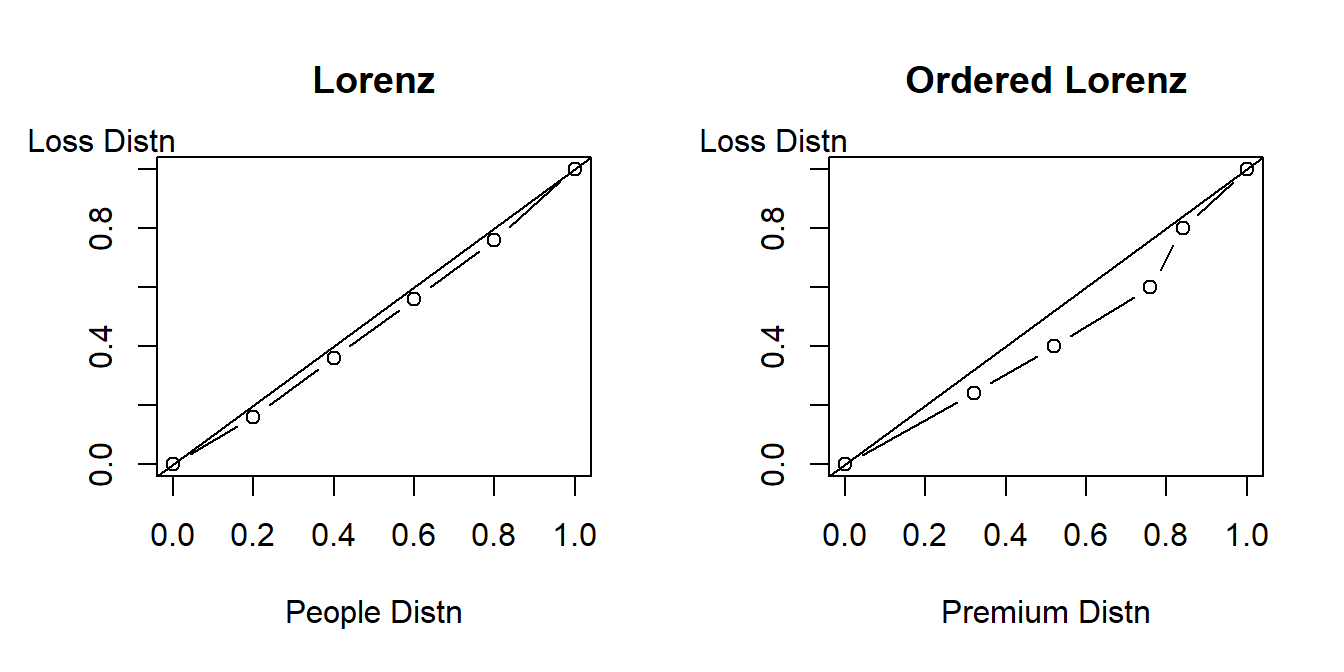}

}

\caption{Lorenz versus Ordered Lorenz Curve}\label{fig:LorenzVsOrdered}
\end{figure}

Figure \ref{fig:LorenzVsOrdered} compares the Lorenz curve to the
ordered version based on this data. The left-hand panel shows the Lorenz
curve. The horizontal axis is the cumulative proportion of policyholders
(0, 0.2, 0.4, and so forth) and the vertical axis is the cumulative
proportion of losses (0, 4/25, 9/25, and so forth). This figure shows
little separation between the distributions of losses and policyholders.

The right-hand panel shows the ordered Lorenz curve. Because
observations are sorted by relativities, the first point after the
origin (reading from left to right) is (8/25, 6/25). The second point is
(13/25, 10/25), with the pattern continuing. For the ordered Lorenz
curve, the horizontal axis uses premium weights, the vertical axis uses
loss weights, and both axes are ordered by relativities. From the
figure, we see that there is greater separation between losses and
premiums when viewed through this relativity.

\subsubsection*{Gini Index}\label{gini-index}
\addcontentsline{toc}{subsubsection}{Gini Index}

Specifically, the Gini index can be calculated as follows. Suppose that
the empirical ordered Lorenz curve is given by
\(\{ (a_0=0, b_0=0), (a_1, b_1), \ldots,\) \((a_n=1, b_n=1) \}\) for a
sample of \(n\) observations. Here, we use \(a_j = \hat{F}_P(R_j)\) and
\(b_j = \hat{F}_{L}(R_j)\). Then, the empirical Gini index is

\begin{eqnarray}
\widehat{Gini} &=&  2\sum_{j=0}^{n-1} (a_{j+1} - a_j) \left \{
\frac{a_{j+1}+a_j}{2} - \frac{b_{j+1}+b_j}{2} \right\} \nonumber \\
&=& 1 - \sum_{j=0}^{n-1} (a_{j+1} - a_j) (b_{j+1}+b_j) .\label{eq:GiniDefn}
\end{eqnarray}

\textbf{Example -- Loss Distribution: Continued.} In the figure, the
Gini index for the left-hand panel is 5.6\%. In contrast, the Gini index
for the right-hand panel is 14.9\%. \(~~\Box\)

\subsection*{TS A.3. Out-of-Sample
Validation}\label{ts-a.3.-out-of-sample-validation}
\addcontentsline{toc}{subsection}{TS A.3. Out-of-Sample Validation}

The Gini statistics based on an ordered Lorenz curve can be used for
out-of-sample validation. The procedure follows:

\begin{enumerate}
\def\labelenumi{\arabic{enumi}.}
\tightlist
\item
  Use an in-sample data set to estimate several competing models.
\item
  Designate an out-of-sample, or validation, data set of the form
  \(\{(y_i, \mathbf{x}_i), i=1,\ldots,n\}\).
\item
  Establish one of the models as the base model. Use this estimated
  model and explanatory variables from the validation sample to form
  premiums of the form \(P(\mathbf{x}_i))\).
\item
  Use an estimated competing model and validation sample explanatory
  variables to form scores of the form \(S(\mathbf{x}_i))\).
\item
  From the premiums and scores, develop relativities
  \(R_i =S(\mathbf{x}_i)/P(\mathbf{x}_i)\).
\item
  Use the validation sample outcomes \(y_i\) to compute the Gini
  statistic.
\end{enumerate}

\textbf{Example -- Out-of-Sample Validation.}

Suppose that we have experience from 25 states. For each state, we have
available 500 observations that can be used to predict future losses.
For this illustration, we have generated losses using a gamma
distrbution with common shape parameter equal to 5 and a scale parameter
that varies by state, from a low of 20 to 66.

Determine the ordered Lorenz curve and the corresponding Gini statistic
to compare the two rate procedures.

Show Example Solution

\hypertarget{toggleExampleLor}{}
For our base premium, we simply use the maximum likelihood estimate
assuming a common distribution among all states. For the gamma
distribution, this turns out to be simply the average which for our
simulation is \textbf{P}=219.96. You can think of this common premium as
based on a \emph{community rating} principle. As an alternative, we use
averages that are state-specific. Because this illustration uses means
that vary by states, we anticipate this alternative rating procedure to
be preferred to the community rating procedure. (Recall for the gamma
distribution that the mean equals the shape times the scale or, 5 times
the scale parameter, for our example.)

Out of sample claims were generated from the same gamma distribution,
with 200 observations for each state. In the following, we have the
ordered Lorenz curve.

\begin{figure}

{\centering \includegraphics[width=0.5\linewidth]{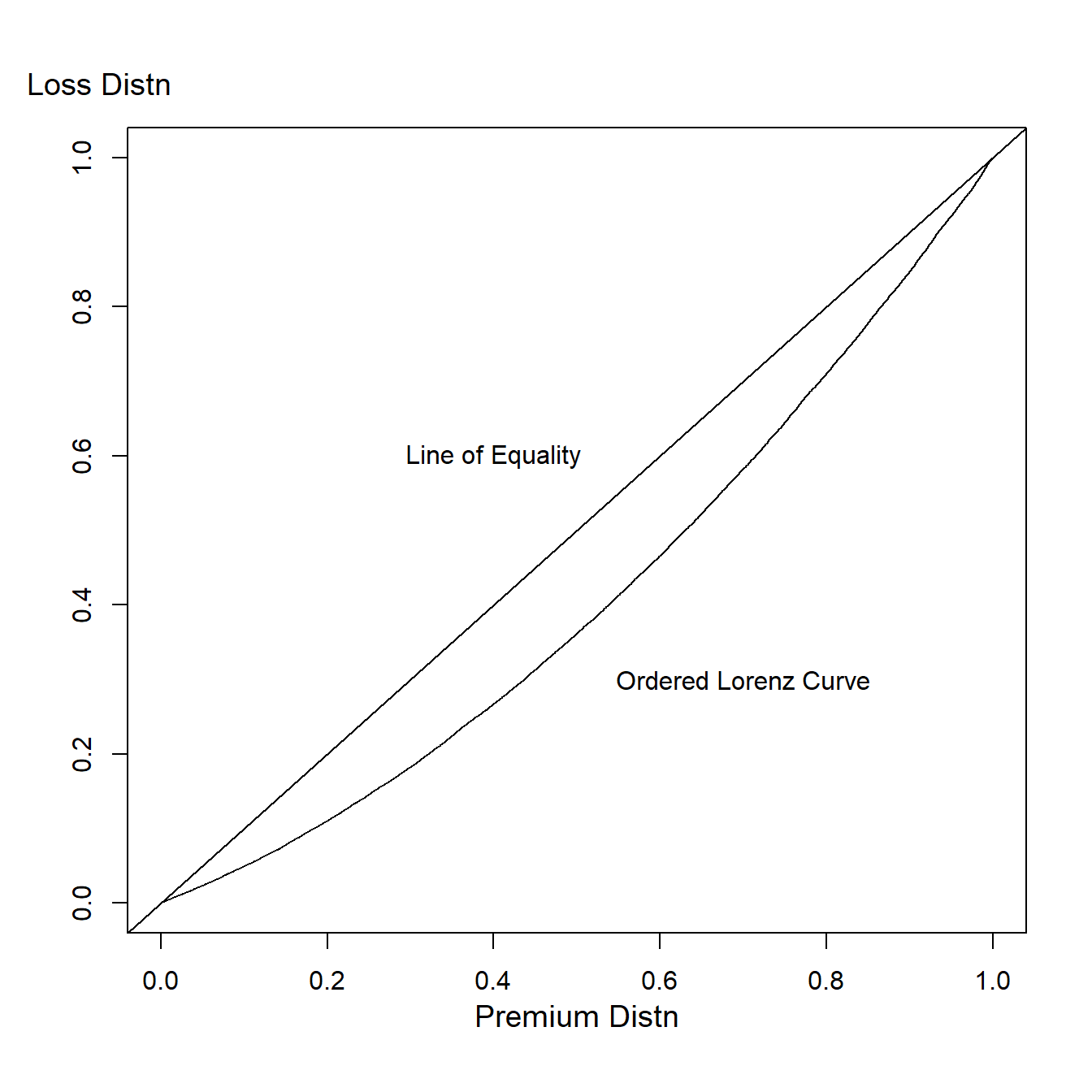}

}

\end{figure}

For these data, the Gini index is 0.187 with a standard error equal to
0.00381.

\subsubsection*{Discussion}\label{discussion}
\addcontentsline{toc}{subsubsection}{Discussion}

In insurance claims modeling, standard out-of-sample validation measures
are not the most informative due to the high proportions of zeros
(corresponding to no claim) and the skewed fat-tailed distribution of
the positive values. The Gini index can be motivated by the economics of
insurance. Intuitively, the Gini index measures the negative covariance
between a policy's ``profit'' (\(P-y\), premium minus loss) and the rank
of the relativity (\textbf{R}, score divided by premium). That is, the
close approximation

\[\widehat{Gini} \approx - \frac{2}{n} \widehat{Cov} \left( (P-y), rank(R) \right) .\]

This observation leads an insurer to seek an ordering that produces to a
large Gini index. Thus, the Gini index and associated ordered Lorenz
curve are useful for identifying profitable blocks of insurance
business.

Unlike classical measures of association, the Gini index assumes that a
premium base \textbf{P} is currently in place and seeks to assess
vulnerabilities of this structure. This approach is more akin to
hypothesis testing (when compared to goodness of fit) where one
identifies a ``null hypothesis'' as the current state of the world and
uses decision-making criteria/statistics to compare this with an
``alternative hypothesis.''

The insurance version of the Gini statistic was developed by
\citep{frees2011summarizing} and \citep{frees2014insurance} where you
can find formulas for the standard errors and other additional
background information.

\chapter{Aggregate Loss Models}\label{C:AggLossModels}

\emph{Chapter Preview}. This chapter introduces probability models for
describing the aggregate claims that arise from a portfolio of insurance
contracts. We presents two standard modeling approaches, the individual
risk model and the collective risk model. Further, we discuss strategies
for computing the distribution of the aggregate claims. Finally, we
examine the effects of individual policy modifications on the aggregate
loss distribution.

\section{Introduction}\label{introduction}

The objective of this chapter is to build a probability model to
describe the aggregate claims by an insurance system occurring in a
fixed time period. The insurance system could be a single policy, a
group insurance contract, a business line, or an entire book of an
insurance's business. In the chapter, aggregate claims refers to either
the number or the amount of claims from a portfolio of insurance
contracts. However, the modeling framework is readily to apply in the
more general setup.

Consider an insurance portfolio of \(n\) individual contracts, and let
\(S\) denote the aggregate losses of the portfolio in a given time
period. There are two approaches to modeling the aggregate losses \(S\),
the individual risk model and the collective risk model. The individual
risk model emphasizes the loss from each individual contract and
represents the aggregate losses as: \[\begin{aligned}
S=X_1 +X_2 +\cdots+X_n,
\end{aligned}\] where \(X_i~(i=1,\ldots,n)\) is interpreted as the loss
amount from the \(i\)th contract. It is worth stressing that \(n\)
denotes the number of contracts in the portfolio and thus is a fixed
number rather than a random variable. For the individual risk model, one
usually assumes \(X_{i}\)'s are independent, i.e., \(X_{i}\perp X_{j}\)
\(\forall\) \(i,j\). Because of different contract features such as
coverage and exposure, \(X_{i}\)'s are not necessarily identically
distributed. A notable feature of the distribution of each \(X_i\) is
the probability mass at zero corresponding to the event of no claims.

The collective risk model represents the aggregate losses in terms of a
frequency distribution and a severity distribution: \[\begin{aligned}
S=X_1 +X_2 +\cdots+X_N.
\end{aligned}\] Here one thinks of a random number of claims \(N\) that
may represent either the number of losses or the number of payments. In
contrast, in the individual risk model, we use a fixed number of
contracts \(n\). We think of \(X_1, X_2, \ldots, X_N\) as representing
the amount of each loss. Each loss may or may not corresponding to a
unique contract. For instance, there may be multiple claims arising from
a single contract. It is natural to think about \(X_i>0\) because if
\(X_i=0\) then no claim has occurred. Typically we assume that
conditional on \(N=n\), \(X_{1},X_{2},\cdots ,X_{n}\) are \emph{iid}
random variables. The distribution of \(N\) is known as the frequency
distribution, and the common distribution of \(X\) is known as the
severity distribution. We further assume \(N\) and \(X\) are
independent. With the collective risk model, we may decompose the
aggregate losses into the frequency (\(N\)) process and the severity
(\(X\)) process. This flexibility allows the analyst to comment on these
two separate components. For example, sales growth due to lower
underwriting standards could lead to higher frequency of losses but
might not affect severity. Similarly, inflation or other economic forces
could have an impact on severity but not on frequency.

\section{Individual Risk Model}\label{individual-risk-model}

As noted earlier, for the individual risk model, we think of \(X_i\) as
the loss from \(i^{th}\) contract and interpret

\begin{eqnarray*}
S_n=X_1 +X_2 +\cdots+X_n
\end{eqnarray*}

to be the aggregate loss from all contracts in a portfolio or group of
contracts. Under the independence assumption on \(X_i's\), it is
straightforward to show \[\begin{aligned}
    {\rm E}(S_n) &= \sum_{i=1}^{n} {\rm E}(X_i),~~~~
    {\rm Var}(S_n) = \sum_{i=1}^{n} {\rm Var}(X_i)\\
    P_{S_n}(z) &= \prod_{i=1}^{n}P_{X_i}(z), ~~~~
    M_{S_n}(t) = \prod_{i=1}^{n}M_{X_i}(t) \\
 \end{aligned}\] where \(P_S(\cdot)\) and \(M_S(\cdot)\) are probability
generating function and moment generating function of \(S\),
respectively. The distribution of each \(X_i\) contains mass at zero,
corresponding to the event of no claim. One strategy to incorporate the
zero mass in the distribution is using the two-part framework:
\[\begin{aligned}
X_i = I_i\times B_i = \left\{\begin{array}{ll}
                               0 & I_i=0 \\
                               B_i & I_i=1
                             \end{array}
             \right.
\end{aligned}\] Here \(I_i\) is a Bernoulli variable indicating whether
or not a loss occurs for the \(i\)th contract, and \(B_i\), a r.v. with
nonnegative support, represents the amount of losses of the contract
given loss occurrence. Assume that \(I_1 ,\ldots,I_n ,B_1 ,\ldots,B_n\)
are mutually independent. Denote \({\rm Pr} (I_i =1)=q_i\),
\(\mu_i={\rm E}(B_i)\), and \(\sigma_i^2={\rm Var}(B_i)\). One can show
\[\begin{aligned}
\mathrm{E}(S_n)& =\sum_{i=1}^n ~q_i  ~\mu _j \\
\mathrm{Var}(S_n) & =\sum_{i=1}^n \left( q_i \sigma _i^2+q_i (1-q_j)\mu_i^2 \right)\\
P_{S_n}(z) & =\prod_{i=1}^n \left( 1-q_i+q_i P_{B_i}(z) \right)\\
M_{S_n}(t) & =\prod_{i=1}^n \left( 1-q_i+q_i M_{B_i}(t) \right)
\end{aligned}\] A special case of the above model is when \(B_i\)
follows a degenerate distribution with \(\mu_i=b_i\) and
\(\sigma^2_i=0\). One example is term life insurance or a pure endowment
insurance where \(b_i\) represents the amount of insurance of the
\(i\)th contract.

Another strategy to accommodate zero mass in the distribution of \(X_i\)
is a collective risk model, i.e. \(X_i=Z_{i1}+\cdots+Z_{iN_i}\) where
\(X_i=0\) when \(N_i=0\). The collective risk model will be discussed in
detail in the next section.

\textbf{Example 5.2.1. SOA Exam Question.} An insurance company sold 300
fire insurance policies as follows:

\[\begin{matrix}
    \begin{array}{c c c} \hline
        \text{Number of} & \text{Policy} & \text{Probability of}\\
        \text{Policies} &  \text{Maximum} &  \text{Claim Per Policy}\\ \hline
        100 & 400 & 0.05\\
        200 & 300 & 0.06\\ \hline
    \end{array}
\end{matrix}\]

You are given:\\
(i) The claim amount for each policy is uniformly distributed between
\(0\) and the policy maximum.\\
(ii) The probability of more than one claim per policy is \(0\).\\
(iii) Claim occurrences are independent.

Calculate the mean \(\mathrm{E~}S_n\) and variance \(\mathrm{Var~}S_n\)
of the aggregate claims. How would these results change if every claim
is equal to the policy maximum?

Show Example Solution

\hypertarget{toggleExampleAggLoss.2.1}{}
\textbf{Solution.} The aggregate claims are
\(S_{300} = X_1+\cdots+X_{300}\). Policy claims amounts are uniformly
distributed on \((0,PolMax)\), so the mean claim amount is \(PolMax/2\)
and the variance is\(PolMax^2/12\). Thus, for policy \(i=1,...,300\), we
have \[\begin{matrix}
    \begin{array}{ccccc} \hline
        \text{Number of} & \text{Policy} & \text{Probability of} & \text{Mean} & \text{Variance}\\
        \text{Policies} &  \text{Maximum} &  \text{Claim Per Policy} & \text{Amount} & \text{Amount}\\
        & & (q_{i}) & (\mu_{i}) & (\sigma_{i}^2) \\ \hline
        100 & 400 & 0.05 & 200 & 400^2/12\\
        200 & 300 & 0.06 & 150 & 300^2/12 \\ \hline
    \end{array}
\end{matrix}\]

The mean of the aggregate claims is
\[\mathrm{E~} S_{300} = \sum_{i=1}^{300} q_i \mu_i = 100\left\{0.05(200)\right\} + 200\left\{0.06 (150) \right\} = 1,000+1,824 =
2,824\]

The variance of the aggregate claims is

\begin{eqnarray*}
    \mathrm{Var~}S_{300} &=& \sum_{i=1}^{300} \left( q_i \sigma _i^2+q_i (1-q_i
    )\mu_i^2 \right) \\
    &=& 100\left\{ 0.05 \left(\frac{400^2}{12}\right) +0.05 (1-0.05 )200^2 \right\}+
    200\left\{
    0.06 \left(\frac{300^2}{12}\right) +0.06 (1-0.06 )150^2 \right\}\\
    &=& 600,466.67 .
\end{eqnarray*}

\begin{center}\rule{0.5\linewidth}{\linethickness}\end{center}

\emph{Follow-Up.} Now suppose everybody receives the policy maximum if a
claim occurs. What is the expected aggregate loss and variance of the
aggregate loss? Each policy claim amount \(B_i\) is now fixed at
\(PolMax\) instead of random, so \(\sigma_i^2 = \mathrm{Var~} B_i = 0\)
and \(\mu_i = PolMax\). \[\begin{aligned}
\mathrm{E~}S^X &= \sum_{i=1}^{300} q_i \mu_i = 100 \left\{0.05(400) \right\} + 200 \left\{ 0.06(300) \right\} = 5,648
\end{aligned}\]\\
\[\begin{aligned}
\mathrm{Var~}S^X &= \sum_{i=1}^{300} \left( q_i \sigma _i^2+q_i (1-q_i
)\mu_i^2 \right) = \sum_{i=1}^{300} \left( q_i (1-q_i) \mu_i^2 \right) \\
&= 100 \left\{(0.05) (1-0.05) 400^2\right\} +
200 \left\{(0.06) (1-0.06)300^2\right\} \\
&= 76,000 + 101,520 = 177,520
\end{aligned}\]

\begin{center}\rule{0.5\linewidth}{\linethickness}\end{center}

The individual risk model can also be used for claim frequency. If
\(X_i\) denotes the number of claims from the \(i\)th contract, and
\(S_n\) is interpreted as the total number of claims from the portfolio.
In this case, the above two-part framework still applies. Assume \(X_i\)
belongs to the \((a,b,0)\) class with pmf denoted by \(p_{ik}\). Let
\(X_i^{T}\) denote the associated zero-truncated distribution in the
\((a,b,1)\) class with the pmf \(p_{ik}^T=p_{ik}/(1-p_{i0})\) for
\(k=1,2,\ldots\). Using the relationship between their generating
functions: \[\begin{aligned}
P_{X_i}(z) = p_{i0} +(1-p_{i0}) P_{X_i^{T}}(z),
\end{aligned}\] we can write \(X_i=I_i\times B_i\) with
\(q_i={\rm Pr}(I_i=1)={\rm Pr}(X_i>0)=1-p_{i0}\) and \(B_i=X_i^T\).

\textbf{Example 5.2.2.} An insurance company sold a portfolio of 100
independent homeowners insurance policies, each of which has claim
frequency following a zero-modified Poisson distribution, as follows:

\[\begin{matrix}
    \begin{array}{cccc} \hline
        \text{Type of} & \text{Number of}  & \text{Probability of} & \lambda \\
        \text{Policy} & \text{Policies}  &  \text{At Least 1 Claim} &  \\ \hline
        \text{Low-risk} & 40 & 0.03 & 1 \\
        \text{High-risk} & 60 & 0.05 & 2 \\ \hline
    \end{array}
\end{matrix}\] Find the expected value and variance of the claim
frequency for the entire portfolio.

Show Example Solution

\hypertarget{toggleExampleAggLoss.2.2}{}
\textbf{Solution.} For each policy, we can write the zero-modified
Poisson claim frequency \(N_i\) as \(N_i = I_i \times B_i\), where
\[q_i = \Pr(I_i = 1) = \Pr(N_i > 0) = 1-p_{i0}\] For the low-risk
policies, we have \(q_i = 0.03\) and for the high-risk policies, we have
\(q_i=0.05\). Further, \(B_i = N_i^T\), the zero-truncated version of
\(N_i\). Thus, we have \[\begin{aligned}
\mu_i &={\rm E}(B_i) = {\rm E}(N_i^T) = \frac{\lambda}{1-e^{-\lambda}} \\
\sigma_i^2 &={\rm Var}(B_i) = {\rm Var}(N_i^T) = \frac{\lambda [1-(\lambda+1)e^{-\lambda}]}{(1-e^{-\lambda})^2}
\end{aligned}\] Let the portfolio claim frequency be
\(S_n = \sum_{i=1}^n N_i\). Using the formulas above, the expected claim
frequency of the portfolio is \[\begin{aligned}
    \mathrm{E~} S_n &= \sum_{i=1}^{100} q_i \mu_i \\
    & = 40\left[0.03 \left(\frac{1}{1-e^{-1}} \right) \right] + 60 \left[0.05 \left( \frac{2}{1-e^{-2}} \right) \right] \\
    &= 40(0.03)(1.5820) + 60(0.05)(2.3130) = 8.8375
\end{aligned}\] The variance of the claim frequency of the portfolio is
\[\begin{aligned}
    \mathrm{Var~}S_n &= \sum_{i=1}^{100} \left( q_i \sigma _i^2+q_i (1-q_i
    )\mu_i^2 \right) \\
    &= 40 \left[ 0.03 \left(\frac{1-2e^{-1}}{(1-e^{-1})^2} \right) + 0.03(0.97)(1.5820^2) \right] + 60 \left[0.05 \left( \frac{2[1-3e^{-2}]}{ (1-e^{-2})^2} \right) + 0.05(0.95)(2.3130^2) \right] \\
    &= 23.7214
\end{aligned}\] Note that equivalently, we could have calculated the
mean and variance of an individual policy directly using the
relationship between the zero-modified and zero-truncated Poisson
distributions.

\begin{center}\rule{0.5\linewidth}{\linethickness}\end{center}

To understand the distribution of the aggregate loss, one could use
central limit theorem to approximate the distribution of \(S_n\). Denote
\(\mu_S={\rm E}(S)\) and \(\sigma^2_S={\rm Var}(S)\), the cdf of \(S_n\)
is: \[\begin{aligned}
 F_{S_n}(s)={\rm Pr}({S_n}\leq s) = \Phi \left(\frac{s-\mu_S}{\sigma_S}\right).
\end{aligned}\]

\textbf{Example 5.2.3. SOA Exam Question - Follow-Up.} As in the
original example earlier, an insurance company sold 300 fire insurance
policies, with claim amounts uniformly distributed between 0 and the
policy maximum. Using the normal approximation, calculate the
probability that the aggregate claim amount exceeds \(\$3,500\).

Show Example Solution

\hypertarget{toggleExampleAggLoss.2.3}{}
\textbf{Solution.} We have seen earlier that
\(\mathrm{E~} S_{300}=2,824\) and \(\mathrm{Var~}S_{300} = 600,466.67\).
Then \[\begin{aligned}
{\rm Pr}(S_{300} > 3,500) &= 1 - {\rm Pr}(S_{300} \leq 3,500) \\
&= 1- \Phi \left( \frac{3,500-2,824}{\sqrt{600,466.67}} \right) = 1 - \Phi \left( 0.87237 \right) \\
&= 1 - 0.8085 = 0.1915
\end{aligned}\]

\begin{center}\rule{0.5\linewidth}{\linethickness}\end{center}

For small \(n\), the distribution of \(S_n\) is likely skewed, and the
normal approximation would be a poor choice. To examine the aggregate
loss distribution, we go back to the basics and first principles.
Specifically, the distribution can be derived recursively. Define
\(S_k=X_1 + \cdots + X_k, k=1,\ldots,n\), we have: For \(k=1\):
\[F_{S_1}(s) = {\rm Pr}(S_1\leq s) = {\rm Pr}(X_1\leq s) = F_{X_1}(s).\]
For \(k=2,\ldots,n\), \[\begin{aligned}
    F_{S_k}(s)&={\Pr}(X_1+\cdots+X_k\leq s) ={\Pr}(S_{k-1}+X_k\leq s) \\
    &={\rm E}_{X_k}\left[{\rm Pr}(S_{k-1}\leq s-X_k|X_k)\right]= {\rm E}_{X_k}\left[F_{S_{k-1}}(s-X_k)\right].
   \end{aligned}\]

There are some simple cases where the \(S_n\) has a closed form.
Examples include

\[\begin{matrix}
\text{Table of Closed Form Partial Sum Distributions}\\
    \begin{array}{c|c} \hline
        \text{Distribution of} & \text{Distribution of}   \\
        X_i   &  S_n  \\ \hline
N(\mu_i,\sigma_i^2) & N(\sum_{i=1}^{n}\mu_i,\sum_{i=1}^{n}\sigma_i^2) \\
Exponential(\theta)& Gamma(n,\theta)\\
Gamma(\alpha_i,\theta)& Gamma(\sum_{i=1}^n\alpha_i,\theta)\\
Poisson(\lambda_i)& Poisson(\sum_{i=1}^{n}\lambda_i)\\
Bin(q,m_i)& Bin(q,\sum_{i=1}^n m_i)\\
 Geometric(\beta)& NegBin(\beta,n)\\
NegBin(\beta,r_i)& NegBin(\beta,\sum_{i=1}^n r_i)\\
\hline
    \end{array}
\end{matrix}\]

A special case is when \(X_i's\) are identically distributed. Let
\(F_X(x)={\Pr}(X\leq x)\) be the common distribution of \(X_i\)
\((i=1,\ldots,n)\), we define
\[F^{*n}_X(x)={\Pr}(X_1+\cdots+X_n\leq x)\] the \(n\)-fold convolution
of \(F_X\).

\textbf{Example 5.2.4. Gamma Distribution.} For an easy case, assume
that \(X_i \sim\) gamma with parameters \((\alpha, \theta)\). As we
know, the moment generating function (\emph{mgf}) is
\(M_{X}(t) = (1 - \theta t)^{- \alpha}\). Thus, the \emph{mgf} of the
sum \(S_n = X_1 + \cdots + X_n\) is

\begin{eqnarray*}
M_{S_n}(t) = \mathrm{E~} \exp(t(X_1 + \cdots + X_n)) = (1 - \theta t)^{-n \alpha} ,
\end{eqnarray*}

Thus, \(S_n\) has a gamma distribution with parameters
\((n \alpha, \theta)\). This makes it easy to compute
\(F^{\ast n}(x) = \Pr(S_n \le x).\) This property is known as ``closed
under convolution''.

\begin{center}\rule{0.5\linewidth}{\linethickness}\end{center}

\textbf{Example 5.2.5. Negative Binomial Distribution.} Assume
\(X_i \sim NegBin(\beta, r_i)\). The probability generating function
(\emph{pgf}) is \(P_X(z) = \left[1-\beta(z-1) \right]^{-r}\). Thus, the
\emph{pgf} of the sum \(S_n =X_1+\cdots+X_n\) is

\[\begin{aligned}
P_{S_n}(z) &= \mathrm{E~}\left[ z^{S_n} \right] = \mathrm{E~}\left[ z^{X_1+\cdots+X_n} \right] = \mathrm{E~}\left[ z^{X_1} z^{X_2} \cdots z^{X_n} \right] = \mathrm{E~}\left[z^{X_1}\right] \cdots \mathrm{E~}\left[z^{X_n}\right] \\
&= \prod_{i=1}^n P_{X_i}(z) = \prod_{i=1}^n \left[1-\beta(z-1) \right]^{-r_i} = \left[1-\beta(z-1) \right]^{-\sum_{i=1}^n r_i}
\end{aligned}\]

Thus, \(S_n\) has a negative binomial distribution with parameters
\((\beta, \sum_{i=1}^n r_i)\).

\begin{center}\rule{0.5\linewidth}{\linethickness}\end{center}

More generally, we can compute \(F^{\ast n}\) recursively. Begin the
recursion at \(n=1\) using \(F^{\ast 1} \left(x \right) =F(x)\). Next,
for \(n=2\), we have

\begin{eqnarray*}
F^{\ast 2} \left(x \right) &=& \Pr(X_1 + X_2 \le x) = \mathrm{E}_{X_2} \Pr(X_1 \le x - X_2|X_2)\\
&=& \mathrm{E}_X F(x - X)\\
&=&\left\{\begin{array}{ll}
\int_{0}^{x} F(x-y) f(y) dy & \text{continuous}\\
\sum_{y \le x} F(x-y) f(y) & \text{discrete}\\
\end{array}\right.
\end{eqnarray*}

Recall \(F(0) = 0\).

Similarly, let \(S_n = X_1 + X_2 + \cdots + X_n\)

\begin{eqnarray*}
F^{\ast n}\left(x\right) &=& \Pr(S_n \le x) = \Pr(S_{n-1} + X_n \le x)\\
&=&\mathrm{E}_{X_n}\Pr(S_{n-1} \le x - X_n|X_n)\\
&=&\mathrm{E}_X F^{\ast(n-1)}(x - X)\\
&=&
\left\{\begin{array}{ll}
\int_{0}^{x} F^{\ast(n-1)}(x-y)f(y)dy & \text{continuous}\\
\sum_{y \le x} F^{\ast(n-1)}(x-y)f(y) & \text{discrete}\\
\end{array}\right.
\end{eqnarray*}

\textbf{Example 5.2.6. SOA Exam Question (modified).} The annual number
of doctor visits for each individual in a family of 4 has geometric
distribution with mean 1.5. The annual numbers of visits for the family
members are mutually independent. An insurance pays 100 per doctor visit
beginning with the 4th visit per family. Calculate the probability that
the family will receive an insurance payment this year.

Show Example Solution

\hypertarget{toggleExampleAggLoss.2.6}{}
\textbf{Solution.} Let \(X_i \sim Geometric(\beta=1.5)\) be the number
of doctor visits for one individual in the family and
\(S_4 = X_1 + X_2 + X_3 + X_4\) be the number of doctor visits for the
family. The sum of 4 independent geometric distributions each with mean
1.5 follows a negative binomial distribution, i.e.
\(S_4 \sim NegBin(\beta=1.5, r=4)\).

If the insurance pays 100 per visit beginning with the 4th visit for the
family, then the family will not receive an insurance payment if they
have less than 4 claims. This probability is \[\begin{aligned}
    \Pr(S_4 < 4) &= \Pr(S_4 = 0) + \Pr(S_4 = 1) + \Pr(S_4 = 2) +\Pr(S_4 = 3) \\
    &= (1+1.5)^{-4} + \frac{4(1.5)}{(1+1.5)^5} + \frac{4(5)(1.5^2)}{2(1+1.5)^6} + \frac{4(5)(6)(1.5^3)}{3!(1+1.5)^7}\\
    &= 0.0256 + 0.0614 + 0.0922 + 0.1106 = 0.2898
\end{aligned}\]

\begin{center}\rule{0.5\linewidth}{\linethickness}\end{center}

\section{Collective Risk Model}\label{collective-risk-model}

\subsection{Moments and Distribution}\label{moments-and-distribution}

Under the collective risk model \(S=X_1+\cdots+X_N\), \(\{X_i\}\) are
\emph{iid}, and independent of \(N\). Let
\(\mu = {\rm E}\left( X_{i}\right)\) and
\(\sigma ^{2} = {\rm Var}\left(X_{i}\right)\) \(\forall\) \(i\). Using
the law of iterated expectations, the mean is

\begin{eqnarray*}
{\rm E}(S)={\rm E}_N[{\rm E}_S(S|N)] = {\rm E}_N[N\mu] = \mu {\rm E}(N).
\end{eqnarray*}

Using the law of total variation, the variance is \[\begin{aligned}
{\rm Var}(S) &= {\rm E}_N[{\rm Var}_S(S|N)] + {\rm Var}_N[{\rm E}_S(S|N)] \\
&={\rm E}_N[N\sigma^2] + {\rm Var}_N[N\mu] \\
&=\sigma^2{\rm E}[N] + \mu^2{\rm Var}[N]
\end{aligned}\]

\textbf{Special Case: Poisson Distributed Frequency.} If
\(N \sim Poisson (\lambda)\), then

\begin{eqnarray*}
\mathrm{E~}N &=& \mathrm{Var~}N = \lambda\\
\mathrm{Var~}S &=& \lambda (\sigma^2 + \mu^2) = \lambda ~\mathrm{E~} X^2 .
\end{eqnarray*}

\begin{center}\rule{0.5\linewidth}{\linethickness}\end{center}

\textbf{Example 5.3.1. SOA Exam Question.} The number of accidents
follows a Poisson distribution with mean 12. Each accident generates 1,
2, or 3 claimants with probabilities 1/2, 1/3, and 1/6 respectively.

Calculate the variance in the total number of claimants.

Show Example Solution

\hypertarget{toggleExampleAggLoss.3.1}{}
\textbf{Solution.}
\[\mathrm{E~}X^2 = 1^2 \left( \frac{1}{2}\right) + 2^2\left(\frac{1}{3} \right) + 3^2\left(\frac{1}{6}\right)
= \frac{10}{3}\]
\[\mathrm{Var~}S = \lambda \ \mathrm{E~}X^2 = 12\left(\frac{10}{3}\right) = 40\]

Alternatively, using the general approach,
\(\mathrm{Var~}S = \sigma^2 \mathrm{E~}N + \mu^2 \mathrm{Var~}N\), where

\[\mathrm{E~}N = \mathrm{Var~}N = 12\]

\[\mu = \mathrm{E~}X = 1\left(\frac{1}{2}\right) + 2\left(\frac{1}{3}\right) + 3\left(\frac{1}{6}\right)
= \frac{5}{3}\]

\[\sigma^2 = \mathrm{E~}X^2 - (\mathrm{E~}X)^2 = \frac{10}{3} - \frac{25}{9}
= \frac{5}{9}\]

\[\Rightarrow \ \mathrm{Var~}S = \left(\frac{5}{9}\right)\left(12\right) + \left(\frac{5}{3}\right)^2\left(12\right) = 40 .\]

\begin{center}\rule{0.5\linewidth}{\linethickness}\end{center}

In general, the moments of \(S\) can be derived from its moment
generating function (\emph{mgf}). Because \(\{X_i\}\) are \emph{iid}, we
denote the \emph{mgf} of \(X\) as \(M_{X}(t) = \mathrm{E~}(e^{tX})\).
Using the law of iterated expectations, the \emph{mgf} of \(S\) is

\begin{eqnarray*}
M_{S}(t) &=& \mathrm{E}(e^{St})=\mathrm{E}[\mathrm{E}(e^{St}|N)]\\
&=& \mathrm{E~}[(M_{X}(t))^N]
\end{eqnarray*}

where we use the relation
\(\mathrm{E}[e^{t(X_1+\cdots+X_n)}] = \mathrm{E}(e^{tX_1})\cdots\mathrm{E}(e^{tX_n}) = (M_{X}(t))^n\).
Now, recall that the probability generating function (\emph{pgf}) of
\(N\) is \(P(z) = \mathrm{E}(z^N)\). Denote \(M_{X}(t)=z\), it is shown

\begin{eqnarray*}
M_{S}(t) = \mathrm{E~}(z^N)  = P_{N}(z) = P_{N}[M_{X}(t)].
\end{eqnarray*}

Similarly, if \(S\) a discrete, one can show the \emph{pgf} of \(S\) is:

\begin{eqnarray*}
P_{S}(z) = P_{N}[P_{X}(z)].
\end{eqnarray*}

To get \(\mathrm{E~}S = M_{S}'(0)\), we use the chain rule \[
M_{S}'(t) = \frac{\partial}{\partial t} P_{N}(M_{X}(t)) = P_{N}'(M_{X}(t)) M_{X}'(t)\\
\] and recall
\(M_{X}(0) = 1, M_{X}'(0) = \mathrm{E~}X = \mu, P_{N}'(1) = \mathrm{E~}N\).
So,

\begin{eqnarray*}
\mathrm{E~}S = M_{S}'(0) = P_{N}'(M_{X}'(0)) M_{X}'(0) = \mu {\rm E}(X)
\end{eqnarray*}

Similarly, one could use relation \[
\mathrm{E~}S^2 = M_{S}''(0)
\] to get
\[\mathrm{Var~}S = \sigma^2 \mathrm{E~}N + \mu^2 \mathrm{Var~}N.\]

\textbf{Special Case. Poisson Frequency.} Let
\(N \sim Poisson (\lambda)\). Thus, the \emph{pgf} of \(N\) is
\(P_N (z) =\exp[\lambda(z-1)]\), and the \emph{mgf} of \(S\) is

\begin{eqnarray*}
M_{S}(t) &=&\exp[\lambda(M_{X}(t) - 1)].
\end{eqnarray*}

Taking derivatives yield

\begin{eqnarray*}
M_{S}(t) &=&\exp(\lambda(M_{X}(t) - 1))\\
M_{S}'(t) &=&\exp(\lambda(M_{X}(t) - 1)) \lambda M_{X}'(t)\\
&=& M_{S}(t) \lambda M_{X}'(t)\\
M_{S}''(t) &=& M_{S}(t) \lambda M_{X}''(t) + \{M_{S}(t) \lambda M_{X}'(t)\} \lambda M_{X}'(t)
\end{eqnarray*}

Evaluating these at \(t=0\) yields

\begin{eqnarray*}
M_{S}''(0) &=& \lambda \mathrm{E}(X^2) + \lambda^2 \mu^2\\
\mathrm{Var~}S &=& \lambda \mathrm{E}(X^2) + \lambda^2 \mu^2 - (\lambda \mu)^2\\
&=& \lambda \mathrm{E}(X^2).
\end{eqnarray*}

\begin{center}\rule{0.5\linewidth}{\linethickness}\end{center}

\textbf{Example 5.3.2. SOA Exam Question.} You are the producer of a
television quiz show that gives cash prizes. The number of prizes,
\(N\), and prize amount, \(X\), have the following distributions:

\[\begin{matrix}
\begin{array}{ccccc}\hline
    n & \Pr(N=n) & & x & \Pr(X=x)\\ \hline
    1 & 0.8 & & 0 & 0.2 \\
    2 & 0.2 & & 100 & 0.7 \\
       &       & & 1000 & 0.1\\\hline
  \end{array}
\end{matrix}\]

Your budget for prizes equals the expected aggregate cash prizes plus
the standard deviation of aggregate cash prizes. Calculate your budget.

Show Example Solution

\hypertarget{toggleExampleAggLoss.3.2}{}
\textbf{Solution.} We need to calculate the mean and standard deviation
of the aggregate (sum) of cash prizes. The moments of the frequency
distribution \(N\) are

\begin{eqnarray*}
\mathrm{E~}N &=& 1 (0.8) + 2 (0.2) =1.2\\
\mathrm{E~}N^2 &=&  1^2 (0.8) + 2^2 (0.2) =1.6\\
\mathrm{Var~}N &=& \mathrm{E~}N^2 - \left( \mathrm{E~}N \right)^2= 0.16
\end{eqnarray*}

The moments of the severity distribution \(X\) are

\begin{eqnarray*}
\mathrm{E~}X &=& 0 (0.2) + 100 (0.7) + 1000 (0.1) = 170 = \mu\\
\mathrm{E~}X^2 &=& 0^2 (0.2) + 100^2 (0.7) + 1000^2 (0.1) = 107,000\\
\mathrm{Var~}X &=& \mathrm{E~}X^2 - \left( \mathrm{E~}X \right)^2=78,100 = \sigma^2
\end{eqnarray*}

Thus, the mean and variance of the aggregate cash prize are

\begin{eqnarray*}
\mathrm{E~}S  &=& \mu \mathrm{E~}N = 170 (1.2) = 204 \\
\mathrm{Var~}S &=& \sigma^2 \mathrm{E~}N + \mu^2 \mathrm{Var~}N\\
&=& 78,100 (1.2) + 170^2 (0.16) = 98,344
\end{eqnarray*}

This gives the following required budget

\begin{eqnarray*}
Budget &=& \mathrm{E~}S + \sqrt{\mathrm{Var~}S} \\
&=& 204 + \sqrt{98,344} = 517.60 .
\end{eqnarray*}

\begin{center}\rule{0.5\linewidth}{\linethickness}\end{center}

The distribution of \(S\) is called a compound distribution, and it can
be derived based on the convolution of \(F_X\) as follows:

\begin{eqnarray*}
F_{S}(s) &=& \Pr \left(X_1 + \cdots + X_N \le s \right) \\
&=&  \mathrm{E} \left[ \Pr \left(X_1 + \cdots + X_N  \le s|N=n \right) \right]\\
&=&  \mathrm{E} \left[ F_{X}^{\ast N}(s) \right] \\
&=&  p_0 + \sum_{n=1}^{\infty }p_n F_{X}^{\ast n}(s)
\end{eqnarray*}

\textbf{Example 5.3.3. SOA Exam Question.} The number of claims in a
period has a geometric distribution with mean \(4\). The amount of each
claim \(X\) follows \(\Pr(X=x) = 0.25, \ x=1,2,3,4.\) The number of
claims and the claim amounts are independent. Let \(S\) denote the
aggregate claim amount in the period. Calculate \(F_{S}(3)\).

Show Example Solution

\hypertarget{toggleExampleAggLoss.3.3}{}
\textbf{Solution.} By definition, we have \[\begin{aligned}
F_{S}\left(3 \right) &= {\rm Pr}\left(\sum_{i=1}^N X_i \leq 3\right) = \sum_{n=0}^\infty {\rm Pr}\left(\sum_{i=1}^n X_i\leq 3|N=n\right){\rm Pr}(N=n) \\
&= \sum_n F^{\ast n} \left(3 \right) p_n = \sum_{n=0}^3 F^{\ast n}(3) p_n \\
&= p_0 + F^{\ast 1}(3) \ p_1 + F^{\ast 2}(3) \ p_2 + F^{\ast 3}(3) \ p_3
\end{aligned}\] Because \(N\) has a geometric distribution with mean 4,
we know that \[\begin{aligned}
p_n &= \frac{1}{1+\beta}
\left(\frac{\beta}{1+ \beta} \right)^n = \frac{1}{5} \left(\frac{4}{5} \right)^n
\end{aligned}\] For the claim severity distribution, recursively, we
have \[\begin{aligned}
F^{\ast 1}(3) &= \Pr(X \le 3) = \frac{3}{4} \\
F^{\ast 2}(3) &= \sum_{y \le 3} F^{\ast 1} (3-y) f(y) = F^{\ast 1}(2)f(1) + F^{\ast 1}(1)f(2) \\
&= \frac{1}{4}\left[F^{\ast 1} (2) + F^{\ast 1}(1)\right] = \frac{1}{4}\left[{\rm Pr}(X\leq 2) + {\rm Pr}(X \leq 1) \right] \\
&= \frac{1}{4} \left(\frac{2}{4} + \frac{1}{4} \right) = \frac{3}{16}\\
F^{\ast 3}(3) &= \Pr(X_1+X_2 + X_3 \le 3) = \Pr(X_1=X_2=X_3=1) = \left(\frac{1}{4} \right)^3
\end{aligned}\] Notice that we did not need to recursively calculate
\(F^{\ast 3}(3)\) by recognizing that each \(X \in \{1,2,3,4\}\), so the
only way of obtaining \(X_1+X_2+X_3 \leq 3\) is to have
\(X_1=X_2=X_3=1\). Additionally, for \(n \geq 4\), \(F^{\ast n} (3)=0\)
since it is impossible for the sum of 4 or more \(X\)'s to be less than
3. For \(n=0\), \(F^{\ast 0}(3) = 1\) since the sum of 0 \(X\)'s is 0,
which is always less than 3. Laying out the probabilities
systematically,

\[\begin {matrix}
\begin{array}{c c c c}\hline
    x & F^{\ast 1}(x) & F^{\ast 2}(x) & F^{\ast 3}(x)\\ \hline
    0 & & & \\
    1 & \frac{1}{4} & 0 & \\
    2 & \frac{2}{4} & \left( \frac{1}{4} \right)^2 & \\
    3 & \frac{3}{4} & \frac{3}{16} & \left( \frac{1}{4} \right)^3 \\ \hline
\end{array}
\end{matrix}\]

Finally, \[\begin{aligned}
F_{S}(3) &= p_0 + F^{\ast 1}(3) \ p_1 + F^{\ast 2}(3) \ p_2 + F^{\ast 3}(3) \ p_3 \\
&= \frac{1}{5} + \frac{3}{4}\left(\frac{4}{25} \right) + \frac{3}{16} \left( \frac{16}{125} \right) + \frac{1}{64} \left( \frac{64}{625}\right) = 0.3456\\
\end{aligned}\]

\begin{center}\rule{0.5\linewidth}{\linethickness}\end{center}

When \(\mathrm{E}(N)\), one may also use the central limit theorem to
approximate the distribution of \(S\) as in the individual risk model.
That is, \(\frac{S - \mathrm{E}(S)}{\sqrt{\mathrm{Var}(S)}}\)
approximately follows \(N(0,1)\).

\textbf{Example 5.3.4. SOA Exam Question..} You are given:

\[\begin{matrix}
  \begin{array}{ c | c  c }
    \hline
      & \text{Mean} & \text{Standard Deviation}\\ \hline
    \text{Number of Claims} & 8 & 3\\
    \text{Individual Losses} & 10,000 & 3,937\\
    \hline
  \end{array}
\end{matrix}\] Using the normal approximation, determine the probability
that the aggregate loss will exceed 150\(\%\) of the expected loss.

Show Example Solution

\hypertarget{toggleExampleAggLoss.3.4}{}
\textbf{Solution.} To use the normal approximation, we must first find
the mean and variance of the aggregate loss \(S\)

\begin{eqnarray*}
\mathrm{E~}S &=& \mu \ \mathrm{E~}N = 10,000(8) = 80,000\\
\mathrm{Var~}S &=& \sigma^2 \ \mathrm{E~}N + \mu^2 \ \mathrm{Var~}N\\
&=& 3937^2(8) + 10000^2 (3^2) = 1,023,999,752\\
\sqrt{\mathrm{Var~}S} &=& 31,999.996 \approx 32,000
\end{eqnarray*}

Then under the normal approximation, aggregate loss \(S\) is
approximately normal with mean 80,000 and standard deviation 32,000. The
probability that \(S\) will exceed 150\(\%\) of the expected aggregate
loss is therefore \[\begin{aligned}
\Pr(S>1.5 \mathrm{E~}S) &= \Pr \left( \frac{S - \mathrm{E~} S}{\sqrt{\mathrm{Var~}S}} > \frac{1.5 \mathrm{E~}S - \mathrm{E~} S}{\sqrt{\mathrm{Var~}S}} \right) \\
&= \Pr \left( N(0,1) > \frac{0.5 \mathrm{E~}S}{\sqrt{\mathrm{Var~}S} } \right) \\
&= \Pr \left( N(0,1) > \frac{0.5(80,000)}{32,000} \right) = \Pr( N(0,1) > 1.25) \\
&= 1-\Phi(1.25) = 0.1056
\end{aligned}\]

\begin{center}\rule{0.5\linewidth}{\linethickness}\end{center}

\textbf{Example 5.3.5. SOA Exam Question.} For an individual over
\(65\):\\
(i) The number of pharmacy claims is a Poisson random variable with mean
\(25\).\\
(ii) The amount of each pharmacy claim is uniformly distributed between
\(5\) and \(95\).\\
(iii) The amounts of the claims and the number of claims are mutually
independent.\\
Determine the probability that aggregate claims for this individual will
exceed \(2000\) using the normal approximation.

Show Example Solution

\hypertarget{toggleExampleAggLoss.3.5}{}
\textbf{Solution.} We have claim frequency
\(N \sim Poisson (\lambda = 25)\) and claim severity
\(X \sim U \left(5, 95 \right)\). To use the normal approximation, we
need to find the mean and variance of the aggregate claims \(S\). Note

\[\begin{matrix}
\begin{array}{lll}
\mathrm{E~} N = 25 & & \mathrm{Var~} N = 25\\
\mathrm{E~}X = \frac{5+95}{2} = 50 = \mu & & \mathrm{Var~}X = \frac{(95-5)^2}{12} = 675 = \sigma^2\\
\end{array}
\end{matrix}\] Then for \(S\),

\begin{eqnarray*}
\mathrm{E~}S &=& \mu \ \mathrm{E~} N = 50(25) = 1,250\\
\mathrm{Var~}S &=& \sigma^2 \ \mathrm{E~}N + \mu^2 \ \mathrm{Var~}N\\
&=& 675 (25) + 50^2 (25) = 79,375
\end{eqnarray*}

Using the normal approximation, \(S\) is approximately normal with mean
1,250 and variance 79,375. The probability that \(S\) exceeds 2,000 is
\[\begin{aligned}
\Pr(S>2,000) &= \Pr \left(\frac{S - \mathrm{E~} S}{\sqrt{\mathrm{Var~} S}} > \frac{2,000- \mathrm{E~} S}{\sqrt{\mathrm{Var~} S}} \right) \\
&= \Pr\left( N(0,1) > \frac{2,000-1,250}{\sqrt{79,375}} \right) \\
&= \Pr (N(0,1) > 2.662) = 1-\Phi(2.662) = 0.003884
\end{aligned}\]

\begin{center}\rule{0.5\linewidth}{\linethickness}\end{center}

\subsection{Stop-loss Insurance}\label{stop-loss-insurance}

Insurance on the aggregate loss \(S\), subjected to a deductible \(d\),
is called \textit{net stop-loss insurance}. The quantity

\begin{eqnarray*}
\mathrm{E}[(S-d)_+]
\end{eqnarray*}

is known as the \emph{net stop-loss premium}.

To calculate the net stop-loss premium, we have

\begin{eqnarray*}
\mathrm{E}(S-d)_+  & =& \int_{d}^{\infty} \left(1-F_S(s) \right) ds\\
&=&
\left\{\begin{array}{ll}
\int_{d}^{\infty}(s-d) f_{S}(s) ds& \text{continuous}\\
 \sum_{s>d}(s-d) f_{S}(s) ds & \text{discrete}\\
 \end{array}\right.\\
&=& \mathrm{E}(S) - \mathrm{E}(S\wedge d)\\
\end{eqnarray*}

\textbf{Example 5.3.6. SOA Exam Question.} In a given week, the number
of projects that require you to work overtime has a geometric
distribution with \(\beta=2\). For each project, the distribution of the
number of overtime hours in the week is as follows:

\[\begin{matrix}
\begin{array}{ccc} \hline
    x &  & f(x)\\ \hline
    5 &  & 0.2 \\
    10 & & 0.3 \\
    20 & & 0.5\\ \hline
  \end{array}
\end{matrix}\]

The number of projects and the number of overtime hours are independent.
You will get paid for overtime hours in excess of 15 hours in the week.
Calculate the expected number of overtime hours for which you will get
paid in the week.

Show Example Solution

\hypertarget{toggleExampleAggLoss.3.6}{}
\textbf{Solution.} The number of projects in a week requiring overtime
work has distribution \(N \sim Geometric(\beta=2)\), while the number of
overtime hours worked per project has distribution \(X\) as described
above. The aggregate number of overtime hours in a week is \(S\) and we
are therefore looking for
\[\mathrm{E~}(S-15)_+ = \mathrm{E~}S - \mathrm{E~}(S \wedge 15).\]

To find \(\mathrm{E~}S = \mathrm{E~}X \ \mathrm{E~}N\), we have

\begin{eqnarray*}
\mathrm{E~}X &=& 5(0.2) + 10(0.3)+ 20(0.5)= 14 \\
\mathrm{E~}N &=& 2 \\
\Rightarrow \ \mathrm{E~}S &=& \mathrm{E~}X \ \mathrm{E~}N = 14(2) = 28
\end{eqnarray*}

To find
\(\mathrm{E~} (S \wedge 15) = 0 \Pr (S=0) + 5 \Pr(S=5) + 10 \Pr(S=10) + 15 \Pr(S \geq 15)\),
we have

\begin{eqnarray*}
\Pr(S=0) &=& \Pr(N=0) = \frac{1}{1+\beta} = \frac{1}{3} \\
\Pr(S=5) &=& \Pr(X=5, \ N=1) = 0.2 \left(\frac{2}{9} \right)= \frac{0.4}{9}\\
\Pr(S=10) &=& \Pr(X=10, \ N=1) + \Pr(X_1=X_2=5, N=2) \\
&=& 0.3 \left(\frac{2}{9} \right) + (0.2)(0.2) \left( \frac{4}{27} \right)= 0.0726 \\
\Pr(S \geq 15) &=& 1 - \left(\frac{1}{3} + \frac{0.4}{9} + 0.0726 \right) = 0.5496\\
\Rightarrow \mathrm{E~}(S \wedge 15) &=& 0 \Pr (S=0) + 5 \Pr(S=5) + 10 \Pr(S=10) + 15 \Pr(S \geq 15) \\
&=& 0 \left( \frac{1}{3} \right) + 5
\left( \frac{0.4}{9} \right) + 10 (0.0726) + 15 (0.5496) = 9.193\\
\end{eqnarray*}

Therefore,

\begin{eqnarray*}
\mathrm{E~}(S-15)_+ &=& \mathrm{E~}S - \mathrm{E~}(S \wedge 15) \\
&=& 28 - 9.193 = 18.807
\end{eqnarray*}

\begin{center}\rule{0.5\linewidth}{\linethickness}\end{center}

\textbf{Recursive Net Stop-Loss Premium Calculation}. For the discrete
case, this can be computed recursively as

\begin{eqnarray*}
\mathrm{E~}\left[ S-(j+1)h \right] _{+}=\mathrm{E~}\left[ ( S-jh )_{+} \right] -h \left( 1-F_S(jh)
\right) .
\end{eqnarray*}

This assumes that the support of \(S\) is equally spaced over units of
\(h\).

To establish this, we assume that \(h=1\). Now, on the left-hand side,
we have
\(\mathrm{E~}\left[ S-(j+1) \right] _{+}=\mathrm{E~}S - \mathrm{E~}S\wedge (j+1)\).
We can write

\begin{eqnarray*}
\mathrm{E~}S\wedge (j+1) = \sum_{x=0}^{j}xf_S(x) + (j+1)\Pr(S \ge j+1).
\end{eqnarray*}

Similarly

\begin{eqnarray*}
\mathrm{E~}S\wedge j = \sum_{x=0}^{j}xf_S(x) + j\Pr(S\ge j+1).
\end{eqnarray*}

With these, expressions, we have

\begin{eqnarray*}
\mathrm{E~}\left[ S-(j+1) \right] _{+} &-& \mathrm{E~}\left[ ( S-j )_{+} \right]  \\
&=&\left\{\mathrm{E~}S - \mathrm{E~}S\wedge (j+1) \right\}
-\left\{\mathrm{E~}S - \mathrm{E~}S\wedge j \right\} \\
&=&\left\{ \sum_{x=0}^{j}xf_S(x) + j\Pr(S\ge j+1) \right\}
- \left\{ \sum_{x=0}^{j}xf_S(x) + (j+1)\Pr(S \ge j+1) \right\} \\
&=& -\Pr(S\ge j+1) = -\{1 - F_{S}(j)\},
\end{eqnarray*}

as required.

\begin{center}\rule{0.5\linewidth}{\linethickness}\end{center}

\textbf{Exercise 5.3.7. Exam M, Fall 2005, 19 - Continued.} Recall that
the goal of this question was to calculate \(\mathrm{E~}(S-15)_+\). Note
that the support of \(S\) is equally spaced over units of 5, so this
question can also be done recursively, using steps of \(h=5\):

\begin{itemize}
\item
  Step 1:\\
  \[\begin{aligned}
  \mathrm{E~}(S-5)_+ &= \mathrm{E~}S - 5 [1-\Pr(S \leq 0) ]\\ 
  &= 28 - 5 \left(1 - \frac{1}{3}\right) = \frac{74}{3}=24.6667
  \end{aligned}\]
\item
  Step 2:\\
  \[\begin{aligned}
  \mathrm{E~}(S-10)_+ &= \mathrm{E~}(S-5)_+ - 5 [1-\Pr(S \leq 5)]\\ 
  &= \frac{74}{3} - 5\left( 1 - \frac{1}{3} - \frac{0.4}{9}\right) = 21.555
  \end{aligned}\]
\item
  Step 3: \[\begin{aligned}
  \mathrm{E~}(S-15)_+ &= \mathrm{E~}(S-10)_+ - 5 [1-\Pr(S \leq 10)] \\ 
  &= \mathrm{E~}(S-10)_+ - 5\Pr (S\ge 15) \\
  &= 21.555 - 5 (0.5496) = 18.887
  \end{aligned}\]
\end{itemize}

\begin{center}\rule{0.5\linewidth}{\linethickness}\end{center}

\subsection{Analytic Results}\label{analytic-results}

There are few combinations of claim frequency and severity distributions
that result in an easy-to-compute distribution for aggregate losses.
This section gives some simple examples. Analysts view these examples as
too simple to be used in practice.

\textbf{Example 5.3.8.} One has a closed-form expression for the
aggregate loss distribution by assuming a geometric frequency
distribution and an exponential severity distribution.

Assume that claim count \(N\) is geometric with parameter \(\beta\) such
that \(\mathrm{E}(N)=\beta\), and that claim amount \(X\) is exponential
with parameter \(\theta\) such that \(\mathrm{E}(X)=\theta\). Recall
that the \emph{pgf} of \(N\) and the \emph{mgf} of \(X\) are:
\[\begin{aligned}
P_N (z) &=\frac{1}{1- \beta (z-1)},\\
M_{X}(t) &=\frac{1}{1-\theta t}.
\end{aligned}\] Thus, the \emph{mgf} of aggregate loss \(S\) is

\begin{eqnarray}
M_{S}(t) &=& P_N [M_{X}(t)] = \frac{1}{1 - \beta \left( \frac{1}{1-\theta t} + 1\right)} \nonumber\\
&=& 1+ \frac{1}{1+\beta} ([1-\theta(1+\beta)z]^{-1}-1)...(1)\\
&=& \frac{1}{1+\beta}(1) +\frac{\beta}{1+\beta}
\left( \frac{1}{1-\theta (1+\beta)t}\right)...(2)
\end{eqnarray}

From (1), we note that \(S\) is equivalent to the compound distribution
of \(S=X^{*}_1+\cdots+X^{*}_{N^{*}}\), where \(N^{*}\) is a Bernoulli
with mean \(\beta/(1+\beta)\) and \(x^{*}\) is an exponential with mean
\(\theta(1+\beta)\). To see this, we examine the \emph{mgf} of \(S\):
\[\begin{aligned}
M_{S}(t) = P_N [M_{X}(t)] = P_{N^{*}} [M_{X^{*}}(t)],
\end{aligned}\] where \[\begin{aligned}
P_{N^*} (z) &=1+ \color{blue}{\frac{\beta}{1+ \beta}} (z-1),\\
M_{X^*} (t) &=\frac{1}{1- {\color{blue}{\theta(1+\beta)}} t}.
\end{aligned}\]

From (2), we note that \(S\) is also equivalent to a 2-point mixture of
0 and \(X^{*}\). Specifically,

\begin{eqnarray*}
S &=&
\left\{
\begin{array}{cl}
0 & {\rm with~ probability ~Pr}(N^*=0) = 1/(1+\beta) \\
Y^{*} & {\rm with~ probability ~Pr}(N^*=1) = \beta/(1+\beta)
\end{array}
\right..
\end{eqnarray*}

The distribution function of \(S\) is:

\begin{eqnarray*}
\Pr(S=0) &=& \frac{1}{1+\beta}\\
\Pr(S>s) &=& \Pr(X^*>s) =\frac{\beta}{1+\beta} \exp\left( -\frac{s}{
\theta (1+\beta)}\right)
\end{eqnarray*}

with \emph{pdf}

\begin{eqnarray*}
f_{S}(s) = \frac{\beta}{\theta (1+\beta)^2}\exp\left( -\frac{s}{
\theta (1+\beta)}\right).
\end{eqnarray*}

\begin{center}\rule{0.5\linewidth}{\linethickness}\end{center}

\textbf{Example 5.3.9.} Consider a collective risk model with an
exponential severity and an arbitrary frequency distribution. Recall
that if If \(X_i\sim Exponential(\theta)\), then the sum of \emph{iid}
exponential, \(S_n=X_1+\cdots+X_n\), has a Gamma distribution, i.e.
\(S_n\sim Gamma(n,\theta)\). This has cdf:

\begin{eqnarray*}
F_{X}^{\ast n}(s) &=& \Pr (S_n \le s) = \int_{0}^{s} \frac{1}{\Gamma(n)\theta^n}s^{n-1}\exp\left(-\frac{s}{\theta}\right) ds\\
&=& 1-\sum_{j=0}^{n-1}\frac{1}{j!}\left( \frac{s}{\theta}\right)^j e^{-s/\theta } .
\end{eqnarray*}

The last equality is derived by integration by parts.

For the aggregate loss distribution, we can interchange order of
summations to get

\begin{eqnarray*}
F_{S}\left(s\right) &=& p_{0}+\sum_{n=1}^{\infty }p_n F_{X}^{\ast n}\left(s\right)\\
&=& 1 - \sum_{n=1}^{\infty }p_n \sum_{j=0}^{n-1}\frac{1}{j!}
\left( \frac{s}{\theta}\right)^j e^{-s/\theta }\\
&=& 1-e^{-s/\theta}\sum_{j=0}^{\infty} \frac{1}{j!}
\left( \frac{s}{\theta} \right)^j \overline{P}_j
\end{eqnarray*}

where \(\overline{P}_j =p_{j+1}+p_{j+2}+\cdots = \Pr (N>j),\) the
``survival function'' of the claims count distribution.

\begin{center}\rule{0.5\linewidth}{\linethickness}\end{center}

\subsection{Tweedie Distribution}\label{tweedie-distribution}

In this section, we examine a particular compound distribution where the
number of claims is a Poisson distribution and the amount of claims is a
Gamma distribution. This specification leads to what is known as a
Tweedie distribution. The Tweedie distribution has a mass probability at
zero and a continuous component for positive values. Because of this
feature, it is widely used in insurance claims modeling, where the zero
mass is interpreted as no claims and the positive component as the
amount of claims.

Specifically, consider the collective risk model \(S=X_1+\cdots+X_N\).
Suppose that \(N\) has a Poisson distribution with mean \(\lambda\), and
each \(X_i\) has a Gamma distribution shape parameter \(\alpha\) and
scale parameter \(\gamma\). The Tweedie distribution is derived as the
Poisson sum of gamma variables. To understand the distribution of \(S\),
we first examine the mass probability at zero. It is straightforward to
see that the aggregate loss is zero when there is no claims occurred,
thus: \[f_S(0)={\rm Pr}(S=0)= {\rm Pr}(N=0)=e^{-\lambda}.\] In addition,
one notes that that \(S\) conditional on \(N_i=n\), denoted by
\(S_n=X_1+\cdots+X_n\), follows a gamma distribution with shape
\(n\alpha\) and scale \(\gamma_i\). Thus, for \(s>0\), the density of a
Tweedie distribution can be calculated as \[\begin{aligned}
f_S(s)&=\sum_{n=1}^{\infty} p_n f_{S_n}(s)\\
&=\sum_{n=1}^{\infty}e^{-\lambda_i}\frac{(\lambda_i)^n}{n!}\frac{1}{\gamma^{n\alpha}}y^{n\alpha-1}e^{-y\gamma}
\end{aligned}\] Thus, the Tweedie distribution can be thought of a
mixture of zero and a positive valued distribution, which makes it a
convenient tool for modeling insurance claims and for calculating pure
premiums. The mean and variance of the Tweedie compound Poisson model
are:
\[{\rm E} (S)=\lambda\frac{\alpha}{\gamma}~~~~{\rm and}~~~~{\rm Var} (S)=\lambda\frac{\alpha(1+\alpha)}{\gamma^2}.\]

As another important feature, the Tweedie distribution is a special case
of exponential dispersion models, a class of models used to describe the
random component in generalized linear models. To see this, we consider
the following reparameterizations:

\begin{equation*}
\lambda=\frac{\mu^{2-p}}{\phi(2-p)},~~~~\alpha=\frac{2-p}{p-1},~~~~\gamma=\phi(p-1)\mu^{p-1}
\end{equation*}

With the above relationships, one can show that the distribution of
\(S\) is
\[f_S(s)=\exp\left[\frac{1}{\phi}\left(\frac{-s}{(p-1)\mu^{p-1}}-\frac{\mu^{2-p}}{2-p}\right)+C(s;\phi)\right]\]
where

\begin{equation*}
C(s;\phi/\omega_i)=\left\{\begin{array}{ll}
                    \displaystyle 0 & {\rm if}~ y=0 \\
                   \displaystyle \ln \sum\limits_{n\ge 1} \left\{\frac{(1/\phi)^{1/(p-1)}y^{(2-p)/(p-1)}}{(2-p)(p-1)^{(2-p)/(p-1)}}\right\}^{n}\frac{1}{n!\Gamma(n(2-p)/(p-1))s} & {\rm if}~ y>0
                  \end{array}\right.
\end{equation*}

Hence, the distribution of \(S\) belongs to the exponential family with
parameters \(\mu\), \(\phi\), and \(p\in(1,2)\), and we have
\[{\rm E} (S)=\mu~~~~{\rm and}~~~~{\rm Var} (S)=\phi\mu^{p}\] It is also
worth mentioning the two limiting cases of the Tweedie model:
\(p\rightarrow 1\) results in the Poisson distribution and
\(p\rightarrow 2\) results in the gamma distribution. The Tweedie
compound Poisson model accommodates the situations in between.

\section{Computing the Aggregate Claims
Distribution}\label{computing-the-aggregate-claims-distribution}

Computing the distribution of aggregate losses is a difficult, yet
important, problem. As we have seen, for both individual risk model and
collective risk model, computing the distribution involves the
evaluation of a \(n\)-fold convolution. To make the problem tractable,
one strategy is to use a distribution that is easy to evaluate to
approximate the aggregate loss distribution. For instance, normal
distribution is a natural choice based on central limit theorem where
parameters of the normal distribution can be estimated by matching the
moments. This approach has its strength and limitations. The main
advantage is the ease of computation. The disadvantage are: first, the
size and direction of approximation error are unknown; second, the
approximation may fail to capture some special features of the aggregate
loss such as mass point at zero.

This section discusses two practical approaches to computing the
distribution of aggregate loss, the recursive method and the simulation.

\subsection{Recursive Method}\label{recursive-method}

The recursive method applies to compound models where the frequency
component \(N\) belongs to either \((a,b,0)\) or \((a,b,1)\) class and
the severity component \(X\) has a discrete distribution. For continuous
\(X\), a common practice is to first discretize the severity
distribution and then the recursive method is ready to apply.

Assume that \(N\) is in the \((a,b,1)\) class so that
\(p_{k}=\left( a+\frac{b}{k} \right) p_{k-1}, k = 2,3,\ldots\). Further
assume that the support of \(X\) is \(\{0,1,\ldots,m\}\), discrete and
finite. Then, the probability function of \(S\) is: \[\begin{aligned}
f_{S}(s)&=\Pr (S=s) \\
&=\frac{1}{1-af_{X}(0)}\left\{ \left[ p_1 -(a+b)p_{0}\right]
f_X (s)+\sum_{x=1}^{s\wedge m}\left( a+\frac{bx}{s} \right) f_X (x)f_{S}(s-x)\right\}.
\end{aligned}\] If \(N\) is in the \((a,b,0)\) class, then
\(p_1=(a+b)p_0\) and so \[
f_S(s)=\frac{1}{1-af_X (0)}\left\{ \sum_{x=1}^{s\wedge m}\left( a+\frac{bx
}{s}\right) f_X (x)f_{S}(s-x)\right\}.
\] \emph{Special Case}: If \(N \sim\) Poisson with mean \(\lambda\),
then \(a=0\) and \(b=\lambda\), thus \[
f_{S}(s)=\frac{\lambda }{s}\left\{ \sum_{x=1}^{s \wedge
m} x f_X (x) f_S (s-x)\right\} .
\]

\textbf{Example 5.4.1. SOA Exam Question.} The number of claims in a
period \(N\) has a geometric distribution with mean 4. The amount of
each claim \(X\) follows \({\rm Pr} (X = x) = 0.25\), for
\(x = 1,2,3,4\). The number of claims and the claim amount are
independent. \(S\) is the aggregate claim amount in the period.
Calculate \(F_S(3)\).

Show Example Solution

\hypertarget{toggleExampleAggLoss.4.1}{}
\textbf{Solution.} The severity distribution \(X\) follows
\[f_X (x) = \frac{1}{4}, \ \ x=1, 2, 3, 4.\] The frequency distribution
\(N\) is geometric with mean 4, which is a member of the \((a,b,0)\)
class with \(b=0\), \(a=\frac{\beta}{1+\beta} = \frac{4}{5}\), and
\(p_0 = \frac{1}{1+\beta} = \frac{1}{5}\). Thus, we can use the
recursive method

\begin{eqnarray*}
f_S (x) &=& 1 \sum_{y=1}^{x\wedge m} (a+0) f_X (y) f_S (x-y) \\
&=& \frac{4}{5} \sum_{y=1}^{x\wedge m} f_X (y) f_S (x-y)
\end{eqnarray*}

Specifically, we have

\begin{eqnarray*}
f_S (0) &=& \Pr(N=0) = p_0=\frac{1}{5}\\
f_S (1) &=& \frac{4}{5}\sum_{y=1}^{1} f_X (y) f_S (1-y) = \frac{4}{5} f_X(1) f_S(0)\\
&=& \frac{4}{5}\left( \frac{1}{4}\right)\left(\frac{1}{5} \right) = \frac{1}{25}\\
f_S (2) &=&  \frac{4}{5}\sum_{y=1}^{2} f_X (y) f_S (2-y) = \frac{4}{5} \left[ f_X(1)f_S(1) + f_X(2) f_S(0) \right] \\
&=& \frac{4}{5}\left[ \frac{1}{4} \left( \frac{1}{25} + \frac{1}{5}\right) \right] =
\frac{4}{5}\left( \frac{6}{100}\right) = \frac{6}{125}\\
f_S (3) &=& \frac{4}{5} \left[ f_X(1) f_S(2) + f_X(2)f_S(1) + f_X(3) f_S(0) \right]\\
&=& \frac{4}{5}\left[ \frac{1}{4} \left( \frac{1}{25} + \frac{1}{5} +
\frac{6}{125}\right) \right] = \frac{1}{5}\left( \frac{5+25+6}{125}\right) = 0.0576\\
\Rightarrow \ F_S (3) &=& f_S (0)+f_S (1)+f_S (2)+f_S (2)+f_S (3) = 0.3456
\end{eqnarray*}

\begin{center}\rule{0.5\linewidth}{\linethickness}\end{center}

\subsection{Simulation}\label{simulation}

The distribution of aggregate loss can be evaluated using Monte Carlo
simulation. The idea is one can calculate the empirical distribution of
\(S\) using a random sample. Blow we summarize the simulation procedures
for the aggregate loss models.

\begin{enumerate}
\def\labelenumi{\arabic{enumi}.}
\tightlist
\item
  Individual Risk Model \(S_n=X_1+\cdots+X_n\)
\end{enumerate}

\begin{itemize}
\item
  For each \(X_i\), \(i=1,\ldots,n\), generate random sample of size
  \(m\), denoted by \(x_{ij}~(j=1,\ldots,m)\);
\item
  Calculate the aggregate loss \(s_j=x_{1j}+\ldots+x_{nj}\) for
  \(j=1,\ldots,m\);
\item
  We obtain a random sample of \(S\), i.e. \(\{s_1,\ldots,s_m\}\).
\end{itemize}

\begin{enumerate}
\def\labelenumi{\arabic{enumi}.}
\setcounter{enumi}{1}
\tightlist
\item
  Collective Risk Model \(S=Y_1+\cdots+Y_N\)
\end{enumerate}

\begin{itemize}
\item
  Generate the number of claims \(n_j\) from frequency distribution
  \(N\);
\item
  Given \(n_j\), generate the amount of claims for each claim
  independently from \(Y\), denoted by \(y_{1},\ldots,y_{n_j}\);
\item
  Calculate the aggregate loss \(s_j=y_{1j}+\ldots+y_{n_j}\);
\item
  Repeat the above three steps for \(j=1,\ldots,m\);
\item
  We obtain a random sample of \(S\), i.e. \(\{s_1,\ldots,s_m\}\).
\end{itemize}

Given the random sample of \(S\), the empirical distribution can be
calculated as \[\hat{F}_S(s)=\frac{1}{m}\sum_{i=1}^{m}I(s_i\leq s),\]
where \(I(\cdot)\) is an indicator function. The empirical distribution
\(\hat{F}_S(s)\) will converge to \({F}_S(s)\) almost surely as
\(m\rightarrow \infty\).

The above procedure assumes that the parameters of the frequency and
severity distributions are known. In practice, one would need to
estimate these parameters from the data. For instance, the assumptions
in the collective risk model suggest a two-stage estimation where a
model is developed for the number of claims \(N\) from the data on claim
counts and a model is developed for the severity of claims \(X\) from
the data on the amount of claims.

\section{Effects of Coverage
Modifications}\label{effects-of-coverage-modifications}

\subsection{Impact of Exposure on
Frequency}\label{impact-of-exposure-on-frequency}

This section focuses on an individual risk model for claim counts.
Consider the number of claims from a group of \(n\) policies:
\[S=X_1+\cdots+X_n\] where we assume \(X_i\) are \emph{iid} representing
the number of claims from policy \(i\). In this case, the exposure for
the portfolio is \(b\) using policy as exposure base. The \emph{pgf} of
\(S\) is \[\begin{aligned}
P_S(z)&={\rm E}(z^S)={\rm E}\left(z^{\sum_{i=1}^nX_i}\right)\\
&=\prod_{i=1}^n{\rm E}(z^{X_i})=[P_X(z)]^n
\end{aligned}\]

\emph{\textbf{Special Case} Poisson. If \(X_i\sim Poisson(\lambda)\),
its }pgf* is \(P_X(z)=e^{\lambda(z-1)}\). Then the \emph{pgf} of \(S\)
is \[P_S(z)=[e^{\lambda(z-1)}]^n=e^{n\lambda(z-1)}.\] So
\(S\sim Poisson(n\lambda)\).

\begin{center}\rule{0.5\linewidth}{\linethickness}\end{center}

\textbf{Special Case} Negative binomial. If \(X_i\sim NegBin(\beta,r)\),
its \emph{pgf} is \(P_X(z)=[1-\beta(z-1)]^{-r}\). Then the \emph{pgf} of
\(S\) is \[P_S(z)=[[1-\beta(z-1)]^{-r}]^n=[1-\beta(z-1)]^{-nr}.\] So
\(S\sim NB(\beta,nr)\).

\begin{center}\rule{0.5\linewidth}{\linethickness}\end{center}

\textbf{Example 5.5.1.} Assume that the number of claims for each
vehicle is Poisson with mean \(\lambda\). Given the following data on
the observed number of claims for each household, calculate the MLE of
\(\lambda\).

\[\begin{matrix}
\begin{array}{c|c|c}
  \hline
  \text{Household ID} & \text{Number of vehicles} & \text{Number of claims} \\
  \hline
  1 & 2 & 0 \\
  2 & 1 & 2 \\
  3 & 3 & 2 \\
  4 & 1 & 0 \\
  5 & 1 & 1 \\
  \hline
\end{array}
\end{matrix}\]

Show Example Solution

\hypertarget{toggleExampleAggLoss.5.1}{}
\textbf{Solution.} Each of the 5 households has number of exposures
\(b_j\) (number of vehicles) and number of claims \(S_j\),
\(j=1,...,5\). Note for each household, the number of claims
\(S_j \sim Poisson (b_j \lambda)\). The likelihood function is\\
\[\begin{aligned}
L(\lambda) &= \prod_{j=1}^5 \Pr(S_j=s_j) = \prod_{j=1}^5 \frac{e^{-b_j\lambda} (b_j \lambda)^{s_j}}{s_j!} \\
&= \left(\frac{e^{-2\lambda} (2 \lambda)^{0}}{0!} \right)
\left(\frac{e^{-1\lambda} (1 \lambda)^{2}}{2!} \right)
\left(\frac{e^{-3\lambda} (3 \lambda)^{2}}{2!} \right)
\left(\frac{e^{-1\lambda} (1 \lambda)^{0}}{0!} \right)
\left(\frac{e^{-1\lambda} (1 \lambda)^{1}}{1!} \right) \\
&\propto e^{-8\lambda} \lambda^5
\end{aligned}\] Taking the log-likehood, we have \[\begin{aligned}
l(\lambda) = \log L(\lambda) = -8\lambda + 5\log(\lambda)
\end{aligned}\] Setting the first derivative of the log-likehood to 0,
we have \[\begin{aligned}
&l'(\lambda) = -8 + \frac{5}{\lambda} = 0 \\
\Rightarrow \ & 8 = \frac{5}{\hat{\lambda}} \ \Rightarrow \ \hat{\lambda} = \frac{5}{8}
\end{aligned}\]

\begin{center}\rule{0.5\linewidth}{\linethickness}\end{center}

If the exposure of the portfolio change from \(n_1\) to \(n_2\), we can
establish the following relation between the aggregate claim counts:
\[P_{S_2}(z)=[P_X(z)]^{n_2}=[P_X(z)^{n_1}]^{n_2/n_1}=P_{S_1}(z)^{n_2/n_1}.\]

\subsection{Impact of Deductibles on Claim
Frequency}\label{impact-of-deductibles-on-claim-frequency}

This section examine the effect of deductible on claim frequency.
Intuitively, there will be fewer claims filed when a policy deductible
is imposed because a loss below deductible might not result in a claim.
Even if an insured does file a claim, this may not result in a payment
by the policy, since the claim may be denied or the loss amount may
ultimately be determined to be below deductible. Let \(N^L\) denote the
number of loss (i.e.~the number of claims with no deductible), and
\(N^P\) denote the number of payments when a deductible \(d\) is
imposed. Our goal is to identify the distribution of \(N^P\) given the
distribution of \(N^L\). We show below that the relationship between
\(N^L\) and \(N^P\) can be established within an aggregate risk model
framework.

Note that sometimes changes in deductible will affect policyholder
behavior. We assume that this is not the case, i.e.~the distribution of
losses for both frequency and severity remain unchanged when the
deductible changes.

Given there are \(N^L\) losses, let \(X_1,X_2\ldots,X_{N^L}\) be the
associated amount of losses. For \(j=1,\ldots,N^L\), define

\begin{eqnarray*}
I_j&=&
\left \{
\begin{array}{cc}
1 & \text{if} ~X_j>d\\
0 & \text{otherwise}\\
\end{array}
\right..
\end{eqnarray*}

Then we establish \[N^P=I_1+I_2+\cdots+I_{N_L}.\]

Note that conditioning on \(N^L\), the distribution of
\(N^P \sim Binomial (N^L, v)\), where \(v=\Pr(X>d)\). Thus, given
\(N^L\),

\begin{eqnarray*}
\mathrm{E}\left(z^{N^P}|N^L\right)&=&\left[ 1+v(z-1)\right]^{N^L}
\end{eqnarray*}

So the \emph{pgf} of \(N^P\) is

\begin{eqnarray*}
P_{N^P}(z)&=&\mathrm{E}_{N^P}\left(z^{N^P}\right)=\mathrm{E}_{N^L}\left[\mathrm{E}_{N^P}\left(z^{N^P}|N^L\right)\right]\\
&=&\mathrm{E}_{N^L}\left[(1+v(z-1))^{N^L}\right]\\
&=&P_{N^L}\left(1+v(z-1)\right)
\end{eqnarray*}

Thus, we can write the \emph{pgf} of \(N^P\) as the \emph{pgf} of
\(N^L\), evaluated at a new argument \(z^* = 1+v(z-1)\), that is,
\(P_{N^P}(z)=P_{N^L}(z^*)\).

\textbf{Special Cases:}

\begin{itemize}
\tightlist
\item
  \(N^L\sim Poisson (\lambda)\). The \emph{pgf} of \(N^L\) is
  \(P_{N^L}=\exp(\lambda(z-1))\). Thus the \emph{pgf} of \(N^P\) is

  \begin{eqnarray*}
  P_{N^P}(z)&=&\exp\left( \lambda(1+v(z-1)-1)\right)\\
  &=&\exp(\lambda v(z-1))\sim Poisson (\lambda v)
  \end{eqnarray*}
\end{itemize}

So the payment number has the same distribution as the loss number but
with the expected number of payments equal to
\(\lambda v = \lambda \Pr(X>d)\).

\begin{itemize}
\tightlist
\item
  \(N^L \sim NegBin(\beta, r)\). The \emph{pgf} of \(N^L\) is
  \(P_{N^{L}}\left( z\right) =\left[ 1-\beta \left( z-1\right)\right]^{-r}\).

  \begin{eqnarray*}
  P_{N^P}(z)&=&\left( 1-\beta (1+v(z-1)-1)\right)^{-r}\\
  &=&\left( 1-\beta v(z-1)\right)^{-r} \:\:\:\sim NegBin(\beta v, r)
  \end{eqnarray*}
\end{itemize}

So the payment number has the same distribution as the loss number but
with parameters \(\beta v\) and \(r\).

\textbf{Example 5.5.2.} Suppose that loss amounts
\(X_i\sim Pareto(\alpha=4,\ \theta=150)\). You are given that the loss
frequency is \(N^L\sim Poisson(\lambda)\) and the payment frequency
distribution \(N^{P_1}\sim Poisson (0.4)\) with \(d_1=30\). Find the
distribution of \(N^{P_2}\) with \(d_2=100\).

Show Example Solution

\hypertarget{toggleExampleAggLoss.5.2}{}
\textbf{Solution.} Because the loss frequency \(N^L\) is Poisson, we can
relate the means of the loss distribution \(N^L\) and the first payment
distribution \(N^{P_1}\) as \(0.4 = v_1 \lambda\), where
\[\begin{aligned}
&v_1 = \Pr(X > 30) = \left( \frac{150}{30+150}\right)^4=\left( \frac{5}{6}\right)^4 \\
\Rightarrow \ & \lambda = 0.4 \left( \frac{6}{5} \right)^4
\end{aligned}\] With this, we can assess the second payment distribution
\(N^{P_2}\) as being Poisson with mean \(\lambda_2 = \lambda v_2\),
where \[\begin{aligned}
& v_2 = \Pr(X>100)=\left( \frac{150}{100+150}\right)^4=\left( \frac{3}{5}\right)^4 \\
\Rightarrow \ & \lambda_2 = \lambda v_2 = 0.4\left( \frac{6}{5} \right)^4 \left( \frac{3}{5} \right)^4 = 0.1075
\end{aligned}\]

\begin{center}\rule{0.5\linewidth}{\linethickness}\end{center}

\textbf{Example 5.5.3. Follow-Up.} Now suppose instead that the loss
frequency is \(N^L \sim NegBin(\beta,\ r)\) and for deductible
\(d_1=30\), the payment frequency \(N^{P_1}\) is negative binomial with
mean \(0.4\). Find the mean of the payment frequency \(N^{P_2}\) with
deductible \(d_2=100\).

Show Example Solution

\hypertarget{toggleExampleAggLoss.5.3}{}
\textbf{Solution.} Because the loss frequency \(N^L\) is negative
binomial, we can relate the parameter \(\beta\) of the \(N^L\)
distribution and the parameter \(\beta_1\) of the first payment
distribution \(N^{P_1}\) using \(\beta_1 = \beta v_1\), where
\[v_1 = \Pr(X > 30) = \left( \frac{5}{6} \right)^4\] Thus, the mean of
\(N^{P_1}\) and the mean of \(N^L\) are related \[\begin{aligned}
&0.4 =  r \beta_1 = r \left(\beta v_1\right) \\
\Rightarrow \ & r\beta = \frac{0.4}{v_1} = 0.4 \left(\frac{6}{5} \right)^4
\end{aligned}\] Note that
\(v_2 = \Pr(X > 100) = \left( \frac{3}{5}\right)^4\) as in the original
question. Then the second payment frequency distribution is
\(N^{P_2} \sim NegBin(\beta v_2, \ r)\) with mean \[\begin{aligned}
r (\beta v_2) = (r \beta) v_2 = 0.4 \left( \frac{6}{5}\right)^4 \left( \frac{3}{5} \right)^4 = 0.1075
\end{aligned}\]

\begin{center}\rule{0.5\linewidth}{\linethickness}\end{center}

Next we examine the more general case where \(N^L\) is a zero-modified
distribution. Recall that a modified distribution is defined in terms of
an unmodified one. That is, \[\begin{aligned}
p_k^M = c~p_k^0, {~\rm for~} k=1,2,3,\ldots,  {~\rm with~}c = \frac{1-p_0^M}{1-p_0^0}.
\end{aligned}\] In the case that \(p_0^M=0\), we call this a
``truncated'' distribution at zero, or \(ZT\). For other arbitrary
values of \(p_0^M\), this is a zero-modified, or \(ZM\), distribution.
The \emph{pgf} for the modified distribution is shown as

\begin{eqnarray*}
P^M(z) = 1-c+c~P^0(z).
\end{eqnarray*}

When \(N^L\) follows a zero-modified distribution, the distribution of
\(N^P\) is established using the same relation
\(P_{N^P}(z)=P_{N^L}\left(1+v(z-1)\right)\).

\textbf{Special Cases:}

\begin{itemize}
\item
  \(N^{L}\) is a ZM-Poisson with parameters \(\lambda\) and \(p_0^{M}\).
  The \emph{pgf} of \(N^L\) is
  \[P_{N^{L}}(z)=1-\cfrac{1-p_0^{M}}{1-\exp(-\lambda)}+\cfrac{1-p_0^{M}}{1-\exp(-\lambda)}\exp[\lambda(z-1)].\]
  Thus the \emph{pgf} of \(N^P\) is
  \[P_{N^{L}}(z)=1-\cfrac{1-p_0^{M}}{1-\exp(-\lambda)}+\cfrac{1-p_0^{M}}{1-\exp(-\lambda)}\exp[\lambda v(z-1)].\]
  So the number of payments is also a ZM-Poisson distribution with
  parameters \(\lambda v\) and \(p_0^{M}\). The probability at zero can
  be evaluated using \({\rm Pr}(N^P=0) = P_{N^P}(0)\).
\item
  \(N^{L}\) is a ZM-NegBin with parameters \(\beta\), \(r\), and
  \(p_0^{M}\). The \emph{pgf} of \(N^L\) is
  \[P_{N^{L}}(z)=1-\cfrac{1-p_0^{M}}{1-(1+\beta)^{-r}}+\cfrac{1-p_0^{M}}{1-(1+\beta)^{-r}}\left[ 1-\beta \left( z-1\right)\right]^{-r}.\]
  Thus the \emph{pgf} of \(N^P\) is
  \[P_{N^{L}}(z)=1-\cfrac{1-p_0^{M}}{1-(1+\beta)^{-r}}+\cfrac{1-p_0^{M}}{1-(1+\beta)^{-r}}\left[ 1-\beta v\left( z-1\right)\right]^{-r}.\]
  So the number of payments is also a ZM-NegBin distribution with
  parameters \(\beta v\), \(r\), and \(p_0^{M}\). Similarly, the
  probability at zero can be evaluated using
  \({\rm Pr}(N^P=0) = P_{N^P}(0)\).
\end{itemize}

\textbf{Example 5.5.4.} Aggregate losses are modeled as follows:\\
(i) The number of losses follows a zero-modified Poisson distribution
with \(\lambda=3\) and \(p_0^M = 0.5\).\\
(ii) The amount of each loss has a Burr distribution with
\(\alpha=3, \theta=50, \gamma=1\).\\
(iii) There is a deductible of \(d=30\) on each loss.\\
(iv) The number of losses and the amounts of the losses are mutually
independent.

Calculate \(\mathrm{E~} N^P\) and \(\mathrm{Var~} N^P\).

Show Example Solution

\hypertarget{toggleExampleAggLoss.5.4}{}
\textbf{Solution.} Since \(N^L\) follows a ZM-Poisson distribution with
parameters \(\lambda\) and \(p_0^M\), we know that \(N^P\) also follows
a ZM-Poisson distribution, but with parameters \(\lambda v\) and
\(p_0^M\), where

\[v = \Pr(X>30) = \left( \frac{1}{1+(30/50)} \right)^3 = 0.2441\]

Thus, \(N^P\) follows a ZM-Poisson distribution with parameters
\(\lambda^\ast = \lambda v= 0.7324\) and \(p_0^M = 0.5\). Finally,
\[\begin{aligned}
\mathrm{E~} N^P &= (1-p_0^M) \frac{\lambda^\ast}{1-e^{-\lambda^\ast}} = 0.5 \left( \frac{0.7324}{1-e^{-0.7324}} \right) \\
&= 0.7053 \\
\mathrm{Var~} N^P &= (1-p_0^M) \left( \frac{\lambda^\ast[1-(\lambda^\ast + 1) e^{-\lambda^\ast}]}{(1-e^{-\lambda^\ast})^2} \right) + p_0^M(1-p_0^M) \left(\frac{\lambda^\ast}{1-e^{-\lambda^\ast}} \right)^2 \\
&= 0.5 \left( \frac{0.7324(1-1.7324 e^{-0.7324})}{(1-e^{-0.7324})^2} \right) + 0.5^2 \left( \frac{0.7324}{1-e^{-0.7324}} \right)^2 \\
&= 0.7244
\end{aligned}\]

\begin{center}\rule{0.5\linewidth}{\linethickness}\end{center}

\subsection{Impact of Policy Modifications on Aggregate
Claims}\label{impact-of-policy-modifications-on-aggregate-claims}

In this section, we examine how the change in deductibles affect
aggregate payments from an insurance portfolio. We assume that policy
limits, coinsurance, and inflation have no effect on the frequency of
payments made by an insurer. As in the previous section, we further
assume that deductible changes do not impact the distribution of losses
for both frequency and severity.

Recall the notation \(N^L\) for the number of losses. With ground-up
loss \(X\) and policy deductible \(d\), we use
\(N^P = I(X_1>d) + \cdots + I(X_{N^L}>d)\) for the number of payments.
Also, define the amount of payment on a per-loss basis as

\begin{eqnarray*}
    X^{L}&=\left\{
      \begin{array}{cc}
        0 & X<\cfrac{d}{1+r} \\
        \alpha[(1+r)X-d] & \cfrac{d}{1+r}\leq X<\cfrac{u}{1+r} \\
        \alpha(u-d) &  X \ge \cfrac{u}{1+r}\\
      \end{array}
\right.,
\end{eqnarray*}

and the the amount of payment on a per-payment basis as

\begin{eqnarray*}
    X^{P}&=\left\{
      \begin{array}{cc}
        {\rm Undefined} & X<\cfrac{d}{1+r} \\
        \alpha[(1+r)X-d] & \cfrac{d}{1+r}\leq X<\cfrac{u}{1+r} \\
        \alpha(u-d) &  X \ge \cfrac{u}{1+r}\\
      \end{array}
\right..
\end{eqnarray*}

In the above, \(r\), \(u\), \(\alpha\) represents the inflation rate,
policy limit, and coinsurance, respectively. Hence, aggregate costs
(payment amounts) can be expressed either on a per loss or per payment
basis:

\begin{eqnarray*}
S &=& X^L_1 + \cdots + X^L_{N^L} \\
&=&X^P_1 + \cdots + X^P_{N^P} .
\end{eqnarray*}

The fundamentals regarding collective risk models are ready to apply.
For instance, we have: \[\begin{aligned}
  {\rm E}(S) &= {\rm E}\left(N^L\right) {\rm E}\left(X^L\right) = {\rm E}\left(N^P\right) {\rm E}\left(X^P\right)\\
  {\rm Var}(S) &= {\rm E}\left(N^L\right) {\rm Var}\left(X^L\right) + \left[{\rm E}\left(X^L\right)\right]^2 {\rm Var}(N^L) \\
  &= {\rm E}\left(N^P\right) {\rm Var}\left(X^P\right) + \left[{\rm E}\left(X^P\right)\right]^2 {\rm Var}(N^P)\\
  M_S(z)&=P_{N^L}\left[M_{X^L}(z)\right]=P_{N^P}\left[M_{X^P}(z)\right]
\end{aligned}\]

\textbf{Example 5.5.5. SOA Exam Question.} A group dental policy has a
negative binomial claim count distribution with mean 300 and variance
800. Ground-up severity is given by the following table:

\[\begin{matrix}
  \begin{array}{ c | c }
    \hline
      \text{Severity} & \text{Probability}\\ \hline
    40 & 0.25\\
    80 & 0.25\\
    120 & 0.25\\
    200 & 0.25\\
    \hline
  \end{array}
\end{matrix}\]

You expect severity to increase 50\% with no change in frequency. You
decide to impose a per claim deductible of 100. Calculate the expected
total claim payment after these changes.

Show Example Solution

\hypertarget{toggleExampleAggLoss.5.5}{}
\textbf{Solution.} The cost per loss with a 50\% increase in severity
and a 100 deductible per claim is

\begin{eqnarray*}
Y^L &=&
\left\{
\begin{array}{cc}
0 & 1.5x<100 \\
1.5x-100 & 1.5x\ge 100\\
\end{array}
\right.
\end{eqnarray*}

This has expectation \[\begin{aligned}
\mathrm{E~}Y^L &= \frac{1}{4} \left[ \left(1.5(40)-100\right)_+ +
\left(1.5(80)-100\right)_+ + \left(1.5(120)-100\right)_+ +
\left(1.5(200)-100\right)_+ \right]  \\
&= \frac{1}{4}\left[ (60-100)_+ + (120-100)_+ + (180-100)_+ + (300-100)_+\right] \\
&= \frac{1}{4}\left[ 0 + 20 + 80 + 200 \right] = 75
\end{aligned}\] Thus, the expected aggregate loss is
\[\mathrm{E~}S=(\mathrm{E~}N) \left( \mathrm{E~}Y^L \right)= 300 (75) = 22,500
.\]

\begin{center}\rule{0.5\linewidth}{\linethickness}\end{center}

\textbf{Example 5.5.6. Follow-Up.} What is \(\mathrm{Var~}S\)?

Show Example Solution

\hypertarget{toggleExampleAggLoss.5.6}{}
\textbf{Solution.} On a per loss basis, we have \[\begin{aligned}
\mathrm{Var~}S &= \left(\mathrm{E~}N \right) \left( \mathrm{Var~} Y^L \right) + \left[ \mathrm{E~} Y^L \right]^2 \left(\mathrm{Var~} N \right)
\end{aligned}\] where \(\mathrm{E~}N = 300\) and
\(\mathrm{Var~} N = 800\). We find \[\begin{aligned}
&\mathrm{E} \left[ (Y^L)^2 \right] = \frac{1}{4} \left[ 0^2 + 20^2 + 80^2 + 200^2 \right] = 11,700 \\
\Rightarrow \ & \mathrm{Var~} Y^L = \mathrm{E} \left[ (Y^L)^2 \right] - \left( \mathrm{E~}Y^L \right)^2 = 11,700 - 75^2 = 6,075
\end{aligned}\] Thus, the variance of the aggregate claim payment is

\begin{eqnarray*}
\mathrm{Var~}S &=& 300(6,075) + 75^2 (800) = 6,322,500
\end{eqnarray*}

\begin{center}\rule{0.5\linewidth}{\linethickness}\end{center}

\emph{Alternative Method: Using the Per Payment Basis.} Previously, we
calculated the expected total claim payment by multiplying the expected
number of losses by the expected payment \emph{per loss}. Recall that we
can also multiply the expected number of payments by the expected
payment \emph{per payment}. In this case, we have
\[S=Y_1^P + \cdots + Y_{N_P}^P \] The probability of a payment is
\[v=\Pr(1.5X \ge 100)=\Pr(X \ge 66.\bar{6})=\frac{3}{4} .\] Thus, the
number of payments, \(N^P\) has a negative binomial distribution with
mean \[\mathrm{E~}N^P=300 \left(\frac{3}{4} \right)=225\] The cost per
payment is

\begin{eqnarray*}
Y^P &=&
\left\{
\begin{array}{cc}
\text{undefined} & 1.5x<100 \\
1.5x-100 & 1.5x\ge 100\\
\end{array}
\right.
\end{eqnarray*}

This has expectation
\[\mathrm{E~}Y^P=\frac{\mathrm{E~}Y^L}{\Pr(1.5X > 100)}=
\frac{\mathrm{E~}Y^L}{v}=\frac{75}{(3/4)}=100\] Thus, as before, the
expected aggregate loss is
\[\mathrm{E~}S=\left(\mathrm{E~}Y^P\right) \left(\mathrm{E~}N^P\right) =
100(225)=22,500\]

\begin{center}\rule{0.5\linewidth}{\linethickness}\end{center}

\textbf{Example 5.5.7. SOA Exam Question.} A company insures a fleet of
vehicles. Aggregate losses have a compound Poisson distribution. The
expected number of losses is 20. Loss amounts, regardless of vehicle
type, have exponential distribution with \(\theta=200\). To reduce the
cost of the insurance, two modifications are to be made:\\
(i) A certain type of vehicle will not be insured. It is estimated that
this will reduce loss frequency by 20\(\%\).\\
(ii) A deductible of 100 per loss will be imposed.

Calculate the expected aggregate amount paid by the insurer after the
modifications.

Show Example Solution

\hypertarget{toggleExampleAggLoss.5.7}{}
\textbf{Solution.} On a per loss basis, we have a 100 deductible. Thus,
the expectation per loss is \[\begin{aligned}
\mathrm{E~} Y^L &= E[(X-100)_+] = E(X) - E(X\wedge 100) \\
&= 200 - 200(1-e^{-100/200}) = 121.31
\end{aligned}\] Loss frequency has been reduced by 20\(\%\), resulting
in an expected number of losses \[\mathrm{E~}N^L = 0.8(20) = 16\] Thus,
the expected aggregate amount paid after the modifications is
\[\mathrm{E~}S = \left(\mathrm{E~}Y^L \right) \left( \mathrm{E~} N^L\right) = 121.31(16) = 1,941\]

\begin{center}\rule{0.5\linewidth}{\linethickness}\end{center}

\emph{Alternative Method: Using the Per Payment Basis.} We can also use
the per payment basis to find the expected aggregate amount paid after
the modifications. For the per payment severity, \[\begin{aligned}
\mathrm{E~} Y^P = \frac{\mathrm{E~} Y^L}{\Pr(X > 100)} = \frac{200 - 200(1-e^{-100/200})}{e^{-100/200}} = 200
\end{aligned}\] This is not surprising -- recall that the exponential
distribution is memoryless, so the expected claim amounts paid in excess
of 100 is still exponential with mean 200. Now we look at the payment
frequency. With the deductible of 100, the probability that a payment
occurs is \(\Pr(X > 100) = e^{-100/200}\) Thus,
\[\mathrm{E~} N^P = 16 e^{-100/200} = 9.7\] Putting this together, we
produce the same answer using the per payment basis as the per loss
basis from earlier
\[\mathrm{E~}S = \left( \mathrm{E~} Y^P \right) \left( \mathrm{E~} N^P \right) = 200(9.7) = 1,941\]

\begin{center}\rule{0.5\linewidth}{\linethickness}\end{center}

\section{Further Resources and
Contributors}\label{AL-further-reading-and-resources}

\subsubsection*{Exercises}\label{exercises-3}
\addcontentsline{toc}{subsubsection}{Exercises}

Here are a set of exercises that guide the viewer through some of the
theoretical foundations of \textbf{Loss Data Analytics}. Each tutorial
is based on one or more questions from the professional actuarial
examinations, typically the Society of Actuaries Exam C.

\href{https://www.ssc.wisc.edu/~jfrees/loss-data-analytics/aggregate-loss-guided-tutorials/}{Aggregate
Loss Guided Tutorials}

\subsubsection*{Contributors}\label{contributors-3}
\addcontentsline{toc}{subsubsection}{Contributors}

\begin{itemize}
\tightlist
\item
  \textbf{Peng Shi} and \textbf{Lisa Gao}, University of
  Wisconsin-Madison, are the principal authors of the initital version
  of this chapter. Email:
  \href{mailto:pshi@bus.wisc.edu}{\nolinkurl{pshi@bus.wisc.edu}} for
  chapter comments and suggested improvements.
\end{itemize}

\chapter{Simulation}\label{simulation-1}

Simulation is a computer-based, computationally intensive, method of
solving difficult problems, such as analyzing business processes.
Instead of creating physical processes and experimenting with them in
order to understand their operational characteristics, a simulation
study is based on a computer representation - it considers various
hypothetical conditions as inputs and summarizes the results. Through
simulation, a vast number of hypothetical conditions can be quickly and
inexpensively examined. Performing the same analysis with a physical
system is not only expensive and time-consuming but, in many cases,
impossible. A drawback of simulation is that computer models are not
perfect representations of business processes.

There are three basic steps for producing a simulation study:

\begin{itemize}
\item
  Generating approximately independent realizations that are uniformly
  distributed
\item
  Transforming the uniformly distributed realizations to observations
  from a probability distribution of interest
\item
  With the generated observations as inputs, designing a structure to
  produce interesting and reliable results.
\end{itemize}

Designing the structure can be a difficult step, where the degree of
difficulty depends on the problem being studied. There are many
resources, including this tutorial, to help the actuary with the first
two steps.

\section{Generating Independent Uniform
Observations}\label{generating-independent-uniform-observations}

We begin with a historically prominent method.

\textbf{Linear Congruential Generator.} To generate a sequence of random
numbers, start with \(B_0\), a starting value that is known as a
``seed.'' Update it using the recursive relationship
\[B_{n+1} = a B_n + c  \text{ modulo }m, ~~ n=0, 1, 2, \ldots .\] This
algorithm is called a \emph{linear congruential generator}. The case of
\(c=0\) is called a \emph{multiplicative} congruential generator; it is
particularly useful for really fast computations.

For illustrative values of \(a\) and \(m\), Microsoft's Visual Basic
uses \(m=2^{24}\), \(a=1,140,671,485\), and \(c = 12,820,163\) (see
\url{http://support.microsoft.com/kb/231847}). This is the engine
underlying the random number generation in Microsoft's Excel program.

The sequence used by the analyst is defined as \(U_n=B_n/m.\) The
analyst may interpret the sequence \{\(U_{i}\)\} to be (approximately)
identically and independently uniformly distributed on the interval
(0,1). To illustrate the algorithm, consider the following.

\textbf{Example.} Take \(m=15\), \(a=3\), \(c=2\) and \(B_0=1\). Then we
have:

\begin{longtable}[]{@{}clc@{}}
\toprule
step \(n\) & \(B_n\) & \(U_n\)\tabularnewline
\midrule
\endhead
0 & \(B_0=1\) &\tabularnewline
1 & \(B_1 =\mod(3 \times 1 +2) = 5\) &
\(U_1 = \frac{5}{15}\)\tabularnewline
2 & \(B_2 =\mod(3 \times 5 +2) = 2\) &
\(U_2 = \frac{2}{15}\)\tabularnewline
3 & \(B_3 =\mod(3 \times 2 +2) = 8\) &
\(U_3 = \frac{8}{15}\)\tabularnewline
4 & \(B_4 =\mod(3 \times 8 +2) = 11\) &
\(U_4 = \frac{11}{15}\)\tabularnewline
\bottomrule
\end{longtable}

Sometimes computer generated random results are known as
\emph{pseudo}-random numbers to reflect the fact that they are machine
generated and can be replicated. That is, despite the fact that
\{\(U_{i}\)\} appears to be i.i.d, it can be reproduced by using the
same seed number (and the same algorithm). The ability to replicate
results can be a tremendous tool as you use simulation while trying to
uncover patterns in a business process.

The linear congruential generator is just one method of producing
pseudo-random outcomes. It is easy to understand and is (still) widely
used. The linear congruential generator does have limitations, including
the fact that it is possible to detect long-run patterns over time in
the sequences generated (recall that we can interpret ``independence''
to mean a total lack of functional patterns). Not surprisingly, advanced
techniques have been developed that address some of this method's
drawbacks.

\section{Inverse Transform}\label{inverse-transform}

With the sequence of uniform random numbers, we next transform them to a
distribution of interest. Let \(F\) represent a distribution function of
interest. Then, use the \emph{inverse transform}
\[X_i=F^{-1}\left( U_i \right) .\] The result is that the sequence
\{\(X_{i}\)\} is approximately i.i.d. with distribution function \(F\).

To interpret the result, recall that a distribution function, \(F\), is
monotonically increasing and so the inverse function, \(F^{-1}\), is
well-defined. The inverse distribution function (also known as the
\emph{quantile function}), is defined as \[\begin{aligned}
F^{-1}(y) = \inf_x \{ F(x) \ge y \} ,\end{aligned}\] where ``\(\inf\)''
stands for ``infimum'', or the greatest lower bound.

\textbf{Inverse Transform Visualization.} Here is a graph to help you
visualize the inverse transform. When the random variable is continuous,
the distribution function is strictly increasing and we can readily
identify a unique inverse at each point of the distribution.

\begin{figure}

{\centering \includegraphics[width=0.6\linewidth]{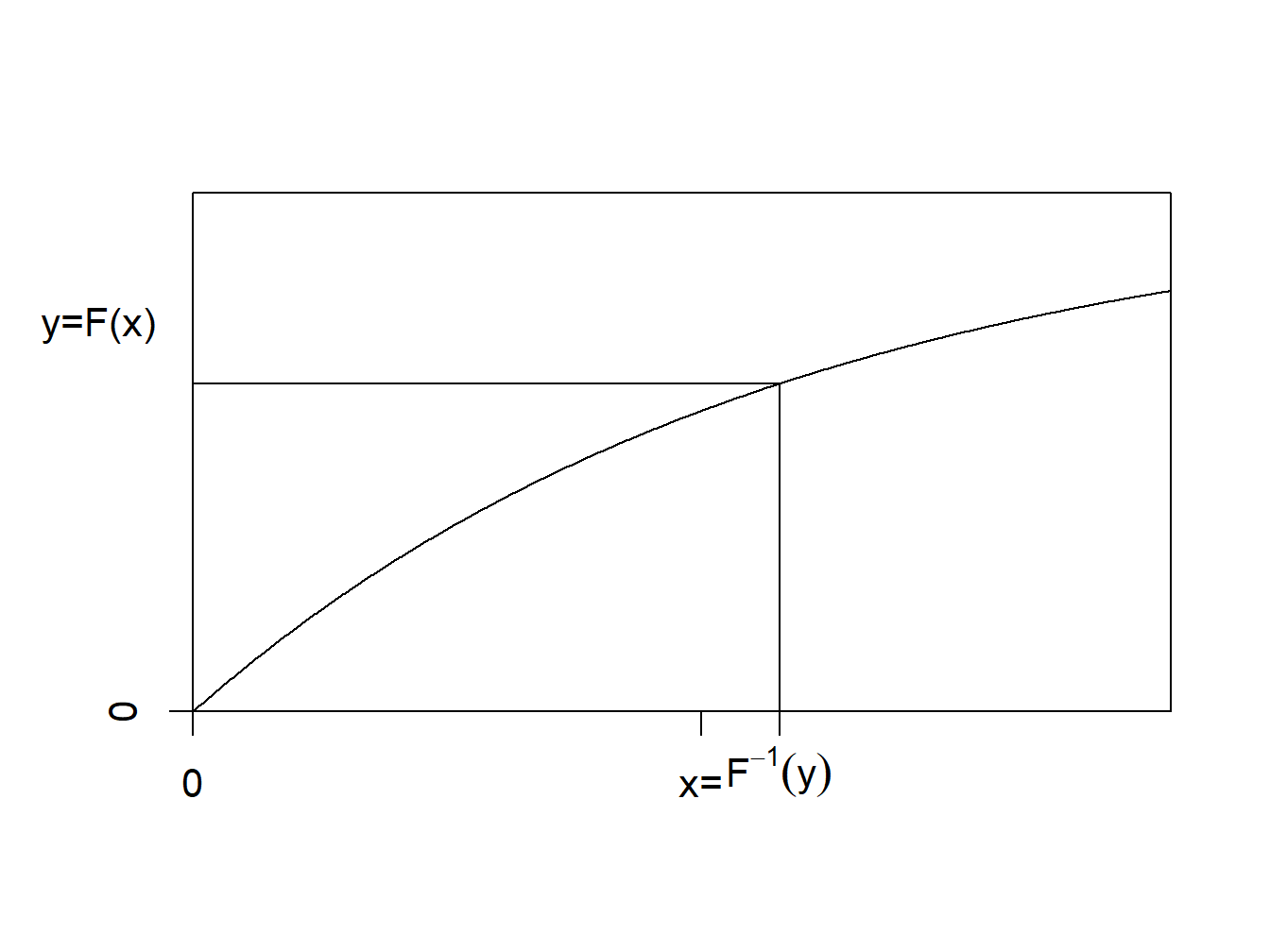}

}

\caption{Inverse of a Distribution Function}\label{fig:InverseDF}
\end{figure}

The inverse transform result is available when the underlying random
variable is continuous, discrete or a mixture. Here is a series of
examples to illustrate its scope of applications.

\textbf{Exponential Distribution Example.} Suppose that we would like to
generate observations from an exponential distribution with scale
parameter \(\theta\) so that \(F(x) = 1 - e^{-x/\theta}\). To compute
the inverse transform, we can use the following steps: \[\begin{aligned}
 y = F(x) &\Leftrightarrow  y = 1-e^{-x/\theta} \\
  &\Leftrightarrow -\theta \ln(1-y) = x = F^{-1}(y) .\end{aligned}\]
Thus, if \(U\) has a uniform (0,1) distribution, then
\(X = -\theta \ln(1-U)\) has an exponential distribution with parameter
\(\theta\).

\emph{Some Numbers.} Take \(\theta = 10\) and generate three random
numbers to get

\begin{longtable}[]{@{}lrrr@{}}
\toprule
\(U\) & 0.26321364 & 0.196884752 & 0.897884218\tabularnewline
\(X = -10\ln(1-U)\) & 1.32658423 & 0.952221285 &
9.909071325\tabularnewline
\bottomrule
\end{longtable}

\textbf{Pareto Distribution Example.} Suppose that we would like to
generate observations from a Pareto distribution with parameters
\(\alpha\) and \(\theta\) so that
\(F(x) = 1 - \left(\frac{\theta}{x+\theta} \right)^{\alpha}\). To
compute the inverse transform, we can use the following steps:
\[\begin{aligned}
 y = F(x) &\Leftrightarrow 1-y = \left(\frac{\theta}{x+\theta} \right)^{\alpha} \\
  &\Leftrightarrow \left(1-y\right)^{-1/\alpha} = \frac{x+\theta}{\theta} = \frac{x}{\theta} +1 \\
    &\Leftrightarrow \theta \left((1-y)^{-1/\alpha} - 1\right) = x = F^{-1}(y) .\end{aligned}\]
Thus, \(X = \theta \left((1-U)^{-1/\alpha} - 1\right)\) has a Pareto
distribution with parameters \(\alpha\) and \(\theta\).

\textbf{Inverse Transform Justification.} Why does the random variable
\(X = F^{-1}(U)\) have a distribution function ``\(F\)''?

This is easy to establish in the continuous case. Because \(U\) is a
Uniform random variable on (0,1), we know that \(\Pr(U \le y) = y\), for
\(0 \le y \le 1\). Thus, \[\begin{aligned}
\Pr(X \le x) &= \Pr(F^{-1}(U) \le x) \\
 &= \Pr(F(F^{-1}(U)) \le F(x)) \\
&= \Pr(U \le F(x)) = F(x)\end{aligned}\] as required. The key step is
that \$ F(F\^{}\{-1\}(u)) = u\$ for each \(u\), which is clearly true
when \(F\) is strictly increasing.

\textbf{Bernoulli Distribution Example.} Suppose that we wish to
simulate random variables from a Bernoulli distribution with parameter
\(p=0.85\). A graph of the cumulative distribution function shows that
the quantile function can be written as

\begin{figure}

{\centering \includegraphics[width=0.5\linewidth]{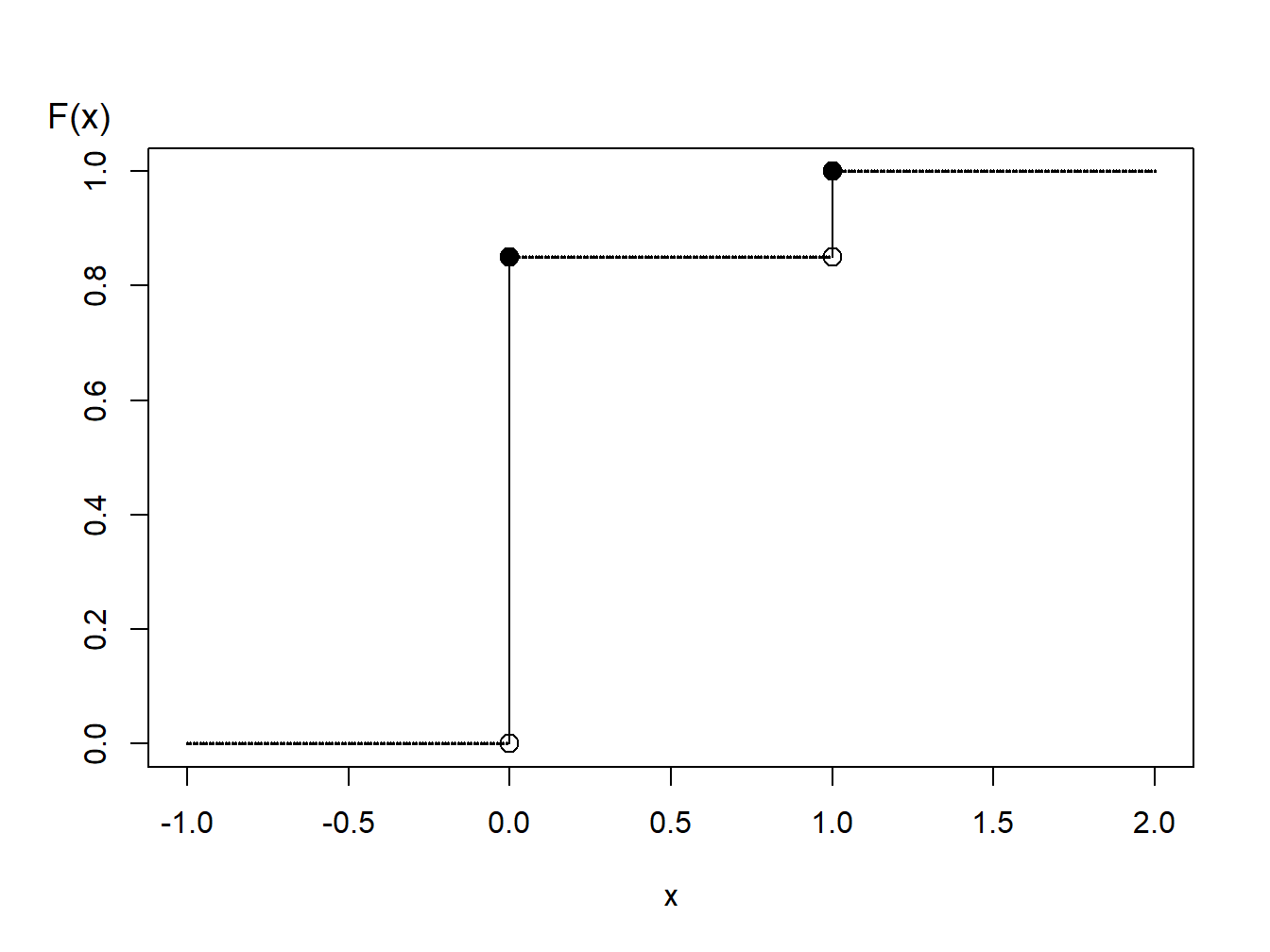}

}

\caption{Distribution Function of a Binary Random Variable}\label{fig:BinaryDF}
\end{figure}

\[\begin{aligned}
F^{-1}(y) = \left\{ \begin{array}{cc}
              0 & 0<y \leq 0.85 \\
              1 & 0.85 < y  \leq  1.0 .
            \end{array} \right.\end{aligned}\]

Thus, with the inverse transform we may define \[\begin{aligned}
X = \left\{ \begin{array}{cc}
              0 & 0<U \leq 0.85  \\
              1 &  0.85 < U  \leq  1.0
            \end{array} \right.\end{aligned}\] \emph{Some Numbers.}
Generate three random numbers to get

\begin{longtable}[]{@{}lrrr@{}}
\toprule
\(U\) & 0.26321364 & 0.196884752 & 0.897884218\tabularnewline
\(X =F^{-1}(U)\) & 0 & 0 & 1\tabularnewline
\bottomrule
\end{longtable}

\textbf{Discrete Distribution Example.} Consider the time of a machine
failure in the first five years. The distribution of failure times is
given as:

\begin{longtable}[]{@{}lrrrrr@{}}
\toprule
Time (\(x\)) & 1 & 2 & 3 & 4 & 5\tabularnewline
probability & 0.1 & 0.2 & 0.1 & 0.4 & 0.2\tabularnewline
\(F(x)\) & 0.1 & 0.3 & 0.4 & 0.8 & 1.0\tabularnewline
\bottomrule
\end{longtable}

\begin{figure}

{\centering \includegraphics[width=0.8\linewidth]{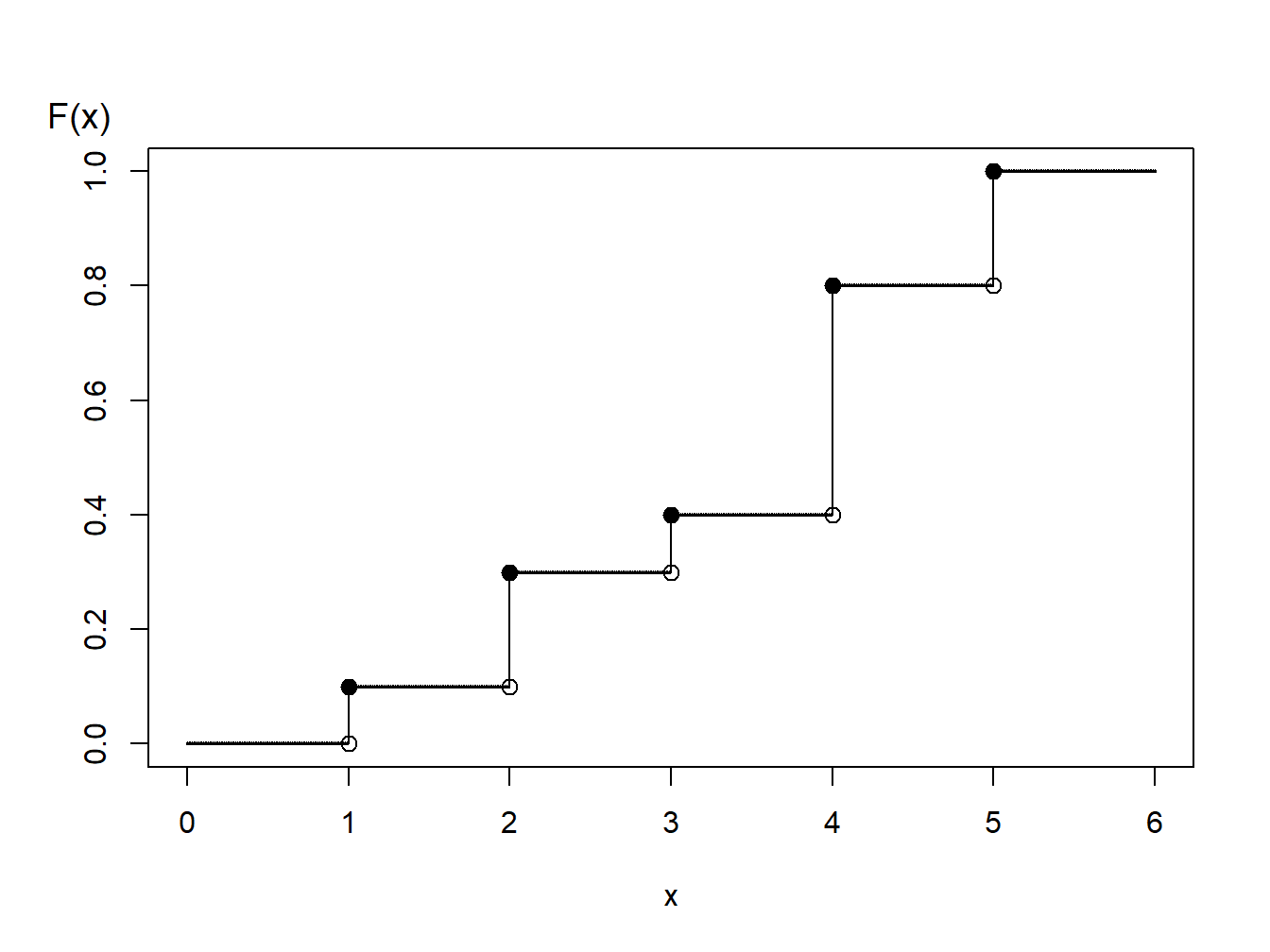}

}

\caption{Distribution Function of a Discrete Random Variable}\label{fig:DiscreteDF}
\end{figure}

Using the graph of the distribution function, with the inverse transform
we may define \[\begin{aligned}
X = \left\{ \begin{array}{cc}
              1 &   0<U  \leq 0.1  \\
              2 &  0.1 < U  \leq  0.3\\
              3 &  0.3 < U  \leq  0.4\\
              4 &  0.4 < U  \leq  0.8  \\
              5 &  0.8 < U  \leq  1.0     .
            \end{array} \right.\end{aligned}\]

For general discrete random variables there may not be an ordering of
outcomes. For example, a person could own one of five types of life
insurance products and we might use the following algorithm to generate
random outcomes:

\[\begin{aligned}
X = \left\{ \begin{array}{cc}
  \textrm{whole life} &   0<U  \leq 0.1  \\
 \textrm{endowment} &  0.1 < U  \leq  0.3\\
\textrm{term life} &  0.3 < U  \leq  0.4\\
  \textrm{universal life} &  0.4 < U  \leq  0.8  \\
  \textrm{variable life} &  0.8 < U  \leq  1.0 .
            \end{array} \right.\end{aligned}\]

Another analyst may use an alternative procedure such as:

\[\begin{aligned}
X = \left\{ \begin{array}{cc}
  \textrm{whole life} &   0.9<U<1.0  \\
 \textrm{endowment} &  0.7 \leq U < 0.9\\
\textrm{term life} &  0.6 \leq U < 0.7\\
  \textrm{universal life} &  0.2 \leq U < 0.6  \\
  \textrm{variable life} &  0 \leq U < 0.2 .
            \end{array} \right.\end{aligned}\]

Both algorithms produce (in the long-run) the same probabilities, e.g.,
\(\Pr(\textrm{whole life})=0.1\), and so forth. So, neither is
incorrect. You should be aware that there is ``more than one way to skin
a cat.'' (What an old expression!) Similarly, you could use an
alterative algorithm for ordered outcomes (such as failure times 1, 2,
3, 4, or 5, above).

\textbf{Mixed Distribution Example.} Consider a random variable that is
0 with probability 70\% and is exponentially distributed with parameter
\(\theta= 10,000\) with probability 30\%. In practice, this might
correspond to a 70\% chance of having no insurance claims and a 30\%
chance of a claim - if a claim occurs, then it is exponentially
distributed. The distribution function is given as

\[\begin{aligned}
F(y) = \left\{ \begin{array}{cc}
              0 &  x<0  \\
              1 - 0.3 \exp(-x/10000) & x \ge 0 .
            \end{array} \right.\end{aligned}\]

\begin{figure}

{\centering \includegraphics[width=0.6\linewidth]{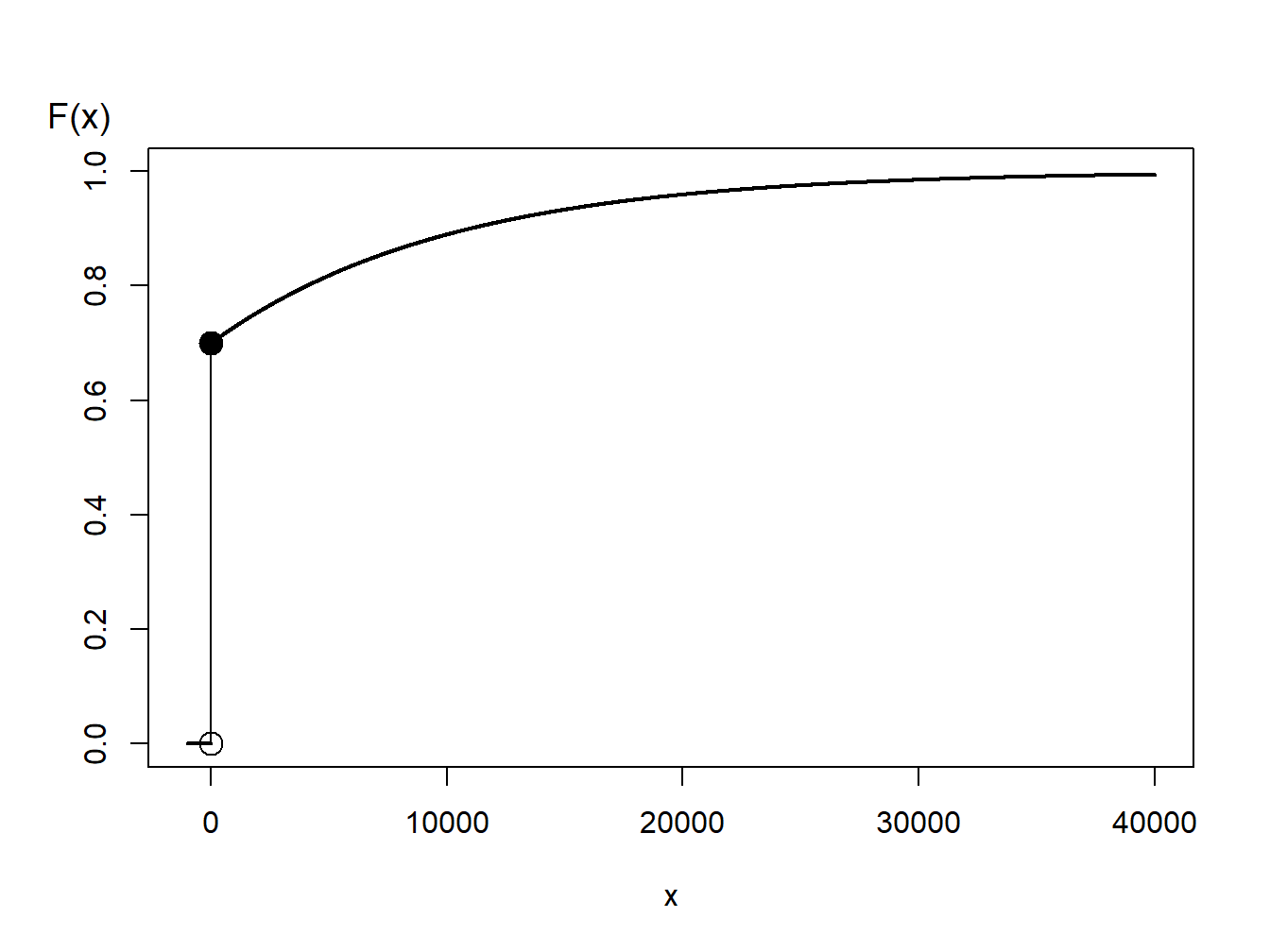}

}

\caption{Distribution Function of a Hybrid Random Variable}\label{fig:MixedDF}
\end{figure}

From the graph, we can see that the inverse transform for generating
random variables with this distribution function is

\[\begin{aligned}
X = F^{-1}(U) = \left\{ \begin{array}{cc}
              0 &  0< U  \leq  0.7  \\
              -1000 \ln (\frac{1-U}{0.3}) & 0.7 < U < 1 .
            \end{array} \right.\end{aligned}\]

As you have seen, for the discrete and mixed random variables, the key
is to draw a graph of the distribution function that allows you to
visualize potential values of the inverse function.

\section{How Many Simulated Values?}\label{how-many-simulated-values}

There are many topics to be described in the study of simulation (and
fortunately many good sources to help you). The best way to appreciate
simulation is to experience it. One topic that inevitably comes up is
the number of simulated trials needed to rid yourself of sampling
variability so that you may focus on patterns of interest.

How many simulated values are recommended? 100? 1,000,000? We can use
the central limit theorem to respond to this question. Suppose that we
wish to use simulation to calculate \(\mathrm{E~}h(X)\), where
\(h(\cdot)\) is some known function. Then, based on \(R\) simulations
(replications), we get \$ X\_1,\ldots,X\_R\$. From this simulated
sample, we calculate a sample average
\[\overline{h}_R=\frac{1}{R}\sum_{i=1}^{R} h(X_i)\] and a sample
standard deviation
\[s_{h,R}^2 = \frac{1}{R} \sum_{i=1}^{R}\left( h(X_i) -\overline{h}_R
\right) ^2.\] So, \(\overline{h}_R\) is your best estimate of
\(\mathrm{E~}h(X)\) and \(s_{h,R}^2\) provides an indication of the
uncertainty of your estimate. As one criterion for your confidence in
the result, suppose that you wish to be within 1\% of the mean with 95\%
certainty. According to the central limit theorem, your estimate should
be approximately normally distributed. Thus, you should continue your
simulation until \[\frac{.01\overline{h}_R}{s_{h,R}/\sqrt{R}}\geq 1.96\]
or equivalently \[R \geq 38,416\frac{s_{h,R}^2}{\overline{h}_R^2}.\]
This criterion is a direct application of the approximate normality
(recall that 1.96 is the 97.5th percentile of the standard normal
curve). Note that \(\overline{h}_R\) and \(s_{h,R}\) are not known in
advance, so you will have to come up with estimates as you go
(sequentially), either by doing a little pilot study in advance or by
interrupting your procedure intermittently to see if the criterion is
satisfied.

\chapter{Premium Calculation Fundamentals}\label{C:PremCalc}

This is a placeholder file

\chapter{Risk Classification}\label{C:RiskClass}

\emph{Chapter Preview.} This chapter motivates the use of risk
classification in insurance pricing and introduces readers to the
Poisson regression as a prominent example of risk classification. In
Section \ref{S:RC:Introduction} we explain why insurers need to
incorporate various risk characteristics, or rating factors, of
individual policyholders in pricing insurance contracts. We then
introduce Section \ref{S:RC:PoissonRegression} the Poisson regression as
a pricing tool to achieve such premium differentials. The concept of
exposure is also introduced in this section. As most rating factors are
categorical, we show in Section \ref{S:CatVarMultiTarriff} how the
multiplicative tariff model can be incorporated in the Poisson
regression model in practice, along with numerical examples for
illustration.

\section{Introduction}\label{S:RC:Introduction}

\begin{center}\rule{0.5\linewidth}{\linethickness}\end{center}

In this section, you learn:

\begin{itemize}
\item
  Why premiums should vary across policyholders with different risk
  characteristics.
\item
  The meaning of the adverse selection spiral.
\item
  The need for risk classification.
\end{itemize}

\begin{center}\rule{0.5\linewidth}{\linethickness}\end{center}

Through insurance contracts, the policyholders effectively transfer
their risks to the insurer in exchange for premiums. For the insurer to
stay in business, the premium income collected from a pool of
policyholders must at least equal to the benefit outgo. Ignoring the
frictional expenses associated with the administrative cost and the
profit margin, the net premium charged by the insurer thus should be
equal to the expected loss occurring from the risk that is transferred
from the policyholder.

If all policyholders in the insurance pool have identical risk profiles,
the insurer simply charges the same premium for all policyholders
because they have the same expected loss. In reality however the
policyholders are hardly homogeneous. For example, mortality risk in
life insurance depends on the characteristics of the policyholder, such
as, age, sex and life style. In auto insurance, those characteristics
may include age, occupation, the type or use of the car, and the area
where the driver resides. The knowledge of these characteristics or
variables can enhance the ability of calculating fair premiums for
individual policyholders as they can be used to estimate or predict the
expected losses more accurately.

Indeed, if the insurer do not differentiate the risk characteristics of
individual policyholders and simply charges the same premium to all
insureds based on the average loss in the portfolio, the insurer would
face adverse selection, a situation where individuals with a higher
chance of loss are attracted in the portfolio and low-risk individuals
are repelled. For example, consider a health insurance industry where
smoking status is an important risk factor for mortality and morbidity.
Most health insurers in the market require different premiums depending
on smoking status, so smokers pay higher premiums than non-smokers, with
other characteristics being identical. Now suppose that there is an
insurer, we will call EquitabAll, that offers the same premium to all
insureds regardless of smoking status, unlike other competitors. The net
premium of EquitabAll is natually an average mortality loss accounting
for both smokers and non-smokers. That is, the net premium is a weighted
average of the losses with the weights being the proportion of smokers
and non-smokers, respectively. Thus it is easy to see that that a smoker
would have a good incentive to purchase insurance from EquitabAll than
from other insurers as the offered premium by EquitabAll is relatively
lower. At the same time non-smokers would prefer buying insurance from
somewhere else where lower premiums, computed from the non-smoker group
only, are offered. As a result, there will be more smokers and less
non-smokers in the EquitabAll's portfolio, which leads to
larger-than-expected losses and hence a higher premium for insureds in
the next period to cover the higher costs. With the raised new premium
in the next period, non-smokers in EquitabAll will have even greater
incentives to switch the insurer. As this cycle continues over time,
EquitabAll would gradually retain more smokers and less non-smokers in
its portfolio with the premium continually raised, eventually leading to
a collapsing of business. In the literature this phenomenon is known as
the \emph{adverse selection spiral} or death spiral. Therefore,
incorporating and differentiating important risk characteristics of
individuals in the insurance pricing process are a pertinent component
for both the determination of fair premium for individual policyholders
and the long term sustainability of insurers.

In order to incorporate relevant risk characteristics of policyholders
in the pricing process, insurers maintain some classification system
that assigns each policyholder to one of the risk classes based on a
relatively small number of risk characteristics that are deemed most
relevant. These characteristics used in the classification system are
called the \emph{rating factors}, which are \emph{a priori} variables in
the sense that they are known before the contract begins (e.g., sex,
health status, vehicle type, etc, are known during the underwriting).
All policyholders sharing identical risk factors thus are assigned to
the same risk class, and are considered homogeneous from the pricing
viewpoint; the insurer consequently charge them the same premium.

An important task in any risk classification is to construct a
quantitative model that can determine the expected loss given various
rating factors of a policyholder. The standard approach is to adopt a
statistical regression model which produces the expected loss as the
output when the relevant risk factors are given as the inputs. In this
chapter we learn the Poisson regression, which can be used when the loss
is a count variable, as a prominent example of an insurance pricing
tool.

\section{Poisson Regression Model}\label{S:RC:PoissonRegression}

The Poisson regression model has been successfully used in a wide range
of applications and has an advantage of allowing closed-form expressions
for important quantities, which provides a informative intuition and
interpretation. In this section we introduce the Poisson regression as a
natural extension of the Poisson distribution.

\begin{center}\rule{0.5\linewidth}{\linethickness}\end{center}

In this section you will:

\begin{itemize}
\item
  Understand Poisson regressions as convenient tool to combine
  individual Poisson distributions in a unified fashion.
\item
  Learn the concept of exposure and its importance.
\item
  Formally learn how to formulate the Poisson regression model using
  indicator variables when the explanatory variables are categorical.
\end{itemize}

\begin{center}\rule{0.5\linewidth}{\linethickness}\end{center}

\subsection{Need for Poisson Regression}\label{S:RC:Need.Poi.reg}

\textbf{Poisson Distribution}

To introduce the Poisson regression, let us consider a hypothetical
health insurance portfolio where all policyholders are of the same age
and only one risk factor, smoking status, is relevant. Smoking status
thus is a categorical variable containing two different types: smoker
and non-smoker. In the statistical literature different types in a given
categorical variable are commonly called \emph{levels}. As there are two
levels for the smoking status, we may denote smoker and non-smoker by
level 1 and 2, respectively. Here the numbering is arbitrary and
nominal. Suppose now that we are interested in pricing a health
insurance where the premium for each policyholder is determined by the
number of outpatient visits to doctor's office during a year. The amount
of medical cost for each visit is assumed to be the same regardless of
the smoking status for simplicity. Thus if we believe that smoking
status is a valid risk factor in this health insurance, it is natural to
consider the data separately for each smoking status. In
\protect\hyperlink{tab:8.1}{Table 8.1} we present the data for this
portfolio.

\[\begin{matrix}
\begin{array}{cc|cc|cc}
\hline
\text{Smoker} & \text{(level 1)}  & \text{Non-smoker}&\text{(level 2)}  & & \text{Both}\\
  \text{Count} & \text{Observed} &  \text{Count} & \text{Observed}  &   \text{Count} & \text{Observed} \\ \hline
0 & 2213 &   0 & 6671 &  0 & 8884 \\
1 & 178  &   1 & 430  &  1 & 608 \\
2 & 11   &   2 & 25   &  2 & 36 \\
3 & 6    &   3 & 9    &  3 & 15 \\
4 & 0    &   4 & 4    &  4 & 4 \\
5 & 1    &   5 & 2    &  5 & 3 \\ \hline
\text{Total} & 2409  &   \text{Total} & 7141 & \text{Total} & 9550 \\
\text{Mean} & 0.0926 &   \text{Mean} & 0.0746 & \text{Mean} & 0.0792 \\
\hline
    \end{array}
\end{matrix}\]

\protect\hyperlink{tab:8.1}{Table 8.1} : Number of visits to doctor's
office in last year

As this dataset contains random counts we try to fit a Poisson
distribution for each level.

The \emph{pmf} of the Poisson with mean \(\mu\) is given by

\begin{equation}
\Pr(Y=y)=\frac{\mu^y e^{-\mu}}{y!},\qquad y=0,1,2, \ldots
\label{eq:Pois-pmf}
\end{equation}

and \(\mathrm{E~}{(Y)}=\mathrm{Var~}{(Y)}=\mu\). Furthermore, the
\emph{mle} of the Poisson distribution is given by the sample mean. Thus
if we denote the Poisson mean parameter for each level by \(\mu_{(1)}\)
(smoker) and \(\mu_{(2)}\) (non-smoker), we see from
\protect\hyperlink{tab:8.1}{Table 8.1} that \(\hat{\mu}_{(1)}=0.0926\)
and \(\hat{\mu}_{(2)}=0.0746\). This simple example shows the basic idea
of risk classification. Depending on the smoking status a policyholder
will have a different risk characteristic and it can be incorporated
through varying Poisson parameter in computing the fair premium. In this
example the ratio of expected loss frequencies is
\(\hat{\mu}_{(1)}/\hat{\mu}_{(2)}=1.2402\), implying that smokers tend
to visit doctor's office 24.02\(\%\) times more frequently compared to
non-smokers.

It is also informative to note that if the insurer charges the same
premium to all policyholders regardless of the smoking status, based on
the average characteristic of the portfolio, as was the case for
EquitabAll described in Introduction, the expected frequency (or the
premium) \(\hat{\mu}\) is 0.0792, obtained from the last column of
\protect\hyperlink{tab:8.1}{Table 8.1}. It is easily verified that

\begin{equation}
\hat{\mu} = \left(\frac{n_1}{n_1+n_2}\right)\hat{\mu}_{(1)}+\left(\frac{n_2}{n_1+n_2}\right)\hat{\mu}_{(2)}=0.0792,
\label{eq:coll-prem-avg}
\end{equation}

where \(n_i\) is the number of observations in each level. Clearly, this
premium is a weighted average of the premiums for each level with the
weight equal to the proportion of the insureds in that level.

\textbf{A simple Poisson regression}\\
In the example above, we have fitted a Poisson distribution for each
level separately, but we can actually combine them together in a unified
fashion so that a single Poisson model can encompass both smoking and
non-smoking statuses. This can be done by relating the Poisson mean
parameter with the risk factor. In other words, we make the Poisson
mean, which is the expected loss frequency, respond to the change in the
smoking status. The conventional approach to deal with a categorical
variable is to adopt indicator or dummy variables that take either 1 or
0, so that we turn the switch on for one level and off for others.
Therefore we may propose to use

\begin{equation}
\mu=\beta_0+\beta_1 x_1
\label{eq:lin-mu}
\end{equation}

or, more commonly, a log linear form

\begin{equation}
\log \mu=\beta_0+\beta_1 x_1,
\label{eq:log-lin-mu}
\end{equation}

where \(x_1\) is an indicator variable with

\begin{equation}
x_1=
\begin{cases}
     1 & \text{if smoker}, \\
     0 & \text{otherwise}.
\end{cases}
\label{eq:dummy-x1}
\end{equation}

We generally prefer the log linear relation \eqref{eq:log-lin-mu} to the
linear one in \eqref{eq:lin-mu} to prevent undesirable events of producing
negative \(\mu\) values, which may happen when there are many different
risk factors and levels. The setup \eqref{eq:log-lin-mu} and
\eqref{eq:dummy-x1} then results in different Poisson frequency parameters
depending on the level in the risk factor:

\begin{equation}
\log \mu=
\begin{cases}
     \beta_0+\beta_1 \\
     \beta_0
\end{cases}
\quad \text{or equivalently,}\qquad \mu= \begin{cases}
     e^{\beta_0+\beta_1} & \text{if smoker (level 1)}, \\
     e^{\beta_0} & \text{if non-smoker (level 2)},
\end{cases}
\label{eq:ind-mu}
\end{equation}

achieving what we aim for. This is the simplest form of the Poisson
regression. Note that we require a single indicator variable to model
two levels in this case. Alternatively, it is also possible to use two
indicator variables through a different coding scheme. This scheme
requires dropping the intercept term so that \eqref{eq:log-lin-mu} is
modified to

\begin{equation}
\log \mu=\beta_1 x_1+\beta_2 x_2,
\label{eq:log-lin-mu-2}
\end{equation}

where \(x_2\) is the second indicator variable with

\begin{equation}
x_2=
\begin{cases}
     1 & \text{if non-smoker}, \\
     0 & \text{otherwise}.
\end{cases}
\label{eq:dummy-x-2}
\end{equation}

Then we have, from \eqref{eq:log-lin-mu-2},

\begin{equation}
\log \mu=
\begin{cases}
     \beta_1 \\
     \beta_2
\end{cases}
\quad \text{or}\qquad \mu= \begin{cases}
     e^{\beta_1} & \text{if smoker (level 1)}, \\
     e^{\beta_2} & \text{if non-smoker (level 2)}.
\end{cases}
\label{eq:ind-mu-2}
\end{equation}

The numerical result of \eqref{eq:ind-mu} is the same as \eqref{eq:ind-mu-2}
as all the coefficients are given as numbers in actual estimation, with
the former setup more common in most texts; we also stick to the former.

With this Poisson regression model we can easily understand how the
coefficients \(\beta_0\) and \(\beta_1\) are linked to the expected loss
frequency in each level. According to \eqref{eq:ind-mu}, the Poisson mean
of the smokers, \(\mu_{(1)}\), is given by

\begin{equation}
\mu_{(1)}=e^{\beta_0+\beta_1}=\mu_{(2)} \,e^{\beta_1} \quad \text{or}\quad  \mu_{(1)}/\mu_{(2)} =e^{\beta_1}
\label{eq:no-label}
\end{equation}

where \(\mu_{(2)}\) is the Poisson mean for the non-smokers. This
relation between the smokers and non-smokers suggests a useful way to
compare the risks embedded in different levels of a given risk factor.
That is, the proportional increase in the expected loss frequency of the
smokers compared to that of the non-smokers is simply given by a
multiplicative factor \(e^{\beta_1}\). Putting another way, if we set
the expected loss frequency of the non-smokers as the base value, the
expected loss frequency of the smokers is obtained by applying
\(e^{\beta_1}\) to the base value.

\textbf{Dealing with multi-level case}\\
We can readily extend the two-level case to a multi-level one where
\(l\) different levels are involved for a single rating factor. For this
we generally need \(l-1\) indicator variables to formulate

\begin{equation}
\log \mu=\beta_0+\beta_1 x_1+\cdots+\beta_{l-1} x_{l-1},
\label{eq:log-lin-mu-1}
\end{equation}

where \(x_k\) is an indicator variable that takes 1 if the policy
belongs to level \(k\) and 0 otherwise, for \(k=1,2, \ldots, l-1\). By
omitting the indicator variable associated with the last level in
\eqref{eq:log-lin-mu-1} we effectively chose level \(l\) as the base case,
but this choice is arbitrary and does not matter numerically. The
resulting Poisson parameter for policies in level \(k\) then becomes,
from \eqref{eq:log-lin-mu-1},

\begin{equation}
\nonumber
\mu= \begin{cases}
     e^{\beta_0+\beta_k} & \text{if the policy belongs to level k (k=1,2, ..., l-1)}, \\
     e^{\beta_0} & \text{if the policy belongs to level l}.
\end{cases}
\end{equation}

Thus if we denote the Poisson parameter for policies in level \(k\) by
\(\mu_{(k)}\), we can relate the Poisson parameter for different levels
through \(\mu_{(k)}=\mu_{(l)}\, e^{\beta_k}\), \(k=1,2, \ldots, l-1\).
This indicates that, just like the two-level case, the expected loss
frequency of the \(k\)th level is obtained from the base value
multiplied by the relative factor \(e^{\beta_k}\). This relative
interpretation becomes more powerful when there are many risk factors
with multi-levels, and leads us to a better understanding of the
underlying risk and more accurate prediction of future losses. Finally,
we note that the varying Poisson mean is completely driven by the
coefficient parameters \(\beta_k\)'s, which are to be estimated from the
dataset; the procedure of the parameter estimation will be discussed
later in this chapter.

\subsection{Poisson Regression}\label{poisson-regression}

We now describe the Poisson regression in a formal and more general
setting. Let us assume that there are \(n\) independent policyholders
with a set of rating factors characterized by a \(k\)-variate
vector\footnote{For example, if there are 3 risk factors each of which
  the number of levels are 2, 3 and 4, respectively, we have
  \(k=(2-1)\times(3-1)\times (4-1)=6\).}. The \(i\)th policyholder's
rating factor is thus denoted by vector
\(\mathbf{ x}_i=(1, x_{i1}, \ldots, x_{ik})^{\prime}\), and the
policyholder has recorded the loss count \(y_i \in \{0,1,2, \ldots \}\)
from the last period of loss observation, for \(i=1, \ldots, n\). In the
regression literature, the values \(x_{i1}, \ldots, x_{ik}\) are
generally known as the \emph{explanatory variables}, as these are
measurements providing information about the variable of interest
\(y_i\). In essence, regression analysis is a method to quantify the
relationship between a variable of interest and explanatory variables.

We also assume, for now, that all policyholders have the same one unit
period for loss observation, or equal exposure of 1, to keep things
simple; we will discuss more details on the exposure in the following
subsection.

As done before, we describe the Poisson regression through its mean
function. For this we first denote \(\mu_i\) to be the expected loss
count of the \(i\)th policyholder under the Poisson specification
\eqref{eq:Pois-pmf}:

\begin{equation}
\mu_i=\mathrm{E~}{(y_i|\mathbf{ x}_i)}, \qquad y_i \sim Pois(\mu_i), \, i=1, \ldots, n.
\label{eq:mui-glm}
\end{equation}

The condition inside the expectation operation in \eqref{eq:mui-glm}
indicates that the loss frequency \(\mu_i\) is the model output
responding to the given set of risk factors or explanatory variables. In
principle the conditional mean \(\mathrm{E~}{(y_i|\mathbf{ x}_i)}\) in
\eqref{eq:mui-glm} can take different forms depending on how we specify
the relationship between \(\mathbf{ x}\) and \(y\). The standard choice
for the Poisson regression is to adopt the exponential function, as we
mentioned previously, so that

\begin{equation}
\mu_i=\mathrm{E~}{(y_i|\mathbf{ x}_i)}=e^{\mathbf{ x}^{\prime}_i\beta}, \qquad y_i \sim Pois(\mu_i), \, i=1, \ldots, n.
\label{eq:mean-ft-Pois}
\end{equation}

Here \(\beta=(\beta_0, \ldots, \beta_k)^{\prime}\) is the vector of
coefficients so that
\(\mathbf{ x}^{\prime}_i\beta=\beta_0+\beta_1x_{i1} +\ldots+\beta_k x_{ik}\).
The exponential function in \eqref{eq:mean-ft-Pois} ensures that
\(\mu_i >0\) for any set of rating factors \(\mathbf{ x}_i\). Often
\eqref{eq:mean-ft-Pois} is rewritten as a log linear form

\begin{equation}
\log \mu_i=\log \mathrm{E~}{(y_i|\mathbf{ x}_i)}=\mathbf{ x}^{\prime}_i\beta, \qquad y_i \sim Pois(\mu_i), \, i=1, \ldots, n
\label{eq:mean-ft-Pois-2}
\end{equation}

to reveal the relationship when the right side is set as the linear
form, \(\mathbf{ x}^{\prime}_i\beta\). Again, we see that the mapping
works well as both sides of \eqref{eq:mean-ft-Pois-2}, \(\log \mu_i\) and
\(\mathbf{ x}_i\beta\), can now cover the entire real values. This is
the formulation of the Poisson regression, assuming that all
policyholders have the same unit period of exposure. When the exposures
differ among the policyholders, however, as is the case in most
practical cases, we need to revise this formulation by adding exposure
component as an additional term in \eqref{eq:mean-ft-Pois-2}.

\subsection{Incorporating Exposure}\label{incorporating-exposure}

\textbf{Concept of Exposure}

In order to determine the size of potential losses in any type of
insurance, one must always know the corresponding exposure. The concept
of exposure is an extremely important ingredient in insurance pricing,
though we usually take it for granted. For example, when we say the
expected claim frequency of a health insurance policy is 0.2, it does
not mean much without the specification of the exposure such as, in this
case, per month or per year. In fact, all premiums and losses need the
exposure precisely specified and must be quoted accordingly; otherwise
all subsequent statistical analyses and predictions will be distorted.

In the previous section we assumed the same unit of exposure across all
policyholders, but this is hardly realistic in practice. In health
insurance, for example, two different policyholders with different
lengths of insurance coverage (e.g., 3 months and 12 months,
respectively) could have recorded the same number of claim counts. As
the expected number of claim counts would be proportional to the length
of coverage, we should not treat these two policyholders' loss
experiences identically in the modelling process. This motivates the
need of the concept of \emph{exposure} in the Poisson regression.

The Poisson distribution in \eqref{eq:Pois-pmf} is parametrised via its
mean. To understand the exposure, we alternatively parametrize the
Poisson \emph{pmf} in terms of the \emph{rate} parameter \(\lambda\),
based on the definition of the Poisson process:

\begin{equation}
\Pr(Y=y)=\frac{(\lambda t)^y e^{-\lambda t}}{y!},\qquad y=0,1,2, \ldots
\label{eq:Pois-pmf-2}
\end{equation}

with \(\mathrm{E~}{(Y)}=\mathrm{Var~}{(Y)}=\lambda t\). Here \(\lambda\)
is known as the rate or intensity per unit period of the Poisson process
and \(t\) represents the length of time or \emph{exposure}, a known
constant value. For given \(\lambda\) the Poisson distribution
\eqref{eq:Pois-pmf-2} produces a larger expected loss count as the
exposure \(t\) gets larger. Clearly, \eqref{eq:Pois-pmf-2} reduces to
\eqref{eq:Pois-pmf} when \(t=1\), which means that the mean and the rate
become the same for the unit exposure, the case we considered in the
previous subsection.

In principle the exposure does not need to be measured in units of time
and may represent different things depending the problem at hand. For
example,

\begin{enumerate}
\def\labelenumi{\arabic{enumi}.}
\item
  In health insurance, the rate may be the occurrence of a specific
  disease per 1,000 people and the exposure is the number of people
  considered in the unit of 1,000.
\item
  In auto insurance, the rate may be the number of accidents per year of
  a driver and the exposure is the length of the observed period for the
  driver in the unit of year.
\item
  For workers compensation, the rate may be the probability of injury in
  the course of employment per dollar and the exposure is the payroll
  amount in dollar.
\item
  In marketing, the rate may be the number of customers who enter a
  store per hour and the exposure is the number of hours observed.
\item
  In civil engineering, the rate may be the number of major cracks on
  the paved road per 10 kms and the exposure is the length of road
  considered in the unit of 10 kms.
\end{enumerate}

\begin{enumerate}
\def\labelenumi{\arabic{enumi}.}
\setcounter{enumi}{5}
\tightlist
\item
  In credit risk modelling, the rate may be the number of default events
  per 1000 firms and the exposure is the number of firms under
  consideration in the unit of 1,000.
\end{enumerate}

Actuaries may be able to use different exposure bases for a given
insurable loss. For example, in auto insurance, both the number of
kilometres driven and the number of months coved by insurance can be
used as exposure bases. Here the former is more accurate and useful in
modelling the losses from car accidents, but more difficult to measure
and manage for insurers. Thus, a good exposure base may not be the
theoretically best one due to various practical constraints. As a rule,
an exposure base must be easy to determine, accurately measurable,
legally and socially acceptable, and free from potential manipulation by
policyholders.

\textbf{Incorporating exposure in Poisson regression}\\
As exposures affect the Poisson mean, constructing Poisson regressions
requires us to carefully separate the rate and exposure in the modelling
process. Focusing on the insurance context, let us denote the rate of
the loss event of the \(i\)th policyholder by \(\lambda_i\), the known
exposure (the length of coverage) by \(m_i\) and the expected loss count
under the given exposure by \(\mu_i\). Then the Poisson regression
formulation in \eqref{eq:mean-ft-Pois} and \eqref{eq:mean-ft-Pois-2} should
be revised in light of \eqref{eq:Pois-pmf-2} as

\begin{equation}
\mu_i=\mathrm{E~}{(y_i|\mathbf{ x}_i)}=m_i \,\lambda_i=m_i \, e^{\mathbf{ x}^{\prime}_i\beta}, \qquad y_i \sim Pois(\mu_i), \, i=1, \ldots, n,
\label{eq:mean-ft-Pois-6}
\end{equation}

which gives

\begin{equation}
\log \mu_i=\log m_i+\mathbf{ x}^{\prime}_i\beta, \qquad y_i \sim Pois(\mu_i), \, i=1, \ldots,
\label{eq:mean-ft-Pois-7}
\end{equation}

Adding \(\log m_i\) in \eqref{eq:mean-ft-Pois-7} does not pose a problem
in fitting as we can always specify this as an extra explanatory
variable, as it is a known constant, and fix its coefficient to 1. In
the literature the log of exposure, \(\log m_i\), is commonly called the
\textbf{offset}.

\subsection{Exercises}\label{exercises-4}

\begin{enumerate}
\def\labelenumi{\arabic{enumi}.}
\item
  Regarding \protect\hyperlink{tab:8.1}{Table 8.1} answer the
  followings.

  \begin{enumerate}
  \def\labelenumii{(\alph{enumii})}
  \item
    Verify the mean values in the table.
  \item
    Verify the number in equation \eqref{eq:coll-prem-avg}.
  \item
    Produce the fitted Poisson counts for each smoking status in the
    table.
  \end{enumerate}
\item
  In the Poisson regression formulation \eqref{eq:mui-glm}, consider using
  \(\mu_i=\mathrm{E~}{(y_i|\mathbf{ x}_i)}=({\mathbf{ x}^{\prime}_i\beta})^2\),
  for \(i=1, \ldots, n\), instead of the exponential function What
  potential issue would you have?
\item
  Verify equation \eqref{eq:Inf-mtx-Poi} by differentiating the
  log-likelihood \eqref{eq:ll-Poi-reg}.
\end{enumerate}

\section{Categorical Variables and Multiplicative
Tariff}\label{S:CatVarMultiTarriff}

\begin{center}\rule{0.5\linewidth}{\linethickness}\end{center}

In this section you will learn:

\begin{itemize}
\item
  The multiplicative tariff model when the rating factors are
  categorical.
\item
  How to construct the Poisson regression model based on the
  multiplicative tariff structure.
\end{itemize}

\begin{center}\rule{0.5\linewidth}{\linethickness}\end{center}

\subsection{Rating Ractors and Tariff}\label{rating-ractors-and-tariff}

In practice most rating factors in insurance are \emph{categorical
variables}, meaning that they take one of the pre-determined number of
possible values. Examples of categorical variables include sex, type of
cars, the driver's region of residence and occupation. Continuous
variables, such as age or auto mileage, can also be grouped by bands and
treated as categorical variables. Thus we can imagine that, with a small
number of rating factors, there will be many policyholders falling into
the same risk class, charged with the same premium. For the remaining of
this chapter we assume that all rating factors are categorical
variables.

To illustrate how categorical variables are used in the pricing process,
we consider a hypothetical auto insurance with only two rating factors:

\begin{itemize}
\tightlist
\item
  Type of vehicle: Type A (personally owned) and B (owned by
  corporations). We use index \(j=1\) and \(2\) to respectively
  represent each level of this rating factor.\\
\item
  Age band of the driver: Young (age \(<\) 25), middle (25 \(\le\) age
  \(<\) 60) and old age (age \(\ge\) 60). We use index \(k=1, 2\) and
  \(3\), respectively, for this rating factor.
\end{itemize}

From this classification rule, we may create an organized table or list,
such as the one shown in \protect\hyperlink{tab:8.2}{Table 8.2},
collected from all policyholders. Clearly there are \(2 \times 3=6\)
different risk classes in total. Each row of the table shows a
combination of different risk characteristics of individual
policyholders. Our goal is to compute six different premiums for each of
these combinations. Once the premium for each row has been determined
using the given exposure and claim counts, the insurer can replace the
last two columns in \protect\hyperlink{tab:8.2}{Table 8.2} with a single
column containing the computed premiums. This new table then can serve
as a manual to determine the premium for a new policyholder given the
rating factors during the underwriting process. In non-life insurance, a
table (or a set of tables) or list that contains each set of rating
factors and the associated premium is referred to as a \emph{tariff}.
Each unique combination of the rating factors in a tariff is called a
\emph{tariff cell}; thus, in \protect\hyperlink{tab:8.2}{Table 8.2} the
number of tariff cells is six, same as the number of risk classes.

\[\begin{matrix}
\begin{array}{ccrrc}
 \hline
\text{Rating} &\text{factors}  &   \text{Exposure} & \text{Claim count} \\
\text{Type }(j) & \text{Age }(k) &  \text{in year} & \text{observed}\\
\hline \hline
j=1 & k=1 &  89.1 & 9\\
1 & 2   & 208.5& 8\\
1 & 3  & 155.2 & 6  \\
2  & 1  & 19.3 & 1 \\
2  & 2  & 360.4 & 13 \\
2   & 3  & 276.7 & 6 \\ \hline
\end{array}
\end{matrix}\]

\protect\hyperlink{tab:8.2}{Table 8.2} : Loss record of the illustrative
auto insurer

Let us now look at the loss information in
\protect\hyperlink{tab:8.2}{Table 8.2} more closely. The exposure in
each row represents the sum of the length of insurance coverages, or
in-force times, in the unit of year, of all the policyholders in that
tariff cell. Similarly the claim counts in each row is the number of
claims at each cell. Naturally the exposures and claim counts vary due
to the different number of drivers across the cells, as well as
different in-force time periods among the drivers within each cell.

In light of the Poisson regression framework, we denote the exposure and
claim count of cell \((j,k)\) as \(m_{jk}\) and \(y_{jk}\),
respectively, and define the claim count per unit exposure as

\begin{equation}
\nonumber
z_{jk}= \frac{y_{jk}}{ m_{jk}}, \qquad j=1,2;\, k=1, 2,3.
\end{equation}

For example, \(z_{12}=8/208.5=0.03837\), meaning that a policyholder in
tariff cell (1,2) would have 0.03837 accidents if insured for a full
year on average. The set of \(z_{ij}\) values then corresponds to the
rate parameter in the Poisson distribution \eqref{eq:Pois-pmf-2} as they
are the event occurrence rates per unit exposure. That is, we have
\(z_{jk}=\hat{\lambda}_{jk}\) where \({\lambda}_{jk}\) is the Poisson
rate parameter. Producing \(z_{ij}\) values however does not do much
beyond comparing the average loss frequencies across risk classes. To
fully exploit the dataset, we will construct a pricing model from
\protect\hyperlink{tab:8.2}{Table 8.2} using the Poisson regression, for
the remaining part of the chapter.

We comment that actual loss records used by insurers typically include
much more risk factors, in which case the number of cells grows
exponentially. The tariff would then consist of a set of tables, instead
of one, separated by some of the basic rating factors, such as sex or
territory.

\subsection{Multiplicative Tariff
Model}\label{multiplicative-tariff-model}

In this subsection, we introduce the multiplicative tariff model, a
popular pricing structure that can be naturally used within the Poisson
regression framework. The developments here is based on
\protect\hyperlink{tab:8.2}{Table 8.2}. Recall that the loss count of a
policyholder is described by the Poisson regression model with rate
\(\lambda\) and the exposure \(m\), so that the expected loss count
becomes \(m\lambda\). As \(m\) is a known constant, we are essentially
concerned with modelling \(\lambda\), so that it responds to the change
in the rating factors. Among other possible functional forms, we
commonly choose the multiplicative\footnote{Preferring the
  multiplicative form to others (e.g., additive one) was already hinted
  in \eqref{eq:log-lin-mu}.} relation to model the Poisson rate
\(\lambda_{jk}\) for rating factor (\(j,k\)):

\begin{equation}
\lambda_{jk}= f_0 \times f_{1j} \times f_{2k}, \qquad j=1,2;\, k=1, 2,3.
\label{eq:multiplicative-tarrif}
\end{equation}

Here \(\{ f_{1j}, j=1,2\}\) are the parameters associated with the two
levels in the first rating factor, car type, and
\(\{ f_{2k}, k=1,2,3\}\) associated with the three levels in the age
band, the second rating factor. For instance, the Poisson rate for a
mid-aged policyholder with a Type B vehicle is given by
\(\lambda_{22}=f_0 \times f_{12} \times f_{22}\). The first term \(f_0\)
is some base value to be discussed shortly. Thus these six parameters
are understood as numerical representations of the levels within each
rating factor, and are to be estimated from the dataset.

The multiplicative form \eqref{eq:multiplicative-tarrif} is easy to
understand and use, because it clearly shows how the expected loss count
(per unit exposure) changes as each rating factor varies. For example,
if \(f_{11}=1\) and \(f_{12}=1.2\), then the expected loss count of a
policyholder with a vehicle of type B would be 20\(\%\) larger than type
A, when the other factors are the same. In non-life insurance, the
parameters \(f_{1j}\) and \(f_{2k}\) are known as \emph{relativities} as
they determine how much expected loss should change relative to the base
value \(f_0\). The idea of relativity is quite convenient in practice,
as we can decide the premium for a policyholder by simply multiplying a
series of corresponding relativities to the base value.

Dropping an existing rating factor or adding a new one is also
transparent with this multiplicative structure. In addition, the insurer
may easily adjust the overall premium for all policyholders by
controlling the base value \(f_0\) without changing individual
relativities. However, by adopting the multiplicative form, we
implicitly assume that there is no serious interaction among the risk
factors.

When the multiplicative form is used we need to address an
identification issue. That is, for any \(c>0\), we can write

\begin{equation}
\lambda_{jk}= f_0 \times \frac{f_{1j}}{c} \times c\,f_{2k}.
\end{equation}

By comparing with \eqref{eq:multiplicative-tarrif}, we see that the
identical rate parameter \(\lambda_{jk}\) can be obtained for very
different individual relativities. This over-parametrization, meaning
that many different sets of parameters arrive at the identical model,
obviously calls for some restriction on \(f_{1j}\) and \(f_{2k}\). The
standard practice is to make one relativity in each rating factor equal
to one. This can be made arbitrarily, so we will assume that
\(f_{11}=1\) and \(f_{21}=1\) for our purpose. This way all other
relativities are uniquely determined. The tariff cell \((j,k)=(1,1)\) is
then called the \emph{base tariff cell}, where the rate simply becomes
\(\lambda_{11}=f_0\), corresponding to the base value according to
\eqref{eq:multiplicative-tarrif}. Thus the base value \(f_0\) is generally
interpreted as the Poisson rate of the base tariff cell.

Again, \eqref{eq:multiplicative-tarrif} is log-transformed and rewritten
as

\begin{equation}
\log \lambda_{jk}= \log f_0 + \log f_{1j} + \log f_{2k},
\label{eq:log-linear-tariff}
\end{equation}

as it is easier to work with in estimating process, similar to
\eqref{eq:mean-ft-Pois-2}. This log linear form makes the log relativities
of the base level in each rating factor equal to zero, i.e.,
\(\log f_{11}=\log f_{21}=0\), and leads to the following alternative,
more explicit expression for \eqref{eq:log-linear-tariff}:

\begin{equation}
\log \lambda=\begin{cases}
      \log f_0 + \quad 0 \quad \,\,+ \quad 0 \quad \,\,& \text{for a policy in cell $(1,1)$}, \\
            \log f_0+ \quad 0 \quad \,\,+\log f_{22}& \text{for a policy in cell $(1,2)$}, \\
                  \log f_0+ \quad 0 \quad \,\,+\log f_{23}& \text{for a policy in cell $(1,3)$}, \\
                        \log f_0+\log f_{12}+ \quad 0 \quad \,\,& \text{for a policy in cell $(2,1)$}, \\
                              \log f_0+\log f_{12}+\log f_{22}& \text{for a policy in cell $(2,2)$}, \\
                                    \log f_0+\log f_{12}+\log f_{23}& \text{for a policy in cell $(2,3)$}. \\
\end{cases}
\label{eq:log-rate-Poi-tariff-3}
\end{equation}

This clearly shows that the Poisson rate parameter \(\lambda\) varies
across different tariff cells, with the same log linear form used in the
Poisson regression framework. In fact the reader may see that
\eqref{eq:log-rate-Poi-tariff-3} is an extended version of the early
expression \eqref{eq:ind-mu} with multiple risk factors and that the log
relativities now play the role of \(\beta_i\) parameters. Therefore all
the relativities can be readily estimated via fitting a Poisson
regression with a suitably chosen set of indicator variables.

\subsection{Poisson Regression for Multiplicative
Tariff}\label{poisson-regression-for-multiplicative-tariff}

\textbf{Indicator Variables for Tariff Cells}

We now explain how the relativities can be incorporated in the Poisson
regression. As seen early in this chapter we use indicator variables to
deal with categorial variables. For our illustrative auto insurer,
therefore, we define an indicator variable for the first rating factor
as

\begin{equation}
x_1=
\begin{cases}
      1 & \text{ for vehicle type B}, \\
      0 & \text{ otherwise}.
\end{cases}
\end{equation}

For the second rating factor, we employ two indicator variables for the
age band, that is,

\begin{equation}
x_2=
\begin{cases}
     1 & \text{for age band 2}, \\
     0 & \text{otherwise}.
\end{cases}
\end{equation}

and

\begin{equation}
x_3=
\begin{cases}
     1 & \text{for age band 3}, \\
     0 & \text{otherwise}.
\end{cases}
\end{equation}

The triple \((x_1, x_2, x_3)\) then can effectively and uniquely
determine each risk class. By observing that the indicator variables
associated with Type A and Age band 1 are omitted, we see that tariff
cell \((j,k)=(1,1)\) plays the role of the base cell. We emphasize that
our choice of the three indicator variables above has been carefully
made so that it is consistent with the choice of the base levels in the
multiplicative tariff model in the previous subsection (i.e.,
\(f_{11}=1\) and \(f_{21}=1\)).\\

With the proposed indicator variables we can rewrite the log rate
\eqref{eq:log-linear-tariff} as

\begin{equation}
\log \lambda_{}= \log f_0+ \log f_{12}  \times x_1 + \log f_{22} \times x_2 +\log f_{23} \times x_3,
\label{eq:log-linear-tariff-3}
\end{equation}

which is identical to \eqref{eq:log-rate-Poi-tariff-3} when each triple
value is actually applied. For example, we can verify that the base
tariff cell \((j,k)=(1,1)\) corresponds to \((x_1, x_2,x_3)=(0, 0, 0)\),
and in turn produces \(\log \lambda=\log f_0\) or \(\lambda= f_0\) in
\eqref{eq:log-linear-tariff-3} as required.

\textbf{Poisson regression for the tariff model\}}\\
Under this specification, let us consider \(n\) policyholders in the
portfolio with the \(i\)th policyholder's risk characteristic given by a
vector of explanatory variables
\(\mathbf{ x}_i=(x_{i1}, x_{i2},x_{i3})^{\prime}\), for
\(i=1, \ldots, n\). We then recognize \eqref{eq:log-linear-tariff-3} as

\begin{equation}
\log \lambda_{i}= \beta_0+ \beta_1 \, x_{i1} + \beta_{2} \, x_{i2} +\beta_3  \, x_{i3}=\mathbf{ x}^{\prime}_i\beta, \qquad i=1, \ldots, n,
\end{equation}

where \(\beta_0, \ldots, \beta_3\) can be mapped to the corresponding
log relativities in \eqref{eq:log-linear-tariff-3}. This is exactly the
same setup as in \eqref{eq:mean-ft-Pois-7} except for the exposure
component. Therefore, by incorporating the exposure in each risk class,
the Poisson regression model for this multiplicative tariff model
finally becomes

\begin{equation}
\log \mu_i=\log \lambda_{i}+\log m_i= \log m_i+ \beta_0+ \beta_1 \, x_{i1} + \beta_{2} \, x_{i2} +\beta_3  \, x_{i3}=\log m_i+\mathbf{ x}^{\prime}_i\beta,
\end{equation}

for \(i=1, \ldots, n\). As a result, the relativities are given by

\begin{equation}
{f}_0=e^{\beta_0}, \quad {f}_{12}=e^{\beta_1}, \quad {f}_{22}=e^{\beta_2} \quad \text{and}\quad {f}_{23}=e^{\beta_3},
\label{eq:relativity-1}
\end{equation}

with \(f_{11}=1\) and \(f_{21}=1\) from the original construction. For
the actual dataset, \(\beta_i\), \(i=0,1, 2, 3\), is replaced with the
\emph{mle} \(b_i\) using the method in the technical supplement at the
end of this chapter (Section \ref{S:RC:mle-Pois-reg}).

\subsection{Numerical Examples}\label{numerical-examples}

We present two numerical examples of the Poisson regression. In the
first example we construct a Poisson regression model from
\protect\hyperlink{tab:8.2}{Table 8.2}, which is a dataset of a
hypothetical auto insurer. The second example uses an actual industry
dataset with more risk factors. As our purpose is to show how the
Poisson regression model can be used under a given classification rule,
we are not concerned with the quality of the Poisson model fit in this
chapter.

\textbf{Example 8.1: Poisson regression for the illustrative auto
insurer}

In the last few subsections we considered a dataset of a hypothetical
auto insurer with two risk factors, as given in
\protect\hyperlink{tab:8.2}{Table 8.2}. We now apply the Poisson
regression model to this dataset. As done before, we have set
\((j,k)=(1,1)\) as the base tariff cell, so that \(f_{11}=f_{21}=1\).
The result of the regression gives the coefficient estimates
\((b_0, b_1,b_2,b_3)=(-2.3359, -0.3004, -0.7837, -1.0655 )\), which in
turn produces the corresponding relativities

\begin{equation}
\nonumber
{f}_0=0.0967, \quad {f}_{12}=  0.7405, \quad {f}_{22}=0.4567 \quad \text{and}\quad {f}_{23}=0.3445.
\end{equation}

from the relation given in \eqref{eq:relativity-1}. The R script and the
output are as follows.

Show R Code

\hypertarget{toggleCodeRiskClass.1}{}
\begin{Shaded}
\begin{Highlighting}[]
\NormalTok{> mydat1<- read.csv("eg1_v1a.csv")}
\NormalTok{> mydat1}
\NormalTok{  Vtype Agebnd Expsr Claims}
\NormalTok{1     1      1  89.1      9}
\NormalTok{2     1      2 208.5      8}
\NormalTok{3     1      3 155.2      6}
\NormalTok{4     2      1  19.3      1}
\NormalTok{5     2      2 360.4     13}
\NormalTok{6     2      3 276.7      6}
\NormalTok{> VtypeF <- relevel(factor(Vtype), ref="1") # treat Vtype as factors with 1 as base.}
\NormalTok{> AgebndF <- relevel(factor(Agebnd), ref="1") # treat Age band as factors.}
\NormalTok{> Pois_reg1 = glm(Claims ~ VtypeF + AgebndF,}
\NormalTok{                    data = mydat1, family = poisson(link = log), offset = log(Expsr) )}
\NormalTok{> Pois_reg1}

\NormalTok{Coefficients:}
\NormalTok{(Intercept)      VtypeF2     AgebndF2     AgebndF3  }
\NormalTok{    -2.3359      -0.3004      -0.7837      -1.0655  }

\NormalTok{Degrees of Freedom: 5 Total (i.e. Null);  2 Residual}
\NormalTok{Null Deviance:      8.774 }
\NormalTok{Residual Deviance: 0.6514   AIC: 30.37}
\end{Highlighting}
\end{Shaded}

\begin{center}\rule{0.5\linewidth}{\linethickness}\end{center}

\textbf{Example 8.2. Poisson regression for Singapore insurance claims
data}

This actual data is a subset of the data used by
\citep{frees2008hierarchical}. The data is from the General Insurance
Association of Singapore, an organisation consisting of non-life
insurers in Singapore. The data contains the number of car accidents for
\(n=7,483\) auto insurance policies with several categorical explanatory
variables and the exposure for each policy. The explanatory variables
include four risk factors: the type of the vehicle insured (either
automobile (A) or other (O), denoted by \(\tt{Vtype}\)), the age of the
vehicle in years (\(\tt{Vage}\)), gender of the policyholder
(\(\tt{Sex}\)) and the age of the policyholder (in years, grouped into
seven categories, denoted \(\tt{Age}\)).

Based on the data description, there are several things to remember
before constructing a model (May need the table from the Jed's pdf
file). First, there are 3,842 policies with vehicle type A (automobile)
and 3,641 policies with other vehicle types. However, age and sex
information is available for the policies of vehicle type A only; the
drivers of all other types of vehicles are recorded to be aged 21 or
less with sex unspecified, except for one policy, indicating that no
driver information has been collected for non-automobile vehicles.
Second, type A vehicles are all classified as private vehicles and all
the other types are not.

When we include these risk factors, we assume that all unspecified sex
to be male. As the age information is only applicable to type A
vehicles, we set the model accordingly. That is, we apply the age
variable only to vehicles of type A. Also we used five vehicle age
bands, simplifying the original seven bands, by combining vehicle ages
0,1 and 2; the combined band is marked as level 2\footnote{corresponding
  to \(\texttt{VAgecat1}\)} in the data file\}. Thus our Poisson model
has the following explicit form:

\begin{align*}
\log \mu_i= \mathbf{ x}^{\prime}_i\beta+&\log m_i=\beta_0+\beta_1 I(Sex_i=M)+ \sum_{t=2}^6 \beta_t\, I(Vage_i=t+1) \\
&+  \sum_{t=7}^{13} \beta_t \,I(Vtype_i=A)\times I(Age_i=t-7)+\log m_i.
\end{align*}

The fitting result is given in Table \protect\hyperlink{tab:8.3}{Table
8.3}, for which we have several comments.

\begin{itemize}
\item
  The claim frequency is higher for male by 17.3\%, when other rating
  factors are held fixed. However, this may have been affected by the
  fact that all unspecified sex has been assigned to male.
\item
  Regarding the vehicle age, the claim frequency gradually decreases as
  the vehicle gets old, when other rating factors are held fixed. The
  level starts from 2 for this variable but, again, the numbering is
  nominal and does not affect the numerical result.
\item
  The policyholder age variable only applies to type A (automobile)
  vehicle, and there is no policy in the first age band. We may
  speculate that younger drivers less than age 21 drive their parents'
  cars rather than having their own because of high insurance premiums
  or related regulations. The missing relativity may be estimated by
  some interpolation or the professional judgement of the actuary. The
  claim frequency is the lowest for age band 3 and 4, but gets
  substantially higher for older age bands, a reasonable pattern seen in
  many auto insurance loss datasets.
\end{itemize}

*We also note that there is no base level in the policyholder age
variable, in the sense that no relativity is equal to 1. This is because
the variable is only applicable to vehicle type A. This does not cause a
problem numerically, but one may set the base relativity as follows if
necessary for other purposes. Since there is no policy in age band 0, we
consider band 1 as the base case. Specifically, we treat its relativity
as a product of 0.918 and 1, where the former is the common relativity
(that is, the common premium reduction) applied to all policies with
vehicle type A and the latter is the base value for age band 1. Then the
relativity of age band 2 can be seen as \(0.917=0.918 \times 0.999\),
where 0.999 is understood as the relativity for age band 2. The
remaining age bands can be treated similarly.

\[\begin{matrix}
\begin{array}{clcc}
\hline
\text{Rating factor} & \text{Level} & \text{Relativity in the tariff} & \text{Note}\\ \hline\hline
\text{Base value}  &  & 0.167 & f_0\\ \hline
\text{Sex} & 1 (F) & 1.000 & \text{Base level}\\
 & 2 (M) & 1.173 &\\\hline
 \text{Vehicle age} & 2 (0-2\text{ yrs}) & 1.000 & \text{Base level}\\
  & 3 (3-5\text{ yrs}) & 0.843 \\
  & 4 (6-10\text{ yrs}) & 0.553 \\
  & 5 (11-15\text{ yrs}) & 0.269 \\
  & 6 (16+\text{ yrs}) & 0.189 &\\\hline
  \text{Policyholder age} & 0 (0-21) & \text{N/A} & \text{No policy} \\
  \text{(Only applicable to} & 1 (22-25) & 0.918 \\
 \text{vehicle type A)}  & 2 (26-35) & 0.917 \\
  & 3 (36-45) & 0.758 \\
  & 4 (46-55) & 0.632 \\
  & 5 (56-65) &  1.102\\
  & 6 (65+) & 1.179\\ \hline \hline
\end{array}
\end{matrix}\]

\protect\hyperlink{tab:8.3}{Table 8.3} : Singapore insurance claims data

Let us try several examples based on \protect\hyperlink{tab:8.3}{Table
8.3}. Suppose a male policyholder aged 40 who owns a 7-year-old vehicle
of type A. The expected claim frequency for this policyholder is then
given by

\begin{equation}
\lambda=0.167 \times 1.173 \times 0.553 \times 0.758 = 0.082.
\end{equation}

As another example consider a female policyholder aged 60 who owns a
3-year-old vehicle of type O. The expected claim frequency for this
policyholder is

\begin{equation}
\lambda=0.167 \times 1 \times 0.843  = 0.141.
\end{equation}

Note that for this policy the age band variable is not used as the
vehicle type is not A. The R script is given as follows.

Show R Code

\hypertarget{toggleCodeRiskClass.2}{}
\begin{Shaded}
\begin{Highlighting}[]
\NormalTok{mydat <- read.csv("SingaporeAuto.csv",  quote = "", header = TRUE)}
\NormalTok{attach(mydat)}

\NormalTok{# create vehicle type as factor}
\NormalTok{TypeA = 1 * (VehicleType == "A")}
\NormalTok{table(VehicleType)}
\NormalTok{VtypeF <- as.character(VehicleType)}
\NormalTok{VtypeF[VtypeF != "A"] <- "O"}
\NormalTok{VtypeF = relevel(factor(VtypeF), ref="A")}

\NormalTok{# create gender as factor}
\NormalTok{Female = 1 * (SexInsured == "F" )}
\NormalTok{Sex = as.character(SexInsured)}
\NormalTok{Sex[Sex != "F"] <- "M"}
\NormalTok{SexF = relevel(factor(Sex), ref = "F")}

\NormalTok{# create driver age as factor}
\NormalTok{AgeCat = pmax(AgeCat - 1, 0)}
\NormalTok{AgeCatF = relevel(factor(AgeCat), ref = "0")}
\NormalTok{table(AgeCatF) # No policy in the first age band}

\NormalTok{# create vehicle age as factor}
\NormalTok{VAgeCatF = relevel( factor(VAgeCat), ref = "0" )}
\NormalTok{VAgecat1 = factor(VAgecat1, labels = }
\NormalTok{                    c("Vage0-2", "Vage3-5", "Vage6-10", "Vage11-15", "Vage15+") )}
\NormalTok{VAgecat1F = relevel( factor(VAgecat1), ref = "Vage0-2" )}

\NormalTok{# Poisson reg model}
\NormalTok{Pois_reg2 = glm(Clm_Count ~ SexF + TypeA:AgeCatF + VAgecat1F, }
\NormalTok{                   offset = LNWEIGHT, poisson(link = log) )}
\NormalTok{summary(Pois_reg2) }

\NormalTok{# compute relativities}
\NormalTok{exp(Pois_reg2}\SpecialStringTok{$coefficients)}

\SpecialStringTok{detach(mydat)}
\end{Highlighting}
\end{Shaded}

\begin{center}\rule{0.5\linewidth}{\linethickness}\end{center}

\section{Contributors and Further
Resources}\label{RC:further-reading-and-resources}

\subsubsection*{Further Reading and
References}\label{further-reading-and-references}
\addcontentsline{toc}{subsubsection}{Further Reading and References}

The Poisson regression is a special member of a more general regression
model class known as the generalized linear model (glm). The glm
develops a unified regression framework for datasets when the response
valuables are continuous, binary or discrete. The classical linear
regression model with normal error is also a member of the glm. There
are many standard statistical texts dealing with the glm, including
\citep{mccullagh1989generalized}. More accessible texts are
\citep{dobson2008introduction}, \citep{agresti1996introduction} and
\citep{faraway2016extending}. For actuarial and insurance applications
of the glm see \citep{frees2009regression}, \citep{de2008generalized}.
Also, \citep{ohlsson2010non} discusses the glm in non-life insurance
pricing context with tariff analyses.

\subsubsection*{Contributor}\label{contributor-1}
\addcontentsline{toc}{subsubsection}{Contributor}

\begin{itemize}
\tightlist
\item
  \textbf{Joseph H. T. Kim}, Yonsei University, is the principal author
  of the initital version of this chapter. Email:
  \href{mailto:jhtkim.yu@gmail.com}{\nolinkurl{jhtkim.yu@gmail.com}} for
  chapter comments and suggested improvements.
\end{itemize}

\section{Technical Supplement -- Estimating Poisson Regression
Models}\label{S:RC:mle-Pois-reg}

\textbf{Maximum Likelihood Estimation for Individual Data}

In the Poisson regression the varying Poisson mean is determined by
parameters \(\beta_i\)'s, as shown in \eqref{eq:mean-ft-Pois-7}. In this
subsection we use the maximum likelihood method to estimate these
parameters. Again, we assume that there are \(n\) policyholders and the
\(i\)th policyholder is characterized by
\(\mathbf{ x}_i=(1, x_{i1}, \ldots, x_{ik})^{\prime}\) with the observed
loss count \(y_i\). Then, from \eqref{eq:mean-ft-Pois-6} and
\eqref{eq:mean-ft-Pois-7}, the log-likelihood function of vector
\(\beta=(\beta_0, \dots, \beta_k)\) is given by

\begin{align}
\nonumber \log L(\beta)    &= l(\beta)=\sum^n_{i=1} \left( -\mu_i +y_i \, \log \mu_i -\log y_i! \right)  \\
    &  = \sum^n_{i=1} \left( -m_i \exp(\mathbf{ x}^{\prime}_i\beta) +y_i \,(\log m_i+\mathbf{ x}^{\prime}_i\beta)  -\log y_i! \right)
\label{eq:ll-Poi-reg}
\end{align}

To obtain the \emph{mle} of
\(\beta=(\beta_0, \ldots, \beta_k)^{\prime}\), we
differentiate\footnote{We use matrix derivative here.} \(l(\beta)\) with
respect to vector \(\beta\) and set it to zero:

\begin{equation}
\frac{\partial}{\partial \beta}l(\beta)\Bigg{|}_{\beta=\mathbf{b}}=\sum^n_{i=1} \left(y_i -m_i \exp(\mathbf{ x}^{\prime}_i \mathbf{ b}) \right)\mathbf{ x}_i=\mathbf{ 0}.
\label{eq:score-ft-Poi}
\end{equation}

Numerically solving this equation system gives the \emph{mle} of
\(\beta\), denoted by \(\mathbf{ b}=(b_0, b_1, \ldots, b_k)^{\prime}\).
Note that, as \(\mathbf{ x}_i=(1, x_{i1}, \ldots, x_{ik})^{\prime}\) is
a column vector, equation \eqref{eq:score-ft-Poi} is a system of \(k+1\)
equations with both sides written as column vectors of size \(k+1\). If
we denote \(\hat{\mu}_i=m_i \exp(\mathbf{ x}^{\prime}_i \mathbf{ b})\),
we can rewrite \eqref{eq:score-ft-Poi} as

\begin{equation}
\sum^n_{i=1} \left(y_i -\hat{\mu}_i \right)\mathbf{ x}_i=\mathbf{ 0}.
\end{equation}

Since the solution \(\mathbf{ b}\) satisfies this equation, it follows
that the first among the array of \(k+1\) equations, corresponding to
the first constant element of \(\mathbf{ x}_i\), yields

\begin{equation}
\sum^n_{i=1}\left( y_i -\hat{\mu}_i \right)\times 1={ 0},
\end{equation}

which implies that we must have

\begin{equation}
n^{-1}\sum_{i=1}^n y_i =\bar{y}=n^{-1}\sum_{i=1}^n \hat{\mu}_i.
\end{equation}

This is an interesting property saying that the average of the
individual losses, \(\bar{y}\), is same as the average of the estimated
values. That is, the sample mean is preserved under the fitted Poisson
regression model.

\textbf{Maximum Likelihood Estimation for Grouped Data}

Sometimes the data is not available at the individual policy level. For
example, \protect\hyperlink{tab:8.2}{Table 8.2} provides collective loss
information for each risk class after grouping individual policies. When
this is the case, \(y_i\) and \(m_i\), the quantities needed for the
\emph{mle} calculation in \eqref{eq:score-ft-Poi}, are unavailable for
each \(i\). However this does not pose a problem as long as we have the
total loss counts and total exposure for each risk class.

To elaborate, let us assume that there are \(K\) different risk classes,
and further that, in the \(k\)th risk class, we have \(n_k\) policies
with the total exposure \(m_{(k)}\) and the average loss count
\(\bar{y}_{(k)}\), for \(k=1, \ldots, K\); the total loss count for the
\(k\)th risk class is then \(n_k\, \bar{y}_{(k)}\). We denote the set of
indices of the policies belonging to the \(k\)th class by \(C_k\). As
all policies in a given risk class share the same risk characteristics,
we may denote \(\mathbf{ x}_i=\mathbf{ x}_{(k)}\) for all \(i \in C_k\).
With this notation, we can rewrite \eqref{eq:score-ft-Poi} as

\begin{align}
\nonumber \sum^n_{i=1} \left(y_i -m_i \exp(\mathbf{ x}^{\prime}_i \mathbf{ b}) \right)\mathbf{ x}_i &= \sum^K_{k=1}\Big{\{}\sum_{i \in C_k} \left(y_i -m_i \exp(\mathbf{ x}^{\prime}_i \mathbf{ b}) \right)\mathbf{ x}_i  \Big{\}} \\
\nonumber     &  =\sum^K_{k=1}\Big{\{} \sum_{i \in C_k} \left(y_i -m_i \exp(\mathbf{ x}^{\prime}_{(k)} \mathbf{ b}) \right)\mathbf{ x}_{(k)}  \Big{\}} \\
\nonumber     &  =\sum^K_{k=1}\Big{\{}  \Big(\sum_{i \in C_k}y_i -\sum_{i \in C_k}m_i \exp(\mathbf{ x}^{\prime}_{(k)} \mathbf{ b}) \Big)\mathbf{ x}_{(k)}  \Big{\}} \\
      &  =\sum^K_{k=1} \Big(n_k\, \bar{y}_{(k)}-m_{(k)} \exp(\mathbf{ x}^{\prime}_{(k)} \mathbf{ b}) \Big)\mathbf{ x}_{(k)} =0.
\label{eq:score-ft-Poi-2}
\end{align}

Since \(n_k\, \bar{y}_{(k)}\) in \eqref{eq:score-ft-Poi-2} represents the
total loss count for the \(k\)th risk class and \(m_{(k)}\) is its total
exposure, we see that for the Poisson regression the \emph{mle}
\(\mathbf{ b}\) is the same whether if we use the individual data or the
grouped data.

\textbf{Information matrix}\\
Taking second derivatives to \eqref{eq:ll-Poi-reg} gives the information
matrix of the \emph{mle} estimators,

\begin{equation}
\mathbf{ I}(\beta)=-\mathrm{E~}{\left( \frac{\partial^2}{\partial \beta\partial \beta^{\prime}}l(\beta) \right)}=\sum^n_{i=1}m_i \exp(\mathbf{ x}^{\prime}_i \mathbf{ \beta})\mathbf{ x}_i \mathbf{ x}_i^{\prime}=\sum^n_{i=1} {\mu}_i \mathbf{ x}_i \mathbf{ x}_i^{\prime}.
\label{eq:Inf-mtx-Poi}
\end{equation}

For actual datasets, \({\mu}_i\) in \eqref{eq:Inf-mtx-Poi} is replaced
with \(\hat{\mu}_i=m_i \exp(\mathbf{ x}^{\prime}_i \mathbf{ b})\) to
estimate the relevant variances and covariances of the \emph{mle}
\(\mathbf{ b}\) or its functions.

For grouped datasets, we have

\begin{equation}
\mathbf{ I}(\beta)=\sum^K_{k=1} \Big{\{}\sum_{i \in C_k}m_i \exp(\mathbf{ x}^{\prime}_i \mathbf{ \beta})\mathbf{ x}_i \mathbf{ x}_i^{\prime} \Big{\}}=\sum^K_{k=1} m_{(k)} \exp(\mathbf{ x}^{\prime}_{(k)} \mathbf{ \beta})\mathbf{ x}_{(k)} \mathbf{ x}_{(k)}^{\prime}.
\end{equation}

\chapter{Experience Rating Using Credibility
Theory}\label{experience-rating-using-credibility-theory}

\emph{Chapter Preview.} This chapter introduces credibility theory which
is an important actuarial tool for estimating pure premiums,
frequencies, and severities for individual risks or classes of risks.
Credibility theory provides a convenient framework for combining the
experience for an individual risk or class with other data to produce
more stable and accurate estimates. Several models for calculating
credibility estimates will be discussed including limited fluctuation,
Bühlmann, Bühlmann-Straub, and nonparametric and semiparametric
credibility methods. The chapter will also show a connection between
credibility theory and Bayesian estimation which was introduced in
Chapter \ref{C:ModelSelection}.

\section{Introduction to Applications of Credibility
Theory}\label{introduction-to-applications-of-credibility-theory}

What premium should be charged to provide insurance? The answer depends
upon the exposure to the risk of loss. A common method to compute an
insurance premium is to rate an insured using a classification rating
plan. A classification plan is used to select an insurance rate based on
an insured's rating characteristics such as geographic territory, age,
etc. All classification rating plans use a limited set of criteria to
group insureds into a ``class'' and there will be variation in the risk
of loss among insureds within the class.

An experience rating plan attempts to capture some of the variation in
the risk of loss among insureds within a rating class by using the
insured's own loss experience to complement the rate from the
classification rating plan. One way to do this is to use a credibility
weight \(Z\) with \(0\leq Z \leq 1\) to compute

\begin{equation*}
\hat{R}=Z\bar{X}+(1-Z)M,
\end{equation*}

\begin{eqnarray*}
\hat{R}&=&\textrm{credibility weighted rate for risk,}\\
           \bar{X}&=&\textrm{average loss for the risk over a specified time period,}\\
                  M&=&\textrm{the rate for the classification group, often called the manual rate.}\\
\end{eqnarray*}

For a large risk whose loss experience is stable from year to year,
\(Z\) might be close to 1. For a smaller risk whose losses vary widely
from year to year, \(Z\) may be close to 0.

Credibility theory is also used for computing rates for individual
classes within a classification rating plan. When classification plan
rates are being determined, some or many of the groups may not have
sufficient data to produce stable and reliable rates. The actual loss
experience for a group will be assigned a credibility weight \(Z\) and
the complement of credibility \(1-Z\) may be given to the average
experience for risk across all classes. Or, if a class rating plan is
being updated, the complement of credibility may be assigned to the
current class rate. Credibility theory can also be applied to the
calculation of expected frequencies and severities.

Computing numeric values for \(Z\) requires analysis and understanding
of the data. What are the variances in the number of losses and sizes of
losses for risks? What is the variance between expected values across
risks?

\section{Limited Fluctuation
Credibility}\label{limited-fluctuation-credibility}

\begin{center}\rule{0.5\linewidth}{\linethickness}\end{center}

In this section, you learn how to:

\begin{itemize}
\tightlist
\item
  Calculate full credibility standards for number of claims, average
  size of claims, and aggregate losses.
\item
  Learn how the relationship between means and variances of underlying
  distributions affects full credibility standards.
\item
  Determine credibility-weight \(Z\) using the square-root partial
  credibility formula.
\end{itemize}

\begin{center}\rule{0.5\linewidth}{\linethickness}\end{center}

Limited fluctuation credibility, also called ``classical credibility'',
was given this name because the method explicitly attempts to limit
fluctuations in estimates for claim frequencies, severities, or losses.
For example, suppose that you want to estimate the expected number of
claims for a group of risks in an insurance rating class. How many risks
are needed in the class to ensure that a specified level of accuracy is
attained in the estimate? First the question will be considered from the
perspective of how many claims are needed.

\subsection{Full Credibility for Claim Frequency}\label{S:frequency}

Let \(N\) be a random variable representing the number of claims for a
group of risks. The observed number of claims will be used to estimate
\(\mu_N=\mathrm{E}[N]\), the expected number of claims. How big does
\(\mu_N\) need to be to get a good estimate? One way to quantify the
accuracy of the estimate would be a statement like: ``The observed value
of \(N\) should be within 5\(\%\) of \(\mu_N\) at least 90\(\%\) of the
time." Writing this as a mathematical expression would give
\(\Pr[0.95\mu_N\leq N \leq1.05\mu_N] \geq 0.90\). Generalizing this
statement by letting \(k\) replace 5\(\%\) and probability \(p\) replace
0.90 produces a confidence interval

\begin{equation}
\Pr[(1-k)\mu_N\leq N \leq(1+k)\mu_N] \geq p.
\label{eq:confidence-interval}
\end{equation}

The expected number of claims required for the probability on the
left-hand side of \eqref{eq:confidence-interval} to equal \(p\) is called
the \textbf{full credibility} standard.

If the expected number of claims is greater than or equal to the full
credibility standard then full credibility can be assigned to the data
so \(Z=1\). Usually the expected value \(\mu_N\) is not known so full
credibility will be assigned to the data if the actual observed value of
\(N\) is greater than or equal to the full credibility standard. The
\(k\) and \(p\) values must be selected and the actuary may rely on
experience, judgment, and other factors in making the choices.

Subtracting \(\mu_N\) from each term in \eqref{eq:confidence-interval} and
dividing by the standard deviation \(\sigma_N\) of \(N\) gives

\begin{equation}
\Pr\left[\frac{-k\mu_N}{\sigma_N}\leq \frac{N-\mu_N}{\sigma_N} \leq \frac{k\mu_N}{\sigma_N}\right] \geq p.
\label{eq:normalized-interval}
\end{equation}

For large values of \(\mu_N=\mathrm{E}[N]\) it may be reasonable to
approximate the distribution for \(Z=(N-\mu_N)/\sigma_N\) with the
standard normal distribution.

Let \(y_p\) be the value such that
\(\Pr[-y_p\leq Z \leq y_p]=\Phi(y_p)-\Phi(-y_p)=p\) where \(\Phi( )\) is
the cumulative standard normal distribution. Because
\(\Phi(-y_p)=1-\Phi(y_p)\), the equality can be rewritten as
\(2\Phi(y_p)-1=p\). Solving for \(y_p\) gives \(y_p=\Phi^{-1}((p+1)/2)\)
where \(\Phi^{-1}( )\) is the inverse of the cumulative normal.

Equation \eqref{eq:normalized-interval} will be satisfied if
\(k\mu_N/\sigma_N \geq y_p\) assuming the normal approximation. First we
will consider this inequality for the case when \(N\) has a Poisson
distribution: \(\Pr[N=n] = \lambda^n\textrm{e}^{\lambda}/n!\). Because
\(\lambda=\mu_N=\sigma_N^2\) for the Poisson, taking square roots yields
\(\mu_N^{1/2}=\sigma_N\). So, \(k\mu_N/\mu_N^{1/2} \geq y_p\) which is
equivalent to \(\mu_N \geq (y_p/k)^2\). Let's define \(\lambda_{kp}\) to
be the value of \(\mu_N\) for which equality holds. Then the full
credibility standard for the Poission distribution is

\begin{equation}
\lambda_{kp} = \left(\frac{y_p}{k}\right)^2 \textrm{with } y_p=\Phi^{-1}((p+1)/2).
\label{eq:full-credibility-Poisson}
\end{equation}

If the expected number of claims \(\mu_N\) is greater than or equal to
\(\lambda_{kp}\) then equation \eqref{eq:confidence-interval} is assumed
to hold and full credibility can be assigned to the data. As noted
previously, because \(\mu_N\) is usually unknown, full credibility is
given if the observed value of \(N\) satisfies \(N \geq \lambda_{kp}.\)

\textbf{Example 9.2.1.} The full credibility standard is set so that the
observed number of claims is to be within 5\% of the expected value with
probability \(p=0.95\). If the number of claims has a Poisson
distribution find the number of claims needed for full credibility.

Show Example Solution

\hypertarget{toggleExampleCred.2.1}{}
\textbf{Solution} Referring to a normal table,
\(y_p=\Phi^{-1}((p+1)/2)=\Phi^{-1}((0.95+1)/2)\)=\(\Phi^{-1}(0.975)=1.960\).
Using this value and \(k=.05\) then
\(\lambda_{kp} = (y_p/k)^{2}=(1.960/0.05)^{2}=1,536.64\). After rounding
up the full credibility standard is 1,537.

\begin{center}\rule{0.5\linewidth}{\linethickness}\end{center}

If claims are not Poisson distributed then equation
\eqref{eq:normalized-interval} does not imply
\eqref{eq:full-credibility-Poisson}. Setting the upper bound of \(Z\) in
\eqref{eq:normalized-interval} equal to \(y_p\) gives
\(k\mu_N/\sigma_N=y_p\). Squaring both sides and moving everything to
the right side except for one of the \(\mu_N\)'s gives
\(\mu_N=(y_p/k)^2(\sigma_N^2/\mu_N)\). This is the full credibility
standard for frequency and will be denoted by \(n_f\),

\begin{equation}
n_f=\left(\frac{y_p}{k}\right)^2\left(\frac{\sigma_N^2}{\mu_N}\right)=\lambda_{kp}\left(\frac{\sigma_N^2}{\mu_N}\right).
\label{eq:full-credibility-frequency}
\end{equation}

This is the same equation as the Poisson full credibility standard
except for the \((\sigma_N^2/\mu_N)\) multiplier. When the claims
distribution is Poisson this extra term is one because the variance
equals the mean.

\textbf{Example 9.2.2.} The full credibility standard is set so that the
total number of claims is to be within 5\(\%\) of the observed value
with probability \(p=0.95\). The number of claims has a negative
binomial distribution

\begin{equation*}
\Pr(N=x)={x+r-1\choose x} \left(\frac{1}{1+\beta}\right)^r \left(\frac{\beta}{1+\beta}\right)^x
\end{equation*}

with \(\beta=1\). Calculate the full credibility standard.

Show Example Solution

\hypertarget{toggleExampleCred.2.2}{}
\textbf{Solution} From the prior example, \(\lambda_{kp} =1,536.64\).
The mean and variance for the negative binomial are
\(\mathrm{E}(N)=r\beta\) and \(\mathrm{Var}(N)=r\beta(1+\beta)\) so
\((\sigma_N^2/\mu_N)=(r\beta(1+\beta)/(r\beta))=1+\beta\) which equals 2
when \(\beta=1\). So,
\(n_f=\lambda_{kp}(\sigma_N^2/\mu_N)=1,536.64(2)=3,073.28\) and rounding
up gives a full credibility standard of 3,074.

\begin{center}\rule{0.5\linewidth}{\linethickness}\end{center}

We see that the negative binomial distribution with
\((\sigma_N^2/\mu_N)>1\) requires more claims for full credibility than
a Poission distribution for the same \(k\) and \(p\) values. The next
example shows that a binomial distribution which has
\((\sigma_N^2/\mu_N)<1\) will need fewer claims for full credibility.

\textbf{Example 9.2.3.} The full credibility standard is set so that the
total number of claims is to be within 5\(\%\) of the observed value
with probability \(p=0.95\). The number of claims has a binomial
distribution

\begin{equation*}
\Pr(N=x)={m\choose x}q^x(1-q)^{m-x}.
\end{equation*}

Calculate the full credibility standard for \(q=1/4\).

Show Example Solution

\hypertarget{toggleExampleCred.2.3}{}
\textbf{Solution} From the first example in this section
\(\lambda_{kp} =1,536.64\). The mean and variance for a binomial are
\(\mathrm{E}(N)=mq\) and \(\mathrm{Var}(N)=mq(1-q)\) so
\((\sigma_N^2/\mu_N)=(mq(1-q)/(mq))=1-q\) which equals 3/4 when
\(q=1/4\). So,
\(n_f=\lambda_{kp}(\sigma_N^2/\mu_N)=1,536.64(3/4)=1,152.48\) and
rounding up gives a full credibility standard of 1,153.

\begin{center}\rule{0.5\linewidth}{\linethickness}\end{center}

Rather than use expected number of claims to define the full credibility
standard, the number of exposures can be used for the full credibility
standard. An exposure is a measure of risk. For example, one car insured
for a full year would be one car-year. Two cars each insured for exactly
one-half year would also result in one car-year. Car-years attempt to
quantify exposure to loss. Two car-years would be expected to generate
twice as many claims as one car-year if the vehicles have the same risk
of loss. To translate a full credibility standard denominated in terms
of number of claims to a full credibility standard denominated in
exposures one needs a reasonable estimate of the expected number of
claims per exposure.

\textbf{Example 9.2.4.} The full credibility standard should be selected
so that the observed number of claims will be within 5\(\%\) of the
expected value with probability \(p=0.95\). The number of claims has a
Poisson distribution. If one exposure is expected to have about 0.20
claims per year, find the number of exposures needed for full
credibility.

Show Example Solution

\hypertarget{toggleExampleCred.2.4}{}
\textbf{Solution} With \(p=0.95\) and \(k=.05\),
\(\lambda_{kp} = (y_p/k)^{2}=(1.960/0.05)^{2}=1,536.64\) claims are
required for full credibility. The claims frequency rate is 0.20
claims/exposures. To convert the full credibility standard to a standard
denominated in exposures the calculation is: (1,536.64 claims)/(0.20
claims/exposures) = 7,683.20 exposures. This can be rounded up to 7,684.

\begin{center}\rule{0.5\linewidth}{\linethickness}\end{center}

Frequency can be defined as the number of claims per exposure. Letting
\(m\) represent number of exposures then the observed claim frequency is
\(N/m\) which is used to estimate \(\mathrm{E}(N/m)\):

\begin{equation*}
\Pr[(1-k)\mathrm{E}(N/m)\leq N/m \leq(1+k)\mathrm{E}(N/m)] \geq p.
\end{equation*}

.

Because the number of exposures is not a random variable,
\(\mathrm{E}(N/m)=\mathrm{E}(N)/m=\mu_N/m\) and the prior equation
becomes

\begin{equation*}
\Pr\left[(1-k)\frac{\mu_N}{m}\leq \frac{N}{m} \leq(1+k)\frac{\mu_N}{m}\right] \geq p.
\end{equation*}

Mulitplying through by \(m\) results in equation
\eqref{eq:confidence-interval} at the beginning of the section. The full
credibility standards that were developed for estimating expected number
of claims also apply to frequency.

\subsection{Full Credibility for Aggregate Losses and Pure
Premium}\label{full-credibility-for-aggregate-losses-and-pure-premium}

Aggregate losses are the total of all loss amounts for a risk or group
of risks. Letting \(S\) represent aggregate losses then

\begin{equation*}
S=X_1+X_2+\cdots+X_N.
\end{equation*}

The random variable \(N\) represents the number of losses and random
variables \(X_1, X_2,\ldots,X_N\) are the individual loss amounts. In
this section it is assumed that \(N\) is independent of the loss amounts
and that \(X_1, X_2,\ldots,X_N\) are \emph{iid}.

The mean and variance of \(S\) are

\begin{equation*}
\mu_S=\mathrm{E}(S)=\mathrm{E}(N)\mathrm{E}(X)=\mu_N\mu_X\textrm{  and}
\end{equation*}

\begin{equation*}
\sigma^{2}_S=\mathrm{Var}(S)=\mathrm{E}(N)\mathrm{Var}(X)+[\mathrm{E}(X)]^{2}\mathrm{Var}(N)=\mu_N\sigma^{2}_X+\mu^{2}_X\sigma^{2}_N.
\end{equation*}

where \(X\) is the amount of a single loss.

Observed losses \(S\) will be used to estimate expected losses
\(\mu_S=\mathrm{E}(S)\). As with the frequency model in the previous
section, the observed losses must be close to the expected losses as
quantified in the equation

\begin{equation*}
\Pr[(1-k)\mu_S\leq S \leq(1+k)\mu_S] \geq p.
\end{equation*}

\noindent After subtracting the mean and dividing by the standard
deviation,

\begin{equation*}
\Pr\left[\frac{-k\mu_S}{\sigma_S}\leq Z \leq \frac{k\mu_S}{\sigma_S}\right] \geq p
\end{equation*}

with \(Z = (S-\mu_S)/\sigma_S\). As done in the previous section the
distribution for \(Z\) is assumed to be normal and
\(k\mu_S/\sigma_S=y_p=\Phi^{-1}((p+1)/2)\). This equation can be
rewritten as \(\mu_S^2=(y_p/k)^2\sigma_S^2\). Using the prior formulas
for \(\mu_S\) and \(\sigma_{S}^2\) gives
\((\mu_N\mu_X)^2=(y_p/k)^2(\mu_N\sigma^{2}_X+\mu^{2}_X\sigma^{2}_N)\).
Dividing both sides by \(\mu_N\mu_X^2\) and reordering terms on the
right side results in a full credibility standard \(n_S\) for aggregate
losses

\begin{equation}
n_S=\left(\frac{y_p}{k}\right)^2\left[\left(\frac{\sigma_N^2}{\mu_N}\right)+\left(\frac{\sigma_X}{\mu_X}\right)^2\right]=\lambda_{kp}\left[\left(\frac{\sigma_N^2}{\mu_N}\right)+\left(\frac{\sigma_X}{\mu_X}\right)^2\right].
\label{eq:full-credibility-losses}
\end{equation}

\textbf{Example 9.2.5.} The number of claims has a Poisson distribution.
Individual loss amounts are independently and identically distributed
with a Pareto distribution \(F(x)=1-[\theta/(x+\theta)]^{\alpha}\). The
number of claims and loss amounts are independent. If observed aggregate
losses should be within 5\(\%\) of the expected value with probability
\(p=0.95\), how many losses are required for full credibility?

Show Example Solution

\hypertarget{toggleExampleCred.2.5}{}
\textbf{Solution} Because the number of claims is Poission,
\((\sigma_N^2/\mu_N)=1\). The mean of the Pareto is
\(\mu_X=\theta/(\alpha-1)\) and the variance is
\(\sigma_X^2=\theta^{2}\alpha/[(\alpha-1)^{2}(\alpha-2)]\) so
\((\sigma_X/\mu_X)^2=\alpha/(\alpha-2)\). Combining the frequency and
severity terms gives
\([(\sigma_N^2/\mu_N)+(\sigma_X/\mu_X)^2]=2(\alpha-1)/(\alpha-2)\). From
a normal table \(y_p=\Phi^{-1}((0.95+1)/2)=1.960\). The full credibility
standard is
\(n_S=(1.96/0.05)^{2}[2(\alpha-1)/(\alpha-2)]=3,073.28(\alpha-1)/(\alpha-2)\).
Suppose \(\alpha=3\) then \(n_S=6,146.56\) for a full credibility
standard of 6,147. Note that considerably more claims are needed for
full credibility for aggregate losses that frequency alone.

\begin{center}\rule{0.5\linewidth}{\linethickness}\end{center}

When the number of claims are Poisson distributed then equation
\eqref{eq:full-credibility-losses} can be simplified using
\((\sigma_N^2/\mu_N)=1\). It follows that
\([(\sigma_N^2/\mu_N)+(\sigma_X/\mu_X)^2]=[1+(\sigma_X/\mu_X)^2]=[(\mu_x^2+\sigma_X^2)/\mu_X^2]=\mathrm{E}(X^2)/\mathrm{E}(X)^2\)
using the relationship \(\mu_X^2+\sigma_X^2=\mathrm{E}(X^2)\). The full
credibility standard is
\(n_S=\lambda_{kp}\mathrm{E}(X^2)/\mathrm{E}(X)^2\).

The pure premium \(PP\) is equal to aggregate losses \(S\) divided by
exposures \(m\): \(PP=S/m\). The full credibility standard for pure
premium will require

\begin{equation*}
\Pr\left[(1-k)\mu_{PP}\leq PP \leq(1+k)\mu_{PP}\right] \geq p.
\end{equation*}

\noindent The number of exposures \(m\) is assumed fixed and not a
random variable so \(\mu_{PP}=\mathrm{E}(S/m)=\mathrm{E}(S)/m=\mu_S/m\).

\begin{equation*}
\Pr\left[(1-k)\left(\frac{\mu_S}{m}\right)\leq \left(\frac{S}{m}\right) \leq(1+k)\left(\frac{\mu_S}{m}\right)\right] \geq p.
\end{equation*}

\noindent Multiplying through by exposures \(m\) returns the confidence
interval for losses

\begin{equation*}
\Pr[(1-k)\mu_S\leq S \leq(1+k)\mu_S] \geq p.
\end{equation*}

\noindent This means that the full credibility standard \(n_{PP}\) for
the pure premium is the same as that for aggregate losses

\begin{equation*}
n_{PP}=n_S=\lambda_{kp}\left[\left(\frac{\sigma_N^2}{\mu_n}\right)+\left(\frac{\sigma_X}{\mu_X}\right)^2\right].
\end{equation*}

\subsection{Full Credibility for
Severity}\label{full-credibility-for-severity}

Let \(X\) be a random variable representing the size of one claim. Claim
severity is \(\mu_X=\mathrm{E}(X)\). Suppose that
\({X_1,X_2, \ldots, X_n}\) is a random sample of \(n\) claims that will
be used to estimate claim severity \(\mu_X\). The claims are assumed to
be \emph{iid}. The average value of the sample is

\begin{equation*}
\bar{X}=\frac{1}{n}\left(X_1+X_2+\cdots+X_n\right).
\end{equation*}

How big does \(n\) need to be to get a good estimate? Note that \(n\) is
not a random variable whereas it is in the aggregate loss model.

In Section \ref{S:frequency} the accuracy of an estimator was defined in
terms of a confidence interval. For severity this confidence interval is

\begin{equation*}
\Pr[(1-k)\mu_X\leq \bar{X} \leq(1+k)\mu_X ]\geq p
\end{equation*}

\noindent where \(k\) and \(p\) need to be specified. Following the
steps in Section \ref{S:frequency}, mean claim severity \(\mu_X\) is
subtracted from each term and the standard deviation of the claim
severity estimator \(\sigma_{\bar{X}}\) is divided into each term
yielding

\begin{equation*}
\Pr\left[\frac{-k\mu_X}{\sigma_{\bar{X}}}\leq Z \leq \frac{k\mu_X}{\sigma_{\bar{X}}}\right] \geq p
\end{equation*}

with \(Z = (\bar{X}-\mu_X)/\sigma_X\). As in prior sections, it is
assumed that \(Z\) is approximately normally distributed and the prior
equation is satistifed if \(k\mu_X/\sigma_{\bar{X}}\geq y_p\) with
\(y_p=\Phi^{-1}((p+1)/2)\). Because \(\bar{X}\) is the average of
individual claims \(X_1, X_2,\dots, X_n\), its standard deviation is
equal to the standard deviation of an individual claim divided by
\(\sqrt{n}\): \(\sigma_{\bar{X}}=\sigma_X/\sqrt{n}\). So,
\(k\mu_X/(\sigma_X/\sqrt{n})\geq y_p\) and with a little algebra this
can be rewritten as \(n \geq (y_p/k)^2(\sigma_X/\mu_X)^2\). The full
credibility standard for severity is

\begin{equation}
n_X=\left(\frac{y_p}{k}\right)^2\left(\frac{\sigma_X}{\mu_X}\right)^2=\lambda_{kp}\left(\frac{\sigma_X}{\mu_X}\right)^2.
\label{eq:full-credibility-frequency}
\end{equation}

Note that the term \(\sigma_X/\mu_X\) is the coefficient of variation
for an individual claim. Even though \(\lambda_{kp}\) is the full
credibility standard for frequency given a Poisson distribution, there
is no assumption about the distribution for the number of claims.

\textbf{Example 9.2.6.} Individual loss amounts are independently and
identically distributed with a Pareto distribution
\(F(x)=1-[\theta/(x+\theta)]^{\alpha}\). How many claims are required
for the average severity of observed claims to be within 5\(\%\) of the
expected severity with probability \(p=0.95\)?

Show Example Solution

\hypertarget{toggleExampleCred.2.6}{}
\textbf{Solution} The mean of the Pareto is \(\mu_X=\theta/(\alpha-1)\)
and the variance is
\(\sigma_X^2=\theta^{2}\alpha/[(\alpha-1)^{2}(\alpha-2)]\) so
\((\sigma_X/\mu_X)^2=\alpha/(\alpha-2)\). From a normal table
\(y_p=\Phi^{-1}((0.95+1)/2)=1.960\). The full credibility standard is
\(n_X=(1.96/0.05)^{2}[\alpha/(\alpha-2)]=1,536.64\alpha/(\alpha-2)\).
Suppose \(\alpha=3\) then \(n_X=4,609.92\) for a full credibility
standard of 4,610.

\begin{center}\rule{0.5\linewidth}{\linethickness}\end{center}

\subsection{Partial Credibility}\label{partial-credibility}

In prior sections full credibility standards were calculated for
estimating frequency (\(n_f\)), pure premium (\(n_{PP}\)), and severity
(\(n_X\)) - in this section these full credibility standards will be
denoted by \(n_{0}\). In each case the full credibility standard was the
expected number of claims required to achieve a defined level of
accuracy when using empirical data to estimate an expected value. If the
observed number of claims is greater than or equal to the full
credibility standard then a full credibility weight \(Z=1\) is given to
the data.

In limited fluctuation credibility, credibility weights \(Z\) assigned
to data are

\begin{equation*}
Z=\quad \sqrt{\frac{n}{n_{0}}} \quad \textrm{if} \quad   n < n_{0} \quad \textrm{and}  \quad Z=\quad 1 \quad \textrm{for} \quad   n \geq n_{0}
\end{equation*}

where \(n_0\) is the full credibility standard. The quantity \(n\) is
the number of claims for the data that is used to estimate the expected
frequency, severity, or pure premium.

\textbf{Example 9.2.7.} The number of claims has a Poisson distribution.
Individual loss amounts are independently and identically distributed
with a Pareto distribution \(F(x)=1-[\theta/(x+\theta)]^{\alpha}\).
Assume that \(\alpha=3\). The number of claims and loss amounts are
independent. The full credibility standard is that the observed pure
premium should be within 5\(\%\) of the expected value with probability
\(p=0.95\). What credibility \(Z\) is assigned to a pure premium
computed from 1,000 claims?

Show Example Solution

\hypertarget{toggleExampleCred.2.7}{}
\textbf{Solution} Because the number of claims is Poisson,

\[
\frac{\mathrm{E}(X^2)}{[\mathrm{E}~X]^2}
=\frac{\sigma_N^2}{\mu_N}+\left(\frac{\sigma_X}{\mu_X}\right)^2.
\]

The mean of the Pareto is \(\mu_X=\theta/(\alpha-1)\) and the second
moment is \(\mathrm{E}(X^2)=2\theta^{2}/[(\alpha-1)(\alpha-2)]\) so
\(\mathrm{E}(X^2)/[\mathrm{E}~X]^2=2(\alpha-1)/(\alpha-2)\). From a
normal table \(y_p=\Phi^{-1}((0.95+1)/2)=1.960\). The full credibility
standard is

\[n_{PP}=(1.96/0.05)^{2}[2(\alpha-1)/(\alpha-2)]=3,073.28(\alpha-1)/(\alpha-2)\]
and if \(\alpha=3\) then \(n_0=n_{PP}=6,146.56\) or 6,147 if rounded up.
The credibility assigned to 1,000 claims is
\(Z=(1,000/6,147)^{1/2}=0.40\).

\begin{center}\rule{0.5\linewidth}{\linethickness}\end{center}

Limited fluctuation credibility uses the formula \(Z=\sqrt{n/n_0}\) to
limit the fluctuation in the credibility-weighted estimate to match the
fluctuation allowed for data with expected claims at the full
credibility standard. Variance or standard deviation is used as the
measure of fluctuation. Rather than derive the square-root formula an
example is shown

Suppose that average claim severity is being estimated from a sample of
size \(n\) that is less that the full credibility standard \(n_0=n_X\).
Applying credibility theory the estimate \(\hat{\mu}_X\) would be

\begin{equation*}
\hat{\mu}_X=Z\bar{X}+(1-Z)M_X
\end{equation*}

with \(\bar{X}=(X_1+X_2+\cdots+X_n)/n\) and independent random variables
\(X_i\) representing the sizes of individual claims. The complement of
credibility is applied to \(M_X\) which could be last year's estimated
average severity adjusted for inflation, the average severity for a much
larger pool of risks, or some other relevant quantity selected by the
actuary. It is assumed that the variance of \(M_X\) is zero or
negligible. With this assumption

\begin{equation*}
\mathrm{Var}(\hat{\mu}_X)=\mathrm{Var}(Z\bar{X})=Z^2\mathrm{Var}(\bar{X})=\frac{n}{n_0}\mathrm{Var}(\bar{X}).
\end{equation*}

Because \(\bar{X}=(X_1+X_2+\cdots+X_n)/n\) it follows that
\(\mathrm{Var}(\bar{X})=\mathrm{Var}(X)/n\) where random variable \(X\)
is one claim. So,

\begin{equation*}
\mathrm{Var}(\hat{\mu}_X)=\frac{n}{n_0}\mathrm{Var}(\bar{X})=\frac{n}{n_0}\frac{\mathrm{Var}(X)}{n}=\frac{\mathrm{Var}(X)}{n_0}.
\end{equation*}

The last term is exactly the variance of a sample mean \(\bar{X}\) when
the sample size is equal to the full credibility standard \(n_0=n_X\).

\section{Bühlmann Credibility}\label{buhlmann-credibility}

\begin{center}\rule{0.5\linewidth}{\linethickness}\end{center}

In this section, you learn how to:

\begin{itemize}
\tightlist
\item
  Compute a credibility-weighted estimate for the expected loss for a
  risk or group of risks.
\item
  Determine the credibility \(Z\) assigned to observations.
\item
  Calculate the values required in Bühlmann credibility including the
  Expected Value of the Process Variance (\emph{EPV}), Variance of the
  Hypothetical Means (\emph{VHM}) and collective mean \(\mu\).
\item
  Recognize situations when the Bühlmann model is appropriate.
\end{itemize}

\begin{center}\rule{0.5\linewidth}{\linethickness}\end{center}

A classification rating plan groups policyholders together into classes
based on risk characteristics. Although policyholders within a class
have similarities, they are not identical and their expected losses will
not be exactly the same. An experience rating plan can supplement a
class rating plan by credibility weighting an individual policyholder's
loss experience with the class rate to produce a more accurate rate for
the policyholder.

In the presentation of Bühlmann credibility it is convenient to assign a
risk parameter \(\theta\) to each policyholder. Losses \(X\) for the
policyholder will have a common distribution function \(F_{\theta}(x)\)
with mean \(\mu(\theta)=\mathrm{E}(X|\theta)\) and variance
\(\sigma^2(\theta)=\mathrm{Var}(X|\theta)\). In the prior sentence
\emph{losses} can represent pure premiums, aggegrate losses, number of
claims, claim severities, or some other measure of loss. Parameter
\(\theta\) can be continuous, discrete, or multivariate depending on the
model.

If the policyholder had losses \(x_1, \ldots, x_n\) during \(n\)
observation periods then we want to find
E(\(\mu(\theta)|x_1,\ldots,\ldots, x_n)\), the conditional expectation
of \(\mu(\theta)\) given \(x_1,\ldots, x_n\). Another way to view this
is to consider random variable \(X_{n+1}\) which is the observation
during period \(n+1\). Finding E\((X_{n+1}|x_1, x_2,\ldots, x_n)\) is
equivalent to finding E(\(\mu(\theta)|x_1, x_2,\ldots, x_n)\) assuming
that \(X_1,\ldots, X_n, X_{n+1}\) are \emph{iid}.

The Bühlmann credibility-weighted estimate for
E(\(\mu(\theta)|X_1,\ldots, X_n)\) for the policyholder is

\begin{equation}
\hat{\mu}(\theta)=Z\bar{X}+(1-Z)\mu
\label{eq:buhlcred}
\end{equation}

with

\begin{eqnarray*}
\theta&=&\textrm{a risk parameter that identifies a policyholder's risk level}\\
\hat{\mu}(\theta)&=&\textrm{estimated expected loss for a policyholder with parameter }\theta\\
 & & \textrm{and loss experience } \bar{X}\\
\bar{X}&=&(X_1+\cdots+X_n)/n \textrm{ is the average of $n$ observations of the policyholder } \\
 Z&=&\textrm{credibility assigned to $n$ observations } \\
\mu&=&\textrm{the expected loss for a randomly chosen policyholder in the class.}\\
\end{eqnarray*}

Random variables \(X_j\) are assumed to be \emph{iid} for
\(j=1,\ldots,n\). The quantity \(\bar{X}\) is the average of \(n\)
observations and
\(\mathrm{E}(\bar{X}|\theta)=\mathrm{E}(X_j|\theta)=\mu(\theta)\).

If a policyholder is randomly chosen from the class and there is no loss
information about the risk then it's expected loss is
\(\mu=\mathrm{E}(\mu(\theta))\) where the expectation is taken over all
\(\theta\)'s in the class. In this situation \(Z=0\) and the expected
loss is \(\hat\mu(\theta)=\mu\) for the risk. The quantity \(\mu\) can
also be written as \(\mu=\mathrm{E}(X_j)\) or
\(\mu=\mathrm{E}(\bar{X})\) and is often called the overall mean or
collective mean. Note that E(\(X_j\)) is evaluated with the ``law of
total expectation'': E(\(X_j\))=E(E(\(X_j|\theta)\)).

\textbf{Example 9.3.1.} The number of claims \(X\) for an insured in a
class has a Poisson distribution with mean \(\theta>0\). The risk
parameter \(\theta\) is exponentially distributed within the class with
\emph{pdf} \(f(\theta)=e^{-\theta}\). What is the expected number of
claims for an insured chosen at random from the class?

Show Example Solution

\hypertarget{toggleExampleCred.3.1}{}
\textbf{Solution} Random variable \(X\) is Poisson with parameter
\(\theta\) and E\((X|\theta)=\theta\). The expected number of claims for
a randomly chosen insured is
\(\mu=\mathrm{E}(\mu(\theta))=\mathrm{E}(\mathrm{E}(X|\theta))=\)E\((\theta)=\int_{0}^{\infty}\theta e^{-\theta} d\theta\).
Integration by parts gives \(\mu=1\).

\begin{center}\rule{0.5\linewidth}{\linethickness}\end{center}

The prior example has risk parameter \(\theta\) as a positive real
number but the risk parameter can be a categorical variable as shown in
the next example.

\textbf{Example 9.3.2.} For any risk (policyholder) in a population the
number of losses \(N\) in a year has a Poisson distribution with
parameter \(\lambda\). Individual loss amounts \(X_i\) for a risk are
independent of \(N\) and are \emph{iid} with Pareto distribution
\(F(x)=1-[\theta/(x+\theta)]^{\alpha}\). There are three types of risks
in the population as follows:

\[\begin{matrix}
\begin{array}{|c|c|c|c|}
\hline
\text{Risk } & \text{Percentage} & \text{Poisson} & \text{Pareto} \\
\text{Type} & \text{of Population} & \text{Parameter} & \text{Parameters} \\
\hline
A & 50\% & \lambda=0.5 & \theta=1000, \alpha=2.0 \\
B & 30\% & \lambda=1.0 & \theta=1500, \alpha=2.0 \\
C & 20\% & \lambda=2.0 & \theta=2000, \alpha=2.0 \\
\hline
\end{array}
\end{matrix}\] If a risk is selected at random from the population, what
is the expected aggregate loss in a year?

Show Example Solution

\hypertarget{toggleExampleCred.3.2}{}
\textbf{Solution} The expected number of claims for a risk is
E(\(N\))=\(\lambda\). The expected value for a Pareto distributed random
variable is E(\(X\))=\(\theta/(\alpha-1)\). The expected value of the
aggregate loss random variable \(S=X_1+\cdots+X_N\) for a risk is
E(\(S\))=E(\(N\))E(\(X\))=\(\lambda\theta/(\alpha-1)\). The expected
aggregate loss for a risk of type A is
E(\(S_{\textrm{A}}\))=(0.5)(1000)/(2-1)=500. The expected aggregate loss
for a risk selected at random from the population is
E(\(S\))=0.5{[}(0.5)(1000){]}+0.3{[}(1.0)(1500){]}+0.2{[}(2.0)(2000){]}=1500.

\begin{center}\rule{0.5\linewidth}{\linethickness}\end{center}

Although formula \eqref{eq:buhlcred} was introduced using experience
rating as an example, the Bühlmann credibility model has wider
application. Suppose that a rating plan has multiple classes.
Credibility formula \eqref{eq:buhlcred} can be used to determine
individual class rates. The overall mean \(\mu\) would be the average
loss for all classes combined, \(\bar{X}\) would be the experience for
the individual class, and \(\hat{\mu}(\theta)\) would be the estimated
loss for the class.

\subsection{\texorpdfstring{Credibility Z, \emph{EPV}, and
\emph{VHM}}{Credibility Z, EPV, and VHM}}\label{S:EPV-VHM-Z}

When computing the credibility estimate
\(\hat{\mu}(\theta)=Z\bar{X}+(1-Z)\mu\), how much weight \(Z\) should go
to experience \(\bar{X}\) and how much weight \((1-Z)\) to the overall
mean \(\mu\)? In Bühlmann credibility there are three factors that need
to be considered:

\begin{itemize}
\tightlist
\item
  How much variation is there in a single observation \(X_j\) for a
  selected risk? With \(\bar{X}=(X_1+\cdots+X_n)/n\) and assuming that
  the observations are \emph{iid}, it follows that
  Var(\(\bar{X}|\theta)\)=Var(\(X_j|\theta)/n\). For larger
  Var(\(\bar{X}|\theta)\) less credibility weight \(Z\) should be given
  to experience \(\bar{X}\). The Expected Value of the Process Variance,
  abbreviated \emph{EPV}, is the expected value of Var(\(X_j|\theta\))
  across all risks:
\end{itemize}

\begin{equation*}
EPV=\mathrm{E}(\mathrm{Var}(X_j|\theta)).
\end{equation*}

Because Var(\(\bar{X}|\theta)\)=Var(\(X_j|\theta)/n\) it follows that
E(Var(\(\bar{X}|\theta)\))=EPV/\(n\).

\begin{itemize}
\tightlist
\item
  How homogeneous is the population of risks whose experience was
  combined to compute the overall mean \(\mu\)? If all the risks are
  similar in loss potential then more weight \((1-Z)\) would be given to
  the overall mean \(\mu\) because \(\mu\) is the average for a group of
  similar risks whose means \(\mu(\theta)\) are not far apart. The
  homogeneity or heterogeneity of the population is measured by the
  Variance of the Hypothetical Means with abbreviation \emph{VHM}:
\end{itemize}

\begin{equation*}
VHM=\mathrm{Var}(\mathrm{E}(X_j|\theta))=\mathrm{Var}(\mathrm{E}(\bar{X}|\theta)).
\end{equation*}

Note that we used \(\mathrm{E}(\bar{X}|\theta)=\mathrm{E}(X_j|\theta)\)
for the second equality. *How many observations \(n\) were used to
compute \(\bar{X}\)? More observations would infer a larger \(Z\).

\textbf{Example 9.3.3.} The number of claims \(N\) in a year for a risk
in a population has a Poisson distribution with mean \(\lambda>0\). The
risk parameters \(\lambda\) for the population are uniformly distributed
over the interval (0,2). Calculate the EPV and \emph{VHM} for the
population.

Show Example Solution

\hypertarget{toggleExampleCred.3.3}{}
\textbf{Solution} Random variable \(N\) is Poisson with parameter
\(\lambda\) so Var\((N|\lambda)=\lambda\). The Expected Value of the
Process variance is
EPV=E(Var(\(N|\lambda\)))=E\((\lambda)=\int_{0}^{2}\lambda \frac{1}{2} d\lambda=1\).
The Variance of the Hypothetical Means is
\emph{VHM}=Var(E(N\(|\lambda\)))=
Var(\(\lambda\))=E(\(\lambda^2)-(\mathrm{E}(\lambda))^2=\int_{0}^{2}\lambda^2 \frac{1}{2} d\lambda-(1)^2=\frac{1}{3}\).

\begin{center}\rule{0.5\linewidth}{\linethickness}\end{center}

The Bühlmann credibility formula includes values for \(n\), \emph{EPV},
and \emph{VHM}:

\begin{equation}
Z=\frac{n}{n+K} \quad , \quad K =\frac{EPV}{VHM}.
\label{eq:buhlZ}
\end{equation}

If \(n\) increases then so does \(Z\). If the \emph{VHM} increases then
\(Z\) increases. If the \emph{EPV} increases then \(Z\) gets smaller.
Unlike limited fluctuation credibility where \(Z=1\) when the expected
number of claims is greater than the full credibility standard, \(Z\)
can approach but not equal 1 as the number of observations \(n\) goes to
infinity.

If you multiply the numerator and denominator of the \(Z\) formula by
(\emph{VHM}/\(n\)) then \(Z\) can be rewritten as

\begin{equation*}
Z=\frac{VHM}{VHM+(EPV/n)} .
\end{equation*}

The number of observations \(n\) is captured in the term
(\emph{EPV}/\(n\)). As shown in bullet (1) at the beginning of the
section, E(Var(\(\bar{X}|\theta)\))=\emph{EPV}/\(n\). As the number of
observations get larger, the expected variance of \(\bar{X}\) gets
smaller and credibility \(Z\) increases so that more weight gets
assigned to \(\bar{X}\) in the credibility-weighted estimate
\(\hat{\mu}(\theta)\).

\textbf{Example 9.3.4.} Use the ``law of total variance" to show that
Var(\(\bar{X}\)) = \emph{VHM} + (\emph{EPV}/n) and derive a formula for
\(Z\) in terms of \(\bar{X}\).

Show Example Solution

\hypertarget{toggleExampleCred.3.4}{}
\textbf{Solution} The quantity Var(\(\bar{X}\)) is called the
unconditional variance or the total variance of \(\bar{X}\). The law of
total variance says

\begin{equation*}
\mathrm{Var}(\bar{X})=\textrm{E(Var}(\bar{X}|\theta))+\textrm{Var(E}(\bar{X}|\theta)).
\end{equation*}

In bullet (1) at the beginning of this section we showed
E(Var(\(\bar{X}|\theta)\))=\emph{EPV}/\(n\). In the second bullet (2),
Var(E(\(\bar{X}|\theta\)))=\emph{VHM}. Reordering the right hand side
gives Var(\(\bar{X}\))= \emph{VHM} +(\emph{EPV}/\(n\)). Another way to
write the formula for credibility \(Z\) is
\(Z\)=Var(E(\(\bar{X}|\theta\)))/Var(\(\bar{X}\)). This implies
\((1-Z)\)=E(Var(\(\bar{X}|\theta\)))/Var(\(\bar{X}\)).

\begin{center}\rule{0.5\linewidth}{\linethickness}\end{center}

The following long example and solution demonstrates how to compute the
credibility-weighted estimate with frequency and severity data.

\textbf{Example 9.3.5.} For any risk in a population the number of
losses \(N\) in a year has a Poisson distribution with parameter
\(\lambda\). Individual loss amounts \(X\) for a selected risk are
independent of \(N\) and are \emph{iid} with exponential distribution
\(F(x)=1-e^{x/\beta}\). There are three types of risks in the population
as shown below. A risk was selected at random from the population and
all losses were recorded over a five-year period. The total amount of
losses over the five-year period was 5,000. Use Bühlmann credibility to
estimate the annual expected aggregate loss for the risk.\\
\[\begin{matrix}
\begin{array}{|c|c|c|c|}
\hline
\text{Risk } & \text{Percentage} & \text{Poisson} & \text{Exponential} \\
\text{Type} & \text{of Population} & \text{Parameter} & \text{Parameter} \\
\hline
A & 50\% & \lambda=0.5 & \beta=1000 \\
B & 30\% & \lambda=1.0 & \beta=1500 \\
C & 20\% & \lambda=2.0 & \beta=2000 \\
\hline
\end{array}
\end{matrix}\]

Show Example Solution

\hypertarget{toggleExampleCred.3.5}{}
\textbf{Solution} Because individual loss amounts \(X\) are
exponentially distributed, E(\(X\))=\(\beta\) and
Var(\(X\))=\(\beta^2\). For aggregate loss \(S=X_1+\cdots+X_N\), the
mean is E(\(S\))=E(\(N\))E(\(X\)) and process variance is
Var(\(S\))=E(\(N\))Var(\(X\))+{[}E(\(X\)){]}\(^2\)Var(\(N\)). With
Poisson frequency and exponentially distributed loss amounts,
E(\(S\))=\(\lambda\beta\) and
Var(\(S\))=\(\lambda\beta^2+\beta^2\lambda=2\lambda\beta^2\).\\
\textbf{Population mean \(\mu\)}: Risk means are
\(\mu\)(A)=0.5(1000)=500; \(\mu\)(B)=1.0(1500)=1500;
\(\mu\)(C)=2.0(2000)=4000; and
\(\mu\)=0.50(500)+0.30(1500)+0.20(4000)=1,500.\\
\textbf{VHM}:
\emph{VHM}=\(0.50(500-1500)^2+0.30(1500-1500)^2+0.20(4000-1500)^2\)=1,750,000.\\
\textbf{EPV}: Process variances are
\(\sigma^2(A)=2(0.5)(1000)^2=1,000,000\);
\(\sigma^2(B)=2(1.0)(1500)^2=4,500,000\);
\(\sigma^2(C)=2(2.0)(2000)^2=16,000,000\); and
\emph{EPV}=0.50(1,000,000)+0.30(4,500,000)+0.20(16,000,000)=5,050,000.\\
\textbf{\(\mathbf{\bar{X}}\)}: \(\bar{X}_5=5,000/5\)=1,000.\\
\textbf{\(\mathbf{K}\)}: \(K=5,050,000/1,750,000\)=2.89.\\
\textbf{\(\mathbf{Z}\)}: There are five years of observations so
\(n=5\). \(Z=5/(5+2.89)\)=0.63.\\
\textbf{\(\boldsymbol{\hat{\mu}(\theta)}\)}:
\(\hat{\mu}(\theta)=0.63(1,000)+(1-0.63)1,500=\boxed{\mathbf{1,185.00}}\).

\begin{center}\rule{0.5\linewidth}{\linethickness}\end{center}

In real world applications of Bühlmann credibility the value of
\(K=EPV/VHM\) must be estimated. Sometimes a value for \(K\) is selected
using judgment. A smaller \(K\) makes estimator \(\hat{\mu}(\theta)\)
more responsive to actual experience \(\bar{X}\) whereas a larger \(K\)
produces a more stable estimate by giving more weight to \(\mu\).
Judgment may be used to balance responsiveness and stability. A later
section in this chapter will discuss methods for determining \(K\) from
data.

For a policyholder with risk parameter \(\theta\), Bühlmann credibility
uses a linear approximation \(\hat{\mu}(\theta)=Z\bar{X}+(1-Z)\mu\) to
estimate E(\(\mu(\theta)|X_1,\ldots,X_n\)), the expected loss for the
policyholder given prior losses \(X_1,\ldots, X_n\). We can rewrite this
as \(\hat{\mu}(\theta)=a+b\bar{X}\) which makes it obvious that the
credibility estimate is a linear function of \(\bar{X}\).

If E(\(\mu(\theta)|X_1,\ldots,X_n\)) is approximated by the linear
function \(a+b\bar{X}\) and constants \(a\) and \(b\) are chosen so that
E{[}(E(\(\mu(\theta)|X_1,\ldots,X_n)-(a+b\bar{X}))^2\){]} is minimized,
what are \(a\) and \(b\)? The answer is \(b=n/(n+K)\) and \(a=(1-b)\mu\)
with \(K=EPV/VHM\) and \(\mu=E(\mu(\theta))\). More detail can be found
in references \citep{buhlmann}, \citep{buhlmanngisler},
\citep{klugman2012}, and \citep{tse}.

Bühlmann credibility is also called least-squares credibility, greatest
accuracy credibility, or Bayesian credibility.

\section{Bühlmann-Straub Credibility}\label{buhlmann-straub-credibility}

\begin{center}\rule{0.5\linewidth}{\linethickness}\end{center}

In this section, you learn how to:

\begin{itemize}
\tightlist
\item
  Compute a credibility-weighted estimate for the expected loss for a
  risk or group of risks using the Bühlmann-Straub model.
\item
  Determine the credibility \(Z\) assigned to observations.
\item
  Calculate required values including the Expected Value of the Process
  Variance (\emph{EPV}), Variance of the Hypothetical Means (\emph{VHM})
  and collective mean \(\mu\).
\item
  Recognize situations when the Bühlmann-Straub model is appropriate.
\end{itemize}

\begin{center}\rule{0.5\linewidth}{\linethickness}\end{center}

With standard Bühlmann or least-squares credibility as described in the
prior section, losses \(X_1,\ldots,X_n\) for a policyholder are assumed
to be \emph{iid}. If the subscripts indicate year 1, year 2 and so on up
to year \(n\), then the \emph{iid} assumption means that the
policyholder has the same exposure to loss every year. For commercial
insurance this assumption is frequently violated.

Consider a commercial policyholder that uses a fleet of vehicles in its
business. In year 1 there are \(m_1\) vehicles in the fleet, \(m_2\)
vehicles in year 2, .., and \(m_n\) vehicles in year \(n\). The exposure
to loss from ownership and use of this fleet is not constant from year
to year. The annual losses for the fleet are not \emph{iid}.

Define \(Y_{jk}\) to be the loss for the \(k^{th}\) vehicle in the fleet
for year \(j\). Then, the total losses for the fleet in year \(j\) are
\(Y_{j1}+\cdots+Y_{jm_j}\) where we are adding up the losses for each of
the \(m_j\) vehicles. In the Bühlmann-Straub model it is assumed that
random variables \(Y_{jk}\) are \emph{iid} across all vehicles and years
for the policyholder. With this assumption the means
E(\(Y_{jk}|\theta)=\mu(\theta)\) and variances
Var(\(Y_{jk}|\theta)=\sigma^2(\theta)\) are the same for all vehicles
and years. The quantity \(\mu(\theta)\) is the expected loss and
\(\sigma^2(\theta)\) is the variance in the loss for one year for one
vehicle for a policyholder with risk parameter \(\theta\).

If \(X_j\) is the average loss per unit of exposure in year \(j\),
\(X_j=(Y_1+\cdots+Y_{m_j})/m_j\), then E(\(X_j)=\mu(\theta)\) and
Var(\(X_j)=\sigma^2(\theta)/m_j\) for policyholder with risk parameter
\(\theta\). The average loss per vehicle for the entire \(n\)-year
period is

\begin{equation*}
\bar{X}= \frac{1}{m} \sum_{j=1}^{n} m_j X_{j} \quad , \quad  m=\sum_{j=1}^{n}  m_j.
\end{equation*}

It follows that E\((\bar{X}|\theta)=\mu(\theta)\) and
Var\((\bar{X}|\theta)=\sigma^2(\theta)/m\) where \(\mu(\theta)\) and
\(\sigma^2(\theta)\) are the mean and variance for a single vehicle for
one year for the policyholder.

\textbf{Example 9.4.1.} Prove that
Var\((\bar{X}|\theta)=\sigma^2(\theta)/m\) for a risk with risk
parameter \(\theta\).

Show Example Solution

\hypertarget{toggleExampleCred.4.1}{}
\textbf{Solution}

\begin{eqnarray*}
\mathrm{Var}(\bar{X}|\theta)&=&\mathrm{Var}\left(\frac{1}{m} \sum_{j=1}^{n} m_j X_j|\theta \right)\\
                                  &=&\frac{1}{m^2}\sum_{j=1}^{n} \mathrm{Var}(m_j X_{j}|\theta)=\frac{1}{m^2}\sum_{j=1}^{n} m_j^2 \mathrm{Var}(X_j|\theta)\\
                                  &=&\frac{1}{m^2}\sum_{j=1}^{n} m_j^2 (\sigma^2(\theta)/m_j)=\frac{\sigma^2(\theta)}{m^2}\sum_{j=1}^{n} m_j=\sigma^2(\theta)/m.\\
\end{eqnarray*}

\begin{center}\rule{0.5\linewidth}{\linethickness}\end{center}

The Bühlmann-Straub credibility estimate is:

\begin{equation}\hat{\mu}(\theta)=Z\bar{X}+(1-Z)\mu
\label{eq:bscred}
\end{equation}

with

\begin{eqnarray*}
\theta&=&\textrm{a risk parameter that identifies a policyholder's risk level}\\
\hat{\mu}(\theta)&=&\textrm{estimated expected loss for one exposure for the policyholder}\\
 & & \textrm{with loss experience } \bar{X}\\
\bar{X}&=& \frac{1}{m} \sum_{j=1}^{n} m_j X_j \textrm{ is the average loss per exposure for m exposures } \\
Z&=&\textrm{credibility assigned to $m$ exposures } \\
 \mu&=&\textrm{expected loss for one exposure for randomly chosen}\\
 & & \textrm{ policyholder from population.}\\
\end{eqnarray*}

Note that \(\hat{\mu}(\theta)\) is the estimator for the expected loss
for one exposure. If the policyholder has \(m_j\) exposures then the
expected loss is \(m_j\hat{\mu}(\theta)\).

In an example in the prior section it was shown that
\(Z\)=Var(E(\(\bar{X}|\theta\)))/Var(\(\bar{X}\)) where \(\bar{X}\) is
the average loss for \(n\) observations. In equation \eqref{eq:bscred} the
\(\bar{X}\) is the average loss for \(m\) exposures and the same \(Z\)
formula can be used:

\begin{equation*}
Z=\frac{\mathrm{Var}(\mathrm{E}(\bar{X}))}{\mathrm{Var}(\bar{X})}=
\frac{\mathrm{Var}(\mathrm{E}(\bar{X}))}{\mathrm{E}(\mathrm{Var}(\bar{X}|\theta))+\mathrm{Var}(\mathrm{E}(\bar{X}|\theta))}.
\end{equation*}

The denominator was expanded using ``the law of total variance." As
noted above \(\mathrm{E}(\bar{X}|\theta)=\mu(\theta)\) so
\(\mathrm{Var}(\mathrm{E}(\bar{X}|\theta))=\mathrm{Var}(\mu(\theta))=VHM\).
Because Var\((\bar{X}|\theta)=\sigma^2(\theta)/m\) it follows that
E(Var(\(\bar{X}|\theta\)))=E(\(\sigma^2(\theta))/m\)=\emph{EPV}/m.
Making these substitutions and a little algebra gives

\begin{equation}
Z=\frac{m}{m+K} \quad , \quad K =\frac{EPV}{VHM}.
\label{eq:bsZ}
\end{equation}

This is the same \(Z\) as for Bühlmann credibility except number of
exposures \(m\) replaces number of years or observations \(n\).

\textbf{Example 9.4.2.}\\
A commercial automobile policyholder had the following exposures and
claims over a three-year period: \[\begin{matrix}
\begin{array}{|c|c|c|}
\hline
\text{Year} & \text{Number of Vehicles} & \text{Number of Claims} \\
\hline
1 &   9 &  5  \\
2 & 12 &  4  \\
3 & 15 &  4  \\
\hline
\end{array}
\end{matrix}\]

\begin{itemize}
\tightlist
\item
  The number of claims in a year for each vehicle in the policyholder's
  fleet is Poisson distributed with the same mean (parameter)
  \(\lambda\).
\item
  Parameter \(\lambda\) is distributed among the policyholders in the
  population with \emph{pdf} \(f(\lambda)=6\lambda(1-\lambda)\) with
  \(0<\lambda<1\).
\end{itemize}

The policyholder has 18 vehicles in its fleet in year 4. Use
Bühlmann-Straub credibility to estimate the expected number of
policyholder claims in year 4.

Show Example Solution

\hypertarget{toggleExampleCred.4.2}{}
\textbf{Solution} The expected number of claims for one vehicle for a
randomly chosen policyholder is
\(\mu=\mathrm{E}(\lambda)=\int_{0}^{1} \lambda[6\lambda(1-\lambda)] d\lambda=1/2\).
The average number of claims per vehicle for the policyholder is
\(\bar{X}\)=13/36. The Expected Value of the Process Variance for a
single vehicle is \emph{EPV}=E(\(\lambda)=1/2\). The Variance of the
Hypothetical Means across policyholders is
\emph{VHM}=Var(\(\lambda\))=E(\(\lambda^2\))-\((\mathrm{E}(\lambda))^2=\int_{0}^{1} \lambda^2[6\lambda(1-\lambda)] d\lambda-(1/2)^2=(3/10)-(1/4)=(6/20)-(5/20)=1/20\).
So, \(K\)=\emph{EPV}/\emph{VHM}=(1/2)/(1/20)=10. The number of exposures
in the experience period is \(m=9+12+15=36\). The credibility is
\(Z=36/(36+10)=18/23\). The credibility-weighted estimate for the number
of claims for one vehicle is
\(\hat{\mu}(\theta)=Z\bar{X}+(1-Z)\mu\)=(18/23)(13/36)+(5/23)(1/2)=9/23.
With 18 vehicles in the fleet in year 4 the expected number of claims is
18(9/23)=162/23=7.04 .

\begin{center}\rule{0.5\linewidth}{\linethickness}\end{center}

\section{Bayesian Inference and
Bühlmann}\label{bayesian-inference-and-buhlmann}

\begin{center}\rule{0.5\linewidth}{\linethickness}\end{center}

In this section, you learn how to:

\begin{itemize}
\tightlist
\item
  Use Bayes Theorem to determine a formula for the expected loss of a
  risk when given a likelihood and prior distribution.
\item
  Determine the posterior distributions for the Gamma-Poisson and
  Beta-Binomial Bayesian models and compute expected values.
\item
  Understand the connection between the Bühlmann and Bayesian estimates
  for the Gamma-Poisson and Beta-Binomial models.
\end{itemize}

\begin{center}\rule{0.5\linewidth}{\linethickness}\end{center}

Section \ref{S:MS:BayesInference} reviews Bayesian inference and it is
assumed that the reader is familiar with that material. This section
will compare Bayesian inference and Bühlmann credibility and show
connections between the two models.

A risk with risk parameter \(\theta\) has expected loss
\(\mu(\theta)=E(X|\theta)\) with random variable \(X\) representing pure
premium, aggegrate loss, number of claims, claim severity, or some other
measure of loss. If the risk had \(n\) losses \(x_1,\ldots, x_n\) then
E(\(\mu(\theta)|x_1,\ldots, x_n)\) is the conditional expectation of
\(\mu(\theta)\). The Bühlmann credibility formula
\(\hat{\mu}(\theta)=Z\bar{X}+(1-Z)\mu\) is a linear function of
\(\bar{X}=(x_1+\cdots+x_n)/n\) used to estimate
\(E(\mu(\theta)|x_1,\ldots,x_n)\).

Expectation \(E(\mu(\theta)|x_1,\ldots,x_n)\) can be calculated from the
conditional density function \(f(x|\theta)\) and the posterior
distribution \(\pi(\theta|x_1,\ldots,x_n)\):

\begin{eqnarray*}
\mathrm{E}(\mu(\theta)|x_1,\ldots,x_n)&=&\int \mu(\theta) \pi(\theta|x_1,..x_n) d\theta \\
                           \mu(\theta)&=&\mathrm{E}(X|\theta)=\int  xf(x|\theta) dx .\\
\end{eqnarray*}

The posterior distribution comes from Bayes theorem

\begin{equation*}
\pi(\theta|x_1,\ldots,x_n)=\frac{\prod_{j=1}^{n} f(x_j|\theta)}{f(x_1,..x_n)}\pi({\theta}).
\end{equation*}

The conditional density function \(f(x|\theta)\) and the prior
distribution \(\pi(\theta)\) must be specified. The numerator on the
right-hand side is called the likelihood.

\subsection{Gamma-Poisson Model}\label{gamma-poisson-model}

In the Gamma-Poisson model the number of claims \(X\) has a Poisson
distribution Pr(\(X=x|\lambda)=\lambda^xe^{-\lambda}/x!\) for a risk
with risk parameter \(\lambda\). The prior distribution for \(\lambda\)
is gamma with
\(\pi(\lambda)=\beta^\alpha\lambda^{\alpha-1}e^{-\beta\lambda}/\Gamma(\alpha)\).
(Note that a rate parameter \(\beta\) is being used in the gamma
distribution rather than a scale parameter.) The mean of the gamma is
E(\(\lambda)=\alpha/\beta\) and the variance is
Var(\(\lambda)=\alpha/\beta^2\). In this section we will assume that
\(\lambda\) is the expected number of claims per year though we could
have chosen another time interval.

If a risk is selected at random from the population then the expected
number of claims in a year is
E(\(N\))=E(E(\(N|\lambda\)))=E(\(\lambda\))=\(\alpha/\beta\). If we had
no observations for the selected risk then the expected number of claims
for the risk is \(\alpha/\beta\).

During \(n\) years the following number of claims by year was observed
for the randomly selected risk: \(x_1,\ldots,x_n\). From Bayes theorem
the posterior distribution is

\begin{equation*}
\pi(\lambda|x_1,\ldots,x_n)=\frac{\prod_{j=1}^{n} (\lambda^{x_j}e^{-\lambda}/x_j!)}{\Pr(x_1,\ldots,x_n)}\beta^\alpha\lambda^{\alpha-1}e^{-\beta\lambda}/\Gamma(\alpha).
\end{equation*}

\noindent Combining terms that have a \(\lambda\) and putting all other
terms into constant \(C\) gives

\begin{equation*}
\pi(\lambda|x_1,\ldots,x_n)=C\lambda^{(\alpha+\sum_{j=1}^{n}x_j)-1}e^{-(\beta+n)\lambda}.
\end{equation*}

This is a gamma distribution with parameters
\(\alpha'=\alpha+\sum_{j=1}^{n}x_i\) and \(\beta'=\beta+n\). The
constant must be \(C={\beta'}^{\alpha'}/\Gamma(\alpha')\) so that
\(\int_{0}^{\infty}\pi(\lambda|x_1,\ldots,x_n) d\lambda=1\) though we do
not need to know \(C\). As explained in chapter four the gamma
distribution is a conjugate prior for the Poisson distribution so the
posterior distribution is also gamma.

Because the posterior distribtution is gamma the expected number of
claims for the selected risk is

\begin{equation*}
\mathrm{E}(\lambda|x_1,\ldots,x_n) = \frac{\alpha+\sum_{j=1}^{n}x_j}{\beta+n}=\frac{\alpha + \textrm{number of claims}}{\beta+\textrm{number of years}}.
\end{equation*}

This formula is slightly different from chapter four because \(\beta\)
is multiplied times \(\lambda\) in the exponential of the gamma
\emph{pdf} whereas in chapter four \(\lambda\) is divided by parameter
\(\theta\).

Now we will compute the Bühlmann credibility estimate for the
Gamma-Poisson model. The variance for a Poisson distribution with
parameter \(\lambda\) is \(\lambda\) so
\emph{EPV}=E(Var(\(X|\lambda\)))=E(\(\lambda\))=\(\alpha/\beta\). The
mean number claims for the risk is \(\lambda\) so
\emph{VHM}=Var(E(\(X|\lambda\)))=Var(\(\lambda\))=\(\alpha/\beta^2\).
The credibility parameter is
\(K\)=\emph{EPV}/\emph{VHM}=\((\alpha/\beta)/(\alpha/\beta^2)=\beta\).
The overall mean is E(E(\(X|\lambda\)))=E(\(\lambda\))=\(\alpha/\beta\).
The sample mean is \(\bar{X}=(\sum_{j=1}^{n}x_j)/n\). The
credibility-weighted estimate for the expected number of claims for the
risk is

\begin{equation*}
\hat{\mu}=\frac{n}{n+\beta}\frac{\sum_{j=1}^{n}x_j}{n} +(1-\frac{n}{n+\beta})\frac{\alpha}{\beta}=\frac{\alpha+\sum_{j=1}^{n}x_j}{\beta+n}
\end{equation*}

For the Gamma-Poisson model the Bühlmann credibility estimate equals the
Bayesian analysis answer.

\subsection{Exact Credibility}\label{exact-credibility}

For the Gamma-Poisson claims model the Bühlmann credibility estimate for
the expected number of claims exactly matches the Bayesian answer. The
term \emph{exact credibility} is applied in this situation. Exact
credibility may occur if the probability distribution for \(X_j\) is in
the linear exponential family and the prior distribution is a conjugate
prior. Besides the Gamma-Poisson model other examples include
Gamma-Exponential, Normal-Normal, and Beta-Binomial. More information
about exact credibility can be found in \citep{buhlmanngisler},
\citep{klugman2012}, and \citep{tse}.

The beta-binomial model is useful for modeling the probability of an
event. Assume that random variable \(X\) is the number of successes in
\(n\) trials and that \(X\) has a binomial distribution
Pr(\(X=x|p)=\binom{n}{x}p^x(1-p)^{n-x}\). In the beta-binomial model the
prior distribution for probability \(p\) is a beta distribution with
\emph{pdf}

\begin{equation*}
\pi(p)=\frac{\Gamma(\alpha+\beta)}{\Gamma(\alpha)\Gamma(\beta)}p^{\alpha-1}(1-p)^{\beta-1} , \quad  0<p<1, \alpha>0, \beta>0.
\end{equation*}

The posterior distribution for \(p\) given outcome \(x\) is

\begin{equation*}
\pi(p|x)=\frac{\binom{n}{x}p^x(1-p)^{n-x}}{\Pr(x)}\frac{\Gamma(\alpha+\beta)}{\Gamma(\alpha)\Gamma(\beta)}p^{\alpha-1}(1-p)^{\beta-1}.
\end{equation*}

Combining terms that have a \(p\) and putting everything else into the
constant \(C\) yields

\begin{equation*}
\pi(p| x)=Cp^{\alpha+x-1}(1-p)^{\beta+(n-x)-1}.
\end{equation*}

This is a beta distribtuion with new parameters
\(\alpha^\prime=\alpha+x\) and \(\beta^\prime=\beta+(n-x)\). The
constant must be
\(C=\frac{\Gamma(\alpha+\beta+n)}{\Gamma(\alpha+x)\Gamma(\beta+n-x)}\).

The mean for the beta distribution with parameters \(\alpha\) and
\(\beta\) is E(\(p)=\frac{\alpha}{\alpha+\beta}\). Given \(x\) successes
in \(n\) trials in the beta-binomial model the mean of the posterior
distribution is E(\(p|x)=\frac{\alpha+x}{\alpha+\beta+n}\). As the
number of trials \(n\) and successes \(x\) increase, the expected value
of \(p\) approaches \(x/n\). The Bühlmann credibility estimate for
E(\(p|x\)) is exactly the same as shown in the following example.

\textbf{Example 9.5.1} The probability that a coin toss will yield heads
is \(p\). The prior distribution for probability \(p\) is beta with
parameters \(\alpha\) and \(\beta\). On \(n\) tosses of the coin there
were exactly \(x\) heads. Use Bühlmann credibility to estimate the
expected value of \(p\).

Show Example Solution

\hypertarget{toggleExampleCred.5.1}{}
\textbf{Solution} Define random variables \(Y_j\) such that \(Y_j=1\) if
the \(j^{th}\) coin toss is heads and \(Y_j=0\) if tails for
\(j=1,\ldots, n\). Random variables \(Y_j\) are \emph{iid} with
Pr\([Y=1|p]=p\) and Pr\([Y=0|p]=1-p\) The number of heads in \(n\)
tosses can be represented by the random variable \(X=Y_1+\cdots+Y_n\).
We want to estimate \(p=E[Y_j]\) using Bühlmann credibility:
\(\hat{p} = Z\bar{Y} +(1-Z)\mu\). The overall mean is
\(\mu=E(E(Y_j|p))=E(p)=\alpha/(\alpha+\beta)\). The sample mean is
\(\bar{y}=x/n\). The credibility is \(Z=n/(n+K)\) and
K=\emph{EPV}/\emph{VHM}. With Var\((Y_j|p)=p(1-p)\) it follows that
\emph{EPV}=E(Var\((Y_j|p)\))=E(\(p(1-p)\)). Because E\((Y_j)=p\) then
\emph{VHM}=Var\((E(Y_j|p))\)=Var(\(p\)). For the beta distribution

\begin{equation*}
\mathrm{E}(p)=\frac{\alpha}{\alpha+\beta}, \mathrm{E}(p^2)=\frac{\alpha(\alpha+1)}{(\alpha+\beta)(\alpha+\beta+1)}, \textrm{ and } \mathrm{Var}(p)=\frac{\alpha\beta}{(\alpha+\beta)^2(\alpha+\beta+1)}.
\end{equation*}

\begin{center}\rule{0.5\linewidth}{\linethickness}\end{center}

Parameter
\(K\)=\emph{EPV}/\emph{VHM}={[}E(\(p\))-E(\(p^2\)){]}/Var(\(p\)). With
some algebra this reduces to \(K=\alpha+\beta\). The Bühlmann
credibility-weighted estimate is

\begin{align*}
  \hat{p} &= \frac{n}{n+\alpha+\beta}\left(\frac{x}{n}\right)+\left(1-\frac{n}{n+\alpha+\beta}\right)\frac{\alpha}{\alpha+\beta} \\
  \hat{p} & =\frac{\alpha+x}{\alpha+\beta+n}\\
\end{align*}

which is the same as the Bayesian posterior mean.

\section{Estimating Credibility
Parameters}\label{estimating-credibility-parameters}

\begin{center}\rule{0.5\linewidth}{\linethickness}\end{center}

In this section, you learn how to:

\begin{itemize}
\tightlist
\item
  Perform nonparametric estimation with the Bühlmann and Bühlmann-Straub
  credibility models.
\item
  Identify situations when semiparametric estimation is appropriate.
\item
  Use data to approximate the \emph{EPV} and \emph{VHM}.
\item
  Balance credibility-weighted estimates.
\end{itemize}

\begin{center}\rule{0.5\linewidth}{\linethickness}\end{center}

The examples in this chapter have provided assumptions for calculating
credibility parameters. In actual practice the actuary must use real
world data and judgment to determine credibility parameters.

\subsection{Full Credibility Standard for Limited Fluctuation
Credibility}\label{full-credibility-standard-for-limited-fluctuation-credibility}

Limited-fluctuation credibility requires a full credibility standard.
The general formula for aggregate losses or pure premium is

\begin{equation*}
n_S=\left(\frac{y_p}{k}\right)^2\left[\left(\frac{\sigma_N^2}{\mu_N}\right)+\left(\frac{\sigma_X}{\mu_X}\right)^2\right]
\end{equation*}

\noindent with \(N\) representing number of claims and \(X\) the size of
claims. If one assumes \(\sigma_X=0\) then the full credibility standard
for frequency results. If \(\sigma_N=0\) then the full credibility
formula for severity follows. Probability \(p\) and \(k\) value are
often selected using judgment and experience.

In practice it is often assumed that the number of claims is Poisson
distributed so that \(\sigma_N^2/\mu_N=1\). In this case the formula can
be simplified to

\begin{equation*}
n_S=\left(\frac{y_p}{k}\right)^2\left[\frac{\mathrm{E}(X^2)}{(\mathrm{E}(X))^2}\right].
\end{equation*}

An empirical mean and second moment for the sizes of individual claim
losses can be computed from past data, if available.

\subsection{Nonparametric Estimation for Bühlmann and Bühlmann-Straub
Models}\label{nonparametric-estimation-for-buhlmann-and-buhlmann-straub-models}

Bayesian analysis as described previously requires assumptions about a
prior distribution and likelihood. It is possible to produce estimates
without these assumptions and these methods are often referred to as
empirical Bayes methods. Bühlmann and Bühlmann-Straub credibility with
parameters estimated from the data are included in category of empirical
Bayes methods.

\textbf{Bühlmann Model} First we will address the simpler Bühlmann
model. Assume that there are \(r\) risks in a population. For risk \(i\)
with risk parameter \(\theta_i\) the losses for \(n\) periods are
\(X_{i1},\ldots, X_{in}\). The losses for a risk are \emph{iid} across
periods as assumed in the Bühlmann model. For risk \(i\) the sample mean
is \(\bar{X}_i=\sum_{j=1}^{n}X_{ij}/n\) and the unbiased sample process
variance is \(s_i^2=\sum_{j=1}^{n}(X_{ij}-\bar{X}_i)^2/(n-1)\). An
unbiased estimator for the \emph{EPV} can be calculated by taking the
average of \(s_i^2\) for the \(r\) risks in the population:

\begin{equation}
\widehat{EPV}=\frac{1}{r}\sum_{i=1}^{r} s_i^2 = \frac{1}{r(n-1)} \sum_{i=1}^{r} \sum_{j=1}^{n}(X_{ij}-\bar{X}_i)^2 .
\label{eq:EPV-estimate}
\end{equation}

The individual risk means \(\bar{X}_i\) for \(i=1,\ldots, r\) can be
used to estimate the \emph{VHM}. An unbiased estimator of
Var(\(\bar{X}_i\)) is

\begin{equation*}
\widehat{\mathrm{Var}}(\bar{X}_i)=\frac{1}{r-1} \sum_{i=1}^{r}(\bar{X}_i-\bar{X})^2 \textrm{  and  }  \bar{X}=\frac{1}{r}\sum_{i=1}^{r} \bar{X}_i,
\end{equation*}

but Var(\(\bar{X}_i\)) is not the \emph{VHM}. The total variance formula
is

\begin{equation*}
\mathrm{Var}(\bar{X}_i)=\textrm{E(Var}(\bar{X}_i|\Theta=\theta_i))+\textrm{Var(E}(\bar{X}_i|\Theta=\theta_i)).
\end{equation*}

The \emph{VHM} is the second term on the right because
\(\mu(\theta_i)=\mathrm{E}(\bar{X}_i|\Theta=\theta_i)\) is the
hypothetical mean for risk \(i\). So,

\begin{equation*}
VHM=\textrm{Var(E}(\mu(\theta_i)) = \mathrm{Var}(\bar{X}_i) - \textrm{E(Var}(\bar{X}_i|\Theta=\theta_i)).
\end{equation*}

As discussed previously in Section \ref{S:EPV-VHM-Z}, \emph{EPV}/n =
E(Var(\(\bar{X}_i|\Theta=\theta_i\))) and using the above estimators
gives an unbiased estimator for the \emph{VHM}:

\begin{equation}
\widehat{VHM} = \frac{1}{r-1} \sum_{i=1}^{r}(\bar{X}_i-\bar{X})^2 - \frac{\widehat{EPV}}{n} .
\label{eq:VHM-estimate}
\end{equation}

Although the expected loss for a risk with parameter \(\theta_i\) is
\(\mu(\theta_i)\)=E(\(\bar{X}_i|\Theta=\theta_i\)), the variance of the
sample mean \(\bar{X}_i\) is greater that the variance of the
hypothetical means: Var(\(\bar{X}_i)\geq\)Var(\(\mu(\theta_i)\)). The
variance in the sample means Var(\(\bar{X}_i\)) includes both the
variance in the hypothetical means plus a process variance term because
for individual observations \(X_{ij}\),
\(Var(X_{ij}|\Theta=\theta_i)>0\).

In some cases formula \eqref{eq:VHM-estimate} can produce a negative value
for \(\widehat{VHM}\) because of the subtraction of \(\widehat{EPV}/n\),
but a variance cannot be negative. The process variance within risks is
so large that it overwhelms the measurement of the variance in means
between risks. We cannot use this method to determine the values needed
for Bühlmann credibility.

\textbf{Example 9.6.1.} Two policyholders had claims over a three-year
period as shown in the table below. Estimate the expected number of
claims for each policyholder using Bühlmann credibility and calculating
necessary parameters from the data.

\[\begin{matrix}
\begin{array}{|c|c|c|}
\hline
\text{Year} & \text{Risk A} & \text{Risk B} \\
\hline
1 & 0 &  2 \\
2 & 1 &  1  \\
3 & 0 &  2  \\
\hline
\end{array}
\end{matrix}\]

Show Example Solution

\hypertarget{toggleExampleCred.6.1}{}
\textbf{Solution} \(\bar{x}_A=\frac{1}{3}(0+1+0)=\frac{1}{3}\),
\(\bar{x}_B=\frac{1}{3}(2+1+2)=\frac{5}{3}\),
\(\bar{x}=\frac{1}{2}(\frac{1}{3}+\frac{5}{3})=1\)\\
\(s_A^2=\frac{1}{3-1}\left[(0-\frac{1}{3})^2+(1-\frac{1}{3})^2+(0-\frac{1}{3})^2\right]=\frac{1}{3}\)\\
\(s_B^2=\frac{1}{3-1}\left[(2-\frac{5}{3})^2+(1-\frac{5}{3})^2+(2-\frac{5}{3})^2\right]=\frac{1}{3}\)\\
\(\widehat{EPV}=\frac{1}{2}\left(\frac{1}{3}+\frac{1}{3}\right)=\frac{1}{3}\),
\(\widehat{VHM}=\frac{1}{2-1}\left[(\frac{1}{3}-1)^2+(\frac{5}{3}-1)^2\right]-\frac{1/3}{3}=\frac{7}{9}\)\\
\(K=\frac{1/3}{7/9}=\frac{3}{7}\),
\(Z=\frac{3}{3+(3/7))}=\frac{7}{8}\)\\
\(\hat{\mu}_A=\frac{7}{8}\left(\frac{1}{3}\right)+(1-\frac{7}{8})1=\frac{5}{12}\),
\(\hat{\mu}_B=\frac{7}{8}\left(\frac{5}{3}\right)+(1-\frac{7}{8})1=\frac{19}{12}\)

\begin{center}\rule{0.5\linewidth}{\linethickness}\end{center}

\textbf{Example9.6.2.} Two policyholders had claims over a three-year
period as shown in the table below. Calculate the nonparametric estimate
for the \emph{VHM}.

\[\begin{matrix}
\begin{array}{|c|c|c|}
\hline
\text{Year} & \text{Risk A} & \text{Risk B} \\
\hline
1 & 3 &  3 \\
2 & 0 &  0  \\
3 & 0 &  3  \\
\hline
\end{array}
\end{matrix}\]

Show Example Solution

\hypertarget{toggleExampleCred.6.2}{}
\textbf{Solution} \(\bar{x}_A=\frac{1}{3}(3+0+0)=1\),
\(\bar{x}_B=\frac{1}{3}(3+0+3)=2\),
\(\bar{x}=\frac{1}{2}(1+2)=\frac{3}{2}\)\\
\(s_A^2=\frac{1}{3-1}\left[(3-1)^2+(0-1)^2+(0-1)^2\right]=3\)\\
\(s_B^2=\frac{1}{3-1}\left[(3-2)^2+(0-2)^2+(3-2)^2\right]=3\)\\
\(\widehat{EPV}=\frac{1}{2}(3+3)=3\)\\
\(\widehat{VHM}=\frac{1}{2-1}\left[(1-\frac{3}{2})^2+(2-\frac{3}{2})^2\right]-\frac{3}{3}=-\frac{1}{2}.\)\\
The process variance is so large that it is not possible to estimate the
\emph{VHM}.

\begin{center}\rule{0.5\linewidth}{\linethickness}\end{center}

\textbf{Bühlmann-Straub Model} Empirical formulas for \emph{EPV} and
\emph{VHM} in the Bühlmann-Straub model are more complicated because a
risk's number of exposures can change from one period to another. Also,
the number of experience periods does not have to be constant across the
population because exposure rather than time measures loss potential.
First some definitions:

\begin{itemize}
\tightlist
\item
  \(X_{ij}\) is the losses per exposure for risk \(i\) in period \(j\).
  Losses can refer to number of claims or amount of loss. There are
  \(r\) risks so \(i=1,\ldots,r\).
\item
  \(n_i\) is the number of observation periods for risk \(i\)
\item
  \(m_{ij}\) is the number of exposures for risk \(i\) in period \(j\)
  for \(j=1,\ldots,n_i\)
\end{itemize}

Risk \(i\) with risk parameter \(\theta_i\) has \(m_{ij}\) exposures in
period \(j\) which means that the losses per exposure random variable
can be written as \(X_{ij}=(Y_{i1}+\cdots+Y_{im_{ij}})/m_{ij}\). Random
variable \(Y_{ik}\) is the loss for one exposure. For risk \(i\) losses
\(Y_{ik}\) are \emph{iid} with mean E(\(Y_{ik}\))=\(\mu(\theta_i)\) and
process variance Var(\(Y_{ik}\))=\(\sigma^2(\theta_i)\). It follows that
Var(\(X_{ij})\)=\(\sigma^2(\theta_i)/m_{i,j}\).

Two more important definitions are:

\begin{itemize}
\tightlist
\item
  \(\bar{X}_i=\frac{1}{m_i}\sum_{j=1}^{n_i} m_{ij}X_{ij}\) with
  \(m_i = \sum_{j=1}^{n_i} m_{ij}\). \(\bar{X}_i\) is the average loss
  per exposure for risk \(i\) for all observation periods combined.
\item
  \(\bar{X}=\frac{1}{m}\sum_{i=1}^{r} m_i \bar{X}_i\) with
  \(m=\sum_{i=1}^r m_i\). \(\bar{X}\) is the average loss per exposure
  for all risks for all observation periods combined.
\end{itemize}

Random variable \(\bar{X}_i\) is the average loss for all \(m_i\)
exposures for risk \(i\) for all years combined. Random variable
\(\bar{X}\) is the average loss for all exposures for all risks for all
years combined.

An unbiased estimator for the process variance \(\sigma^2(\theta_i)\) of
one exposure for risk \(i\) is

\begin{equation*}
{s_i}^2=\frac{\sum_{j=1}^{n_i} m_{ij}(X_{ij}-\bar{X}_i)^2}{n_i-1}. \end{equation*}

The \(m_{ij}\) weights are applied to the squared differences because
the \(X_{ij}\) are the averages of \(m_{ij}\) exposures. The weighted
average of the sample variances \({s_i}^2\) for each risk \(i\) in the
population with weights proportional to the number of \((n_i-1)\)
observation periods will produce the expected value of the process
variance (\emph{EPV}) estimate

\begin{equation*}
\widehat{EPV}=\frac{\sum_{i=1}^r  (n_i-1){s_i}^2}{\sum_{i=1}^r (n_i-1)}=\frac{\sum_{i=1}^r \sum_{j=1}^{n_i} m_{ij}(X_{ij}-\bar{X}_i)^2}{\sum_{i=1}^r (n_i-1)}.
\end{equation*}

The quantity \(\widehat{*EPV*}\) is an unbiased estimator for the
process variance of one exposure for a risk chosen at random from the
population.

To calculate an estimator for the variance in the hypothetical means
(\emph{VHM}) the squared differences of the individual risk sample means
\(\bar{X}_i\) and population mean \(\bar{X}\) are used. An unbiased
estimator for the \emph{VHM} is

\begin{equation*}
\widehat{VHM}=\frac{\sum_{i=1}^r m_i(\bar{X}_i-\bar{X})^2 - (r-1)\widehat{*EPV*}}{m-\frac{1}{m}\sum_{i=1}^r m_i^2}.
\end{equation*}

This complicated formula is necessary because of the varying number of
exposures. Proofs that the \emph{EPV} and \emph{VHM} estimators shown
above are unbiased can be found in several references mentioned at the
end of this chapter including \citep{buhlmanngisler},
\citep{klugman2012}, and \citep{tse}.

\textbf{Example 9.6.3.} Two policyholders had claims shown in the table
below. Estimate the expected number of claims for each policyholder
using Bü hlmann-Straub credibility and calculating parameters from the
data.

\[\begin{matrix}
\begin{array}{|c|c|c|c|c|c|}
\hline
\text{Policyholder} &  & \text{Year 1} & \text{Year 2} & \text{Year 3} & \text{Year 4} \\
\hline
\text{A} & \text{Number of claims} & 0 & 2 & 2 & 3 \\
\hline
\text{A} & \text{Insured vehicles} &  1 & 2 & 2 & 2\\
\hline
 & & & & & \\
\hline
\text{B} & \text{Number of claims} & 0 & 0 & 1 & 2\\
\hline
\text{B} & \text{Insured vehicles} &  0 & 2 & 3 & 4\\
\hline
\end{array}
\end{matrix}\]

Show Example Solution

\hypertarget{toggleExampleCred.6.3}{}
\textbf{Solution} \(\bar{x}_A=\frac{0+2+2+3}{1+2+2+2}=1\),
\(\bar{x}_B=\frac{0+1+2}{2+3+4}=\frac{1}{3}\),
\(\bar{x}=\frac{7(1)+9(1/3)}{7+9}=\frac{5}{8}\)\\
\(s_A^2=\frac{1}{4-1}\left[1(0-1)^2+2(1-1)^2+2(1-1)^2+2(\frac{3}{2}-1)^2\right]=\frac{1}{2 }\)\\
\(s_B^2=\frac{1}{3-1}\left[2(0-\frac{1}{3})^2+3(\frac{1}{3}-\frac{1}{3})^2+4(\frac{1}{2}-\frac{1}{3})^2\right]=\frac{1}{6}\)\\
\(\widehat{EPV}=\left[3\left(\frac{1}{2}\right)+2\left(\frac{1}{6}\right)\right]/(3+2)=\frac{11}{30}=0.3667\)\\
\(\widehat{VHM}=\left[(7(1-\frac{5}{8})^2+9(\frac{1}{3}-\frac{5}{8})^2-(2-1)\frac{11}{30}\right]/\left[16-\left(\frac{1}{16}\right)(7^2+9^2)\right]=0.1757\)\\
\(K=\frac{0.3667}{0.1757}=2.0871\)\\
\(Z_A=\frac{7}{7+2.0871)}=0.7703\),
\(Z_B=\frac{9}{9+2.0871)}=0.8118\),\\
\(\hat{\mu}_A=0.7703(1)+(1-0.7703)(5/8)=0.9139\)\\
\(\hat{\mu}_B=0.8118(1/3)+(1-0.8118)(5/8)=0.3882\)

\begin{center}\rule{0.5\linewidth}{\linethickness}\end{center}

\subsection{Semiparametric Estimation for Bühlmann and Bühlmann-Straub
Models}\label{semiparametric-estimation-for-buhlmann-and-buhlmann-straub-models}

In the prior section on nonparametric estimation, there were no
assumptions about the distribution of the losses per exposure random
variables \(X_{ij}\). Assuming that the \(X_{ij}\) have a particular
distribution and using properties of the distribution along with the
data to determine credibility parameters is referred to as
semiparametric estimation.

An example of semiparametric estimation would be the assumption of a
Poisson distribution when estimating claim frequencies. The Poisson
distribution has the property that the mean and variance are identical
and this property can simplify calculations. The following simple
example comes from the prior section but now includes a Poisson
assumption about claim frequencies.

\textbf{Example 9.6.4.} Two policyholders had claims over a three-year
period as shown in the table below. Assume that the number of claims for
each risk has a Poisson distribution. Estimate the expected number of
claims for each policyholder using Bühlmann credibility and calculating
necessary parameters from the data. \[\begin{matrix}
\begin{array}{|c|c|c|}
\hline
\text{Year} & \text{Risk A} & \text{Risk B} \\
\hline
1 & 0 &  2 \\
2 & 1 &  1  \\
3 & 0 &  2  \\
\hline
\end{array}
\end{matrix}\]

Show Example Solution

\hypertarget{toggleExampleCred.6.4}{}
\textbf{Solution} \(\bar{x}_A=\frac{1}{3}(0+1+0)=\frac{1}{3}\),
\(\bar{x}_B=\frac{1}{3}(2+1+2)=\frac{5}{3}\),
\(\bar{x}=\frac{1}{2}(\frac{1}{3}+\frac{5}{3})=1\)\\
With Poisson assumption the estimated variance for risk A is
\(\hat\sigma_A^2=\bar{x}_A=\frac{1}{3}\)\\
Similarly, \(\hat\sigma_B^2=\bar{x}_B=\frac{5}{3}\)\\
\(\widehat{EPV}=\frac{1}{2}(\frac{1}{3})+\frac{1}{2}(\frac{5}{3})=1\).
This is also \(\bar{x}\) because of Poisson assumption.\\
\(\widehat{VHM}=\frac{1}{2-1}\left[(\frac{1}{3}-1)^2+(\frac{5}{3}-1)^2\right]-\frac{1}{3}=\frac{5}{9}\)\\
\(K=\frac{1}{5/9}=\frac{9}{5}\),
\(Z_A=Z_B=\frac{3}{3+(9/5)}=\frac{5}{8}\)\\
\(\hat{\mu}_A=\frac{5}{8}\left(\frac{1}{3}\right)+(1-\frac{5}{8})1=\frac{7}{12}\),
\(\hat{\mu}_B=\frac{5}{8}\left(\frac{5}{3}\right)+(1-\frac{5}{8})1=\frac{17}{12}.\)

\begin{center}\rule{0.5\linewidth}{\linethickness}\end{center}

We did not have to make the Poisson assumption in the prior example
because there was enough data to use nonparametric estimation but the
following example is commonly used to demonstrate a situation where
semiparametric estimation is needed. There is insufficient data for
nonparametric estimation but with the Poisson assumption estimates can
be calculated.

\textbf{Example 9.6.5.} A portfolio of 2,000 policyholders generated the
following claims profile during a five-year period: \[\begin{matrix}
\begin{array}{|c|c|}
\hline
\text{Number of Claims} &   \\
\text{In 5 Years}           &  \text{Number of policies}\\
\hline
 0 &  923 \\
 1 &  682 \\
 2 &  249 \\
 3 &  70   \\
 4 &  51   \\
 5 &  25   \\
\hline
\end{array}
\end{matrix}\] In your model you assume that the number of claims for
each policyholder has a Poisson distribution and that a policyholder's
expected number of claims is constant through time. Use Bühlmann
credibility to estimate the annual expected number of claims for
policyholders with 3 claims during the five-year period.

Show Example Solution

\hypertarget{toggleExampleCred.6.5}{}
\textbf{Solution} Let \(\theta_i\) be the risk parameter for the
\(i^{th}\) risk in the portfolio with mean \(\mu(\theta_i)\) and
variance \(\sigma^2(\theta_i)\). With the Poisson assumption
\(\mu(\theta_i)=\sigma^2(\theta_i)\). The expected value of the process
variance is EPV=E(\(\sigma^2(\theta_i)\)) where the expectation is taken
across all risks in the population. Because of the Poisson assumption
for all risks it follows that
EPV=E(\(\sigma^2(\theta_i)\))=E(\(\mu(\theta_i)\)). An estimate for the
annual expected number of claims is \(\hat{\mu}(\theta_i)\)= (observed
number of claims)/5. This can also serve as the estimate for the process
variance for a risk. Weighting the process variance estimates (or means)
by the number of policies in each group gives the estimators

\begin{equation*}
\widehat{EPV}=\bar{x}=\frac{923(0)+682(1)+249(2)+70(3)+51(4)+25(5)}{(5)(2000)}=0.1719.
\end{equation*}

The \emph{VHM} estimator is

\begin{eqnarray*}
\hat{VHM}&=&\frac{1}{2000-1}[923(0-0.1719)^2+682(0.20-0.1719)^2+249(0.40-0.1719)^2\\
                            &   &+70(0.60-0.1719)^2+51(0.80-0.1719)^2+25(1-0.1719)^2]-\frac{0.1719}{5}\\
                            &=& 0.0111\\
               \hat{K}  &=& \hat{*EPV*}/\hat{VHM}=0.1719/0.0111=15.49\\
               \hat{Z}  &=& \frac{5}{5+15.49}=0.2440\\
               \hat{\mu}_{3 \textrm{ claims}}& = & 0.2440(3/5)+(1-0.2440)0.1719=0.2764 .\\
\end{eqnarray*}

\begin{center}\rule{0.5\linewidth}{\linethickness}\end{center}

\subsection{Balancing Credibility
Estimators}\label{balancing-credibility-estimators}

The estimated loss for risk \(i\) in a credibility weighted model is
\(\hat{\mu}(\theta_i)=Z_i\bar{X}_i+(1-Z_i)\bar{X}\) where \(\bar{X}_i\)
is the loss per exposure for risk \(i\) and \(\bar{X}\) is loss per
exposure for the population. The overall mean in the Bühlmann-Straub
model is \(\bar{X}=\sum_{i=1}^r(m_i/m) \bar{X}_i\) where \(m_i\) and
\(m\) are number of exposures for risk \(i\) and population,
respectively. The same formula works for the simpler Bühlmann model by
setting \(m_i=1\) and \(m=r\) where \(r\) is the number of risks.

For the credilility weighted estimators to be in balance we want

\begin{equation*}
\bar{X}=\sum_{i=1}^r(m_i/m) \bar{X}_i=\sum_{i=1}^r(m_i/m) \hat{\mu}(\theta_i).
\end{equation*}

If this equation is satisfied then the estimated losses for each risk
will add up to the population total, an important goal in ratemating,
but this may not happen if \(\bar{X}\) is used for the complement of
credibility.

In order to find a complement of credibility that will bring the
credibility-weighted estimators into balance we will set \(\hat{\mu}\)
as the complement of crediblity:

\begin{equation*}
\sum_{i=1}^r(m_i/m) \bar{X}_i=\sum_{i=1}^r(m_i/m) (Z_i\bar{X}_i+(1-Z_i)\hat{\mu}) .
\end{equation*}

A little algebra gives

\begin{equation*}
\sum_{i=1}^r m_i \bar{X}_i=\sum_{i=1}^r m_i Z_i\bar{X}_i + \hat{\mu}\sum_{i=1}^r m_i(1-Z_i),
\end{equation*}

and

\begin{equation*}
\hat{\mu}=\frac{\sum_{i=1}^r m_i(1-Z_i)\bar{X}_i}{\sum_{i=1}^r m_i(1-Z_i)}. \end{equation*}

This can be simplified using the following relationship

\begin{equation*}
m_i(1-Z_i)=m_i\left(1-\frac{m_i}{m_i+K}\right)=m_i\left(\frac{(m_i+K)-m_i}{m_i+K}\right)=KZ_i .
\end{equation*}

A complement of credibility that will bring the credibility-weighed
estimators into balance with the overall mean loss per exposure is

\begin{equation*}
\hat{\mu}=\frac{\sum_{i=1}^r  Z_i \bar{X}_i}{\sum_{i=1}^r  Z_i}. \end{equation*}

\textbf{Example 9.6.6.} An example from the nonparametric
Bühlmann-Straub section had the following data for two risks. Find the
complement of credibility \(\hat{\mu}\) that will produce
crediblity-weighted estimates that are in balance.

\[\begin{matrix}
\begin{array}{|c|c|c|c|c|c|}
\hline
\text{Policyholder} &  & \text{Year 1} & \text{Year 2} & \text{Year 3} & \text{Year 4} \\
\hline
\text{A} & \text{Number of claims} & 0 & 2 & 2 & 3 \\
\hline
\text{A} & \text{Insured vehicles} &  1 & 2 & 2 & 2\\
\hline
 & & & & & \\
\hline
\text{B} & \text{Number of claims} & 0 & 0 & 1 & 2\\
\hline
\text{B} & \text{Insured vehicles} &  0 & 2 & 3 & 4\\
\hline
\end{array}
\end{matrix}\]

Show Example Solution

\hypertarget{toggleExampleCred.6.6}{}
\textbf{Solution} The credibilities from the prior example are
\(Z_A=\frac{7}{7+2.0871)}=0.7703\) and
\(Z_B=\frac{9}{9+2.0871)}=0.8118\). The sample means are \(\bar{X}_A=1\)
and \(\bar{X}_B=1/3\). The balanced complement of credibility is

\begin{equation*}
\hat{\mu}=\frac{0.7703(1)+0.8118(1/3)}{0.7703+0.8118}=0.6579.
\end{equation*}

The updated credibility estimates are
\(\hat{\mu}_A=0.7703(1)+(1-0.7703)(.6579)=0.9214\) versus the previous
0.9139 and \(\hat{\mu}_B=0.8118(1/3)+(1-0.8118)(.6579)=0.3944\) versus
previous 0.3882. Checking the balance on the new estimators:
(7/16)(0.9214)+(9/16)0.3944)=0.6250. This exactly matches
\(\bar{X}=10/16=0.6250\).

\begin{center}\rule{0.5\linewidth}{\linethickness}\end{center}

\section{Further Resources and
Contributors}\label{Cred-further-reading-and-resources}

\subsubsection*{Exercises}\label{exercises-5}
\addcontentsline{toc}{subsubsection}{Exercises}

Here are a set of exercises that guide the viewer through some of the
theoretical foundations of \textbf{Loss Data Analytics}. Each tutorial
is based on one or more questions from the professional actuarial
examinations, typically the Society of Actuaries Exam C.

\href{https://www.ssc.wisc.edu/~jfrees/loss-data-analytics/loss-data-analyticscredibility-guided-tutorials/}{Credibility
Guided Tutorials}

\subsubsection*{Contributors}\label{contributors-4}
\addcontentsline{toc}{subsubsection}{Contributors}

\begin{itemize}
\tightlist
\item
  \textbf{Gary Dean}, Ball State University is the author of the
  initital version of this chapter. Email:
  \href{mailto:cgdean@bsu.edu}{\nolinkurl{cgdean@bsu.edu}} for chapter
  comments and suggested improvements.
\end{itemize}

\chapter{Portfolio Management including Reinsurance}\label{C:PortMgt}

\emph{Chapter Preview}. Define \(S\) to be (random) obligations that
arise from a collection (portfolio) of insurance contracts.

\begin{itemize}
\item
  We are particularly interested in probabilities of large outcomes and
  so formalize the notion of a heavy-tail distribution in Section
  \ref{S:Tails}.
\item
  How much in assets does an insurer need to retain to meet obligations
  arising from the random \(S\)? A study of risk measures in Section
  \ref{S:RiskMeasure} helps to address this question.
\item
  As with policyholders, insurers also seek mechanisms in order to
  spread risks. A company that sells insurance to an insurance company
  is known as a reinsurer, studied in Section \ref{S:Reinsurance}.
\end{itemize}

\section{Tails of Distributions}\label{S:Tails}

In 1998 freezing rains fell on eastern Ontario, south-western Quebec and
lasted for six days. The event doubled the amount of precipitation in
the area experienced in any prior ice storm, and resulted in a
catastrophe that produced excess of 840,000 cases of insurance claims.
This number is 20\(\%\) more than that of the claims caused by the
Hurricane Andrew - one of the largest natural disasters in the history
of North America. After all, the catastrophe caused approximately 1.44
billion Canadian dollars insurance settlements which is the highest loss
burden in the history of Canada. More examples of similar catastrophic
events that caused extremal insurance losses are Hurricanes Harvey and
Sandy, the 2011 Japanese earthquake and tsunami, and so forth.

In the context of insurance, a few large losses hitting a portfolio and
then converting into claims usually represent the greatest part of the
indemnities paid by insurance companies. The aforementioned losses, also
called `extremes', are quantitatively modelled by the tails of the
associated probability distributions. From the quantitative modelling
standpoint, relying on probabilistic models with improper tails is
rather daunting. For instance, periods of financial stress may appear
with higher frequency than expected, and insurance losses may occur with
worse severity. Therefore, the study of probabilistic behavior in the
tail portion of actuarial models is of utmost importance in the modern
framework of quantitative risk management. For this reason, this section
is devoted to the introduction of a few mathematical notions that
characterize the tail weight of random variables (\emph{rv}'s). The
applications of these notions will benefit us in the construction and
selection of appropriate models with desired mathematical properties in
the tail portion, that are suitable for a given task.

Formally, define \(X\) to be the (random) obligations that arise from a
collection (portfolio) of insurance contracts. We are particularly
interested in studying the right tail of the distribution of \(X\),
which represents the occurrence of large losses. Speaking plainly, a
\emph{rv} is said to be heavier-tailed if higher probabilities are
assigned to larger values. Unwelcome outcomes are more likely to occur
for an insurance portfolio that is described by a loss \emph{rv}
possessing heavier (right) tail. Tail weight can be an absolute or a
relative concept. Specifically, for the former, we may consider a
\emph{rv} to be heavy-tailed if certain mathematical properties of the
probability distribution are met. For the latter, we can say the tail of
one distribution is heavier than the other if some tail measures are
larger/smaller.

In the statistics and probability literature, there are several
quantitative approaches have been proposed to classify and compare tail
weight. Among most of these approaches, the survival functions serve as
the building block. In what follows, we are going to introduce two
simple yet useful tail classification methods, in which the basic idea
is to study the quantities that are closely related to behavior of the
survival function of \(X\).

\subsection{Classification Based on
Moments}\label{classification-based-on-moments}

One possible way of classifying the tail weight of distribution is by
assessing the existence of raw moments. Since our major interest lies in
the right tails of distributions, we henceforth assume the
obligation/loss \emph{rv}~\(X\) to be positive. At the outset, let us
recall that the \(k-\)th raw moment of a continuous \emph{rv}~\(X\), for
\(k\geq 0\), can be computed via

\begin{eqnarray*}
    \mu_k' &=& k \int_0^{\infty} x^{k-1} S(x) dx, \\
\end{eqnarray*}

where \(S(\cdot)\) denotes the survival function of \(X\). It is a
simple matter to see that the existence of the raw moments depends on
the asymptotic behavior of the survival function at infinity. Namely,
the faster the survival function decays to zero, the higher the order of
finite moment the associated \emph{rv} possesses. Hence the maximal
order of finite moment, denoted by
\(k^{\ast}:=\sup\{k\in \mathbf{R}_+:\mu_k'<\infty \}\), can be
considered as an indicator of tail weight. This observation leads us to
the moment-based tail weight classification method, which is defined
formally next.

\textbf{Definition 10.1.} \emph{For a positive loss random variable
\(X\), if all the positive raw moments exist, namely the maximal order
of finite moment \(k^{\ast}=\infty\), then \(X\) is said to be
light-tailed based on the moment method. If
\(k^{\ast}=a \in (0,\infty)\), then \(X\) is said to be heavy-tailed
based on the moment method. Moreover, for two positive loss random
variables \(X_1\) and \(X_2\) with maximal orders of moment
\(k^{\ast}_1\) and \(k^{\ast}_2\) respectively, we say \(X_1\) has a
heavier (right) tail than \(X_2\) if \(k^{\ast}_1\leq k^{\ast}_2\).}

It is noteworthy that the first part of Definition 10.1 is an absolute
concept of tail weight, while the second part is a relative concept of
tail weight which compares the (right) tails between two distributions.
Next, we are going to present a few examples that illustrate the
applications of the moment-based method for comparing tail weight. Some
of these examples are borrowed from \citet{klugman2012}.

\textbf{Example 10.1.1. Finiteness of gamma moments.} Let
\(X\sim Gamma(\alpha,\theta)\), with \(\alpha>0\) and \(\theta>0\), then
for all \(k>0\), show that \(\mu_k' < \infty\).

Show Example Solution

\hypertarget{toggleExamplePortMgt.1.1}{}
\textbf{Solution.}

\begin{eqnarray*}
    \mu_k' &=& \int_0^{\infty} x^k \frac{x^{\alpha-1} e^{-x/\theta}}{\Gamma(\alpha) \theta^{\alpha}} dx \\
    &=& \int_0^{\infty} (y\theta)^k  \frac{(y\theta)^{\alpha-1} e^{-y}}{\Gamma(\alpha) \theta^{\alpha}} \theta dy \\
    &=& \frac{\theta^k}{\Gamma(\alpha)} \Gamma(\alpha+k) < \infty.
\end{eqnarray*}

Since all the positive moments exist, i.e., \(k^{\ast}=\infty\), in
accordance with the moment-based classification method in Definition
10.1, the gamma distribution is light-tailed.

\begin{center}\rule{0.5\linewidth}{\linethickness}\end{center}

\textbf{Example 10.1.2. Finiteness of Weibull moments.} Let
\(X\sim Weibull(\theta,\tau)\), with \(\theta>0\) and \(\tau>0\), then
for all \(k>0\), show that \(\mu_k' < \infty\).

Show Example Solution

\hypertarget{toggleExamplePortMgt.1.2}{}
\textbf{Solution.}

\begin{eqnarray*}
    \mu_k' &=& \int_0^{\infty} x^k \frac{\tau x^{\tau-1} }{\theta^{\tau}} e^{-(x/\theta)^{\tau}}dx \\
    &=& \int_0^{\infty}  \frac{ y^{k/\tau} }{\theta^{\tau}} e^{-y/\theta^{\tau}}dy \\
    &=& \theta^{k} \Gamma(1+k/\tau) < \infty.
\end{eqnarray*}

\begin{center}\rule{0.5\linewidth}{\linethickness}\end{center}

Again, due to the existence of all the positive moments, the Weibull
distribution is light-tailed.

We notice in passing that the gamma and Weibull distributions have been
used quite intensively in the actuarial practice nowadays. Applications
of these two distributions are vast which include, but are not limited
to, insurance claim severity modelling, solvency assessment, loss
reserving, aggregate risk approximation, reliability engineering and
failure analysis. We have thus far seen two examples of using the
moment-based method to analyze light-tailed distributions. We document a
heavy-tailed example in what follows.

\textbf{Example 10.1.3. Heavy tail nature of the Pareto distribution.}
Let \(X\sim Pareto(\alpha,\theta)\), with \(\alpha>0\) and \(\theta>0\),
then for \(k>0\)

\begin{eqnarray*}
    \mu_k^{'} &=& \int_0^{\infty} x^k \frac{\alpha \theta^{\alpha}}{(x+\theta)^{\alpha+1}} dx \\
    &=& \alpha \theta^{\alpha} \int_{\theta}^{\infty} (y-\theta)^k {y^{-(\alpha+1)}} dy.
\end{eqnarray*}

Consider a similar integration:

\begin{eqnarray*}
  g_k:=\int_{\theta}^{\infty} {y^{k-\alpha-1}} dy=\left\{
  \begin{array}{ll}
    <\infty, & \hbox{for } k<\alpha;\\
    =\infty, & \hbox{for } k\geq \alpha.
  \end{array}
\right.
\end{eqnarray*}

Meanwhile,

\[\lim_{y\rightarrow \infty} \frac{(y-\theta)^k {y^{-(\alpha+1)}}}{y^{k-\alpha-1}}=\lim_{y\rightarrow \infty}
(1-\theta/y)^{k}=1.\]

Application of the limit comparison theorem for improper integrals
yields \(\mu_k'\) is finite if and only if \(g_k\) is finite. Hence we
can conclude that the raw moments of Pareto \emph{rv}'s exist only up to
\(k<\alpha\), i.e., \(k^{\ast}=\alpha\), and thus the distribution is
heavy-tailed. What is more, the maximal order of finite moment depends
only on the shape parameter \(\alpha\) and it is an increasing function
of \(\alpha\). In other words, based on the moment method, the tail
weight of Pareto \emph{rv}'s is solely manipulated by \(\alpha\) -- the
smaller the value of \(\alpha\), the heavier the tail weight becomes.
Since \(k^{\ast}<\infty\), the tail of Pareto distribution is heavier
than those of the gamma and Weibull distributions.

\begin{center}\rule{0.5\linewidth}{\linethickness}\end{center}

We are going to conclude this current section by an open discussion on
the limitations of the moment-based method. Despite its simple
implementation and intuitive interpretation, there are certain
circumstances in which the application of the moment-based method is not
suitable. First, for more complicated probabilistic models, the \(k\)-th
raw moment may not be simple to derive, and thus the identification of
the maximal order of finite moment can be challenging. Second, the
moment-based method does not well comply with main body of the well
established heavy tail theory in literature. Specifically, the existence
of moment generating functions is arguably the most popular method for
classifying heavy tail versus light tail within the community of
academic actuaries. However, for some \emph{rv}'s such as the lognormal
\emph{rv}'s, their moment generating functions do not exist even that
all the positive moments are finite. In these cases, applications of the
moment-based methods can lead to different tail weight assessment.
Third, when we need to compare the tail weight between two light-tailed
distributions both having all positive moments exist, the moment-based
method is no longer informative (see, e.g., Examples 10.1 and 10.2).

\subsection{Comparison Based on Limiting Tail
Behavior}\label{comparison-based-on-limiting-tail-behavior}

In order to resolve the aforementioned issues of the moment-based
classification method, an alternative approach for comparing tail weight
is to directly study the limiting behavior of the survival functions.

\textbf{Definition 10.2.} For two \emph{rv}'s \(X\) and \(Y\), let

\[
\gamma:=\lim_{t\rightarrow \infty}\frac{S_X(t)}{S_Y(t)}.
\] We say that

\begin{itemize}
\tightlist
\item
  \(X\) has a \textbf{heavier right tail} than \(Y\) if
  \(\gamma=\infty\);\\
\item
  \(X\) and \(Y\) are \textbf{proportionally equivalent in the right
  tail} if \(\gamma =c\in \mathbf{R}_+\);
\item
  \(X\) has a \textbf{lighter right tail} than \(Y\) if \(\gamma=0\).
\end{itemize}

\textbf{Example 10.4. Comparison of Pareto to Weibull distributions.}
Let \(X\sim Pareto(\alpha, \theta)\) and
\(Y\sim Weibull(\tau, \theta)\), for \(\alpha>0\), \(\tau>0\), and
\(\theta>0\). Show that the Pareto has a heavier right tail than the
Weibull.

Show Example Solution

\hypertarget{toggleExamplePortMgt.1.4}{}
\textbf{Solution.}

\begin{eqnarray*}
    \lim_{t\rightarrow \infty}\frac{S_X(t)}{S_Y(t)} &=& \lim_{t\rightarrow \infty}\frac{(1+t/\theta)^{-\alpha}}{\exp\{-(t/\theta)^{\tau}\}} \\
    &=& \lim_{t\rightarrow \infty}\frac{\exp\{t/\theta^{\tau} \}}{(1+t^{1/\tau}/\theta)^{\alpha}} \\
    &=& \lim_{t\rightarrow \infty}\frac{\sum_{i=0}^{\infty}\left(\frac{t}{\theta^{\tau}}\right)^{i}/i!}{(1+t^{1/\tau}/\theta)^{\alpha}}\\
    &=& \lim_{t\rightarrow \infty} \sum_{i=0}^{\infty} \left(t^{-i/\alpha}+\frac{t^{(1/\tau-i/\alpha)}}{\theta} \right)^{-\alpha}/\theta^{\tau i}i!\\
    &=& \infty.
\end{eqnarray*}

Therefore, the Pareto distribution has a heavier tail than the Weibull
distribution. One may also realize that exponentials go to infinity
faster than polynomials, thus the aforementioned limit must be infinite.

\begin{center}\rule{0.5\linewidth}{\linethickness}\end{center}

For some distributions of which the survival functions do not admit
explicit expressions, we may find the following alternative formula
useful:

\begin{eqnarray*}
    \lim_{t\to \infty} \frac{S_X(t)}{S_Y(t)} &=& \lim_{t \to \infty} \frac{S_X^{'}(t)}{S_Y^{'}(t)} \\
    &=& \lim_{t \to \infty} \frac{-f_X(t)}{-f_Y(t)}\\
 &=& \lim_{t\to \infty} \frac{f_X(t)}{f_Y(t)}.
\end{eqnarray*}

given that the density functions exist.

\textbf{Example 10.1.5. Comparison of Pareto to gamma distributions.}
Let \(X\sim Pareto(\alpha, \theta)\) and
\(Y\sim Gamma(\alpha, \theta)\), for \(\alpha>0\) and \(\theta>0\). Show
that the Pareto has a heavier right tail than the gamma.

Show Example Solution

\hypertarget{toggleExamplePortMgt.1.5}{}
\textbf{Solution.}

\begin{eqnarray*}
    \lim_{t\to \infty} \frac{f_{X}(t)}{f_{Y}(t)} &=& \lim_{t \to \infty} \frac{\alpha \theta^{\alpha} (t+ \theta)^{-\alpha-1}}{t^{\tau-1} e^{-t/\lambda} \lambda^{-\tau} \Gamma(\tau)^{-1}} \\
 &\propto&  \lim_{t\to \infty} \frac{e^{t/\lambda}}{(t+\theta)^{\alpha+1} t^{\tau-1}} \\
    &=& \infty,
\end{eqnarray*}

as exponentials go to infinity faster than polynomials.

\begin{center}\rule{0.5\linewidth}{\linethickness}\end{center}

\section{Risk Measures}\label{S:RiskMeasure}

In this previous section, we studied two methods for classifying the
weight of distribution tails. We may claim that the risk associated with
one distribution is more dangerous (asymptotically) than the others if
the tail is heavier. However, knowing one risk is more dangerous
(asymptotically) than the others may not provide sufficient information
for a sophisticated risk management purpose, and in addition, one is
also interested in quantifying how much more. In fact, the magnitude of
risk associated with a given loss distribution is an essential input for
many insurance applications, such as actuarial pricing, reserving,
hedging, insurance regulatory oversight, and so forth.

The literature on risk measures has been growing rapidly in popularity
and importance. In the succeeding twp subsections, we introduce two
indices which have recently earned unprecedented amount of interest
among theoreticians, practitioners, and regulators. They are namely the
\emph{Value-at-Risk} (\emph{VaR}) and the \emph{Tail Value-at-Risk}
(\emph{TVaR}) measures. The economic rationale behind these two popular
risk measures is similar to that for the tail classification methods
introduced in the previous section, with which we hope to capture the
risk of extremal losses represented by the distribution tails. Following
this is a broader discussion of desirable properties of risk measures.

\subsection{Value-at-Risk}\label{value-at-risk}

At the outset, we offer the formal definition of \emph{VaR}.

\textbf{Definition 10.3.} Consider an insurance loss random variable
\(X\). The Value-at-Risk measure of \(X\) with confidence level
\(q\in [0,1]\) is formulated as

\begin{eqnarray}
VaR_q[X]:=\inf\{x\in \mathbf{R}:F_X(x)\geq q\}.
\label{eq:Value-at-Risk}
\end{eqnarray}

The \emph{VaR} measure outputs the smallest value of \(X\) such that the
associated \emph{cdf} first excesses or equates to \(q\). In the fields
of probability and statistics, the \emph{VaR} is also known as the
percentiles.

Here is how we should interpret \emph{VaR} in the lingo of actuarial
mathematics. The \emph{VaR} is a forecast of the `maximal' probable loss
for a insurance product/portfolio or a risky investment occurring
\(q\times 100\%\) of times, over a specific time horizon (typically, one
year). For instance, let \(X\) be the annual loss \emph{rv} of an
insurance product, \(VaR_{0.95}[X]=100\) million means that there is a
\(5\%\) chance that the loss will exceed 100 million over a given year.
Owing to the meaningful interpretation, \emph{VaR} has become the
industrial standard to measuring financial and insurance risks since
1990's. Financial conglomerates, regulators, and academics often utilize
\emph{VaR} to price insurance products, measure risk capital, ensure the
compliance with regulatory rules, and disclose the financial positions.

Next, we are going to present a few examples about the computation of
\emph{VaR}.

\textbf{Example 10.2.1. \emph{VaR} for the exponential distribution.}
Consider an insurance loss \emph{rv} \(X\sim Exp(\theta)\) for
\(\theta>0\), then the \emph{cdf}~of \(X\) is given by \[
F_X(x)=1-e^{-x/\theta}, \text{ for } x>0.
\] Give a closed-form expression for the \emph{VaR}.

Show Example Solution

\hypertarget{toggleExamplePortMgt.2.1}{}
\textbf{Solution.}

Because exponential distribution is a continuous distribution, the
smallest value such that the \emph{cdf}~first exceeds or equates to
\(q \in [0,1]\) must be at the point \(x_q\) satisfying \[
q=F_X(x_q)=1-\exp\{-x_q/\theta \}.
\] Thus \[
VaR_q[X]=F_X^{-1}(q)=-\theta[\log(1-q)].
\]

\begin{center}\rule{0.5\linewidth}{\linethickness}\end{center}

The result reported in Example 10.6 can be generalized to any continuous
\emph{rv}'s having strictly increasing \emph{cdf}. Specifically, the
\emph{VaR} of any continuous \emph{rv}'s is simply the inverse of the
corresponding \emph{cdf} Let us consider another example of continuous
\emph{rv} which has the support from negative infinity to positive
infinity.

\textbf{Example 10.2.2. \emph{VaR} for the normal distribution.}
Consider an insurance loss \emph{rv} \(X\sim Normal(\mu,\sigma^2)\) with
\(\mu\in \mathbf{R}\) and \(\sigma>0\). In this case, one may interpret
the negative values of \(X\) as profit or revenue. Give a closed-form
expression for the \emph{VaR}.

Show Example Solution

\hypertarget{toggleExamplePortMgt.2.2}{}
\textbf{Solution.}

Because normal distribution is a continuous distribution, the \emph{VaR}
of \(X\) must satisfy

\begin{eqnarray*}
 q &=& F_X(VaR_q[X])\\
&=&\Pr\left[(X-\mu)/\sigma\leq (VaR_q[X]-\mu)/\sigma\right]\\
&=&\Phi((VaR_q[X]-\mu)/\sigma).
\end{eqnarray*}

Therefore, we have \[
VaR_q[X]=\Phi^{-1}(q)\ \sigma+\mu.
\]

\begin{center}\rule{0.5\linewidth}{\linethickness}\end{center}

In many insurance applications, we have to deal with transformations of
\emph{rv}'s. For instance, in Example 10.7, the loss
\emph{rv}~\(X\sim Normal(\mu, \sigma^2)\) can be viewed as a linear
transformation of a standard normal \emph{rv}~\(Z\sim Normal(0,1)\),
namely \(X=Z\sigma+\mu\). By setting \(\mu=0\) and \(\sigma=1\), it is
straightforward for us to check \(VaR_q[Z]=\Phi^{-1}[q].\) A useful
finding revealed from Example 10.7 is that the \emph{VaR} of a linear
transformation of the normal \emph{rv}'s is equivalent to the linear
transformation of the \emph{VaR} of the original \emph{rv}'s. This
finding can be further generalized to any \emph{rv}'s as long as the
transformations are strictly increasing. The next example highlights the
usefulness of the abovementioned finding.

\textbf{Example 10.2.3. \emph{VaR} for transformed variables.} Consider
an insurance loss \emph{rv}~\(Y\sim lognormal(\mu,\sigma^2)\), for
\(\mu\in \mathbf{R}\) and \(\sigma>0\). Give an expression of the
\(VaR\) of \(Y\) in terms of the standard normal inverse \emph{cdf}.

Show Example Solution

\hypertarget{toggleExamplePortMgt.2.3}{}
\textbf{Solution.}

Note that \(\log Y\sim Normal(\mu,\sigma^2)\), or equivalently let
\(X\sim Normal(\mu,\sigma^2)\), then \(Y\overset{d}{=}e^{X}\) which is
strictly increasing transformation. Here, the notation
`\(\overset{d}{=}\)' means equality in distribution. The \emph{VaR} of
\(Y\) is thus given by the exponential transformation of the \emph{VaR}
of \(X\). Precisely, for \(q\in [0,1]\), \[
VaR_{q}[Y]= e^{VaR_q[X]}=\exp\{\Phi^{-1}(q)\ \sigma+\mu\}.
\]

\begin{center}\rule{0.5\linewidth}{\linethickness}\end{center}

We have thus far seen a number of examples about the \emph{VaR} for
continuous \emph{rv}'s, let us consider an example concerning the
\emph{VaR} for a discrete \emph{rv}

\textbf{Example 10.2.4. \emph{VaR} for a discrete random variable.}
Consider an insurance loss \emph{rv} with the following probability
distribution: \[
\Pr[X=x]=\left\{
                  \begin{array}{ll}
                    1, & \hbox{with probability $0.75$;} \\
                    3, & \hbox{with probability $0.20$;} \\
                    4, & \hbox{with probability $0.05$.}
                  \end{array}
                \right.
\] Determine the \emph{VaR} at \(q = 0.6, 0.9, 0.95, 0.95001\).

Show Example Solution

\hypertarget{toggleExamplePortMgt.2.4}{}
\textbf{Solution.}

The corresponding \emph{cdf} of \(X\) is \[
F_X(x)=\left\{
         \begin{array}{ll}
           0, & \hbox{ $x<1$;} \\
           0.75, & \hbox{ $1\leq x<3$;} \\
           0.95, & \hbox{ $3\leq x<4$;} \\
           1, & \hbox{ $4\leq x$.}
         \end{array}
       \right.
\] By the definition of \emph{VaR}, we thus have then

\begin{itemize}
\tightlist
\item
  \emph{\(VaR_{0.6}[X]=1\);}
\item
  \emph{\(VaR_{0.9}[X]=3\);}
\item
  \emph{\(VaR_{0.95}[X]=3\);}
\item
  \emph{\(VaR_{0.950001}[X]=4\).}
\end{itemize}

\begin{center}\rule{0.5\linewidth}{\linethickness}\end{center}

Let us now conclude this current subsection by an open discussion of the
\emph{VaR} measure. Some advantages of utilizing \emph{VaR} include

\begin{itemize}
\tightlist
\item
  possessing a practically meaningful interpretation;
\item
  relatively simple to compute for many distributions with closed-form
  distribution functions;
\item
  no additional assumption is required for the computation of
  \emph{VaR}.
\end{itemize}

On the other hand, the limitations of \emph{VaR} can be particularly
pronounced for some risk management practices. We report some of them
herein:

\begin{itemize}
\tightlist
\item
  the selection of the confidence level \(q\in [0,1]\) is highly
  subjective, while the \emph{VaR} can be very sensitive to the choice
  of \(q\) (e.g., in Example 10.9, \(VaR_{0.95}[X]=3\) and
  \(VaR_{0.950001}[X]=4\));
\item
  the scenarios/loss information that are above the
  \((1-p)\times 100\%\) worst event, are completely neglected;
\item
  \emph{VaR} is not a coherent risk measure (specifically, the
  \emph{VaR} measure does not satisfy the subadditivity axiom, meaning
  that diversification benefits may not be fully reflected).
\end{itemize}

\subsection{Tail Value-at-Risk}\label{tail-value-at-risk}

Recall that the \emph{VaR} represents the \((1-p)\times100\%\) chance
maximal loss. As we mentioned in the previous section, one major
drawback of the \emph{VaR} measure is that it does not reflect the
extremal losses occurring beyond the \((1-p)\times100\%\) chance worst
scenario. For an illustration purpose, let us consider the following
slightly unrealistic yet inspiring example.

\textbf{Example 10.2.5.} Consider two loss \emph{rv}'s
\(X\sim Uniform [0,100]\), and \(Y\sim Exp(31.71)\). We use \emph{VaR}
at \(95\%\) confidence level to measure the riskiness of \(X\) and
\(Y\). Simple calculation yields (see, also, Example 10.6), \[
VaR_{0.95}[X]=VaR_{0.95}[Y]=95,
\] and thus these two loss distributions have the same level of risk
according to \(VaR_{0.95}\). However, it is clear that \(Y\) is more
risky than \(X\) if extremal losses are of major concern since \(X\) is
bounded above while \(Y\) is unbounded. Simply quantifying risk by using
\emph{VaR} at a specific confidence level could be misleading and may
not reflect the true nature of risk.

As a remedy, the \emph{Tail Value-at-Risk} (\emph{TVaR}) was proposed to
measure the extremal losses that are above a given level of \emph{VaR}
as an average. We document the definition of \emph{TVaR} in what
follows. For the sake of simplicity, we are going to confine ourselves
to continuous positive \emph{rv}'s only, which are more frequently used
in the context of insurance risk management. We refer the interested
reader to \citet{hardy2006} for a more comprehensive discussion of
\emph{TVaR} for both discrete and continuous \emph{rv}'s.

\textbf{Definition 10.4.} Fix \(q\in [0,1]\), the Tail Value-at-Risk of
a (continuous) \emph{rv} \(X\) is formulated as

\begin{eqnarray*}
  TVaR_q[X] &:=& \mathrm{E}[X|X>VaR_q[X]],
\end{eqnarray*}

given that the expectation exists.

In light of Definition 10.4, the computation of \emph{TVaR} typically
consists of two major components - the VaR and the average of losses
that are above the VaR. The \emph{TVaR} can be computed via a number of
formulas. Consider a continuous positive \emph{rv}~\(X\), for notional
convenience, henceforth let us write \(\pi_q:=VaR_q[X]\). By definition,
the \emph{TVaR} can be computed via

\begin{eqnarray}
TVaR_{q}[X]=\frac{1}{(1-q)}\int_{\pi_q}^{\infty}xf_X(x)dx.
\label{eq:cte-pdf}
\end{eqnarray}

\textbf{Example 10.2.6. \emph{TVaR} for a normal distribution.} Consider
an insurance loss \emph{rv}~\(X\sim Normal (\mu,\sigma^2)\) with
\(\mu\in \mathbf{R}\) and \(\sigma>0\). Give an expression for
\emph{TVaR}.

Show Example Solution

\hypertarget{toggleExamplePortMgt.2.6}{}
\textbf{Solution.}

Let \(Z\) be the standard normal \emph{rv}. For \(q\in[0,1]\), the
\emph{TVaR} of \(X\) can be computed via

\begin{eqnarray*}
  TVaR_q[X] &=& \mathrm{E}[X|X>VaR_q[X]]\\
&=&\mathrm{E}[\sigma Z+\mu|\sigma Z+\mu>VaR_q[X]]\\
&=& \sigma\mathrm{E}[Z|Z>(VaR_q[X]-\mu)/\sigma]+\mu\\
&\overset{(1)}{=}& \sigma\mathrm{E}[Z|Z>VaR_q[Z]]+\mu,
\end{eqnarray*}

where `\(\overset{(1)}{=}\)' holds because of the results reported in
Example 10.7. Next, we turn to study
\(TVaR_q[Z]=\mathrm{E}[Z|Z>VaR_q[Z]]\). Let
\(\omega(q)=(\Phi^{-1}(q))^2/2\), we have

\begin{eqnarray*}
  (1-q)\ TVaR_q[Z] &=& \int_{\Phi^{-1}(q)}^{\infty} z \frac{1}{\sqrt{2\pi}} e^{-z^2/2}dz\\
&=& \int_{\omega(q)}^{\infty}  \frac{1}{\sqrt{2\pi}} e^{-x}dx\\
&=& \frac{1}{\sqrt{2\pi}} e^{-\omega(q)}\\
&=& \phi(\Phi^{-1}(q)).
\end{eqnarray*}

Thus, \[
TVaR_q[X]=\sigma\frac{\phi(\Phi^{-1}(q))}{1-q}+\mu.
\]

\begin{center}\rule{0.5\linewidth}{\linethickness}\end{center}

We mentioned earlier in the previous subsection that the \emph{VaR} of a
strictly increasing function of \emph{rv} is equal to the function of
VaR of the original \emph{rv}. Motivated by the results in Example
10.11, one can show that the \emph{TVaR} of a strictly increasing linear
transformation of \emph{rv} is equal to the function of VaR of the
original \emph{rv} This is due to the linearity property of
expectations. However, the aforementioned finding can not be extended to
non-linear functions. The following example of lognormal \emph{rv}
serves as a counter example.

\textbf{Example 10.2.7. \emph{TVaR} of a lognormal distribution.}
Consider an insurance loss \emph{rv}~\(X\sim lognormal (\mu,\sigma^2)\),
with \(\mu\in \mathbf{R}\) and \(\sigma>0\). Show that

\begin{eqnarray*}
  TVaR_q[X] &=& \frac{e^{\mu+\sigma^2/2}}{(1-q)} \Phi(\Phi^{-1}(q)-\sigma).
\end{eqnarray*}

Show Example Solution

\hypertarget{toggleExamplePortMgt.2.7}{}
\textbf{Solution.}

Recall that the \emph{pdf} of lognormal distribution is formulated as \[
f_X(x)=\frac{1}{\sigma\sqrt{2\pi} x}\exp\{-(\ln x-\mu )^2/2\sigma^2 \}, \text{ for } x>0.
\] Fix \(q\in[0,1]\), then the \emph{TVaR} of \(X\) can be computed via

\begin{eqnarray}
  TVaR_q[X] &=& \frac{1}{(1-q)} \int_{\pi_q}^{\infty} x f_X(x)dx \nonumber\\
&=&\frac{1}{(1-q)} \int_{\pi_q}^{\infty} \frac{1}{\sigma \sqrt{2\pi}} \exp\left\{ -\frac{(\log x-\mu)^2}{2\sigma^2}
\right\}dx\nonumber\\
&\overset{(1)}{=}&\frac{1}{(1-q)} \int_{\omega(q)}^{\infty} \frac{1}{\sqrt{2\pi}} e^{ -\frac{1}{2}w^2+\sigma w+\mu}dw\nonumber\\
&=&\frac{e^{\mu+\sigma^2/2}}{(1-q)} \int_{\omega(q)}^{\infty} \frac{1}{\sqrt{2\pi}} e^{ -\frac{1}{2}(w-\sigma)^2}dw\nonumber\\
&=&\frac{e^{\mu+\sigma^2/2}}{(1-q)} \Phi(\omega(q)-\sigma),
\label{eq:cte-normal}
\end{eqnarray}

where `\(\overset{(1)}{=}\)' holds by applying change of variable
\(w=(\log x-\mu)/\sigma\), and \(\omega(q)=(\log \pi_q-\mu)/\sigma\).
Evoking the formula of VaR for lognormal \emph{rv}~reported in Example
10.7, we can simplify the expression \eqref{eq:cte-normal} into

\begin{eqnarray*}
  TVaR_q[X] &=& \frac{e^{\mu+\sigma^2/2}}{(1-q)} \Phi(\Phi^{-1}(q)-\sigma).
\end{eqnarray*}

\begin{center}\rule{0.5\linewidth}{\linethickness}\end{center}

Clearly, the \emph{TVaR} of lognormal \emph{rv} is not the exponential
of the \emph{TVaR} of normal \emph{rv}.

For distributions of which the distribution functions are more tractable
to work with, we may apply integration by parts technique to rewrite
equation \eqref{eq:cte-pdf} as

\begin{eqnarray*}
TVaR_{q}[X]&=&\left[-x S_X(x)\big |_{\pi_q}^{\infty}+\int_{\pi_q}^{\infty}S_X(x)dx\right]\frac{1}{(1-q)}\\
&=& \pi_q +\frac{1}{(1-q)}\int_{\pi_q}^{\infty}S_X(x)dx.
\end{eqnarray*}

\textbf{Example 10.2.8. \emph{TVaR} of an exponential distribution.}
Consider an insurance loss \emph{rv}~\(X\sim Exp(\theta)\) for
\(\theta>0\). Give an expression for the \emph{TVaR}.

Show Example Solution

\hypertarget{toggleExamplePortMgt.2.8}{}
\textbf{Solution.}

We have seen from the previous subsection that \[
\pi_q=-\theta[\log(1-q)].
\] Let us now consider the \emph{TVaR}:

\begin{eqnarray*}
  TVaR_q[X] &=& \pi_q+\int_{\pi_q}^{\infty} e^{-x/\theta}dx/(1-q)\\
&=& \pi_q+\theta e^{-\pi_q/\theta}dx/(1-q)\\
&=& \pi_q+\theta.
\end{eqnarray*}

\begin{center}\rule{0.5\linewidth}{\linethickness}\end{center}

It can also be helpful to express the \emph{TVaR} in terms of limited
expected values. Specifically, we have

\begin{eqnarray}
  TVaR_q[X] &=& \int_{\pi_q}^{\infty} (x-\pi_q+\pi_q)f_X(x)dx/(1-q) \nonumber\\
&=& \pi_q+\frac{1}{(1-q)}\int_{\pi_q}^{\infty} (x-\pi_q)f_X(x)dx\nonumber\\
&=& \pi_q+e_X(\pi_q)\nonumber\\
&=& \pi_q +\frac{\left({\mathrm{E}[X]-\mathrm{E}[X\wedge\pi_q]}\right)}{(1-q)},
\label{eq:cte-expectation}
\end{eqnarray}

where \(e_X(d):=\mathrm{E}[X-d|X>d]\) for \(d>0\) denotes the \emph{mean
excess loss function}. For many commonly used parametric distributions,
the formulas for calculating \(\mathrm{E}[X]\) and
\(\mathrm{E}[X\wedge\pi_q]\) can be found in a table of distributions.

\textbf{Example 10.2.9. \emph{TVaR} of the Pareto distribution.}
Consider a loss \emph{rv}~\(X\sim Pareto(\theta,\alpha)\) with
\(\theta>0\) and \(\alpha>0\). The \emph{cdf}~of \(X\) is given by \[
F_X(x)=1-\left(\frac{\theta}{\theta+x} \right)^{\alpha}, \text{ for } x>0 .
\]

Fix \(q\in [0,1]\) and set \(F_X(\pi_q)=q\), we readily obtain

\begin{eqnarray}
\pi_q=\theta\left[(1-q)^{-1/\alpha}-1 \right].
\label{eq:var-pareto}
\end{eqnarray}

According to the distribution table provided in the SOA exam C, we know
\[
\mathrm{E}[X]=\frac{\theta}{\alpha-1},
\] and \[
\mathrm{E}[X\wedge \pi_q]=\frac{\theta}{\alpha-1}\left[
1-\left(\frac{\theta}{\theta+\pi_q}\right)^{\alpha-1}
\right].
\] Evoking equation \eqref{eq:cte-expectation} yields

\begin{eqnarray*}
  TVaR_q[X] &=& \pi_q+\frac{\theta}{\alpha-1} \frac{(\theta/(\theta+\pi_q))^{\alpha-1}}
{(\theta/(\theta+\pi_q))^{\alpha}}\\
&=&\pi_q +\frac{\theta}{\alpha-1}\left( \frac{\pi_q+\theta}{\theta} \right)\\
&=& \pi_q+\frac{\pi_q+\theta}{\alpha-1},
\end{eqnarray*}

where \(\pi_q\) is given by \eqref{eq:var-pareto}.

Via a change of variables, we can also rewrite equation \eqref{eq:cte-pdf}
as

\begin{eqnarray}
  TVaR_{q}[X] &=& \frac{1}{(1-q)}\int_{q}^{1} VaR_{\alpha}[X]\ d\alpha.
  \label{eq:cte-var}
\end{eqnarray}

What this alternative formula \eqref{eq:cte-var} tells is that \emph{TVaR}
in fact is the average of \(VaR_{\alpha}[X]\) with varying degree of
confidence level over \(\alpha\in [q,1]\). Therefore, the \emph{TVaR}
effectively resolves most of the limitations of \emph{VaR} outlined in
the previous subsection. First, due to the averaging effect, the
\emph{TVaR} may be less sensitive to the change of confidence level
compared with \emph{VaR}. Second, all the extremal losses that are above
the \((1-q)\times 100\%\) worst probable event are taken in account.

In this respect, it is a simple matter for us to see that for any given
\(q\in [0,1]\) \[
TVaR_q[X]\geq VaR_q[X].
\] Third and perhaps foremost, \emph{TVaR} is a coherent risk measure
and thus is able to more accurately capture the diversification effects
of insurance portfolio. Herein, we do not intend to provide the proof of
the coherent feature for \emph{TVaR}, which is considered to be
challenging technically.

\subsection{Properties of risk
measures}\label{properties-of-risk-measures}

To compare the magnitude of risk in a practically convenient manner, we
aim to seek a function that maps the loss \emph{rv} of interest to a
numerical value indicating the level of riskiness, which is termed the
risk measure. Putting mathematically, denoted by \(\mathcal{X}\) a set
of insurance loss \emph{rv}'s, a risk measure is a functional map
\(H:\mathcal{X}\rightarrow \mathbf{R}_+\). In principle, risk measures
can admit an unlimited number of functional formats. Classical examples
of risk measures include the mean \(\mathrm{E}[X]\), the standard
deviation \(\mathrm{SD}(X):=\sqrt{\mathrm{Var}(X)}\), the standard
deviation principle

\begin{equation}
H_{\mathrm{SD}}(X):=\mathrm{E}[X]+\alpha \mathrm{SD}(X),\text{ for } \alpha\geq 0,
\label{eq:SD-principle}
\end{equation}

and the variance principle \[
H_{\mathrm{Var}}(X):=\mathrm{E}[X]+\alpha \mathrm{Var}(X),\text{ for } \alpha\geq 0.
\] It is a simple matter to check that all the aforementioned functions
are risk measures in which we input the loss \emph{rv} and the functions
output a numerical value. On a different note, the function
\(H^{\ast}(X):=\alpha X^{\beta}\) for any real-valued
\(\alpha,\beta\neq 0\), is not a risk measure since \(H^{\ast}\)
produces another \emph{rv} rather than a single numerical value.

Since risk measures are scalar measures which aim to use a single
numerical value to describe the stochastic nature of loss \emph{rv}'s,
it should not be surprising to us that there is no risk measures can
capture all the risk information of the associated \emph{rv}'s.
Therefore, when seeking useful risk measures, it is important for us to
keep in mind that the measures should be at least

\begin{itemize}
\tightlist
\item
  interpretable practically;\\
\item
  computable conveniently; and\\
\item
  being able to reflect the most critical information of risk
  underpinning the loss distribution.
\end{itemize}

A vast number of risk measures have been developed in the literature of
actuarial mathematics. Unfortunately, there is no best risk measure that
can outperform the others, and the selection of appropriate risk measure
depends mainly on the application questions at hand. In this respect, it
is imperative to emphasize that `risk' is a subjective concept, and thus
even given the same problem, there are multifarious approaches to assess
risk. However, for many risk management applications, there is a wide
agreement that economically sounded risk measures should satisfy four
major axioms which we are going to describe them in detail next. Risk
measures that satisfy these axioms are termed \emph{coherent} risk
measures.

Consider in what follows a risk measure \(H(\cdot)\), then \(H\) is a
coherent risk measure if the following axioms are satisfied.

\begin{itemize}
\tightlist
\item
  \textbf{Axiom 1.} \emph{Subadditivity:} \(H(X+Y)\leq H(X)+H(Y)\). The
  economic implication of this axiom is that diversification benefits
  exist if different risks are combined.\\
\item
  \textbf{Axiom 2.} \emph{Monotonicity:} if \(\Pr[X\leq Y]=1\), then
  \(H(X)\leq H(Y)\). Recall that \(X\) and \(Y\) are \emph{rv}'s
  representing losses, the underlying economic implication is that
  higher losses essentially leads to a higher level of risk.\\
\item
  \textbf{Axiom 3.} \emph{Positive homogeneity:} \(H(cX)=cH(X)\) for any
  positive constant \(c\). A potential economic implication about this
  axiom is that risk measure should be independent of the monetary units
  in which the risk is measured. For example, let \(c\) be the currency
  exchange rate between the US and Canadian dollars, then the risk of
  random losses measured in terms of US dollars (i.e., X) and Canadian
  dollars (i.e., cX) should be different only up to the exchange rate
  \(c\) (i.e., \(cH(x)=H(cX)\)).\\
\item
  \textbf{Axiom 4.} \emph{Translation invariance:} \(H(X+c)=H(X)+c\) for
  any positive constant \(c\). If the constant \(c\) is interpreted as
  risk-free cash, this axiom tells that no additional risk is created
  for adding cash to an insurance portfolio, and injecting risk-free
  capital of \(c\) can only reduce the risk by the same amount.
\end{itemize}

Verifying the coherent properties for some risk measures can be quite
straightforward, but it can be very challenging sometimes. For example,
it is a simple matter to check that the mean is a coherent risk measure
since for any pair of \emph{rv}'s \(X\) and \(Y\) having finite means
and constant \(c>0\),

\begin{itemize}
\tightlist
\item
  validation of \emph{subadditivity}:
  \(\mathrm{E}[X+Y]=\mathrm{E}[X]+\mathrm{E}[Y]\);
\item
  validation of \emph{monotonicity}: if \(\Pr[X\leq Y]=1\), then
  \(\mathrm{E}[X]\leq \mathrm{E}[Y]\);
\item
  validation of \emph{positive homogeneity}:
  \(\mathrm{E}[cX]=c\mathrm{E}[X]\);
\item
  validation of \emph{translation invariance}:
  \(\mathrm{E}[X+c]=\mathrm{E}[X]+c\)
\end{itemize}

On a different note, the standard deviation is not a coherent risk
measure. Specifically, one can check that the standard deviation
satisfies

\begin{itemize}
\tightlist
\item
  validation of \emph{subadditivity}:
\end{itemize}

\begin{eqnarray*}
\mathrm{SD}[X+Y]&=&\sqrt{\mathrm{Var}(X)+\mathrm{Var}(Y)+2\mathrm{Cov}(X,Y)}\\
      &\leq& \sqrt{\mathrm{SD}(X)^2+\mathrm{SD}(Y)^2+2\mathrm{SD}(X)\mathrm{SD}(Y)}\\
      &=& \mathrm{SD}(X)+\mathrm{SD}(Y);
\end{eqnarray*}

\begin{itemize}
\tightlist
\item
  validation of \emph{positive homogeneity}:
  \(\mathrm{SD}[cX]=c~\mathrm{SD}[X]\).
\end{itemize}

However, the standard deviation does not comply with translation
invariance property as for any positive constant \(c\), \[
\mathrm{SD}(X+c)=\mathrm{SD}(X)<\mathrm{SD}(X)+c.
\] Moreover, the standard deviation also does not satisfy the
monotonicity property. To see this, consider the following two
\emph{rv}'s

\begin{eqnarray}
X=\left\{
    \begin{array}{ll}
      0, & \hbox{with probability $0.25$;} \\
      4, & \hbox{with probability $0.75$,}
    \end{array}
  \right.
\label{eq:special-x}
\end{eqnarray}

and \(Y\) is a degenerate \emph{rv} such that

\begin{eqnarray}
\Pr[Y = 4] = 1.
\label{eq:special-y}
\end{eqnarray}

It is easy to check that \(\Pr[X\leq Y]=1\), but
\(\mathrm{SD}(X)=\sqrt{4^2\cdot 0.25\cdot 0.75}=\sqrt{3}>\mathrm{SD}(Y)=0\).

We have so far checked that \(\mathrm{E}[\cdot]\) is a coherent risk
measure, but not \(\mathrm{SD}(\cdot)\). Let us now proceed to study the
coherent property for the standard deviation principle
\eqref{eq:SD-principle} which is a linear combination of two coherent and
incoherent risk measures. To this end, for a given \(\alpha>0\), we
check the four axioms for \(H_{\mathrm{SD}}(X+Y)\) one by one:

\begin{itemize}
\tightlist
\item
  validation of \emph{subadditivity:}
\end{itemize}

\begin{eqnarray*}
  H_{\mathrm{SD}}(X+Y) &=& \mathrm{E}[X+Y]+\alpha \mathrm{SD}(X+Y) \\
  &\leq& \mathrm{E}[X]+\mathrm{E}[Y]+\alpha [\mathrm{SD}(X) +\mathrm{SD}(Y)]\\
  &=& H_{\mathrm{SD}}(X)+ H_{\mathrm{SD}}(Y);
\end{eqnarray*}

\begin{itemize}
\tightlist
\item
  validation of \emph{positive homogeneity:} \[
  H_{\mathrm{SD}}(cX)=c\mathrm{E}[X]+c\alpha\mathrm{SD}(X)=cH_{\mathrm{SD}}(X);
  \]
\item
  validation of \emph{translation invariance:} \[
  H_{\mathrm{SD}}(X+c)=\mathrm{E}[X]+c+\alpha\mathrm{SD}(X)=H_{\mathrm{SD}}(X)+c.
  \]
\end{itemize}

It only remains to verify the monotonicity property, which may or may
not be satisfied depending on the value of \(\alpha\). To see this,
consider again the setup of \eqref{eq:special-x} and \eqref{eq:special-y} in
which \(\Pr[X\leq Y]=1\). Let \(\alpha=0.1\cdot \sqrt{3}\), then
\(H_{\mathrm{SD}}(X)=3+0.3=3.3< H_{\mathrm{SD}}(Y)=4\) and the
monotonicity condition is met. On the other hand, let
\(\alpha=\sqrt{3}\), then
\(H_{\mathrm{SD}}(X)=3+3=6> H_{\mathrm{SD}}(Y)=4\) and the monotonicity
condition is not satisfied. More precisely, by setting

\[
  H_{\mathrm{SD}}(X) = 3+\alpha\sqrt{3} \leq4= H_{\mathrm{SD}}(Y),
\]

we find that the monotonicity condition is only satisfied for
\(0\leq\alpha\leq 1/\sqrt{3}\), and thus the standard deviation
principle \(H_{\mathrm{SD}}\) is coherent. This result appears to be
very intuitive to us since the standard deviation principle
\(H_{\mathrm{SD}}\) is a linear combination two risk measures of which
one is coherent and the other is incoherent. If
\(\alpha\leq 1/\sqrt{3}\), then the coherent measure dominates the
incoherent one, thus the resulting measure \(H_{\mathrm{SD}}\) is
coherent and vice versa.

\section{Reinsurance}\label{S:Reinsurance}

Recall that \emph{reinsurance} is simply insurance purchased by an
insurer. Insurance purchased by non-insurers is sometimes known as
\emph{primary} insurance to distinguish it from reinsurance. Reinsurance
differs from personal insurance purchased by individuals, such as auto
and homeowners insurance, in contract flexibility. Like insurance
purchased by major corporations, reinsurance programs are generally
tailored more closely to the buyer. For contrast, in personal insurance
buyers typically cannot negotiate on the contract terms although they
may have a variety of different options (contracts) from which to
choose.

The two broad types are \emph{proportional} and \emph{non-proportional}
reinsurance. A proportional reinsurance contract is an agreement between
a reinsurer and a \emph{ceding} company (also known as the
\emph{reinsured}) in which the reinsurer assumes a given percent of
losses and premium. A reinsurance contract is also known as a
\emph{treaty}. Non-proportional agreements are simply everything else.
As examples of non-proportional agreements, this chapter focuses on
\emph{stop-loss} and \emph{excess of loss} contracts. For all types of
agreements, we split the total risk \(S\) into the portion taken on by
the reinsurer, \(Y_{reinsurer}\), and that retained by the insurer,
\(Y_{insurer}\), that is, \(S= Y_{insurer}+Y_{reinsurer}\).

The mathematical structure of a basic reinsurance treaty is the same as
the coverage modifications of personal insurance introduced in Chapter
3. For a proportional reinsurance, the transformation
\(Y_{insurer} = c S\) is identical to a coinsurance adjustment in
personal insurance. For stop-loss reinsurance, the transformation
\(Y_{reinsurer} = \max(0,S-M)\) is the same as an insurer's payment with
a deductible \(M\) and \(Y_{insurer} = \min(S,M) = S \wedge M\) is
equivalent to what a policyholder pays with deductible \(M\). For
practical applications of the mathematics, in personal insurance the
focus is generally upon the expectation as this is a key ingredient used
in pricing. In constrast, for reinsurance the focus is on the entire
distribution of the risk, as the extreme events are a primary concern of
the financial stability of the insurer and reinsurer.

This chapter describes the foundational and most basic of reinsurance
treaties: Section \ref{S:ProportionalRe} for proportional and Section
\ref{S:NonProportionalRe} for non-proportional. Section
\ref{S:AdditionalRe} gives a flavor of more complex contracts.

\subsection{Proportional Reinsurance}\label{S:ProportionalRe}

The simplest example of a proportional treaty is called \emph{quota
share}.

\begin{itemize}
\item
  In a quota share treaty, the reinsurer receives a flat percent, say
  50\%, of the premium for the book of business reinsured.
\item
  In exchange, the reinsurer pays 50\% of losses, including allocated
  loss adjustment expenses
\item
  The reinsurer also pays the ceding company a ceding commission which
  is designed to reflect the differences in underwriting expenses
  incurred.
\end{itemize}

The amounts paid by the direct insurer and the reinsurer are summarized
as

\[
Y_{insurer} = c S \ \ \text{and} \ \ \ Y_{reinsurer} = (1-c) S.
\]

Note that \(Y_{insurer}+Y_{reinsurer}=S\).

\textbf{Example 10.3.1. Distribution of losses under quota share.} To
develop intuition for the effect of quota-share agreement on the
distribution of losses, the following is a short \texttt{R}
demonstration using simulation. Note the relative shapes of the
distributions of total losses, the retained portion (of the insurer),
and the reinsurer's portion.

\begin{center}\includegraphics{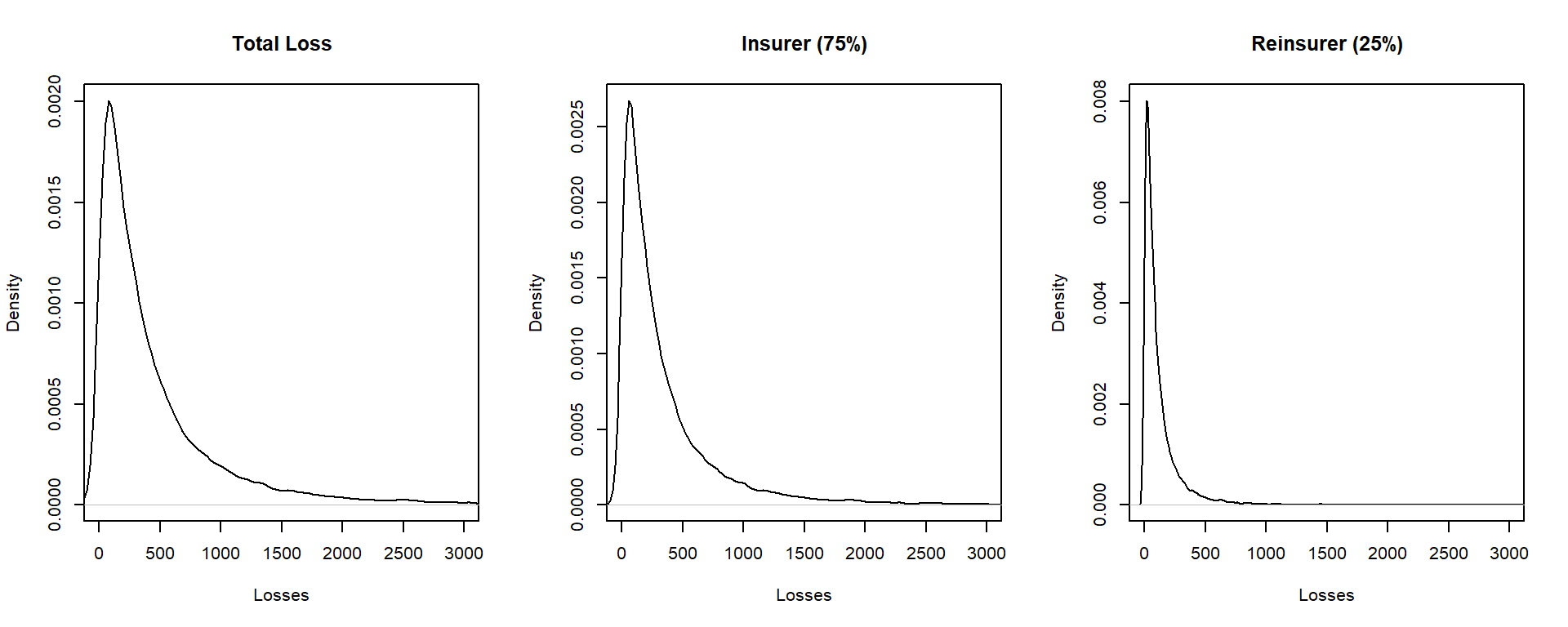} \end{center}

Show the R Code

\hypertarget{toggleQuotaShare}{}
\begin{Shaded}
\begin{Highlighting}[]
\KeywordTok{set.seed}\NormalTok{(}\DecValTok{2018}\NormalTok{)}
\NormalTok{theta =}\StringTok{ }\DecValTok{1000}
\NormalTok{alpha =}\StringTok{ }\DecValTok{3}
\NormalTok{nSim =}\StringTok{ }\DecValTok{10000}
\KeywordTok{library}\NormalTok{(actuar)}
\NormalTok{S <-}\StringTok{  }\KeywordTok{rpareto}\NormalTok{(nSim, }\DataTypeTok{shape =}\NormalTok{ alpha, }\DataTypeTok{scale =}\NormalTok{ theta)}

\KeywordTok{par}\NormalTok{(}\DataTypeTok{mfrow=}\KeywordTok{c}\NormalTok{(}\DecValTok{1}\NormalTok{,}\DecValTok{3}\NormalTok{))}
\KeywordTok{plot}\NormalTok{(}\KeywordTok{density}\NormalTok{(S), }\DataTypeTok{xlim=}\KeywordTok{c}\NormalTok{(}\DecValTok{0}\NormalTok{,}\DecValTok{3}\OperatorTok{*}\NormalTok{theta), }\DataTypeTok{main=}\StringTok{"Total Loss"}\NormalTok{, }\DataTypeTok{xlab=}\StringTok{"Losses"}\NormalTok{)}
\KeywordTok{plot}\NormalTok{(}\KeywordTok{density}\NormalTok{(}\FloatTok{0.75}\OperatorTok{*}\NormalTok{S), }\DataTypeTok{xlim=}\KeywordTok{c}\NormalTok{(}\DecValTok{0}\NormalTok{,}\DecValTok{3}\OperatorTok{*}\NormalTok{theta), }\DataTypeTok{main=}\StringTok{"Insurer (75%)"}\NormalTok{, }\DataTypeTok{xlab=}\StringTok{"Losses"}\NormalTok{)}
\KeywordTok{plot}\NormalTok{(}\KeywordTok{density}\NormalTok{(}\FloatTok{0.25}\OperatorTok{*}\NormalTok{S), }\DataTypeTok{xlim=}\KeywordTok{c}\NormalTok{(}\DecValTok{0}\NormalTok{,}\DecValTok{3}\OperatorTok{*}\NormalTok{theta), }\DataTypeTok{main=}\StringTok{"Reinsurer (25%)"}\NormalTok{, }\DataTypeTok{xlab=}\StringTok{"Losses"}\NormalTok{)}
\end{Highlighting}
\end{Shaded}

\subsubsection{Quota Share is Desirable for
Reinsurers}\label{quota-share-is-desirable-for-reinsurers}

The quota share contract is particularly desirable for the reinsurer. To
see this, suppose that an insurer and reinsurer wish to enter a contract
to share total losses \(S\) such that
\[Y_{insurer}=g(S) \ \ \ \text{and} \ \ \ \ Y_{reinsurer}=S-g(S),\] for
some generic function \(g(\cdot)\) (known as the \emph{retention}
function). Suppose further that the insurer only cares about the
variability of retained claims and is indifferent to the choice of \(g\)
as long as \(Var~Y_{insurer}\) stays the same and equals, say, \(Q\).
Then, the following result shows that the quota share reinsurance treaty
minimizes the reinsurer's uncertainty as measured by
\(Var~Y_{reinsurer}\).

\textbf{Proposition}. Suppose that \(Var~Y_{insurer}=Q.\) Then,
\(Var ((1-c)S) \le Var(g(S))\) for all \(g(.)\).

Show the Justification of the Proposition

\hypertarget{toggleProof}{}
\textbf{Proof of the Proposition}. With
\(Y_{reinsurer} = S - Y_{insurer}\) and the law of total variation

\[
\begin{array}{ll}
Var (Y_{reinsurer}) &= Var (S-Y_{insurer}) \\
&= Var (S) + Var (Y_{insurer})  - 2 Cov (S,Y_{insurer}) \\
&=Var (S) + Q - 2 Corr (S,Y_{insurer}) \times \sqrt{Q} \sqrt{Var (S)}
\end{array}
\] In this expression, we see that \(Q\) and \(Var(S)\) do not change
with the choice of \(g\). Thus, we can minimize \(Var (Y_{reinsurer})\)
by maximizing the correlation \(Corr (S,Y_{insurer})\). If we use a
quota share reinsurance agreement, then
\(Corr (S,Y_{insurer})=Corr (S,(1-c)S)=1\), the maximum possible
correlation. This establishes the proposition.

\(\Box\)`

The proposition is intuitively appealing - with quota share insurance,
the reinsurer shares the responsibility for very large claims in the
tail of the distribution. This is in contrast to non-proportional
agreements where reinsurers take responsibility for the very large
claims.

\subsubsection{Optimizing Quota Share Agreements for
Insurers}\label{optimizing-quota-share-agreements-for-insurers}

Now assume \(n\) risks in the porfolio, \(X_1, \ldots, X_n,\) so that
the portfolio sum is \(S= X_1 + \cdots + X_n\). For simplicity, we focus
on the case of independent risks. Let us consider a variation of the
basic quota share agreement where the amount retained by the insurer may
vary with each risk, say \(c_i\). Thus, the insurer's portion of the
portfolio risk is \(Y_{insurer} = \sum_{i=1}^n c_i X_i\). What is the
best choice of the proportions \(c_i\)?

To formalize this question, we seek to find those values of \(c_i\) that
minimize \(Var ~Y_{insurer}\) subject to the constraint that
\(E ~Y_{insurer} = K.\) The requirement that \(E ~Y_{insurer} = K\)
suggests that the insurers wishes to retain a revenue in at least the
amount of the constant \(K\). Subject to this revenue constraint, the
insurer wishes to minimize uncertainty of the retained risks as measured
by the variance.

Show the Optimal Retention Proportions

\hypertarget{toggleDerivationProof}{}
\textbf{The Optimal Retention Proportions}

Minimizing \(Var ~Y_{insurer}\) subject to \(E ~Y_{insurer} = K\) is a
constrained optimization problem - we can use the method of Lagrange
multipliers, a calculus technique, to solve this. To this end, define
the Lagrangian

\[
\begin{array}{ll}
L &= Var (Y_{insurer}) - \lambda (E ~Y_{insurer} - K) \\
&= \sum_{i=1}^n c_i^2 ~Var ~X_i - \lambda (\sum_{i=1}^n c_i ~E ~X_i - K)
\end{array}
\] Taking a partial derivative with respect to \(\lambda\) and setting
this equal simply means that the constraint, \(E ~Y_{insurer} = K\), is
enforced and we have to choose the proportions \(c_i\) to satisfy this
constraint. Moreover, taking the partial derivative with respect to each
proportion \(c_i\) yields \[
\frac{\partial}{\partial c_i} L = 2 c_i ~Var~ X_i - \lambda ~E ~X_i = 0
\]

so that

\[
c_i  =  \frac{\lambda}{2} \frac{E ~X_i}{Var ~X_i} .
\]

From the math, it turns out that the constant for the \(i\)th risk,
\(c_i\) is proportional to \(\frac{E ~X_i}{Var ~X_i}\). This is
intuitively appealing. Other things being equal, a higher revenue as
measured by \(E ~X_i\) means a higher value of \(c_i\). In the same way,
a higher value of uncertainty as measured by \(Var ~X_i\) means a lower
value of \(c_i\). The proportional scaling factor is determined by the
revenue requirement \(E ~Y_{insurer} = K\). The following example helps
to develop a feel for this relationship.

\textbf{Example 10.3.2. Three Pareto risks.} Consider three risks that
have a Pareto distribution. Provide a graph, and supporting code, that
give values of \(c_1\), \(c_2\), and \(c_3\) for a required revenue
\(K\). Note that these values increase linearly with \(K\).

Show an Example with Three Pareto Risks

\hypertarget{toggleParetoRisksProp}{}
\begin{Shaded}
\begin{Highlighting}[]
\NormalTok{theta1 =}\StringTok{ }\DecValTok{1000}\NormalTok{;theta2 =}\StringTok{ }\DecValTok{2000}\NormalTok{;theta3 =}\StringTok{ }\DecValTok{3000}\NormalTok{;}
\NormalTok{alpha1 =}\StringTok{ }\DecValTok{3}\NormalTok{;alpha2 =}\StringTok{ }\DecValTok{3}\NormalTok{;alpha3 =}\StringTok{ }\DecValTok{4}\NormalTok{;}
\KeywordTok{library}\NormalTok{(actuar)}
\NormalTok{propnfct <-}\StringTok{ }\ControlFlowTok{function}\NormalTok{(alpha,theta)\{}
\NormalTok{  mu    <-}\StringTok{ }\KeywordTok{mpareto}\NormalTok{(}\DataTypeTok{shape=}\NormalTok{alpha, }\DataTypeTok{scale=}\NormalTok{theta, }\DataTypeTok{order=}\DecValTok{1}\NormalTok{)}
\NormalTok{  var   <-}\StringTok{ }\KeywordTok{mpareto}\NormalTok{(}\DataTypeTok{shape=}\NormalTok{alpha, }\DataTypeTok{scale=}\NormalTok{theta, }\DataTypeTok{order=}\DecValTok{2}\NormalTok{) }\OperatorTok{-}\StringTok{ }\NormalTok{mu}\OperatorTok{^}\DecValTok{2}
\NormalTok{  ratio <-}\StringTok{ }\NormalTok{mu}\OperatorTok{/}\NormalTok{var}
\NormalTok{  ratio}
\NormalTok{\}}
\NormalTok{c1 <-}\StringTok{ }\KeywordTok{propnfct}\NormalTok{(alpha1, theta1)}
\NormalTok{c2 <-}\StringTok{ }\KeywordTok{propnfct}\NormalTok{(alpha2, theta2)}
\NormalTok{c3 <-}\StringTok{ }\KeywordTok{propnfct}\NormalTok{(alpha3, theta3)}
\NormalTok{summeans =}\StringTok{ }\KeywordTok{mpareto}\NormalTok{(}\DataTypeTok{shape=}\NormalTok{alpha1, }\DataTypeTok{scale=}\NormalTok{theta1, }\DataTypeTok{order=}\DecValTok{1}\NormalTok{)}\OperatorTok{+}
\StringTok{           }\KeywordTok{mpareto}\NormalTok{(}\DataTypeTok{shape=}\NormalTok{alpha2, }\DataTypeTok{scale=}\NormalTok{theta2, }\DataTypeTok{order=}\DecValTok{1}\NormalTok{)}\OperatorTok{+}
\StringTok{           }\KeywordTok{mpareto}\NormalTok{(}\DataTypeTok{shape=}\NormalTok{alpha3, }\DataTypeTok{scale=}\NormalTok{theta3, }\DataTypeTok{order=}\DecValTok{1}\NormalTok{)  }
\NormalTok{temp =}\StringTok{ }\NormalTok{c1}\OperatorTok{*}\KeywordTok{mpareto}\NormalTok{(}\DataTypeTok{shape=}\NormalTok{alpha1, }\DataTypeTok{scale=}\NormalTok{theta1, }\DataTypeTok{order=}\DecValTok{1}\NormalTok{)}\OperatorTok{+}
\StringTok{       }\NormalTok{c2}\OperatorTok{*}\KeywordTok{mpareto}\NormalTok{(}\DataTypeTok{shape=}\NormalTok{alpha2, }\DataTypeTok{scale=}\NormalTok{theta2, }\DataTypeTok{order=}\DecValTok{1}\NormalTok{)}\OperatorTok{+}
\StringTok{       }\NormalTok{c3}\OperatorTok{*}\KeywordTok{mpareto}\NormalTok{(}\DataTypeTok{shape=}\NormalTok{alpha3, }\DataTypeTok{scale=}\NormalTok{theta3, }\DataTypeTok{order=}\DecValTok{1}\NormalTok{)  }
\NormalTok{KVec =}\StringTok{ }\KeywordTok{seq}\NormalTok{(}\DecValTok{100}\NormalTok{,summeans,}\DataTypeTok{length.out=}\DecValTok{20}\NormalTok{)}
\NormalTok{c1Vec <-}\StringTok{ }\NormalTok{c2Vec <-c3Vec <-}\StringTok{ }\DecValTok{0}\OperatorTok{*}\NormalTok{KVec }
\ControlFlowTok{for}\NormalTok{ (j }\ControlFlowTok{in} \DecValTok{1}\OperatorTok{:}\DecValTok{20}\NormalTok{) \{}
\NormalTok{  c1Vec[j] =}\StringTok{ }\NormalTok{c1 }\OperatorTok{*}\StringTok{ }\NormalTok{KVec[j]}\OperatorTok{/}\NormalTok{temp}
\NormalTok{  c2Vec[j] =}\StringTok{ }\NormalTok{c2 }\OperatorTok{*}\StringTok{ }\NormalTok{KVec[j]}\OperatorTok{/}\NormalTok{temp}
\NormalTok{  c3Vec[j] =}\StringTok{ }\NormalTok{c3 }\OperatorTok{*}\StringTok{ }\NormalTok{KVec[j]}\OperatorTok{/}\NormalTok{temp}
\NormalTok{  \}}
\KeywordTok{plot}\NormalTok{(KVec,c1Vec, }\DataTypeTok{type=}\StringTok{"l"}\NormalTok{, }\DataTypeTok{ylab=}\StringTok{"proportion"}\NormalTok{, }\DataTypeTok{xlab=}\StringTok{"required revenue"}\NormalTok{, }\DataTypeTok{ylim=}\KeywordTok{c}\NormalTok{(}\DecValTok{0}\NormalTok{,}\DecValTok{1}\NormalTok{))}
\KeywordTok{lines}\NormalTok{(KVec,c2Vec)}
\KeywordTok{lines}\NormalTok{(KVec,c3Vec)}
\KeywordTok{text}\NormalTok{(}\DecValTok{1200}\NormalTok{,.}\DecValTok{8}\NormalTok{,}\KeywordTok{expression}\NormalTok{(c[}\DecValTok{1}\NormalTok{]))}
\KeywordTok{text}\NormalTok{(}\DecValTok{2000}\NormalTok{,.}\DecValTok{75}\NormalTok{,}\KeywordTok{expression}\NormalTok{(c[}\DecValTok{2}\NormalTok{]))}
\KeywordTok{text}\NormalTok{(}\DecValTok{1500}\NormalTok{,.}\DecValTok{3}\NormalTok{,}\KeywordTok{expression}\NormalTok{(c[}\DecValTok{3}\NormalTok{]))}
\end{Highlighting}
\end{Shaded}

\begin{center}\includegraphics{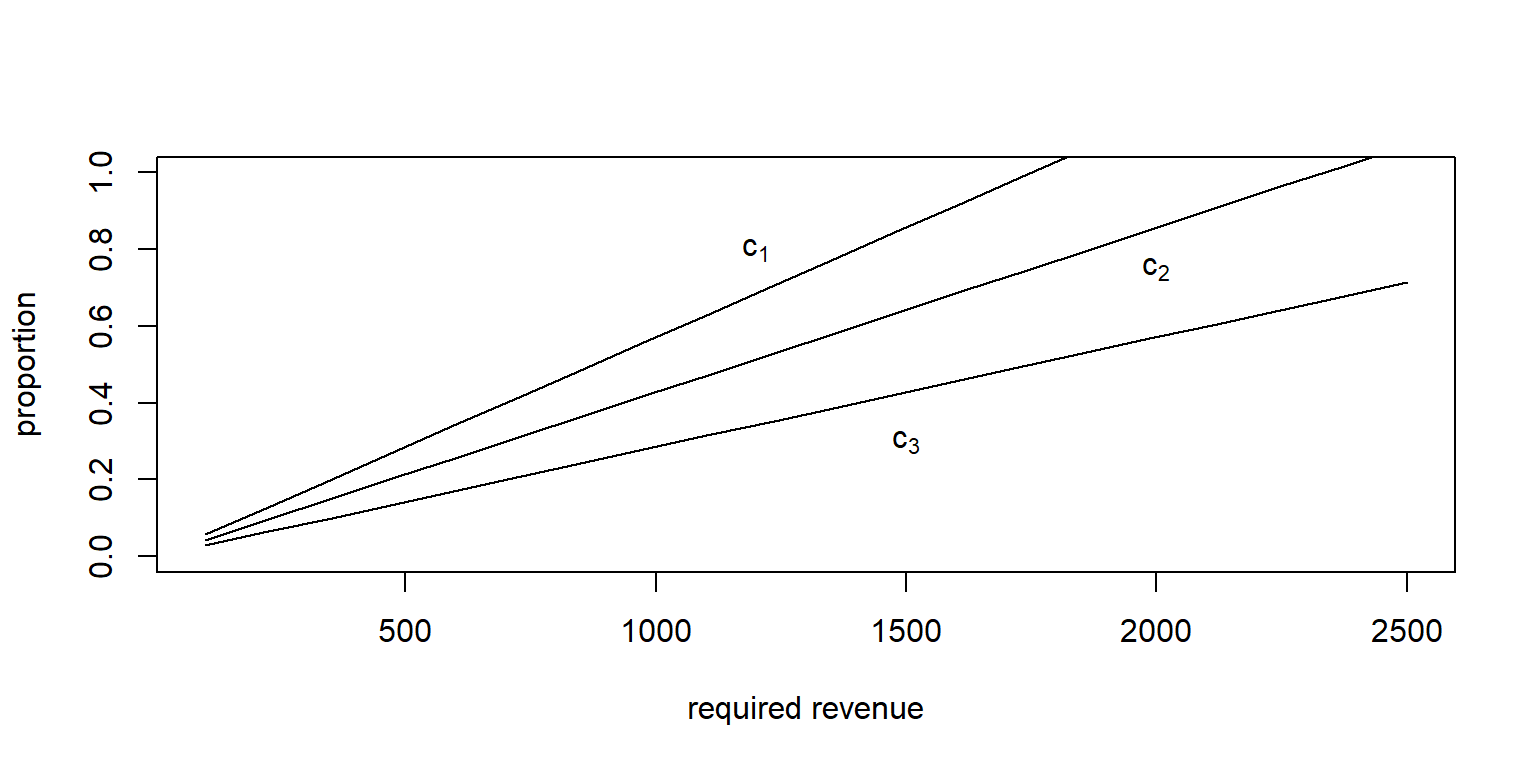} \end{center}

\subsection{Non-Proportional Reinsurance}\label{S:NonProportionalRe}

\subsubsection{The Optimality of Stop Loss
Insurance}\label{the-optimality-of-stop-loss-insurance}

Under a \textbf{stop loss} arrangement, the insurer sets a retention
level \(M (>0)\) and pays in full total claims for which \(S \le M\).
Thus, the insurer retains an amount \(M\) of the risk. Further, for
claims for which \(S > M\), the direct insurer pays \(M\) and the
reinsurer pays the remaining amount \(S-M\). Summarizing, the amounts
paid by the direct insurer and the reinsurer are

\[
Y_{insurer} =
\begin{cases}
S & \text{for } S \le M\\
M & \text{for } S >M \\
\end{cases} \ \ \ \ = \min(S,M) = S \wedge M
\]

and

\[
Y_{reinsurer} =
\begin{cases}
0 & \text{for } S \le M\\
S- M &  \text{for } S >M \\
\end{cases} \ \ \ \  = \max(0,S-M) .
\]

As before, note that \(Y_{insurer}+Y_{reinsurer}=S\).

The stop loss type of contract is particularly desirable for the
insurer. Similar to earlier, suppose that an insurer and reinsurer wish
to enter a contract so that \(Y_{insurer}=g(S)\) and
\(Y_{reinsurer}=S-g(S)\) for some generic retention function
\(g(\cdot)\). Suppose further that the insurer only cares about the
variability of retained claims and is indifferent to the choice of \(g\)
as long as \(Var~Y_{insurer}\) can be minimized. Again, we impose the
constraint that \(E ~Y_{insurer} = K\); the insurer needs to retain a
revenue \(K\). Subject to this revenue constraint, the insurer wishes to
minimize uncertainty of the retained risks (as measured by the
variance). Then, the following result shows that the stop loss
reinsurance treaty minimizes the reinsurer's uncertainty as measured by
\(Var~Y_{reinsurer}\).

\textbf{Proposition}. Suppose that \(E~Y_{insurer}=K.\) Then,
\(Var (S \wedge M) \le Var(g(S))\) for all \(g(.)\).

Show the Justification of the Proposition

\hypertarget{toggleProofStopLoss}{}
\textbf{Proof of the Proposition}. Add and subtract a constant \(M\) and
expand the square to get \[
\begin{array}{ll}
Var~ g(S) &= E (g(S) - K)^2 = E (g(S) -M +M- K)^2 \\
&= E (g(S) -M)^2 +  (M- K)^2 +2 E (g(S) -M)(M- K) \\
&= E (g(S) -M)^2 -  (M- K)^2 ,
\end{array}
\] because \(E ~g(S)= K.\)

Now, for any retention function, we have \(g(S) \le S\), that is, the
insurer's retained claims are less than or equal to total claims. Using
the notation \(g_{SL}(S) = S \wedge M\) for stop loss insurance, we have

\[
\begin{array}{ll}
M- g_{SL}(S) &= M-(S \wedge M) \\
&= (M-S) \wedge 0 \\
&\le (M-g(S)) \wedge 0 .
\end{array}
\] Squaring each side yields
\[(M- g_{SL}(S))^2 \le (M-g(S))^2 \wedge 0 \le (M-g(S))^2.\]

Returning to our expression for the variance, we have \[
\begin{array}{ll}
Var~ g_{SL}(S) &= E (g_{SL}(S) -M)^2 -  (M- K)^2 \\
&\le E (g_{SL}(S) -M)^2 -  (M- K)^2 = Var~ g(S) ,
\end{array}
\] for any retention function \(g\). This establishes the proposition.

\(\Box\)`

The proposition is intuitively appealing - with stop loss insurance, the
reinsurer takes the responsibility for very large claims in the tail of
the distribution, not the insurer.

\subsubsection{Excess of Loss}\label{excess-of-loss}

A closely related form of non-proportional reinsurance is the
\textbf{excess of loss} coverage. Under this contract, we assume that
the total risk \(S\) can be thought of as composed as \(n\) separate
risks \(X_1, \ldots, X_n\) and that each of these risks are subject to
upper limit, say, \(M_i\). So the insurer retains

\[
Y_{i,insurer} = X_i \wedge M_i \ \ \ \ Y_{insurer} = \sum_{i=1}^n Y_{i,insurer}
\] and the reinsurer is responsible for the excess,
\(Y_{reinsurer}=S - Y_{insurer}\). The retention limits may vary by risk
or may be the same for all risks, \(M_i =M\), for all \(i\).

\subsubsection{Optimal Choice for Excess of Loss Retention
Limits}\label{optimal-choice-for-excess-of-loss-retention-limits}

What is the best choice of the excess of loss retention limits \(M_i\)?
To formalize this question, we seek to find those values of \(M_i\) that
minimize \(Var ~Y_{insurer}\) subject to the constraint that
\(E ~Y_{insurer} = K.\) Subject to this revenue constraint, the insurer
wishes to minimize uncertainty of the retained risks (as measured by the
variance).

Show the Optimal Retention Proportions

\hypertarget{toggleDerivationProofExcess}{}
\textbf{The Optimal Retention Limits}

Minimizing \(Var ~Y_{insurer}\) subject to \(E ~Y_{insurer} = K\) is a
constrained optimization problem - we can use the method of Lagrange
multipliers, a calculus technique, to solve this. As before, define the
Lagrangian \[
\begin{array}{ll}
L &= Var (Y_{insurer}) - \lambda (E ~Y_{insurer} - K) \\
&= \sum_{i=1}^n ~Var (X_i \wedge M_i) - \lambda (\sum_{i=1}^n ~E(X_i \wedge M_i)- K)
\end{array}
\]

We first recall the relationships

\[
E~S \wedge M = \int_0^M ~(1- F(S))dx
\] and

\[
E~(S \wedge M)^2 = 2\int_0^M ~x(1- F(x))dx
\]

Taking a partial derivative with respect to \(\lambda\) and setting this
equal simply means that the constraint, \(E ~Y_{insurer} = K\), is
enforced and we have to choose the limits \(M_i\) to satisfy this
constraint. Moreover, taking the partial derivative with respect to each
limit \(M_i\) yields

\[
\begin{array}{ll}
\frac{\partial}{\partial M_i} L
&= \frac{\partial}{\partial M_i}  ~Var~ (X_i \wedge M_i)  - \lambda \frac{\partial}{\partial M_i} ~E ~(X_i \wedge M_i) \\
&= \frac{\partial}{\partial M_i} \left(E~ (X_i \wedge M_i)^2 -(E ~(X_i \wedge M_i))^2\right) - \lambda (1-F_i(M_i)) \\
&= 2 M_i (1-F_i(M_i)) - 2 E ~(X_i \wedge M_i) (1-F_i(M_i))-
\lambda (1-F_i(M_i)).
\end{array}
\]

Setting \(\frac{\partial}{\partial M_i} L =0\) and solving for
\(\lambda\), we get

\[
\lambda = 2 (M_i - E ~(X_i \wedge M_i)) .
\]

From the math, it turns out that the retention limit less the expected
insurer's claims, \(M_i - E ~(X_i \wedge M_i)\), is the same for
\emph{all} risks. This is intuitively appealing.

\textbf{Example 10.3.3. Excess of loss for three Pareto risks.} Consider
three risks that have a Pareto distribution, each having a different set
of parameters (so they are independent but non-identical). Show
numerically that the optimal retention limits \(M_1\), \(M_2\), and
\(M_3\) resulting retention limit minus expected insurer's claims,
\(M_i - E ~(X_i \wedge M_i)\), is the same for all risks, as we derived
theoretically. Further, graphically compare the distribution of total
risks to that retained by the insurer and by the reinsurer.

Show an Example with Three Pareto Risks

\hypertarget{toggleParetoRisksExcess}{}
We first optimize the Lagrangian using the \texttt{R} package
\texttt{alabama} for \emph{Augmented Lagrangian Adaptive Barrier
Minimization Algorithm}.

\begin{Shaded}
\begin{Highlighting}[]
\NormalTok{theta1 =}\StringTok{ }\DecValTok{1000}\NormalTok{;theta2 =}\StringTok{ }\DecValTok{2000}\NormalTok{;theta3 =}\StringTok{ }\DecValTok{3000}\NormalTok{;}
\NormalTok{alpha1 =}\StringTok{ }\DecValTok{3}\NormalTok{;   alpha2 =}\StringTok{ }\DecValTok{3}\NormalTok{;   alpha3 =}\StringTok{ }\DecValTok{4}\NormalTok{;}
\NormalTok{Pmin <-}\StringTok{ }\DecValTok{2000}
\KeywordTok{library}\NormalTok{(actuar)}
\NormalTok{VarFct <-}\StringTok{ }\ControlFlowTok{function}\NormalTok{(M)\{}
\NormalTok{  M1=M[}\DecValTok{1}\NormalTok{];M2=M[}\DecValTok{2}\NormalTok{];M3=M[}\DecValTok{3}\NormalTok{]}
\NormalTok{  mu1    <-}\StringTok{ }\KeywordTok{levpareto}\NormalTok{(}\DataTypeTok{limit=}\NormalTok{M1,}\DataTypeTok{shape=}\NormalTok{alpha1, }\DataTypeTok{scale=}\NormalTok{theta1, }\DataTypeTok{order=}\DecValTok{1}\NormalTok{)}
\NormalTok{  var1   <-}\StringTok{ }\KeywordTok{levpareto}\NormalTok{(}\DataTypeTok{limit=}\NormalTok{M1,}\DataTypeTok{shape=}\NormalTok{alpha1, }\DataTypeTok{scale=}\NormalTok{theta1, }\DataTypeTok{order=}\DecValTok{2}\NormalTok{)}\OperatorTok{-}\NormalTok{mu1}\OperatorTok{^}\DecValTok{2}
\NormalTok{  mu2    <-}\StringTok{ }\KeywordTok{levpareto}\NormalTok{(}\DataTypeTok{limit=}\NormalTok{M2,}\DataTypeTok{shape=}\NormalTok{alpha2, }\DataTypeTok{scale=}\NormalTok{theta2, }\DataTypeTok{order=}\DecValTok{1}\NormalTok{)}
\NormalTok{  var2   <-}\StringTok{ }\KeywordTok{levpareto}\NormalTok{(}\DataTypeTok{limit=}\NormalTok{M2,}\DataTypeTok{shape=}\NormalTok{alpha2, }\DataTypeTok{scale=}\NormalTok{theta2, }\DataTypeTok{order=}\DecValTok{2}\NormalTok{)}\OperatorTok{-}\NormalTok{mu2}\OperatorTok{^}\DecValTok{2}
\NormalTok{  mu3    <-}\StringTok{ }\KeywordTok{levpareto}\NormalTok{(}\DataTypeTok{limit=}\NormalTok{M3,}\DataTypeTok{shape=}\NormalTok{alpha3, }\DataTypeTok{scale=}\NormalTok{theta3, }\DataTypeTok{order=}\DecValTok{1}\NormalTok{)}
\NormalTok{  var3   <-}\StringTok{ }\KeywordTok{levpareto}\NormalTok{(}\DataTypeTok{limit=}\NormalTok{M3,}\DataTypeTok{shape=}\NormalTok{alpha3, }\DataTypeTok{scale=}\NormalTok{theta3, }\DataTypeTok{order=}\DecValTok{2}\NormalTok{)}\OperatorTok{-}\NormalTok{mu3}\OperatorTok{^}\DecValTok{2}
\NormalTok{  varFct <-}\StringTok{ }\NormalTok{var1 }\OperatorTok{+}\NormalTok{var2}\OperatorTok{+}\NormalTok{var3}
\NormalTok{  meanFct <-}\StringTok{ }\NormalTok{mu1}\OperatorTok{+}\NormalTok{mu2}\OperatorTok{+}\NormalTok{mu3}
  \KeywordTok{c}\NormalTok{(meanFct,varFct)}
\NormalTok{  \}}
\NormalTok{f <-}\StringTok{ }\ControlFlowTok{function}\NormalTok{(M)\{}\KeywordTok{VarFct}\NormalTok{(M)[}\DecValTok{2}\NormalTok{]\}}
\NormalTok{h <-}\StringTok{ }\ControlFlowTok{function}\NormalTok{(M)\{}\KeywordTok{VarFct}\NormalTok{(M)[}\DecValTok{1}\NormalTok{] }\OperatorTok{-}\StringTok{ }\NormalTok{Pmin\}}
\KeywordTok{library}\NormalTok{(alabama)}
\NormalTok{par0=}\KeywordTok{rep}\NormalTok{(}\DecValTok{1000}\NormalTok{,}\DecValTok{3}\NormalTok{)}
\NormalTok{op <-}\StringTok{ }\KeywordTok{auglag}\NormalTok{(}\DataTypeTok{par=}\NormalTok{par0,}\DataTypeTok{fn=}\NormalTok{f,}\DataTypeTok{hin=}\NormalTok{h,}\DataTypeTok{control.outer=}\KeywordTok{list}\NormalTok{(}\DataTypeTok{trace=}\OtherTok{FALSE}\NormalTok{))}
\end{Highlighting}
\end{Shaded}

The optimal retention limits \(M_1\), \(M_2\), and \(M_3\) resulting
retention limit minus expected insurer's claims,
\(M_i - E ~(X_i \wedge M_i)\), is the same for all risks, as we derived
theoretically.

\begin{Shaded}
\begin{Highlighting}[]
\NormalTok{M1star =}\StringTok{ }\NormalTok{op}\OperatorTok{$}\NormalTok{par[}\DecValTok{1}\NormalTok{];M2star =}\StringTok{ }\NormalTok{op}\OperatorTok{$}\NormalTok{par[}\DecValTok{2}\NormalTok{];M3star =}\StringTok{ }\NormalTok{op}\OperatorTok{$}\NormalTok{par[}\DecValTok{3}\NormalTok{]}
\NormalTok{M1star }\OperatorTok{-}\KeywordTok{levpareto}\NormalTok{(M1star,}\DataTypeTok{shape=}\NormalTok{alpha1, }\DataTypeTok{scale=}\NormalTok{theta1,}\DataTypeTok{order=}\DecValTok{1}\NormalTok{)}
\end{Highlighting}
\end{Shaded}

\begin{verbatim}
[1] 1344.135
\end{verbatim}

\begin{Shaded}
\begin{Highlighting}[]
\NormalTok{M2star }\OperatorTok{-}\KeywordTok{levpareto}\NormalTok{(M2star,}\DataTypeTok{shape=}\NormalTok{alpha2, }\DataTypeTok{scale=}\NormalTok{theta2,}\DataTypeTok{order=}\DecValTok{1}\NormalTok{)}
\end{Highlighting}
\end{Shaded}

\begin{verbatim}
[1] 1344.133
\end{verbatim}

\begin{Shaded}
\begin{Highlighting}[]
\NormalTok{M3star }\OperatorTok{-}\KeywordTok{levpareto}\NormalTok{(M3star,}\DataTypeTok{shape=}\NormalTok{alpha3, }\DataTypeTok{scale=}\NormalTok{theta3,}\DataTypeTok{order=}\DecValTok{1}\NormalTok{)}
\end{Highlighting}
\end{Shaded}

\begin{verbatim}
[1] 1344.133
\end{verbatim}

We graphically compare the distribution of total risks to that retained
by the insurer and by the reinsurer.

\begin{Shaded}
\begin{Highlighting}[]
\KeywordTok{set.seed}\NormalTok{(}\DecValTok{2018}\NormalTok{)}
\NormalTok{nSim =}\StringTok{ }\DecValTok{10000}
\KeywordTok{library}\NormalTok{(actuar)}
\NormalTok{Y1 <-}\StringTok{ }\KeywordTok{rpareto}\NormalTok{(nSim, }\DataTypeTok{shape =}\NormalTok{ alpha1, }\DataTypeTok{scale =}\NormalTok{ theta1)}
\NormalTok{Y2 <-}\StringTok{ }\KeywordTok{rpareto}\NormalTok{(nSim, }\DataTypeTok{shape =}\NormalTok{ alpha2, }\DataTypeTok{scale =}\NormalTok{ theta2)}
\NormalTok{Y3 <-}\StringTok{ }\KeywordTok{rpareto}\NormalTok{(nSim, }\DataTypeTok{shape =}\NormalTok{ alpha3, }\DataTypeTok{scale =}\NormalTok{ theta3)}
\NormalTok{YTotal <-}\StringTok{ }\NormalTok{Y1 }\OperatorTok{+}\StringTok{ }\NormalTok{Y2 }\OperatorTok{+}\StringTok{ }\NormalTok{Y3}
\NormalTok{Yinsur <-}\StringTok{  }\KeywordTok{pmin}\NormalTok{(Y1,M1star)}\OperatorTok{+}\KeywordTok{pmin}\NormalTok{(Y2,M2star)}\OperatorTok{+}\KeywordTok{pmin}\NormalTok{(Y3,M3star)}
\NormalTok{Yreinsur <-}\StringTok{ }\NormalTok{YTotal }\OperatorTok{-}\StringTok{ }\NormalTok{Yinsur}

\KeywordTok{par}\NormalTok{(}\DataTypeTok{mfrow=}\KeywordTok{c}\NormalTok{(}\DecValTok{1}\NormalTok{,}\DecValTok{3}\NormalTok{))}
\KeywordTok{plot}\NormalTok{(}\KeywordTok{density}\NormalTok{(YTotal),   }\DataTypeTok{xlim=}\KeywordTok{c}\NormalTok{(}\DecValTok{0}\NormalTok{,}\DecValTok{10000}\NormalTok{), }\DataTypeTok{main=}\StringTok{"Total Loss"}\NormalTok{, }\DataTypeTok{xlab=}\StringTok{"Losses"}\NormalTok{)}
\KeywordTok{plot}\NormalTok{(}\KeywordTok{density}\NormalTok{(Yinsur),   }\DataTypeTok{xlim=}\KeywordTok{c}\NormalTok{(}\DecValTok{0}\NormalTok{,}\DecValTok{10000}\NormalTok{), }\DataTypeTok{main=}\StringTok{"Insurer"}\NormalTok{,    }\DataTypeTok{xlab=}\StringTok{"Losses"}\NormalTok{)}
\KeywordTok{plot}\NormalTok{(}\KeywordTok{density}\NormalTok{(Yreinsur), }\DataTypeTok{xlim=}\KeywordTok{c}\NormalTok{(}\DecValTok{0}\NormalTok{,}\DecValTok{10000}\NormalTok{), }\DataTypeTok{main=}\StringTok{"Reinsurer"}\NormalTok{,  }\DataTypeTok{xlab=}\StringTok{"Losses"}\NormalTok{)}
\end{Highlighting}
\end{Shaded}

\begin{center}\includegraphics{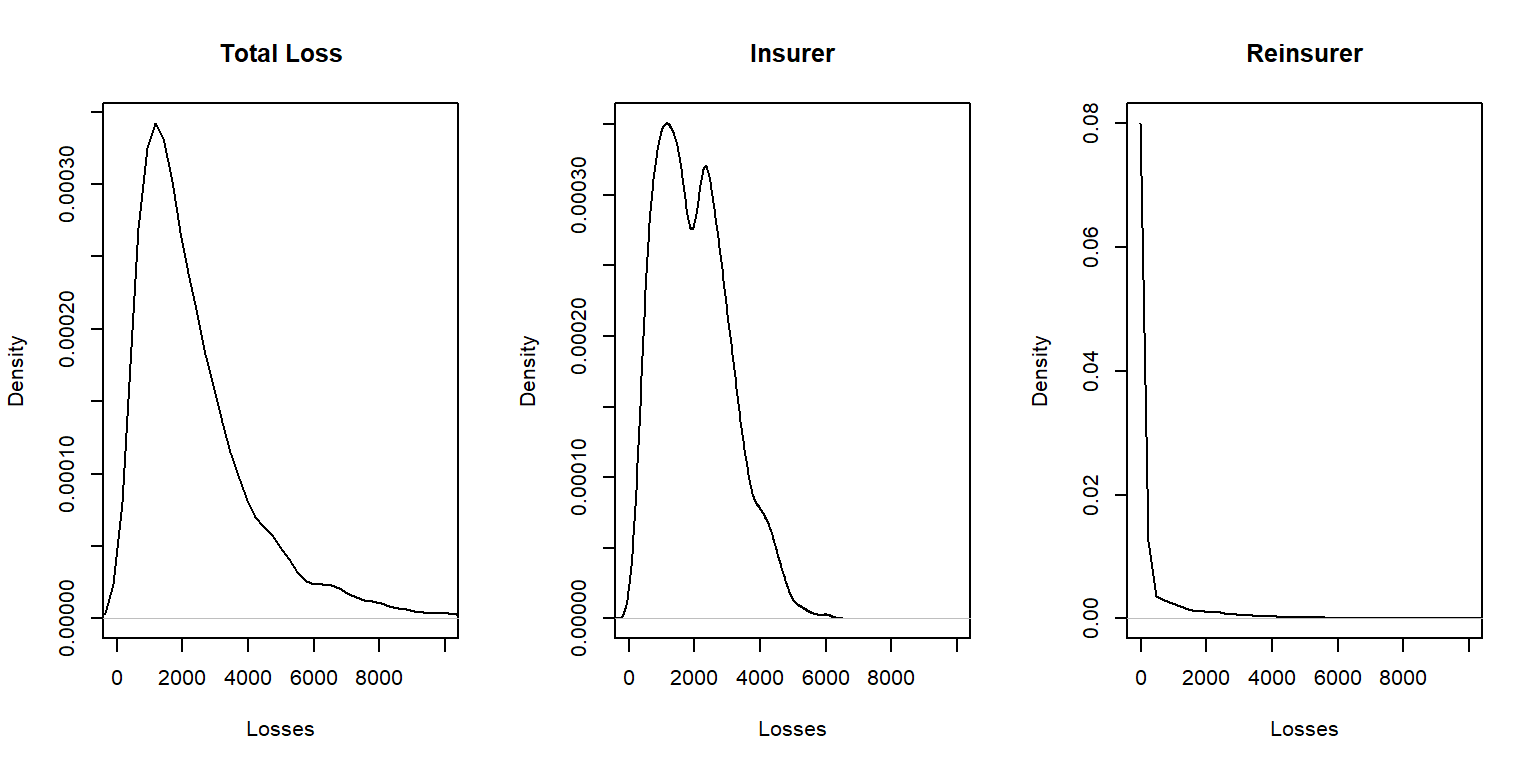} \end{center}

\begin{center}\rule{0.5\linewidth}{\linethickness}\end{center}

\subsection{Additional Reinsurance Treaties}\label{S:AdditionalRe}

\subsubsection{Surplus Share Proportional
Treaty}\label{surplus-share-proportional-treaty}

Another proportional treaty is known as \textbf{surplus share}; this
type of contract is common in commercial property insurance.

\begin{itemize}
\tightlist
\item
  A surplus share treaty allows the reinsured to limit its exposure on
  any one risk to a given amount (the \emph{retained line}).
\item
  The reinsurer assumes a part of the risk in proportion to the amount
  that the insured value exceeds the retained line, up to a given limit
  (expressed as a multiple of the retained line, or number of lines).
\end{itemize}

For example, let the retained line be \$100,000 and let the given limit
be 4 lines (\$400,000). Then, if \(S\) is the loss, the reinsurer's
portion is \(\min(400000, (S-100000)_+)\).

\subsubsection{Layers of Coverage}\label{layers-of-coverage}

One can also extend non-proportional stop loss treaties by introducing
additional parties to the contract. For example, instead of simply an
insurer and reinsurer or an insurer and a policyholder, think about the
situation with all three parties, a policyholder, insurer, and
reinsurer, who agree on how to share a risk. More generally, we consider
\(k\) parties. If \(k=4\), it could be an insurer and three different
reinsurers.

\textbf{Example 10.3.4. Layers of coverage for three parties.}

\begin{itemize}
\item
  Suppose that there are \(k=3\) parties. The first party is responsible
  for the first 100 of claims, the second responsible for claims from
  100 to 3000, and the third responsible for claims above 3000.
\item
  If there are four claims in the amounts 50, 600, 1800 and 4000, then
  they would be allocated to the parties as follows:
\end{itemize}

\begin{longtable}[]{@{}lccccl@{}}
\toprule
Layer & Claim 1 & Claim 2 & Claim 3 & Claim 4 & Total\tabularnewline
\midrule
\endhead
(0, 100{]} & 50 & 100 & 100 & 100 & 350\tabularnewline
(100, 3000{]} & 0 & 500 & 1700 & 2900 & 5100\tabularnewline
(3000, \(\infty\)) & 0 & 0 & 0 & 1000 & 1000\tabularnewline
Total & 50 & 600 & 1800 & 4000 & 6450\tabularnewline
\bottomrule
\end{longtable}

\begin{center}\rule{0.5\linewidth}{\linethickness}\end{center}

To handle the general situation with \(k\) groups, partition the
positive real line into \(k\) intervals using the cut-points
\[0 = M_0 < M_1 < \cdots < M_{k-1} < M_k = \infty.\]

Note that the \(j\)th interval is \((M_{j-1}, M_j]\). Now let \(Y_j\) be
the amount of risk shared by the \(j\)th party. To illustrate, if a loss
\(x\) is such that \(M_{j-1} <x \le M_j\), then \[\left(\begin{array}{c}
    Y_1\\ Y_2 \\ \vdots \\ Y_j \\Y_{j+1} \\ \vdots \\Y_k
    \end{array}\right)
    =\left(\begin{array}{c}
    M_1-M_0 \\ M_2-M_1  \\ \vdots \\ x-M_{j-1}  \\ 0 \\ \vdots \\0
    \end{array}\right)\]

More succinctly, we can write \[Y_j = \min(S,M_j) - \min(S,M_{j-1}) .\]

With the expression \(Y_j = \min(S,M_j) - \min(S,M_{j-1})\), we see that
the \(j\)th party is responsible for claims in the interval
\((M_{j-1}, M_j].\) With this, it is easy to check that
\(S = Y_1 + Y_2 + \cdots + Y_k.\) As emphasized in the following
example, we also remark that the parties need not be different.

\textbf{Example 10.3.5.} - Suppose that a policyholder is responsible
for the first 500 of claims and all claims in excess of 100,000. The
insurer takes claims between 100 and 100,000. - Then, we would use
\(M_1 = 100\), \(M_2 =100000\). - The policyholder is responsible for
\(Y_1 =\min(S,100)\) and
\(Y_3 = S - \min(S,100000) = \max(0, S-100000)\).

For additional reading, see the
\href{https://sites.google.com/a/wisc.edu/local-government-property-insurance-fund/home/reinsurance}{Wisconsin
Property Fund site} for more info on layers of reinsurance.

\subsubsection{Portfolio Management
Example}\label{portfolio-management-example}

Many other variations of the foundational contracts are possible. For
one more illustration, consider the following.

\textbf{Example. 10.3.6. Portfolio management.} You are the Chief Risk
Officer of a telecommunications firm. Your firm has several property and
liabililty risks. We will consider:

\begin{itemize}
\tightlist
\item
  \(X_1\) - buildings, modeled using a gamma distribution with mean 200
  and scale parameter 100.
\item
  \(X_2\) - motor vehicles, modeled using a gamma distribution with mean
  400 and scale parameter 200.
\item
  \(X_3\) - directors and executive officers risk, modeled using a
  Pareto distribution with mean 1000 and scale parameter 1000.
\item
  \(X_4\) - cyber risks, modeled using a Pareto distribution with mean
  1000 and scale parameter 2000.
\end{itemize}

Denote the total risk as \[S = X_1 + X_2 + X_3 + X_4 .\]

For simplicity, you assume that these risks are independent.

To manage the risk, you seek some insurance protection. You wish to
manage internally small building and motor vehicles amounts, up to
\(M_1\) and \(M_2\), respectively. You seek insurance to cover all other
risks. Specifically, the insurer's portion is
\[ Y_{insurer} = (X_1 - M_1)_+ + (X_2 - M_2)_+ + X_3 + X_4 ,\] so that
your retained risk is
\(Y_{retained}= S- Y_{insurer} = \min(X_1,M_1) + \min(X_2,M_2)\). Using
deductibles \(M_1=\) 100 and \(M_2=\) 200:

\begin{enumerate}
\def\labelenumi{\alph{enumi}.}
\tightlist
\item
  Determine the expected claim amount of (i) that retained, (ii) that
  accepted by the insurer, and (iii) the total overall amount.
\item
  Determine the 80th, 90th, 95th, and 99th percentiles for (i) that
  retained, (ii) that accepted by the insurer, and (iii) the total
  overall amount.
\item
  Compare the distributions by plotting the densities for (i) that
  retained, (ii) that accepted by the insurer, and (iii) the total
  overall amount.
\end{enumerate}

Show Example Solution with R Code

\hypertarget{togglePortMgtExample}{}
In preparation, here is the code needed to set the parameters.

\begin{Shaded}
\begin{Highlighting}[]
\CommentTok{# For the gamma distributions, use}
\NormalTok{alpha1 <-}\StringTok{ }\DecValTok{2}\NormalTok{;      theta1 <-}\StringTok{ }\DecValTok{100}
\NormalTok{alpha2 <-}\StringTok{ }\DecValTok{2}\NormalTok{;      theta2 <-}\StringTok{ }\DecValTok{200}
\CommentTok{# For the Pareto distributions, use}
\NormalTok{alpha3 <-}\StringTok{ }\DecValTok{2}\NormalTok{;      theta3 <-}\StringTok{ }\DecValTok{1000}
\NormalTok{alpha4 <-}\StringTok{ }\DecValTok{3}\NormalTok{;      theta4 <-}\StringTok{ }\DecValTok{2000}
\CommentTok{# Limits}
\NormalTok{M1     <-}\StringTok{ }\DecValTok{100}
\NormalTok{M2     <-}\StringTok{ }\DecValTok{200}
\end{Highlighting}
\end{Shaded}

With these parameters, we can now simulate realizations of the portfolio
risks.

\begin{Shaded}
\begin{Highlighting}[]
\CommentTok{# Simulate the risks}
\NormalTok{nSim <-}\StringTok{ }\DecValTok{10000}  \CommentTok{#number of simulations}
\KeywordTok{set.seed}\NormalTok{(}\DecValTok{2017}\NormalTok{) }\CommentTok{#set seed to reproduce work }
\NormalTok{X1 <-}\StringTok{ }\KeywordTok{rgamma}\NormalTok{(nSim,alpha1,}\DataTypeTok{scale =}\NormalTok{ theta1)  }
\NormalTok{X2 <-}\StringTok{ }\KeywordTok{rgamma}\NormalTok{(nSim,alpha2,}\DataTypeTok{scale =}\NormalTok{ theta2)  }
\CommentTok{# For the Pareto Distribution, use}
\KeywordTok{library}\NormalTok{(actuar)}
\NormalTok{X3 <-}\StringTok{ }\KeywordTok{rpareto}\NormalTok{(nSim,}\DataTypeTok{scale=}\NormalTok{theta3,}\DataTypeTok{shape=}\NormalTok{alpha3)}
\NormalTok{X4 <-}\StringTok{ }\KeywordTok{rpareto}\NormalTok{(nSim,}\DataTypeTok{scale=}\NormalTok{theta4,}\DataTypeTok{shape=}\NormalTok{alpha4)}
\CommentTok{# Portfolio Risks}
\NormalTok{S         <-}\StringTok{ }\NormalTok{X1 }\OperatorTok{+}\StringTok{ }\NormalTok{X2 }\OperatorTok{+}\StringTok{ }\NormalTok{X3 }\OperatorTok{+}\StringTok{ }\NormalTok{X4}
\NormalTok{Yretained <-}\StringTok{ }\KeywordTok{pmin}\NormalTok{(X1,M1) }\OperatorTok{+}\StringTok{ }\KeywordTok{pmin}\NormalTok{(X2,M2)}
\NormalTok{Yinsurer  <-}\StringTok{ }\NormalTok{S }\OperatorTok{-}\StringTok{ }\NormalTok{Yretained}
\end{Highlighting}
\end{Shaded}

\textbf{(a)} Here is the code for the expected claim amounts.

\begin{Shaded}
\begin{Highlighting}[]
\CommentTok{# Expected Claim Amounts}
\NormalTok{ExpVec <-}\StringTok{ }\KeywordTok{t}\NormalTok{(}\KeywordTok{as.matrix}\NormalTok{(}\KeywordTok{c}\NormalTok{(}\KeywordTok{mean}\NormalTok{(Yretained),}\KeywordTok{mean}\NormalTok{(Yinsurer),}\KeywordTok{mean}\NormalTok{(S))))}
\KeywordTok{colnames}\NormalTok{(ExpVec) <-}\StringTok{ }\KeywordTok{c}\NormalTok{(}\StringTok{"Retained"}\NormalTok{, }\StringTok{"Insurer"}\NormalTok{,}\StringTok{"Total"}\NormalTok{)}
\KeywordTok{round}\NormalTok{(ExpVec,}\DataTypeTok{digits=}\DecValTok{2}\NormalTok{)}
\end{Highlighting}
\end{Shaded}

\begin{verbatim}
     Retained Insurer   Total
[1,]   269.05 2274.41 2543.46
\end{verbatim}

\textbf{(b)} Here is the code for the quantiles.

\begin{Shaded}
\begin{Highlighting}[]
\CommentTok{# Quantiles}
\NormalTok{quantMat <-}\StringTok{ }\KeywordTok{rbind}\NormalTok{(}
  \KeywordTok{quantile}\NormalTok{(Yretained, }\DataTypeTok{probs=}\KeywordTok{c}\NormalTok{(}\FloatTok{0.80}\NormalTok{, }\FloatTok{0.90}\NormalTok{, }\FloatTok{0.95}\NormalTok{, }\FloatTok{0.99}\NormalTok{)),}
  \KeywordTok{quantile}\NormalTok{(Yinsurer,  }\DataTypeTok{probs=}\KeywordTok{c}\NormalTok{(}\FloatTok{0.80}\NormalTok{, }\FloatTok{0.90}\NormalTok{, }\FloatTok{0.95}\NormalTok{, }\FloatTok{0.99}\NormalTok{)),}
  \KeywordTok{quantile}\NormalTok{(S       ,  }\DataTypeTok{probs=}\KeywordTok{c}\NormalTok{(}\FloatTok{0.80}\NormalTok{, }\FloatTok{0.90}\NormalTok{, }\FloatTok{0.95}\NormalTok{, }\FloatTok{0.99}\NormalTok{)))}
\KeywordTok{rownames}\NormalTok{(quantMat) <-}\StringTok{ }\KeywordTok{c}\NormalTok{(}\StringTok{"Retained"}\NormalTok{, }\StringTok{"Insurer"}\NormalTok{,}\StringTok{"Total"}\NormalTok{)}
\KeywordTok{round}\NormalTok{(quantMat,}\DataTypeTok{digits=}\DecValTok{2}\NormalTok{)}
\end{Highlighting}
\end{Shaded}

\begin{verbatim}
             80%     90%     95%      99%
Retained  300.00  300.00  300.00   300.00
Insurer  3075.67 4399.80 6172.69 11859.02
Total    3351.35 4675.04 6464.20 12159.02
\end{verbatim}

\textbf{(c)} Here is the code for the density plots of the retained,
insurer, and total portfolio risk.

\begin{Shaded}
\begin{Highlighting}[]
\KeywordTok{par}\NormalTok{(}\DataTypeTok{mfrow=}\KeywordTok{c}\NormalTok{(}\DecValTok{1}\NormalTok{,}\DecValTok{3}\NormalTok{))}
\KeywordTok{plot}\NormalTok{(}\KeywordTok{density}\NormalTok{(Yretained), }\DataTypeTok{xlim=}\KeywordTok{c}\NormalTok{(}\DecValTok{0}\NormalTok{,}\DecValTok{500}\NormalTok{), }\DataTypeTok{main=}\StringTok{"Retained Portfolio Risk"}\NormalTok{, }\DataTypeTok{xlab=}\StringTok{"Loss (Note the different horizontal scale)"}\NormalTok{)}
\KeywordTok{plot}\NormalTok{(}\KeywordTok{density}\NormalTok{(Yinsurer), }\DataTypeTok{xlim=}\KeywordTok{c}\NormalTok{(}\DecValTok{0}\NormalTok{,}\DecValTok{15000}\NormalTok{), }\DataTypeTok{main=}\StringTok{"Insurer Portfolio Risk"}\NormalTok{, }\DataTypeTok{xlab=}\StringTok{"Loss"}\NormalTok{)}
\KeywordTok{plot}\NormalTok{(}\KeywordTok{density}\NormalTok{(S), }\DataTypeTok{xlim=}\KeywordTok{c}\NormalTok{(}\DecValTok{0}\NormalTok{,}\DecValTok{15000}\NormalTok{), }\DataTypeTok{main=}\StringTok{"Total Portfolio Risk"}\NormalTok{, }\DataTypeTok{xlab=}\StringTok{"Loss"}\NormalTok{)}
\end{Highlighting}
\end{Shaded}

\begin{center}\includegraphics{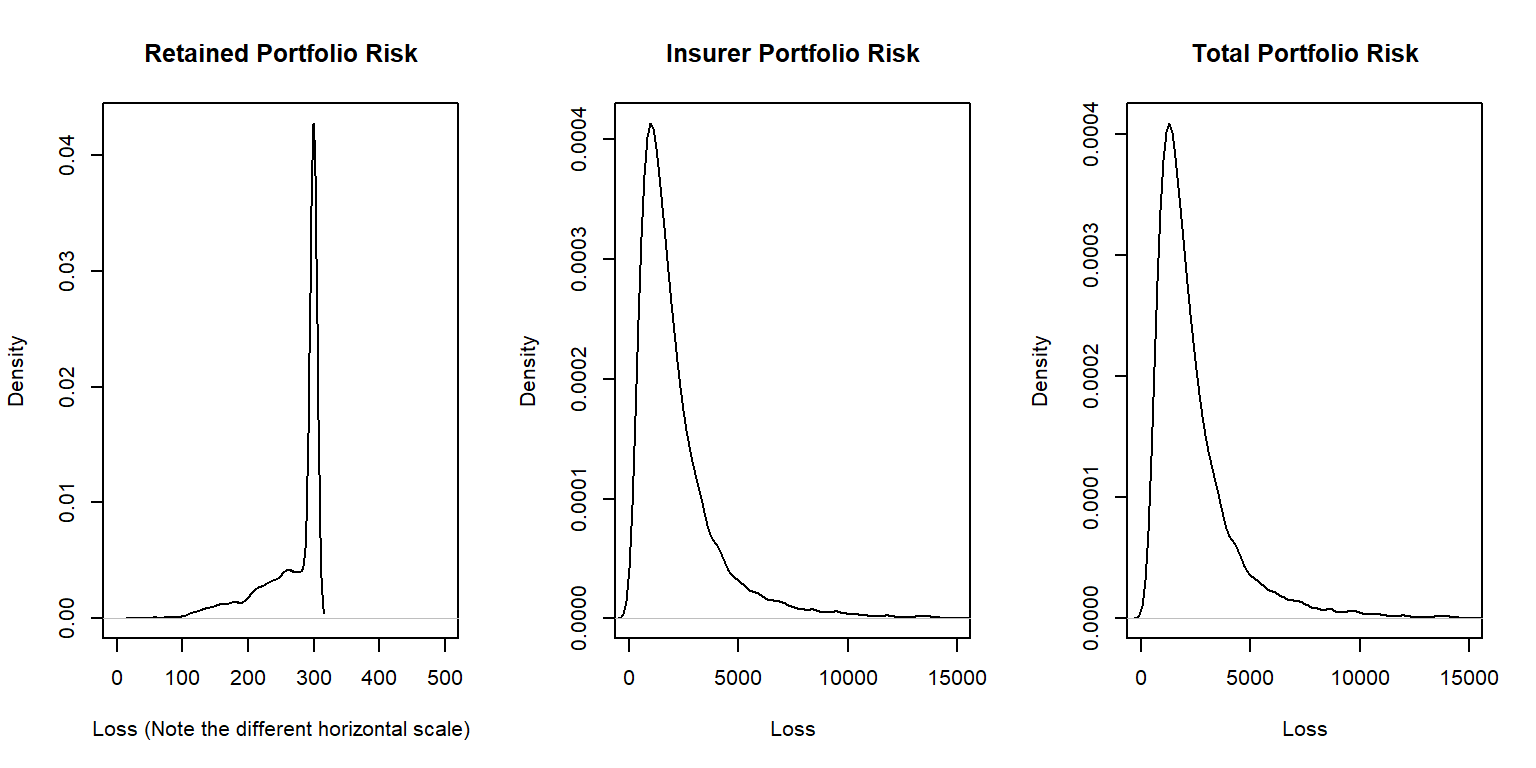} \end{center}

\chapter{Loss Reserving}\label{C:LossReserves}

This is a placeholder file

\chapter{Experience Rating using Bonus-Malus}\label{C:BonusMalus}

This is a placeholder file

\textbf{Bonus-Malus}

Bonus-malus system, which is used interchangeably as ``no-fault
discount'', ``merit rating'', ``experience rating'' or ``no-claim
discount'' in different countries, is based on penalizing insureds who
are responsible for one or more claims by a premium surcharge, and
awarding insureds with a premium discount if they do not have any claims
(Frangos and Vrontos, 2001). Insurers use bonus-malus systems for two
main purposes; firstly, to encourage drivers to drive more carefully in
a year without any claims, and secondly, to ensure insureds to pay
premiums proportional to their risks which are based on their claims
experience.

\textbf{NCD and Experience Rating}

No Claim Discount (NCD) system is an experience rating system commonly
used in motor insurance. NCD system represents an attempt to categorize
insureds into homogeneous groups who pay premiums based on their claims
experience. Depending on the rules in the scheme, new policyholders may
be required to pay full premium initially, and obtain discounts in the
future years as a results of claim-free years.

\textbf{Hunger for Bonus }

An NCD system rewards policyholders for not making any claims during a
year, or in other words, it grants a bonus to a careful driver. This
bonus principle may affect policy holders' decisions whether to claim or
not to claim, especially when involving accidents with slight damages,
which is known as `hunger for bonus' phenomenon (Philipson, 1960). The
option of `hunger for bonus' implemented on insureds under an NCD system
may reduce insurers' claim costs, and may be able to offset the expected
decrease in premium income.

\chapter{Data Systems}\label{C:DataSystems}

\emph{Chapter Preview}. This chapter covers the learning areas on data
and systems outlined in the IAA (International Actuarial Association)
Education Syllabus published in September 2015.

\section{Data}\label{data}

\subsection{Data Types and Sources}\label{data-types-and-sources}

In terms of how data are collected, data can be divided into two types
\citep{hox2005data}: primary data and secondary data. Primary data are
original data that are collected for a specific research problem.
Secondary data are data originally collected for a different purpose and
reused for another research problem. A major advantage of using primary
data is that the theoretical constructs, the research design, and the
data collection strategy can be tailored to the underlying research
question to ensure that the data collected indeed help to solve the
problem. A disadvantage of using primary data is that data collection
can be costly and time-consuming. Using secondary data has the advantage
of lower cost and faster access to relevant information. However, using
secondary data may not be optimal for the research question under
consideration.

In terms of the degree of organization of the data, data can be also
divided into two types
\citep{inmon2014, leary2013bigdata, hashem2015bigdata, abdullah2013data, pries2015}:
structured data and unstructured data. Structured data have a
predictable and regularly occurring format. In contrast, unstructured
data are unpredictable and have no structure that is recognizable to a
computer. Structured data consists of records, attributes, keys, and
indices and are typically managed by a database management system (DBMS)
such as IBM DB2, Oracle, MySQL, and Microsoft SQL Server. As a result,
most units of structured data can be located quickly and easily.
Unstructured data have many different forms and variations. One common
form of unstructured data is text. Accessing unstructured data is
clumsy. To find a given unit of data in a long text, for example,
sequentially search is usually performed.

In terms of how the data are measured, data can be classified as
qualitative or quantitative. Qualitative data is data about qualities,
which cannot be actually measured. As a result, qualitative data is
extremely varied in nature and includes interviews, documents, and
artifacts \citep{miles2014}. Quantitative data is data about quantities,
which can be measured numerically with numbers. In terms of the level of
measurement, quantitative data can be further classified as nominal,
ordinal, interval, or ratio \citep{gan2011}. Nominal data, also called
categorical data, are discrete data without a natural ordering. Ordinal
data are discrete data with a natural order. Interval data are
continuous data with a specific order and equal intervals. Ratio data
are interval data with a natural zero.

There exist a number of data sources. First, data can be obtained from
university-based researchers who collect primary data. Second, data can
be obtained from organizations that are set up for the purpose of
releasing secondary data for general research community. Third, data can
be obtained from national and regional statistical institutes that
collect data. Finally, companies have corporate data that can be
obtained for research purpose.

While it might be difficult to obtain data to address a specific
research problem or answer a business question, it is relatively easy to
obtain data to test a model or an algorithm for data analysis. In
nowadays, readers can obtain datasets from the Internet easily. The
following is a list of some websites to obtain real-world data:

\begin{itemize}
\item
  \textbf{UCI Machine Learning Repository} This website (url:
  \url{http://archive.ics.uci.edu/ml/index.php}) maintains more than 400
  datasets that can be used to test machine learning algorithms.
\item
  \textbf{Kaggle} The Kaggle website (url:
  \url{https://www.kaggle.com/}) include real-world datasets used for
  data science competition. Readers can download data from Kaggle by
  registering an account.
\item
  \textbf{DrivenData} DrivenData aims at bringing cutting-edge practices
  in data science to solve some of the world's biggest social
  challenges. In its website (url: \url{https://www.drivendata.org/}),
  readers can participate data science competitions and download
  datasets.
\item
  \textbf{Analytics Vidhya} This website (url:
  \url{https://datahack.analyticsvidhya.com/contest/all/}) allows you to
  participate and download datasets from practice problems and hackathon
  problems.
\item
  \textbf{KDD Cup} KDD Cup is the annual Data Mining and Knowledge
  Discovery competition organized by ACM Special Interest Group on
  Knowledge Discovery and Data Mining. This website (url:
  \url{http://www.kdd.org/kdd-cup}) contains the datasets used in past
  KDD Cup competitions since 1997.
\item
  \textbf{U.S. Government's open data} This website (url:
  \url{https://www.data.gov/}) contains about 200,000 datasets covering
  a wide range of areas including climate, education, energy, and
  finance.
\item
  \textbf{AWS Public Datasets} In this website (url:
  \url{https://aws.amazon.com/datasets/}), Amazon provides a centralized
  repository of public datasets, including some huge datasets.
\end{itemize}

\subsection{Data Structures and
Storage}\label{data-structures-and-storage}

As mentioned in the previous subsection, there are structured data as
well as unstructured data. Structured data are highly organized data and
usually have the following tabular format:

\[\begin{matrix}
\begin{array}{lllll} \hline
 & V_1 & V_2 & \cdots & V_d \
\\\hline
\textbf{x}_1 & x_{11} & x_{12} & \cdots & x_{1d} \\
\textbf{x}_2 & x_{21} & x_{22} & \cdots & x_{2d} \\
\vdots & \vdots & \vdots & \cdots & \vdots \\
\textbf{x}_n & x_{n1} & x_{n2} & \cdots & x_{nd} \\
\hline
\end{array}
\end{matrix}
\]

In other words, structured data can be organized into a table consists
of rows and columns. Typically, each row represents a record and each
column represents an attribute. A table can be decomposed into several
tables that can be stored in a relational database such as the Microsoft
SQL Server. The SQL (Structured Query Language) can be used to access
and modify the data easily and efficiently.

Unstructured data do not follow a regular format
\citep{abdullah2013data}. Examples of unstructured data include
documents, videos, and audio files. Most of the data we encounter are
unstructured data. In fact, the term ``big data'' was coined to reflect
this fact. Traditional relational databases cannot meet the challenges
on the varieties and scales brought by massive unstructured data
nowadays. NoSQL databases have been used to store massive unstructured
data.

There are three main NoSQL databases \citep{chen2014b}: key-value
databases, column-oriented databases, and document-oriented databases.
Key-value databases use a simple data model and store data according to
key-values. Modern key-value databases have higher expandability and
smaller query response time than relational databases. Examples of
key-value databases include Dynamo used by Amazon and Voldemort used by
LinkedIn. Column-oriented databases store and process data according to
columns rather than rows. The columns and rows are segmented in multiple
nodes to achieve expandability. Examples of column-oriented databases
include BigTable developed by Google and Cassandra developed by
FaceBook. Document databases are designed to support more complex data
forms than those stored in key-value databases. Examples of document
databases include MongoDB, SimpleDB, and CouchDB. MongoDB is an
open-source document-oriented database that stores documents as binary
objects. SimpleDB is a distributed NoSQL database used by Amazon.
CouchDB is an another open-source document-oriented database.

\subsection{Data Quality}\label{data-quality}

Accurate data are essential to useful data analysis. The lack of
accurate data may lead to significant costs to organizations in areas
such as correction activities, lost customers, missed opportunities, and
incorrect decisions \citep{olson2003}.

Data has quality if it satisfies its intended use, that is, the data is
accurate, timely, relevant, complete, understood, and trusted
\citep{olson2003}. As a result, we first need to know the specification
of the intended uses and then judge the suitability for those uses in
order to assess the quality of the data. Unintended uses of data can
arise from a variety of reasons and lead to serious problems.

Accuracy is the single most important component of high-quality data.
Accurate data have the following properties \citep{olson2003}:

\begin{itemize}
\tightlist
\item
  The data elements are not missing and have valid values.
\item
  The values of the data elements are in the right ranges and have the
  right representations.
\end{itemize}

Inaccurate data arise from different sources. In particular, the
following areas are common areas where inaccurate data occur:

\begin{itemize}
\tightlist
\item
  Initial data entry. Mistakes (including deliberate errors) and system
  errors can occur during the initial data entry. Flawed data entry
  processes can result in inaccurate data.
\item
  Data decay. Data decay, also known as data degradation, refers to the
  gradual corruption of computer data due to an accumulation of
  non-critical failures in a storage device.
\item
  Data moving and restructuring. Inaccurate data can also arise from
  data extracting, cleaning, transforming, loading, or integrating.
\item
  Data using. Faulty reporting and lack of understanding can lead to
  inaccurate data.
\end{itemize}

Reverification and analysis are two approaches to find inaccurate data
elements. To ensure that the data elements are 100\% accurate, we must
use reverification. However, reverification can be time-consuming and
may not be possible for some data. Analytical techniques can also be
used to identify inaccurate data elements. There are five types of
analysis that can be used to identify inaccurate data \citep{olson2003}:
data element analysis, structural analysis, value correlation,
aggregation correlation, and value inspection

Companies can create a data quality assurance program to create
high-quality databases. For more information about data quality issues
management and data profiling techniques, readers are referred to
\citep{olson2003}.

\subsection{Data Cleaning}\label{data-cleaning}

Raw data usually need to be cleaned before useful analysis can be
conducted. In particular, the following areas need attention when
preparing data for analysis \citep{janert2010}:

\begin{itemize}
\item
  \textbf{Missing values} It is common to have missing values in raw
  data. Depending on the situations, we can discard the record, discard
  the variable, or impute the missing values.
\item
  \textbf{Outliers} Raw data may contain unusual data points such as
  outliers. We need to handle outliers carefully. We cannot just remove
  outliers without knowing the reason for their existence. Sometimes the
  outliers are caused by clerical errors. Sometimes outliers are the
  effect we are looking for.
\item
  \textbf{Junk} Raw data may contain junks such as nonprintable
  characters. Junks are typically rare and not easy to get noticed.
  However, junks can cause serious problems in downstream applications.
\item
  \textbf{Format} Raw data may be formated in a way that is inconvenient
  for subsequent analysis. For example, components of a record may be
  split into multiple lines in a text file. In such cases, lines
  corresponding to a single record should be merged before loading to a
  data analysis software such as R.
\item
  \textbf{Duplicate records} Raw data may contain duplicate records.
  Duplicate records should be recognized and removed. This task may not
  be trivial depending on what you consider ``duplicate.''
\item
  \textbf{Merging datasets} Raw data may come from different sources. In
  such cases, we need to merge the data from different sources to ensure
  compatibility.
\end{itemize}

For more information about how to handle data in R, readers are referred
to \citep{forte2015} and \citep{buttrey2017}.

\section{Data Analysis Preliminary}\label{data-analysis-preliminary}

Data analysis involves inspecting, cleansing, transforming, and modeling
data to discover useful information to suggest conclusions and make
decisions. Data analysis has a long history. In 1962, statistician John
Tukey defined data analysis as:

\begin{quote}
procedures for analyzing data, techniques for interpreting the results
of such procedures, ways of planning the gathering of data to make its
analysis easier, more precise or more accurate, and all the machinery
and results of (mathematical) statistics which apply to analyzing data.

--- \citep{tukey1962data}
\end{quote}

Recently, Judd and coauthors defined data analysis as the following
equation\citep{judd2017}:

\[\hbox{Data} = \hbox{Model} + \hbox{Error},\] where Data represents a
set of basic scores or observations to be analyzed, Model is a compact
representation of the data, and Error is simply the amount the model
fails to represent accurately. Using the above equation for data
analysis, an analyst must resolve the following two conflicting goals:

\begin{itemize}
\tightlist
\item
  to add more parameters to the model so that the model represents the
  data better.
\item
  to remove parameters from the model so that the model is simple and
  parsimonious.
\end{itemize}

In this section, we give a high-level introduction to data analysis,
including different types of methods.

\subsection{Data Analysis Process}\label{S:process}

\begin{figure}

{\centering \includegraphics[width=0.8\linewidth]{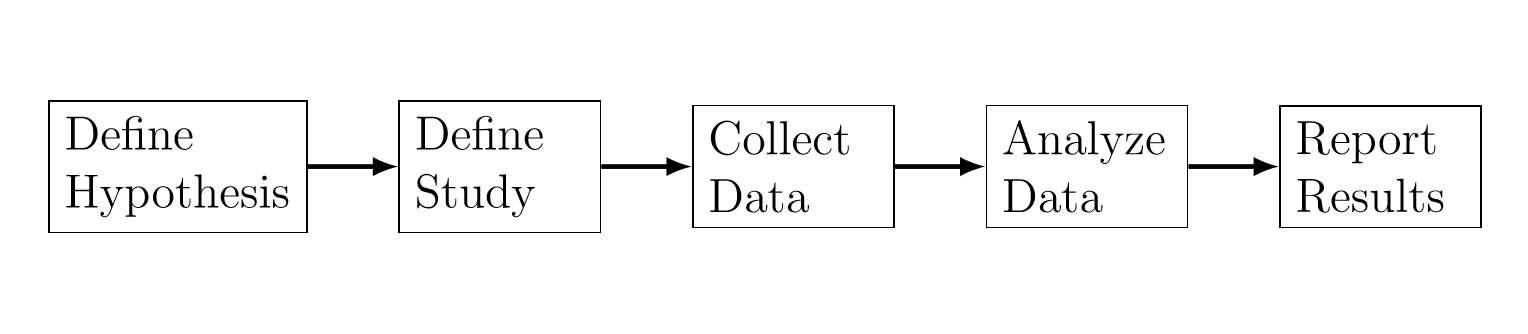}

}

\caption{The process of a typical study in behavioral and social sciences.}\label{fig:study}
\end{figure}

Data analysis is part of an overall study. For example, Figure
\ref{fig:study} shows the process of a typical study in behavioral and
social sciences as described in \citep{albers2017}. The data analysis
part consists of the following steps:

\begin{itemize}
\item
  \textbf{Exploratory analysis} The purpose of this step is to get a
  feel of the relationships with the data and figure out what type of
  analysis for the data makes sense.
\item
  \textbf{Statistical analysis} This step performs statistical analysis
  such as determining statistical significance and effect size.
\item
  \textbf{Make sense of the results} This step interprets the
  statistical results in the context of the overall study.
\item
  \textbf{Determine implications} This step interprets the data by
  connecting it to the study goals and the larger field of this study.
\end{itemize}

The goal of the data analysis as described above focuses on explaining
some phenomenon (See Section \ref{S:expred}).

\begin{figure}

{\centering \includegraphics[width=0.8\linewidth]{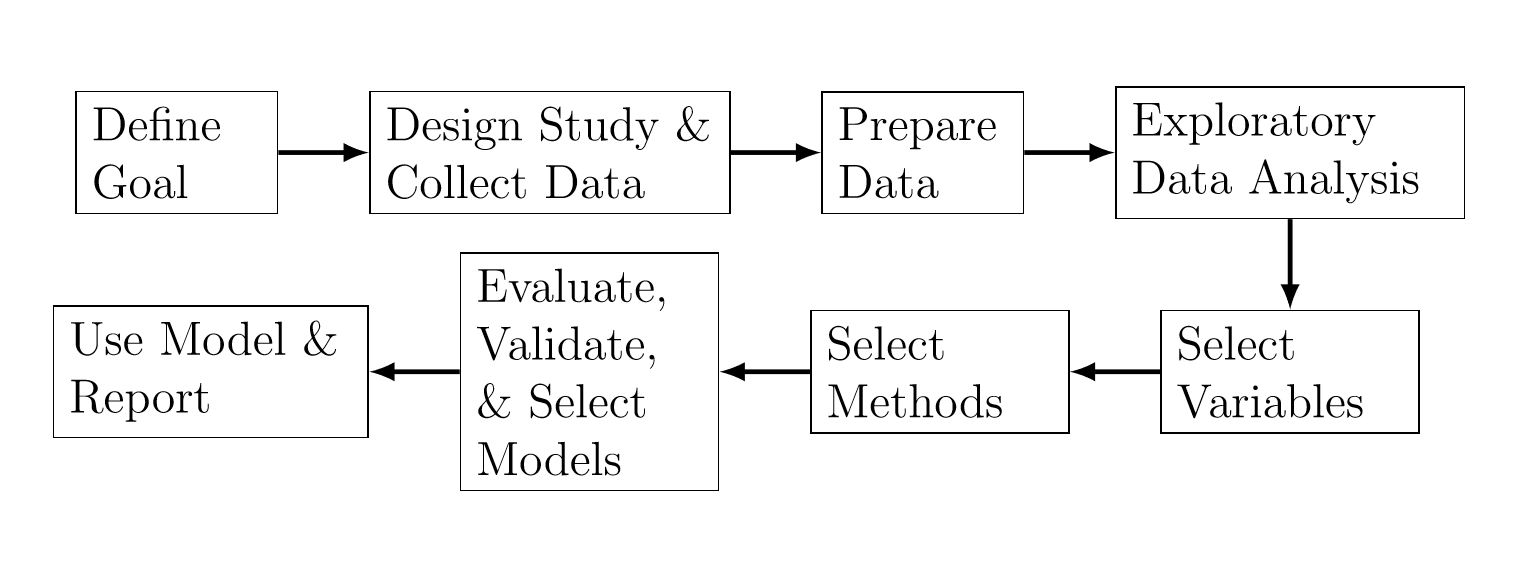}

}

\caption{The process of statistical modeling.}\label{fig:modeling}
\end{figure}

\citet{shmueli2010model} described a general process for statistical
modeling, which is shown in Figure \ref{fig:modeling}. Depending on the
goal of the analysis, the steps differ in terms of the choice of
methods, criteria, data, and information.

\subsection{Exploratory versus
Confirmatory}\label{exploratory-versus-confirmatory}

There are two phases of data analysis \citep{good1983data}: exploratory
data analysis (EDA) and confirmatory data analysis (CDA).
\protect\hyperlink{tab:13.1}{Table 13.1} summarizes some differences
between EDA and CDA. EDA is usually applied to observational data with
the goal of looking for patterns and formulating hypotheses. In
contrast, CDA is often applied to experimental data (i.e., data obtained
by means of a formal design of experiments) with the goal of quantifying
the extent to which discrepancies between the model and the data could
be expected to occur by chance \citep{gelman2004eda}.

\[\begin{matrix}
\begin{array}{lll} \hline
 & \textbf{EDA} & \textbf{CDA} \\\hline
\text{Data} & \text{Observational data} & \text{Experimental data}\\[3mm]
\text{Goal} & \text{Pattern recognition,}  & \text{Hypothesis testing,}  \\
& \text{formulate hypotheses} & \text{estimation, prediction} \\[3mm]
\text{Techniques} & \text{Descriptive statistics,} & \text{Traditional statistical tools of} \\
& \text{visualization, clustering} & \text{inference, significance, and}\\
& & \text{confidence} \\
\hline
\end{array}
\end{matrix}
\] \protect\hyperlink{tab:13.1}{Table 13.1}: Comparison of exploratory
data analysis and confirmatory data analysis.

Techniques for EDA include descriptive statistics (e.g., mean, median,
standard deviation, quantiles), distributions, histograms, correlation
analysis, dimension reduction, and cluster analysis. Techniques for CDA
include the traditional statistical tools of inference, significance,
and confidence.

\subsection{Supervised versus
Unsupervised}\label{supervised-versus-unsupervised}

Methods for data analysis can be divided into two types
\citep{abbott2014, igual2017}: supervised learning methods and
unsupervised learning methods. Supervised learning methods work with
labeled data, which include a target variable. Mathematically,
supervised learning methods try to approximate the following function:
\[
Y = f(X_1, X_2, \ldots, X_p),
\] where \(Y\) is a target variable and \(X_1\), \(X_2\), \(\ldots\),
\(X_p\) are explanatory variables. Other terms are also used to mean a
target variable. \protect\hyperlink{tab:13.2}{Table 13.2} gives a list
of common names for different types of variables \citep{frees2009}. When
the target variable is a categorical variable, supervised learning
methods are called classification methods. When the target variable is
continuous, supervised learning methods are called regression methods.

\[\begin{matrix}
\begin{array}{ll}
\hline
\textbf{Target Variable}  &  \textbf{Explanatory Variable}\\\hline
\text{Dependent variable} & \text{Independent variable}\\
\text{Response} & \text{Treatment} \\
\text{Output} & \text{Input} \\
\text{Endogenous variable} & \text{Exogenous variable} \\
\text{Predicted variable} & \text{Predictor variable} \\
\text{Regressand} & \text{Regressor} \\
\hline
\end{array}
\end{matrix}
\] \protect\hyperlink{tab:13.2}{Table 13.2}: Common names of different
variables.

Unsupervised learning methods work with unlabeled data, which include
explanatory variables only. In other words, unsupervised learning
methods do not use target variables. As a result, unsupervised learning
methods are also called descriptive modeling methods.

\subsection{Parametric versus
Nonparametric}\label{parametric-versus-nonparametric}

Methods for data analysis can be parametric or nonparametric
\citep{abbott2014}. Parametric methods assume that the data follow a
certain distribution. Nonparametric methods do not assume distributions
for the data and therefore are called distribution-free methods.

Parametric methods have the advantage that if the distribution of the
data is known, properties of the data and properties of the method
(e.g., errors, convergence, coefficients) can be derived. A disadvantage
of parametric methods is that analysts need to spend considerable time
on figuring out the distribution. For example, analysts may try
different transformation methods to transform the data so that it
follows a certain distribution.

Since nonparametric methods make fewer assumptions, nonparametric
methods have the advantage that they are more flexible, more robust, and
applicable to non-quantitative data. However, a drawback of
nonparametric methods is that the conclusions drawn from nonparametric
methods are not as powerful as those drawn from parametric methods.

\subsection{Explanation versus Prediction}\label{S:expred}

There are two goals in data analysis
\citep{breiman2001modeling, shmueli2010model}: explanation and
prediction. In some scientific areas such as economics, psychology, and
environmental science, the focus of data analysis is to explain the
causal relationships between the input variables and the response
variable. In other scientific areas such as natural language processing
and bioinformatics, the focus of data analysis is to predict what the
responses are going to be given the input variables.

\citet{shmueli2010model} discussed in detail the distinction between
explanatory modeling and predictive modeling, which reflect the process
of using data and methods for explaining or predicting, respectively.
Explanatory modeling is commonly used for theory building and testing.
However, predictive modeling is rarely used in many scientific fields as
a tool for developing theory.

Explanatory modeling is typically done as follows:

\begin{itemize}
\item
  State the prevailing theory.
\item
  State causal hypotheses, which are given in terms of theoretical
  constructs rather than measurable variables. A causal diagram is
  usually included to illustrate the hypothesized causal relationship
  between the theoretical constructs.
\item
  Operationalize constructs. In this step, previous literature and
  theoretical justification are used to build a bridge between
  theoretical constructs and observable measurements.
\item
  Collect data and build models alongside the statistical hypotheses,
  which are operationalized from the research hypotheses.
\item
  Reach research conclusions and recommend policy. The statistical
  conclusions are converted into research conclusions. Policy
  recommendations are often accompanied.
\end{itemize}

\citet{shmueli2010model} defined predictive modeling as the process of
applying a statistical model or data mining algorithm to data for the
purpose of predicting new or future observations. Predictions include
point predictions, interval predictions, regions, distributions, and
rankings of new observations. Predictive model can be any method that
produces predictions.

\subsection{Data Modeling versus Algorithmic
Modeling}\label{data-modeling-versus-algorithmic-modeling}

\citet{breiman2001modeling} discussed two cultures in the use of
statistical modeling to reach conclusions from data: the data modeling
culture and the algorithmic modeling culture. In the data modeling
culture, the data are assumed to be generated by a given stochastic data
model. In the algorithmic modeling culture, the data mechanism is
treated as unknown and algorithmic models are used.

Data modeling gives the statistics field many successes in analyzing
data and getting information about the data mechanisms. However,
\citet{breiman2001modeling} argued that the focus on data models in the
statistical community has led to some side effects such as

\begin{itemize}
\item
  Produced irrelevant theory and questionable scientific conclusions.
\item
  Kept statisticians from using algorithmic models that might be more
  suitable.
\item
  Restricted the ability of statisticians to deal with a wide range of
  problems.
\end{itemize}

Algorithmic modeling was used by industrial statisticians long time ago.
However, the development of algorithmic methods was taken up by a
community outside statistics \citep{breiman2001modeling}. The goal of
algorithmic modeling is predictive accuracy. For some complex prediction
problems, data models are not suitable. These prediction problems
include speech recognition, image recognition, handwriting recognition,
nonlinear time series prediction, and financial market prediction. The
theory in algorithmic modeling focuses on the properties of algorithms,
such as convergence and predictive accuracy.

\subsection{Big Data Analysis}\label{big-data-analysis}

Unlike traditional data analysis, big data analysis employs additional
methods and tools that can extract information rapidly from massive
data. In particular, big data analysis uses the following processing
methods \citep{chen2014b}:

\begin{itemize}
\item
  \textbf{Bloom filter} A bloom filter is a space-efficient
  probabilistic data structure that is used to determine whether an
  element belongs to a set. It has the advantages of high space
  efficiency and high query speed. A drawback of using bloom filter is
  that there is a certain misrecognition rate.
\item
  \textbf{Hashing} Hashing is a method that transforms data into
  fixed-length numerical values through a hash function. It has the
  advantages of rapid reading and writing. However, sound hash functions
  are difficult to find.
\item
  \textbf{Indexing} Indexing refers to a process of partitioning data in
  order to speed up reading. Hashing is a special case of indexing.
\item
  \textbf{Tries} A trie, also called digital tree, is a method to
  improve query efficiency by using common prefixes of character strings
  to reduce comparison on character strings to the greatest extent.
\item
  \textbf{Parallel computing} Parallel computing uses multiple computing
  resources to complete a computation task. Parallel computing tools
  include MPI (Message Passing Interface), MapReduce, and Dryad.
\end{itemize}

Big data analysis can be conducted in the following levels
\citep{chen2014b}: memory-level, business intelligence (BI) level, and
massive level. Memory-level analysis is conducted when the data can be
loaded to the memory of a cluster of computers. Current hardware can
handle hundreds of gigabytes (GB) of data in memory. BI level analysis
can be conducted when the data surpass the memory level. It is common
for BI level analysis products to support data over terabytes (TB).
Massive level analysis is conducted when the data surpass the
capabilities of products for BI level analysis. Usually Hadoop and
MapReduce are used in massive level analysis.

\subsection{Reproducible Analysis}\label{reproducible-analysis}

As mentioned in Section \ref{S:process}, a typical data analysis
workflow includes collecting data, analyzing data, and reporting
results. The data collected are saved in a database or files. The data
are then analyzed by one or more scripts, which may save some
intermediate results or always work on the raw data. Finally a report is
produced to describe the results, which include relevant plots, tables,
and summaries of the data. The workflow may subject to the following
potential issues \citep[Chapter 2]{mailund2017}:

\begin{itemize}
\item
  The data are separated from the analysis scripts.
\item
  The documentation of the analysis is separated from the analysis
  itself.
\end{itemize}

If the analysis is done on the raw data with a single script, then the
first issue is not a major problem. If the analysis consists of multiple
scripts and a script saves intermediate results that are read by the
next script, then the scripts describe a workflow of data analysis. To
reproduce an analysis, the scripts have to be executed in the right
order. The workflow may cause major problems if the order of the scripts
is not documented or the documentation is not updated or lost. One way
to address the first issue is to write the scripts so that any part of
the workflow can be run completely automatically at any time.

If the documentation of the analysis is synchronized with the analysis,
then the second issue is not a major problem. However, the documentation
may become completely useless if the scripts are changed but the
documentation is not updated.

Literate programming is an approach to address the two issues mentioned
above. In literate programming, the documentation of a program and the
code of the program are written together. To do literate programming in
R, one way is to use the R Markdown and the \(\texttt{knitr}\) package.

\subsection{Ethical Issues}\label{ethical-issues}

Analysts may face ethical issues and dilemmas during the data analysis
process. In some fields, for example, ethical issues and dilemmas
include participant consent, benefits, risk, confidentiality, and data
ownership \citep{miles2014}. For data analysis in actuarial science and
insurance in particular, we face the following ethical matters and
issues \citep{miles2014}:

\begin{itemize}
\item
  \textbf{Worthness of the project} Is the project worth doing? Will the
  project contribute in some significant way to a domain broader than my
  career? If a project is only opportunistic and does not have a larger
  significance, then it might be pursued with less care. The result may
  be looked good but not right.
\item
  \textbf{Competence} Do I or the whole team have the expertise to carry
  out the project? Incompetence may lead to weakness in the analytics
  such as collecting large amounts of data poorly and drawing
  superficial conclusions.
\item
  \textbf{Benefits, costs, and reciprocity} Will each stakeholder gain
  from the project? Are the benefit and the cost equitable? A project
  will likely to fail if the benefit and the cost for a stakeholder do
  not match.
\item
  \textbf{Privacy and confidentiality} How do we make sure that the
  information is kept confidentially? Where raw data and analysis
  results are stored and how will have access to them should be
  documented in explicit confidentiality agreements.
\end{itemize}

\section{Data Analysis Techniques}\label{data-analysis-techniques}

Techniques for data analysis are drawn from different but overlapping
fields such as statistics, machine learning, pattern recognition, and
data mining. Statistics is a field that addresses reliable ways of
gathering data and making inferences based on them
\citep{bandyo2011, bluman2012}. The term machine learning was coined by
Samuel in 1959 \citep{samuel1959ml}. Originally, machine learning refers
to the field of study where computers have the ability to learn without
being explicitly programmed. Nowadays, machine learning has evolved to
the broad field of study where computational methods use experience
(i.e., the past information available for analysis) to improve
performance or to make accurate predictions
\citep{bishop2007, clarke2009, mohri2012, kubat2017}. There are four
types of machine learning algorithms (See
\protect\hyperlink{tab:13.3}{Table 13.3} depending on the type of the
data and the type of the learning tasks.

\[\begin{matrix}
\begin{array}{rll} \hline
& \textbf{Supervised} & \textbf{Unsupervised} \\\hline
\textbf{Discrete Label} & \text{Classification} & \text{Clustering} \\
\textbf{Continuous Label} & \text{Regression} & \text{Dimension reduction} \\
\hline
\end{array}
\end{matrix}
\] \protect\hyperlink{tab:13.3}{Table 13.3}: Types of machine learning
algorithms.

Originating in engineering, pattern recognition is a field that is
closely related to machine learning, which grew out of computer science.
In fact, pattern recognition and machine learning can be considered to
be two facets of the same field \citep{bishop2007}. Data mining is a
field that concerns collecting, cleaning, processing, analyzing, and
gaining useful insights from data \citep{aggarwal2015}.

\subsection{Exploratory Techniques}\label{exploratory-techniques}

Exploratory data analysis techniques include descriptive statistics as
well as many unsupervised learning techniques such as data clustering
and principal component analysis.

\subsection{Descriptive Statistics}\label{descriptive-statistics}

In the mass noun sense, descriptive statistics is an area of statistics
that concerns the collection, organization, summarization, and
presentation of data \citep{bluman2012}. In the count noun sense,
descriptive statistics are summary statistics that quantitatively
describe or summarize data.

\[\begin{matrix}
\begin{array}{ll} \hline
& \textbf{Descriptive Statistics} \\\hline
\text{Measures of central tendency} & \text{Mean, median, mode, midrange}\\
\text{Measures of variation} & \text{Range, variance, standard deviation} \\
\text{Measures of position} & \text{Quantile} \\
\hline
\end{array}
\end{matrix}
\] \protect\hyperlink{tab:13.4}{Table 13.4}: Some commonly used
descriptive statistics.

\protect\hyperlink{tab:13.4}{Table 13.4} lists some commonly used
descriptive statistics. In R, we can use the function
\(\texttt{summary}\) to calculate some of the descriptive statistics.
For numeric data, we can visualize the descriptive statistics using a
boxplot.

In addition to these quantitative descriptive statistics, we can also
qualitatively describe shapes of the distributions \citep{bluman2012}.
For example, we can say that a distribution is positively skewed,
symmetric, or negatively skewed. To visualize the distribution of a
variable, we can draw a histogram.

\subsubsection{Principal Component
Analysis}\label{principal-component-analysis}

Principal component analysis (PCA) is a statistical procedure that
transforms a dataset described by possibly correlated variables into a
dataset described by linearly uncorrelated variables, which are called
principal components and are ordered according to their variances. PCA
is a technique for dimension reduction. If the original variables are
highly correlated, then the first few principal components can account
for most of the variation of the original data.

To describe PCA, let \(X_1,X_2,\ldots,X_d\) be a set of variables. The
first principal component is defined to be the normalized linear
combination of the variables that has the largest variance, that is, the
first principal component is defined as

\[Z_1=w_{11} X_1 + w_{12} X_2 + \cdots + w_{1d} X_d,\] where
\(\textbf{w}_1=(w_{11}, w_{12}, \ldots, w_{1d})'\) is a vector of
loadings such that \(\mathrm{Var~}{(Z_1)}\) is maximized subject to the
following constraint:
\[\textbf{w}_1'\textbf{w}_1 = \sum_{j=1}^d w_{1j}^2 = 1.\]

For \(i=2,3,\ldots,d\), the \(i\)th principal component is defined as

\[Z_i=w_{i1} X_1 + w_{i2} X_2 + \cdots + w_{id} X_d,\] where
\(\textbf{w}_i=(w_{i1}, w_{i2}, \ldots, w_{id})'\) is a vector of
loadings such that \(\mathrm{Var~}{(Z_i)}\) is maximized subject to the
following constraints:
\[\textbf{w}_i'\textbf{w}_i=\sum_{j=1}^d w_{ij}^2 = 1,\]
\[\mathrm{cov~}{(Z_i, Z_j)} = 0,\quad j=1,2,\ldots,i-1.\]

The principal components of the variables are related to the
eigenvectors and eigenvectors of the covariance matrix of the variables.
For \(i=1,2,\ldots,d\), let \((\lambda_i, \textbf{e}_i)\) be the \(i\)th
eigenvalue-eigenvector pair of the covariance matrix \({\Sigma}\) such
that \(\lambda_1\ge \lambda_2\ge \ldots\ge \lambda_d\ge 0\) and the
eigenvectors are normalized. Then the \(i\)th principal component is
given by

\[Z_{i} = \textbf{e}_i' \textbf{X} =\sum_{j=1}^d e_{ij} X_j,\] where
\(\textbf{X}=(X_1,X_2,\ldots,X_d)'\). It can be shown that
\(\mathrm{Var~}{(Z_i)} = \lambda_i\). As a result, the proportion of
variance explained by the \(i\)th principal component is calculated as

\[\dfrac{\mathrm{Var~}{(Z_i)}}{ \sum_{j=1}^{d} \mathrm{Var~}{(Z_j)}} = \dfrac{\lambda_i}{\lambda_1+\lambda_2+\cdots+\lambda_d}.\]

For more information about PCA, readers are referred to
\citep{mirkin2011}.

\subsection{Cluster Analysis}\label{cluster-analysis}

Cluster analysis (aka data clustering) refers to the process of dividing
a dataset into homogeneous groups or clusters such that points in the
same cluster are similar and points from different clusters are quite
distinct \citep{gan2007, gan2011}. Data clustering is one of the most
popular tools for exploratory data analysis and has found applications
in many scientific areas.

During the past several decades, many clustering algorithms have been
proposed. Among these clustering algorithms, the \(k\)-means algorithm
is perhaps the most well-known algorithm due to its simplicity. To
describe the \(k\)-means algorithm, let
\(X=\{\textbf{x}_1,\textbf{x}_2,\ldots,\textbf{x}_n\}\) be a dataset
containing \(n\) points, each of which is described by \(d\) numerical
features. Given a desired number of clusters \(k\), the \(k\)-means
algorithm aims at minimizing the following objective function:

\[P(U,Z) = \sum_{l=1}^k\sum_{i=1}^n u_{il} \Vert \textbf{x}_i-\textbf{z}_l\Vert^2,\]
where \(U=(u_{il})_{n\times k}\) is an \(n\times k\) partition matrix,
\(Z=\{\textbf{z}_1,\textbf{z}_2,\ldots,\textbf{z}_k\}\) is a set of
cluster centers, and \(\Vert\cdot\Vert\) is the \(L^2\) norm or
Euclidean distance. The partition matrix \(U\) satisfies the following
conditions:

\[u_{il}\in \{0,1\},\quad i=1,2,\ldots,n,\:l=1,2,\ldots,k,\]
\[\sum_{l=1}^k u_{il}=1,\quad i=1,2,\ldots,n.\]

The \(k\)-means algorithm employs an iterative procedure to minimize the
objective function. It repeatedly updates the partition matrix \(U\) and
the cluster centers \(Z\) alternately until some stop criterion is met.
When the cluster centers \(Z\) are fixed, the partition matrix \(U\) is
updated as follows:

\[\begin{aligned}u_{il}=\left\{
\begin{array}{ll}
1, & \text{if } \Vert \textbf{x}_i - \textbf{z}_l\Vert = \min_{1\le j\le k} \Vert \textbf{x}_i - \textbf{z}_j\Vert;\\
0, & \text{if otherwise,}
\end{array}
\right.
\end{aligned}\] When the partition matrix \(U\) is fixed, the cluster
centers are updated as follows:

\[z_{lj} = \dfrac{\sum_{i=1}^n u_{il} x_{ij} } { \sum_{i=1}^n u_{il}},\quad l=1,2,\ldots,k,\: j=1,2,\ldots,d,\]
where \(z_{lj}\) is the \(j\)th component of \(\textbf{z}_l\) and
\(x_{ij}\) is the \(j\)th component of \(\textbf{x}_i\).

For more information about \(k\)-means, readers are referred to
\citep{gan2007} and \citep{mirkin2011}.

\subsection{Confirmatory Techniques}\label{confirmatory-techniques}

Confirmatory data analysis techniques include the traditional
statistical tools of inference, significance, and confidence.

\subsubsection{Linear Models}\label{linear-models}

Linear models, also called linear regression models, aim at using a
linear function to approximate the relationship between the dependent
variable and independent variables. A linear regression model is called
a simple linear regression model if there is only one independent
variable. When more than one independent variables are involved, a
linear regression model is called a multiple linear regression model.

Let \(X\) and \(Y\) denote the independent and the dependent variables,
respectively. For \(i=1,2,\ldots,n\), let \((x_i, y_i)\) be the observed
values of \((X,Y)\) in the \(i\)th case. Then the simple linear
regression model is specified as follows \citep{frees2009}:

\[y_i  = \beta_0 + \beta_1 x_i + \epsilon_i,\quad i=1,2,\ldots,n,\]
where \(\beta_0\) and \(\beta_1\) are parameters and \(\epsilon_i\) is a
random variable representing the error for the \(i\)th case.

When there are multiple independent variables, the following multiple
linear regression model is used:

\[y_i = \beta_0 + \beta_1 x_{i1} + \cdots + \beta_k x_{ik} + \epsilon_i,\]
where \(\beta_0\), \(\beta_1\), \(\ldots\), \(\beta_k\) are unknown
parameters to be estimated.

Linear regression models usually make the following assumptions:

\begin{enumerate}
\def\labelenumi{(\alph{enumi})}
\item
  \(x_{i1},x_{i2},\ldots,x_{ik}\) are nonstochastic variables.
\item
  \(\mathrm{Var~}(y_i)=\sigma^2\), where \(\mathrm{Var~}(y_i)\) denotes
  the variance of \(y_i\).
\item
  \(y_1,y_2,\ldots,y_n\) are independent random variables.
\end{enumerate}

For the purpose of obtaining tests and confidence statements with small
samples, the following strong normality assumption is also made:

\begin{enumerate}
\def\labelenumi{(\alph{enumi})}
\setcounter{enumi}{3}
\tightlist
\item
  \(\epsilon_1,\epsilon_2,\ldots,\epsilon_n\) are normally distributed.
\end{enumerate}

\subsubsection{Generalized Linear
Models}\label{generalized-linear-models}

The generalized linear model (GLM) is a wide family of regression models
that include linear regression models as special cases. In a GLM, the
mean of the response (i.e., the dependent variable) is assumed to be a
function of linear combinations of the explanatory variables, i.e.,

\[\mu_i = E[y_i],\]
\[\eta_i = \textbf{x}_i'\boldsymbol{\beta} = g(\mu_i),\] where
\(\textbf{x}_i=(1,x_{i1}, x_{i2}, \ldots, x_{ik})'\) is a vector of
regressor values, \(\mu_i\) is the mean response for the \(i\)th case,
and \(\eta_i\) is a systematic component of the GLM. The function
\(g(\cdot)\) is known and is called the link function. The mean response
can vary by observations by allowing some parameters to change. However,
the regression parameters \(\boldsymbol{\beta}\) are assumed to be the
same among different observations.

GLMs make the following assumptions:

\begin{enumerate}
\def\labelenumi{(\alph{enumi})}
\item
  \(x_{i1},x_{i2},\ldots,x_{in}\) are nonstochastic variables.
\item
  \(y_1,y_2,\ldots,y_n\) are independent.
\item
  The dependent variable is assumed to follow a distribution from the
  linear exponential family.
\item
  The variance of the dependent variable is not assumed to be constant
  but is a function of the mean, i.e.,
\end{enumerate}

\[\mathrm{Var~}{(y_i)} = \phi \nu(\mu_i),\] where \(\phi\) denotes the
dispersion parameter and \(\nu(\cdot)\) is a function.

As we can see from the above specification, the GLM provides a unifying
framework to handle different types of dependent variables, including
discrete and continuous variables. For more information about GLMs,
readers are referred to \citep{dejong2008} and \citep{frees2009}.

\subsubsection{Tree-based Models}\label{tree-based-models}

Decision trees, also known as tree-based models, involve dividing the
predictor space (i.e., the space formed by independent variables) into a
number of simple regions and using the mean or the mode of the region
for prediction \citep{breiman1984}. There are two types of tree-based
models: classification trees and regression trees. When the dependent
variable is categorical, the resulting tree models are called
classification trees. When the dependent variable is continuous, the
resulting tree models are called regression trees.

The process of building classification trees is similar to that of
building regression trees. Here we only briefly describe how to build a
regression tree. To do that, the predictor space is divided into
non-overlapping regions such that the following objective function

\[f(R_1,R_2,\ldots,R_J) = \sum_{j=1}^J \sum_{i=1}^n I_{R_j}(\textbf{x}_i)(y_i - \mu_j)^2\]
is minimized, where \(I\) is an indicator function, \(R_j\) denotes the
set of indices of the observations that belong to the \(j\)th box,
\(\mu_j\) is the mean response of the observations in the \(j\)th box,
\(\textbf{x}_i\) is the vector of predictor values for the \(i\)th
observation, and \(y_i\) is the response value for the \(i\)th
observation.

In terms of predictive accuracy, decision trees generally do not perform
to the level of other regression and classification models. However,
tree-based models may outperform linear models when the relationship
between the response and the predictors is nonlinear. For more
information about decision trees, readers are referred to
\citep{breiman1984} and \citep{mitchell1997}.

\section{Some R Functions}\label{some-r-functions}

R is an open-source software for statistical computing and graphics. The
R software can be downloaded from the R project website at
\url{https://www.r-project.org/}. In this section, we give some R
function for data analysis, especially the data analysis tasks mentioned
in previous sections.

\[\begin{matrix}
\begin{array}{lll} \hline
\text{Data Analysis Task} & \text{R package} & \text{R Function} \\\hline
\text{Descriptive Statistics} & \texttt{base} & \texttt{summary}\\
\text{Principal Component Analysis} & \texttt{stats} & \texttt{prcomp} \\
\text{Data Clustering} & \texttt{stats} & \texttt{kmeans}, \texttt{hclust} \\
\text{Fitting Distributions} & \texttt{MASS} & \texttt{fitdistr} \\
\text{Linear Regression Models} & \texttt{stats} & \texttt{lm} \\
\text{Generalized Linear Models} & \texttt{stats} & \texttt{glm} \\
\text{Regression Trees} & \texttt{rpart} & \texttt{rpart} \\
\text{Survival Analysis} & \texttt{survival} & \texttt{survfit} \\
\hline
\end{array}
\end{matrix}
\] \protect\hyperlink{tab:13.5}{Table 13.5}: Some R functions for data
analysis.

\protect\hyperlink{tab:13.5}{Table 13.5} lists a few R functions for
different data analysis tasks. Readers can read the R documentation for
examples of using these functions. There are also other R functions from
other packages to do similar things. However, the functions listed in
this table provide good start points for readers to conduct data
analysis in R. For analyzing large datasets in R in an efficient way,
readers are referred to \citep{daroczi2015}.

\section{Summary}\label{summary}

In this chapter, we gave a high-level overview of data analysis. The
overview is divided into three major parts: data, data analysis, and
data analysis techniques. In the first part, we introduced data types,
data structures, data storages, and data sources. In particular, we
provided several websites where readers can obtain real-world datasets
to horn their data analysis skills. In the second part, we introduced
the process of data analysis and various aspects of data analysis. In
the third part, we introduced some commonly used techniques for data
analysis. In addition, we listed some R packages and functions that can
be used to perform various data analysis tasks.

\section{Further Resources and
Contributors}\label{DS:further-reading-and-resources}

\subsubsection*{Contributor}\label{contributor-2}
\addcontentsline{toc}{subsubsection}{Contributor}

\begin{itemize}
\tightlist
\item
  \textbf{Guojun Gan}, University of Connecticut, is the principal
  author of the initital version of this chapter. Email:
  \href{mailto:guojun.gan@uconn.edu}{\nolinkurl{guojun.gan@uconn.edu}}
  for chapter comments and suggested improvements.
\end{itemize}

\chapter{Dependence Modeling}\label{C:DependenceModel}

\emph{Chapter Preview}. In practice, there are many types of variables
that one encounter and the first step in dependence modeling is
identifying the type of variable you are dealing with to help direct you
to the appropriate technique.This chapter introduces readers to variable
types and techniques for modeling dependence or association of
multivariate distributions. Section \ref{S:VarTypes} provides an
overview of the types of variables. Section \ref{S:Measures} then
elaborates basic measures for modeling the dependence between variables.

Section \ref{S:Copula} introduces a novel approach to modeling
dependence using Copulas which is reinforced with practical
illustrations in Section \ref{S:CopAppl}. The types of Copula families
and basic properties of Copula functions is explained Section
\ref{S:CopTyp}. The chapter concludes by explaining why the study of
dependence modeling is important in Section \ref{S:CopImp}.

\section{Variable Types}\label{S:VarTypes}

\begin{center}\rule{0.5\linewidth}{\linethickness}\end{center}

In this section, you learn how to:

\begin{itemize}
\tightlist
\item
  Classify variables as qualitative or quantitative.
\item
  Describe multivariate variables.
\end{itemize}

\begin{center}\rule{0.5\linewidth}{\linethickness}\end{center}

People, firms, and other entities that we want to understand are
described in a dataset by numerical characteristics. As these
characteristics vary by entity, they are commonly known as
\emph{variables}. To manage insurance systems, it will be critical to
understand the distribution of each variable and how they are associated
with one another. It is common for data sets to have many variables
(high dimensional) and so it useful to begin by classifying them into
different types. As will be seen, these classifications are not strict;
there is overlap among the groups. Nonetheless, the grouping summarized
in \protect\hyperlink{tab:14.1}{Table 14.1} and explained in the
remainder of this section provide a solid first step in framing a data
set.

\[
{\small \begin{matrix}
\begin{array}{l|l} \hline
\textbf{Variable Type} & \textbf{Example} \\\hline
Qualitative &            \\
    \text{Binary} &        \text{Sex} \\
\text{Categorical (Unordered, Nominal)} & \text{Territory (e.g., state/province) in which an insured resides} \\
\text{Ordered Category (Ordinal)} & \text{Claimant satisfaction (five point scale ranging from 1=dissatisfied} \\
& ~~~ \text{to 5 =satisfied)} \\\hline
Quantitative &            \\
\text{Continuous} & \text{Policyholder's age, weight, income} \\
  \text{Discrete} & \text{Amount of deductible} \\
\text{Count} & \text{Number of insurance claims} \\
\text{Combinations of}  & \text{Policy losses, mixture of 0's (for no loss)}  \\
~~~ \text{Discrete and Continuous} & ~~~\text{and positive claim amount} \\
\text{Interval Variable} & \text{Driver Age: 16-24 (young), 25-54 (intermediate),}  \\
& ~~~\text{55 and over (senior)} \\
\text{Circular Data} & \text{Time of day measures of customer arrival} \\ \hline
Multivariate ~ Variable &            \\
\text{High Dimensional Data} & \text{Characteristics of a firm purchasing worker's compensation} \\
& ~~~\text{insurance (location of plants, industry, number of employees,} \\
&~~~\text{and so on)} \\
\text{Spatial Data} & \text{Longitude/latitude of the location an insurance hailstorm claim} \\
\text{Missing Data} & \text{Policyholder's age (continuous/interval) and "-99" for} \\
&~~~ \text{"not reported," that is, missing} \\
\text{Censored and Truncated Data} & \text{Amount of insurance claims in excess of a deductible} \\
\text{Aggregate Claims} & \text{Losses recorded for each claim in a motor vehicle policy.} \\
\text{Stochastic Process Realizations} & \text{The time and amount of each occurrence of an insured loss} \\ \hline
\end{array}
\end{matrix}}
\]

\protect\hyperlink{tab:14.1}{Table 14.1} : Variable types

In data analysis, it is important to understand what type of variable
you are working with. For example, Consider a pair of random variables
\emph{(Coverage,Claim)} from the LGPIF data introduced in chapter 1 as
displayed in Figure \ref{fig:IntroPlot} below. We would like to know
whether the distribution of \emph{Coverage} depends on the distribution
of \emph{Claim} or whether they are statistically independent. We would
also want to know how the \emph{Claim} distribution depends on the
\emph{EntityType} variable. Because the \emph{EntityType} variable
belongs to a different class of variables, modeling the dependence
between \emph{Claim} and \emph{Coverage} may require a different
technique from that of \emph{Claim} and \emph{EntityType}.

\begin{figure}

{\centering \includegraphics[width=1\linewidth]{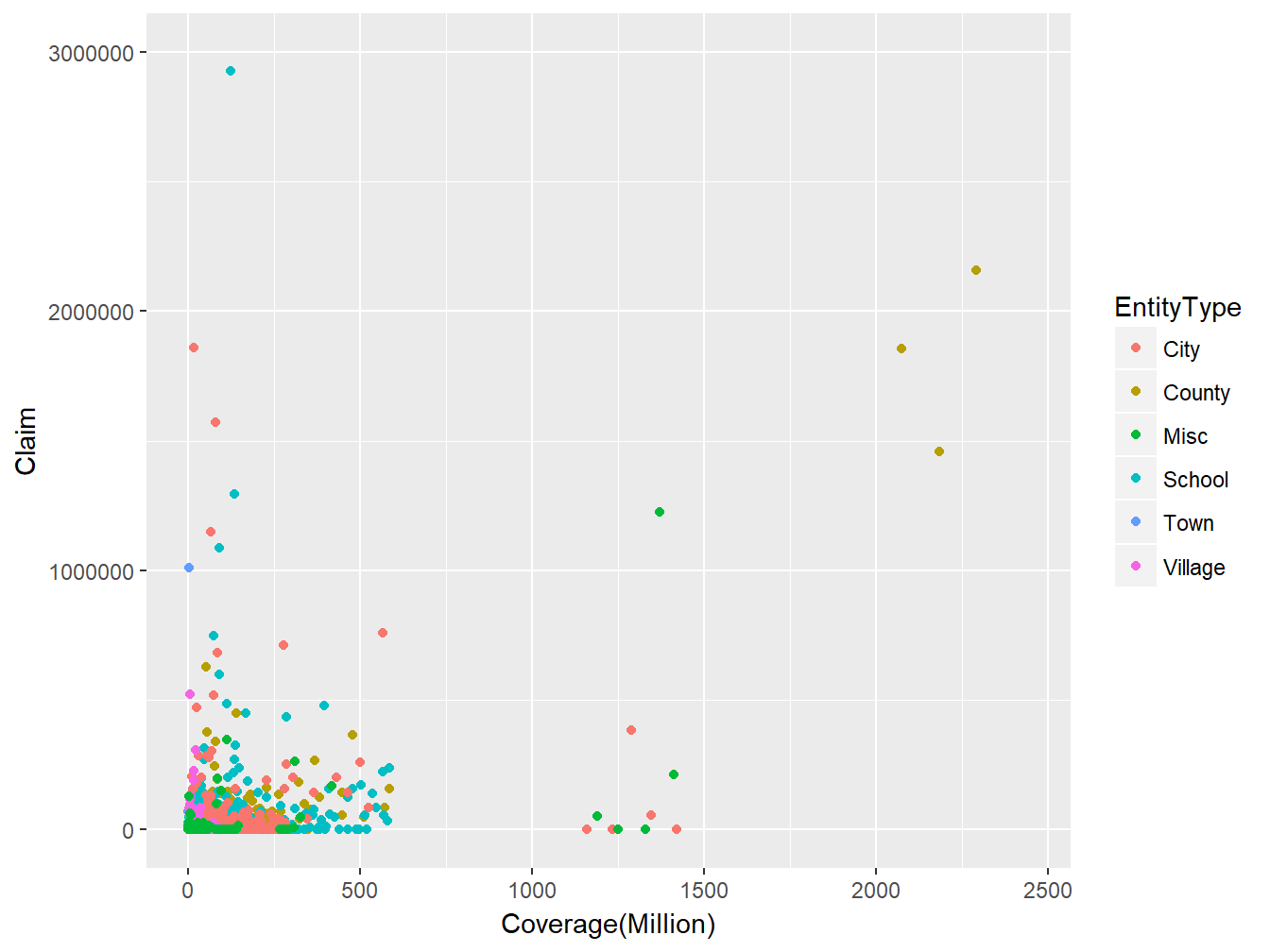}

}

\caption{Scatter plot of *(Coverage,Claim)* from LGPIF data}\label{fig:IntroPlot}
\end{figure}

\subsection{Qualitative Variables}\label{S:QuaVar}

\begin{center}\rule{0.5\linewidth}{\linethickness}\end{center}

In this sub-section, you learn how to:

\begin{itemize}
\tightlist
\item
  Classify qualitative variables as nominal or ordinal
\item
  Describe binary variable
\end{itemize}

\begin{center}\rule{0.5\linewidth}{\linethickness}\end{center}

A \emph{qualitative}, or \emph{categorical}, variable is one for which
the measurement denotes membership in a set of groups, or categories.
For example, if you were coding which area of the country an insured
resides, you might use a 1 for the northern part, 2 for southern, and 3
for everything else. This location variable is an example of a
\emph{nominal} variable, one for which the levels have no natural
ordering. Any analysis of nominal variables should not depend on the
labeling of the categories. For example, instead of using a 1,2,3 for
north, south, other, I should arrive at the same set of summary
statistics if I used a 2,1,3 coding instead, interchanging north and
south.

In contrast, an \emph{ordinal} variable is a type of categorical
variable for which an ordering does exist. For example, with a survey to
see how satisfied customers are with our claims servicing department, we
might use a five point scale that ranges from 1 meaning dissatisfied to
a 5 meaning satisfied. Ordinal variables provide a clear ordering of
levels of a variable but the amount of separation between levels is
unknown.

A \emph{binary} variable is a special type of categorical variable where
there are only two categories commonly taken to be a 0 and a 1. For
example, we might code a variable in a dataset to be a 1 if an insured
is female and a 0 if male.

\subsection{Quantitative Variables}\label{S:QuanVar}

\begin{center}\rule{0.5\linewidth}{\linethickness}\end{center}

In this sub-section, you learn how to:

\begin{itemize}
\tightlist
\item
  Differentiate between continuous and discrete variable
\item
  Use a combination of continuous and discrete variable
\item
  Describe circular data
\end{itemize}

\begin{center}\rule{0.5\linewidth}{\linethickness}\end{center}

Unlike a qualitative variable, a quantitative variable is one in which
numerical level is a realization from some scale so that the distance
between any two levels of the scale takes on meaning. A \emph{continuous
variable} is one that can take on any value within a finite interval.
For example, it is common to represent a policyholder's age, weight, or
income, as a continuous variable. In contrast, a \emph{discrete
variable} is one that takes on only a finite number of values in any
finite interval. Like an ordinal variable, these represent distinct
categories that are ordered. Unlike an ordinal variable, the numerical
difference between levels takes on economic meaning. A special type of
discrete variable is a \emph{count variable}, one with values on the
nonnegative integers. For example, we will be particularly interested in
the number of claims arising from a policy during a given period.

Some variables are inherently a \emph{combination of discrete and
continuous} components. For example, when we analyze the insured loss of
a policyholder, we will encounter a discrete outcome at zero,
representing no insured loss, and a continuous amount for positive
outcomes, representing the amount of the insured loss. Another
interesting variation is an \emph{interval variable}, one that gives a
range of possible outcomes.

\emph{Circular data} represent an interesting category typically not
analyzed by insurers. As an example of circular data, suppose that you
monitor calls to your customer service center and would like to know
when is the peak time of the day for calls to arrive. In this context,
one can think about the time of the day as a variable with realizations
on a circle, e.g., imagine an analog picture of a clock. For circular
data, the distance between observations at 00:15 and 00:45 are just as
close as observations 23:45 and 00:15 (here, we use the convention
\emph{HH:MM} means hours and minutes).

\subsection{Multivariate Variables}\label{multivariate-variables}

\begin{center}\rule{0.5\linewidth}{\linethickness}\end{center}

In this sub-section, you learn how to:

\begin{itemize}
\tightlist
\item
  Differentiate between univariate and multivariate data
\item
  Handle missing variables
\end{itemize}

\begin{center}\rule{0.5\linewidth}{\linethickness}\end{center}

Insurance data typically are \emph{multivariate} in the sense that we
can take many measurements on a single entity. For example, when
studying losses associated with a firm's worker's compensation plan, we
might want to know the location of its manufacturing plants, the
industry in which it operates, the number of employees, and so forth.
The usual strategy for analyzing multivariate data is to begin by
examining each variable in isolation of the others. This is known as a
\emph{univariate} approach.

In contrast, for some variables, it makes little sense to only look at
one dimensional aspects. For example, insurers typically organize
\emph{spatial} data by longitude and latitude to analyze the location of
weather related insurance claims due hailstorms. Having only a single
number, either longitude or latitude, provides little information in
understanding geographical location.

Another special case of a multivariate variable, less obvious, involves
coding for \emph{missing data}. When data are missing, it is better to
think about the variable as two dimensions, one to indicate whether or
not the variable is reported and the second providing the age (if
reported). In the same way, insurance data are commonly \emph{censored}
and \emph{truncated}. We refer you to Chapter 4 for more on censored and
truncated data. \emph{Aggregate claims} can also be coded as another
special type of multivariate variable. We refer you to Chapter 5 for
more Aggregate claims.

Perhaps the most complicated type of multivariate variable is a
\emph{realization of a stochastic process}. You will recall that a
stochastic process is little more than a collection of random variables.
For example, in insurance, we might think about the times that claims
arrive to an insurance company in a one year time horizon. This is a
high dimensional variable that theoretically is infinite dimensional.
Special techniques are required to understand realizations of stochastic
processes that will not be addressed here.

\section{Classic Measures of Scalar Associations}\label{S:Measures}

\begin{center}\rule{0.5\linewidth}{\linethickness}\end{center}

In this section, you learn how to:

\begin{itemize}
\tightlist
\item
  Estimate correlation using Pearson method
\item
  Use rank based measures like Spearman, Kendall to estimate correlation
\item
  Measure dependence using odds ratio,Pearson chi-square and likelihood
  ratio test statistic
\item
  Use normal-based correlations to quantify associations involving
  ordinal variables
\end{itemize}

\begin{center}\rule{0.5\linewidth}{\linethickness}\end{center}

\subsection{Association Measures for Quantitative
Variables}\label{association-measures-for-quantitative-variables}

For this section, consider a pair of random variables \((X,Y)\) having
joint distribution function \(F(\cdot)\) and a random sample
\((X_i,Y_i), i=1, \ldots, n\). For the continuous case, suppose that
\(F(\cdot)\) is absolutely continuous with absolutely continuous
marginals.

\subsubsection{Pearson Correlation}\label{pearson-correlation}

Define the sample covariance function
\(Cov(X,Y) = \frac{1}{n} \sum_{i=1}^n (X_i - \bar{X})(Y_i - \bar{Y})\),
where \(\bar{X}\) and \(\bar{Y}\) are the sample means of \(X\) and
\(Y\), respectively. Then, the product-moment (Pearson) correlation can
be written as

\begin{equation*}
r = \frac{Cov(X,Y)}{\sqrt{Cov(X,X) Cov(Y,Y)}}.
\end{equation*}

The correlation statistic \(r\) is widely used to capture association
between random variables. It is a (nonparametric) estimator of the
correlation parameter \(\rho\), defined to be the covariance divided by
the product of standard deviations. In this sense, it captures
association for any pair of random variables.

This statistic has several important features. Unlike regression
estimators, it is symmetric between random variables, so the correlation
between \(X\) and \(Y\) equals the correlation between \(Y\) and \(X\).
It is unchanged by linear transformations of random variables (up to
sign changes) so that we can multiply random variables or add constants
as is helpful for interpretation. The range of the statistic is
\([-1,1]\) which does not depend on the distribution of either \(X\) or
\(Y\).

Further, in the case of independence, the correlation coefficient \(r\)
is 0. However, it is well known that zero correlation does not imply
independence, except for normally distributed random variables. The
correlation statistic \(r\) is also a (maximum likelihood) estimator of
the association parameter for bivariate normal distribution. So, for
normally distributed data, the correlation statistic \(r\) can be used
to assess independence. For additional interpretations of this
well-known statistic, readers will enjoy \citep{lee1988thirteen}.

You can obtain the correlation statistic \(r\) using the \texttt{cor()}
function in \texttt{R} and selecting the \texttt{pearson} method. This
is demonstrated below by using the \emph{Coverage} rating variable in
millions of dollars and \emph{Claim} amount variable in dollars from the
LGPIF data introduced in chapter 1.

R Code for Pearson Correlation Statistic

\hypertarget{display.pearson.2}{}
\begin{verbatim}
### Pearson correlation between Claim and Coverage ###
r<-cor(Claim,Coverage, method = c("pearson"))
round(r,2)

Output:
[1] 0.31

### Pearson correlation between Claim and log(Coverage) ###
r<-cor(Claim,log(Coverage), method = c("pearson"))
round(r,2)

Output:
[1] 0.1
\end{verbatim}

From \texttt{R} output above, \(r=0.31\) , which indicates a positive
association between \emph{Claim} and \emph{Coverage}. This means that as
the coverage amount of a policy increases we expect claim to increase.

\subsection{Rank Based Measures}\label{rank-based-measures}

\subsubsection{Spearman's Rho}\label{spearmans-rho}

The Pearson correlation coefficient does have the drawback that it is
not invariant to nonlinear transforms of the data. For example, the
correlation between \(X\) and \(\ln Y\) can be quite different from the
correlation between \(X\) and \(Y\). As we see from the \texttt{R} code
for Pearson correlation statistic above, the correlation statistic \(r\)
between \emph{Coverage} rating variable in logarithmic millions of
dollars and \emph{Claim} amounts variable in dollars is \(0.1\) as
compared to \(0.31\) when we calculate the correlation between
\emph{Coverage} rating variable in millions of dollars and \emph{Claim}
amounts variable in dollars. This limitation is one reason for
considering alternative statistics.

Alternative measures of correlation are based on ranks of the data. Let
\(R(X_j)\) denote the rank of \(X_j\) from the sample
\(X_1, \ldots, X_n\) and similarly for \(R(Y_j)\). Let
\(R(X) = \left(R(X_1), \ldots, R(X_n)\right)'\) denote the vector of
ranks, and similarly for \(R(Y)\). For example, if \(n=3\) and
\(X=(24, 13, 109)\), then \(R(X)=(2,1,3)\). A comprehensive introduction
of rank statistics can be found in, for example,
\citep{hettmansperger1984statistical}. Also, ranks can be used to obtain
the empirical distribution function, refer to section 4.1.1 for more on
the empirical distribution function.

With this, the correlation measure of \citep{spearman1904proof} is
simply the product-moment correlation computed on the ranks:

\begin{equation*}
r_S = \frac{Cov(R(X),R(Y))}{\sqrt{Cov(R(X),R(X))Cov(R(Y),R(Y))}}
= \frac{Cov(R(X),R(Y))}{(n^2-1)/12} .
\end{equation*}

You can obtain the Spearman correlation statistic \(r_S\) using the
\texttt{cor()} function in \texttt{R} and selecting the
\texttt{spearman} method. From below, the Spearman correlation between
the \emph{Coverage} rating variable in millions of dollars and
\emph{Claim} amount variable in dollars is \(0.41\).

R Code for Spearman Correlation Statistic

\hypertarget{display.spearman.2}{}
\begin{verbatim}
### Spearman correlation between Claim and Coverage ###
rs<-cor(Claim,Coverage, method = c("spearman"))
round(rs,2)

Output:
[1] 0.41

### Spearman correlation between Claim and log(Coverage) ###
rs<-cor(Claim,log(Coverage), method = c("spearman"))
round(rs,2)

Output:
[1] 0.41
\end{verbatim}

To show that the Spearman correlation statistic is invariate under
strictly increasing transformations , from the \texttt{R} Code for
Spearman correlation statistic above, \(r_S=0.41\) between the
\emph{Coverage} rating variable in logarithmic millions of dollars and
\emph{Claim} amount variable in dollars.

\subsubsection{Kendall's Tau}\label{kendalls-tau}

An alternative measure that uses ranks is based on the concept of
\emph{concordance}. An observation pair \((X,Y)\) is said to be
concordant (discordant) if the observation with a larger value of \(X\)
has also the larger (smaller) value of \(Y\). Then
\(\Pr(concordance) = \Pr[ (X_1-X_2)(Y_1-Y_2) >0 ]\) ,
\(\Pr(discordance) = \Pr[ (X_1-X_2)(Y_1-Y_2) <0 ]\) and

\begin{eqnarray*}
\tau(X,Y)= \Pr(concordance) - \Pr(discordance) = 2\Pr(concordance) - 1 + \Pr(tie).
\end{eqnarray*}

To estimate this, the pairs \((X_i,Y_i)\) and \((X_j,Y_j)\) are said to
be concordant if the product \(sgn(X_j-X_i)sgn(Y_j-Y_i)\) equals 1 and
discordant if the product equals -1. Here, \(sgn(x)=1,0,-1\) as \(x>0\),
\(x=0\), \(x<0\), respectively. With this, we can express the
association measure of \citep{kendall1938new}, known as \emph{Kendall's
tau}, as

\begin{equation*}
\begin{array}{rl}
\tau &= \frac{2}{n(n-1)} \sum_{i<j}sgn(X_j-X_i)sgn(Y_j-Y_i)\\
&= \frac{2}{n(n-1)} \sum_{i<j}sgn(R(X_j)-R(X_i))sgn(R(Y_j)-R(Y_i))
\end{array}.
\end{equation*}

Interestingly, \citep{hougaard2000analysis}, page 137, attributes the
original discovery of this statistic to
\citep{fechnerkollektivmasslehre}, noting that Kendall's discovery was
independent and more complete than the original work.

You can obtain the Kendall's tau, using the \texttt{cor()} function in
\texttt{R} and selecting the \texttt{kendall} method. From below,
\(\tau=0.32\) between the \emph{Coverage} rating variable in millions of
dollars and \emph{Claim} amount variable in dollars.

R Code for Kendall's Tau

\hypertarget{display.kendall.2}{}
\begin{verbatim}
### Kendall's tau correlation between Claim and Coverage ###
tau<-cor(Claim,Coverage, method = c("kendall"))
round(tau,2)

Output:
[1]  0.32

### Kendall's tau correlation between Claim and log(Coverage) ###
tau<-cor(Claim,log(Coverage), method = c("kendall"))
round(tau,2)

Output:
[1] 0.32
\end{verbatim}

Also,to show that the Kendall's tau is invariate under strictly
increasing transformations , \(\tau=0.32\) between the \emph{Coverage}
rating variable in logarithmic millions of dollars and \emph{Claim}
amount variable in dollars.

\subsection{Nominal Variables}\label{nominal-variables}

\subsubsection{Bernoulli Variables}\label{bernoulli-variables}

To see why dependence measures for continuous variables may not be the
best for discrete variables, let us focus on the case of Bernoulli
variables that take on simple binary outcomes, 0 and 1. For notation,
let \(\pi_{jk} = \Pr(X=j, Y=k)\) for \(j,k=0,1\) and let
\(\pi_X=\Pr(X=1)\) and similarly for \(\pi_Y\). Then, the population
version of the product-moment (Pearson) correlation can be easily seen
to be

\begin{eqnarray*}
\rho = \frac{\pi_{11} - \pi_X \pi_Y}{\sqrt{\pi_X(1-\pi_X)\pi_Y(1-\pi_Y)}} .
\end{eqnarray*}

Unlike the case for continuous data, it is not possible for this measure
to achieve the limiting boundaries of the interval \([-1,1]\). To see
this, students of probability may recall the
Fr\(\acute{e}\)chet-H\(\ddot{o}\)effding bounds for a joint distribution
that turn out to be
\(\max\{0, \pi_X+\pi_Y-1\} \le \pi_{11} \le \min\{\pi_X,\pi_Y\}\) for
this joint probability. This limit on the joint probability imposes an
additional restriction on the Pearson correlation. As an illustration,
assume equal probabilities \(\pi_X =\pi_Y = \pi > 1/2\). Then, the lower
bound is

\begin{eqnarray*}
\frac{2\pi - 1 - \pi^2}{\pi(1-\pi)} = -\frac{1-\pi}{\pi} .
\end{eqnarray*}

For example, if \(\pi=0.8\), then the smallest that the Pearson
correlation could be is -0.25. More generally, there are bounds on
\(\rho\) that depend on \(\pi_X\) and \(\pi_Y\) that make it difficult
to interpret this measure.

As noted by \citep{bishop1975discrete} (page 382), squaring this
correlation coefficient yields the Pearson chi-square statistic. Despite
the boundary problems described above, this feature makes the Pearson
correlation coefficient a good choice for describing dependence with
binary data. The other is the odds ratio, described as follows.

As an alternative measure for Bernoulli variables, the \emph{odds ratio}
is given by

\begin{eqnarray*}
OR(\pi_{11}) = \frac{\pi_{11} \pi_{00}}{\pi_{01} \pi_{10}} = \frac{\pi_{11} \left( 1+\pi_{11}-\pi_1 -\pi_2\right)}{(\pi_1-\pi_{11})(\pi_2- \pi_{11})} .
\end{eqnarray*}

Pleasant calculations show that \(OR(z)\) is \(0\) at the lower
Fr\(\acute{e}\)chet-H\(\ddot{o}\)effding bound
\(z= \max\{0, \pi_1+\pi_2-1\}\) and is \(\infty\) at the upper bound
\(z=\min\{\pi_1,\pi_2\}\). Thus, the bounds on this measure do not
depend on the marginal probabilities \(\pi_X\) and \(\pi_Y\), making it
easier to interpret this measure.

As noted by \citep{yule1900association}, odds ratios are invariant to
the labeling of 0 and 1. Further, they are invariant to the marginals in
the sense that one can rescale \(\pi_1\) and \(\pi_2\) by positive
constants and the odds ratio remains unchanged. Specifically, suppose
that \(a_i\), \(b_j\) are sets of positive constants and that

\begin{eqnarray*}
\pi_{ij}^{new} &=& a_i b_j \pi_{ij}
\end{eqnarray*}

and \(\sum_{ij} \pi_{ij}^{new}=1.\) Then,

\begin{eqnarray*}
OR^{new} = \frac{(a_1 b_1 \pi_{11})( a_0 b_0 \pi_{00})}{(a_0 b_1 \pi_{01})( a_1 b_0\pi_{10})}
= \frac{\pi_{11} \pi_{00}}{\pi_{01} \pi_{10}} =OR^{old} .
\end{eqnarray*}

For additional help with interpretation, Yule proposed two transforms
for the odds ratio, the first in \citep{yule1900association},

\begin{eqnarray*}
\frac{OR-1}{OR+1},
\end{eqnarray*}

and the second in \citep{yule1912methods},

\begin{eqnarray*}
\frac{\sqrt{OR}-1}{\sqrt{OR}+1}.
\end{eqnarray*}

Although these statistics provide the same information as is the
original odds ration \(OR\), they have the advantage of taking values in
the interval \([-1,1]\), making them easier to interpret.

In a later section, we will also see that the marginal distributions
have no effect on the Fr\(\acute{e}\)chet-H\(\ddot{o}\)effding of the
tetrachoric correlation, another measure of association, see also,
\citep{joe2014dependence}, page 48.

\[
{\small \begin{matrix}
\begin{array}{l|rr|r}
    \hline
                  & \text{Fire5} & & \\
\text{NoClaimCredit} & 0     & 1     & \text{Total} \\
  \hline
           0  & 1611  & 2175  & 3786 \\
           1  & 897   & 956   & 1853 \\
    \hline
    \text{Total}    & 2508  & 3131  & 5639 \\
   \hline
\end{array}
\end{matrix}}
\]

\protect\hyperlink{tab:14.2}{Table 14.2} : 2 \(\times\) 2 table of
counts for \emph{Fire5} and \emph{NoClaimCredit}

From \protect\hyperlink{tab:14.2}{Table 14.2},
\(OR(\pi_{11})=\frac{1611(956)}{897(2175)}=0.79\). You can obtain the
\(OR(\pi_{11})\), using the \texttt{oddsratio()} function from the
\texttt{epitools} library in \texttt{R}. From the output below,
\(OR(\pi_{11})=0.79\) for the binary variables \emph{NoClaimCredit} and
\emph{Fier5} from the LGPIF data.

R Code for Odds Ratios

\hypertarget{display.wald.2}{}
\begin{verbatim}
library(epitools)
oddsratio(NoClaimCredit, Fire5,method = c("wald"))$measure

Output:
[1]  0.79
\end{verbatim}

\subsubsection{Categorical Variables}\label{categorical-variables}

More generally, let \((X,Y)\) be a bivariate pair having \(ncat_X\) and
\(ncat_Y\) numbers of categories, respectively. For a two-way table of
counts, let \(n_{jk}\) be the number in the \(j\)th row, \(k\) column.
Let \(n_{j\cdot}\) be the row margin total and \(n_{\cdot k}\) be the
column margin total. Define Pearson chi-square statistic as

\begin{eqnarray*}
chi^2 = \sum_{jk} \frac{(n_{jk}- n_{j\cdot}n_{\cdot k}/n)^2}{n_{j\cdot}n_{\cdot k}/n} .
\end{eqnarray*}

The likelihood ratio test statistic is

\begin{eqnarray*}
G^2 = 2 \sum_{jk} n_{jk} \ln\frac{n_{jk}}{n_{j\cdot}n_{\cdot k}/n} .
\end{eqnarray*}

Under the assumption of independence, both \(chi^2\) and \(G^2\) have an
asymptotic chi-square distribution with \((ncat_X-1)(ncat_Y-1)\) degrees
of freedom.

To help see what these statistics are estimating, let
\(\pi_{jk} = \Pr(X=j, Y=k)\) and let \(\pi_{X,j}=\Pr(X=j)\) and
similarly for \(\pi_{Y,k}\). Assuming that \(n_{jk}/n \approx \pi_{jk}\)
for large \(n\) and similarly for the marginal probabilities, we have

\begin{eqnarray*}
\frac{chi^2}{n} \approx \sum_{jk} \frac{(\pi_{jk}- \pi_{X,j}\pi_{Y,k})^2}{\pi_{X,j}\pi_{Y,k}}
\end{eqnarray*}

and

\begin{eqnarray*}
\frac{G^2}{n} \approx 2 \sum_{jk} \pi_{jk} \ln\frac{\pi_{jk}}{\pi_{X,j}\pi_{Y,k}} .
\end{eqnarray*}

Under the null hypothesis of independence, we have
\(\pi_{jk} =\pi_{X,j}\pi_{Y,k}\) and it is clear from these
approximations that we anticipate that these statistics will be small
under this hypothesis.

Classical approaches, as described in \citep{bishop1975discrete} (page
374), distinguish between tests of independence and measures of
associations. The former are designed to detect whether a relationship
exists whereas the latter are meant to assess the type and extent of a
relationship. We acknowledge these differing purposes but also less
concerned with this distinction for actuarial applications.

\[
{\small \begin{matrix}
\begin{array}{l|rr}
    \hline
                  & \text{NoClaimCredit} &  \\
       \text{EntityType} & 0     & 1      \\
  \hline
            \text{City}    & 644  & 149 \\
          \text{County}    & 310  &  18 \\
            \text{Misc}    & 336  & 273 \\
          \text{School}    & 1103 & 494 \\
           \text{Town}     & 492  & 479 \\
         \text{Village}    & 901  & 440 \\
   \hline
\end{array}
\end{matrix}}
\]

\protect\hyperlink{tab:14.3}{Table 14.3} : Two-way table of counts for
\emph{EntityType} and \emph{NoClaimCredit}

You can obtain the Pearson chi-square statistic, using the
\texttt{chisq.test()} function from the \texttt{MASS} library in
\texttt{R}. Here, we test whether the \emph{EntityType} variable is
independent of \emph{NoClaimCredit} variable using
\protect\hyperlink{tab:14.3}{Table 14.3}.

R Code for Pearson Chi-square Statistic

\hypertarget{display.chi.2}{}
\begin{verbatim}
library(MASS)
table = table(EntityType, NoClaimCredit)
chisq.test(table)


Output:
------------------------------------
 Test statistic   df     P value
---------------- ---- --------------
     344.2        5   3.15e-72 * * *
------------------------------------

Table: Pearson's Chi-squared test
\end{verbatim}

As the p-value is less than the .05 significance level, we reject the
null hypothesis that the \emph{EntityType} is independent of
\emph{NoClaimCredit}.

Furthermore, you can obtain the likelihood ratio test statistic , using
the \texttt{likelihood.test()} function from the \texttt{Deducer}
library in \texttt{R}. From below, we test whether the \emph{EntityType}
variable is independent of \emph{NoClaimCredit} variable from the LGPIF
data. Same conclusion is drawn as the Pearson chi-square test.

R Code for Likelihood Ratio Test Statistic

\hypertarget{display.lik.2}{}
\begin{verbatim}
library(Deducer)
likelihood.test(EntityType, NoClaimCredit)

Output:
-----------------------------------------
 Test statistic   X-squared df   P value
---------------- -------------- ---------
     378.7             5         0 * * *
-----------------------------------------

Table: Log likelihood ratio (G-test) test of independence without correction
\end{verbatim}

\subsubsection{Ordinal Variables}\label{ordinal-variables}

As the analyst moves from the continuous to the nominal scale, there are
two main sources of loss of information \citep{bishop1975discrete} (page
343). The first is breaking the precise continuous measurements into
groups. The second is losing the ordering of the groups. So, it is
sensible to describe what we can do with variables that in discrete
groups but where the ordering is known.

As described in Section \ref{S:QuaVar}, ordinal variables provide a
clear ordering of levels of a variable but distances between levels are
unknown. Associations have traditionally been quantified parametrically
using normal-based correlations and nonparametrically using Spearman
correlations with tied ranks.

\subsubsection{Parametric Approach Using Normal Based
Correlations}\label{parametric-approach-using-normal-based-correlations}

Refer to page 60, Section 2.12.7 of \citep{joe2014dependence}. Let
\((y_1,y_2)\) be a bivariate pair with discrete values on
\(m_1, \ldots, m_2\). For a two-way table of ordinal counts, let
\(n_{st}\) be the number in the \(s\)th row, \(t\) column. Let
\((n_{m_1*}, \ldots, n_{m_2*})\) be the row margin total and
\((n_{*m_1}, \ldots, n_{*m_2})\) be the column margin total.

Let \(\hat{\xi}_{1s} = \Phi^{-1}((n_{m_1}+\cdots+n_{s*})/n)\) for
\(s=m_1, \ldots, m_2\) be a cutpoint and similarly for
\(\hat{\xi}_{2t}\). The \emph{polychoric} correlation, based on a
two-step estimation procedure, is

\begin{eqnarray*}
\begin{array}{cr}
  \hat{\rho_N} &=\text{argmax}_{\rho}
  \sum_{s=m_1}^{m_2} \sum_{t=m_1}^{m_2} n_{st} \log\left\{
    \Phi_2(\hat{\xi}_{1s}, \hat{\xi}_{2t};\rho)
    -\Phi_2(\hat{\xi}_{1,s-1}, \hat{\xi}_{2t};\rho) \right.\\
   & \left. -\Phi_2(\hat{\xi}_{1s}, \hat{\xi}_{2,t-1};\rho)
    +\Phi_2(\hat{\xi}_{1,s-1}, \hat{\xi}_{2,t-1};\rho)
    \right\}
\end{array}
\end{eqnarray*}

It is called a tetrachoric correlation for binary variables.

\[
{\small \begin{matrix}
\begin{array}{l|rr}
    \hline
                  & \text{NoClaimCredit} &  \\
\text{AlarmCredit} & 0     & 1      \\
  \hline
          1  & 1669  &  942   \\
          2  &    121 &  118 \\
          3  &  195  &   132 \\
          4 &  1801  &   661 \\
   \hline
\end{array}
\end{matrix}}
\]

\protect\hyperlink{tab:14.4}{Table 14.4} : Two-way table of counts for
\emph{AlarmCredit} and \emph{NoClaimCredit}

You can obtain the polychoric or tetrachoric correlation using the
\texttt{polychoric()} or \texttt{tetrachoric()} function from the
\texttt{psych} library in \texttt{R}. The polychoric correlation is
illustrated using \protect\hyperlink{tab:14.4}{Table 14.4}.
\(\hat{\rho_N}=-0.14\), which means that there is a negative
relationship between \emph{AlarmCredit} and \emph{NoClaimCredit}.

R Code for Polychoric Correlation

\hypertarget{display.poly.2}{}
\begin{verbatim}
library(psych)
AlarmCredit<-as.numeric(ifelse(Insample$AC00==1,"1",
                   ifelse(Insample$AC05==1,"2",
                          ifelse(Insample$AC10==1,"3",
                                 ifelse(Insample$AC15==1,"4",0)))))
x <- table(AlarmCredit,NoClaimCredit)
rhoN<-polychoric(x,correct=FALSE)$rho
round(rhoN,2)

Output:
[1] -0.14
\end{verbatim}

\subsubsection{Interval Variables}\label{interval-variables}

As described in Section \ref{S:QuanVar}, interval variables provide a
clear ordering of levels of a variable and the numerical distance
between any two levels of the scale can be readily interpretable. For
example, a claims count variable is an interval variable.

For measuring association, both the continuous variable and ordinal
variable approaches make sense. The former takes advantage of knowledge
of the ordering although assumes continuity. The latter does not rely on
the continuity but also does not make use of the information given by
the distance between scales.

For applications, one type is a count variable, a random variable on the
discrete integers. Another is a mixture variable, on that has discrete
and continuous components.

\subsubsection{Discrete and Continuous
Variables}\label{discrete-and-continuous-variables}

The polyserial correlation is defined similarly, when one variable
(\(y_1\)) is continuous and the other (\(y_2\)) ordinal. Define \(z\) to
be the normal score of \(y_1\). The polyserial correlation is

\begin{eqnarray*}
\hat{\rho_N} = \text{argmax}_{\rho}
\sum_{i=1}^n \log\left\{ \phi(z_{i1})\left[
\Phi(\frac{\hat{\xi}_{2,y_{i2}} - \rho z_{i1}}
{(1-\rho^2)^{1/2}})
-\Phi(\frac{\hat{\xi}_{2,y_{i2-1}} - \rho z_{i1}}
{(1-\rho^2)^{1/2}})
\right]
\right\}
\end{eqnarray*}

The biserial correlation is defined similarly, when one variable is
continuous and the other binary.

\[
{\small \begin{matrix}
\begin{array}{l|r|r}
    \hline
\text{NoClaimCredit} & \text{Mean}     &\text{Total}       \\
 & \text{Claim}     &\text{Claim}       \\
  \hline
          0  & 22,505  &  85,200,483   \\
          1  &    6,629 &  12,282,618 \\
   \hline
\end{array}
\end{matrix}}
\]

\protect\hyperlink{tab:14.5}{Table 14.5} : Summary of \emph{Claim} by
\emph{NoClaimCredit}

You can obtain the polyserial or biserial correlation using the
\texttt{polyserial()} or \texttt{biserial()} function from the
\texttt{psych} library in \texttt{R}. \protect\hyperlink{tab:14.5}{Table
14.5} gives the summary of \emph{Claim} by \emph{NoClaimCredit} and the
biserial correlation is illustrated using \texttt{R} code below. The
\(\hat{\rho_N}=-0.04\) which means that there is a negative correlation
between \emph{Claim} and \emph{NoClaimCredit}.

R Code for Biserial Correlation

\hypertarget{display.bis.2}{}
\begin{verbatim}
library(psych)
rhoN<-biserial(Claim,NoClaimCredit)
round(rhoN,2)

Output:
[1] -0.04
\end{verbatim}

\section{Introduction to Copulas}\label{S:Copula}

Copula functions are widely used in statistics and actuarial science
literature for dependency modeling.

\begin{center}\rule{0.5\linewidth}{\linethickness}\end{center}

In this section, you learn how to:

\begin{itemize}
\tightlist
\item
  Describe a multivariate distribution function in terms of a copula
  function.
\end{itemize}

\begin{center}\rule{0.5\linewidth}{\linethickness}\end{center}

A \(copula\) is a multivariate distribution function with uniform
marginals. Specifically, let \(U_1, \ldots, U_p\) be \(p\) uniform
random variables on \((0,1)\). Their distribution function
\[{C}(u_1, \ldots, u_p) = \Pr(U_1 \leq u_1, \ldots, U_p \leq u_p),\]

is a copula. We seek to use copulas in applications that are based on
more than just uniformly distributed data. Thus, consider arbitrary
marginal distribution functions \({F}_1(y_1)\),\ldots{},\({F}_p(y_p)\).
Then, we can define a multivariate distribution function using the
copula such that \[{F}(y_1, \ldots, y_p)= {C}({F}_1(y_1), \ldots,
{F}_p(y_p)).\]

Here, \(F\) is a multivariate distribution function in this equation.
Sklar (1959) showed that \(any\) multivariate distribution function
\(F\), can be written in the form of this equation, that is, using a
copula representation.

Sklar also showed that, if the marginal distributions are continuous,
then there is a unique copula representation. In this chapter we focus
on copula modeling with continuous variables. For discrete case, readers
can see \citep{joe2014dependence} and \citep{genest2007methods}.

For bivariate case, \(p=2\) , the distribution function of two random
variables can be written by the bivariate copula function:
\[{C}(u_1, \, u_2) = \Pr(U_1 \leq u_1, \, U_2 \leq
u_2),\]

\[{F}(y_1, \, y_2)= {C}({F}_1(y_1), \,
{F}_p(y_2)).\]

To give an example for bivariate copula, we can look at Frank's (1979)
copula. The equation is

\[{C}(u_1,u_2) = \frac{1}{\theta} \ln \left( 1+ \frac{ (\exp(\theta
u_1) -1)(\exp(\theta u_2) -1)} {\exp(\theta) -1} \right).\]

This is a bivariate distribution function with its domain on the unit
square \([0,1]^2.\) Here \(\theta\) is dependence parameter and the
range of dependence is controlled by the parameter \(\theta\). Positive
association increases as \(\theta\) increases and this positive
association can be summarized with Spearman's rho (\(\rho\)) and
Kendall's tau (\(\tau\)). Frank's copula is one of the commonly used
copula functions in the copula literature. We will see other copula
functions in Section \ref{S:CopTyp}.

\section{Application Using Copulas}\label{S:CopAppl}

\begin{center}\rule{0.5\linewidth}{\linethickness}\end{center}

In this section, you learn how to:

\begin{itemize}
\tightlist
\item
  Discover dependence structure between random variables
\item
  Model the dependence with a copula function
\end{itemize}

\begin{center}\rule{0.5\linewidth}{\linethickness}\end{center}

This section analyzes the insurance \emph{losses } and \emph{expenses }
data with the statistical programming \texttt{R}. This data set was
introduced in \citet{frees1998understanding} and is now readily
available in the \texttt{copula} package. The model fitting process is
started by marginal modeling of two variables (\(loss\) and
\(expense\)). Then we model the joint distribution of these marginal
outcomes.

\subsection{Data Description}\label{data-description}

We start with getting a sample (\(n = 1500\)) from the whole data. We
consider first two variables of the data; \emph{losses} and
\emph{expenses}.

\begin{itemize}
\tightlist
\item
  \emph{losses }: general liability claims from Insurance Services
  Office, Inc. (ISO)
\item
  \emph{expenses }: ALAE, specifically attributable to the settlement of
  individual claims (e.g.~lawyer's fees, claims investigation expenses)
\end{itemize}

\begin{figure}
\centering
\includegraphics{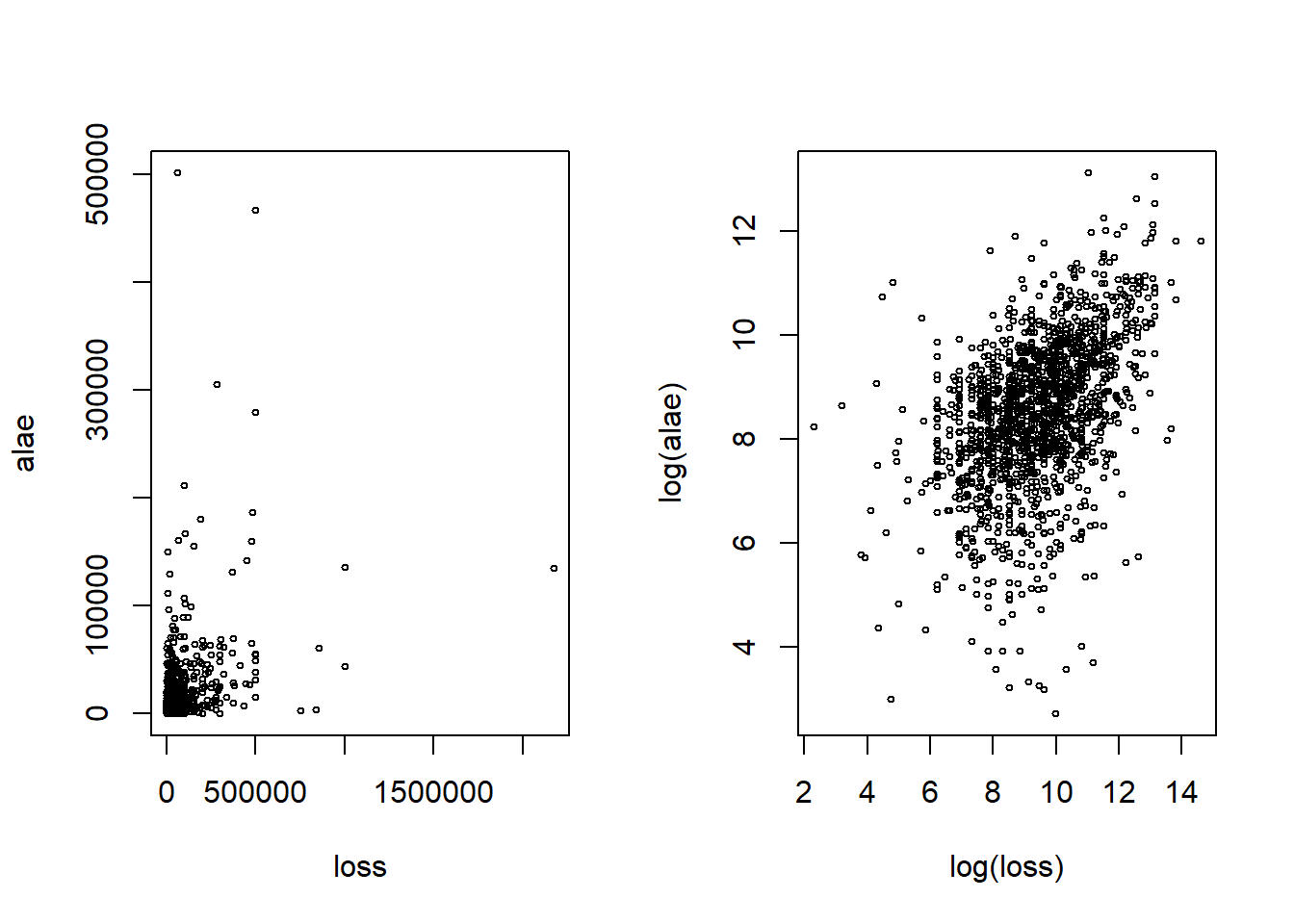}
\caption{\label{fig:Scatter}Scatter plot of Loss and ALAE}
\end{figure}

To visualize the relationship between \emph{losses } and \emph{expenses
} (ALAE), scatterplots in figure \ref{fig:Scatter} are created on the
real dollar scale and on the log scale.

R Code for Scatterplots

\hypertarget{display.ScaHis.2}{}
\begin{verbatim}
library(copula)
data(loss) # loss data
Lossdata <- loss
attach(Lossdata)
loss <- Lossdata$loss
par(mfrow=c(1, 2))
plot(loss,alae, cex=.5) # real dollar scale
plot(log(loss),log(alae),cex=.5) # log scale
par(mfrow=c(1, 2))
\end{verbatim}

\subsection{Marginal Models}\label{marginal-models}

We first examine the marginal distributions of \emph{losses } and
\emph{expenses } before going through the joint modeling. The histograms
show that both \emph{losses } and \emph{expenses } are right-skewed and
fat-tailed.

For marginal distributions of losses and expenses, we consider a
Pareto-type distribution, namely a Pareto type II with distribution
function

\[ F(y)=1- \left( 1 + \frac{y}{\theta} \right) ^{-\alpha},\] where
\(\theta\) is the scale parameter and \(\alpha\) is the shape parameter.

The marginal distributions of losses and expenses are fitted with
maximum likelihood. Specifically, we use the \(vglm\) function from the
\texttt{R\ VGAM} package. Firstly, we fit the marginal distribution of
\emph{expenses }.

R Code for Pareto Fitting

\hypertarget{display.Reg.2}{}
\begin{verbatim}
library(VGAM)

fit = vglm(alae ~ 1, paretoII(location=0, lscale="loge", lshape="loge")) # fit the model by vlgm function
coef(fit, matrix=TRUE) # extract fitted model coefficients, matrix=TRUE gives logarithm of estimated parameters instead of default normal scale estimates
Coef(fit)

Output:
               loge(scale) loge(shape)
 (Intercept)     9.624673   0.7988753

                  scale        shape
 (Intercept)  15133.603598     2.223039
\end{verbatim}

We repeat this procedure to fit the marginal distribution of the
\emph{loss} variable. Because the loss data also seems right-skewed and
heavy-tail data, we also model the marginal distribution with Pareto II
distribution.

R Code for Pareto Fitting

\hypertarget{display.ParFit.2}{}
\begin{verbatim}
fitloss = vglm(loss ~ 1, paretoII, trace=TRUE)
Coef(fit)
summary(fit)


Output:
       scale        shape
15133.603598     2.223039
\end{verbatim}

To visualize the fitted distribution of \emph{expenses } and \emph{loss}
variables, we use the estimated parameters and plot the corresponding
distribution function and density function. For more details on marginal
model selection, see Chapter \ref{C:ModelSelection}.

\subsection{Probability Integral
Transformation}\label{probability-integral-transformation}

The \emph{probability integral transformation } shows that any
continuous variable can be mapped to a \(U(0,1)\) random variable via
its distribution function.

Given the fitted Pareto II distribution, the variable \emph{expenses} is
transformed to the variable \(u_1\), which follows a uniform
distribution on \([0,1]\):

\[u_1 = 1 - \left( 1 + \frac{ALAE}{\hat{\theta}} \right)^{-\hat{\alpha}}.\]

\begin{figure}

{\centering \includegraphics{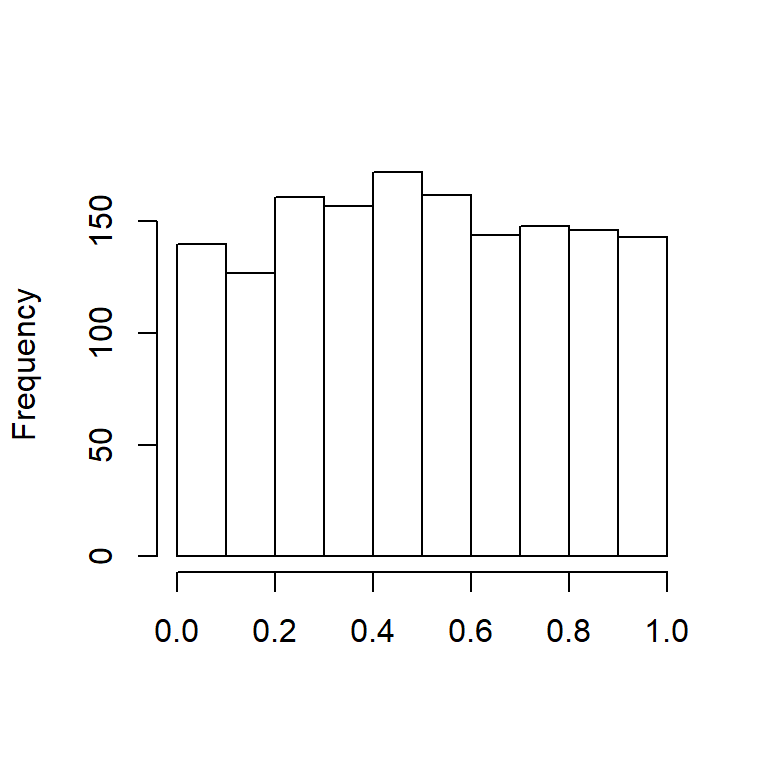}

}

\caption{Histogram of Transformed Alae}\label{fig:Hist}
\end{figure}

After applying the probability integral transformation to \emph{expenses
} variable, we plot the histogram of \emph{Transformed Alae } in Figure
\ref{fig:Hist}.

\begin{figure}
\centering
\includegraphics{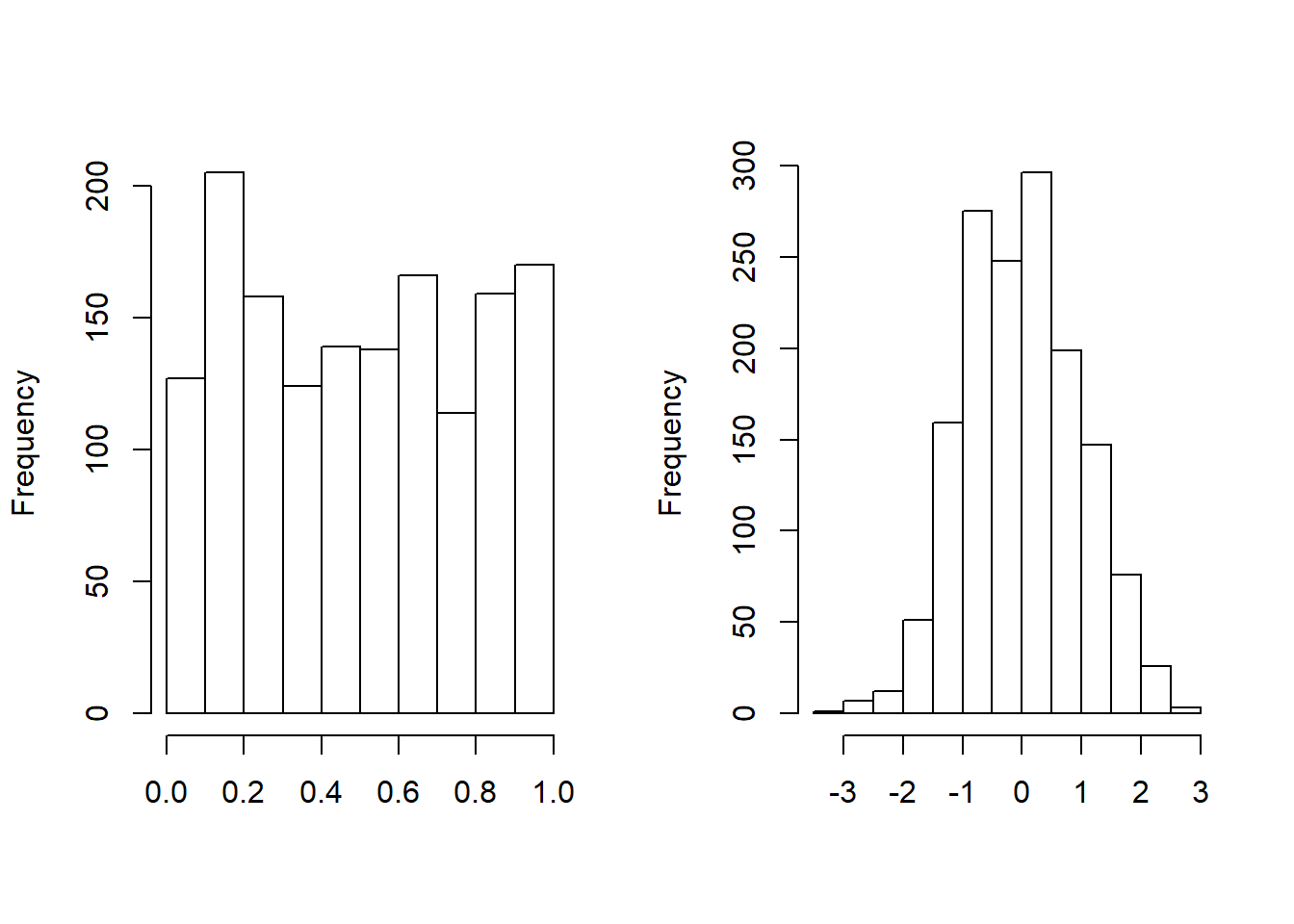}
\caption{\label{fig:Hist2}Histogram of Transformed Loss. The left-hand panel
shows the distribution of probability integral transformed losses. The
right-hand panel shows the distribution for the corresponding normal
scores.}
\end{figure}

After fitting process,the variable \emph{loss} is also transformed to
the variable \(u_2\), which follows a uniform distribution on \([0,1]\).
We plot the histogram of \emph{Transformed Loss }. As an alternative,
the variable \emph{loss} is transformed to \(normal\) \(scores\) with
the quantile function of standard normal distribution. As we see in
Figure \ref{fig:Hist2}, normal scores of the variable \emph{loss} are
approximately marginally standard normal.

R Code for Histograms of Transformed Variables

\hypertarget{display.transalae.2}{}
\begin{verbatim}
u1 <- 1 - (1 + (alae/b))^(-s) # or u1 <- pparetoII(alae, location=0, scale=b, shape=s)
hist(u1, main = "", xlab = "Histogram of Transformed alae")

scaleloss <- Coef(fitloss)[1]
shapeloss <- Coef(fitloss)[2]
u2 <- 1 - (1 + (loss/scaleloss))^(-shapeloss)
par(mfrow = c(1, 2))
hist(u2, main = "", xlab = "Histogram of Transformed Loss")
hist(qnorm(u2), main = "", xlab = "Histogram of qnorm(Loss)")
\end{verbatim}

\subsection{Joint Modeling with Copula
Function}\label{joint-modeling-with-copula-function}

\begin{figure}
\centering
\includegraphics{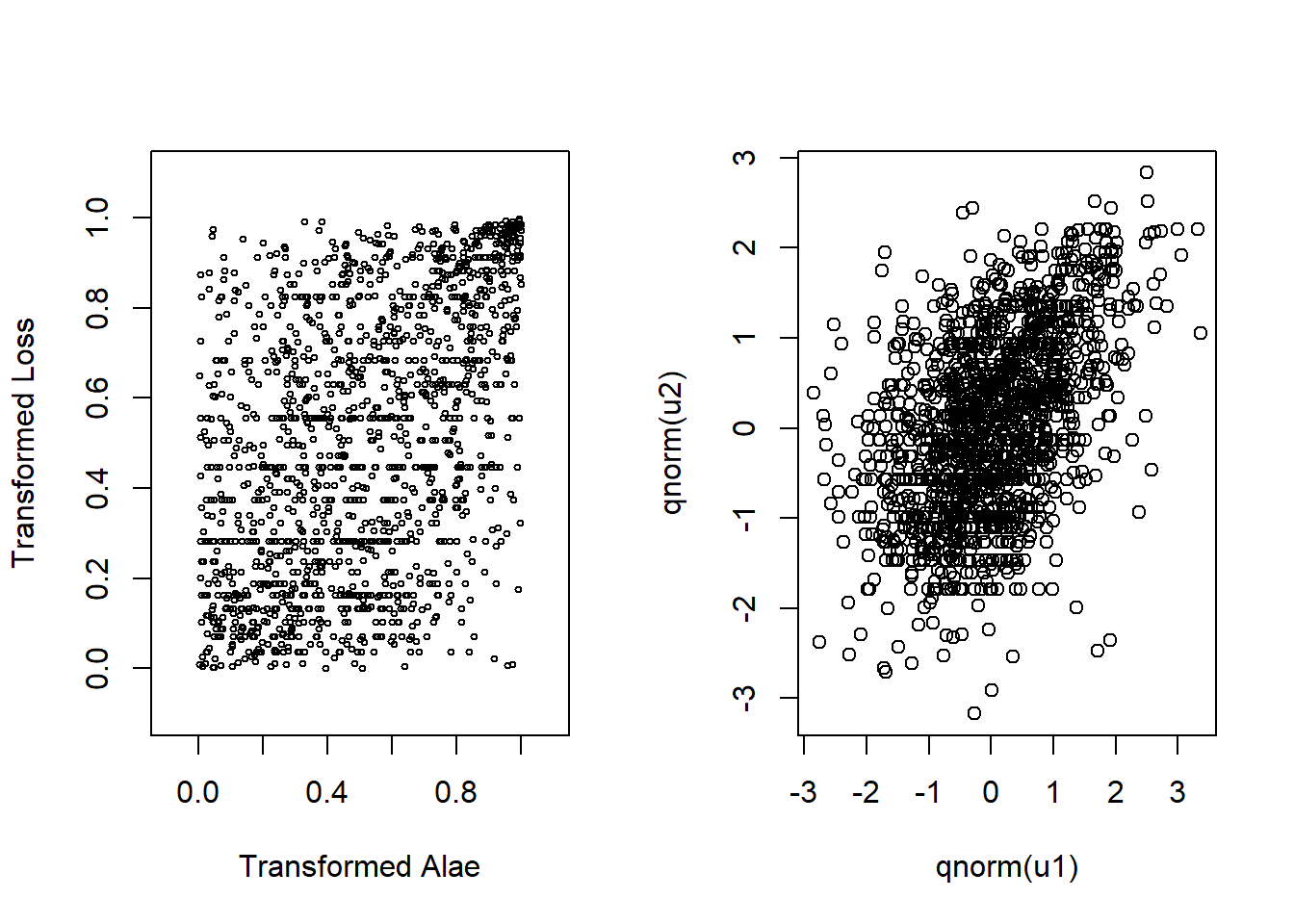}
\caption{\label{fig:Scatter2}Left: Scatter plot for transformed variables.
Right:Scatter plot for normal scores}
\end{figure}

Before jointly modeling losses and expenses, we draw the scatterplot of
transformed variables \((u_1, u_2)\) and the scatterplot of normal
scores in Figure \ref{fig:Scatter2}.

Then we calculate the Spearman's rho between these two uniform random
variables.

R Code for Scatter Plots and Correlation

\hypertarget{display.Cor.2}{}
\begin{verbatim}
par(mfrow = c(1, 2))
plot(u1, u2, cex = 0.5, xlim = c(-0.1,1.1), ylim = c(-0.1,1.1),
     xlab = "Transformed Alae", ylab = "Transformed Loss")
plot(qnorm(u1), qnorm(u2))
cor(u1, u2, method = "spearman")

Output:

[1] 0.451872
\end{verbatim}

Scatter plots and Spearman's rho correlation value (0.451) shows us
there is a positive dependency between these two uniform random
variables. It is more clear to see the relationship with normal scores
in the second graph. To learn more details about normal scores and their
applications in copula modeling, see \citep{joe2014dependence}.

\((U_1, U_2)\), (\(U_1 = F_1(ALAE)\) and \(U_2=F_2(LOSS)\)), is fit to
Frank's copula with maximum likelihood method.

R Code for Modeling with Frank Copula

\hypertarget{display.FrankCopula.2}{}
\begin{verbatim}
uu = cbind(u1,u2)
frank.cop <- archmCopula("frank", param= c(5), dim = 2)
fit.ml <- fitCopula(frank.cop, uu, method="ml", start=c(0.4))
summary(fit.ml)


Output:

Call: fitCopula(copula, data = data, method = "ml", start = ..2)
Fit based on "maximum likelihood" and 1500 2-dimensional observations.
Copula: frankCopula
param
3.114
The maximized loglikelihood is 172.6
Convergence problems: code is 52 see ?optim.
Call: fitCopula(copula, data = data, method = "ml", start = ..2)
Fit based on "maximum likelihood" and 1500 2-dimensional observations.
Frank copula, dim. d = 2
      Estimate Std. Error
param    3.114         NA
The maximized loglikelihood is 172.6
Convergence problems: code is 52 see ?optim.
Number of loglikelihood evaluations:
function gradient
      45       45
\end{verbatim}

The fitted model implies that losses and expenses are positively
dependent and their dependence is significant.

We use the fitted parameter to update the Frank's copula. The Spearman's
correlation corresponding to the fitted copula parameter(3.114) is
calculated with the \texttt{rho} function. In this case, the Spearman's
correlation coefficient is 0.462, which is very close to the sample
Spearman's correlation coefficient; 0.452.

R Code for Spearman's Correlation Using Frank's Copula

\hypertarget{display.fittedCop.2}{}
\begin{verbatim}
(param = fit.ml@estimate)
frank.cop <- archmCopula("frank", param= param, dim = 2)
rho(frank.cop)

Output :
[1] 0.4622722
\end{verbatim}

\begin{figure}
\centering
\includegraphics{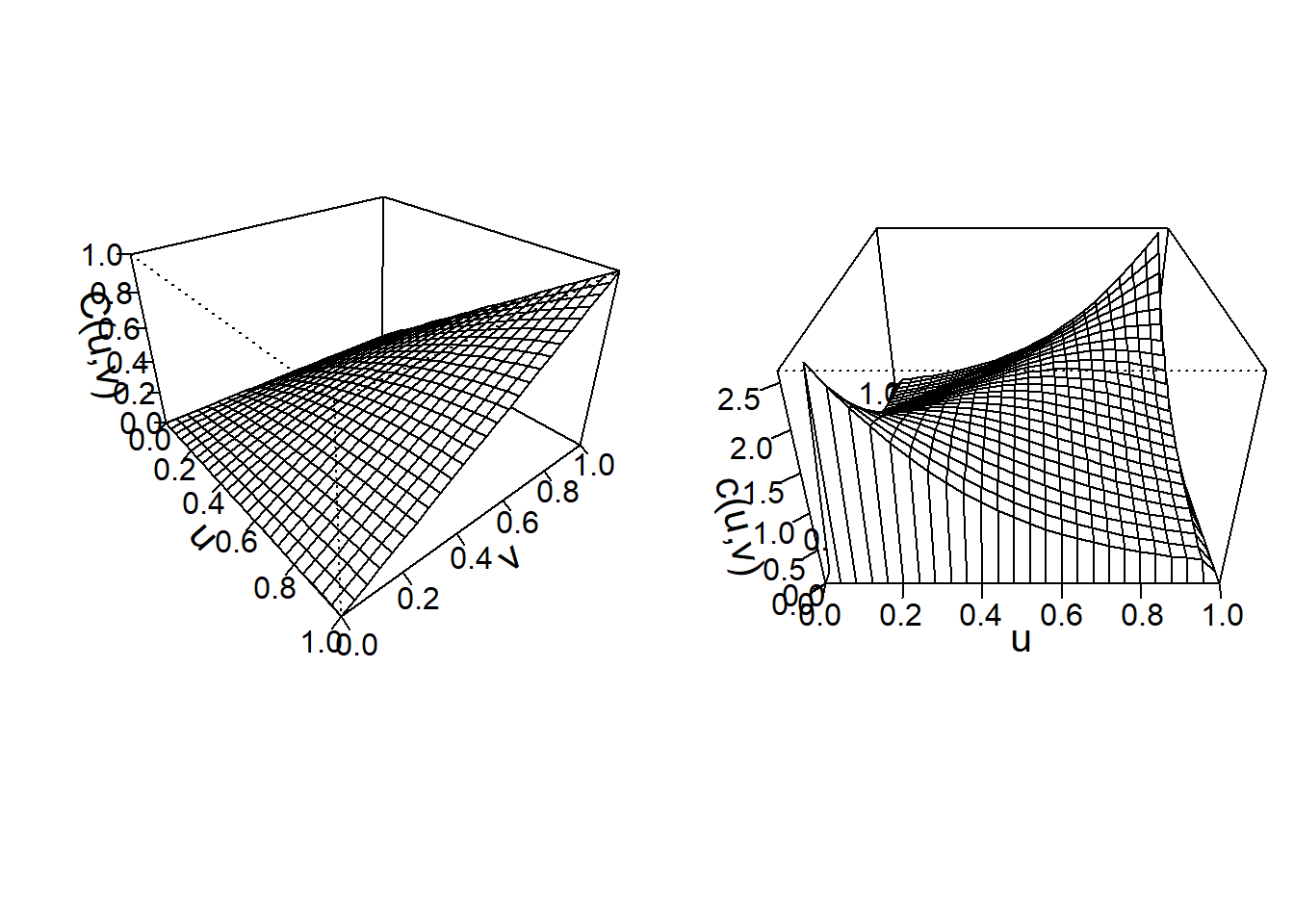}
\caption{\label{fig:FrankCop}Left: Plot for distribution function for Franks
Copula. Right:Plot for density function for Franks Copula}
\end{figure}

To visualize the fitted Frank's copula, the distribution function and
density function perspective plots are drawn in Figure
\ref{fig:FrankCop}.

R Code for Frank's Copula Plots

\hypertarget{display.DistriPlot.2}{}
\begin{verbatim}
par(mar=c(3.2,3,.2,.2),mfrow=c(1,2))
persp(frank.cop, pCopula, theta=50, zlab="C(u,v)",
        xlab ="u", ylab="v", cex.lab=1.3)
persp(frank.cop, dCopula, theta=0, zlab="c(u,v)",
        xlab ="u", ylab="v", cex.lab=1.3)
\end{verbatim}

Frank's copula models positive dependence for this data set, with
\(\theta=3.114\). For Frank's copula, the dependence is related to
values of \(\theta\). That is:

\begin{itemize}
\tightlist
\item
  \(\theta=0\): independent copula
\item
  \(\theta>0\): positive dependence
\item
  \(\theta<0\): negative dependence
\end{itemize}

\section{Types of Copulas}\label{S:CopTyp}

\begin{center}\rule{0.5\linewidth}{\linethickness}\end{center}

In this section, you learn how to:

\begin{itemize}
\tightlist
\item
  Define the basic families of the copula functions
\item
  Calculate the association coefficients by the help of copula functions
\end{itemize}

\begin{center}\rule{0.5\linewidth}{\linethickness}\end{center}

There are several families of copulas have been described in the
literature. Two main families of the copula families are the
\textbf{Archimedian} and \textbf{Elliptical} copulas.

\subsection{Elliptical Copulas}\label{elliptical-copulas}

Elliptical copulas are constructed from elliptical distributions. This
copula decompose (multivariate) elliptical distributions into their
univariate elliptical marginal distributions by Sklar's theorem
\citep{hofertelements}.

Properties of elliptical copulas are typically obtained from the
properties ofcorresponding elliptical distributions
\citep{hofertelements}.

For example, the normal distribution is a special type of elliptical
distribution. To introduce the elliptical class of copulas, we start
with the familiar multivariate normal distribution with probability
density function
\[\phi_N (\mathbf{z})= \frac{1}{(2 \pi)^{p/2}\sqrt{\det \boldsymbol \Sigma}}
\exp\left( -\frac{1}{2} \mathbf{z}^{\prime} \boldsymbol
\Sigma^{-1}\mathbf{z}\right).\]

Here, \(\boldsymbol \Sigma\) is a correlation matrix, with ones on the
diagonal. Let \(\Phi\) and \(\phi\) denote the standard normal
distribution and density functions. We define the Gaussian (normal)
copula density function as

\[{c}_N(u_1,  \ldots, u_p) = \phi_N \left(\Phi^{-1}(u_1), \ldots, \Phi^{-1}(u_p) \right) \prod_{j=1}^p \frac{1}{\phi(\Phi^{-1}(u_j))}.\]

As with other copulas, the domain is the unit cube \([0,1]^p\).

Specifically, a \(p\)-dimensional vector \({z}\) has an \({elliptical}\)
\({distribution}\) if the density can be written as
\[h_E (\mathbf{z})= \frac{k_p}{\sqrt{\det \boldsymbol \Sigma}}
g_p \left( \frac{1}{2} (\mathbf{z}- \boldsymbol \mu)^{\prime}
\boldsymbol \Sigma^{-1}(\mathbf{z}- \boldsymbol \mu) \right).\]

We will use elliptical distributions to generate copulas. Because
copulas are concerned primarily with relationships, we may restrict our
considerations to the case where \(\mu = \mathbf{0}\) and
\(\boldsymbol \Sigma\) is a correlation matrix. With these restrictions,
the marginal distributions of the multivariate elliptical copula are
identical; we use \(H\) to refer to this marginal distribution function
and \(h\) is the corresponding density. This marginal density is
\(h(z) = k_1 g_1(z^2/2).\)

We are now ready to define the \(elliptical\) \(copula\), a function
defined on the unit cube \([0,1]^p\) as

\[{c}_E(u_1,  \ldots, u_p) = h_E \left(H^{-1}(u_1), \ldots,
H^{-1}(u_p) \right) \prod_{j=1}^p \frac{1}{h(H^{-1}(u_j))}.\]

In the elliptical copula family, the function \(g_p\) is known as a
\emph{generator} in that it can be used to generate alternative
distributions.

\[
\small\begin{array}{lc}
\hline & Generator \\
 Distribution &  \mathrm{g}_p(x)  \\
\hline
 \text{Normal distribution} &  e^{-x}\\
 \text{t-distribution with r degrees of freedom} &   (1+2x/r)^{-(p+r)/2}\\
 \text{Cauchy} &  (1+2x)^{-(p+1)/2}\\
\text{Logistic} &  e^{-x}/(1+e^{-x})^2\\
 \text{Exponential power} &   \exp(-rx^s)\\
\hline
\end{array}
\]

\protect\hyperlink{tab:14.6}{Table 14.6} : Distribution and Generator
Functions (\(\mathrm{g}_p(x)\)) for Selected Elliptical Copulas

Most empirical work focuses on the normal copula and \(t\)-copula. That
is, \(t\)-copulas are useful for modeling the dependency in the tails of
bivariate distributions, especially in financial risk analysis
applications.

The \(t\)-copulas with same association parameter in varying the degrees
of freedom parameter show us different tail dependency structures. For
more information on about \(t\)-copulas readers can see
\citep{joe2014dependence}, \citep{hofertelements}.

\subsection{Archimedian Copulas}\label{archimedian-copulas}

This class of copulas are constructed from a \(generator\)
function,which is \(\mathrm{g}(\cdot)\) is a convex, decreasing function
with domain {[}0,1{]} and range \([0, \infty)\) such that
\(\mathrm{g}(0)=0\). Use \(\mathrm{g}^{-1}\) for the inverse function of
\(\mathrm{g}\). Then the function

\[\mathrm{C}_{\mathrm{g}}(u_1, \ldots, u_p) = \mathrm{g}^{-1} \left(\mathrm{g}(u_1)+ \cdots + \mathrm{g}(u_p) \right)\]

is said to be an \emph{Archimedean} copula. The function \(\mathrm{g}\)
is known as the \emph{generator} of the copula
\(\mathrm{C}_{\mathrm{g}}\).

For bivariate case; \(p=2\) , Archimedean copula function can be written
by the function

\[\mathrm{C}_{\mathrm{g}}(u_1, \, u_2) = \mathrm{g}^{-1} \left(\mathrm{g}(u_1) + \mathrm{g}(u_2) \right).\]

Some important special cases of Archimedean copulas are Frank copula,
Clayton/Cook-Johnson copula, Gumbel/Hougaard copula. This copula classes
are derived from different generator functions.

We can remember that we mentioned about Frank's copula with details in
Section \ref{S:Copula} and in Section \ref{S:CopAppl}. Here we will
continue to express the equations for Clayton copula and Gumbel/Hougaard
copula.

\subsubsection{Clayton Copula}\label{clayton-copula}

For \(p=2\), the Clayton copula is parameterized by
\(\theta \in [-1,\infty)\) is defined by
\[C_{\theta}^C(u)=\max\{u_1^{-\theta}+u_2^{-\theta}-1,0\}^{1/\theta}, \quad u\in[0,1]^2.\]

This is a bivariate distribution function of Clayton copula defined in
unit square \([0,1]^2.\) The range of dependence is controlled by the
parameter \(\theta\) as the same as Frank copula.

\subsubsection{Gumbel-Hougaard copula}\label{gumbel-hougaard-copula}

The Gumbel-Hougaarg copula is parametrized by \(\theta \in [1,\infty)\)
and defined by
\[C_{\theta}^{GH}(u)=\exp\left(-\left(\sum_{i=1}^2 (-\log u_i)^{\theta}\right)^{1/\theta}\right), \quad u\in[0,1]^2.\]

Readers seeking deeper background on Archimedean copulas can see
\citet{joe2014dependence}, \citet{frees1998understanding}, and
\citet{genest1986bivariate}.

\subsection{Properties of Copulas}\label{properties-of-copulas}

\subsubsection{Bounds on Association}\label{bounds-on-association}

Like all multivariate distribution functions, copulas are bounded. The
Fr\('{e}\)chet-Hoeffding bounds are

\[\max( u_1 +\cdots+ u_p + p -1, 0) \leq  \mathrm{C}(u_1,  \ldots, u_p) \leq \min (u_1,  \ldots,u_p).\]

To see the right-hand side of the equation, note that
\[\mathrm{C}(u_1,\ldots, u_p) = \Pr(U_1 \leq u_1, \ldots, U_p \leq u_p) \leq  \Pr(U_j \leq u_j)\],
for \(j=1,\ldots,p\). The bound is achieved when \(U_1 = \cdots = U_p\).
To see the left-hand side when \(p=2\), consider \(U_2=1-U_1\). In this
case, if \(1-u_2 < u_1\) then
\(\Pr(U_1 \leq u_1, U_2 \leq u_2) = \Pr ( 1-u_2 \leq U_1 < u_1) =u_1+u_2-1.\)
\citep{nelsen1997introduction}

The product copula is \(\mathrm{C}(u_1,u_2)=u_1u_2\) is the result of
assuming independence between random variables.

The lower bound is achieved when the two random variables are perfectly
negatively related (\(U_2=1-U_1\)) and the upper bound is achieved when
they are perfectly positively related (\(U_2=U_1\)).

\begin{figure}
\centering
\includegraphics{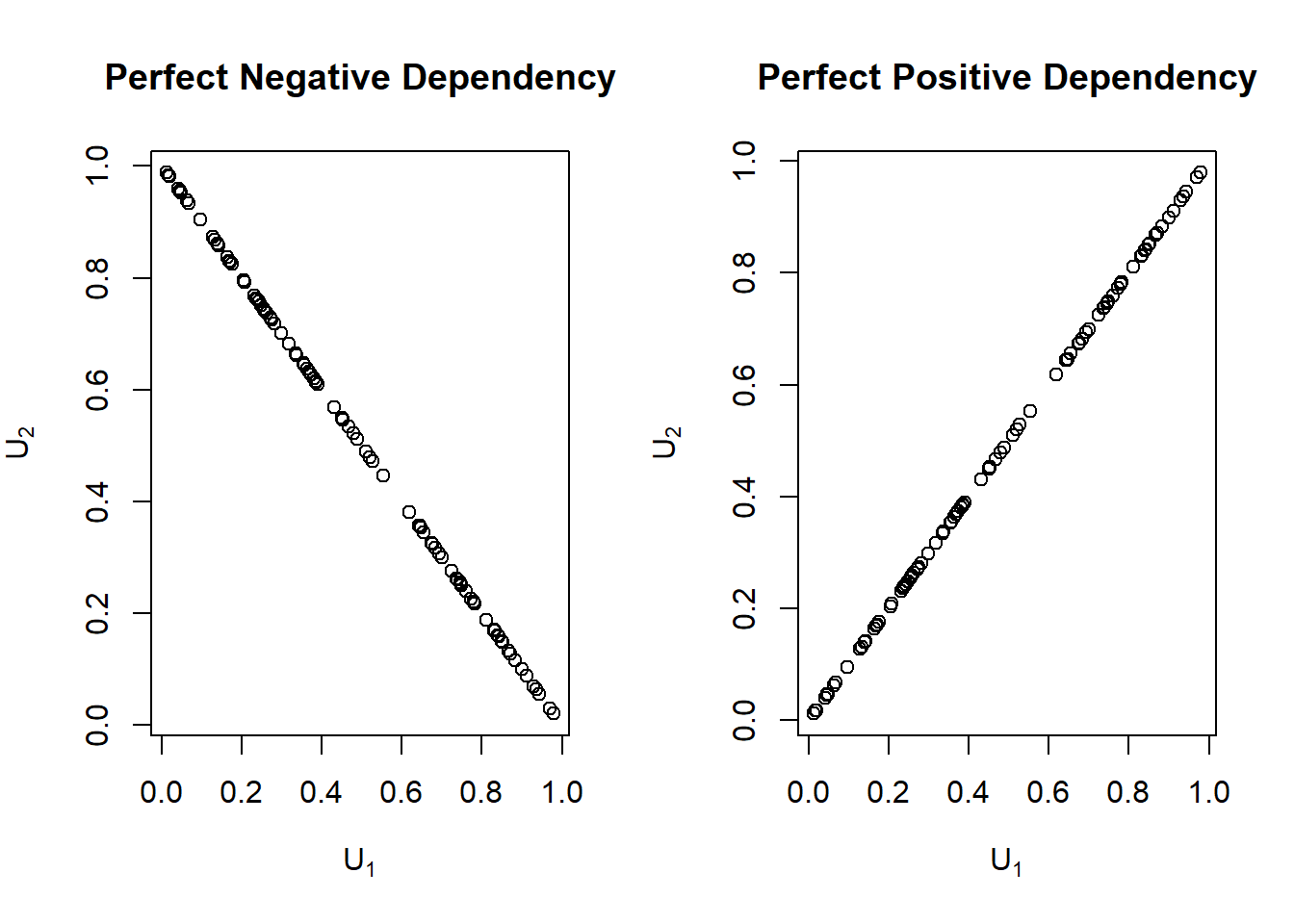}
\caption{\label{fig:Bounds}Perfect Positive and Perfect negative dependence
plots}
\end{figure}

We can see The Frechet-Hoeffding bounds for two random variables in the
Figure \ref{fig:Bounds}.

R Code for Frechet-Hoeffding Bounds for Two Random Variables

\hypertarget{display.plot.2}{}
\begin{verbatim}
library(copula)
n<-100
set.seed(1980)
U<-runif(n)
par(mfrow=c(1, 2))
plot(cbind(U,1-U), xlab=quote(U[1]), ylab=quote(U[2]),main="Perfect Negative Dependency") # W for p=2
plot (cbind(U,U), xlab=quote(U[1]),ylab=quote(U[2]),main="Perfect Positive Dependency")  #M for p=2
\end{verbatim}

\subsubsection{Measures of Association}\label{measures-of-association}

Schweizer and Wolff (1981) established that the copula accounts for all
the dependence between two random variables, \(Y_1\) and \(Y_2\), in the
following sense. Consider m\(_1\) and m\(_2\), strictly increasing
functions. Thus, the manner in which \(Y_1\) and \(Y_2\) ``move
together'' is captured by the copula, regardless of the scale in which
each variable is measured.

Schweizer and Wolff also showed the two standard nonparametric measures
of association could be expressed solely in terms of the copula
function. Spearman's correlation coefficient is given by

\[= 12 \int \int \left\{\mathrm{C}(u,v) - uv \right\} du dv.\]

Kendall's tau is given by

\[= 4 \int \int \mathrm{C}(u,v)d\mathrm{C}(u,v) - 1 .\]

For these expressions, we assume that \(Y_1\) and \(Y_2\) have a jointly
continuous distribution function. Further, the definition of Kendall's
tau uses an independent copy of (\(Y_1\), \(Y_2\)), labeled
(\(Y_1^{\ast}\), \(Y_2^{\ast}\)), to define the measure of
``concordance.'' the widely used Pearson correlation depends on the
margins as well as the copula. Because it is affected by non-linear
changes of scale.

\subsubsection{Tail Dependency}\label{tail-dependency}

There are some applications in which it is useful to distinguish by the
part of the distribution in which the association is strongest. For
example, in insurance it is helpful to understand association among the
largest losses, that is, association in the right tails of the data.

To capture this type of dependency, we use the right-tail concentration
function. The function is

\[R(z) = \frac{\Pr(U_1 >z, U_2 > z)}{1-z} =\Pr(U_1 > z | U_2 > z) =\frac{1 - 2z + \mathrm{C}(z,z)}{1-z} .\]

From this equation , \(R(z)\) will equal to \(z\) under independence.
Joe (1997) uses the term ``upper tail dependence parameter'' for
\(R = \lim_{z \rightarrow 1} R(z)\). Similarly, the left-tail
concentration function is

\[L(z) = \frac{\Pr(U_1 \leq z, U_2 \leq z)}{z}=\Pr(U_1 \leq z | U_2 \leq z) =\frac{ \mathrm{C}(z,z)}{1-z}.\]

Tail dependency concentration function captures the probability of two
random variables both catching up extreme values.

\begin{figure}
\centering
\includegraphics{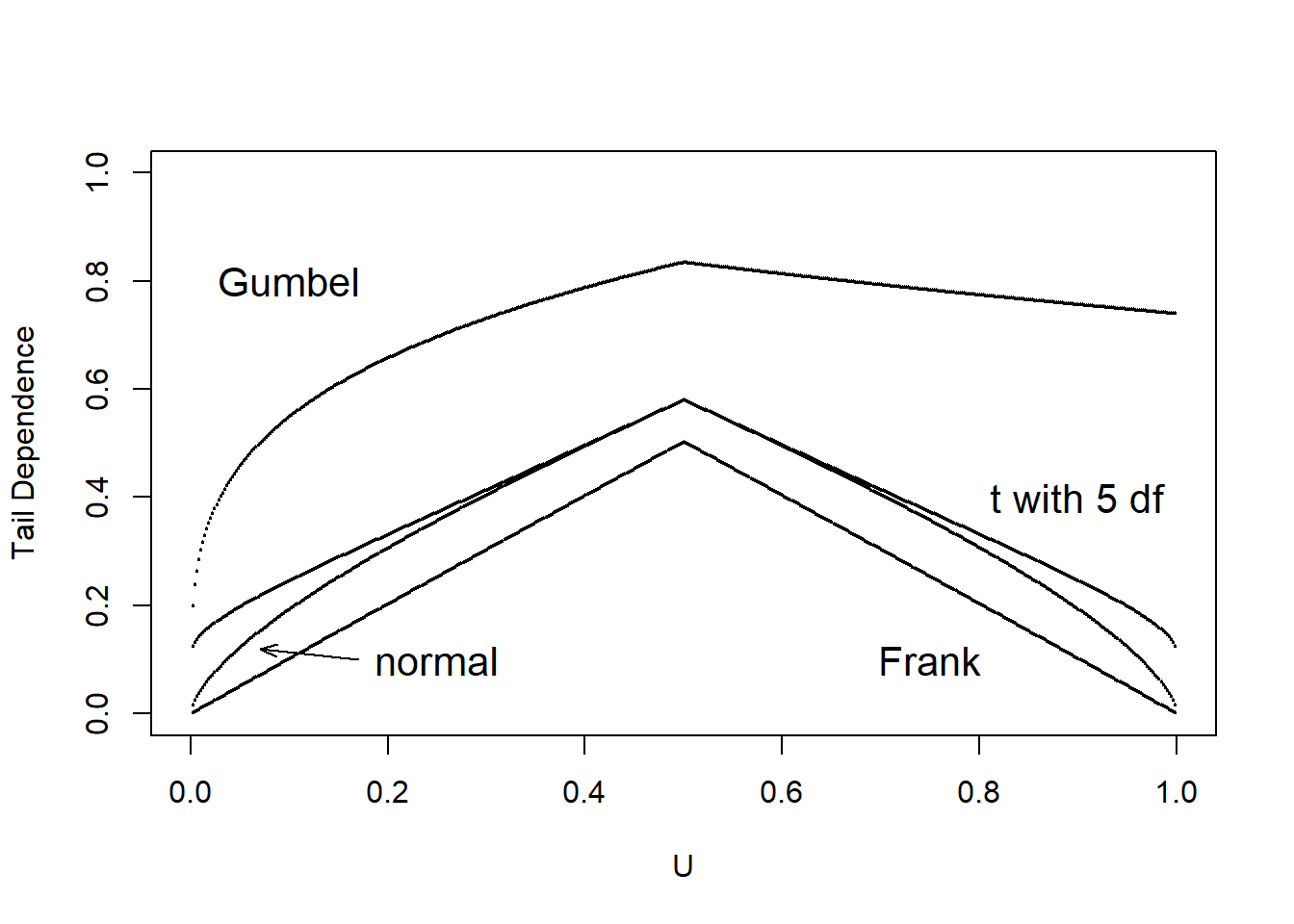}
\caption{\label{fig:DepTails}Tail dependence plots}
\end{figure}

We calculate the left and right tail concentration functions for four
different types of copulas; Normal, Frank,Gumbel and t copula. After
getting tail concentration functions for each copula, we show
concentration function's values for these four copulas in
\protect\hyperlink{tab:14.7}{Table 14.7}. As in \citet{venter2002tails},
we show \(L(z)\) for \(z\leq 0.5\) and \(R(z)\) for \(z>0.5\) in the
tail dependence plot in Figure \ref{fig:DepTails}. We interpret the tail
dependence plot, to mean that both the Frank and Normal copula exhibit
no tail dependence whereas the \(t\) and the Gumbel may do so. The \(t\)
copula is symmetric in its treatment of upper and lower tails.

\[
{\small \begin{matrix}
\begin{array}{l|rr}
    \hline
\text{Copula} & \text{Lower}    & \text{Upper}     \\
\hline
\text{Frank}  & 0  & 0   \\
\text{Gumbel}  & 0   & 0.74    \\
\text{Normal}  & 0   & 0    \\
\text{t}  & 0.10   & 0.10    \\
   \hline
\end{array}
\end{matrix}}
\]

\protect\hyperlink{tab:14.7}{Table 14.7} : Tail concentration function
values for different copulas

R Code for Tail Copula Functions for Different Copulas

\hypertarget{display.Sim.2}{}
\begin{verbatim}

library(copula)
U1 = seq(0,0.5, by=0.002)
U2 = seq(0.5,1, by=0.002)
U = rbind(U1, U2)
TailFunction <- function(Tailcop) {
  lowertail <- pCopula(cbind(U1,U1), Tailcop)/U1
  uppertail <- (1-2*U2 +pCopula(cbind(U2,U2), Tailcop))/(1-U2)
  jointtail <- rbind(lowertail,uppertail)
}
Tailcop1 <- archmCopula(family = "frank", param= c(0.05), dim = 2)
Tailcop2 <- archmCopula(family = "gumbel",param = 3)
Tailcop3 <- ellipCopula("normal", param = c(0.25),dim = 2, dispstr = "un")
Tailcop4 <- ellipCopula("t", param = c(0.25),dim = 2, dispstr = "un", df=5)
jointtail1 <- TailFunction(Tailcop1)
jointtail2 <- TailFunction(Tailcop2)
jointtail3 <- TailFunction(Tailcop3)
jointtail4 <- TailFunction(Tailcop4)
tailIndex(Tailcop1)
tailIndex(Tailcop2)
tailIndex(Tailcop3)
tailIndex(Tailcop4)
\end{verbatim}

R Code for Tail Dependence Plots for Different Copulas

\hypertarget{display.Sim.2}{}
\begin{verbatim}
plot(U,jointtail1, cex=.2, xlim=c(0,1),ylab="Tail Dependence", ylim=c(0,1))
lines(U,jointtail2, type="p",lty=1, cex=.2)
lines(U,jointtail3, type="p",lty=1, cex=.2)
lines(U,jointtail4, type="p",lty=1, cex=.2)
text(0.75, 0.1, "Frank", cex=1.3)        #1
text(0.1, 0.8, "Gumbel", cex=1.3)        #2
text(0.25, 0.1, "normal", cex=1.3)       #3
arrows(.17, 0.1, .07, 0.12,code=2, angle=20, length=0.1)
text(0.9, 0.4, "t with 5 df", cex=1.3)   #4
\end{verbatim}

\section{Why is Dependence Modeling Important?}\label{S:CopImp}

Dependence Modeling is important because it enables us to understand the
dependence structure by defining the relationship between variables in a
dataset. In insurance, ignoring dependence modeling may not impact
pricing but could lead to misestimation of required capital to cover
losses. For instance, from Section \ref{S:CopAppl} , it is seen that
there was a positive relationship between \emph{Loss} and
\emph{Expense}. This means that, if there is a large loss then we expect
expenses to be large as well and ignoring this relationship could lead
to misestimation of reserves.

To illustrate the importance of dependence modeling, we refer you back
to Portfolio Management example in Chapter 6 that assumed that the
property and liability risks are independent. Here, we incorporate
dependence by allowing the 4 lines of business to depend on one another
through a Gaussian copula. In \protect\hyperlink{tab:14.8}{Table 14.8},
we show that dependence affects the portfolio quantiles (\(VaR_q\)),
although not the expect value. For instance , the \(VaR_{0.99}\) for
total risk which is the amount of capital required to ensure, with a
\(99\%\) degree of certainty that the firm does not become technically
insolvent is higher when we incorporate dependence. This leads to less
capital being allocated when dependence is ignored and can cause
unexpected solvency problems.

\[
{\small \begin{matrix}
\begin{array}{l|rrrr}
    \hline
 \text{Independent} &\text{Expected}   & VaR_{0.9}  & VaR_{0.95}  & VaR_{0.99}  \\
                   &\text{Value}      &            &             &             \\
     \hline
\text{Retained}    & 269              &  300       & 300         & 300         \\
\text{Insurer}     & 2,274            &  4,400     & 6,173       & 11,859      \\
\text{Total}       & 2,543            &  4,675     & 6,464       & 12,159      \\
   \hline
\text{Gaussian Copula}&\text{Expected}& VaR_{0.9}  & VaR_{0.95}  & VaR_{0.99}  \\
                      &\text{Value}    &           &             &              \\
     \hline
\text{Retained}       & 269            &  300      & 300         &  300         \\
\text{Insurer}        & 2,340          &  4,988    & 7,339       & 14,905       \\
\text{Total}          & 2,609          &  5,288    & 7,639       & 15,205       \\
   \hline
\end{array}
\end{matrix}}
\]

\protect\hyperlink{tab:14.8}{Table 14.8} : Results for portfolio
expected value and quantiles (\(VaR_q\))

R Code for Simulation Using Gaussian Copula

\hypertarget{display.Sim.2}{}
\begin{verbatim}

# For the gamma distributions, use
alpha1 <- 2;      theta1 <- 100
alpha2 <- 2;      theta2 <- 200
# For the Pareto distributions, use
alpha3 <- 2;      theta3 <- 1000
alpha4 <- 3;      theta4 <- 2000
# Deductibles
d1     <- 100
d2     <- 200


# Simulate the risks
nSim <- 10000  #number of simulations
set.seed(2017) #set seed to reproduce work
X1 <- rgamma(nSim,alpha1,scale = theta1)
X2 <- rgamma(nSim,alpha2,scale = theta2)
# For the Pareto Distribution, use
library(VGAM)
X3 <- rparetoII(nSim,scale=theta3,shape=alpha3)
X4 <- rparetoII(nSim,scale=theta4,shape=alpha4)
# Portfolio Risks
S         <- X1 + X2 + X3 + X4
Sretained <- pmin(X1,d1) + pmin(X2,d2)
Sinsurer  <- S - Sretained

# Expected Claim Amounts
ExpVec <- t(as.matrix(c(mean(Sretained),mean(Sinsurer),mean(S))))
colnames(ExpVec) <- c("Retained", "Insurer","Total")
round(ExpVec,digits=2)

# Quantiles
quantMat <- rbind(
  quantile(Sretained, probs=c(0.80, 0.90, 0.95, 0.99)),
  quantile(Sinsurer,  probs=c(0.80, 0.90, 0.95, 0.99)),
  quantile(S       ,  probs=c(0.80, 0.90, 0.95, 0.99)))
rownames(quantMat) <- c("Retained", "Insurer","Total")
round(quantMat,digits=2)

plot(density(S), main="Density of Total Portfolio Risk S", xlab="S")

### Normal Copula ##
library(VGAM)
library(copula)
library(GB2)
library(statmod)
library(numDeriv)
set.seed(2017)
parm<-c(0.5,0.5,0.5,0.5,0.5,0.5)
nc <- normalCopula(parm, dim = 4, dispstr = "un")
mcc <- mvdc(nc, margins = c("gamma", "gamma","paretoII","paretoII"),
            paramMargins = list(list(scale = theta1, shape=alpha1),
                                list(scale = theta2, shape=alpha2),
                                list(scale = theta3, shape=alpha3),
                                list(scale = theta4, shape=alpha4)))
X <- rMvdc(nSim, mvdc = mcc)

X1<-X[,1]
X2<-X[,2]
X3<-X[,3]
X4<-X[,4]

# Portfolio Risks
S         <- X1 + X2 + X3 + X4
Sretained <- pmin(X1,d1) + pmin(X2,d2)
Sinsurer  <- S - Sretained

# Expected Claim Amounts
ExpVec <- t(as.matrix(c(mean(Sretained),mean(Sinsurer),mean(S))))
colnames(ExpVec) <- c("Retained", "Insurer","Total")
round(ExpVec,digits=2)

# Quantiles
quantMat <- rbind(
  quantile(Sretained, probs=c(0.80, 0.90, 0.95, 0.99)),
  quantile(Sinsurer,  probs=c(0.80, 0.90, 0.95, 0.99)),
  quantile(S       ,  probs=c(0.80, 0.90, 0.95, 0.99)))
rownames(quantMat) <- c("Retained", "Insurer","Total")
round(quantMat,digits=2)

plot(density(S), main="Density of Total Portfolio Risk S", xlab="S")
\end{verbatim}

\section{Further Resources and
Contributors}\label{Dep:further-reading-and-resources}

\subsubsection*{Contributors}\label{contributors-5}
\addcontentsline{toc}{subsubsection}{Contributors}

\begin{itemize}
\tightlist
\item
  \textbf{Edward W. (Jed) Frees} and \textbf{Nii-Armah Okine},
  University of Wisconsin-Madison, and \textbf{Emine Selin Sarıdaş},
  Mimar Sinan University, are the principal authors of the initital
  version of this chapter. Email:
  \href{mailto:jfrees@bus.wisc.edu}{\nolinkurl{jfrees@bus.wisc.edu}} for
  chapter comments and suggested improvements.
\end{itemize}

\section*{Technical Supplement A. Other Classic Measures of Scalar
Associations}\label{technical-supplement-a.-other-classic-measures-of-scalar-associations}
\addcontentsline{toc}{section}{Technical Supplement A. Other Classic
Measures of Scalar Associations}

\subsection*{A.1. Blomqvist's Beta}\label{a.1.-blomqvists-beta}
\addcontentsline{toc}{subsection}{A.1. Blomqvist's Beta}

\citet{blomqvist1950measure} developed a measure of dependence now known
as \emph{Blomqvist's beta}, also called the \emph{median concordance
coefficient} and the \emph{medial correlation coefficient}. Using
distribution functions, this parameter can be expressed as

\begin{equation*}
\beta = 4F\left(F^{-1}_X(1/2),F^{-1}_Y(1/2) \right) - 1.
\end{equation*}

That is, first evaluate each marginal at its median (\(F^{-1}_X(1/2)\)
and \(F^{-1}_Y(1/2)\), respectively). Then, evaluate the bivariate
distribution function at the two medians. After rescaling (multiplying
by 4 and subtracting 1), the coefficient turns out to have a range of
\([-1,1]\), where 0 occurs under independence.

Like Spearman's rho and Kendall's tau, an estimator based on ranks is
easy to provide. First write
\(\beta = 4C(1/2,1/2)-1 = 2\Pr((U_1-1/2)(U_2-1/2))-1\) where
\(U_1, U_2\) are uniform random variables. Then, define

\begin{equation*}
\hat{\beta} = \frac{2}{n} \sum_{i=1}^n I\left( (R(X_{i})-\frac{n+1}{2})(R(Y_{i})-\frac{n+1}{2}) \ge 0 \right)-1 .
\end{equation*}

See, for example, \citep{joe2014dependence}, page 57 or
\citep{hougaard2000analysis}, page 135, for more details.

Because Blomqvist's parameter is based on the center of the
distribution, it is particularly useful when data are censored; in this
case, information in extreme parts of the distribution are not always
reliable. How does this affect a choice of association measures? First,
recall that association measures are based on a bivariate distribution
function. So, if one has knowledge of a good approximation of the
distribution function, then calculation of an association measure is
straightforward in principle. Second, for censored data, bivariate
extensions of the univariate Kaplan-Meier distribution function
estimator are available. For example, the version introduced in
\citep{dabrowska1988kaplan} is appealing. However, because of instances
when large masses of data appear at the upper range of the data, this
and other estimators of the bivariate distribution function are
unreliable. This means that, summary measures of the estimated
distribution function based on Spearman's rho or Kendall's tau can be
unreliable. For this situation, Blomqvist's beta appears to be a better
choice as it focuses on the center of the distribution.
\citep{hougaard2000analysis}, Chapter 14, provides additional
discussion.

You can obtain the Blomqvist's beta, using the \texttt{betan()} function
from the \texttt{copula} library in \texttt{R}. From below,
\(\beta=0.3\) between the \emph{Coverage} rating variable in millions of
dollars and \emph{Claim} amount variable in dollars.

R Code for Blomqvist's Beta

\hypertarget{display.beta.2}{}
\begin{verbatim}
### Blomqvist's beta correlation between Claim and Coverage ###
library(copula)
n<-length(Claim)
U<-cbind(((n+1)/n*pobs(Claim)),((n+1)/n*pobs(Coverage)))
beta<-betan(U, scaling=FALSE)
round(beta,2)

Output:
[1]  0.3

### Blomqvist's beta correlation between Claim and log(Coverage) ###
n<-length(Claim)
Fx<-cbind(((n+1)/n*pobs(Claim)),((n+1)/n*pobs(log(Coverage))))
beta<-betan(Fx, scaling=FALSE)
round(beta,2)

Output:
[1]  0.3
\end{verbatim}

In addition,to show that the Blomqvist's beta is invariate under
strictly increasing transformations , \(\beta=0.3\) between the
\emph{Coverage} rating variable in logarithmic millions of dollars and
\emph{Claim} amount variable in dollars.

\subsection*{A.2. Nonparametric Approach Using Spearman Correlation with
Tied
Ranks}\label{a.2.-nonparametric-approach-using-spearman-correlation-with-tied-ranks}
\addcontentsline{toc}{subsection}{A.2. Nonparametric Approach Using
Spearman Correlation with Tied Ranks}

For the first variable, the average rank of observations in the \(s\)th
row is

\begin{equation*}
r_{1s} = n_{m_1*}+ \cdots+ n_{s-1,*}+ \frac{1}{2} \left(1+ n_{s*}\right)
\end{equation*}

and similarly
\(r_{2t} = \frac{1}{2} \left[(n_{*m_1}+ \cdots+ n_{*,s-1}+1)+ (n_{*m_1}+ \ldots+ n_{*s})\right]\).
With this, we have Spearman's rho with tied rank is

\begin{equation*}
\hat{\rho}_S = \frac{\sum_{s=m_1}^{m_2} \sum_{t=m_1}^{m_2} n_{st}(r_{1s} - \bar{r})(r_{2t} - \bar{r})}
{\left[\sum_{s=m_1}^{m_2}n_{s*}(r_{1s} - \bar{r})^2 \sum_{t=m_1}^{m_2} n_{*t}(r_{2t} - \bar{r})^2
\right]^2}
\end{equation*}

where the average rank is \(\bar{r} = (n+1)/2\).

Click to Show Proof for Special Case: Binary Data.

\hypertarget{display.Thry.2}{}
Special Case: Binary Data. Here, \(m_1=0\) and \(m_2=1\). For the first
variable ranks, we have \(r_{10} = (1+n_{0+})/2\) and
\(r_{11} = (n_{0+}+1+n)/2\). Thus, \(r_{10} -\bar{r}= (n_{0+}-n)/2\) and
\(r_{11}-\bar{r} = n_{0+}/2\). This means that we have
\(\sum_{s=0}^{1}n_{s+}(r_{1s} - \bar{r})^2 = n (n-n_{0+})n_{0+}/4\) and
similarly for the second variable. For the numerator, we have

\begin{eqnarray*}
\sum_{s=0}^{1}  \sum_{t=0}^{1} && n_{st}(r_{1s} - \bar{r})(r_{2t} - \bar{r})\\
&=& n_{00} \frac{n_{0+}-n}{2} \frac{n_{+0}-n}{2}
+n_{01} \frac{n_{0+}-n}{2} \frac{n_{+0}}{2}
+n_{10} \frac{n_{0+}}{2} \frac{n_{+0}-n}{2}
+n_{11} \frac{n_{0+}}{2} \frac{n_{+0}}{2} \\
&=& \frac{1}{4}(n_{00} (n_{0+}-n) (n_{+0}-n)
+(n_{0+}-n_{00}) (n_{0+}-n)n_{+0} \\
&&  ~ ~ ~ +(n_{+0}-n_{00})  n_{0+}(n_{+0}-n)
+(n-n_{+0}-n_{0+}+n_{00}) n_{0+}n_{+0} ) \\
&=& \frac{1}{4}(n_{00} n^2
- n_{0+} (n_{0+}-n)n_{+0} \\
&& ~ ~ ~ +n_{+0}  n_{0+}(n_{+0}-n)
+(n-n_{+0}-n_{0+}) n_{0+}n_{+0} ) \\
&=& \frac{1}{4}(n_{00} n^2
- n_{0+}n_{+0} (n_{0+}-n +n_{+0}-n
+n-n_{+0}-n_{0+}) \\
&=& \frac{n}{4}(n n_{00} - n_{0+}n_{+0}) .
\end{eqnarray*}

This yields

\begin{eqnarray*}
\hat{\rho}_S &=& \frac{n(n n_{00} - n_{0+}n_{+0})}
{4\sqrt{(n (n-n_{0+})n_{0+}/4)(n (n-n_{+0})n_{+0}/4)}} \\
&=& \frac{n n_{00} - n_{0+}n_{+0}}
{\sqrt{ n_{0+} n_{+0}(n-n_{0+}) (n-n_{+0})}} \\
&=& \frac{n_{00} - n (1-\hat{\pi}_X)(1- \hat{\pi}_Y)}
{\sqrt{\hat{\pi}_X(1-\hat{\pi}_X)\hat{\pi}_Y(1-\hat{\pi}_Y) }}
\end{eqnarray*}

where \(\hat{\pi}_X = (n-n_{0+})/n\) and similarly for \(\hat{\pi}_Y\).
Note that this is same form as the Pearson measure. From this, we see
that the joint count \(n_{00}\) drives this association measure.

\bigskip

You can obtain the ties-corrected Spearman correlation statistic \(r_S\)
using the \texttt{cor()} function in \texttt{R} and selecting the
\texttt{spearman} method. From below \(\hat{\rho}_S=-0.09\)

R Code for Ties-corrected Spearman Correlation

\hypertarget{display.spearT.2}{}
\begin{verbatim}
rs_ties<-cor(AlarmCredit,NoClaimCredit, method = c("spearman"))
round(rs_ties,2)

Output:
[1] -0.09
\end{verbatim}

\chapter{Appendix A: Review of Statistical Inference}\label{C:AppA}

\emph{Chapter preview}. The appendix gives an overview of concepts and
methods related to statistical inference on the population of interest,
using a random sample of observations from the population. In the
appendix, Section \ref{S:AppA:BASIC} introduces the basic concepts
related to the population and the sample used for making the inference.
Section \ref{S:AppA:PE} presents the commonly used methods for point
estimation of population characteristics. Section \ref{S:AppA:IE}
demonstrates interval estimation that takes into consideration the
uncertainty in the estimation, due to use of a random sample from the
population. Section \ref{S:AppA:HT} introduces the concept of hypothesis
testing for the purpose of variable and model selection.

\section{Basic Concepts}\label{S:AppA:BASIC}

In this section, you learn the following concepts related to statistical
inference.

\begin{itemize}
\tightlist
\item
  Random sampling from a population that can be summarized using a list
  of items or individuals within the population
\item
  Sampling distributions that characterize the distributions of possible
  outcomes for a statistic calculated from a random sample
\item
  The central limit theorem that guides the distribution of the mean of
  a random sample from the population
\end{itemize}

\textbf{Statistical inference} is the process of making conclusions on
the characteristics of a large set of items/individuals (i.e., the
\textbf{population}), using a representative set of data (e.g., a
\textbf{random sample}) from a list of items or individuals from the
population that can be sampled. While the process has a broad spectrum
of applications in various areas including science, engineering, health,
social, and economic fields, statistical inference is important to
insurance companies that use data from their existing policy holders in
order to make inference on the characteristics (e.g., risk profiles) of
a specific segment of target customers (i.e., the population) whom the
insurance companies do not directly observe.

Show An Empirical Example Using the Wisconsin Property Fund

\hypertarget{EXM:S1:SI}{}
\textbf{Example -- Wisconsin Property Fund.} Assume there are 1,377
\emph{individual} claims from the 2010 experience.

\begin{longtable}[]{@{}rrrrrrrr@{}}
\toprule
& Minimum & First Quartile & Median & Mean & Third Quartile & Maximum &
Standard Deviation\tabularnewline
\midrule
\endhead
Claims & 1 & 788 & 2,250 & 26,620 & 6,171 & 12,920,000 &
368,030\tabularnewline
Logarithmic Claims & 0 & 6.670 & 7.719 & 7.804 & 8.728 & 16.370 &
1.683\tabularnewline
\bottomrule
\end{longtable}

\begin{Shaded}
\begin{Highlighting}[]
\NormalTok{ClaimLev <-}\StringTok{ }\KeywordTok{read.csv}\NormalTok{(}\StringTok{"Data/CLAIMLEVEL.csv"}\NormalTok{, }\DataTypeTok{header=}\OtherTok{TRUE}\NormalTok{)}
\NormalTok{ClaimLevBC10<-}\KeywordTok{subset}\NormalTok{(ClaimLev,Year}\OperatorTok{==}\DecValTok{2010}\NormalTok{); }
\KeywordTok{cat}\NormalTok{(}\StringTok{"Sample size: "}\NormalTok{, }\KeywordTok{nrow}\NormalTok{(ClaimLevBC10), }\StringTok{"}\CharTok{\textbackslash{}n}\StringTok{"}\NormalTok{)}
\KeywordTok{par}\NormalTok{(}\DataTypeTok{mfrow=}\KeywordTok{c}\NormalTok{(}\DecValTok{1}\NormalTok{, }\DecValTok{2}\NormalTok{))}
\KeywordTok{hist}\NormalTok{(ClaimLevBC10}\OperatorTok{$}\NormalTok{Claim, }\DataTypeTok{main=}\StringTok{""}\NormalTok{, }\DataTypeTok{xlab=}\StringTok{"Claims"}\NormalTok{)}
\KeywordTok{hist}\NormalTok{(}\KeywordTok{log}\NormalTok{(ClaimLevBC10}\OperatorTok{$}\NormalTok{Claim), }\DataTypeTok{main=}\StringTok{""}\NormalTok{, }\DataTypeTok{xlab=}\StringTok{"Logarithmic Claims"}\NormalTok{)}
\end{Highlighting}
\end{Shaded}

\begin{figure}

{\centering \includegraphics[width=0.8\linewidth]{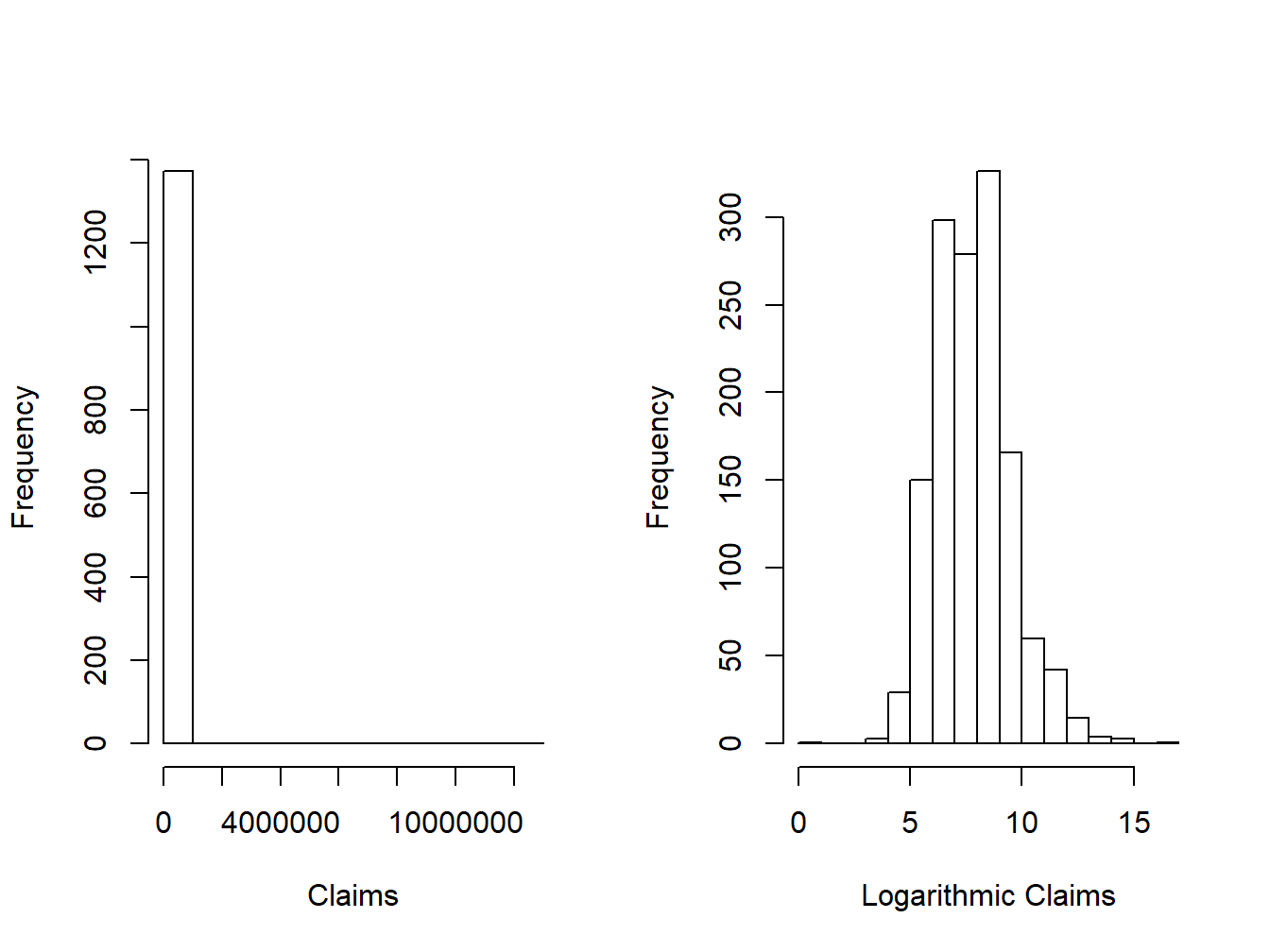}

}

\caption{Distribution of Claims}\label{fig:ClaimDistn1}
\end{figure}

\begin{verbatim}
## Sample size:  1377
\end{verbatim}

Using the 2010 claim experience (the sample), the Wisconsin Property
Fund may be interested in assessing the severity of all claims that
could potentially occur, such as 2010, 2011, and so forth (the
population). This process is important in the contexts of ratemaking or
claim predictive modeling. In order for such inference to be valid, we
need to assume that

\begin{itemize}
\tightlist
\item
  the set of 2010 claims is a \emph{random sample} that is
  representative of the population,
\item
  the \emph{sampling distribution} of the average claim amount can be
  estimated, so that we can quantify the bias and uncertainty in the
  esitmation due to use of a finite sample.
\end{itemize}

\subsection{Random Sampling}\label{random-sampling}

In statistics, a sampling \textbf{error} occurs when the
\textbf{sampling frame}, the list from which the sample is drawn, is not
an adequate approximation of the population of interest. A sample must
be a representative subset of a population, or universe, of interest. If
the sample is not representative, taking a larger sample does not
eliminate bias, as the same mistake is repeated over again and again.
Thus, we introduce the concept for random sampling that gives rise to a
simple \textbf{random sample} that is representative of the population.

We assume that the random variable \(X\) represents a draw from a
population with a distribution function \(F(\cdot)\) with mean
\(\mathrm{E}[X]=\mu\) and variance
\(\mathrm{Var}[X]=\mathrm{E}[(X-\mu)^2]\), where \(E(\cdot)\) denotes
the expectation of a random variable. In \textbf{random sampling}, we
make a total of \(n\) such draws represented by \(X_1, \ldots, X_n\),
each unrelated to one another (i.e., \emph{statistically independent}).
We refer to \(X_1, \ldots, X_n\) as a \textbf{random sample} (\emph{with
replacement}) from \(F(\cdot)\), taking either a parametric or
nonparametric form. Alternatively, we may say that \(X_1, \ldots, X_n\)
are identically and independently distributed (\emph{iid}) with
distribution function \(F(\cdot)\).

\subsection{Sampling Distribution}\label{sampling-distribution}

Using the random sample \(X_1, \ldots, X_n\), we are interested in
making a conclusion on a specific attribute of the population
distribution \(F(\cdot)\). For example, we may be interested in making
an inference on the population mean, denoted \(\mu\). It is natural to
think of the \textbf{sample mean}, \(\bar{X}=\sum_{i=1}^nX_i\), as an
estimate of the population mean \(\mu\). We call the sample mean as a
\textbf{statistic} calculated from the random sample
\(X_1, \ldots, X_n\). Other commonly used summary statistics include
sample standard deviation and sample quantiles.

When using a statistic (e.g., the sample mean \(\bar{X}\)) to make
statistical inference on the population attribute (e.g., population mean
\(\mu\)), the quality of inference is determined by the bias and
uncertainty in the estimation, owing to the use of a sample in place of
the population. Hence, it is important to study the distribution of a
statistic that quantifies the bias and variability of the statistic. In
particular, the distribution of the sample mean, \(\bar{X}\) (or any
other statistic), is called the \textbf{sampling distribution}. The
sampling distribution depends on the sampling process, the statistic,
the sample size \(n\) and the population distribution \(F(\cdot)\). The
central limit theorem gives the large-sample (sampling) distribution of
the sample mean under certain conditions.

\subsection{Central Limit Theorem}\label{central-limit-theorem}

In statistics, there are variations of the central limit theorem (CLT)
ensuring that, under certain conditions, the sample mean will approach
the population mean with its sampling distribution approaching the
normal distribution as the sample size goes to infinity. We give the
Lindeberg--Levy CLT that establishes the asymptotic sampling
distribution of the sample mean \(\bar{X}\) calculated using a random
sample from a universe population having a distribution \(F(\cdot)\).

\textbf{Lindeberg--Levy CLT.} Let \(X_1, \ldots, X_n\) be a random
sample from a population distribution \(F(\cdot)\) with mean \(\mu\) and
variance \(\sigma^2<\infty\). The difference between the sample mean
\(\bar{X}\) and \(\mu\), when multiplied by \(\sqrt{n}\), converges in
distribution to a normal distribution as the sample size goes to
infinity. That is,
\[\sqrt{n}(\bar{X}-\mu)\xrightarrow[]{d}N(0,\sigma).\]

Note that the CLT does not require a parametric form for \(F(\cdot)\).
Based on the CLT, we may perform statistical inference on the population
mean (we \emph{infer}, not \emph{deduce}). The types of inference we may
perform include \textbf{estimation} of the population,
\textbf{hypothesis testing} on whether a null statement is true, and
\textbf{prediction} of future samples from the population.

\section{Point Estimation and Properties}\label{S:AppA:PE}

In this section, you learn how to

\begin{itemize}
\tightlist
\item
  estimate population parameters using method of moments estimation
\item
  estimate population parameters based on maximum likelihood estimation
\end{itemize}

The population distribution function \(F(\cdot)\) can usually be
characterized by a limited (finite) number of terms called
\textbf{parameters}, in which case we refer to the distribution as a
\textbf{parametric distribution}. In contrast, in \textbf{nonparametric}
analysis, the attributes of the sampling distribution are not limited to
a small number of parameters.

For obtaining the population characteristics, there are different
attributes related to the population distribution \(F(\cdot)\). Such
measures include the mean, median, percentiles (i.e., 95th percentile),
and standard deviation. Because these summary measures do not depend on
a specific parametric reference, they are \textbf{nonparametric} summary
measures.

In \textbf{parametric} analysis, on the other hand, we may assume
specific families of distributions with specific parameters. For
example, people usually think of logarithm of claim amounts to be
normally distributed with mean \(\mu\) and standard deviation
\(\sigma\). That is, we assume that the claims have a \emph{lognormal}
distribution with parameters \(\mu\) and \(\sigma\). Alternatively,
insurance companies commonly assume that claim severity follows a gamma
distribution with a shape parameter \(\alpha\) and a scale parameter
\(\theta\). Here, the normal, lognormal, and gamma distributions are
examples of parametric distributions. In the above examples, the
quantities of \(\mu\), \(\sigma\), \(\alpha\), and \(\theta\) are known
as \emph{parameters}. For a given parametric distribution family, the
distribution is uniquely determined by the values of the parameters.

One often uses \(\theta\) to denote a summary attribute of the
population. In parametric models, \(\theta\) can be a parameter or a
function of parameters from a distribution such as the normal mean and
variance parameters. In nonparametric analysis, it can take a form of a
nonparametric summary such as the population mean or standard deviation.
Let \(\hat{\theta} =\hat{\theta}(X_1, \ldots, X_n)\) be a function of
the sample that provides a proxy, or an \textbf{estimate}, of
\(\theta\). It is referred to as a \textbf{statistic}, a function of the
sample \(X_1, \ldots, X_n\).

Show Wisconsin Property Fund Example - Continued

\hypertarget{EXM:S1:PE}{}
\textbf{Example -- Wisconsin Property Fund.} The sample mean 7.804 and
the sample standard deviation 1.683 can be either deemed as
nonparametric estimates of the population mean and standard deviation,
or as parametric estimates of \(\mu\) and \(\sigma\) of the normal
distribution concerning the logarithmic claims. Using results from the
lognormal distribution, we may estimate the expected claim, the
lognormal mean, as 10,106.8 ( \(=\exp(7.804+1.683^2/2)\) ).

For the Wisconsin Property Fund data, we may denote \(\hat{\mu} =7.804\)
and \(\hat{\sigma} = 1.683\), with the hat notation denoting an
\textbf{estimate} of the parameter based on the sample. In particular,
such an estimate is referred to as a \textbf{point estimate}, a single
approximation of the corresponding parameter. For point estimation, we
introduce the two commonly used methods called the method of moments
estimation and maximum likelihood estimation.

\subsection{Method of Moments
Estimation}\label{method-of-moments-estimation}

Before defining the method of moments estimation, we define the the
concept of \textbf{moments}. Moments are population attributes that
characterize the distribution function \(F(\cdot)\). Given a random draw
\(X\) from \(F(\cdot)\), the expectation \(\mu_k=\mathrm{E}[X^k]\) is
called the \textbf{\(k\)th moment} of \(X\), \(k=1,2,3,\cdots\). For
example, the population mean \(\mu\) is the \emph{first} moment.
Furthermore, the expectation \(\mathrm{E}[(X-\mu)^k]\) is called a
\textbf{\(k\)th central moment}. Thus, the variance is the second
central moment.

Using the random sample \(X_1, \ldots, X_n\), we may construct the
corresponding sample moment, \(\hat{\mu}_k=(1/n)\sum_{i=1}^nX_i^k\), for
estimating the population attribute \(\mu_k\). For example, we have used
the sample mean \(\bar{X}\) as an estimator for the population mean
\(\mu\). Similarly, the second central moment can be estimated as
\((1/n)\sum_{i=1}^n(X_i-\bar{X})^2\). Without assuming a parametric form
for \(F(\cdot)\), the sample moments constitute nonparametric estimates
of the corresponding population attributes. Such an estimator based on
matching of the corresponding sample and population moments is called a
\textbf{method of moments estimator} (MME).

While the MME works naturally in a nonparametric model, it can be used
to estimate parameters when a specific parametric family of distribution
is assumed for \(F(\cdot)\). Denote by
\(\boldsymbol{\theta}=(\theta_1,\cdots,\theta_m)\) the vector of
parameters corresponding to a parametric distribution \(F(\cdot)\).
Given a distribution family, we commonly know the relationships between
the parameters and the moments. In particular, we know the specific
forms of the functions \(h_1(\cdot),h_2(\cdot),\cdots,h_m(\cdot)\) such
that
\(\mu_1=h_1(\boldsymbol{\theta}),\,\mu_2=h_2(\boldsymbol{\theta}),\,\cdots,\,\mu_m=h_m(\boldsymbol{\theta})\).
Given the MME \(\hat{\mu}_1, \ldots, \hat{\mu}_m\) from the random
sample, the MME of the parameters
\(\hat{\theta}_1,\cdots,\hat{\theta}_m\) can be obtained by solving the
equations of \[\hat{\mu}_1=h_1(\hat{\theta}_1,\cdots,\hat{\theta}_m);\]
\[\hat{\mu}_2=h_2(\hat{\theta}_1,\cdots,\hat{\theta}_m);\] \[\cdots\]
\[\hat{\mu}_m=h_m(\hat{\theta}_1,\cdots,\hat{\theta}_m).\]

Show Wisconsin Property Fund Example - Continued

\hypertarget{EXM:S1:MME}{}
\textbf{Example -- Wisconsin Property Fund.} Assume that the claims
follow a lognormal distribution, so that logarithmic claims follow a
normal distribution. Specifically, assume \(\ln(X)\) has a normal
distribution with mean \(\mu\) and variance \(\sigma^2\), denoted as
\(\ln(X) \sim N(\mu, \sigma^2)\). It is straightforward that the MME
\(\hat{\mu}=\bar{X}\) and
\(\hat{\sigma}=\sqrt{(1/n)\sum_{i=1}^n(X_i-\bar{X})^2}\). For the
Wisconsin Property Fund example, the method of moments estimates are
\(\hat{\mu} =7.804\) and \(\hat{\sigma} = 1.683\).

\subsection{Maximum Likelihood
Estimation}\label{maximum-likelihood-estimation}

When \(F(\cdot)\) takes a parametric form, the maximum likelihood method
is widely used for estimating the population parameters
\(\boldsymbol{\theta}\). Maximum likelihood estimation is based on the
likelihood function, a function of the parameters given the observed
sample. Denote by \(f(x_i|\boldsymbol{\theta})\) the probability
function of \(X_i\) evaluated at \(X_i=x_i\) \((i=1,2,\cdots,n)\), the
probability mass function in the case of a discrete \(X\) and the
probability density function in the case of a continuous \(X\). Then the
\textbf{likelihood function} of \(\boldsymbol{\theta}\) associated with
the observation \((X_1,X_2,\cdots,X_n)=(x_1,x_2,\cdots,x_n)=\mathbf{x}\)
can be written as
\[L(\boldsymbol{\theta}|\mathbf{x})=\prod_{i=1}^nf(x_i|\boldsymbol{\theta}),\]
with the corresponding \textbf{log-likelihood function} given by
\[l(\boldsymbol{\theta}|\mathbf{x})=\ln(L(\boldsymbol{\theta}|\mathbf{x}))=\sum_{i=1}^n\ln f(x_i|\boldsymbol{\theta}).\]
The maximum likelihood estimator (MLE) of \(\boldsymbol{\theta}\) is the
set of values of \(\boldsymbol{\theta}\) that maximize the likelihood
function (log-likelihood function), given the observed sample. That is,
the MLE \(\hat{\boldsymbol{\theta}}\) can be written as
\[\hat{\boldsymbol{\theta}}={\mbox{argmax}}_{\boldsymbol{\theta}\in\Theta}l(\boldsymbol{\theta}|\mathbf{x}),\]
where \(\Theta\) is the parameter space of \(\boldsymbol{\theta}\), and
\({\mbox{argmax}}_{\boldsymbol{\theta}\in\Theta}l(\boldsymbol{\theta}|\mathbf{x})\)
is defined as the value of \(\boldsymbol{\theta}\) at which the function
\(l(\boldsymbol{\theta}|\mathbf{x})\) reachs its maximum.

Given the analytical form of the likelihood function, the MLE can be
obtained by taking the first derivative of the log-likelihood function
with respect to \(\boldsymbol{\theta}\), and setting the values of the
partial derivatives to zero. That is, the MLE are the solutions of the
equations of
\[\frac{\partial l(\hat{\boldsymbol{\theta}}|\mathbf{x})}{\partial\hat{\theta}_1}=0;\]
\[\frac{\partial l(\hat{\boldsymbol{\theta}}|\mathbf{x})}{\partial\hat{\theta}_2}=0;\]
\[\cdots\]
\[\frac{\partial l(\hat{\boldsymbol{\theta}}|\mathbf{x})}{\partial\hat{\theta}_m}=0,\]
provided that the second partial derivatives are negative.

For parametric models, the MLE of the parameters can be obtained either
analytically (e.g., in the case of normal distributions and linear
estimators), or numerically through iterative algorithms such as the
Newton-Raphson method and its adaptive versions (e.g., in the case of
generalized linear models with a non-normal response variable).

\textbf{Normal distribution.} Assume \((X_1,X_2,\cdots,X_n)\) to be a
random sample from the normal distribution \(N(\mu, \sigma^2)\). With an
observed sample \((X_1,X_2,\cdots,X_n)=(x_1,x_2,\cdots,x_n)\), we can
write the likelihood function of \(\mu,\sigma^2\) as
\[L(\mu,\sigma^2)=\prod_{i=1}^n\left[\frac{1}{\sqrt{2\pi\sigma^2}}e^{-\frac{\left(x_i-\mu\right)^2}{2\sigma^2}}\right],\]
with the corresponding log-likelihood function given by
\[l(\mu,\sigma^2)=-\frac{n}{2}[\ln(2\pi)+\ln(\sigma^2)]-\frac{1}{2\sigma^2}\sum_{i=1}^n\left(x_i-\mu\right)^2.\]

By solving
\[\frac{\partial l(\hat{\mu},\sigma^2)}{\partial \hat{\mu}}=0,\] we
obtain \(\hat{\mu}=\bar{x}=(1/n)\sum_{i=1}^nx_i\). It is straightforward
to verify that
\(\frac{\partial l^2(\hat{\mu},\sigma^2)}{\partial \hat{\mu}^2}\left|_{\hat{\mu}=\bar{x}}\right.<0\).
Since this works for arbitrary \(x\), \(\hat{\mu}=\bar{X}\) is the MLE
of \(\mu\). Similarly, by solving
\[\frac{\partial l(\mu,\hat{\sigma}^2)}{\partial \hat{\sigma}^2}=0,\] we
obtain \(\hat{\sigma}^2=(1/n)\sum_{i=1}^n(x_i-\mu)^2\). Further
replacing \(\mu\) by \(\hat{\mu}\), we derive the MLE of \(\sigma^2\) as
\(\hat{\sigma}^2=(1/n)\sum_{i=1}^n(X_i-\bar{X})^2\).

Hence, the sample mean \(\bar{X}\) and \(\hat{\sigma}^2\) are both the
\emph{MME} and \emph{MLE} for the mean \(\mu\) and variance
\(\sigma^2\), under a normal population distribution \(F(\cdot)\). More
details regarding the properties of the likelihood function, and the
derivation of MLE under parametric distributions other than the normal
distribution are given in Appendix Chapter \ref{C:AppB}.

\section{Interval Estimation}\label{S:AppA:IE}

In this section, you learn how to

\begin{itemize}
\tightlist
\item
  derive the exact sampling distribution of the MLE of the normal mean
\item
  obtain the large-sample approximation of the sampling distribution
  using the large sample properties of the MLE
\item
  construct a confidence interval of a parameter based on the large
  sample properties of the MLE
\end{itemize}

Now that we have introduced the MME and MLE, we may perform the first
type of statistical inference, \textbf{interval estimation} that
quantifies the uncertainty resulting from the use of a finite sample. By
deriving the sampling distribution of MLE, we can estimate an interval
(a confidence interval) for the parameter. Under the frequentist
approach (e.g., that based on maximum likelihood estimation), the
confidence intervals generated from the same random sampling frame will
cover the true value the majority of times (e.g., 95\% of the times), if
we repeat the sampling process and re-calculate the interval over and
over again. Such a process requires the derivation of the sampling
distribution for the MLE.

\subsection{Exact Distribution for Normal Sample
Mean}\label{S:AppA:IE:ED}

Due to the \textbf{additivity} property of the normal distribution
(i.e., a sum of normal random variables that follows a multivariate
normal distribution still follows a normal distribution) and that the
normal distribution belongs to the \textbf{location--scale family}
(i.e., a location and/or scale transformation of a normal random
variable has a normal distribution), the sample mean \(\bar{X}\) of a
random sample from a normal \(F(\cdot)\) has a normal sampling
distribution for any finite \(n\). Given
\(X_i\sim^{iid} N(\mu,\sigma^2)\), \(i=1,\dots,n\), the MLE of \(\mu\)
has an exact distribution
\[\bar{X}\sim N\left(\mu,\frac{\sigma^2}{n}\right).\] Hence, the sample
mean is an unbiased estimator of \(\mu\). In addition, the uncertainty
in the estimation can be quantified by its variance \(\sigma^2/n\), that
decreases with the sample size \(n\). When the sample size goes to
infinity, the sample mean will approach a single mass at the true value.

\subsection{Large-sample Properties of
MLE}\label{large-sample-properties-of-mle}

For the MLE of the mean parameter and any other parameters of other
parametric distribution families, however, we usually cannot derive an
exact sampling distribution for finite samples. Fortunately, when the
sample size is sufficiently large, MLEs can be approximated by a normal
distribution. Due to the general maximum likelihood theory, the MLE has
some nice large-sample properties.

\begin{itemize}
\item
  The MLE \(\hat{\theta}\) of a parameter \(\theta\), is a
  \textbf{consistent} estimator. That is, \(\hat{\theta}\) converges in
  probability to the true value \(\theta\), as the sample size \(n\)
  goes to infinity.
\item
  The MLE has the \textbf{asymptotic normality} property, meaning that
  the estimator will converge in distribution to a normal distribution
  centered around the true value, when the sample size goes to infinity.
  Namely,
  \[\sqrt{n}(\hat{\theta}-\theta)\rightarrow_d N\left(0,\,V\right),\quad \mbox{as}\quad n\rightarrow \infty,\]
  where \(V\) is the inverse of the Fisher Information. Hence, the MLE
  \(\hat{\theta}\) approximately follows a normal distribution with mean
  \(\theta\) and variance \(V/n\), when the sample size is large.
\item
  The MLE is \textbf{efficient}, meaning that it has the smallest
  asymptotic variance \(V\), commonly referred to as the
  \textbf{Cramer--Rao lower bound}. In particular, the Cramer--Rao lower
  bound is the inverse of the Fisher information defined as
  \(\mathcal{I}(\theta)=-\mathrm{E}(\partial^2\ln f(X;\theta)/\partial \theta^2)\).
  Hence, \(\mathrm{Var}(\hat{\theta})\) can be estimated based on the
  observed Fisher information that can be written as
  \(-\sum_{i=1}^n \partial^2\ln f(X_i;\theta)/\partial \theta^2\).
\end{itemize}

For many parametric distributions, the Fisher information may be derived
analytically for the MLE of parameters. For more sophisticated
parametric models, the Fisher information can be evaluated numerically
using numerical integration for continuous distributions, or numerical
summation for discrete distributions.

\subsection{Confidence Interval}\label{confidence-interval}

Given that the MLE \(\hat{\theta}\) has either an exact or an
approximate normal distribution with mean \(\theta\) and variance
\(\mathrm{Var}(\hat{\theta})\), we may take the square root of the
variance and plug-in the estimate to define
\(se(\hat{\theta}) = \sqrt{\mathrm{Var}(\hat{\theta})}\). A
\textbf{standard error} is an estimated standard deviation that
quantifies the uncertainty in the estimation resulting from the use of a
finite sample. Under some regularity conditions governing the population
distribution, we may establish that the statistic
\[\frac{\hat{\theta}-\theta}{se(\hat{\theta})}\] converges in
distribution to a Student-\(t\) distribution with degrees of freedom (a
parameter of the distribution) \({n-p}\), where \(p\) is the number of
parameters in the model other than the variance. For example, for the
normal distribution case, we have \(p=1\) for the parameter \(\mu\); for
a linear regression model with an independent variable, we have \(p=2\)
for the parameters of the intercept and the independent variable. Denote
by \(t_{n-p}(1-\alpha/2)\) the \(100\times(1-\alpha/2)\)-th percentile
of the Student-\(t\) distribution that satisfies
\(\Pr\left[t< t_{n-p}\left(1-{\alpha}/{2}\right) \right]= 1-{\alpha}/{2}\).
We have,
\[\Pr\left[-t_{n-p}\left(1-\frac{\alpha}{2}\right)<\frac{\hat{\theta}-\theta}{se(\hat{\theta})}< t_{n-p}\left(1-\frac{\alpha}{2}\right) \right]= 1-{\alpha},\]
from which we can derive a \textbf{confidence interval} for \(\theta\).
From the above equation we can derive a pair of statistics,
\(\hat{\theta}_1\) and \(\hat{\theta}_2\), that provide an interval of
the form \([\hat{\theta}_1, \hat{\theta}_2]\). This interval is a
\(1-\alpha\) confidence interval for \(\theta\) such that
\(\Pr\left(\hat{\theta}_1 \le \theta \le \hat{\theta}_2\right) = 1-\alpha,\)
where the probability \(1-\alpha\) is referred to as the
\textbf{confidence level}. Note that the above confidence interval is
not valid for small samples, except for the case of the normal mean.

\textbf{Normal distribution.} For the normal population mean \(\mu\),
the MLE has an exact sampling distribution
\(\bar{X}\sim N(\mu,\sigma/\sqrt{n})\), in which we can estimate
\(se(\hat{\theta})\) by \(\hat{\sigma}/\sqrt{n}\). Based on the
\textbf{Cochran's theorem}, the resulting statistic has an exact
Student-\(t\) distribution with degrees of freedom \(n-1\). Hence, we
can derive the lower and upper bounds of the confidence interval as
\[\hat{\mu}_1 = \hat{\mu} - t_{n-1}\left(1-\frac{\alpha}{2}\right)\frac{ \hat{\sigma}}{\sqrt{n}}\]
and
\[\hat{\mu}_2 = \hat{\mu} + t_{n-1}\left(1-\frac{\alpha}{2}\right)\frac{ \hat{\sigma}}{\sqrt{n}}.\]
When \(\alpha = 0.05\), \(t_{n-1}(1-\alpha/2) \approx 1.96\) for large
values of \(n\). Based on the Cochran's theorem, the confidence interval
is valid regardless of the sample size.

\begin{center}\rule{0.5\linewidth}{\linethickness}\end{center}

Show Wisconsin Property Fund Example - Continued

\hypertarget{EXM:S1:CI}{}
\textbf{Example -- Wisconsin Property Fund.} For the lognormal claim
model, (7.715235, 7.893208) is a 95\% confidence interval for \(\mu\).

More details regarding interval estimation based the MLE of other
parameters and distribution families are given in Appendix Chapter
\ref{C:AppC}.

\begin{center}\rule{0.5\linewidth}{\linethickness}\end{center}

\section{Hypothesis Testing}\label{S:AppA:HT}

In this section, you learn how to

\begin{itemize}
\tightlist
\item
  understand the basic concepts in hypothesis testing including the
  level of significance and the power of a test
\item
  perform hypothesis testing such as a Student-\(t\) test based on the
  properties of the MLE
\item
  construct a likelihood ratio test for a single parameter or multiple
  parameters from the same statistical model
\item
  use information criteria such as the Akaike's information criterion or
  the Bayesian information criterion to perform model selection
\end{itemize}

For the parameter(s) \(\boldsymbol{\theta}\) from a parametric
distribution, an alternative type of statistical inference is called
\textbf{hypothesis tesing} that verifies whether a hypothesis regarding
the parameter(s) is true, under a given probability called the
\textbf{level of significance} \(\alpha\) (e.g., 5\%). In hypothesis
testing, we reject the null hypothesis, a restrictive statement
concerning the parameter(s), if the probability of observing a random
sample as extremal as the observed one is smaller than \(\alpha\), if
the null hypothesis were true.

\subsection{Basic Concepts}\label{basic-concepts}

In a statistical test, we are usually interested in testing whether a
statement regarding some parameter(s), a \textbf{null hypothesis}
(denoted \(H_0\)), is true given the observed data. The null hypothesis
can take a general form \(H_0:\theta\in\Theta_0\), where \(\Theta_0\) is
a subset of the parameter space \(\Theta\) of \(\theta\) that may
contain multiple parameters. For the case with a single parameter
\(\theta\), the null hypothesis usually takes either the form
\(H_0:\theta=\theta_0\) or \(H_0:\theta\leq\theta_0\). The opposite of
the null hypothesis is called the \textbf{alternative hypothesis} that
can be written as \(H_a:\theta\neq\theta_0\) or \(H_a:\theta>\theta_0\).
The statistical test on \(H_0:\theta=\theta_0\) is called a
\textbf{two-sided} as the alternative hypothesis contains two
ineqalities of \(H_a:\theta<\theta_0\) or \(\theta>\theta_0\). In
contrast, the statistical test on either \(H_0:\theta\leq\theta_0\) or
\(H_0:\theta\geq\theta_0\) is called a \textbf{one-sided} test.

A statistical test is usually constructed based on a statistic \(T\) and
its exact or large-sample distribution. The test typically rejects a
two-sided test when either \(T > c_1\) or \(T < c_2\), where the two
constants \(c_1\) and \(c_2\) are obtained based on the sampling
distribution of \(T\) at a probability level \(\alpha\) called the
\textbf{level of significance}. In particular, the level of significance
\(\alpha\) satisfies
\[\alpha=\Pr(\mbox{reject }H_0|H_0\mbox{ is true}),\] meaning that if
the null hypothesis were true, we would reject the null hypothesis only
5\% of the times, if we repeat the sampling process and perform the test
over and over again.

Thus, the level of significance is the probability of making a
\textbf{type I error} (error of the first kind), the error of
incorrectly rejecting a true null hypothesis. For this reason, the level
of significance \(\alpha\) is also referred to as the type I error rate.
Another type of error we may make in hypothesis testing is the
\textbf{type II error} (error of the second kind), the error of
incorrectly accepting a false null hypothesis. Similarly, we can define
the \textbf{type II error rate} as the probability of not rejecting
(accepting) a null hypothesis given that it is not true. That is, the
type II error rate is given by
\[\Pr(\mbox{accept }H_0|H_0\mbox{ is false}).\] Another important
quantity concerning the quality of the statistical test is called the
\textbf{power} of the test \(\beta\), defined as the probability of
rejecting a false null hypothesis. The mathematical definition of the
power is \[\beta=\Pr(\mbox{reject }H_0|H_0\mbox{ is false}).\] Note that
the power of the test is typically calculated based on a specific
alternative value of \(\theta=\theta_a\), given a specific sampling
distribution and a given sample size. In real experimental studies,
people usually calculate the required sample size in order to choose a
sample size that will ensure a large chance of obtaining a statistically
significant test (i.e., with a prespecified statistical power such as
85\%).

\subsection{\texorpdfstring{Student-\(t\) test based on
MLE}{Student-t test based on MLE}}\label{student-t-test-based-on-mle}

Based on the results from Section \ref{S:AppA:IE:ED}, we can define a
Student \(t\) test for testing \(H_0:\theta=\theta_0\). In particular,
we define the test statistic as
\[t\text{-stat}=\frac{\hat{\theta}-\theta_0}{se(\hat{\theta})},\] which
has a large-sample distribution of a Student-\(t\) distribution with
degrees of freedom \({n-p}\), when the null hypothesis is true (i.e.,
when \(\theta=\theta_0\)).

For a given \textbf{level of significance} \(\alpha\), say 5\%, we
reject the null hypothesis if the event
\(t\text{-stat}<-t_{n-p}\left(1-{\alpha}/{2}\right)\) or
\(t\text{-stat}> t_{n-p}\left(1-{\alpha}/{2}\right)\) occurs (the
\textbf{rejection region}). Under the null hypothesis \(H_0\), we have
\[\Pr\left[t\text{-stat}<-t_{n-p}\left(1-\frac{\alpha}{2}\right)\right]=\Pr\left[t\text{-stat}> t_{n-p}\left(1-\frac{\alpha}{2}\right) \right]= \frac{\alpha}{2}.\]
In addition to the concept of rejection region, we may reject the test
based on the \(p\)\textbf{-value} defined as \(2\Pr(T>|t\text{-stat}|)\)
for the aforementioned two-sided test, where the random variable
\(T\sim T_{n-p}\). We reject the null hypothesis if \(p\)-value is
smaller than and equal to \(\alpha\). For a given sample, a \(p\)-value
is defined to be the smallest significance level for which the null
hypothesis would be rejected.

Similarly, we can construct a one-sided test for the null hypothesis
\(H_0:\theta\leq\theta_0\) (or \(H_0:\theta\geq\theta_0\)). Using the
same test statistic, we reject the null hypothesis when
\(t\text{-stat}> t_{n-p}\left(1-{\alpha}\right)\) (or
\(t\text{-stat}<- t_{n-p}\left(1-{\alpha}\right)\) for the test on
\(H_0:\theta\geq\theta_0\)). The corresponding \(p\)-value is defined as
\(\Pr(T>|t\text{-stat}|)\) (or \(\Pr(T<|t\text{-stat}|)\) for the test
on \(H_0:\theta\geq\theta_0\)). Note that the test is not valid for
small samples, except for the case of the test on the normal mean.

\textbf{One-sample \(t\) Test for Normal Mean.} For the test on the
normal mean of the form \(H_0:\mu=\mu_0\), \(H_0:\mu\leq\mu_0\) or
\(H_0:\mu\geq\mu_0\), we can define the test statistic as
\[t\text{-stat}=\frac{\bar{X}-\mu_0}{{\hat{\sigma}}/{\sqrt{n}}},\] for
which we have an exact sampling distribution
\(t\text{-stat}\sim T_{n-1}\) from the Cochran's theorem, with
\(T_{n-1}\) denoting a Student-\(t\) distribution with degrees of
freedom \(n-1\). According to the Cochran's theorem, the test is valid
for both small and large samples.

Show Wisconsin Property Fund Example - Continued

\hypertarget{EXM:S1:TST1}{}
\textbf{Example -- Wisconsin Property Fund.} Assume that mean
logarithmic claims have historically been approximately by
\(\mu_0 = \ln(5000)= 8.517\). We might want to use the 2010 data to
assess whether the mean of the distribution has changed (a two-sided
test), or whether it has increased (a one-sided test). Given the actual
2010 average \(\hat{\mu} =7.804\), we may use the one-sample \(t\) test
to assess whether this is a significant departure from \(\mu_0 = 8.517\)
(i.e., in testing \(H_0:\mu=8.517\)). The test statistic
\(t\text{-stat}=(8.517-7.804)/(1.683/\sqrt{1377}) = 15.72>t_{1376}\left(0.975\right)\).
Hence, we reject the two-sided test at \(\alpha=5\%\). Similarly, we
will reject the one-sided test at \(\alpha=5\%\).

Show Wisconsin Property Fund Example - Continued

\hypertarget{EXM:S1:TST2}{}
\textbf{Example -- Wisconsin Property Fund.} For numerical stability and
extensions to regression applications, statistical packages often work
with transformed versions of parameters. The following estimates are
from the \textbf{R} package \textbf{VGAM} (the function). More details
on the MLE of other distribution families are given in Appendix Chapter
\ref{C:AppC}.

\begin{longtable}[]{@{}rrrr@{}}
\toprule
Distribution & Parameter & Standard & \(t\)-stat\tabularnewline
& Estimate & Error &\tabularnewline
Gamma & 10.190 & 0.050 & 203.831\tabularnewline
& -1.236 & 0.030 & -41.180\tabularnewline
Lognormal & 7.804 & 0.045 & 172.089\tabularnewline
& 0.520 & 0.019 & 27.303\tabularnewline
Pareto & 7.733 & 0.093 & 82.853\tabularnewline
& -0.001 & 0.054 & -0.016\tabularnewline
GB2 & 2.831 & 1.000 & 2.832\tabularnewline
& 1.203 & 0.292 & 4.120\tabularnewline
& 6.329 & 0.390 & 16.220\tabularnewline
& 1.295 & 0.219 & 5.910\tabularnewline
\bottomrule
\end{longtable}

\subsection{Likelihood Ratio Test}\label{S:AppA:HT:LRT}

In the previous subsection, we have introduced the Student-\(t\) test on
a single parameter, based on the properties of the MLE. In this section,
we define an alternative test called the \textbf{likelihood ratio test}
(LRT). The LRT may be used to test multiple parameters from the same
statistical model.

Given the likelihood function \(L(\theta|\mathbf{x})\) and
\(\Theta_0 \subset \Theta\), the likelihood ratio test statistic for
testing \(H_0:\theta\in\Theta_0\) against \(H_a:\theta\notin\Theta_0\)
is given by
\[L=\frac{\sup_{\theta\in\Theta_0}L(\theta|\mathbf{x})}{\sup_{\theta\in\Theta}L(\theta|\mathbf{x})},\]
and that for testing \(H_0:\theta=\theta_0\) versis
\(H_a:\theta\neq\theta_0\) is
\[L=\frac{L(\theta_0|\mathbf{x})}{\sup_{\theta\in\Theta}L(\theta|\mathbf{x})}.\]
The LRT rejects the null hypothesis when \(L < c\), with the threshold
depending on the level of significance \(\alpha\), the sample size
\(n\), and the number of parameters in \(\theta\). Based on the
\textbf{Neyman--Pearson Lemma}, the LRT is the \textbf{uniformly most
powerful} (UMP) test for testing \(H_0:\theta=\theta_0\) versis
\(H_a:\theta=\theta_a\). That is, it provides the largest power
\(\beta\) for a given \(\alpha\) and a given alternative value
\(\theta_a\).

Based on the \textbf{Wilks's Theorem}, the likelihood ratio test
statistic \(-2\ln(L)\) converges in distribution to a Chi-square
distribution with the degree of freedom being the difference between the
dimensionality of the parameter spaces \(\Theta\) and \(\Theta_0\), when
the sample size goes to infinity and when the null model is nested
within the alternative model. That is, when the null model is a special
case of the alternative model containing a restricted sample space, we
may approximate \(c\) by \(\chi^2_{p_1 - p_2}(1-\alpha)\), the
\(100\times(1-\alpha)\) th percentile of the Chi-square distribution,
with \(p_1-p_2\) being the degrees of freedom, and \(p_1\) and \(p_2\)
being the numbers of parameters in the alternative and null models,
respectively. Note that the LRT is also a large-sample test that will
not be valid for small samples.

\subsection{Information Criteria}\label{S:AppA:HT:IC}

In real-life applications, the LRT has been commonly used for comparing
two nested models. The LRT approach as a model selection tool, however,
has two major drawbacks: 1) It typically requires the null model to be
nested within the alternative model; 2) models selected from the LRT
tends to provide in-sample over-fitting, leading to poor out-of-sample
prediction. In order to overcome these issues, model selection based on
information criteria, applicable to non-nested models while taking into
consideration the model complexity, is more widely used for model
selection. Here, we introduce the two most widely used criteria, the
Akaike's information criterion and the Bayesian information criterion.

In particular, the \textbf{Akaike's information criterion} (\(AIC\)) is
defined as \[AIC = -2\ln L(\hat{\boldsymbol \theta}) + 2p,\] where
\(\hat{\boldsymbol \theta}\) denotes the MLE of
\({\boldsymbol \theta}\), and \(p\) is the number of parameters in the
model. The additional term \(2 p\) represents a penalty for the
complexity of the model. That is, with the same maximized likelihood
function, the \(AIC\) favors model with less parameters. We note that
the \(AIC\) does not consider the impact from the sample size \(n\).

Alternatively, people use the \textbf{Bayesian information criterion}
(\(BIC\)) that takes into consideration the sample size. The \(BIC\) is
defined as \[BIC = -2\ln L(\hat{\boldsymbol \theta}) + p\,\ln(n).\] We
observe that the \(BIC\) generally puts a higher weight on the number of
parameters. With the same maximized likelihood function, the \(BIC\)
will suggest a more parsimonious model than the \(AIC\).

Show Wisconsin Property Fund Example - Continued

\hypertarget{EXM:S1:AIC}{}
\textbf{Example -- Wisconsin Property Fund.} Both the \(AIC\) and
\(BIC\) statistics suggest that the \emph{GB2} is the best fitting model
whereas gamma is the worst.

\begin{longtable}[]{@{}rrr@{}}
\toprule
Distribution & AIC & BIC\tabularnewline
\midrule
\endhead
Gamma & 28,305.2 & 28,315.6\tabularnewline
Lognormal & 26,837.7 & 26,848.2\tabularnewline
Pareto & 26,813.3 & 26,823.7\tabularnewline
GB2 & 26,768.1 & 26,789.0\tabularnewline
\bottomrule
\end{longtable}

In this graph,

\begin{itemize}
\item
  black represents actual (smoothed) logarithmic claims
\item
  Best approximated by green which is fitted GB2
\item
  Pareto (purple) and Lognormal (lightblue) are also pretty good
\item
  Worst are the exponential (in red) and gamma (in dark blue)
\end{itemize}

\begin{figure}

{\centering \includegraphics[width=0.8\linewidth]{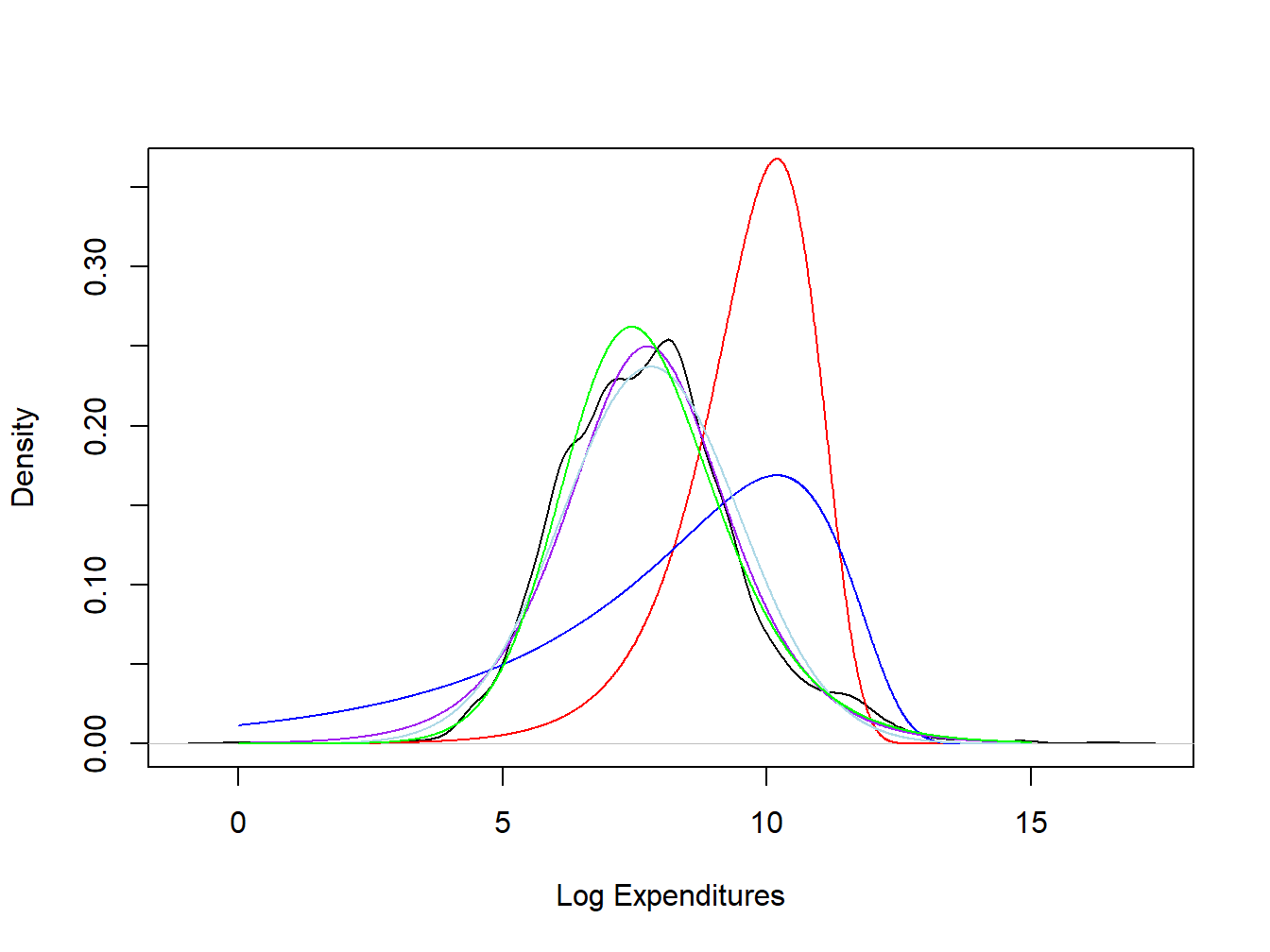}

}

\caption{Fitted Claims Distribution}\label{fig:FitClaimDistn}
\end{figure}

\begin{verbatim}
## Sample size:  6258
\end{verbatim}

Show R Code

\hypertarget{CODE:S1:FIT}{}
R Code for Fitted Claims Distributions

\begin{verbatim}
# R Code to fit several claims distributions
ClaimLev <- read.csv("Data/CLAIMLEVEL.csv", header=TRUE); nrow(ClaimLev)
ClaimData<-subset(ClaimLev,Year==2010);
#Use "VGAM" library for estimation of parameters
library(VGAM)
fit.LN <- vglm(Claim ~ 1, family=lognormal, data = ClaimData)
fit.gamma <- vglm(Claim ~ 1, family=gamma2, data = ClaimData)
  theta.gamma<-exp(coef(fit.gamma)[1])/exp(coef(fit.gamma)[2])
  alpha.gamma<-exp(coef(fit.gamma)[2])
fit.exp <- vglm(Claim ~ 1, exponential, data = ClaimData)
fit.pareto <- vglm(Claim ~ 1, paretoII, loc=0, data = ClaimData)

###################################################
#  Inference assuming a GB2 Distribution - this is more complicated
# The likelihood functon of GB2 distribution (negative for optimization)
likgb2 <- function(param) {
  a1 <- param[1]
  a2 <- param[2]
  mu <- param[3]
  sigma <- param[4]
  yt <- (log(ClaimData$Claim)-mu)/sigma
  logexpyt<-ifelse(yt>23,yt,log(1+exp(yt)))
  logdens <- a1*yt - log(sigma) - log(beta(a1,a2)) - (a1+a2)*logexpyt -log(ClaimData$Claim)
  return(-sum(logdens))
}
#  "optim" is a general purpose minimization function
gb2bop <- optim(c(1,1,0,1),likgb2,method=c("L-BFGS-B"),
                lower=c(0.01,0.01,-500,0.01),upper=c(500,500,500,500),hessian=TRUE)
###################################################
# Plotting the fit using densities (on a logarithmic scale)
plot(density(log(ClaimData$Claim)), ylim=c(0,0.36),main="", xlab="Log Expenditures")
x <- seq(0,15,by=0.01)
fexp_ex = dgamma(exp(x), scale = exp(-coef(fit.exp)), shape = 1)*exp(x)
lines(x,fexp_ex, col="red")
fgamma_ex = dgamma(exp(x), shape = alpha.gamma, scale=theta.gamma)*exp(x)
lines(x,fgamma_ex,col="blue")
fpareto_ex = dparetoII(exp(x),loc=0,shape = exp(coef(fit.pareto)[2]), scale = exp(coef(fit.pareto)[1]))*exp(x)
lines(x,fpareto_ex,col="purple")
flnorm_ex = dlnorm(exp(x), mean = coef(fit.LN)[1], sd = exp(coef(fit.LN)[2]))*exp(x)
lines(x,flnorm_ex, col="lightblue")
# density for GB II
gb2density <- function(x){
  a1 <- gb2bop$par[1]
  a2 <- gb2bop$par[2]
  mu <- gb2bop$par[3]
  sigma <- gb2bop$par[4]
  xt <- (log(x)-mu)/sigma
  logexpxt<-ifelse(xt>23,yt,log(1+exp(xt)))
  logdens <- a1*xt - log(sigma) - log(beta(a1,a2)) - (a1+a2)*logexpxt -log(x)
  exp(logdens)
}
fGB2_ex = gb2density(exp(x))*exp(x)
lines(x,fGB2_ex, col="green")
\end{verbatim}

\chapter{Appendix B: Iterated Expectations}\label{C:AppB}

This appendix introduces the laws related to iterated expectations. In
particular, Section \ref{S:AppB:CD} introduces the concepts of
conditional distribution and conditional expectation. Section
\ref{S:AppB:IE} introduces the Law of Iterated Expectations and the Law
of Total Variance.

In some situations, we only observe a single outcome but can
conceptualize an outcome as resulting from a two (or more) stage
process. Such types of statistical models are called \textbf{two-stage},
or \textbf{hierarchical} models. Some special cases of hierarchical
models include:

\begin{itemize}
\item
  models where the parameters of the distribution are random variables;
\item
  mixture distribution, where Stage 1 represents the draw of a
  sub-population and Stage 2 represents a random variable from a
  distribution that is determined by the sub-population drew in Stage 1;
\item
  an aggregate distribution, where Stage 1 represents the draw of the
  number of events and Stage two represents the loss amount occurred per
  event.
\end{itemize}

In these situations, the process gives rise to a conditional
distribution of a random variable (the Stage 2 outcome) given the other
(the Stage 1 outcome). The Law of Iterated Expectations can be useful
for obtaining the unconditional expectation or variance of a random
variable in such cases.

\section{Conditional Distribution and Conditional
Expectation}\label{S:AppB:CD}

In this section, you learn

\begin{itemize}
\tightlist
\item
  the concepts related to the conditional distribution of a random
  variable given another
\item
  how to define the conditional expectation and variance based on the
  conditional distribution function
\end{itemize}

The iterated expectations are the laws regarding calculation of the
expectation and variance of a random variable using a conditional
distribution of the variable given another variable. Hence, we first
introduce the concepts related to the conditional distribution, and the
calculation of the conditional expectation and variable based on a given
conditional distribution.

\subsection{Conditional Distribution}\label{conditional-distribution}

Here we introduce the concept of conditional distribution respectively
for discrete and continuous random variables.

\subsubsection{Discrete Case}\label{discrete-case}

Suppose that \(X\) and \(Y\) are both discrete random variables, meaning
that they can take a finite or countable number of possible values with
a positive probability. The \textbf{joint probability (mass) function}
of (\(X\), \(Y\)) is defined as \[p(x,y) = \Pr[X=x, Y=y]\].

When \(X\) and \(Y\) are \textbf{independent} (the value of \(X\) does
not depend on that of \(Y\)), we have \[p(x,y)=p(x)p(y),\] with
\(p(x)=\Pr[X=x]\) and \(p(y)=\Pr[Y=y]\) being the \textbf{marginal
probability function} of \(X\) and \(Y\), respectively.

Given the joint probability function, we may obtain the marginal
probability functions of \(Y\) as \[p(y)=\sum_x p(x,y),\] where the
summation is over all possible values of \(x\), and the marginal
probability function of \(X\) can be obtained in a similar manner.

The \textbf{conditional probability (mass) function} of \((Y|X)\) is
defined as \[p(y|x) =\Pr[Y=y|X=x]= \frac{p(x,y)}{\Pr[X=x]},\] where we
may obtain the conditional probability function of \((X|Y)\) in a
similar manner. In particular, the above conditional probability
represents the probability of the event \(Y=y\) given the event \(X=x\).
Hence, even in cases where \(\Pr[X=x]=0\), the function may be given as
a particular form, in real applications.

\subsubsection{Continuous Case}\label{continuous-case}

For continuous random variables \(X\) and \(Y\), we may define their
joint probability (density) function based on the joint cumulative
distribution function. The \textbf{joint cumulative distribution
function} of (\(X\), \(Y\)) is defined as
\[F(x,y) = \Pr[X\leq x, Y\leq y].\]

When \(X\) and \(Y\) are \emph{independent}, we have
\[F(x,y)=F(x)F(y),\] with \(F(x)=\Pr[X\leq x]\) and
\(F(y)=\Pr[Y\leq y]\) being the \textbf{cumulative distribution
function} (cdf) of \(X\) and \(Y\), respectively. The random variable
\(X\) is referred to as a \textbf{continuous} random variable if its cdf
is continuous on \(x\).

When the cdf \(F(x)\) is continuous on \(x\), then we define
\(f(x)=\partial F(x)/\partial x\) as the \textbf{(marginal) probability
density function} (pdf) of \(X\). Similarly, if the joint cdf \(F(x,y)\)
is continuous on both \(x\) and \(y\), we define
\[f(x,y)=\frac{\partial^2 F(x,y)}{\partial x\partial y}\] as the
\textbf{joint probability density function} of (\(X\), \(Y\)), in which
case we refer to the random variables as \textbf{jointly continuous}.

When \(X\) and \(Y\) are \emph{independent}, we have
\[f(x,y)=f(x)f(y).\]

Given the joint density function, we may obtain the marginal density
function of \(Y\) as \[f(y)=\int_x f(x,y)\,dx,\] where the integral is
over all possible values of \(x\), and the marginal probability function
of \(X\) can be obtained in a similar manner.

Based on the joint pdf and the marginal pdf, we define the
\textbf{conditional probability density function} of \((Y|X)\) as

\[f(y|x) = \frac{f(x,y)}{f(x)},\] where we may obtain the conditional
probability function of \((X|Y)\) in a similar manner. Here, the
conditional density function is the density function of \(y\) given
\(X=x\). Hence, even in cases where \(\Pr[X=x]=0\) or when \(f(x)\) is
not defined, the function may be given in a particular form in real
applications.

\subsection{Conditional Expectation and Conditional
Variance}\label{conditional-expectation-and-conditional-variance}

Now we define the conditional expectation and variance based on the
conditional distribution defined in the previous subsection.

\subsubsection{Discrete Case}\label{discrete-case-1}

For a discrete random variable \(Y\), its \textbf{expectation} is
defined as \(\mathrm{E}[Y]=\sum_y y\,p(y)\) if its value is finite, and
its \textbf{variance} is defined as
\(\mathrm{Var}[Y]=\mathrm{E}\{(Y-\mathrm{E}[Y])^2\}=\sum_y y^2\,p(y)-\{\mathrm{E}[Y]\}^2\)
if its value is finite.

For a discrete random variable \(Y\), the \textbf{conditional
expectation} of the random variable \(Y\) given the event \(X=x\) is
defined as \[\mathrm{E}[Y|X=x]=\sum_y y\,p(y|x),\] where \(X\) does not
have to be a discrete variable, as far as the conditional probability
function \(p(y|x)\) is given.

Note that the conditional expectation \(\mathrm{E}[Y|X=x]\) is a fixed
number. When we replace \(x\) with \(X\) on the right hand side of the
above equation, we can define the expectation of \(Y\) given the random
variable \(X\) as \[\mathrm{E}[Y|X]=\sum_y y\,p(y|X),\] which is still a
\emph{random variable}, and the randomness comes from \(X\).

In a similar manner, we can define the \textbf{conditional variance} of
the random variable \(Y\) given the event \(X=x\) as
\[\mathrm{Var}[Y|X=x]=\mathrm{E}[Y^2|X=x]-\{\mathrm{E}[Y|X=x]\}^2=\sum_y y^2\,p(y|x)-\{\mathrm{E}[Y|X=x]\}^2.\]

The variance of \(Y\) given \(X\), \(\mathrm{Var}[Y|X]\) can be defined
by replacing \(x\) by \(X\) in the above equation, and
\(\mathrm{Var}[Y|X]\) is still a random variable and the randomness
comes from \(X\).

\subsubsection{Continuous Case}\label{continuous-case-1}

For a continuous random variable \(Y\), its \textbf{expectation} is
defined as \(\mathrm{E}[Y]=\int_y y\,f(y)dy\) if the integral exists,
and its \textbf{variance} is defined as
\(\mathrm{Var}[Y]=\mathrm{E}\{(X-\mathrm{E}[Y])^2\}=\int_y y^2\,f(y)dy-\{\mathrm{E}[Y]\}^2\)
if its value is finite.

For jointly continuous random variables \(X\) and \(Y\), the
\textbf{conditional expectation} of the random variable \(Y\) given
\(X=x\) is defined as \[\mathrm{E}[Y|X=x]=\int_y y\,f(y|x)dy.\] where
\(X\) does not have to be a continuous variable, as far as the
conditional probability function \(f(y|x)\) is given.

Similarly, the conditional expectation \(\mathrm{E}[Y|X=x]\) is a fixed
number. When we replace \(x\) with \(X\) on the right-hand side of the
above equation, we can define the expectation of \(Y\) given the random
variable \(X\) as \[\mathrm{E}[Y|X]=\int_y y\,p(y|X)\,dy,\] which is
still a \emph{random variable}, and the randomness comes from \(X\).

In a similar manner, we can define the \textbf{conditional variance} of
the random variable \(Y\) given the event \(X=x\) as
\[\mathrm{Var}[Y|X=x]=\mathrm{E}[Y^2|X=x]-\{\mathrm{E}[Y|X=x]\}^2=\int_y y^2\,f(y|x)\,dy-\{\mathrm{E}[Y|X=x]\}^2.\]

The variance of \(Y\) given \(X\), \(\mathrm{Var}[Y|X]\) can then be
defined by replacing \(x\) by \(X\) in the above equation, and similarly
\(\mathrm{Var}[Y|X]\) is also a random variable and the randomness comes
from \(X\).

\section{Iterated Expectations and Total Variance}\label{S:AppB:IE}

In this section, you learn

\begin{itemize}
\tightlist
\item
  the Law of Iterated Expectations for calculating the expectation of a
  random variable based on its conditional distribution given another
  random variable
\item
  the Law of Total Variance for calculating the variance of a random
  variable based on its conditional distribution given another random
  variable
\item
  how to calculate the expectation and variance based on an example of a
  two-stage model
\end{itemize}

\subsection{Law of Iterated
Expectations}\label{law-of-iterated-expectations}

Consider two random variables \(X\) and \(Y\), and \(h(X,Y)\), a random
variable depending on the function \(h\), \(X\) and \(Y\).

Assuming all the expectations exist and are finite, the \textbf{Law of
Iterated Expectations} states that
\[\mathrm{E}[h(X,Y)]= \mathrm{E} \left\{ \mathrm{E} \left[ h(X,Y) | X \right] \right \},\]
where the first (inside) expectation is taken with respect to the random
variable \(Y\) and the second (outside) expectation is taken with
respect to \(X\).

For the Law of Iterated Expectations, the random variables may be
discrete, continuous, or a hybrid combination of the two. We use the
example of discrete variables of \(X\) and \(Y\) to illustrate the
calculation of the unconditional expectation using the Law of Iterated
Expectations. For continuous random variables, we only need to replace
the summation with the integral, as illustrated earlier in the appendix.

Given \(p(y|x)\) the joint pmf of \(X\) and \(Y\), the conditional
expectation of \(h(X,Y)\) given the event \(X=x\) is defined as
\[\mathrm{E} \left[ h(X,Y) | X=x \right] = \sum_y h(x,y) p(y|x),\] and
the conditional expectation of \(h(X,Y)\) given \(X\) being a
\emph{random variable} can be written as
\[\mathrm{E} \left[ h(X,Y) | X \right] = \sum_y h(X,y) p(y|X).\]

The unconditional expectation of \(h(X,Y)\) can then be obtained by
taking the expectation of \(\mathrm{E} \left[ h(X,Y) | X \right]\) with
respect to the random variable \(X\). That is, we can obtain
\(\mathrm{E}[ h(X,Y)]\) as \[\begin{aligned}
     \mathrm{E} \left\{ \mathrm{E} \left[ h(X,Y) | X \right] \right \}
    &= \sum_x  \left\{\sum_y h(x,y) p(y|x) \right \} p(x) \\
    &= \sum_x  \sum_y h(x,y) p(y|x)p(x) \\
    &=  \sum_x  \sum_y h(x,y) p(x,y)
    =  \mathrm{E}[h(X,Y)] \end{aligned}.\]

The Law of Iterated Expectations for the continuous and hybrid cases can
be proved in a similar manner, by replacing the corresponding
summation(s) by integral(s).

\subsection{Law of Total Variance}\label{law-of-total-variance}

Assuming that all the variances exist and are finite, the \textbf{Law of
Total Variance} states that
\[\mathrm{Var}[h(X,Y)]= \mathrm{E} \left\{ \mathrm{Var} \left[h(X,Y) | X \right] \right \}
    +\mathrm{Var} \left\{ \mathrm{E} \left[ h(X,Y) | X \right] \right \},\]
where the first (inside) expectation/variance is taken with respect to
the random variable \(Y\) and the second (outside) expectation/variance
is taken with respect to \(X\). Thus, the unconditional variance equals
to the expectation of the conditional variance plus the variance of the
conditional expectation.

\begin{center}\rule{0.5\linewidth}{\linethickness}\end{center}

Show Technical Detail

\hypertarget{EXM:S2a:LTV}{}
In order to verify this rule, first note that we can calculate a
conditional variance as
\[\mathrm{Var} \left[ h(X,Y) | X \right]  = \mathrm{E} [ h(X,Y)^2 | X ] -\left\{\mathrm{E} \left[ h(X,Y) | X \right] \right\}^2.\]

From this, the expectation of the conditional variance is

\begin{align}
    \mathrm{E}\{\mathrm{Var} \left[ h(X,Y) | X \right] \} &=
    \mathrm{E}\left\{\mathrm{E} \left[ h(X,Y)^2 | X \right] \right\} - \mathrm{E}\left(\left\{\mathrm{E} \left[ h(X,Y) | X \right] \right\}^2\right) \notag \\
    &=\mathrm{E} \left[ h(X,Y)^2\right] - \mathrm{E}\left(\left\{\mathrm{E} \left[ h(X,Y) | X \right] \right\}^2\right).\label{eq:AppBEV1}
\end{align}

Further, note that the conditional expectation,
\(\mathrm{E} \left[ h(X,Y) | X \right]\), is a function of \(X\),
denoted \(g(X)\). Thus, \(g(X)\) is a random variable with mean
\(\mathrm{E}[h(X,Y)]\) and variance

\begin{align}
    \mathrm{Var} \left\{ \mathrm{E} \left[ h(X,Y) | X \right] \right \} &=\mathrm{Var}[g(X)]  \notag \\
    &= \mathrm{E}[g(X)^2]\ - \left\{\mathrm{E}[g(X)]\right\}^2 \nonumber\\
    &= \mathrm{E}\left(\left\{\mathrm{E} \left[ h(X,Y) | X \right] \right\}^2\right)
    - \left\{\mathrm{E}[h(X,Y)]\right\}^2.\label{eq:AppBVE2}
\end{align}

Thus, adding Equations \eqref{eq:AppBEV1} and \eqref{eq:AppBVE2} leads to
the unconditional variance \(\mathrm{Var} \left[ h(X,Y) \right]\).

\begin{center}\rule{0.5\linewidth}{\linethickness}\end{center}

\subsection{Application}\label{application}

To apply the Law of Iterated Expectations and the Law of Total Variance,
we generally adopt the following procedure.

\begin{enumerate}
\def\labelenumi{\arabic{enumi}.}
\item
  Identify the random variable that is being conditioned upon, typically
  a stage 1 outcome (that is not observed).
\item
  Conditional on the stage 1 outcome, calculate summary measures such as
  a mean, variance, and the like.
\item
  There are several results of the step 2, one for each stage 1 outcome.
  Then, combine these results using the iterated expectations or total
  variance rules.
\end{enumerate}

\textbf{Mixtures of Finite Populations.} Suppose that the random
variable \(N_1\) represents a realization of the number of claims in a
policy year from the population of good drivers and \(N_2\) represents
that from the population of bad drivers. For a specific driver, there is
a probability \(\alpha\) that (s)he is a good driver. For a specific
draw \(N\), we have \[N =
    \begin{cases}
    N_1,  &  \text{if (s)he is a good driver;}\\
    N_2,  &   \text{otherwise}.\\
    \end{cases}\]

Let \(T\) be the indicator whether (s)he is a good driver, with \(T=1\)
representing that the driver is a good driver with \(\Pr[T=1]=\alpha\)
and \(T=2\) representing that the driver is a bad driver with
\(\Pr[T=2]=1-\alpha\).

From the Law of Iterated Expectations, we can obtain the expected number
of claims as \[
    \mathrm{E}[N]= \mathrm{E} \left\{ \mathrm{E} \left[ N | T \right] \right \}= \mathrm{E}[N_1] \times \alpha +  \mathrm{E}[N_2] \times (1-\alpha).\]

From the Law of Total Variance, we can obtain the variance of \(N\) as
\[\mathrm{Var}[N]= \mathrm{E} \left\{ \mathrm{Var} \left[ N | T \right] \right \}
    +\mathrm{Var} \left\{ \mathrm{E} \left[ N | T \right] \right \}.\]

To be more concrete, suppose that \(N_j\) follows a Poisson distribution
with the mean \(\lambda_j\), \(j=1,2\). Then we have
\[\mathrm{Var}[N|T=j]= \mathrm{E}[N|T=j] = \lambda_j, \quad j = 1,2.\]

Thus, we can derive the expectation of the conditional variance as
\[\mathrm{E} \left\{ \mathrm{Var} \left[ N | T \right] \right \} = \alpha \lambda_1+ (1-\alpha) \lambda_2\]
and the variance of the conditional expectation as
\[\mathrm{Var} \left\{ \mathrm{E} \left[ N | T \right] \right \} = (\lambda_1-\lambda_2)^2 \alpha (1-\alpha).\]
Note that the later is the variance for a Bernoulli with outcomes
\(\lambda_1\) and \(\lambda_2\), and the binomial probability
\(\alpha\).

Based on the Law of Total Variance, the unconditional variance of \(N\)
is given by
\[\mathrm{Var}[N]= \alpha \lambda_1+ (1-\alpha) \lambda_2 + (\lambda_1-\lambda_2)^2 \alpha (1-\alpha).\]

\chapter{Appendix C: Maximum Likelihood Theory}\label{C:AppC}

\emph{Chapter preview}. Appendix Chapter \ref{C:AppA} introduced the
maximum likelihood theory regarding estimation of parameters from a
parametric family. This appendix gives more specific examples and
expands some of the concepts. Section \ref{S:AppC:LF} reviews the
definition of the likelihood function, and introduces its properties.
Section \ref{S:AppC:MLE} reviews the maximum likelihood estimators, and
extends their large-sample properties to the case where there are
multiple parameters in the model. Section \ref{S:AppC:SI} reviews
statistical inference based on maximum likelihood estimators, with
specific examples on cases with multiple parameters.

\section{Likelihood Function}\label{S:AppC:LF}

\begin{center}\rule{0.5\linewidth}{\linethickness}\end{center}

In this section, you learn

\begin{itemize}
\tightlist
\item
  the definitions of the likelihood function and the log-likelihood
  function
\item
  the properties of the likelihood function.
\end{itemize}

\begin{center}\rule{0.5\linewidth}{\linethickness}\end{center}

From Appendix \ref{C:AppA}, the likelihood function is a function of
parameters given the observed data. Here, we review the concepts of the
likelihood function, and introduces its properties that are bases for
maximum likelihood inference.

\subsection{Likelihood and Log-likelihood
Functions}\label{likelihood-and-log-likelihood-functions}

Here, we give a brief review of the likelihood function and the
log-likelihood function from Appendix \ref{C:AppA}. Let
\(f(\cdot|\boldsymbol\theta)\) be the probability function of \(X\), the
probability mass function (pmf) if \(X\) is discrete or the probability
density function (pdf) if it is continuous. The likelihood is a function
of the parameters (\(\boldsymbol \theta\)) given the data
(\(\mathbf{x}\)). Hence, it is a function of the parameters with the
data being fixed, rather than a function of the data with the parameters
being fixed. The vector of data \(\mathbf{x}\) is usually a realization
of a \emph{random sample} as defined in Appendix \ref{C:AppA}.

Given a realized of a random sample \(\mathbf{x}=(x_1,x_2,\cdots,x_n)\)
of size \(n\), the \textbf{likelihood function} is defined as
\[L(\boldsymbol{\theta}|\mathbf{x})=f(\mathbf{x}|\boldsymbol{\theta})=\prod_{i=1}^nf(x_i|\boldsymbol{\theta}),\]
with the corresponding \textbf{log-likelihood function} given by
\[l(\boldsymbol{\theta}|\mathbf{x})=\ln L(\boldsymbol{\theta}|\mathbf{x})=\sum_{i=1}^n\ln f(x_i|\boldsymbol{\theta}),\]
where \(f(\mathbf{x}|\boldsymbol{\theta})\) denotes the joint
probability function of \(\mathbf{x}\). The log-likelihood function
leads to an additive structure that is easy to work with.

In Appendix \ref{C:AppA}, we have used the normal distribution to
illustrate concepts of the likelihood function and the log-likelihood
function. Here, we derive the likelihood and corresponding
log-likelihood functions when the population distribution is from the
Pareto distribution family.

Show Example

\hypertarget{EXM:S2b:LLK}{}
\textbf{Example -- Pareto Distribution.} Suppose that
\(X_1, \ldots, X_n\) represents a random sample from a single-parameter
Pareto distribution with the \textbf{cumulative distribution function}
given by
\[F(x) = \Pr(X_i\leq x)=1- \left(\frac{500}{x}\right)^{\alpha}, ~~~~ x>500,\]
where the parameter \(\theta = \alpha\).

The corresponding probability density function is
\(f(x) = 500^{\alpha} \alpha x^{-\alpha-1}\) and the log-likelihood
function can be derived as
\[l(\boldsymbol \alpha|\mathbf{x}) = \sum_{i=1}^n \ln f(x_i;\alpha) = n \alpha \ln 500 +n \ln \alpha -(\alpha+1)  \sum_{i=1}^n \ln x_i .\]

\subsection{Properties of Likelihood
Functions}\label{properties-of-likelihood-functions}

In mathematical statistics, the first derivative of the log-likelihood
function with respect to the parameters,
\(u(\boldsymbol\theta)=\partial l(\boldsymbol \theta|\mathbf{x})/\partial \boldsymbol \theta\),
is referred to as the \textbf{score function}, or the \textbf{score
vector} when there are multiple parameters in \(\boldsymbol\theta\). The
score function or score vector can be written as
\[u(\boldsymbol\theta)=\frac{ \partial}{\partial \boldsymbol \theta} l(\boldsymbol \theta|\mathbf{x})
    =\frac{ \partial}{\partial \boldsymbol \theta} \ln \prod_{i=1}^n
    f(x_i;\boldsymbol \theta ) =\sum_{i=1}^n \frac{
    \partial}{\partial \boldsymbol \theta}
    \ln f(x_i;\boldsymbol \theta ),\] where
\(u(\boldsymbol\theta)=(u_1(\boldsymbol\theta),u_2(\boldsymbol\theta),\cdots,u_p(\boldsymbol\theta))\)
when \(\boldsymbol\theta=(\theta_1,\cdots,\theta_p)\) contains \(p>2\)
parameters, with the element
\(u_k(\boldsymbol\theta)=\partial l(\boldsymbol \theta|\mathbf{x})/\partial \theta_k\)
being the partial derivative with respect to \(\theta_k\)
(\(k=1,2,\cdots,p\)).

The likelihood function has the following properties:

\begin{itemize}
\tightlist
\item
  One basic property of the likelihood function is that the expectation
  of the score function with respect to \(\mathbf{x}\) is 0. That is,
  \[\mathrm{E}[u(\boldsymbol\theta)]=\mathrm{E} \left[ \frac{ \partial}{\partial \boldsymbol \theta}
  l(\boldsymbol \theta|\mathbf{x}) \right] = \mathbf 0\]
\end{itemize}

To illustrate this, we have

\[\begin{aligned}
    \mathrm{E} \left[ \frac{ \partial}{\partial \boldsymbol \theta} l(\boldsymbol \theta|\mathbf{x}) \right]
    &= \mathrm{E} \left[ \frac{\frac{\partial}{\partial \boldsymbol \theta}f(\mathbf{x};\boldsymbol \theta)}{f(\mathbf{x};\boldsymbol \theta )}  \right]
    = \int\frac{\partial}{\partial \boldsymbol \theta} f(\mathbf{y};\boldsymbol \theta ) d \mathbf y \\
    &= \frac{\partial}{\partial \boldsymbol \theta} \int f(\mathbf{y};\boldsymbol \theta ) d \mathbf y
    = \frac{\partial}{\partial \boldsymbol \theta} 1 = \mathbf 0.\end{aligned}\]

\begin{itemize}
\item
  Denote by
  \({ \partial^2 l(\boldsymbol \theta|\mathbf{x}) }/{\partial \boldsymbol \theta\partial \boldsymbol \theta^{\prime}}={ \partial^2 l(\boldsymbol \theta|\mathbf{x}) }/{\partial \boldsymbol \theta^{2}}\)
  the second derivative of the log-likelihood function when
  \(\boldsymbol\theta\) is a single parameter, or by
  \({ \partial^2 l(\boldsymbol \theta|\mathbf{x}) }/{\partial \boldsymbol \theta\partial \boldsymbol \theta^{\prime}}=(h_{jk})=({ \partial^2 l(\boldsymbol \theta|\mathbf{x}) }/\partial x_j\partial x_k)\)
  the hessian matrix of the log-likelihood function when it contains
  multiple parameters. Denote
  \([{ \partial l(\boldsymbol \theta|\mathbf{x})}{\partial\boldsymbol \theta}][{ \partial l(\boldsymbol \theta|\mathbf{x})}{\partial\boldsymbol \theta'}]=u^2(\boldsymbol \theta)\)
  when \(\boldsymbol\theta\) is a single parameter, or let
  \([{ \partial l(\boldsymbol \theta|\mathbf{x})}{\partial\boldsymbol \theta}][{ \partial l(\boldsymbol \theta|\mathbf{x})}{\partial\boldsymbol \theta'}]=(uu_{jk})\)
  be a \(p\times p\) matrix when \(\boldsymbol\theta\) contains a total
  of \(p\) parameters, with each element
  \(uu_{jk}=u_j(\boldsymbol \theta)u_k(\boldsymbol \theta)\) and
  \(u_j(\boldsymbol \theta)\) being the \(k\)th element of the score
  vector as defined earlier. Another basic property of the likelihood
  function is that sum of the expectation of the hessian matrix and the
  expectation of the kronecker product of the score vector and its
  transpose is \(\mathbf 0\). That is,
  \[\mathrm{E} \left( \frac{ \partial^2 }{\partial \boldsymbol \theta\partial \boldsymbol \theta^{\prime}} l(\boldsymbol \theta|\mathbf{x}) \right) + \mathrm{E} \left( \frac{ \partial l(\boldsymbol \theta|\mathbf{x})}{\partial\boldsymbol \theta} \frac{ \partial l(\boldsymbol \theta|\mathbf{x})}{\partial\boldsymbol \theta^{\prime}}\right) = \mathbf 0.\]
\item
  Define the \textbf{Fisher information matrix} as \[
  \mathcal{I}(\boldsymbol \theta) = \mathrm{E} \left( \frac{ \partial
  l(\boldsymbol \theta|\mathbf{x})}{\partial \boldsymbol \theta} \frac{ \partial
  l(\boldsymbol \theta|\mathbf{x})}{\partial \boldsymbol \theta^{\prime}}
   \right) = -\mathrm{E} \left( \frac{ \partial^2}{\partial \boldsymbol \theta
  \partial \boldsymbol \theta^{\prime}} l(\boldsymbol \theta|\mathbf{x}) \right).\]
\end{itemize}

As the sample size \(n\) goes to infinity, the score function (vector)
converges in distribution to a \textbf{normal distribution} (or
\textbf{multivariate normal distribution} when \(\boldsymbol \theta\)
contains multiple parameters) with mean \textbf{0} and variance (or
covariance matrix in the multivariate case) given by
\(\mathcal{I}(\boldsymbol \theta)\).

\section{Maximum Likelihood Estimators}\label{S:AppC:MLE}

\begin{center}\rule{0.5\linewidth}{\linethickness}\end{center}

In this section, you learn

\begin{itemize}
\tightlist
\item
  the definition and derivation of the maximum likelihood estimator
  (MLE) for parameters from a specific distribution family
\item
  the properties of maximum likelihood estimators that ensure valid
  large-sample inference of the parameters
\item
  why using the MLE-based method, and what caution that needs to be
  taken.
\end{itemize}

\begin{center}\rule{0.5\linewidth}{\linethickness}\end{center}

In statistics, maximum likelihood estimators are values of the
parameters \(\boldsymbol \theta\) that are most likely to have been
produced by the data.

\subsection{Definition and Derivation of
MLE}\label{definition-and-derivation-of-mle}

Based on the definition given in Appendix \ref{C:AppA}, the value of
\(\boldsymbol \theta\), say \(\hat{\boldsymbol \theta}_{MLE}\), that
maximizes the likelihood function, is called the \emph{maximum
likelihood estimator} (MLE) of \(\boldsymbol \theta\).

Because the log function \(\ln(\cdot)\) is a one-to-one function, we can
also determine \(\hat{\boldsymbol{\theta}}_{MLE}\) by maximizing the
log-likelihood function, \(l(\boldsymbol \theta|\mathbf{x})\). That is,
the MLE is defined as
\[\hat{\boldsymbol \theta}_{MLE}={\mbox{argmax}}_{\boldsymbol{\theta}\in\Theta}l(\boldsymbol{\theta}|\mathbf{x}).\]

Given the analytical form of the likelihood function, the MLE can be
obtained by taking the first derivative of the log-likelihood function
with respect to \(\boldsymbol{\theta}\), and setting the values of the
partial derivatives to zero. That is, the MLE are the solutions of the
equations of
\[\frac{\partial l(\hat{\boldsymbol{\theta}}|\mathbf{x})}{\partial\hat{\boldsymbol{\theta}}}=\mathbf 0.\]

\begin{center}\rule{0.5\linewidth}{\linethickness}\end{center}

Show Example

\hypertarget{EXM:S2b:MLE}{}
\textbf{Example. Course C/Exam 4. May 2000, 21.} You are given the
following five observations: 521, 658, 702, 819, 1217. You use the
single-parameter Pareto with cumulative distribution function:
\[F(x) = 1- \left(\frac{500}{x}\right)^{\alpha}, ~~~~ x>500 .\]
Calculate the maximum likelihood estimate of the parameter \(\alpha\).

Show Solution

\hypertarget{SOL:S2b:MLE}{}
\emph{Solution}. With \(n=5\), the log-likelihood function is
\[l(\alpha|\mathbf{x} ) =  \sum_{i=1}^5 \ln f(x_i;\alpha ) =  5 \alpha \ln 500 + 5 \ln \alpha
-(\alpha+1) \sum_{i=1}^5 \ln x_i.\] Solving for the root of the score
function yields
\[\frac{ \partial}{\partial \alpha } l(\alpha |\mathbf{x}) =    5  \ln 500 + 5 / \alpha -  \sum_{i=1}^5 \ln x_i
=_{set} 0 \Rightarrow \hat{\alpha}_{MLE} = \frac{5}{\sum_{i=1}^5 \ln x_i - 5  \ln 500 } = 2.453 .\]

\begin{center}\rule{0.5\linewidth}{\linethickness}\end{center}

\subsection{Asymptotic Properties of
MLE}\label{asymptotic-properties-of-mle}

From Appendix \ref{C:AppA}, the MLE has some nice large-sample
properties, under certain regularity conditions. We presented the
results for a single parameter in Appendix \ref{C:AppA}, but results are
true for the case when \(\boldsymbol{\theta}\) contains multiple
parameters. In particular, we have the following results, in a general
case when \(\boldsymbol{\theta}=(\theta_1,\theta_2,\cdots,\theta_p)\).

\begin{itemize}
\item
  The MLE of a parameter \(\boldsymbol{\theta}\),
  \(\hat{\boldsymbol{\theta}}_{MLE}\), is a \textbf{consistent}
  estimator. That is, the MLE \(\hat{\boldsymbol{\theta}}_{MLE}\)
  converges in probability to the true value \(\boldsymbol{\theta}\), as
  the sample size \(n\) goes to infinity.
\item
  The MLE has the \textbf{asymptotic normality} property, meaning that
  the estimator will converge in distribution to a multivariate normal
  distribution centered around the true value, when the sample size goes
  to infinity. Namely,
  \[\sqrt{n}(\hat{\boldsymbol{\theta}}_{MLE}-\boldsymbol{\theta})\rightarrow N\left(\mathbf 0,\,\boldsymbol{V}\right),\quad \mbox{as}\quad n\rightarrow \infty,\]
  where \(\boldsymbol{V}\) denotes the asymptotic variance (or
  covariance matrix) of the estimator. Hence, the MLE
  \(\hat{\boldsymbol{\theta}}_{MLE}\) has an approximate normal
  distribution with mean \(\boldsymbol{\theta}\) and variance
  (covariance matrix when \(p>1\)) \(\boldsymbol{V}/n\), when the sample
  size is large.
\item
  The MLE is \textbf{efficient}, meaning that it has the smallest
  asymptotic variance \(\boldsymbol{V}\), commonly referred to as the
  \textbf{Cramer--Rao lower bound}. In particular, the Cramer--Rao lower
  bound is the inverse of the Fisher information (matrix)
  \(\mathcal{I}(\boldsymbol{\theta})\) defined earlier in this appendix.
  Hence, \(\mathrm{Var}(\hat{\boldsymbol{\theta}}_{MLE})\) can be
  estimated based on the observed Fisher information.
\end{itemize}

Based on the above results, we may perform statistical inference based
on the procedures defined in Appendix \ref{C:AppA}.

\begin{center}\rule{0.5\linewidth}{\linethickness}\end{center}

Show Example

\hypertarget{EXM:S2b:COV}{}
\textbf{Example. Course C/Exam 4. Nov 2000, 13.} A sample of ten
observations comes from a parametric family
\(f(x,; \theta_1, \theta_2)\) with log-likelihood function
\[l(\theta_1, \theta_2)= \sum_{i=1}^{10} f(x_i; \theta_1, \theta_2) = -2.5 \theta_1^2 - 3
    \theta_1 \theta_2 - \theta_2^2 + 5 \theta_1 + 2 \theta_2 + k,\]
where \(k\) is a constant. Determine the estimated covariance matrix of
the maximum likelihood estimator, \(\hat{\theta_1}, \hat{\theta_2}\).

Show Solution

\hypertarget{SOL:S2b:COV}{}
\emph{Solution}. Denoting \(l=l(\theta_1, \theta_2)\), the hessian
matrix of second derivatives is \[\left(
\begin{array}{cc}
  \frac{ \partial ^2}{\partial \theta_1 ^2 } l & \frac{ \partial ^2}{\partial \theta_1 \partial \theta_2 } l  \\
  \frac{ \partial ^2}{\partial \theta_1 \partial \theta_2 } l & \frac{ \partial ^2}{\partial \theta_1 ^2 } l
\end{array} \right) =
\left(
\begin{array}{cc}
  -5 & -3  \\
  -3 & -2
\end{array} \right)\] Thus, the information matrix is:
\[\mathcal{I}(\theta_1, \theta_2) = -\mathrm{E} \left( \frac{ \partial^2}{\partial \boldsymbol \theta
\partial \boldsymbol \theta^{\prime}} l(\boldsymbol \theta|\mathbf{x}) \right) = \left(
\begin{array}{cc}
  5 & 3  \\
  3 & 2
\end{array} \right)\] and
\[\mathcal{I}^{-1}(\theta_1, \theta_2) = \frac{1}{5(2) - 3(3)}\left(
\begin{array}{cc}
  2 & -3  \\
  -3 & 5
\end{array} \right) = \left(
\begin{array}{cc}
  2 & -3  \\
  -3 & 5
\end{array} \right) .\]

\begin{center}\rule{0.5\linewidth}{\linethickness}\end{center}

\subsection{Use of Maximum Likelihood
Estimation}\label{use-of-maximum-likelihood-estimation}

The method of maximum likelihood has many advantages over alternative
methods such as the method of moment method introduced in Appendix
\ref{C:AppA}.

\begin{itemize}
\tightlist
\item
  It is a general tool that works in many situations. For example, we
  may be able to write out the closed-form likelihood function for
  censored and truncated data. Maximum likelihood estimation can be used
  for regression models including covariates, such as survival
  regression, generalized linear models and mixed models, that may
  include covariates that are time-dependent.
\item
  From the efficiency of the MLE, it is optimal, the best, in the sense
  that it has the smallest variance among the class of all unbiased
  estimators for large sample sizes.
\item
  From the results on the asymptotic normality of the MLE, we can obtain
  a large-sample distribution for the estimator, allowing users to
  assess the variability in the estimation and perform statistical
  inference on the parameters. The approach is less computationally
  extensive than re-sampling methods that require a large of fittings of
  the model.
\end{itemize}

Despite its numerous advantages, MLE has its drawback in cases such as
generalized linear models when it does not have a closed analytical
form. In such cases, maximum likelihood estimators are computed
iteratively using numerical optimization methods. For example, we may
use the Newton-Raphson iterative algorithm or its variations for
obtaining the MLE. Iterative algorithms require starting values. For
some problems, the choice of a close starting value is critical,
particularly in cases where the likelihood function has local minimums
or maximums. Hence, there may be a convergence issue when the starting
value is far from the maximum. Hence, it is important to start from
different values across the parameter space, and compare the maximized
likelihood or log-likelihood to make sure the algorithms have converged
to a global maximum.

\section{Statistical Inference Based on Maximum Likelhood
Estimation}\label{S:AppC:SI}

\begin{center}\rule{0.5\linewidth}{\linethickness}\end{center}

In this section, you learn how to

\begin{itemize}
\tightlist
\item
  perform hypothesis testing based on MLE for cases where there are
  multiple parameters in \(\boldsymbol\theta\)
\item
  perform likelihood ratio test for cases where there are multiple
  parameters in \(\boldsymbol\theta\)
\end{itemize}

\begin{center}\rule{0.5\linewidth}{\linethickness}\end{center}

In Appendix \ref{C:AppA}, we have introduced maximum likelihood-based
methods for statistical inference when \(\boldsymbol\theta\) contains a
single parameter. Here, we will extend the results to cases where there
are multiple parameters in \(\boldsymbol\theta\).

\subsection{Hypothesis Testing}\label{hypothesis-testing}

In Appendix \ref{C:AppA}, we defined hypothesis testing concerning the
null hypothesis, a statement on the parameter(s) of a distribution or
model. One important type of inference is to assess whether a parameter
estimate is statistically significant, meaning whether the value of the
parameter is zero or not.

We have learned earlier that the MLE \(\hat{\boldsymbol{\theta}}_{MLE}\)
has a large-sample normal distribution with mean \(\boldsymbol \theta\)
and the variance covariance matrix
\(\mathcal{I}^{-1}(\boldsymbol \theta)\). Based on the multivariate
normal distribution, the \(j\)th element of
\(\hat{\boldsymbol{\theta}}_{MLE}\), say \(\hat{\theta}_{MLE,j}\), has a
large-sample univariate normal distribution.

Define \(se(\hat{\theta}_{MLE,j})\), the standard error (estimated
standard deviation) to be the square root of the \(j\)th diagonal
element of \(\mathcal{I}^{-1}(\boldsymbol \theta)_{MLE}\). To assess the
null hypothesis that \(\theta_j=\theta_0\), we define the
\(t\)-statistic or \(t\)-ratio to be
\(t(\hat{\theta}_{MLE,j})=(\hat{\theta}_{MLE,j}-\theta_0)/se(\hat{\theta}_{MLE,j})\).

Under the null hypothesis, it has a Student-\(t\) distribution with
degrees of freedom equal to \(n-p\), with \(p\) being the dimension of
\(\boldsymbol{\theta}\).

For most actuarial applications, we have a large sample size \(n\), so
the \(t\)-distribution is very close to the (standard) normal
distribution. In the case when \(n\) is very large or when the standard
error is known, the \(t\)-statistic can be referred to as a
\(z\)-statistic or \(z\)-score.

Based on the results from Appendix \ref{C:AppA}, if the \(t\)-statistic
\(t(\hat{\theta}_{MLE,j})\) exceeds a cut-off (in absolute value), then
the test for the \(j\) parameter \(\theta_j\) is said to be
statistically significant. If \(\theta_j\) is the regression coefficient
of the \(j\) th independent variable, then we say that the \(j\)th
variable is statistically significant.

For example, if we use a 5\% significance level, then the cut-off value
is 1.96 using a normal distribution approximation for cases with a large
sample size. More generally, using a \(100 \alpha \%\) significance
level, then the cut-off is a \(100(1-\alpha/2)\%\) quantile from a
Student-\(t\) distribution with the degree of freedom being \(n-p\).

Another useful concept in hypothesis testing is the \(p\)-value,
shorthand for probability value. From the mathematical definition in
Appendix \ref{C:AppA}, a \(p\)-value is defined as the smallest
significance level for which the null hypothesis would be rejected.
Hence, the \(p\)-value is a useful summary statistic for the data
analyst to report because it allows the reader to understand the
strength of statistical evidence concerning the deviation from the null
hypothesis.

\subsection{MLE and Model Validation}\label{S:AppC:MLEModelVal}

In addition to hypothesis testing and interval estimation introduced in
Appendix \ref{C:AppA} and the previous subsection, another important
type of inference is selection of a model from two choices, where one
choice is a special case of the other with certain parameters being
restricted. For such two models with one being nested in the other, we
have introduced the likelihood ratio test (LRT) in Appendix
\ref{C:AppA}. Here, we will briefly review the process of performing a
LRT based on a specific example of two alternative models.

Suppose that we have a (large) model under which we derive the maximum
likelihood estimator, \(\hat{\boldsymbol{\theta}}_{MLE}\). Now assume
that some of the \(p\) elements in \(\boldsymbol \theta\) are equal to
zero and determine the maximum likelihood estimator over the remaining
set, with the resulting estimator denoted
\(\hat{\boldsymbol{\theta}}_{Reduced}\).

Based on the definition in Appendix \ref{C:AppA}, the statistic,
\(LRT= 2 \left( l(\hat{\boldsymbol{\theta}}_{MLE}) - l(\hat{\boldsymbol{\theta}}_{Reduced}) \right)\),
is called the likelihood ratio statistic. Under the null hypothesis that
the reduced model is correct, the likelihood ratio has a Chi-square
distribution with degrees of freedom equal to \(d\), the number of
variables set to zero.

Such a test allows us to judge which of the two models is more likely to
be correct, given the observed data. If the statistic \(LRT\) is large
relative to the critical value from the chi-square distribution, then we
reject the reduced model in favor of the larger one. Details regarding
the critical value and alternative methods based on information criteria
are given in Appendix \ref{C:AppA}.

\bibliography{LDAReferenceB}

\end{document}